\documentclass{llncs}

\usepackage[colorinlistoftodos]{todonotes}

\usepackage{graphicx}
\usepackage{siunitx}
\usepackage{enumitem}
\usepackage{subcaption}
\usepackage[colorinlistoftodos]{todonotes}
\usepackage{graphicx}
\usepackage{siunitx}
\usepackage{enumitem}
\usepackage{comment}
\usepackage{markdown}
\usepackage{hyperref}
\usepackage{cleveref}
\usepackage{orcidlink}
\newcommand{\ICBCversion}{}
\begin{document}
\newcommand{\figscale}{0.49}

\title{Analyzing Solana's Blocks and Transactions}

\newcommand{\DvirName}{Dvir David Biton}
\newcommand{\DvirEmail}{dbiton@campus.technion.ac.il}
\newcommand{\DvirORCID}{0009-0009-3440-375X}

\newcommand{\YaronName}{Yaron Hay}
\newcommand{\YaronEmail}{yaron.hay@cs.technion.ac.il}
\newcommand{\YaronORCID}{0009-0006-1263-7318}

\newcommand{\RoyName}{Roy Friedman}
\newcommand{\RoyEmail}{roy@cs.technion.ac.il}
\newcommand{\RoyORCID}{0000-0001-6460-9665}

\newcommand{\AuthorFmt}[1]%
    {\csuse{#1Name}\orcidlink{\csuse{#1ORCID}}}
    
\author{\AuthorFmt{Yaron} \and \AuthorFmt{Dvir} \and \AuthorFmt{Roy}}
\institute{Technion - Israel Institute of Technology}

\maketitle

\begin{abstract}
Solana is one of the most popular blockchains, and is arguably the most widely used blockchain for smart contracts, also known as dApps.
Understanding the types of smart contracts that are being executed by Solana and their interplay is therefore highly beneficial both for designers of modern blockchains and developers of smart contracts.
To that end, in this paper we analyze a million recent Solana blocks.
We report statistics about the size of blocks (number of transactions per block), execution time units for individual transactions and fees, and invoked Solana programs.
Further, based on the declared readset and writeset of each transaction, as mandated by Solana, we analyze the conflicts and corresponding conflict graphs arising within each block.
These latter statistics are important to understand the potential for parallelism in the network, which is one of the main claimed benefits of Solana.
The data \cite{MoreData} and code \cite{Github} are available in open source.
\end{abstract}

\section{Introduction}
\label{sec:intro}

Many modern blockchains support smart contract execution, which enables extending their utility beyond simple asset transfers, and turn them into general purpose decentralized applications execution engines.
A key distinction between blockchains and traditional distributed databases is that blockchain replicas do not mutually trust one another. As a result, each replica. commonly referred to as a miner or validator, must independently execute and verify every transaction. While blockchain performance was historically constrained by inefficient consensus protocols, modern consensus mechanisms are capable of ordering hundreds of thousands of transactions per second~\cite{Shoal++,mysticeti,FireLedger,RedBelly,Tusk,dumbo-ng,Bullshark,raptr,hotstuff}. Consequently, transaction execution and validation have emerged as the primary performance bottlenecks~\cite{diablo}.

Understanding the types of transactions executed on a blockchain, their object access patterns, and the resulting inter-transaction conflicts is key to optimizing performance. 
For example, being able to predict which objects will be accessed allows for pre-warming, which can reduce execution latency.
Parallel execution is another established technique for improving performance~\cite{ParBlockchain,ConcurrencySC,Sui-Parallel,BlockSTM,ColoringSC} \cite{KD04,monad,Blochchain-DB,EmpSdy-Con,sei-2,solana-sealevel}. 
However, to ensure correctness, concurrent execution of transactions within a block must be deterministically serializable~\cite{Blochchain-DB,Schneider1990}; that is, all validators must observe executions equivalent to the same logical sequential order. 
Consequently, the degree of achievable parallelism is inherently constrained by transaction conflicts.
Identifying transaction types or objects responsible for a significant share of these conflicts can therefore inform targeted optimizations, such as redesigning smart contract logic to reduce contention.

In addition, understanding how the distribution of transactions' execution times can help with resources planning at validator nodes.
Understanding the fee distribution can potentially help users optimize their code structure and improve their pricing strategies, as well as expose anomalies. 

In this work, we analyze transactions from the Solana blockchain~\cite{Solana}.
Solana is a highly popular blockchain in terms of transactions volume and market-cap, and arguably the most widely used platform for smart contracts.
Solana avoids sharding, and instead relies on a parallelization engine to expedite local execution~\cite{solana-sealevel}.
In order to facilitate this, Solana also mandates pre-declaration of readsets and writesets, which simplifies the task of detecting potential conflicts between transactions within the same block.
Hence, we have downloaded and analyzed one million recent Solana blocks (at the time of writing), and report our findings below.

\subsubsection*{Summary of Findings}
%
\begin{itemize}
    \item While the execution time of individual transactions seems to be inversely proportional to the total block size, transactions' fees do not correlate to the block size\footnote{
    See explanation in \Cref{sec:exec-time}.}.
    \item Most Solana blocks exhibit a low number of conflicts. This is likely due to its unique mechanism for choosing each block's transactions (as explained in the paper).
    \item While Solana is effective in reducing conflicts, it incurs long conflict chains, This echoes results reported in~\cite{EthSolAnalysis}.
    \item Conflicting transactions tend to form a near clique among each other, and in particular, in large blocks, most of them belong to the same very large (near) clique structure.
    \item The coloring based approach of~\cite{ColoringSC,RHMJFP25} to parallelizing transaction execution can potentially improve block execution time by a factor of 1.2-3 times (up to 10 in extreme cases) compared to the current arbitrary block~order.
    \item The number of transactions that invoke non-standard programs is non-negligible, meaning that many transactions invoke user defined smart contracts.
    \item The are multiple non-standard programs that are invoked by multiple transactions, indicating that some of these user defined smart contracts are fairly popular. Further, this number grows with the conflict rate, indicating that most conflicts are generated by such programs.
    \item On average, each transaction invokes more than one program, but less than three.
    \item Each transaction accesses between 2-10 true data items (not programs) for read only purposes.
\end{itemize}

\section{Related Work}
\label{sec:related}

The work of~\cite{BFH25} investigates conflict graphs arising from transactions bundled within the same Ethereum block and examines how these conflicts affect the potential performance gains achievable through transaction parallelism, and a similar analysis for Sui was given in~\cite{BF25}.
Several of the metrics considered in this paper overlap with those studied in~\cite{BF25,BFH25,KKA26}, though not all.
Additionally,~\cite{VH20} shows that the proportion of conflicting Ethereum transactions increased between 2016 and 2017, substantially limiting the effectiveness of optimistic concurrency control. Further analyses of other aspects of the Ethereum network can be found in, for example,~\cite{EthInfoProp,EthGasSize,RoleReward,ETHLargeBlock,EthPerfAnalysis,XBlock-ETH}.

Further, the work of~\cite{EthSolAnalysis} has analyzed both Ethereum and Solana blocks in terms of their parallelization potential.
Specifically, for each of these two blockchains, they have downloaded 1,000 blocks from each of three historical periods: "old", "mid", and "recent".
They have explored various statistics, and in particular the length of the longest chains of conflicting transactions within a block.
They have discovred that Ethereum blocks frequently achieve high independence, i.e., over 50\% in more than 50\% of blocks, while Solana blocks contain longer conflict chains, comprising $\approx$59\% of the block size compared to $\approx$18\% in Ethereum.
In this work, we have downloaded a significantly larger number of blocks.
Also, we have also measured the execution time units and gas distribution of transactions, as well as the chromatic number of the undirected conflict graphs, which indicates the maximal potential speedup that can be obtained from parallelism~\cite{ColoringSC}.
The difference between the long conflict chains in Solana and the coloring numbers we have found indicates that replacing the reliance on the arbitrary block order with a minimal coloring driven order can substantially reduce the total block execution time in Solana.

Prior work has studied transaction parallelism in blockchains~\cite{ParBlockchain,Blochchain-DB} and in Byzantine fault-tolerant state machine replication systems more generally~\cite{COS,KD04}. Many modern blockchains employ parallel execution, for example~\cite{BlockSTM,monad,sei-2,sui,solana-sealevel}. Most approaches require transactions to declare their read and write sets in advance, while some use optimistic ordered execution instead~\cite{BlockSTM}. In these systems, conflicting transactions must commit in the order in which they appear in the block. In contrast,~\cite{ColoringSC} shows that coloring the conflict graph and ordering transactions by color can yield significant performance gains, proportional to the ratio between the graph’s chromatic number and its longest simple path. We find that this effect also holds for Solana.

Several studies evaluating deterministic concurrency control mechanisms for blockchains rely on synthetic workloads, such as randomized peer-to-peer balance transfers~\cite{ParBlockchain}, randomized multi-input multi-output transfer transactions~\cite{BlockSTM}, and synthetic smart contracts~\cite{ConcurrencySC}. Deterministic databases~\cite{DetOverview,Aria,caracal} are closely related to blockchain systems; however, they typically do not target Byzantine fault tolerance, operate with a small number of replicas, and focus primarily on traditional database benchmarks such as TPC-C~\cite{tpc-c}, TPC-E~\cite{tpc-e}, and YCSB~\cite{YCSB}. The Diablo benchmark~\cite{diablo} can be used to stress-test the raw performance of blockchains, but it does not capture realistic access patterns or transaction conflicts.

Additional works, such as~\cite{ResilientDB,Blochchain-DB}, adopt a database-oriented perspective to design scalable blockchain systems. These studies evaluate the performance of their prototypes using artificial benchmarks and experimental settings. We further note that blockchain performance can be improved by offloading smart contract execution to off-chain components~\cite{Hyperledger,SlimChain}. However, such approaches are orthogonal to our goal of understanding on-chain transaction behavior and leveraging this understanding to improve on-chain performance.

\section{Preliminaries}
\label{sec:prelim}


A block $B$ consists of a set $T$ of $n$ transactions.
Each transaction $tx \in T$ is an atomic execution of a smart contract invocation or a simple transfer, which accesses some of the blockchain's objects; transactions may access such objects for both reading and writing purposes.
In Solana, transactions should pre-declare their readsets and~writesets.

\begin{figure}[t]
    \centering
    \includegraphics[width=0.36\linewidth]{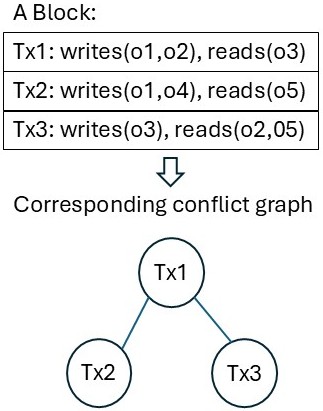}
    \caption{An example of a conflict graph derived from a block's transactions based on the transactions' writesets and readsets.}
    \label{fig:confgraph}
\end{figure}

\paragraph*{Conflicts and Conflict Graphs}
For any transaction $tx \in T$, we define its \emph{read set} as the set of objects accessed for reading during its execution, and its \emph{write set} as the set of objects it write to. Following standard database terminology, two transactions $tx_1 \in T$ and $tx_2 \in T$ are said to \emph{conflict} if they both access a common object and at least one of them performs a write. For a block $B$, we construct the \emph{conflict graph} by adding an undirected edge between every pair of conflicting transactions.
This is illustrated in Figure~\ref{fig:confgraph}.

Conflicts are fundamental in parallel blockchains, which provide replicated state machine semantics with atomic transaction execution~\cite{Schneider1990}. Consequently, parallel execution must preserve deterministic serializability~\cite{DB-textbook,ColoringSC}. One common approach is to disallow concurrent execution of conflicting transactions. Alternatively, optimistic concurrency control can be used, but it may require transaction aborts and re-execution, which are difficult to realize deterministically~\cite{BlockSTM}.

\paragraph*{Graph Properties}
Graphs admit a wide range of structural characterizations. In this work, we analyze key properties of the conflict graphs generated by Solana blocks, with particular focus on those that fundamentally constrain transaction parallelism under serializability requirements~\cite{DB-textbook}.

\begin{description} 
\item[Density:] The ratio of the number of edges to the maximum possible number of edges in an undirected graph. Recall that edges represent conflicts. Hence, low density means few conflicts, while high density indicates many conflicts.
\item[Diameter:] The longest of the shortest paths between each pair of nodes in the graph. 
\item[Max degree:] As its name suggests, the maximum degree among all nodes in the cluster. Together with diameter, can predict other graph properties. For example, a very small diameter with a high max degree is indicative of a star shaped~graph.
\item[Mean degree:] The mean degree of a graph is the average degree across all its nodes. A large gap between the mean and max degree suggests a star-like structure, where few high degree hubs dominate an otherwise sparse graph.
\item[Cluster coefficient:] This measure captures the degree to which nodes in a graph tend to cluster together. For our purposes, suppose transaction $tx_1$ conflicts with $tx_2$, and $tx_2$ conflicts with $tx_3$. A high clustering coefficient would indicate that $tx_1$ is also likely to conflict with $tx_3$.
\item[Assortativity:] The tendency of nodes with similar properties to connect to one another. In this context, degree assortativity captures the extent to which the likelihood of two transactions conflicting is positively correlated with their graph properties, such as node degree and neighborhood.
\item[Chromatic number:] The minimum number of colors required to color the nodes of a graph so that no two adjacent nodes share the same color.
As shown in~\cite{ColoringSC}, the chromatic number of the conflict graph indicates the maximal attainable speedup from parallelism while preserving deterministic serializability.
\item[Clique number:] The size of the largest complete subgraph (clique), in which every pair of nodes is connected by an edge. This quantity provides a lower bound on the chromatic number.
\item[Longest simple path:] The length of the longest path in the graph that does not revisit any node. As shown in~\cite{ColoringSC}, this metric captures the maximum latency required to execute all transactions under na\"ive parallelization strategies. Because computing the longest simple path is NP-hard, we obtain a practical lower bound by performing multiple random traversals and recording the longest path observed~\cite{gnp-dfs}.
\item[Largest connected component:] The size of the largest subset of nodes such that there exists a path between any two nodes in this subset. This serves as an upper bound for the longest simple path.
\end{description}

\section{Solana}
\label{sec:solana}

Solana is a decentralized platform that facilitates P2P transactions while supporting the execution of on-chain programs. 
Solana differs from other modern blockchains by using the account model as well as having an architecture that embraces parallelism, which tends to reduce the number of conflicts between transactions placed within the same block.

\paragraph*{Account Model}
Everything in Solana is addressed as an account, whether it is data or programs.
Each account is idenfified through a unique account address.
An account maintains the following fields:
\begin{itemize}
    \item lamports - the amount of lamports (Solana's smallest currency unit) the account~holds.
    \item data - the account data.
    \item owner - the id of the program that owns the account.
    \item executable - a biary that indicates if this a a program or data item.
\end{itemize}
When the account is a program, its data field either contains the code or the address of the code.
Otherwise, it contains the accounts data, which is a stream of bytes that needs to be deserialized in order to be used.

\paragraph*{Programs}
Smart contracts are called \emph{programs}, which are compiled through LLVM into Solana Bytecode Format (sBPF) files or native.
Programs are stateless and are often written in Rust. 
Transactions include invocations of one or more instructions from such programs, and must declare their readset and writeset in advance.

\paragraph*{Blocks and Slots}
As in many blockchains, transactions in Solana are divided into blocks.
Time is divided into \emph{slots}, with a target duration of 400ms, during which a single validator is designated as the \emph{leader}.
The same validator serves as the leader for consecutive 4 slots.
Issued transactions are forwarded directly to the current or closely scheduled leader rather than being disseminated to all validators (no mempool).
Such transactions are held by the leader until it manages to execute them, unless it is too close to the end of its 4 slots turn as a leader, in which case they are sent to the next scheduled leader.

During each slot, the leader attempts to execute unordered transactions in parallel, while ensuring strict serializability through concurrency control.
Each transaction that terminates its local execution is then inserted into the evolving block that corresponds to the current slot.
At the end of the slot, the block is passed to the consensus protocol to be decided on.
If a leader fails to produce a block within their slot, that slot is skipped, and the network moves to the next leader.
Validators that receive a block from the leader (after the consensus protocol) execute the block's transactions in parallel, but must ensure that the logical serialization order obeys the order in which the transactions appear in the block.

Interestingly, the above mechanism tends to reduce the number of conflicts within a Solana block.
This is because concurrency control often delays the execution of some concurrent conflicting transactions.
This lowers the chances that multiple conflicting transactions would finish executing within the same slot and be inserted into the same block.

\paragraph*{Consensus}
Solana is in the process of migrating from its previous Proof of History (PoH)~\cite{PoH} and Tower BFT consensus protocol~\cite{TowerBFT} to the Alpenglow protocol~\cite{alpenglow}.
For brevity and since the details are not important for the rest of this work, we skip them here.

\section{Methodology}
\label{sec:method}

\subsection{Solana Full Node}

We have rented access time to a Solana node to collect block traces, including their respective transactions, while utilizing the Solana standard RPC interface.
%
Specifically, we have collected transaction traces of 1M blocks, between \#$390$M and \#$391$M, out of a total of 400 million blocks at the time of writing.

\subsection{Solana Transactions Fields} \label{sec:solana-txs}
Using the \texttt{getBlock} RPC method, we can obtain a trace for all transactions in the given block at once.
This method allows specifying how detailed the transaction information should be by using the \texttt{transactionDetails=''full''} field in the the request body. 
In particular, it includes the following fields:


\begin{description}
   \item[\texttt{computeUnitsConsumed}:] Compute unit used at runtime of the transaction.
   \item[\texttt{costUnits}:] Total compute units used by the transaction including execution and all other overheads: signature verifications, accound loadings, write locks, etc.
   \item[\texttt{err}:] Indicates if the transaction was recorded as failed (hard failure).
   \item[\texttt{fee}:] Fee charged for the transaction.
   \item[\texttt{accountKeys}:] List of account the transaction will access. Each account includes a \texttt{writeable} flag, indicating whether the transaction can write to the account data. 
   \item[\texttt{instructions}:] List of programs to be invoked: each instruction contains a pointer to the program that is to be invoked, a parameter, and a subset of accounts that the program may access.
\end{description}


\subsection{Read Sets, Write Sets and Conflicts}
\label{sec:method:readwritesetsandconflicts}

To obtain the write set of a Solana transaction, we analyze the list of account keys in the transaction according to the \texttt{writable} flag.
The read set is extracted as the collection of all account keys with \texttt{writable='false'}, and the write consists of all with \texttt{writable='true'}.
Two transactions are considered conflicting if they both specify at least one common account key and at least one transaction sets \texttt{writable='true'} for it.
Note that Solana considers the \texttt{writable='true'} flag as a requirement for an exclusive lock on the corresponding data account; thus, transactions may read and write when specifying \texttt{writable='true'}.
We are unable to distinguish between true read-write conflicts and true write-write conflicts.
Let us note that the declared objects are indeed accessed by the transactions that declare them, as these declarations are derived from the execution performed by the block producer.

\subsection{ChainGrapher}
\label{sec:chaingrapher}

We extended the ChainGrapher infrastructure~\cite{Github} to support parsing of Solana traces.
The tool supports the collection, retrieval, and processing of execution traces.
It is implemented in Python~3.12, and uses \texttt{networkx}~3.4.2 for conflict graph processing and \texttt{httpx} to interact with the Solana node.

Execution traces are compressed and stored in HDF5 format using \texttt{h5py}. Additional Python libraries, including \texttt{pandas}, \texttt{numpy}, and \texttt{matplotlib}, are used for metric computation and visualization. 

Of the metrics collected and presented below, the following require non-trivial data processing steps:

\begin{description}
\item[Graph Coloring:] To estimate the chromatic number of the conflict graphs, we apply the DSatur greedy coloring algorithm~\cite{dsatur}, since computing an exact minimum graph coloring is NP-hard~\cite{Karp1972}.
\item[Longest Path Estimation:] We approximate the length of the longest path using a Monte Carlo approach. 
We iterate on connected components by size, randomly choosing a set of nodes as starting points. 
From each starting node, we iteratively build a simple path by uniformly selecting an unvisited neighbor at each step, terminating when no unvisited neighbors remain. Connected components with fewer nodes than the current best estimate of the longest path are skipped. 
This approach is known to yield accurate longest-path estimates for random graphs~\cite{gnp-dfs}.
\end{description}

\section{Findings Analysis}
\label{sec:results}

Since the number of transactions in a block can significantly influence most of the metrics we examine, we group blocks into buckets based on their size.
In each graph, we plot one line per bucket: the solid line shows the median across blocks in that bucket, while the shaded region represents the the top and bottom 5\% of observed values.
The legend labels each line by the minimum block size (\#txs) within its corresponding~bucket.

\subsection{Block size}
\Cref{fig:block-size-dist} plots the distribution of block sizes, i.e., the number of transactions per block, in a log-scale histogram. 
Most blocks have around 1,000 transactions, with a long tail reaching 2,000 transactions and more; the largest block size observed is 8,000. Rarely, blocks have a size of 0; in addition, some slots have been completely skipped (yielding empty blocks). Out of the 1M blocks we have downloaded, only $1,498$ were skipped due to consensus protocol failures.

\subsection{Execution Time Units and Fee Distribution}
\label{sec:exec-time}
\Cref{fig:avg-fee} plots the average transaction fee in a block. Note that most transactions have a similar overall fee, except for rare high-spike occasions when the conflict rate is abnormally high compared to all other blocks we have~downloaded.


\Cref{fig:avg-computeUnitsConsumed,fig:avg-costUnits} plot the average number of compute units (CUs) each transaction consume. While the former only considers compute units that are directly accumulated during runtime, the latter includes also all other overheads related to transaction processing apart from execution (see~\cref{sec:solana-txs}). 
Note the similarities between these two graphs.

The execution time units of transactions seem to be inversely proportional to the block size, which is reasonable assuming one aims at obtaining a roughly equal block execution time.
Surprisingly, the transactions fees are not proportional to the block size.
This seems to suggest that a block creator could make more money by packaging many short transactions than a smaller number of longer~transactions.

Note that in Solana, a transaction fee consists of a base fee, equal to 5,000 lamports multiplied by the number of signatures used to validate the transaction, plus a priority fee. The priority fee is not related to the transaction’s execution time. 
Also, Solana imposes limits on block execution time and compute resources. 
As transactions' selection only takes into account absolute fees, a block creator may prefer to include a small number of long transactions that offer slightly higher individual fees, rather than many shorter transactions that each offers a slightly lower fee.
Here, a smarter algorithm could improve the creator's~profit.

Lastly, we look at the average rate of hard failures in \Cref{fig:avg-failed}. Hard failures are transaction that have been executed correctly but encountered a logical error; therefore, they are recorded in blocks rather than omitted.

\begin{figure*}[ht!]
    \centering
    
    \begin{subfigure}{\dimexpr\figscale\linewidth}
        \includegraphics[width=\linewidth]{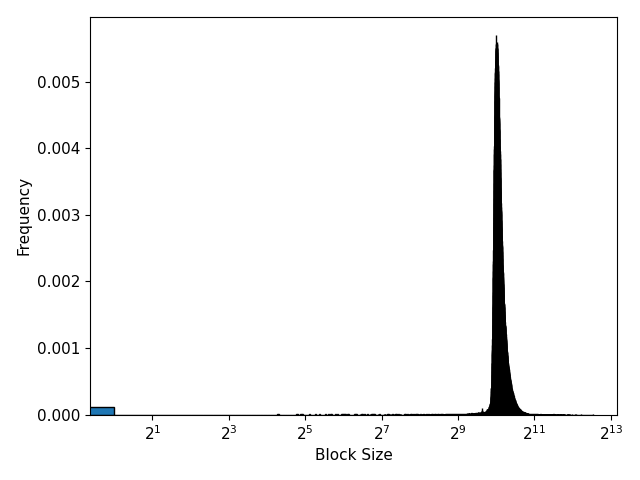}
        \caption{Block Size Distribution}
        \label{fig:block-size-dist}
    \end{subfigure}
    \hfill
    \begin{subfigure}{\dimexpr\figscale\linewidth}
        \includegraphics[width=\linewidth]{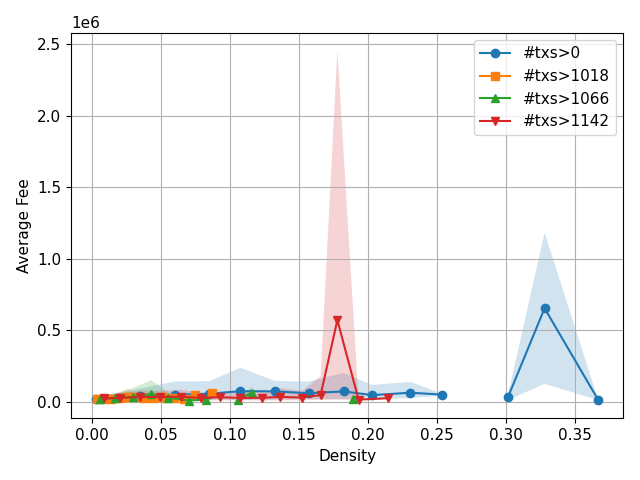}
        \caption{Average Transaction Fee}
        \label{fig:avg-fee}
    \end{subfigure}
    
    \vspace{1em} 

    \begin{subfigure}{\dimexpr\figscale\linewidth}
        \includegraphics[width=\linewidth]{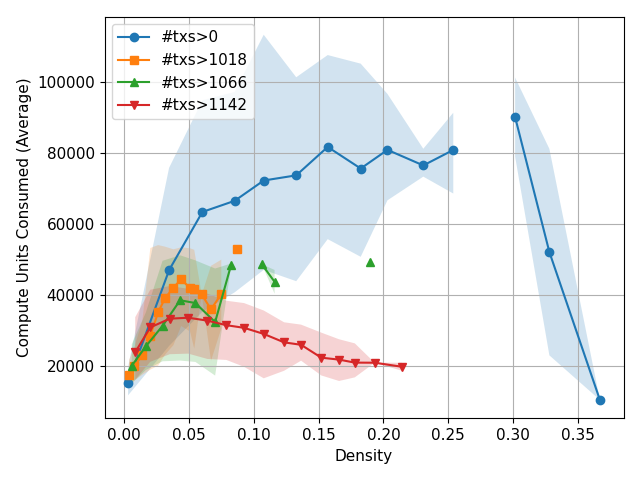}
        \caption{Exec. Compute Units per Tx}
        \label{fig:avg-computeUnitsConsumed}
    \end{subfigure}
    \hfill
    %
    %
    \begin{subfigure}{\dimexpr\figscale\linewidth}
        \includegraphics[width=\linewidth]{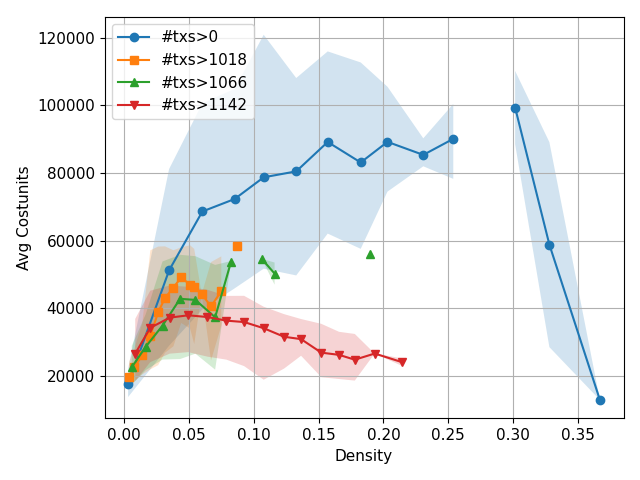}
        \caption{Overall Compute Units per Tx}
        \label{fig:avg-costUnits}
    \end{subfigure}
    \hfill

    \vspace{1em} 

    \begin{subfigure}{\dimexpr\figscale\linewidth}
        \includegraphics[width=\linewidth]{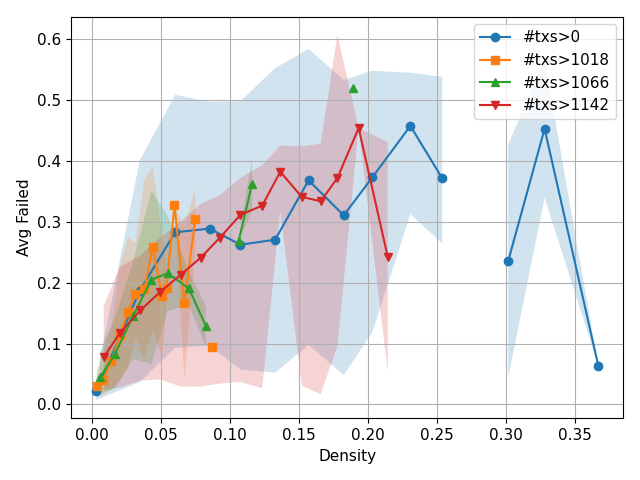}
        \caption{Mean Hard Failures}
        \label{fig:avg-failed}
    \end{subfigure}
    \hfill
    \begin{subfigure}{\dimexpr\figscale\linewidth}
        \includegraphics[width=\linewidth,height=4cm]{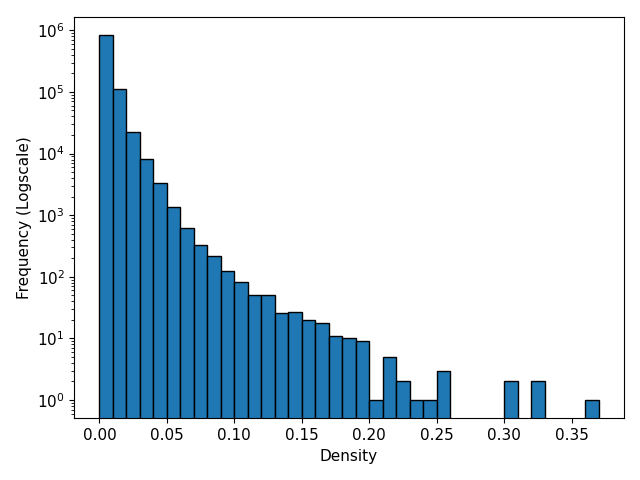}
        \caption{Density Distribution}
        \label{fig:density-dist}
    \end{subfigure}

    \caption{Analysis of Block and Transaction Metrics.}
    \label{fig:tx-metrics-analysis}
\end{figure*}

\subsection{Conflict Graph Properties}
\label{sec:conf-structure}


\subsubsection{Conflict Graph Density}
\Cref{fig:density-dist} exhibits the density distribution of the conflict graphs corresponding to the collected blocks.
As can be seen, the density is quite low, especially compared to what has been reported for Ethereum~\cite{EthSolAnalysis,BFH25,VH20} and Sui~\cite{BF25}, and in correspondence to the findings reported in~\cite{EthSolAnalysis}.
This low density is the result of the block creation in Solana, as reported in \Cref{sec:solana}.
Yet, some graphs reach above 10\% density, and one had even more than 35\%.

\subsubsection{Assortativity and Cluster Coefficient}

Assortativity and cluster coeficient are reported in \Cref{fig:assortativity} and \Cref{fig:cluster-coe}.
The relatively high numbers of these measures indicate that conflicting transactions tend to concentrate is near clique structures or singletons (transactions with no conflicts).

\subsubsection{Degree and Diameter}

\Cref{fig:degree} and \Cref{fig:max-degree} show the average and max degrees of the conflict graphs, while \Cref{fig:diameter} shows the respective diameters.
The fact that the average degree is not too far from the max degree, especially when there are so many singleton nodes (non-conflicting transactions), supports the assumption that conflicting transactions form near clique structures.
The low diameter also corresponds to such~findings.


\begin{figure*}[ht!]
    \centering
    
    \begin{subfigure}{\dimexpr\figscale\linewidth}
        \includegraphics[width=\linewidth]{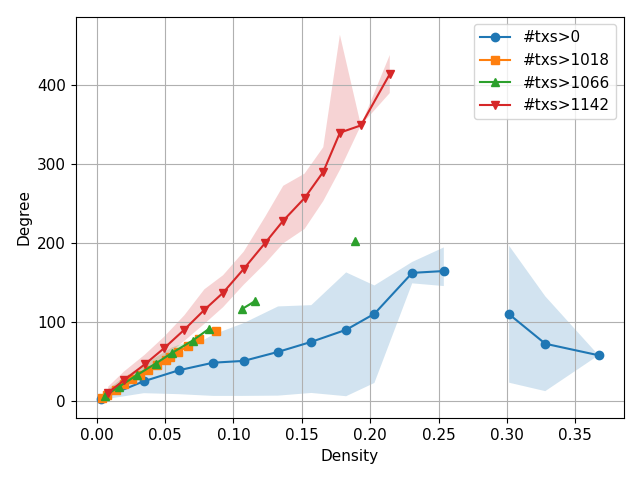}
        \caption{Average Degree}
        \label{fig:degree}
    \end{subfigure}
    \hfill 
    \begin{subfigure}{\dimexpr\figscale\linewidth}
        \includegraphics[width=\linewidth]{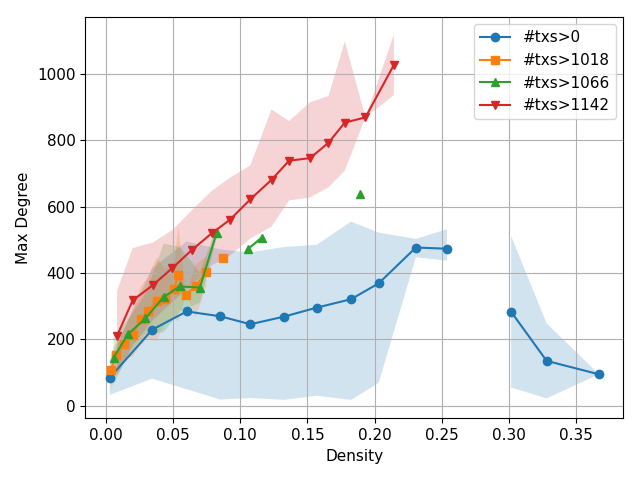}
        \caption{Max Degree}
        \label{fig:max-degree}
    \end{subfigure}
        
    \vspace{1em} 
    
    \begin{subfigure}{\dimexpr\figscale\linewidth}
        \includegraphics[width=\linewidth]{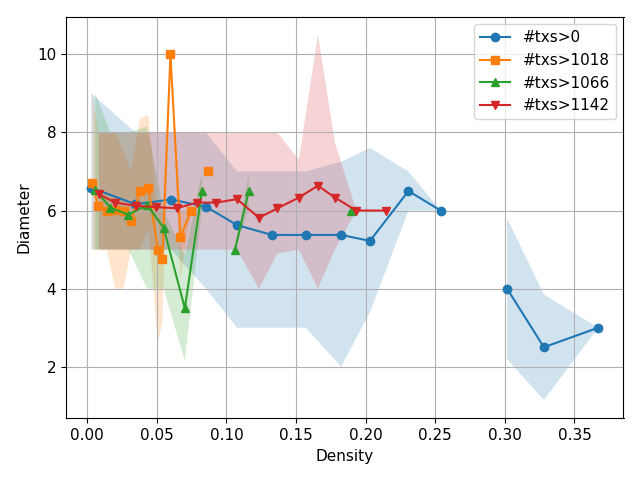}
        \caption{Diameter}
        \label{fig:diameter}
    \end{subfigure}
    \hfill
    %
    %
    \begin{subfigure}{\dimexpr\figscale\linewidth}
        \includegraphics[width=\linewidth]{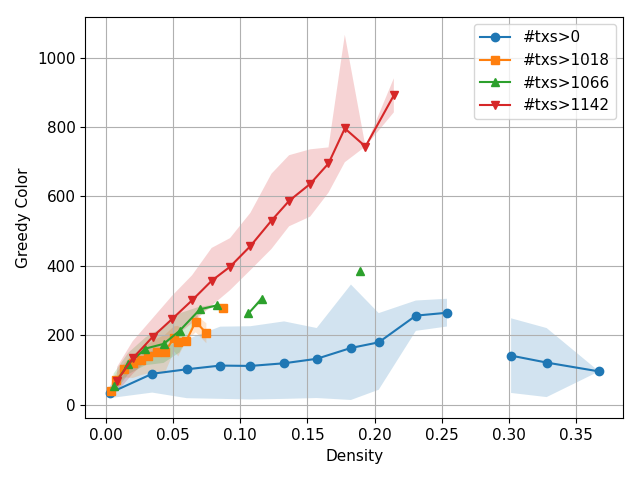}
        \caption{Greedy Chromatic Est.}
        \label{fig:greedy-color}
    \end{subfigure}
        
    \vspace{1em} 
    
    \begin{subfigure}{\dimexpr\figscale\linewidth}
        \includegraphics[width=\linewidth]{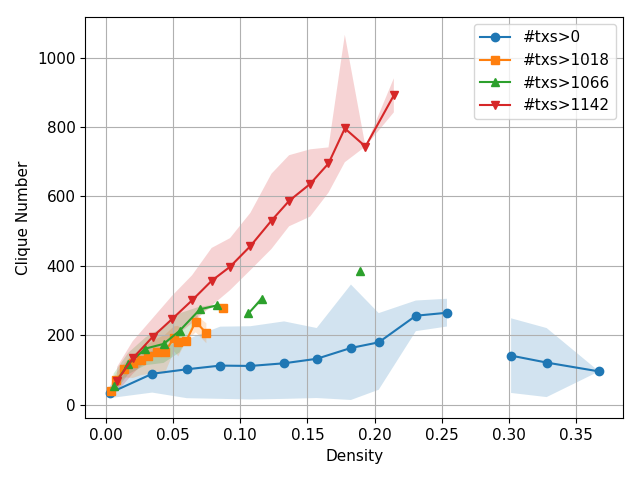}
        \caption{Clique Number}
        \label{fig:clique-number}
    \end{subfigure}
    \hfill
    \begin{subfigure}{\dimexpr\figscale\linewidth}
        \includegraphics[width=\linewidth]{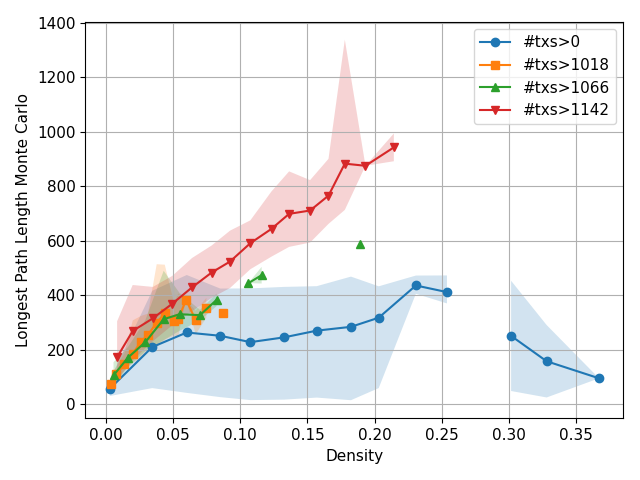}
        \caption{Longest Path Est.}
        \label{fig:longest-path-length-monte-carlo}
    \end{subfigure}
    \caption{Comprehensive analysis of graph metrics - part I.}
    \label{fig:full-analysis-1}
\end{figure*}
    

\begin{figure*}[ht!]
    \centering
       
    \begin{subfigure}{\dimexpr\figscale\linewidth}
        \includegraphics[width=\linewidth]{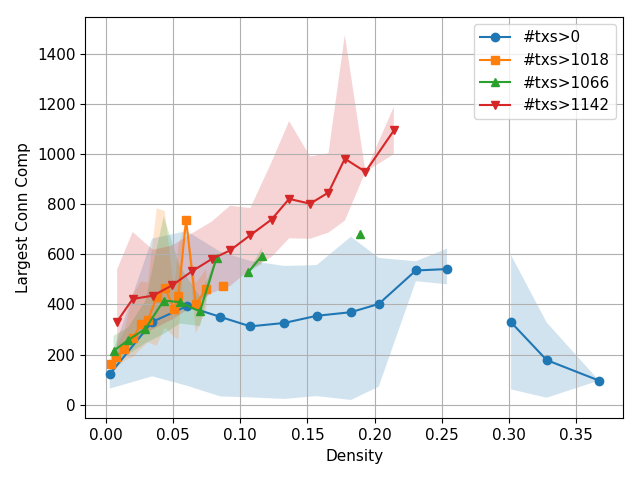}
        \caption{Largest Conn. Comp.}
        \label{fig:largest-conn-comp}
    \end{subfigure}
    \hfill
    \begin{subfigure}{\dimexpr\figscale\linewidth}
        \includegraphics[width=\linewidth]{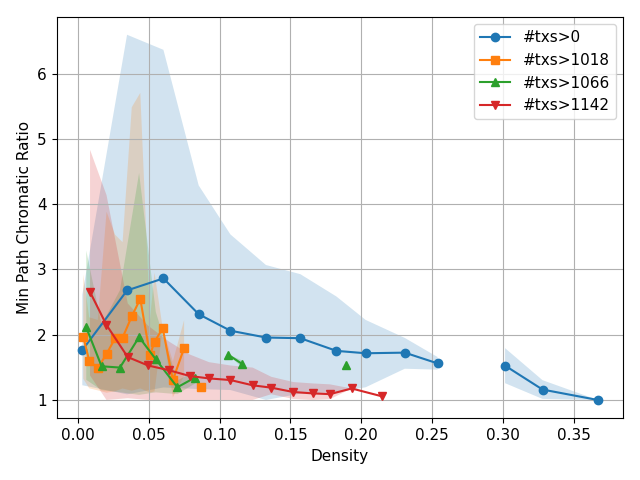}
        \caption{Lower bound ratio}
        \label{fig:min-path-chromatic-ratio}
    \end{subfigure}
    
    \vspace{1em} 
        
    \begin{subfigure}{\dimexpr\figscale\linewidth}
        \includegraphics[width=\linewidth]{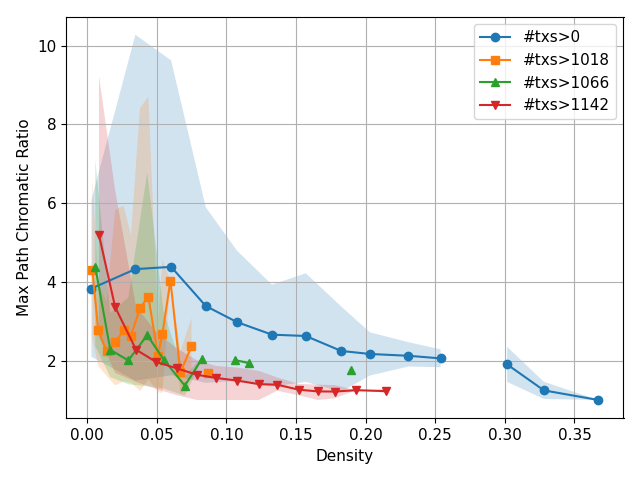}
        \caption{Upper bound ratio}
        \label{fig:max-path-chromatic-ratio}
    \end{subfigure}
    \hfill
    %
    %
    \begin{subfigure}{\dimexpr\figscale\linewidth}
        \includegraphics[width=\linewidth]{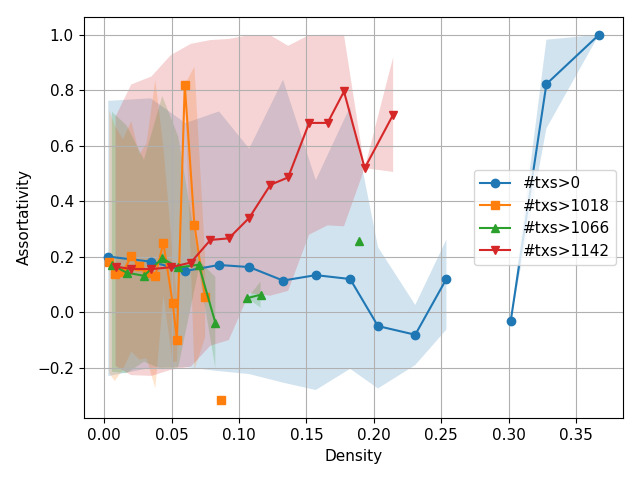}
        \caption{Assortativity}
        \label{fig:assortativity}
    \end{subfigure}
    
    \vspace{1em} 
        
    \begin{subfigure}{\dimexpr\figscale\linewidth}
        \includegraphics[width=\linewidth]{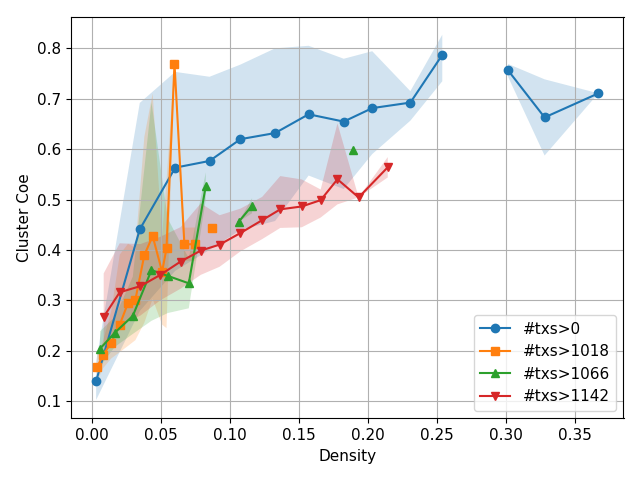}
        \caption{Cluster Coefficient}
        \label{fig:cluster-coe}
    \end{subfigure}
    \caption{Comprehensive analysis of graph metrics - part II.}
    \label{fig:full-analysis-2}
\end{figure*}

\subsubsection{Chromatic Number and Clique Number}

\Cref{fig:greedy-color} exhibits the greedy algorithm's estimations for the chromatic numbers of the graphs, which is an upper bound on the true chromatic number.
\Cref{fig:clique-number} shows the corresponding clique numbers, which serve as lower bounds for the true chromatic number.
The fact that these graphs are almost identical indicates that the DSatur greedy algorithm indeed finds near minimal colorings.

We remind the reader that the size of the block divided by chromatic number is a good indication for the maximal potential speedup due to parallelism.
As can be seen, in most cases parallelism is expected to significantly improve performance.
However, in the denser and larger graphs, the chromatic number is very large, indicating that a significant portion of the graph forms a single clique, or a near clique, structure.
This may indicate a burst of inter-related activities.

\begin{figure*}[t]
    \centering
    \begin{subfigure}{\dimexpr\figscale\linewidth}
        \centerline{\includegraphics[width=\linewidth]{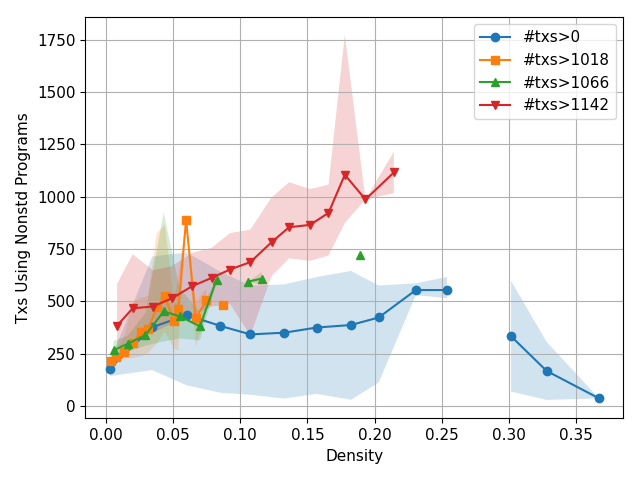}}
        \caption{\#TX Invoking nonstandard Prog.}
        \label{fig:txs-using-nonstd-programs}
    \end{subfigure}
    \hfill
    \begin{subfigure}{\dimexpr\figscale\linewidth}
        \includegraphics[width=\linewidth]{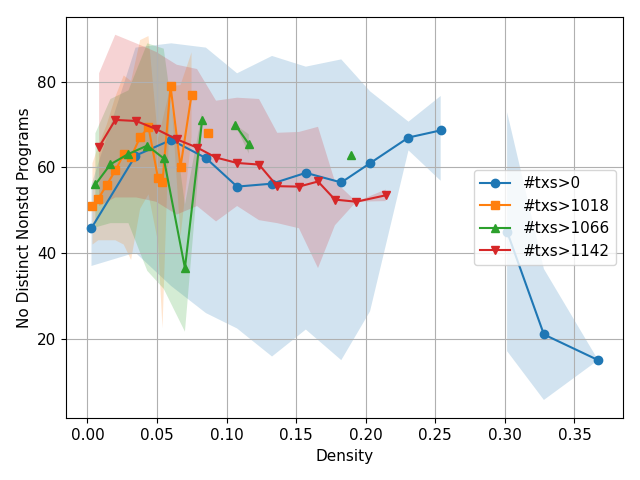}
        \caption{Distinct Non-std Programs/Block}
        \label{fig:no-distinct-nonstd-programs}
    \end{subfigure}
    
    \vspace{1em} 
            
    \begin{subfigure}{\dimexpr\figscale\linewidth}
        \includegraphics[width=\linewidth]{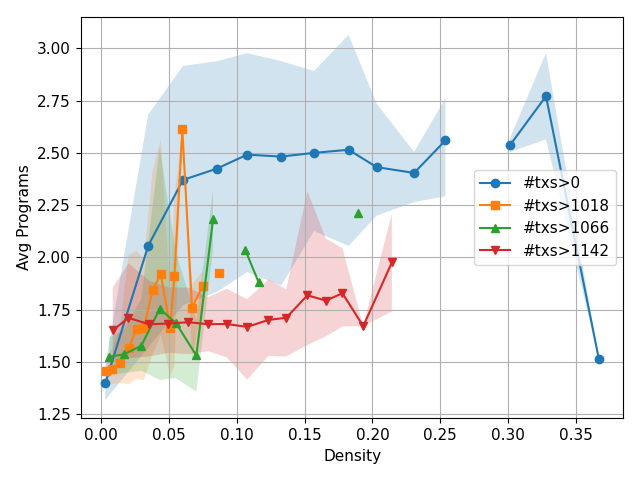}
        \caption{Programs per Transaction}
        \label{fig:avg-programs}
    \end{subfigure}
    \hfill
    \begin{subfigure}{\dimexpr\figscale\linewidth}
    \includegraphics[width=\linewidth]{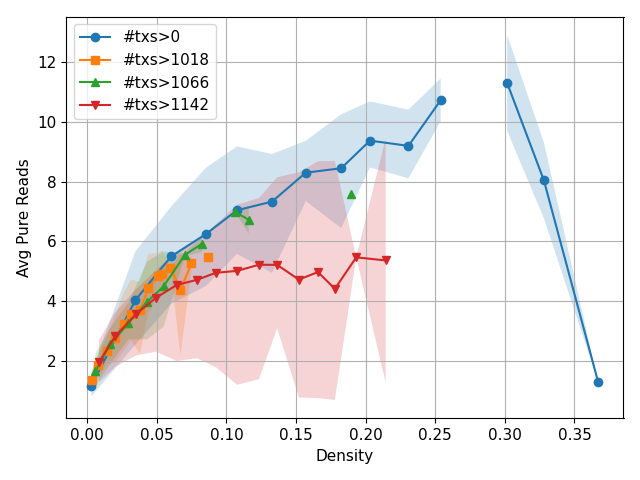}
    \caption{Pure Reads (Data Accounts) per Tx}
    \label{fig:avg-pure-reads}
    \end{subfigure}
    \caption{Analysis of Program Usage and Data Reads.}
    \label{fig:program-read-analysis}
\end{figure*}

\subsubsection{Longest Simple Path and Largest Connected Component}

\Cref{fig:longest-path-length-monte-carlo} shows the longest path estimation as computed by the Monte Carlo based approach, which serves as a lower bound on this figure.
\Cref{fig:largest-conn-comp} exhibits the size of the largest connected component, which is an upper bound on it.
Here, too, the results are not far off.
Recall that this measure serves as a worse case serialization time, which could impede performance when validators must follow a logical serialization order that corresponds to the order in which transactions appear in the block.
The relatively large number in some of the data points echo the findings of~\cite{EthSolAnalysis}, which reported serialization chains as long as 59\% of the block~size.

\subsubsection{Longest Path to Chromatic Number Ratio}

\Cref{fig:min-path-chromatic-ratio} exhibits the ratio between the longest shortest path and the estimated chromic number, which serves as a lower bound for this metric.
\Cref{fig:max-path-chromatic-ratio} exhibits the ratio between the largest connected component and the estimated chromic number, which serves as an upper bound.
Recall that this metric indicates the additional speedup that parallel execution of transactions while obeying a serialization order derived from minimal coloring can have in comparison to an arbitrary derived block order~\cite{ColoringSC}.
We can see that in most cases the coloring approach can improve performance by a factor of 1.5-3, which in extreme cases the improvement can even reach a factor of 6-10.


\subsection{Distribution of Programs}
Finally, we analyze how transactions interact with blockchain programs. We collected 18 publicly known program addresses, including core native programs, SPL programs, loader programs and ecosystem programs, and singled them out. For the purpose of this discussion, we refer to these 18 specific 18 as ``standard programs'' and the rest as ``non-standard''. 
\Cref{fig:txs-using-nonstd-programs} plots the number of transactions that include at least one instruction involving a non-standard program, and \cref{fig:no-distinct-nonstd-programs} counts how many different nonstandard programs are involved in a single block. \Cref{fig:avg-programs} plots the number of programs each transaction uses, and \cref{fig:avg-pure-reads} counts the number of read locks a transaction acquires for pure data accounts (and ignoring program accounts which are also included in the read declarations).
Empirical evidence shows that the two most used  standard programs are Vote1111111111....11 and ComputeBudget111111111....111; however, they are ``kernel''-like programs, rather than user related operations. The top three most used user-related programs are 111111111111.....11111 (system including transfers), TokenkegQfeZyiNwAJbNbGKPFXCWuBvf9Ss623VQ5DA, ATokenGPvbdGVxr1b2hvZbsiqW5xWH25efTNsLJA8knL

\ifdefined\ANON
Additional graphs, as well as the raw data containing per-block metrics, would be made available once the anonymity requirement is lifted.
\else
Additional graphs, as well as the raw data containing per-block metrics, have been made available online~\cite{MoreData,Github}.
\fi

\section{Discussion}
\label{sec:dicussion}

In this paper, we have downloaded and analyzed 1M Solana blocks.
Our analysis focused on transactions execution times, transactions fees, number and types of programs being invoked by each transaction and block, as well as readsets, writesets, and the corresponding conflicts and resulting conflict graph~properties.



An interesting conclusion we derive from analyzing the individual transactions execution time vs. fees is that in Solana, a block creator can make more money by packaging many fast executing transaction than a few slow ones.
We remind the reader that in Solana the transaction fee includes a priority fee part, which is unrelated to its size or execution time.
However, since blocks have a maximum execution time and size, greedily taking the highest paying transactions could result in a suboptimal overall profit for the block, as discussed in \Cref{sec:exec-time}.

Solana uses programs (smart contracts) to manage its own infrastructure.
Still, we have noticed wide usage of user defined (``non-standard'') programs, and that some of these programs are quite popular, and they tend to create most conflicts.
On average, each transaction invokes between 1-3 programs, and transactions in smaller blocks tend to invoke more programs.
This is reasonable since this probably means that their execution time is longer, so fewer of them can fit inside a single block.
Further, transactions tend to access between 2-10 pure read objects (that are not programs).

Despite Solana's successful efforts to reduce the number of conflicts in each blocks, compared e.g., to Ethereum~\cite{BFH25} and Sui~\cite{BF25}, long conflict chains still emerge, as has also been reported in~\cite{EthSolAnalysis}.
Further, our findings suggest that ordering the transactions inside the block in an order that corresponds to a minimal coloring of the conflict graph~\cite{ColoringSC} is likely to reduce the block execution time by anywhere from 20\% and up to a factor of 10 is extreme cases, or if the order is deliberately manipulated to form long chains.
Interestingly, we also found the DSatur greedy graph algorithm~\cite{dsatur} seems to produce near minimal colorings for Solana conflict graphs.

Previous work~\cite{BF25,BFH25,KKA26} have explored somewhat similar metrics related to the Ethereum and Sui blockchains.
When considering the emerging structures inside the corresponding conflicts graphs, it appears that in Solana there is a relatively high number of transactions that do not conflict with each other.
Further, in Solana, conflicting transactions tend to form near cliques, whereas in Ethereum and Sui we often see a hub-and-spoke like structures.
This also means that in Solana conflict chains tend to be longer than in Ethereum and Sui, corresponding also to the findings of~\cite{EthSolAnalysis}.
Further, in Solana, there is a relatively very large number of singleton nodes, i.e., transactions with no conflicts, which results from its unique approach to selecting transactions to be inserted in a created block.
The difference between the characteristics of transactions and blocks in Solana, as we found in this work, compared to previous findings for Ethereum and Sui suggest that it is not possible to design a single representative synthetic workload that would represent all blockchains.
Perhaps, the best one can hope for is a parameterized TPC-like benchmark, whose development is left as an interesting future work.


\noindent\textbf{Acknowledgements:} We would like to thank the anonymous reviewers for their helpful comments and insights that helped improve this paper.

\clearpage
\newpage

\bibliographystyle{splncs04}
\bibliography{citations}

\ifdefined\ICBCversion
\else
\newpage
\begin{figure}
\centerline{\includegraphics[width=\columnwidth]{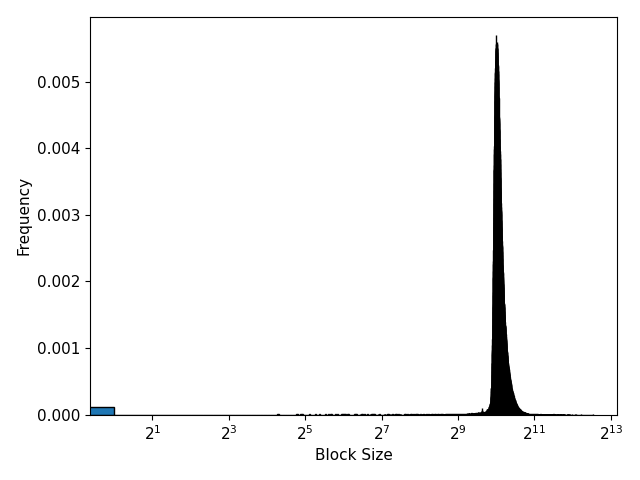}}
\caption{block-size-dist}
\label{fig:block-size-dist}
\end{figure}
\begin{figure}
\centerline{\includegraphics[width=\columnwidth]{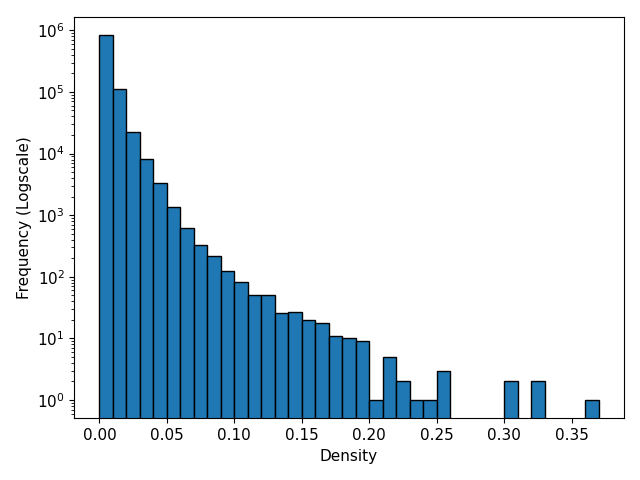}}
\caption{density-dist}
\label{fig:density-dist}
\end{figure}
\begin{figure}
\centerline{\includegraphics[width=\columnwidth]{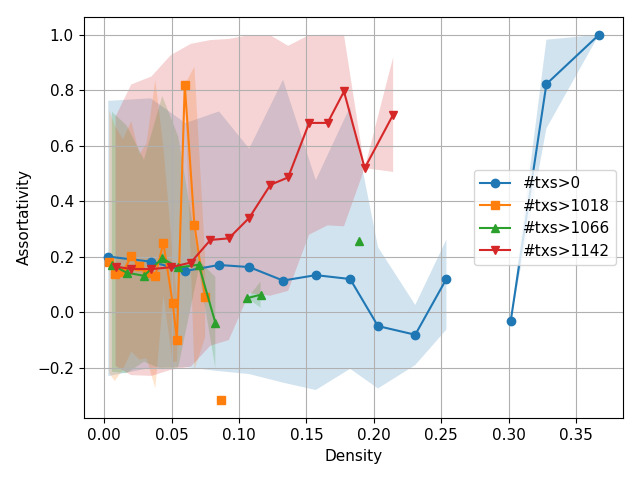}}
\caption{assortativity}
\label{fig:assortativity}
\end{figure}
\begin{figure}
\centerline{\includegraphics[width=\columnwidth]{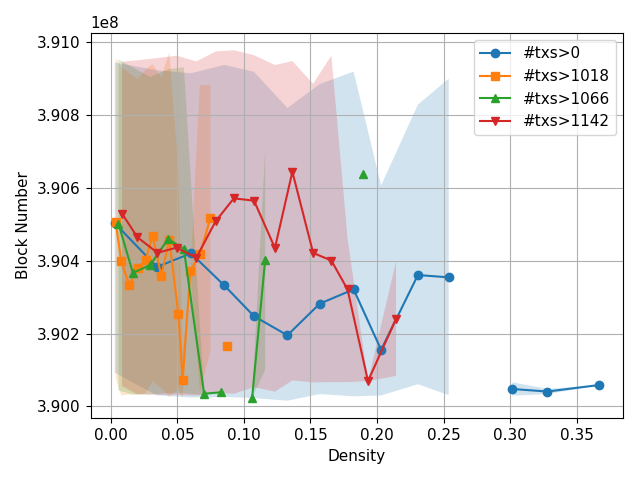}}
\caption{block-number}
\label{fig:block-number}
\end{figure}
\begin{figure}
\centerline{\includegraphics[width=\columnwidth]{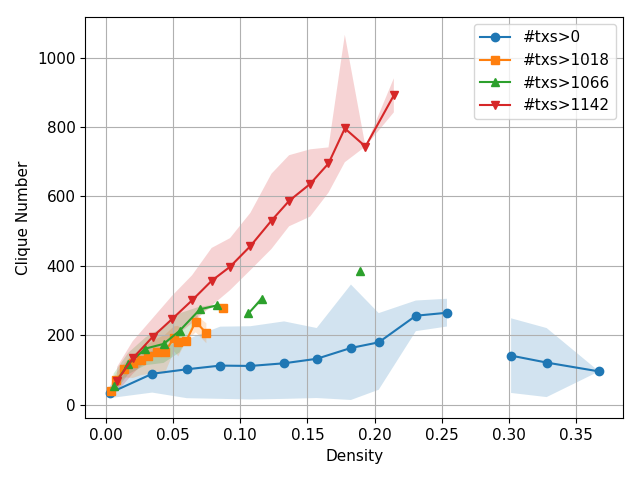}}
\caption{clique-number}
\label{fig:clique-number}
\end{figure}
\begin{figure}
\centerline{\includegraphics[width=\columnwidth]{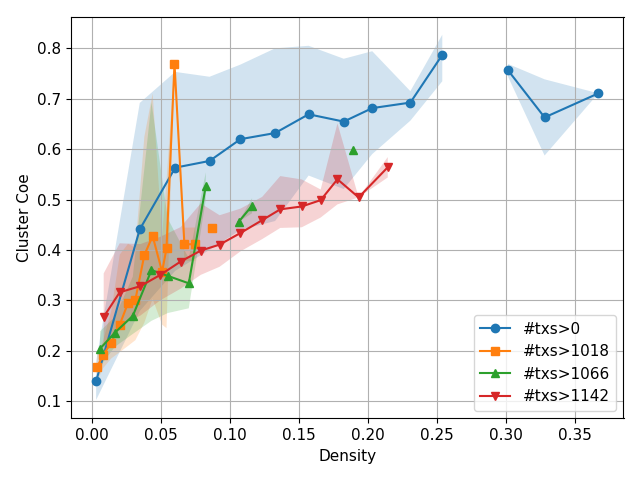}}
\caption{cluster-coe}
\label{fig:cluster-coe}
\end{figure}
\begin{figure}
\centerline{\includegraphics[width=\columnwidth]{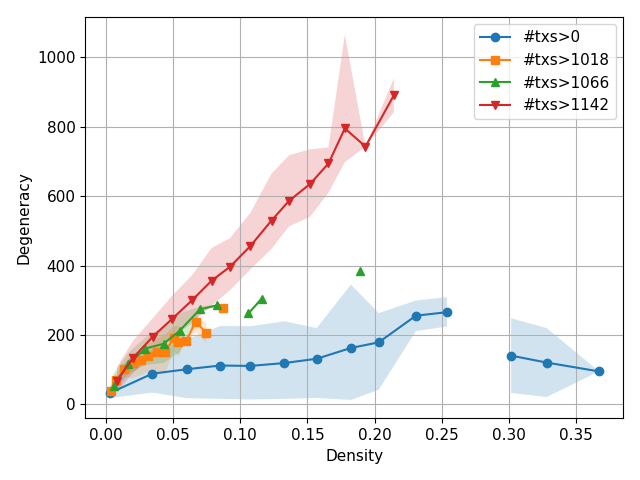}}
\caption{degeneracy}
\label{fig:degeneracy}
\end{figure}
\begin{figure}
\centerline{\includegraphics[width=\columnwidth]{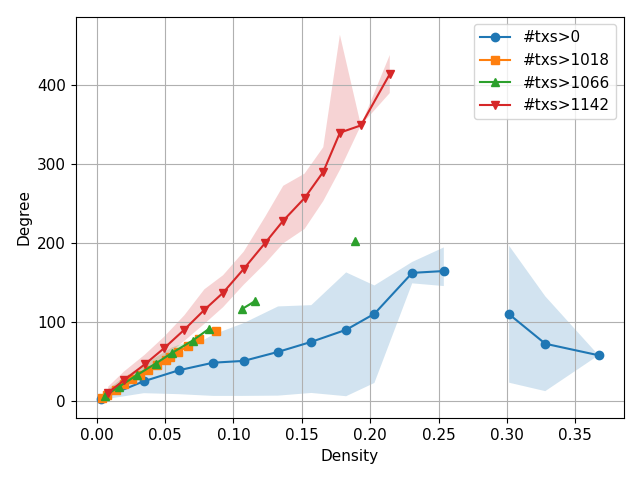}}
\caption{degree}
\label{fig:degree}
\end{figure}
\begin{figure}
\centerline{\includegraphics[width=\columnwidth]{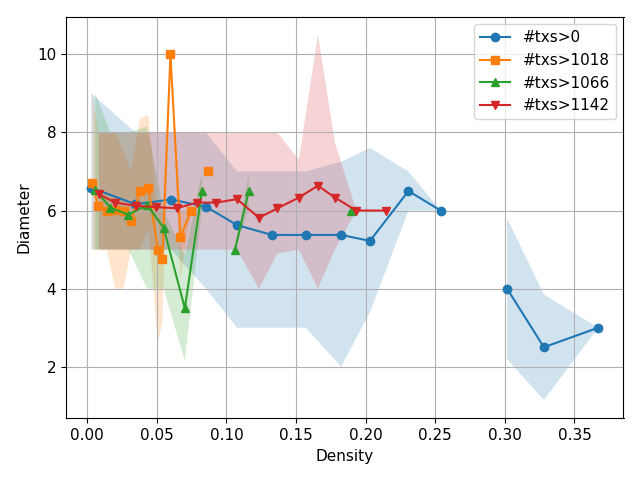}}
\caption{diameter}
\label{fig:diameter}
\end{figure}
\begin{figure}
\centerline{\includegraphics[width=\columnwidth]{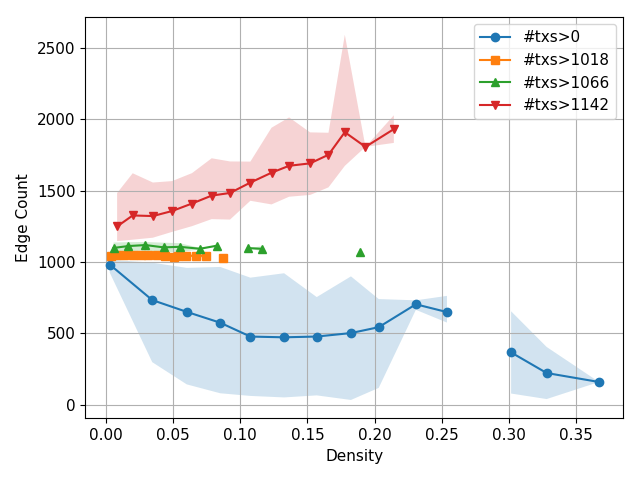}}
\caption{edge-count}
\label{fig:edge-count}
\end{figure}
\begin{figure}
\centerline{\includegraphics[width=\columnwidth]{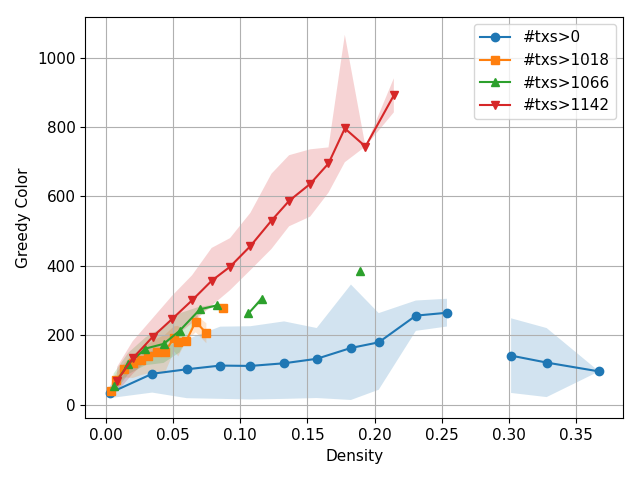}}
\caption{greedy-color}
\label{fig:greedy-color}
\end{figure}
\begin{figure}
\centerline{\includegraphics[width=\columnwidth]{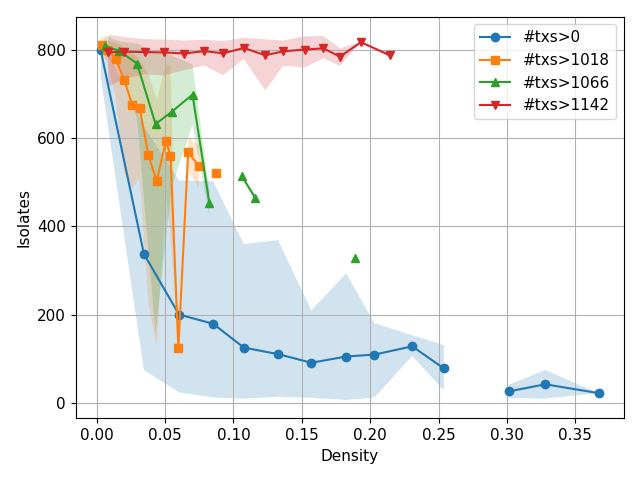}}
\caption{isolates}
\label{fig:isolates}
\end{figure}
\begin{figure}
\centerline{\includegraphics[width=\columnwidth]{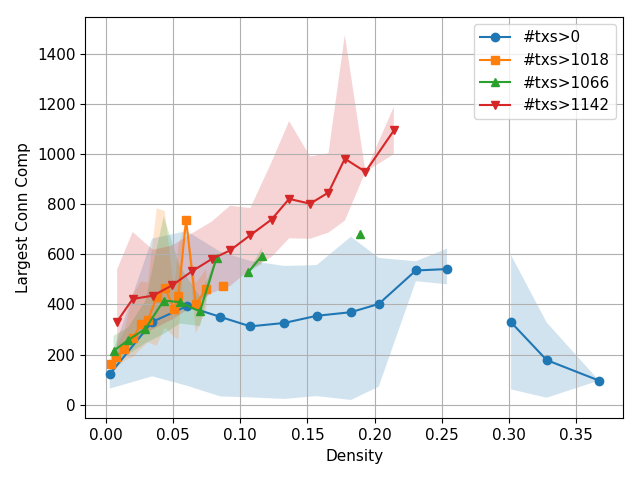}}
\caption{largest-conn-comp}
\label{fig:largest-conn-comp}
\end{figure}
\begin{figure}
\centerline{\includegraphics[width=\columnwidth]{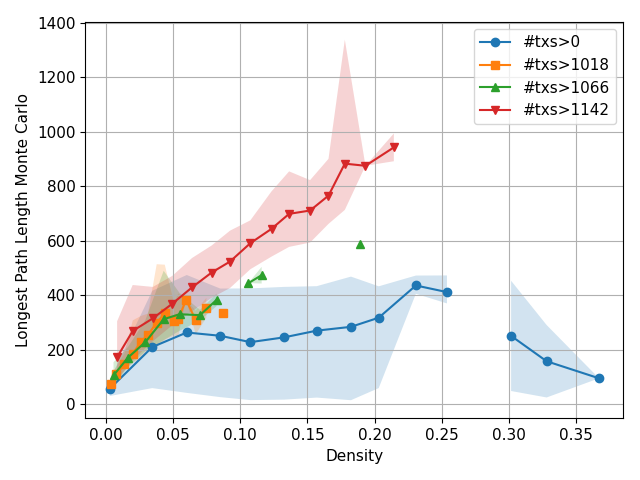}}
\caption{longest-path-length-monte-carlo}
\label{fig:longest-path-length-monte-carlo}
\end{figure}
\begin{figure}
\centerline{\includegraphics[width=\columnwidth]{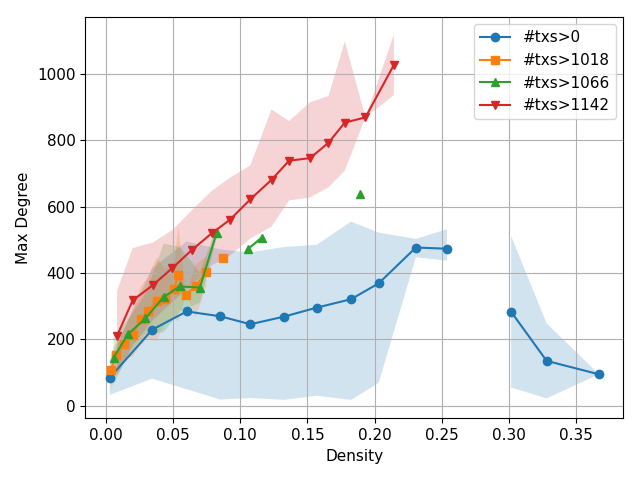}}
\caption{max-degree}
\label{fig:max-degree}
\end{figure}
\begin{figure}
\centerline{\includegraphics[width=\columnwidth]{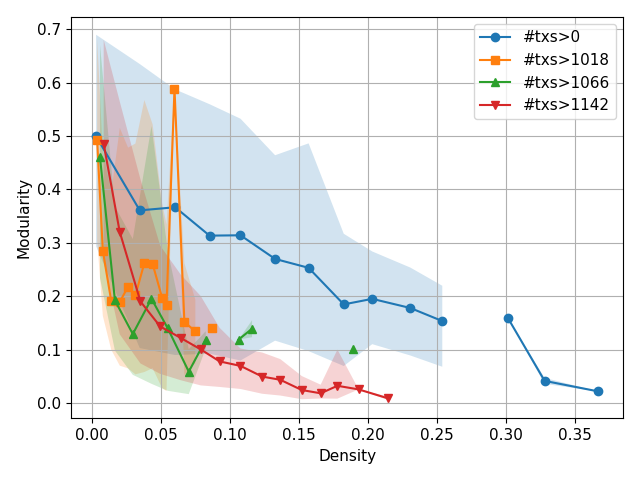}}
\caption{modularity}
\label{fig:modularity}
\end{figure}
\begin{figure}
\centerline{\includegraphics[width=\columnwidth]{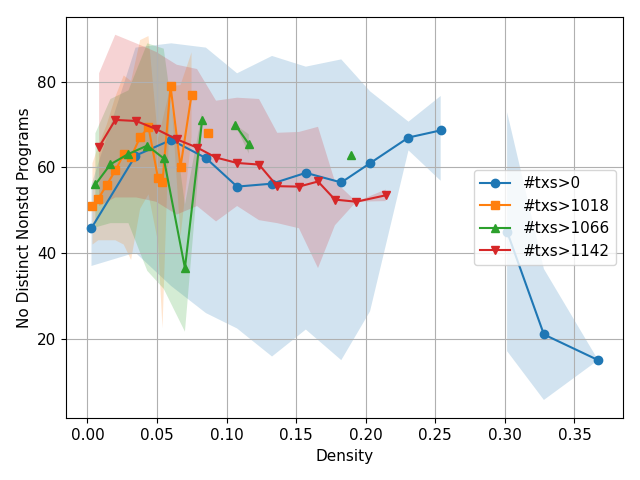}}
\caption{no-distinct-nonstd-programs}
\label{fig:no-distinct-nonstd-programs}
\end{figure}
\begin{figure}
\centerline{\includegraphics[width=\columnwidth]{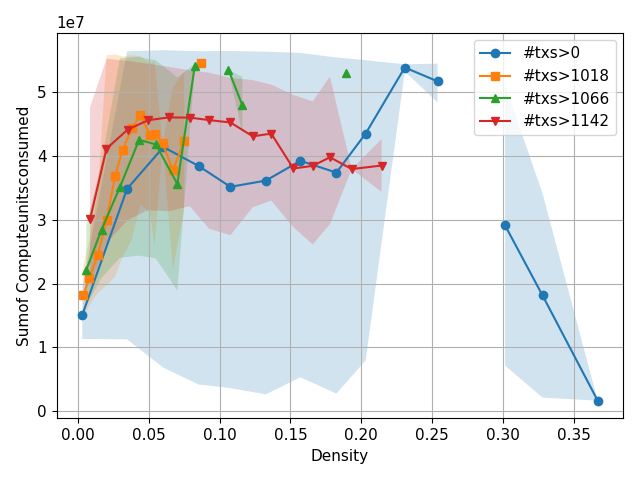}}
\caption{sumof-computeUnitsConsumed}
\label{fig:sumof-computeUnitsConsumed}
\end{figure}
\begin{figure}
\centerline{\includegraphics[width=\columnwidth]{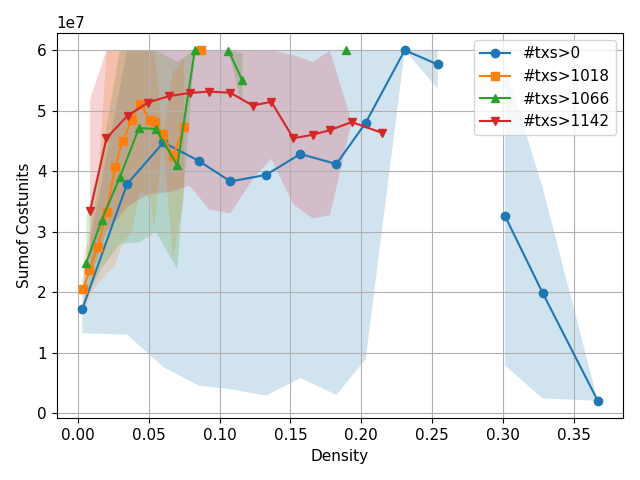}}
\caption{sumof-costUnits}
\label{fig:sumof-costUnits}
\end{figure}
\begin{figure}
\centerline{\includegraphics[width=\columnwidth]{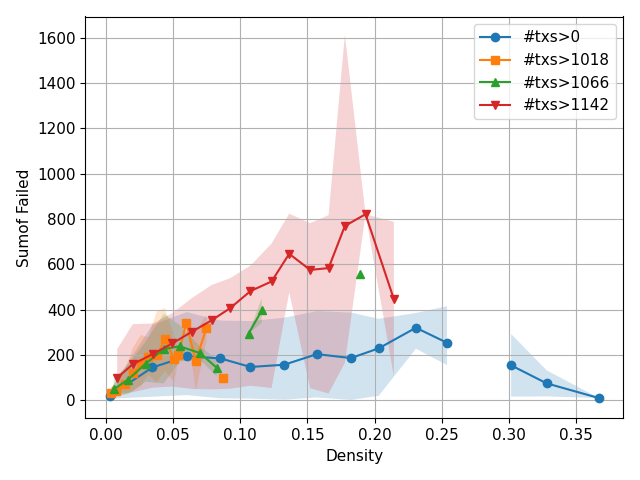}}
\caption{sumof-failed}
\label{fig:sumof-failed}
\end{figure}
\begin{figure}
\centerline{\includegraphics[width=\columnwidth]{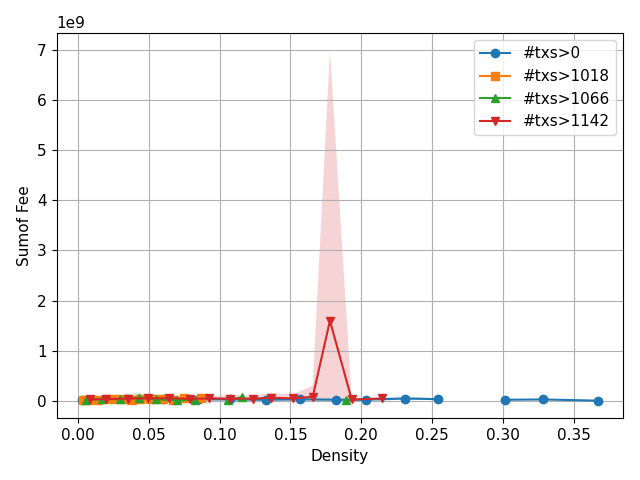}}
\caption{sumof-fee}
\label{fig:sumof-fee}
\end{figure}
\begin{figure}
\centerline{\includegraphics[width=\columnwidth]{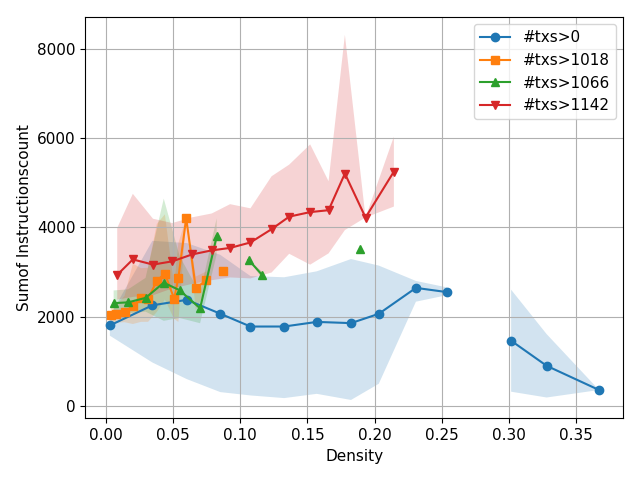}}
\caption{sumof-instructionsCount}
\label{fig:sumof-instructionsCount}
\end{figure}
\begin{figure}
\centerline{\includegraphics[width=\columnwidth]{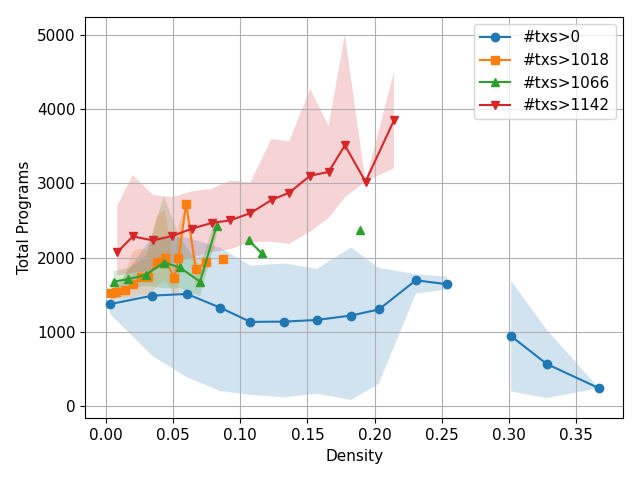}}
\caption{total-programs}
\label{fig:total-programs}
\end{figure}
\begin{figure}
\centerline{\includegraphics[width=\columnwidth]{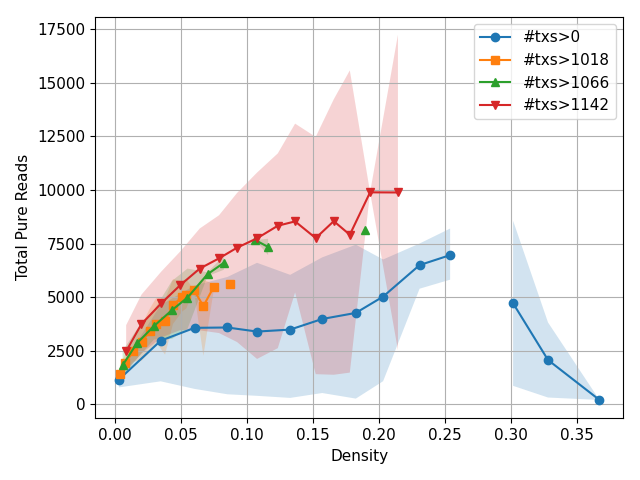}}
\caption{total-pure-reads}
\label{fig:total-pure-reads}
\end{figure}
\begin{figure}
\centerline{\includegraphics[width=\columnwidth]{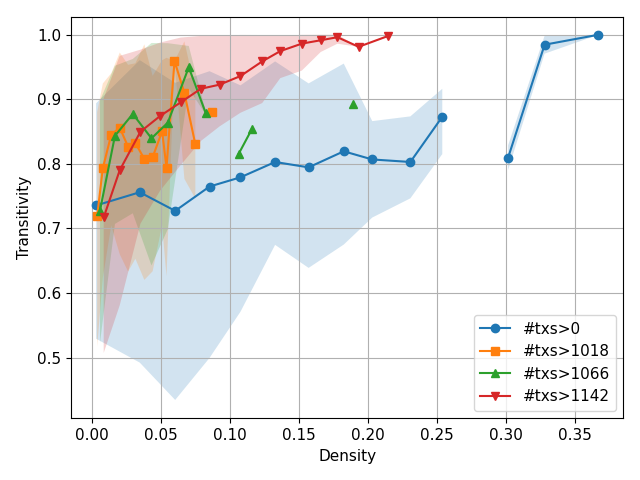}}
\caption{transitivity}
\label{fig:transitivity}
\end{figure}
\begin{figure}
\centerline{\includegraphics[width=\columnwidth]{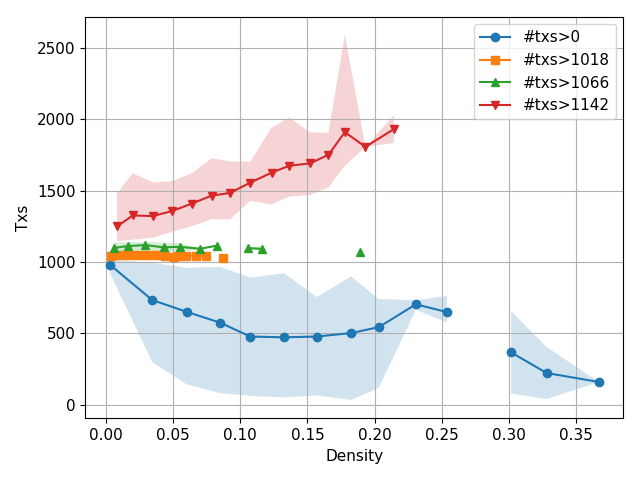}}
\caption{txs}
\label{fig:txs}
\end{figure}
\begin{figure}
\centerline{\includegraphics[width=\columnwidth]{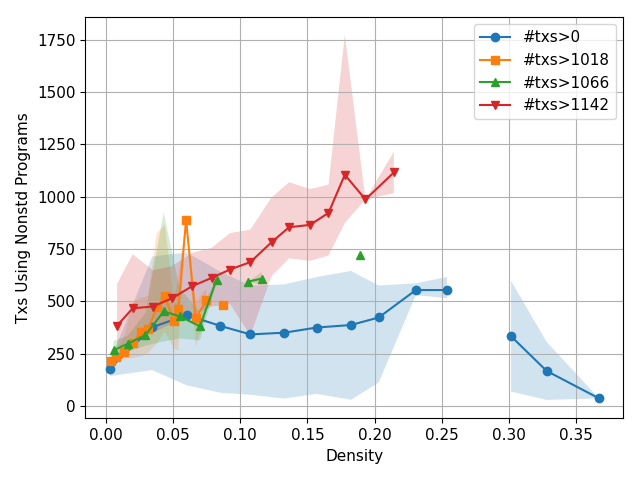}}
\caption{txs-using-nonstd-programs}
\label{fig:txs-using-nonstd-programs}
\end{figure}
\begin{figure}
\centerline{\includegraphics[width=\columnwidth]{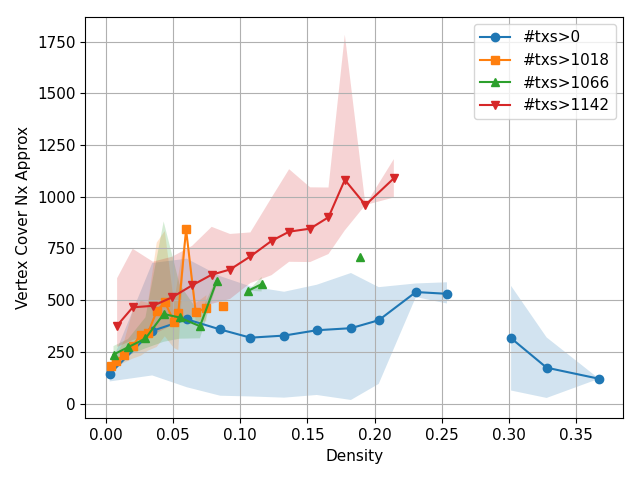}}
\caption{vertex-cover-nx-approx}
\label{fig:vertex-cover-nx-approx}
\end{figure}
\begin{figure}
\centerline{\includegraphics[width=\columnwidth]{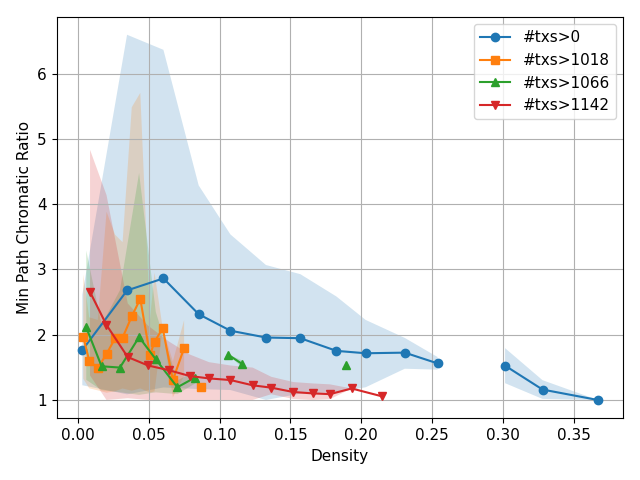}}
\caption{min-path-chromatic-ratio}
\label{fig:min-path-chromatic-ratio}
\end{figure}
\begin{figure}
\centerline{\includegraphics[width=\columnwidth]{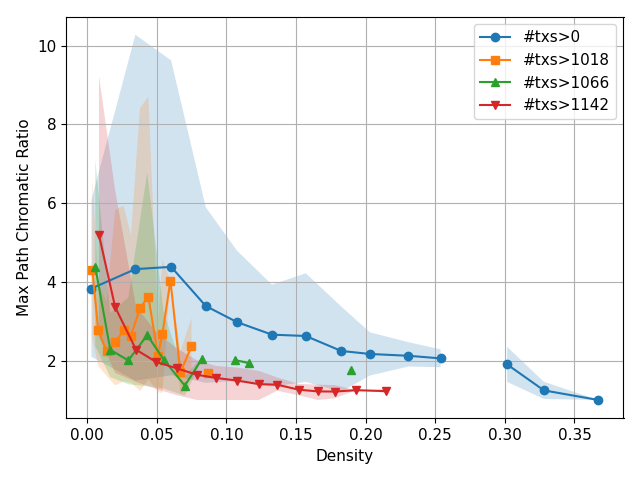}}
\caption{max-path-chromatic-ratio}
\label{fig:max-path-chromatic-ratio}
\end{figure}
\begin{figure}
\centerline{\includegraphics[width=\columnwidth]{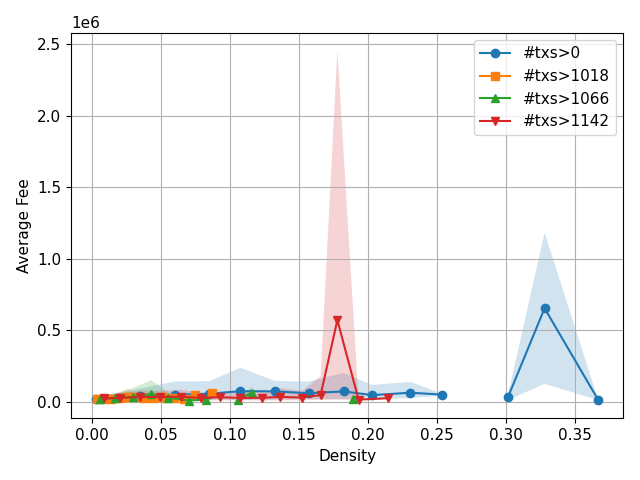}}
\caption{avg-fee}
\label{fig:avg-fee}
\end{figure}
\begin{figure}
\centerline{\includegraphics[width=\columnwidth]{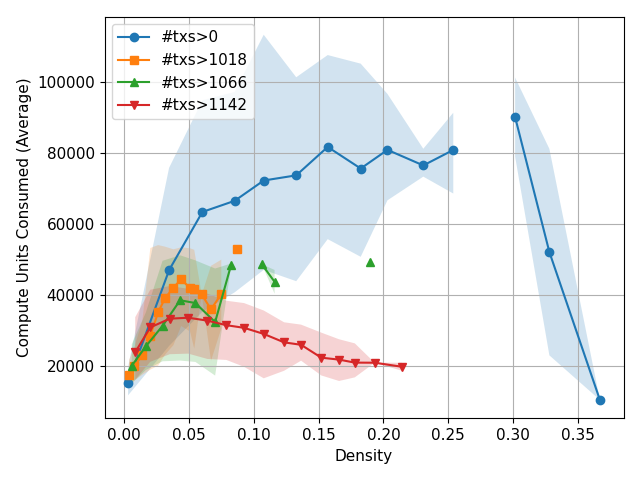}}
\caption{avg-computeUnitsConsumed}
\label{fig:avg-computeUnitsConsumed}
\end{figure}
\begin{figure}
\centerline{\includegraphics[width=\columnwidth]{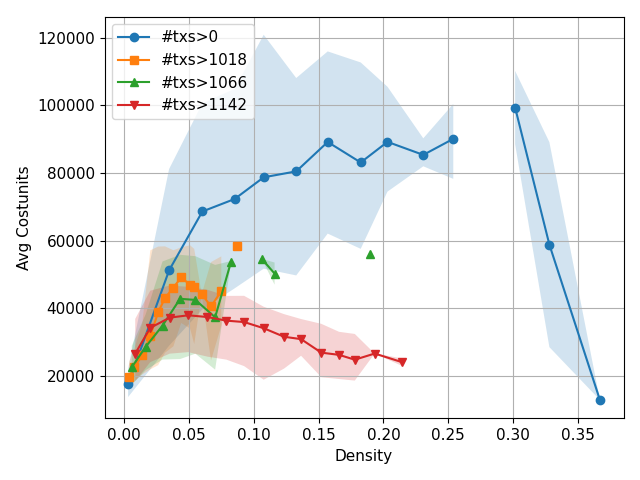}}
\caption{avg-costUnits}
\label{fig:avg-costUnits}
\end{figure}
\begin{figure}
\centerline{\includegraphics[width=\columnwidth]{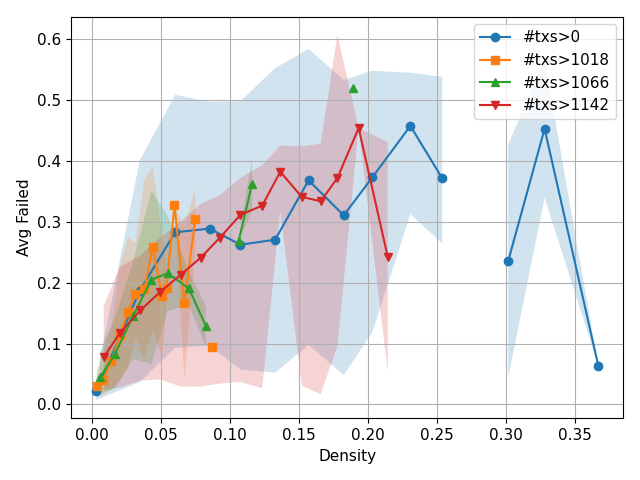}}
\caption{avg-failed}
\label{fig:avg-failed}
\end{figure}
\begin{figure}
\centerline{\includegraphics[width=\columnwidth]{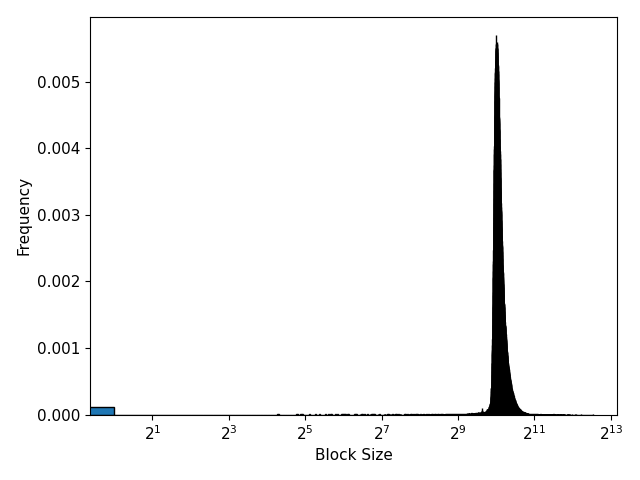}}
\caption{block-size-dist}
\label{fig:block-size-dist}
\end{figure}
\begin{figure}
\centerline{\includegraphics[width=\columnwidth]{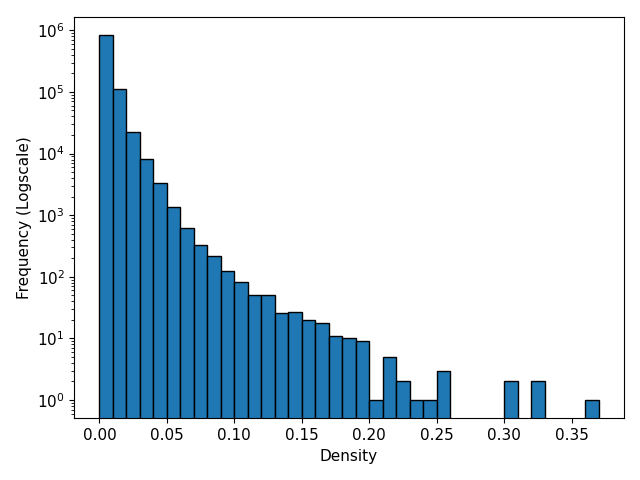}}
\caption{density-dist}
\label{fig:density-dist}
\end{figure}
\begin{figure}
\centerline{\includegraphics[width=\columnwidth]{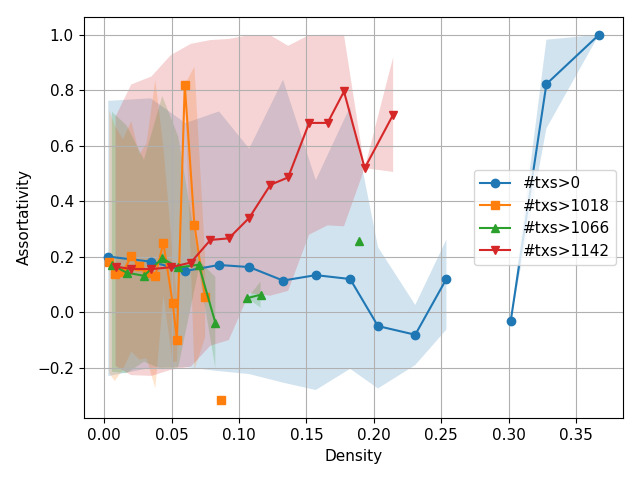}}
\caption{assortativity}
\label{fig:assortativity}
\end{figure}
\begin{figure}
\centerline{\includegraphics[width=\columnwidth]{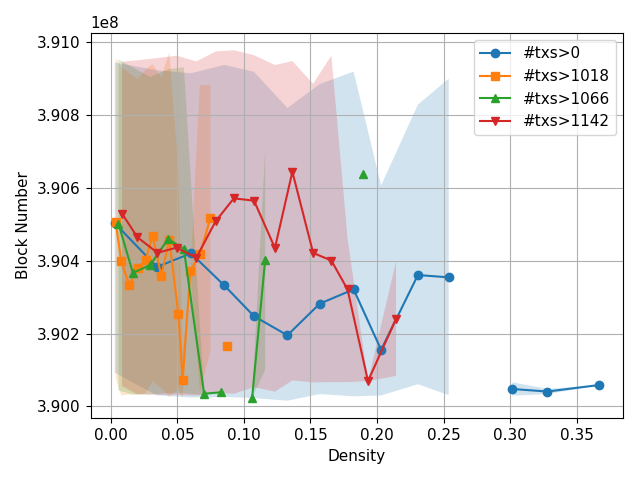}}
\caption{block-number}
\label{fig:block-number}
\end{figure}
\begin{figure}
\centerline{\includegraphics[width=\columnwidth]{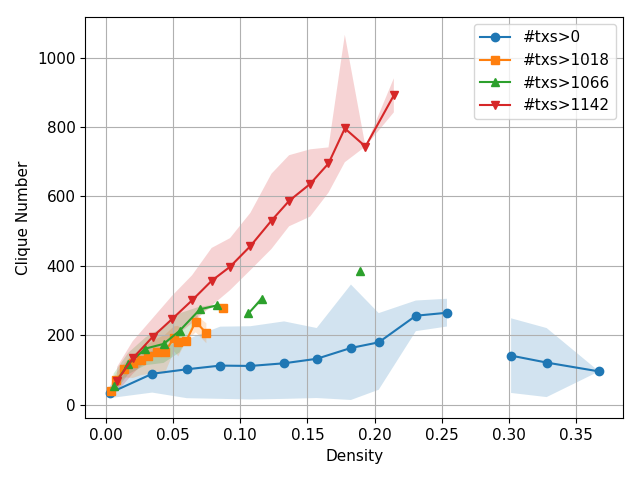}}
\caption{clique-number}
\label{fig:clique-number}
\end{figure}
\begin{figure}
\centerline{\includegraphics[width=\columnwidth]{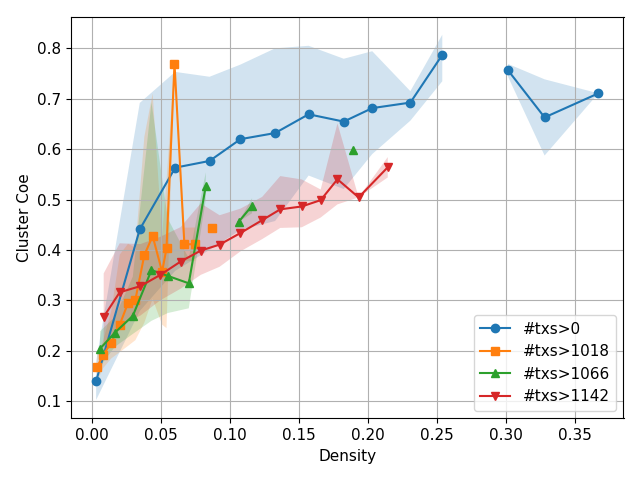}}
\caption{cluster-coe}
\label{fig:cluster-coe}
\end{figure}
\begin{figure}
\centerline{\includegraphics[width=\columnwidth]{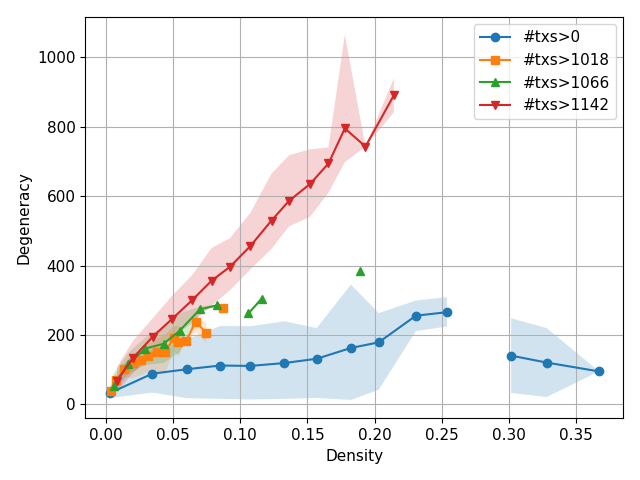}}
\caption{degeneracy}
\label{fig:degeneracy}
\end{figure}
\begin{figure}
\centerline{\includegraphics[width=\columnwidth]{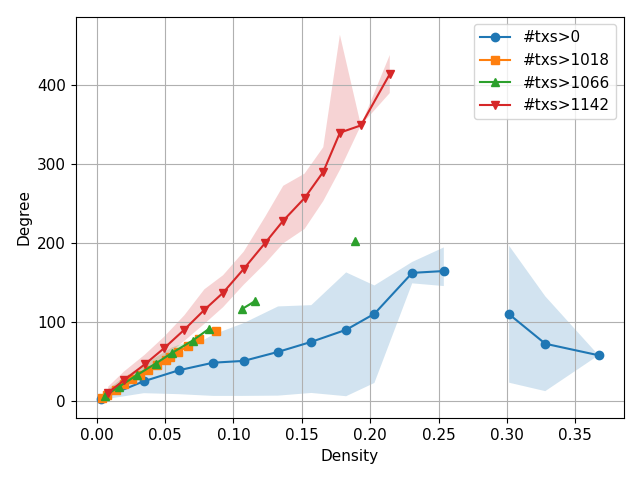}}
\caption{degree}
\label{fig:degree}
\end{figure}
\begin{figure}
\centerline{\includegraphics[width=\columnwidth]{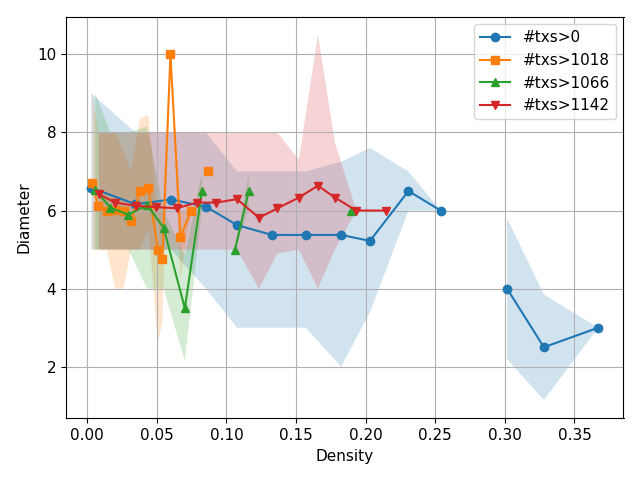}}
\caption{diameter}
\label{fig:diameter}
\end{figure}
\begin{figure}
\centerline{\includegraphics[width=\columnwidth]{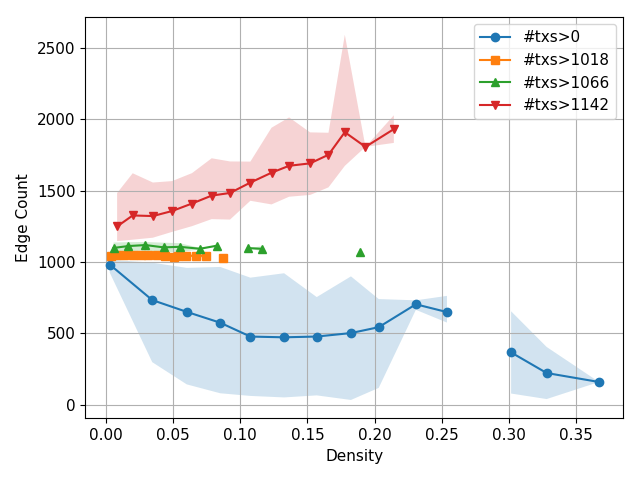}}
\caption{edge-count}
\label{fig:edge-count}
\end{figure}
\begin{figure}
\centerline{\includegraphics[width=\columnwidth]{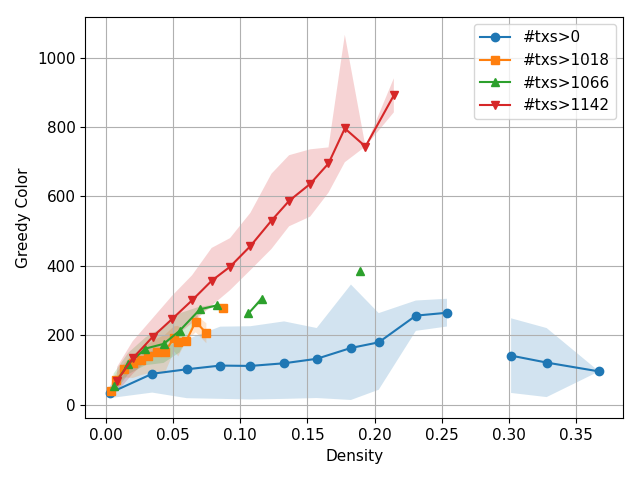}}
\caption{greedy-color}
\label{fig:greedy-color}
\end{figure}
\begin{figure}
\centerline{\includegraphics[width=\columnwidth]{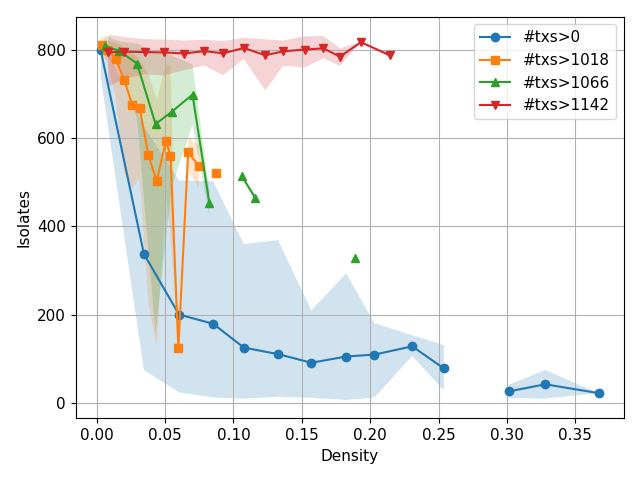}}
\caption{isolates}
\label{fig:isolates}
\end{figure}
\begin{figure}
\centerline{\includegraphics[width=\columnwidth]{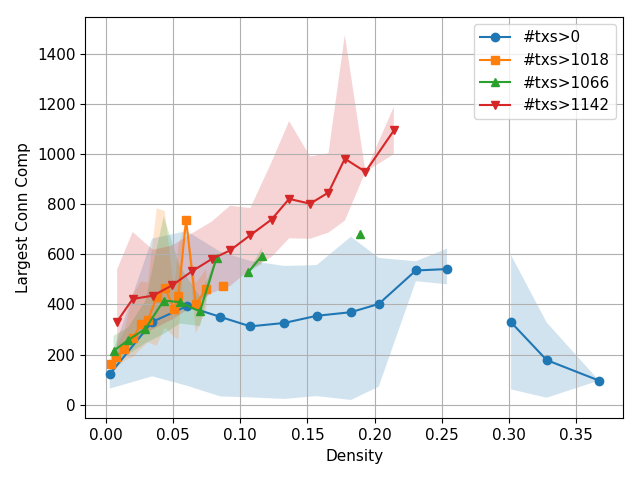}}
\caption{largest-conn-comp}
\label{fig:largest-conn-comp}
\end{figure}
\begin{figure}
\centerline{\includegraphics[width=\columnwidth]{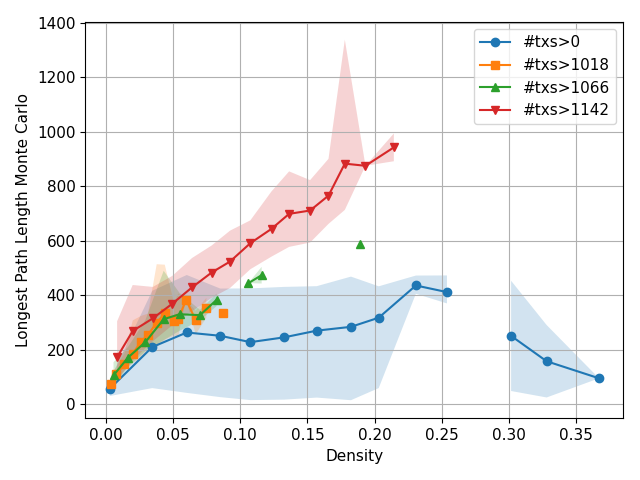}}
\caption{longest-path-length-monte-carlo}
\label{fig:longest-path-length-monte-carlo}
\end{figure}
\begin{figure}
\centerline{\includegraphics[width=\columnwidth]{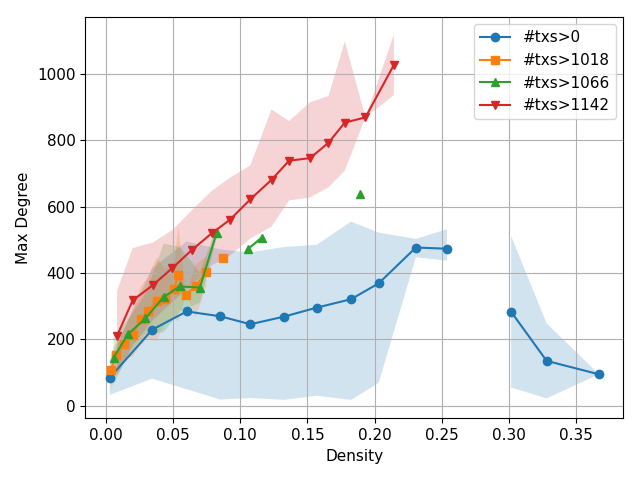}}
\caption{max-degree}
\label{fig:max-degree}
\end{figure}
\begin{figure}
\centerline{\includegraphics[width=\columnwidth]{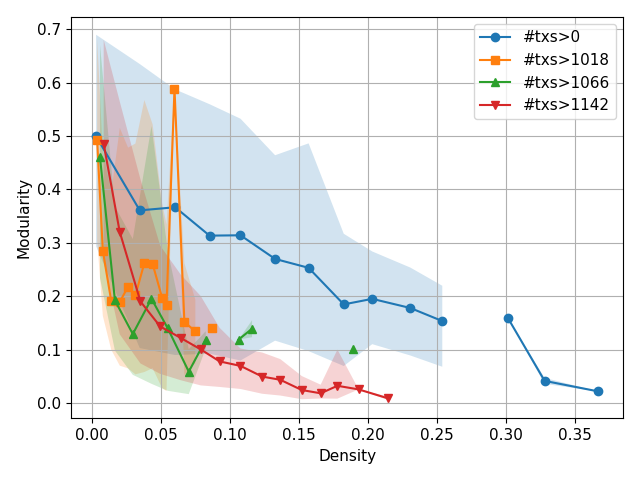}}
\caption{modularity}
\label{fig:modularity}
\end{figure}
\begin{figure}
\centerline{\includegraphics[width=\columnwidth]{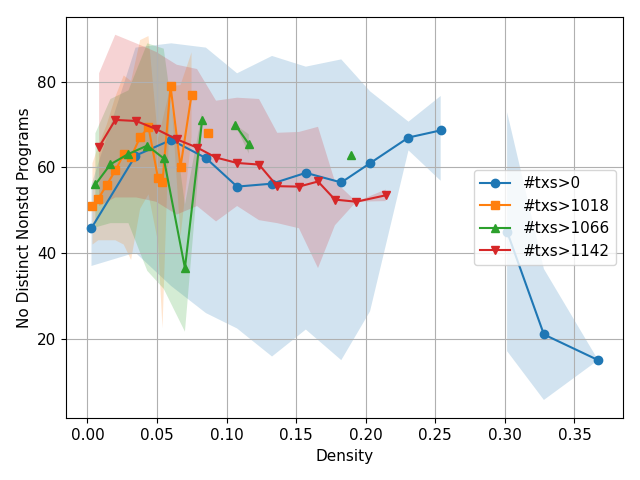}}
\caption{no-distinct-nonstd-programs}
\label{fig:no-distinct-nonstd-programs}
\end{figure}
\begin{figure}
\centerline{\includegraphics[width=\columnwidth]{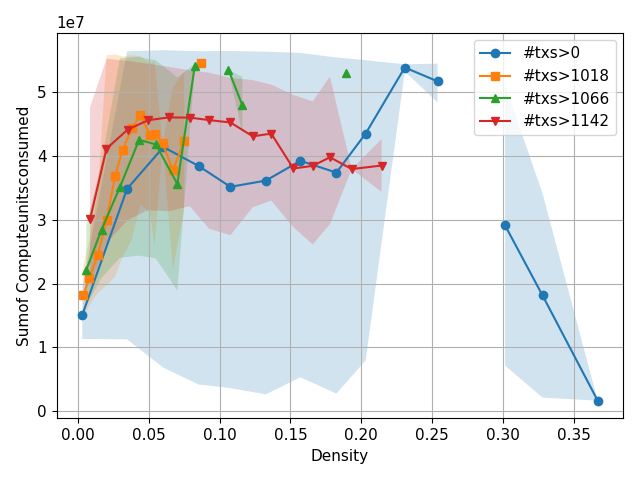}}
\caption{sumof-computeUnitsConsumed}
\label{fig:sumof-computeUnitsConsumed}
\end{figure}
\begin{figure}
\centerline{\includegraphics[width=\columnwidth]{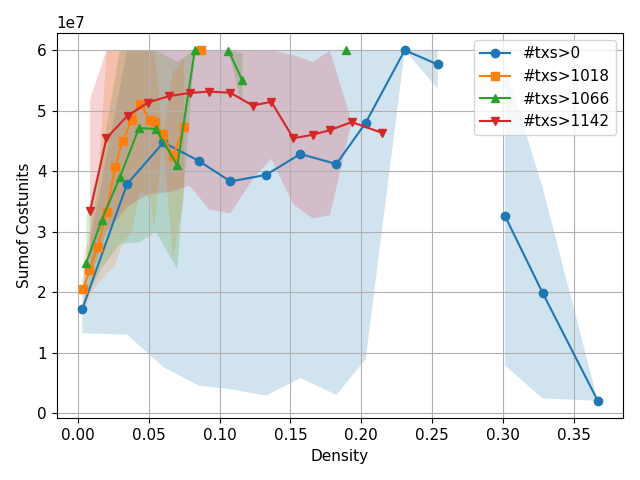}}
\caption{sumof-costUnits}
\label{fig:sumof-costUnits}
\end{figure}
\begin{figure}
\centerline{\includegraphics[width=\columnwidth]{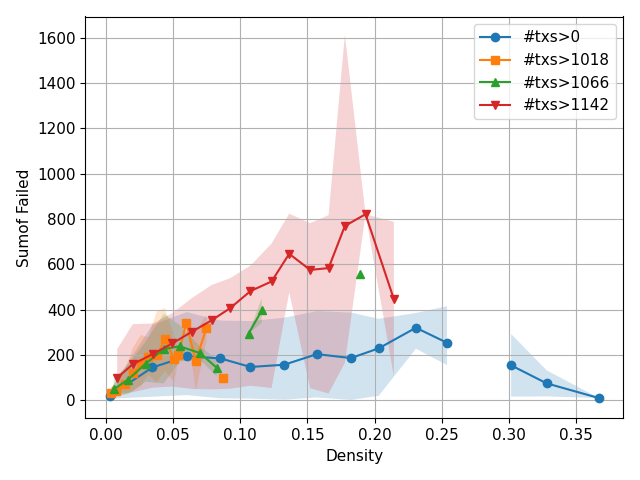}}
\caption{sumof-failed}
\label{fig:sumof-failed}
\end{figure}
\begin{figure}
\centerline{\includegraphics[width=\columnwidth]{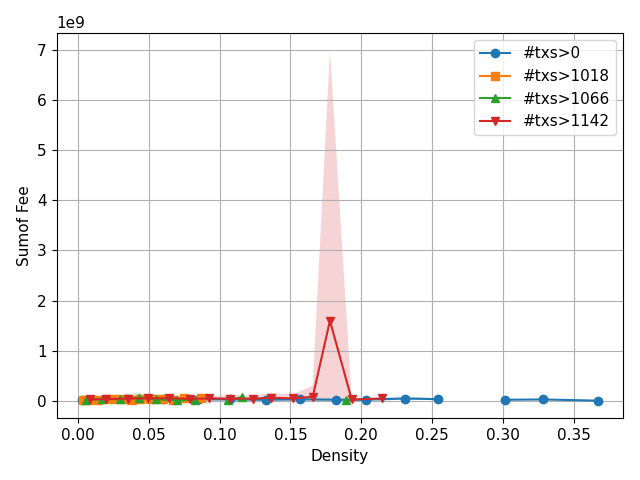}}
\caption{sumof-fee}
\label{fig:sumof-fee}
\end{figure}
\begin{figure}
\centerline{\includegraphics[width=\columnwidth]{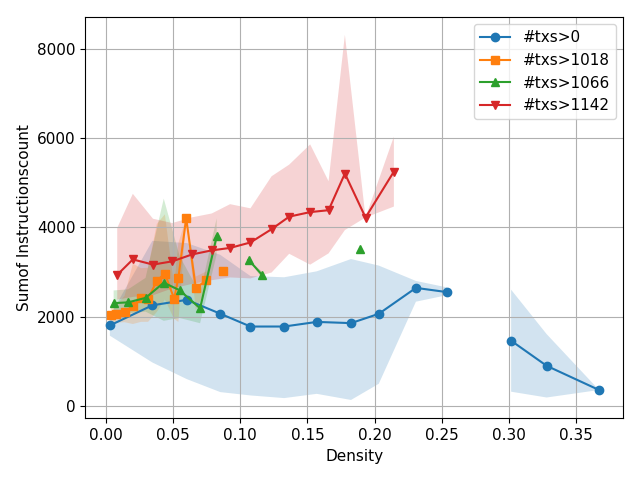}}
\caption{sumof-instructionsCount}
\label{fig:sumof-instructionsCount}
\end{figure}
\begin{figure}
\centerline{\includegraphics[width=\columnwidth]{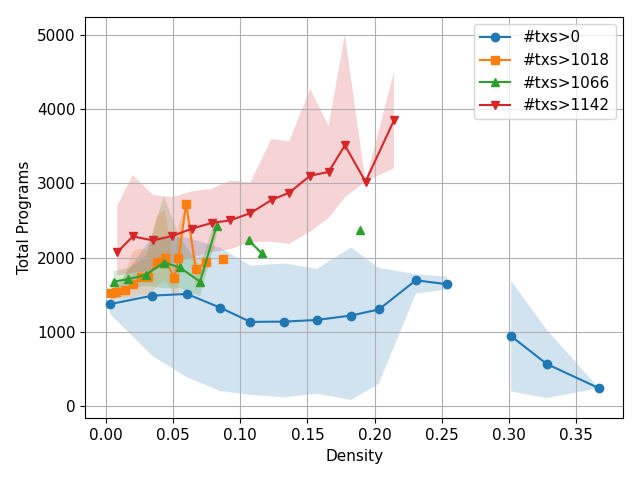}}
\caption{total-programs}
\label{fig:total-programs}
\end{figure}
\begin{figure}
\centerline{\includegraphics[width=\columnwidth]{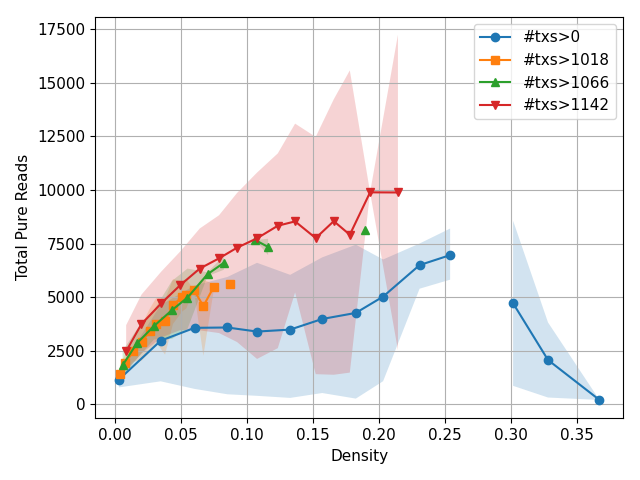}}
\caption{total-pure-reads}
\label{fig:total-pure-reads}
\end{figure}
\begin{figure}
\centerline{\includegraphics[width=\columnwidth]{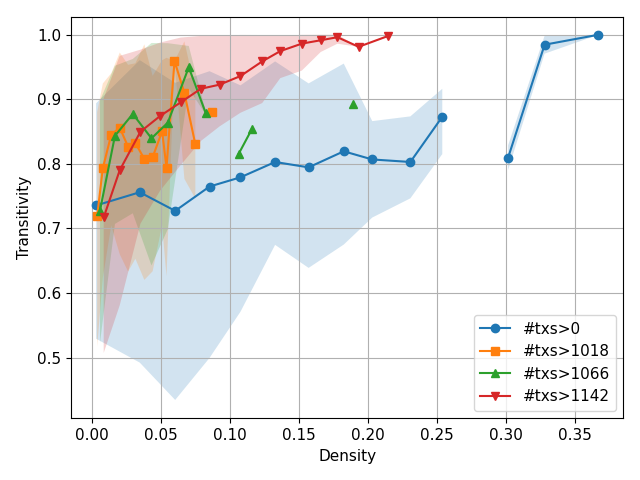}}
\caption{transitivity}
\label{fig:transitivity}
\end{figure}
\begin{figure}
\centerline{\includegraphics[width=\columnwidth]{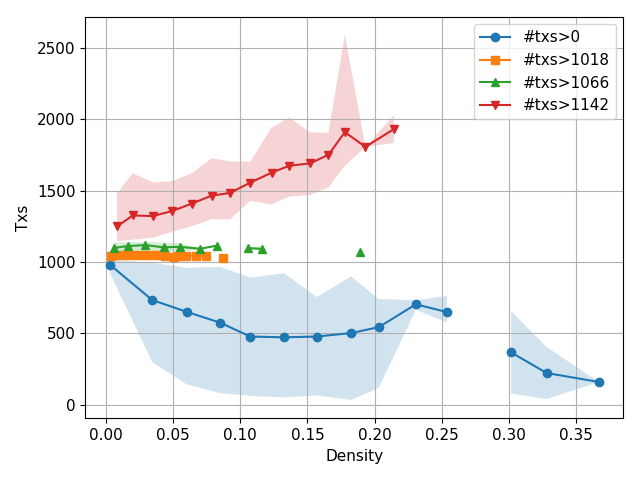}}
\caption{txs}
\label{fig:txs}
\end{figure}
\begin{figure}
\centerline{\includegraphics[width=\columnwidth]{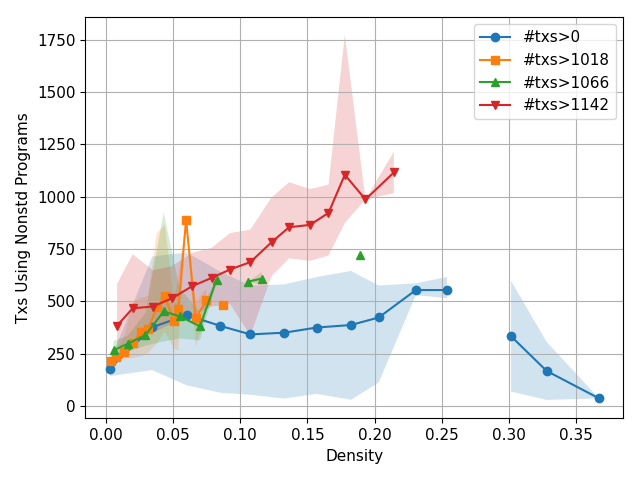}}
\caption{txs-using-nonstd-programs}
\label{fig:txs-using-nonstd-programs}
\end{figure}
\begin{figure}
\centerline{\includegraphics[width=\columnwidth]{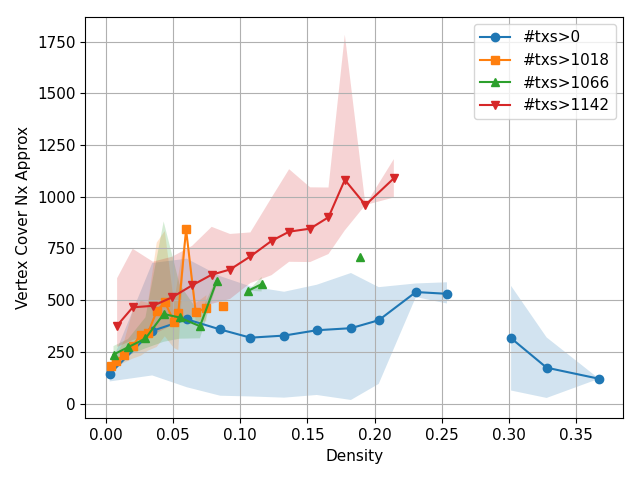}}
\caption{vertex-cover-nx-approx}
\label{fig:vertex-cover-nx-approx}
\end{figure}
\begin{figure}
\centerline{\includegraphics[width=\columnwidth]{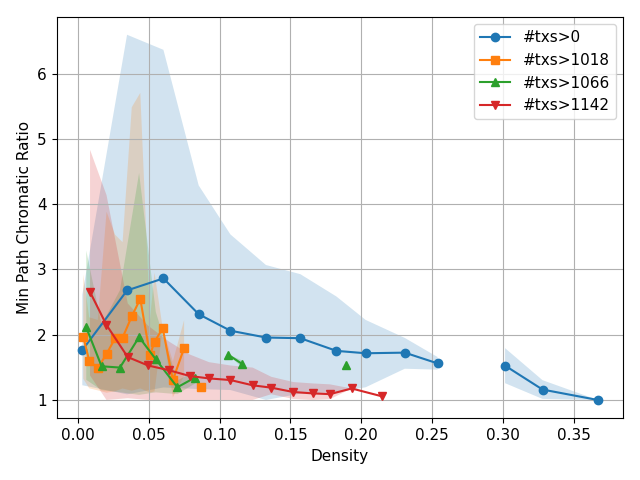}}
\caption{min-path-chromatic-ratio}
\label{fig:min-path-chromatic-ratio}
\end{figure}
\begin{figure}
\centerline{\includegraphics[width=\columnwidth]{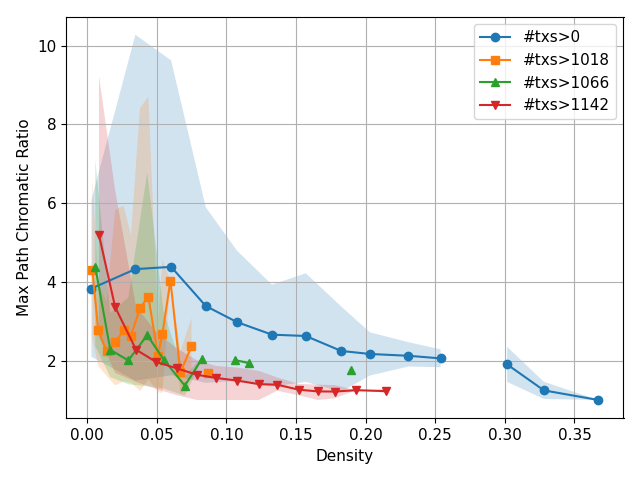}}
\caption{max-path-chromatic-ratio}
\label{fig:max-path-chromatic-ratio}
\end{figure}
\begin{figure}
\centerline{\includegraphics[width=\columnwidth]{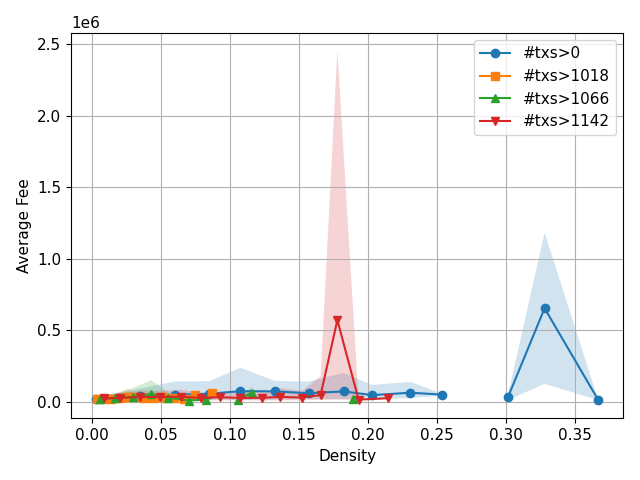}}
\caption{avg-fee}
\label{fig:avg-fee}
\end{figure}
\begin{figure}
\centerline{\includegraphics[width=\columnwidth]{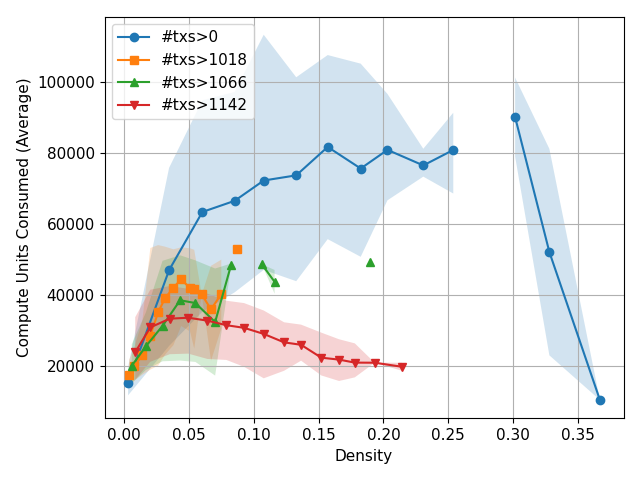}}
\caption{avg-computeUnitsConsumed}
\label{fig:avg-computeUnitsConsumed}
\end{figure}
\begin{figure}
\centerline{\includegraphics[width=\columnwidth]{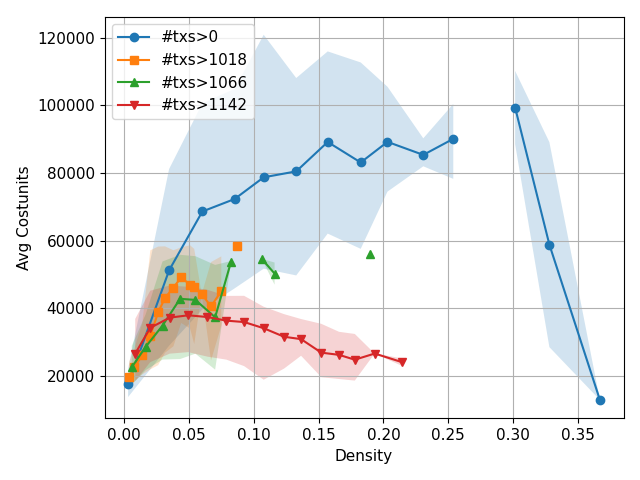}}
\caption{avg-costUnits}
\label{fig:avg-costUnits}
\end{figure}
\begin{figure}
\centerline{\includegraphics[width=\columnwidth]{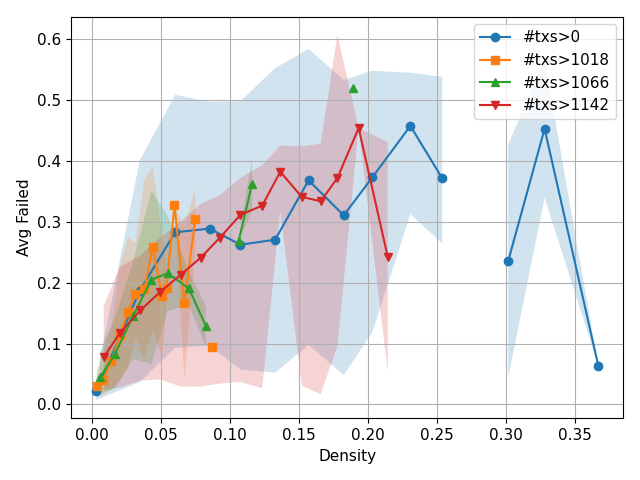}}
\caption{avg-failed}
\label{fig:avg-failed}
\end{figure}
\begin{figure}
\centerline{\includegraphics[width=\columnwidth]{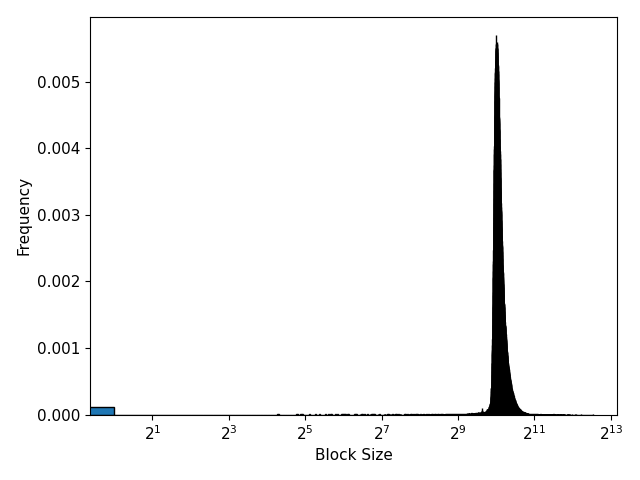}}
\caption{block-size-dist}
\label{fig:block-size-dist}
\end{figure}
\begin{figure}
\centerline{\includegraphics[width=\columnwidth]{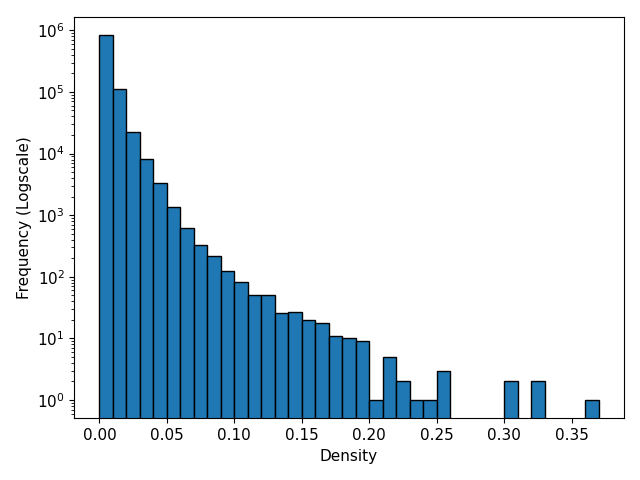}}
\caption{density-dist}
\label{fig:density-dist}
\end{figure}
\begin{figure}
\centerline{\includegraphics[width=\columnwidth]{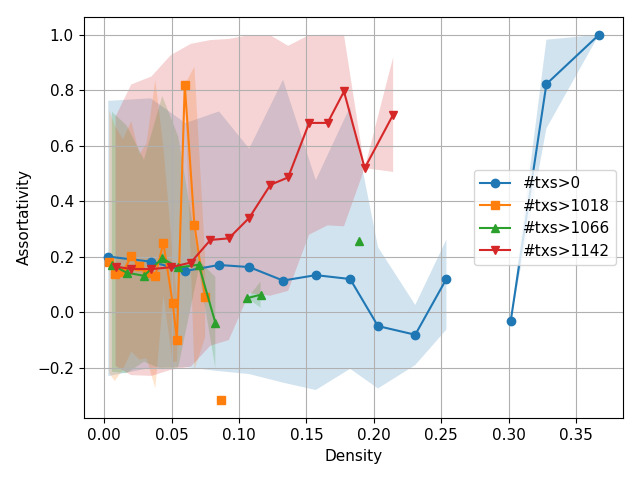}}
\caption{assortativity}
\label{fig:assortativity}
\end{figure}
\begin{figure}
\centerline{\includegraphics[width=\columnwidth]{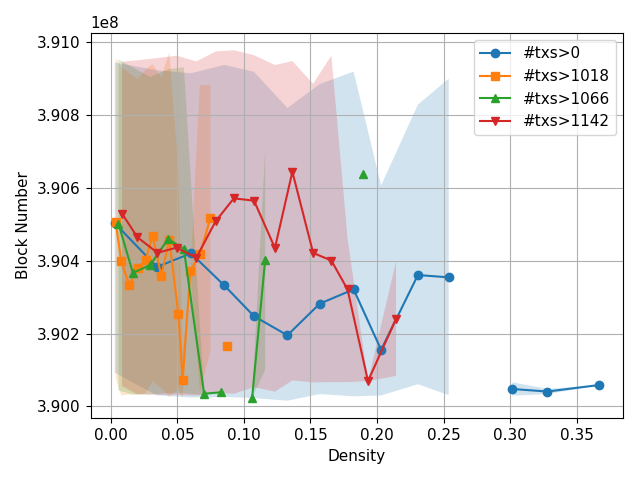}}
\caption{block-number}
\label{fig:block-number}
\end{figure}
\begin{figure}
\centerline{\includegraphics[width=\columnwidth]{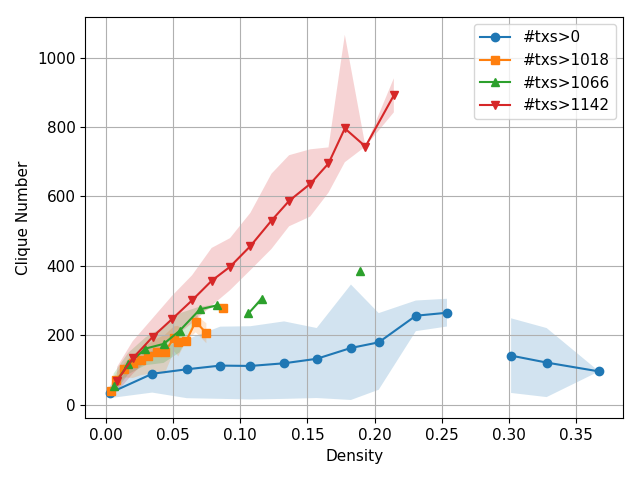}}
\caption{clique-number}
\label{fig:clique-number}
\end{figure}
\begin{figure}
\centerline{\includegraphics[width=\columnwidth]{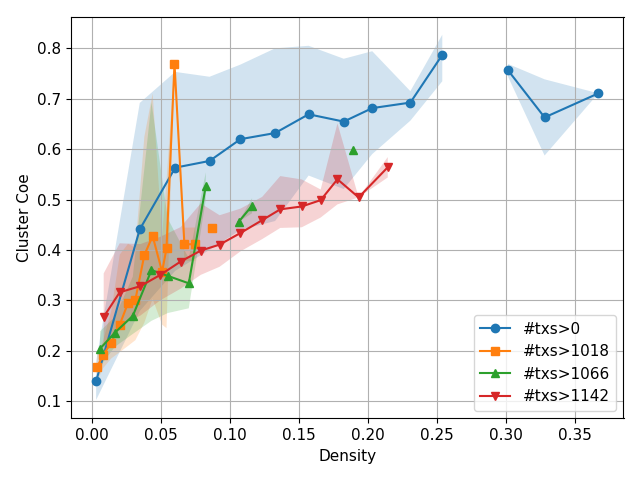}}
\caption{cluster-coe}
\label{fig:cluster-coe}
\end{figure}
\begin{figure}
\centerline{\includegraphics[width=\columnwidth]{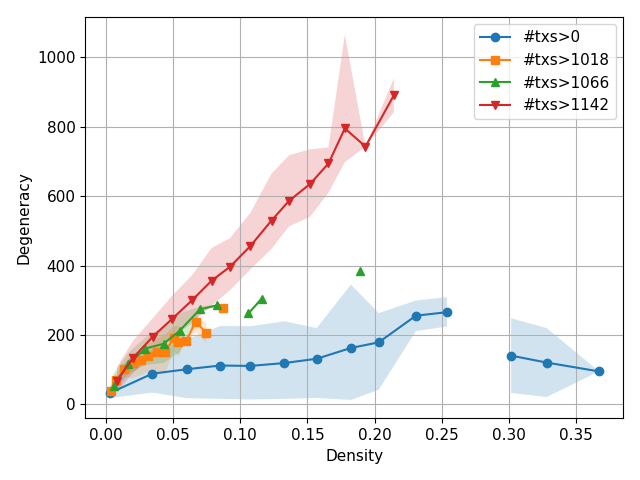}}
\caption{degeneracy}
\label{fig:degeneracy}
\end{figure}
\begin{figure}
\centerline{\includegraphics[width=\columnwidth]{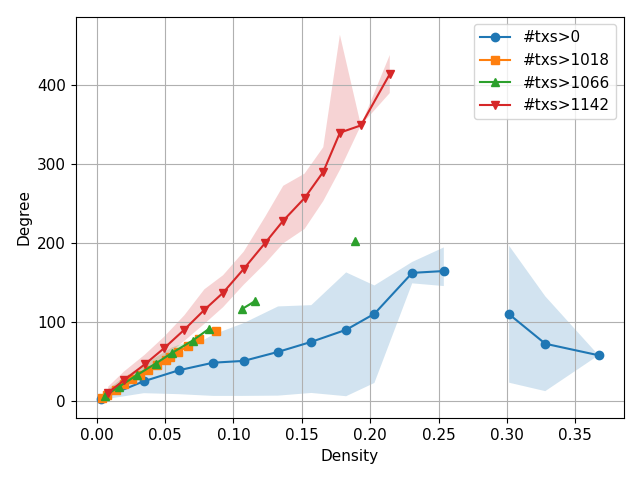}}
\caption{degree}
\label{fig:degree}
\end{figure}
\begin{figure}
\centerline{\includegraphics[width=\columnwidth]{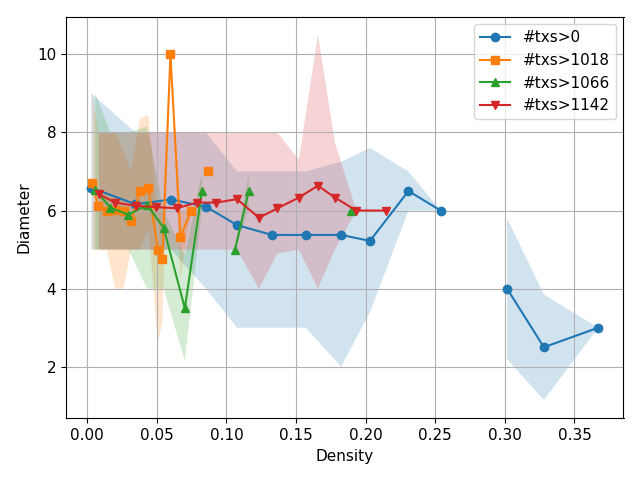}}
\caption{diameter}
\label{fig:diameter}
\end{figure}
\begin{figure}
\centerline{\includegraphics[width=\columnwidth]{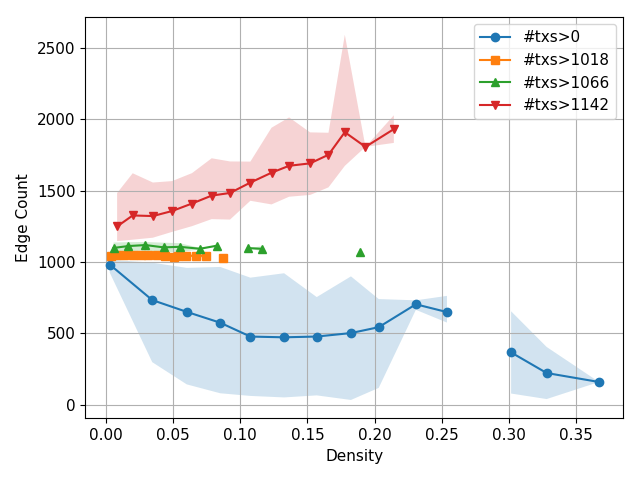}}
\caption{edge-count}
\label{fig:edge-count}
\end{figure}
\begin{figure}
\centerline{\includegraphics[width=\columnwidth]{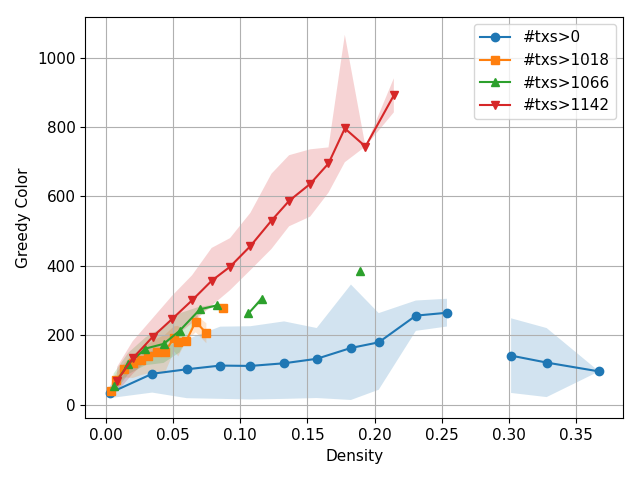}}
\caption{greedy-color}
\label{fig:greedy-color}
\end{figure}
\begin{figure}
\centerline{\includegraphics[width=\columnwidth]{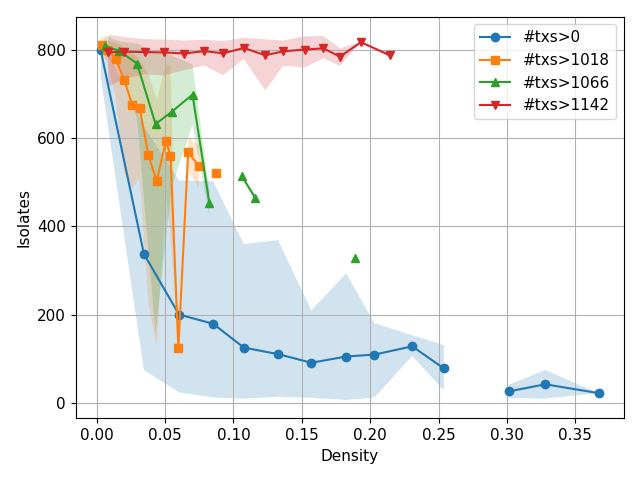}}
\caption{isolates}
\label{fig:isolates}
\end{figure}
\begin{figure}
\centerline{\includegraphics[width=\columnwidth]{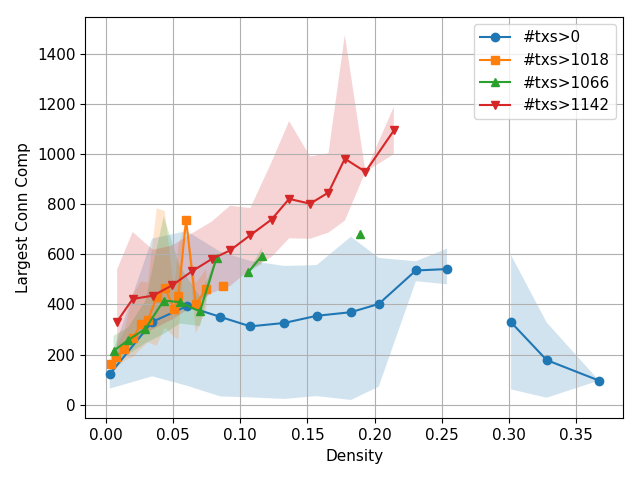}}
\caption{largest-conn-comp}
\label{fig:largest-conn-comp}
\end{figure}
\begin{figure}
\centerline{\includegraphics[width=\columnwidth]{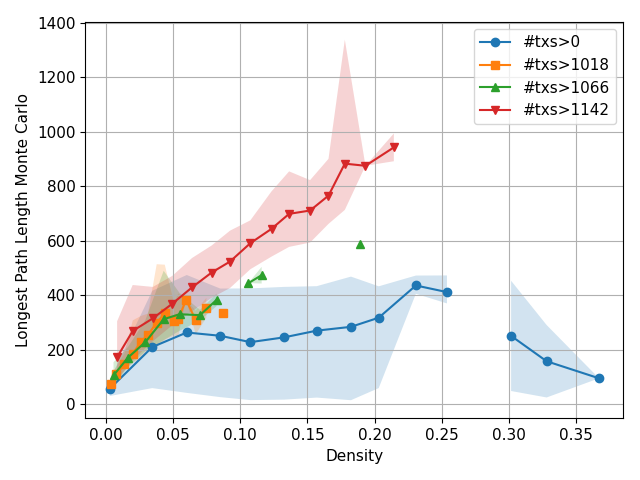}}
\caption{longest-path-length-monte-carlo}
\label{fig:longest-path-length-monte-carlo}
\end{figure}
\begin{figure}
\centerline{\includegraphics[width=\columnwidth]{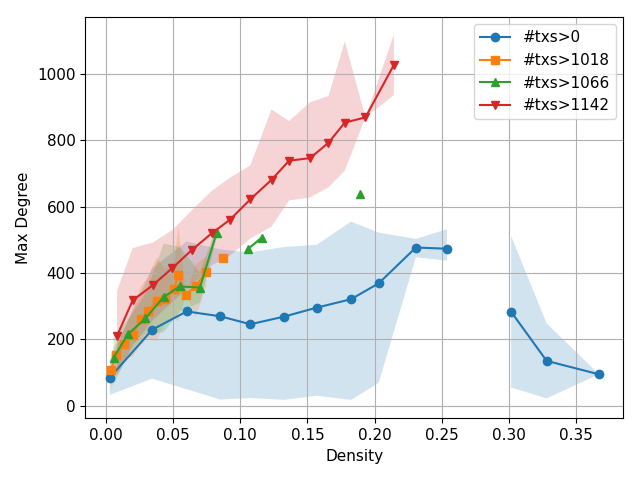}}
\caption{max-degree}
\label{fig:max-degree}
\end{figure}
\begin{figure}
\centerline{\includegraphics[width=\columnwidth]{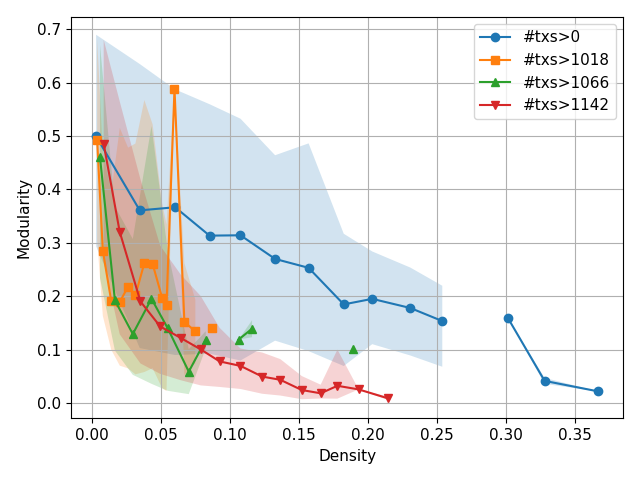}}
\caption{modularity}
\label{fig:modularity}
\end{figure}
\begin{figure}
\centerline{\includegraphics[width=\columnwidth]{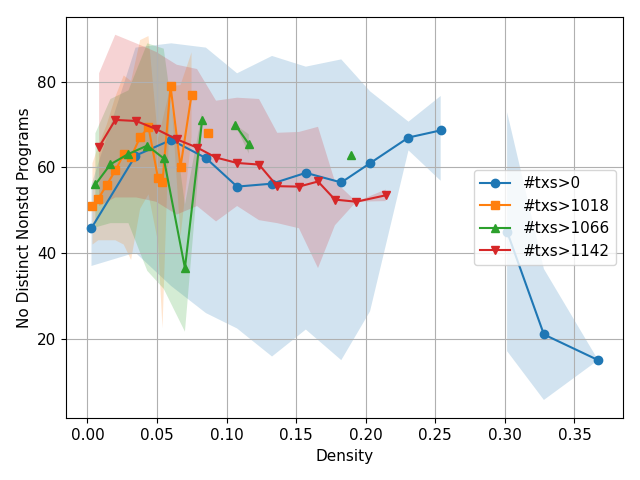}}
\caption{no-distinct-nonstd-programs}
\label{fig:no-distinct-nonstd-programs}
\end{figure}
\begin{figure}
\centerline{\includegraphics[width=\columnwidth]{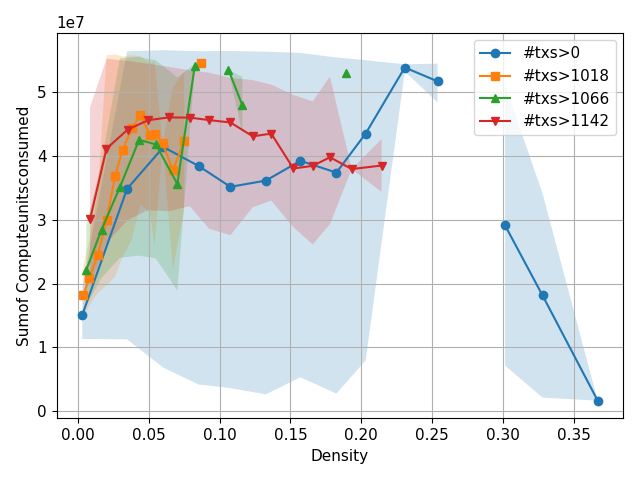}}
\caption{sumof-computeUnitsConsumed}
\label{fig:sumof-computeUnitsConsumed}
\end{figure}
\begin{figure}
\centerline{\includegraphics[width=\columnwidth]{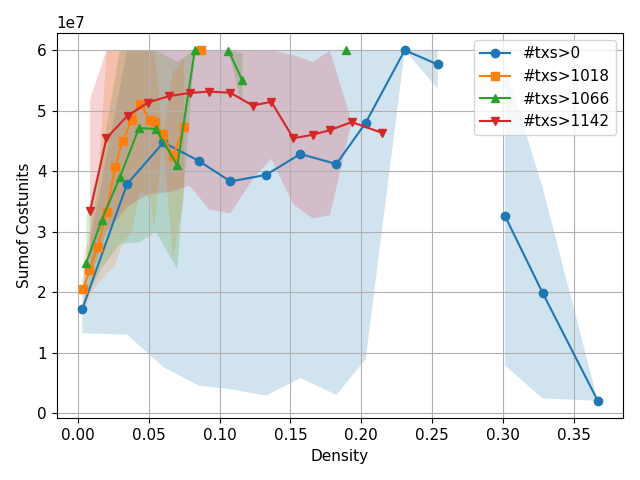}}
\caption{sumof-costUnits}
\label{fig:sumof-costUnits}
\end{figure}
\begin{figure}
\centerline{\includegraphics[width=\columnwidth]{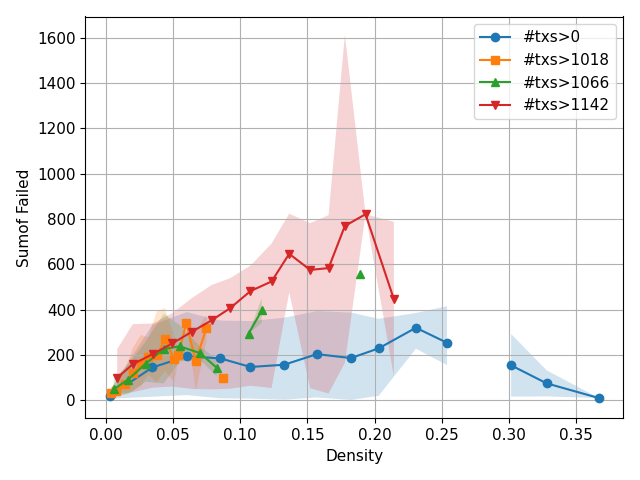}}
\caption{sumof-failed}
\label{fig:sumof-failed}
\end{figure}
\begin{figure}
\centerline{\includegraphics[width=\columnwidth]{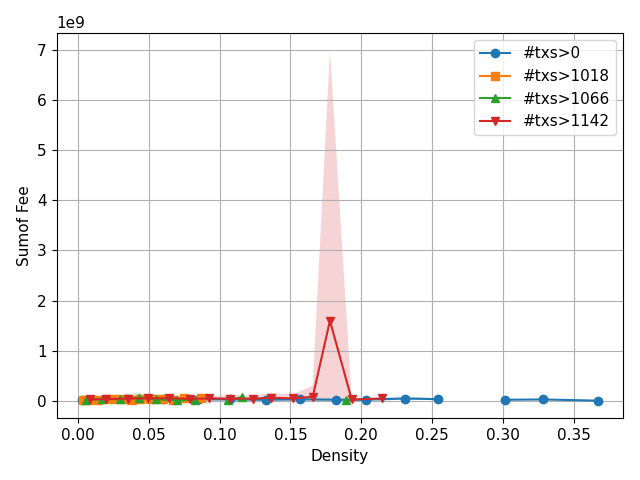}}
\caption{sumof-fee}
\label{fig:sumof-fee}
\end{figure}
\begin{figure}
\centerline{\includegraphics[width=\columnwidth]{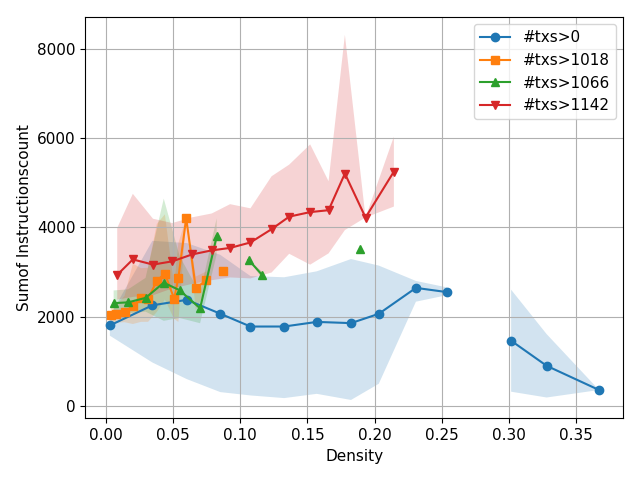}}
\caption{sumof-instructionsCount}
\label{fig:sumof-instructionsCount}
\end{figure}
\begin{figure}
\centerline{\includegraphics[width=\columnwidth]{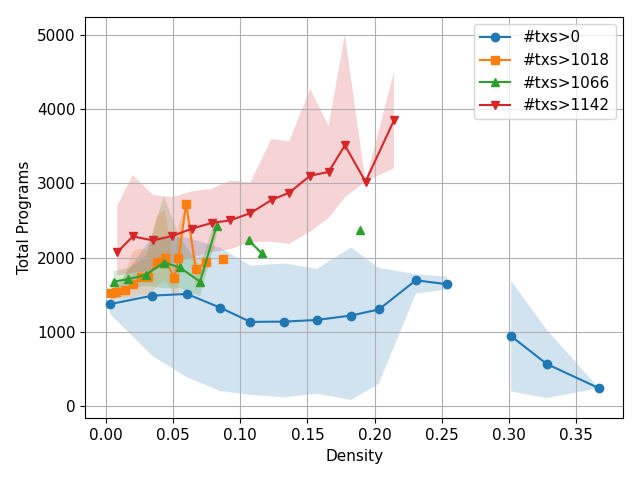}}
\caption{total-programs}
\label{fig:total-programs}
\end{figure}
\begin{figure}
\centerline{\includegraphics[width=\columnwidth]{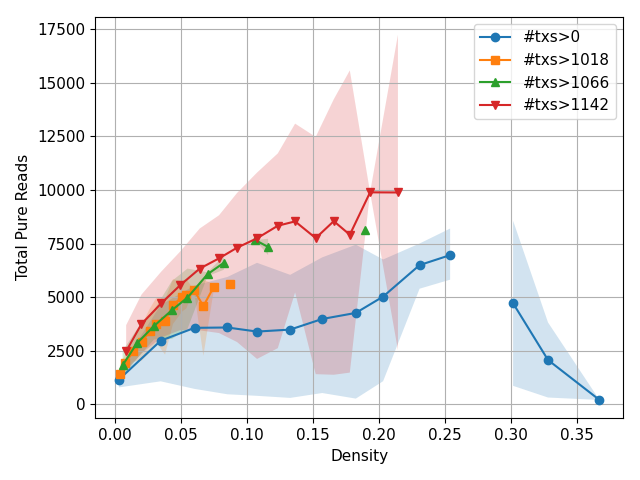}}
\caption{total-pure-reads}
\label{fig:total-pure-reads}
\end{figure}
\begin{figure}
\centerline{\includegraphics[width=\columnwidth]{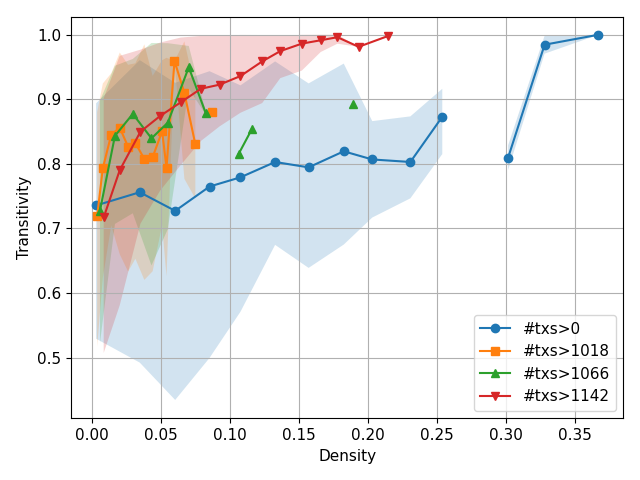}}
\caption{transitivity}
\label{fig:transitivity}
\end{figure}
\begin{figure}
\centerline{\includegraphics[width=\columnwidth]{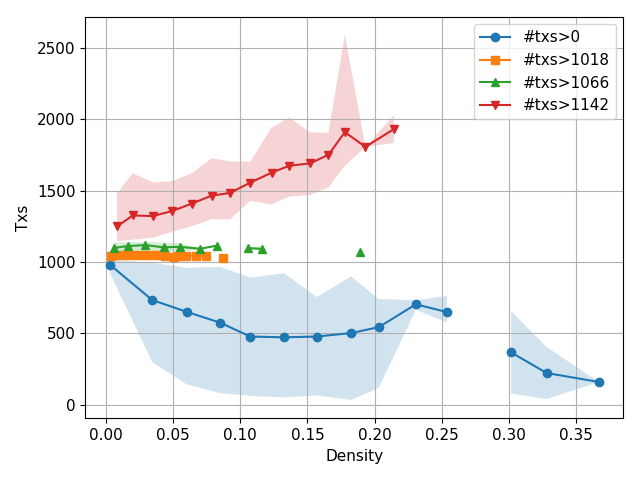}}
\caption{txs}
\label{fig:txs}
\end{figure}
\begin{figure}
\centerline{\includegraphics[width=\columnwidth]{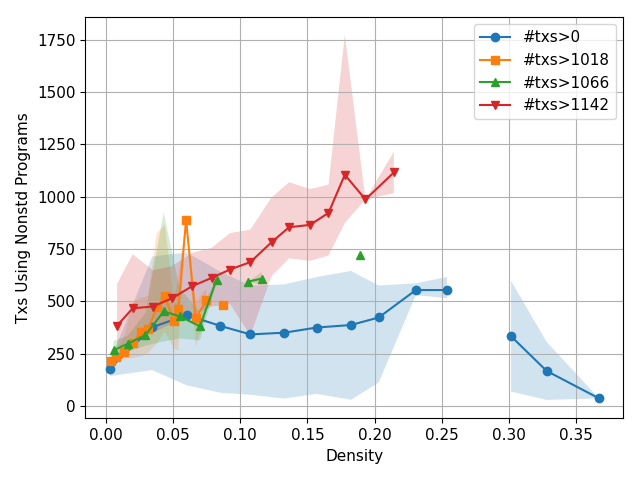}}
\caption{txs-using-nonstd-programs}
\label{fig:txs-using-nonstd-programs}
\end{figure}
\begin{figure}
\centerline{\includegraphics[width=\columnwidth]{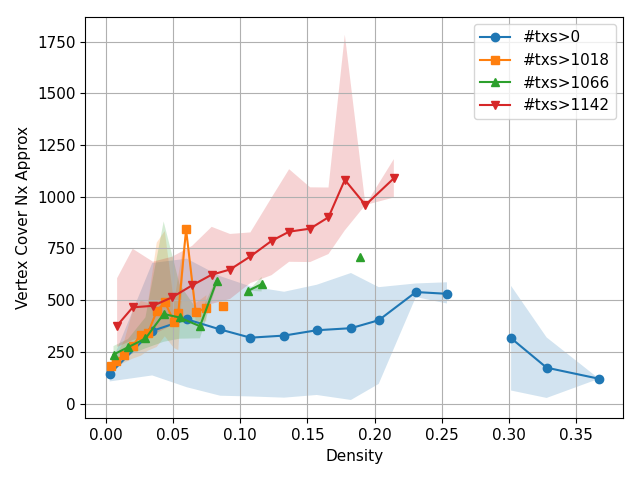}}
\caption{vertex-cover-nx-approx}
\label{fig:vertex-cover-nx-approx}
\end{figure}
\begin{figure}
\centerline{\includegraphics[width=\columnwidth]{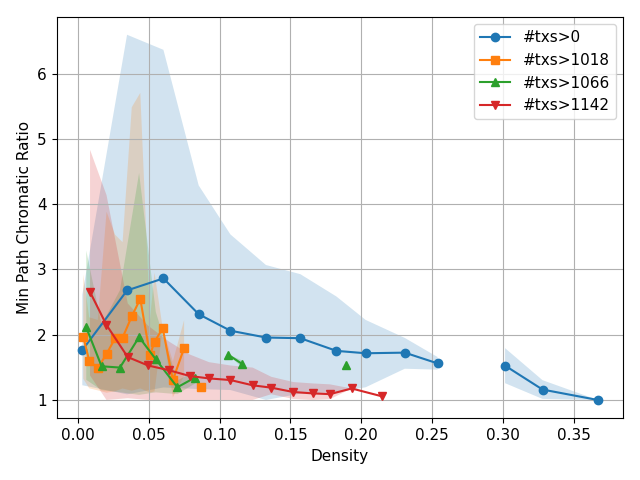}}
\caption{min-path-chromatic-ratio}
\label{fig:min-path-chromatic-ratio}
\end{figure}
\begin{figure}
\centerline{\includegraphics[width=\columnwidth]{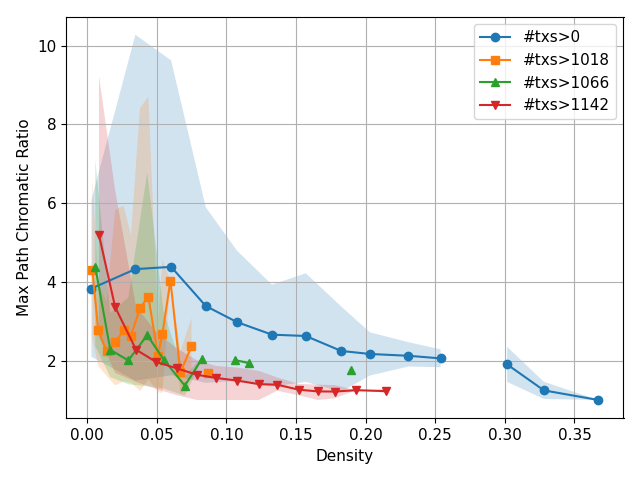}}
\caption{max-path-chromatic-ratio}
\label{fig:max-path-chromatic-ratio}
\end{figure}
\begin{figure}
\centerline{\includegraphics[width=\columnwidth]{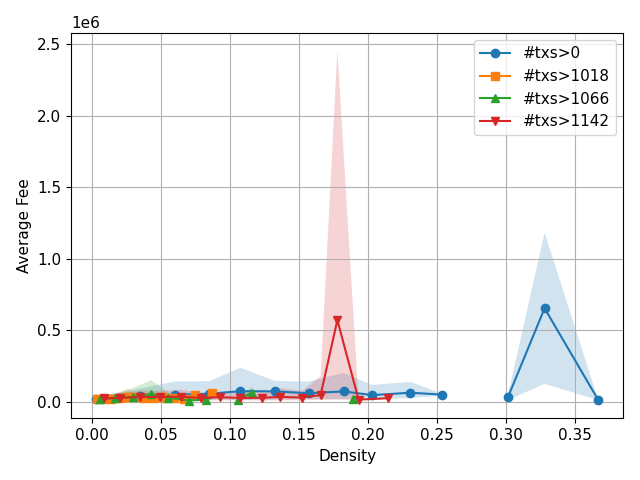}}
\caption{avg-fee}
\label{fig:avg-fee}
\end{figure}
\begin{figure}
\centerline{\includegraphics[width=\columnwidth]{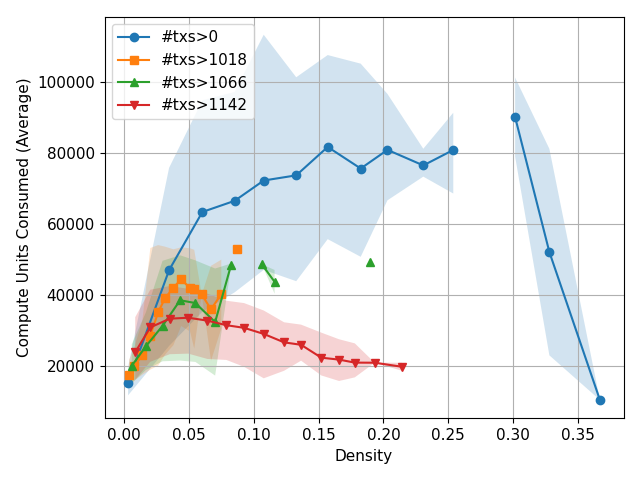}}
\caption{avg-computeUnitsConsumed}
\label{fig:avg-computeUnitsConsumed}
\end{figure}
\begin{figure}
\centerline{\includegraphics[width=\columnwidth]{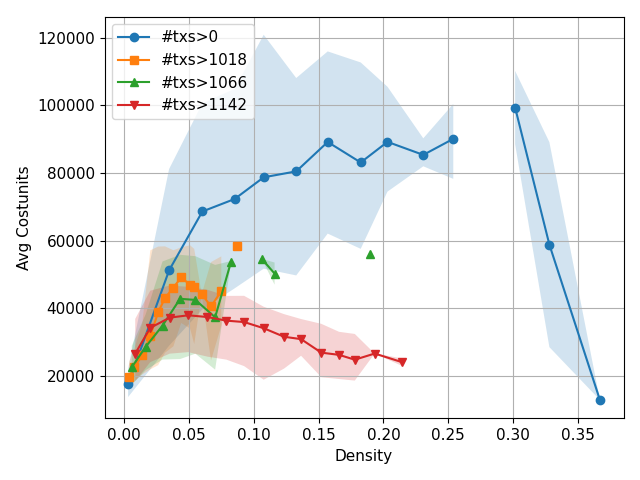}}
\caption{avg-costUnits}
\label{fig:avg-costUnits}
\end{figure}
\begin{figure}
\centerline{\includegraphics[width=\columnwidth]{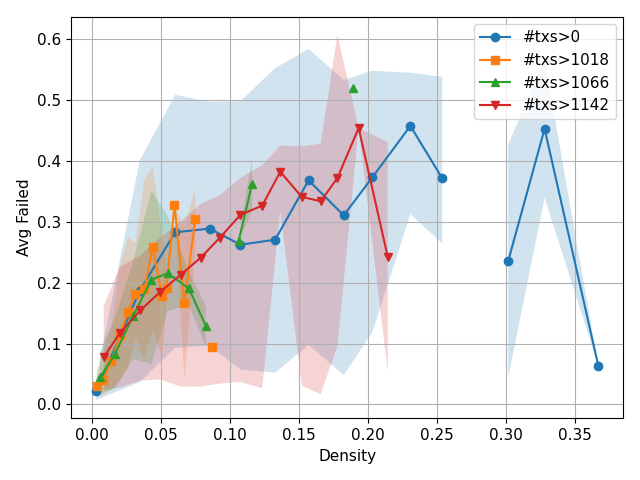}}
\caption{avg-failed}
\label{fig:avg-failed}
\end{figure}
\begin{figure}
\centerline{\includegraphics[width=\columnwidth]{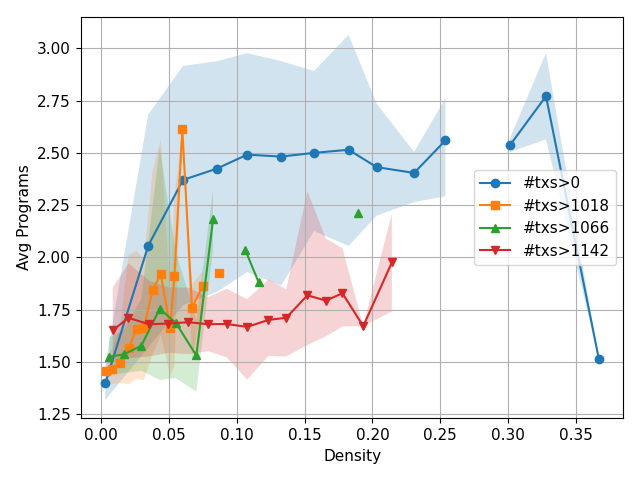}}
\caption{avg-programs}
\label{fig:avg-programs}
\end{figure}
\begin{figure}
\centerline{\includegraphics[width=\columnwidth]{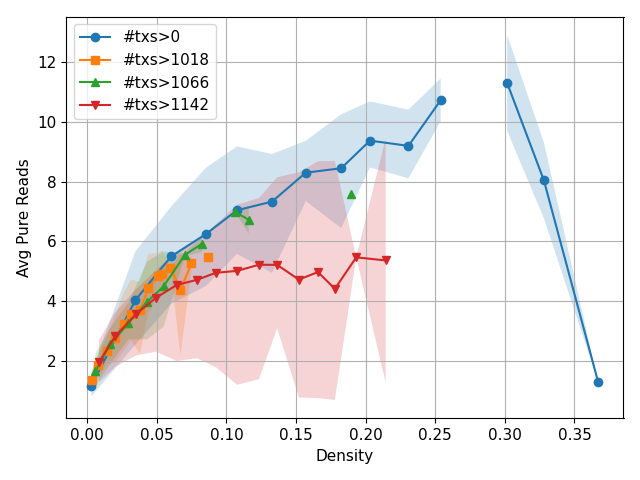}}
\caption{avg-pure-reads}
\label{fig:avg-pure-reads}
\end{figure}
\begin{figure}
\centerline{\includegraphics[width=\columnwidth]{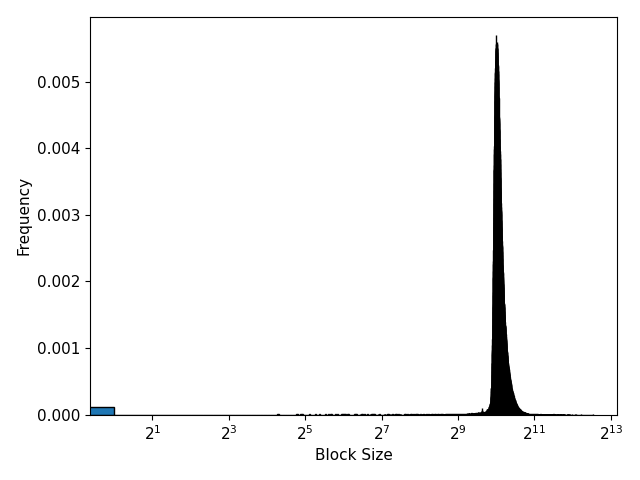}}
\caption{block-size-dist}
\label{fig:block-size-dist}
\end{figure}
\begin{figure}
\centerline{\includegraphics[width=\columnwidth]{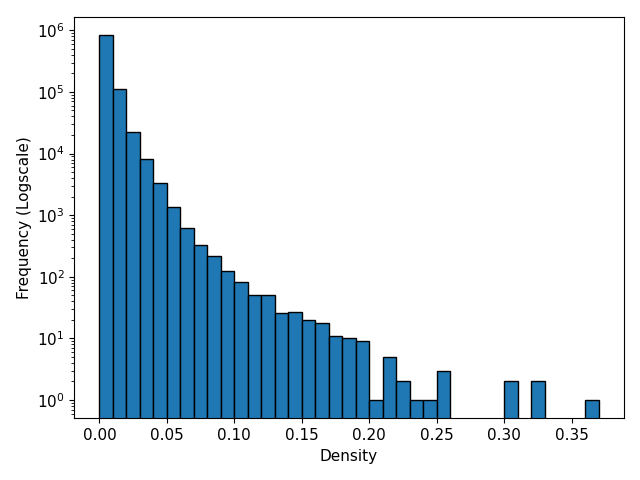}}
\caption{density-dist}
\label{fig:density-dist}
\end{figure}
\begin{figure}
\centerline{\includegraphics[width=\columnwidth]{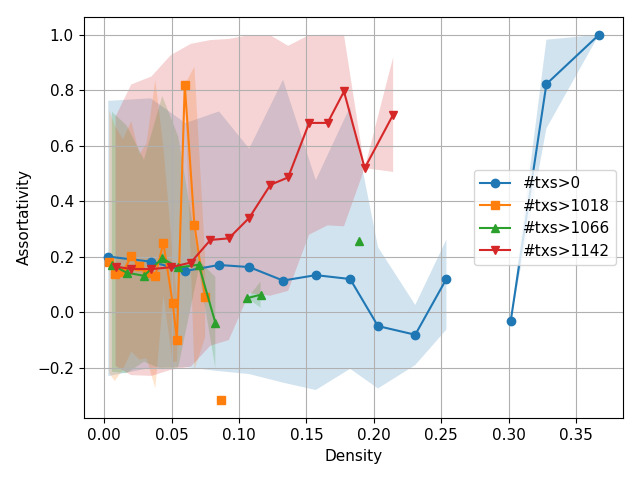}}
\caption{assortativity}
\label{fig:assortativity}
\end{figure}
\begin{figure}
\centerline{\includegraphics[width=\columnwidth]{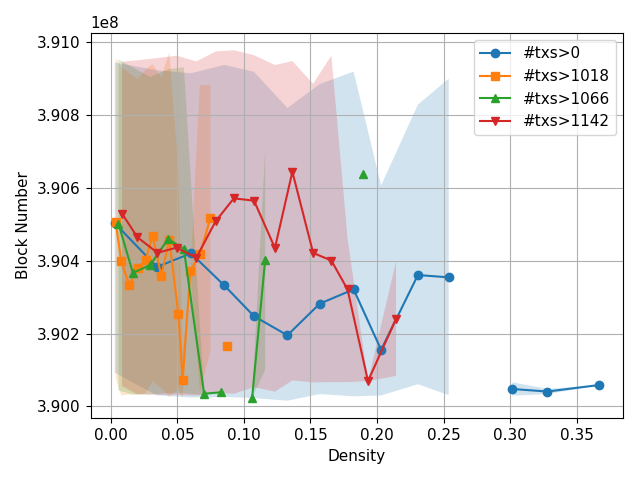}}
\caption{block-number}
\label{fig:block-number}
\end{figure}
\begin{figure}
\centerline{\includegraphics[width=\columnwidth]{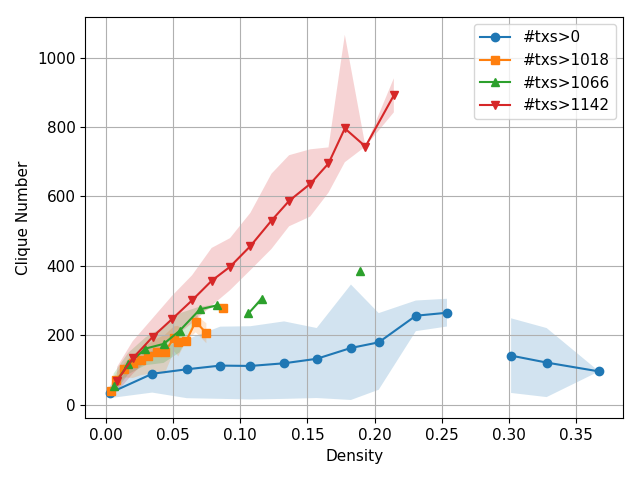}}
\caption{clique-number}
\label{fig:clique-number}
\end{figure}
\begin{figure}
\centerline{\includegraphics[width=\columnwidth]{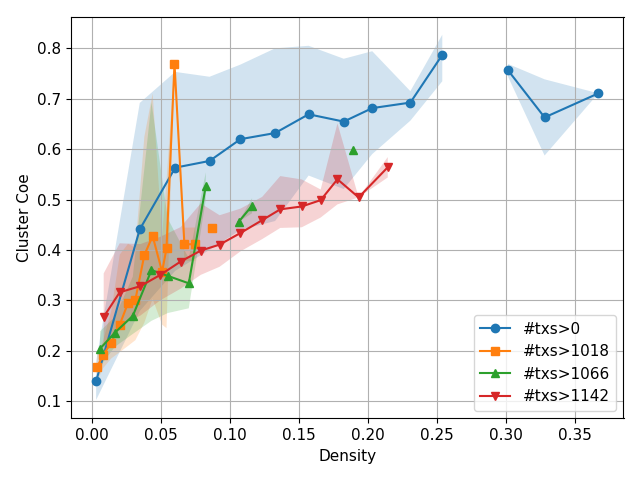}}
\caption{cluster-coe}
\label{fig:cluster-coe}
\end{figure}
\begin{figure}
\centerline{\includegraphics[width=\columnwidth]{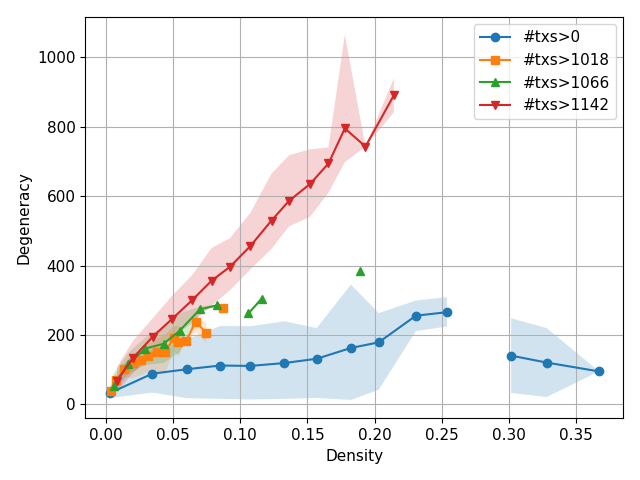}}
\caption{degeneracy}
\label{fig:degeneracy}
\end{figure}
\begin{figure}
\centerline{\includegraphics[width=\columnwidth]{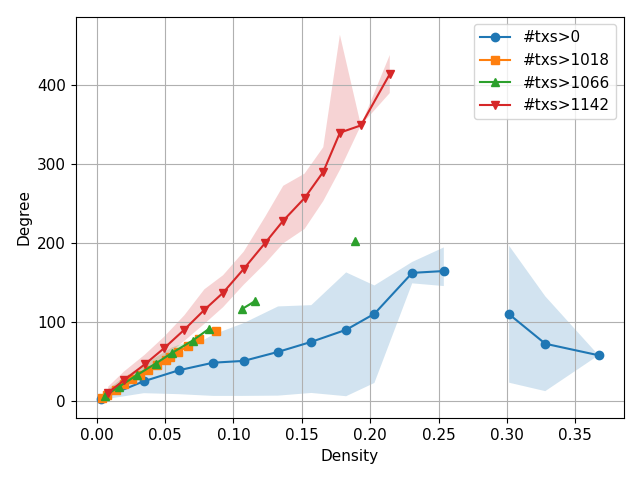}}
\caption{degree}
\label{fig:degree}
\end{figure}
\begin{figure}
\centerline{\includegraphics[width=\columnwidth]{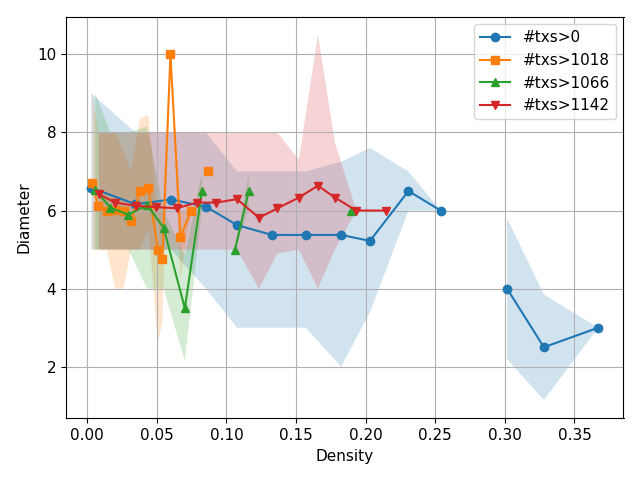}}
\caption{diameter}
\label{fig:diameter}
\end{figure}
\begin{figure}
\centerline{\includegraphics[width=\columnwidth]{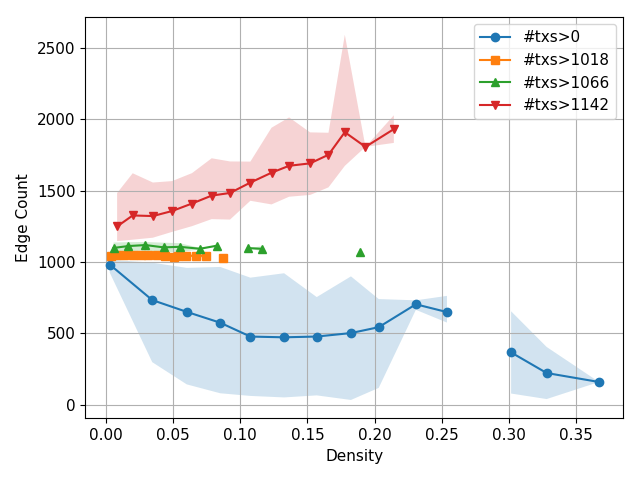}}
\caption{edge-count}
\label{fig:edge-count}
\end{figure}
\begin{figure}
\centerline{\includegraphics[width=\columnwidth]{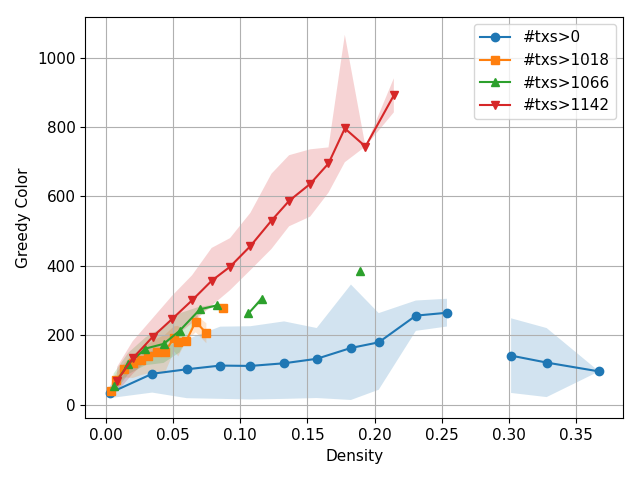}}
\caption{greedy-color}
\label{fig:greedy-color}
\end{figure}
\begin{figure}
\centerline{\includegraphics[width=\columnwidth]{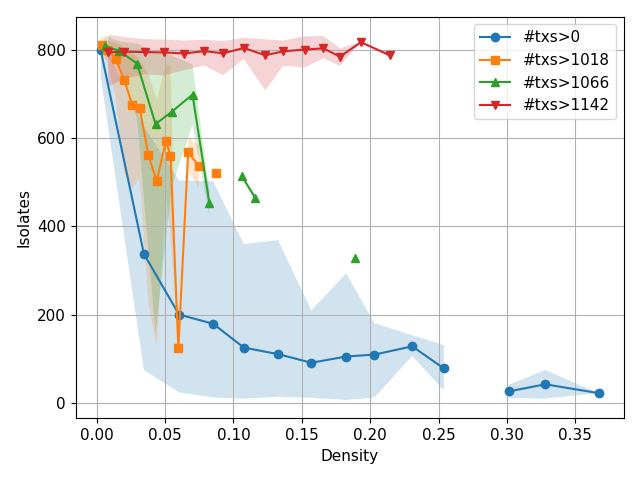}}
\caption{isolates}
\label{fig:isolates}
\end{figure}
\begin{figure}
\centerline{\includegraphics[width=\columnwidth]{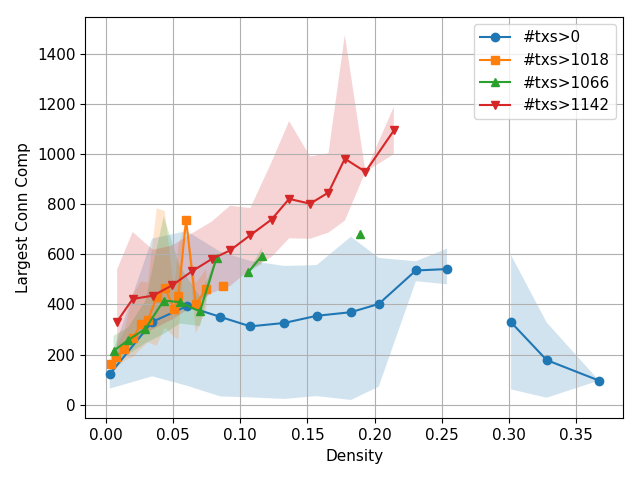}}
\caption{largest-conn-comp}
\label{fig:largest-conn-comp}
\end{figure}
\begin{figure}
\centerline{\includegraphics[width=\columnwidth]{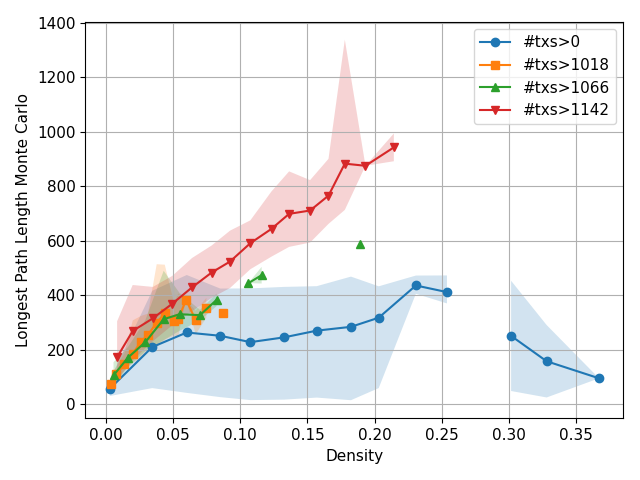}}
\caption{longest-path-length-monte-carlo}
\label{fig:longest-path-length-monte-carlo}
\end{figure}
\begin{figure}
\centerline{\includegraphics[width=\columnwidth]{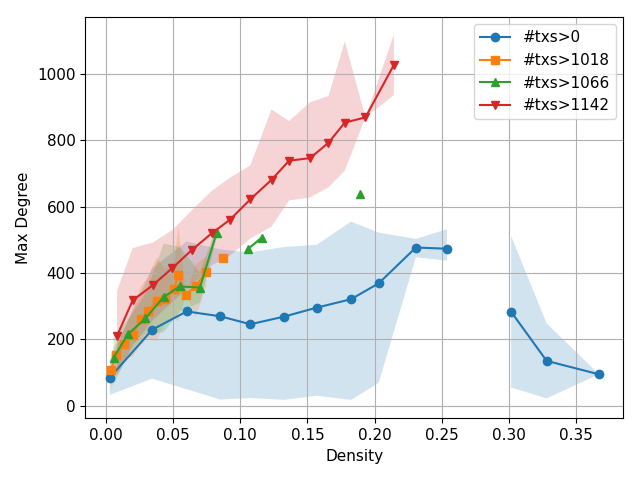}}
\caption{max-degree}
\label{fig:max-degree}
\end{figure}
\begin{figure}
\centerline{\includegraphics[width=\columnwidth]{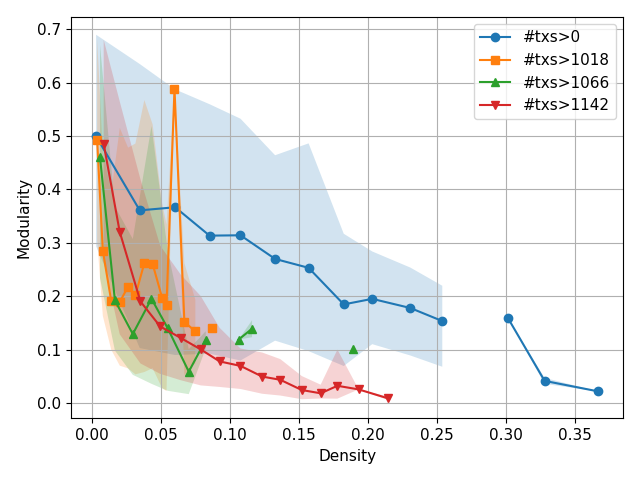}}
\caption{modularity}
\label{fig:modularity}
\end{figure}
\begin{figure}
\centerline{\includegraphics[width=\columnwidth]{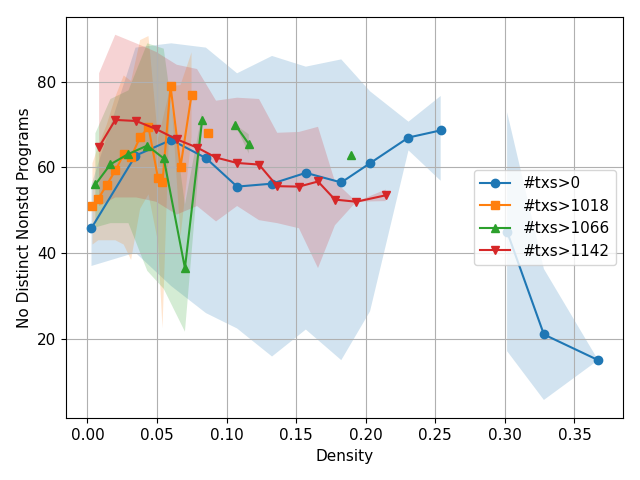}}
\caption{no-distinct-nonstd-programs}
\label{fig:no-distinct-nonstd-programs}
\end{figure}
\begin{figure}
\centerline{\includegraphics[width=\columnwidth]{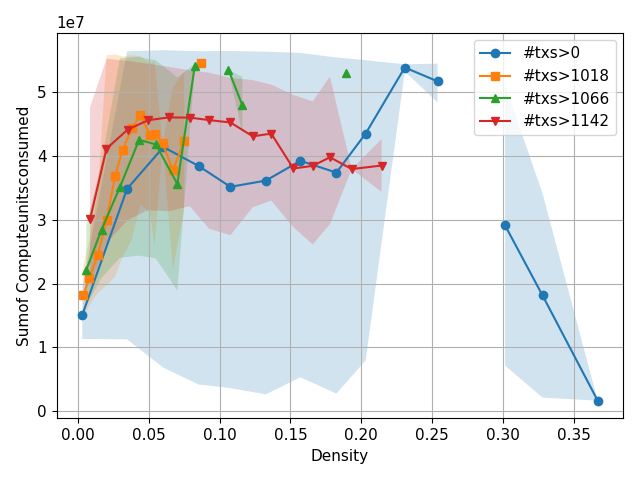}}
\caption{sumof-computeUnitsConsumed}
\label{fig:sumof-computeUnitsConsumed}
\end{figure}
\begin{figure}
\centerline{\includegraphics[width=\columnwidth]{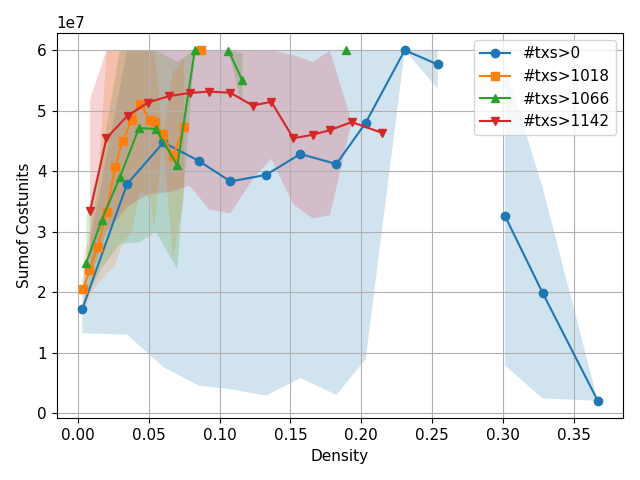}}
\caption{sumof-costUnits}
\label{fig:sumof-costUnits}
\end{figure}
\begin{figure}
\centerline{\includegraphics[width=\columnwidth]{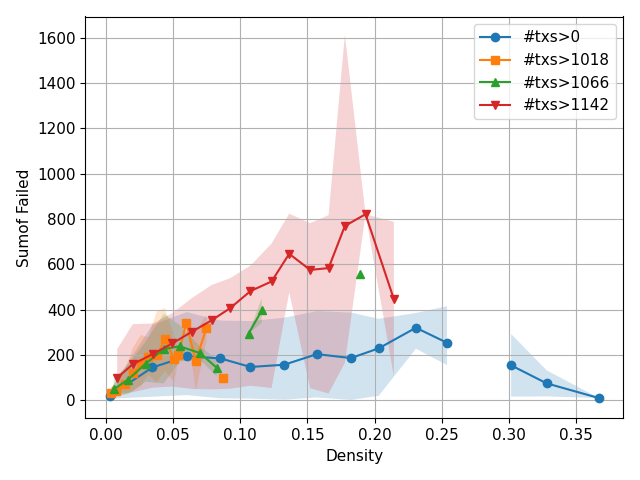}}
\caption{sumof-failed}
\label{fig:sumof-failed}
\end{figure}
\begin{figure}
\centerline{\includegraphics[width=\columnwidth]{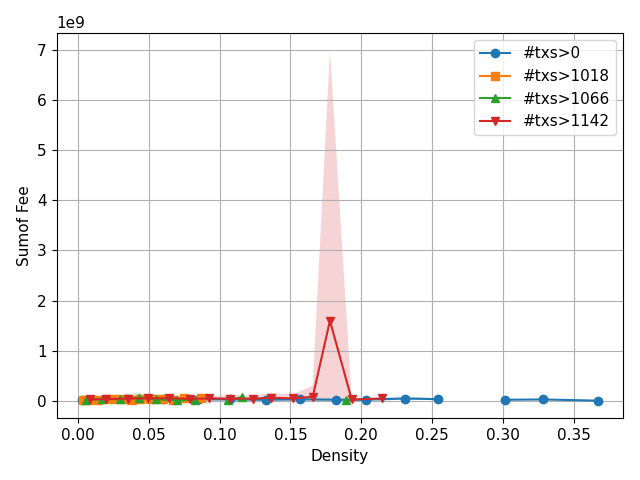}}
\caption{sumof-fee}
\label{fig:sumof-fee}
\end{figure}
\begin{figure}
\centerline{\includegraphics[width=\columnwidth]{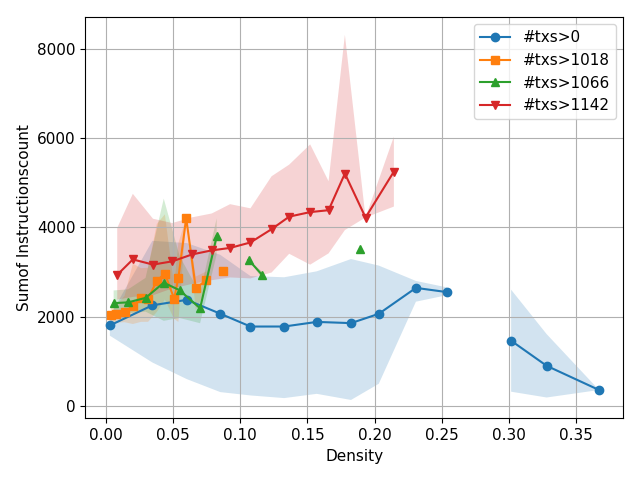}}
\caption{sumof-instructionsCount}
\label{fig:sumof-instructionsCount}
\end{figure}
\begin{figure}
\centerline{\includegraphics[width=\columnwidth]{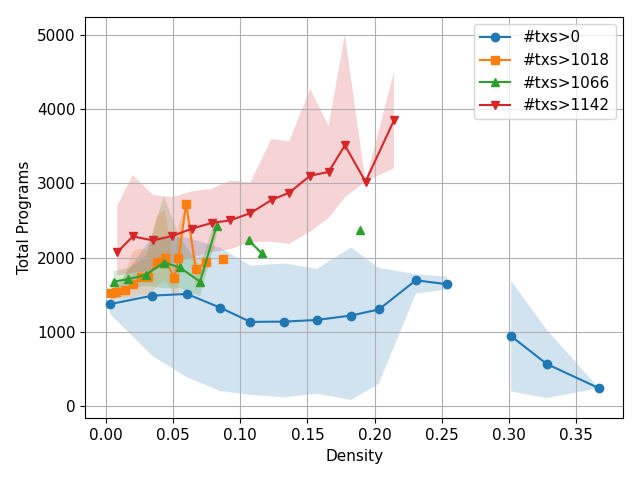}}
\caption{total-programs}
\label{fig:total-programs}
\end{figure}
\begin{figure}
\centerline{\includegraphics[width=\columnwidth]{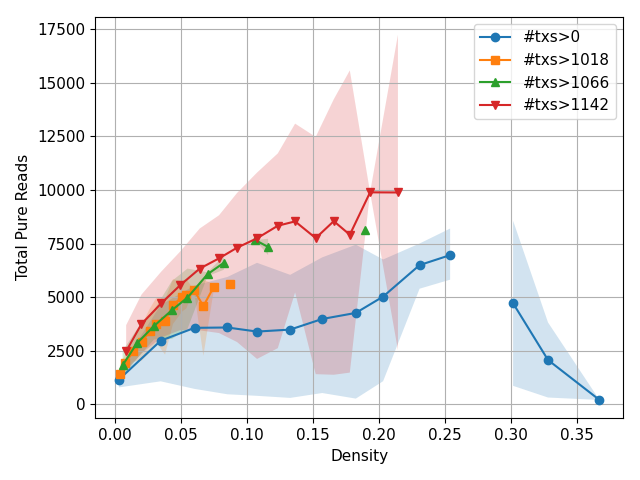}}
\caption{total-pure-reads}
\label{fig:total-pure-reads}
\end{figure}
\begin{figure}
\centerline{\includegraphics[width=\columnwidth]{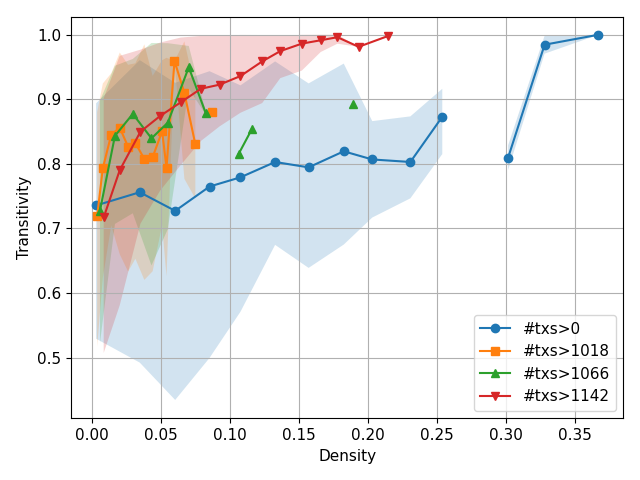}}
\caption{transitivity}
\label{fig:transitivity}
\end{figure}
\begin{figure}
\centerline{\includegraphics[width=\columnwidth]{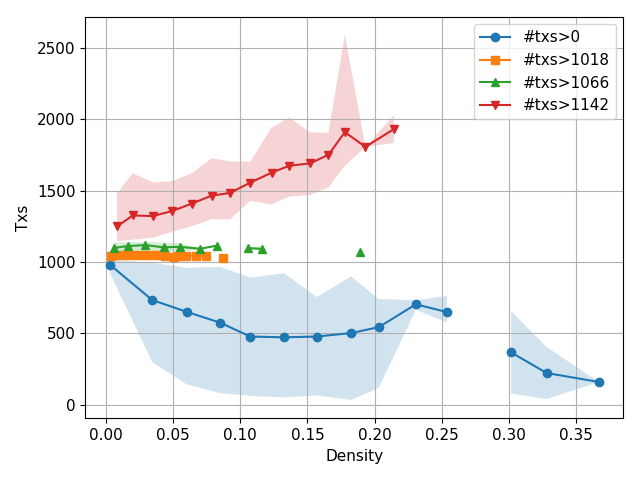}}
\caption{txs}
\label{fig:txs}
\end{figure}
\begin{figure}
\centerline{\includegraphics[width=\columnwidth]{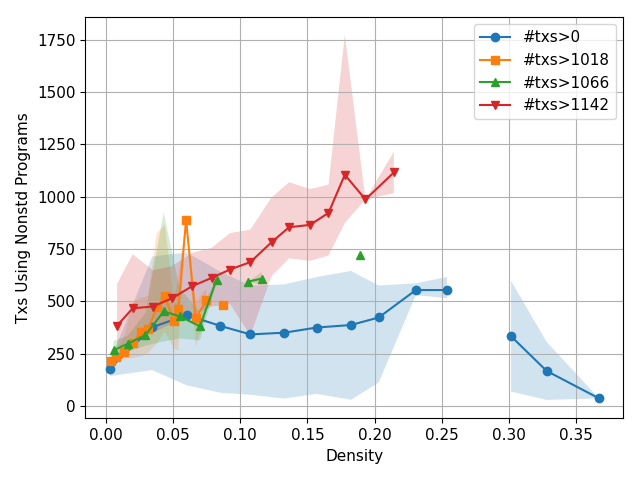}}
\caption{txs-using-nonstd-programs}
\label{fig:txs-using-nonstd-programs}
\end{figure}
\begin{figure}
\centerline{\includegraphics[width=\columnwidth]{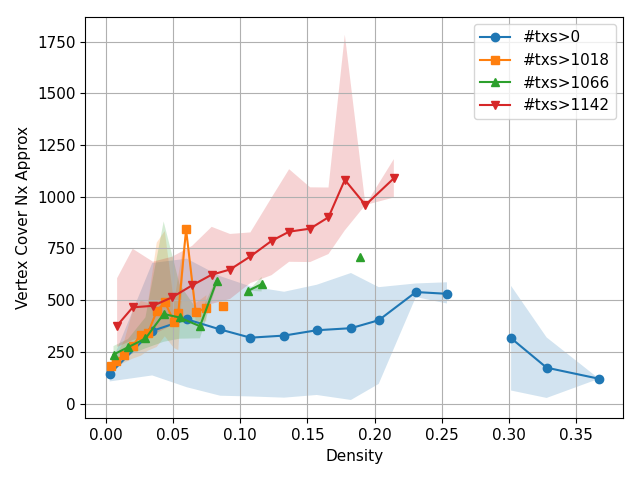}}
\caption{vertex-cover-nx-approx}
\label{fig:vertex-cover-nx-approx}
\end{figure}
\begin{figure}
\centerline{\includegraphics[width=\columnwidth]{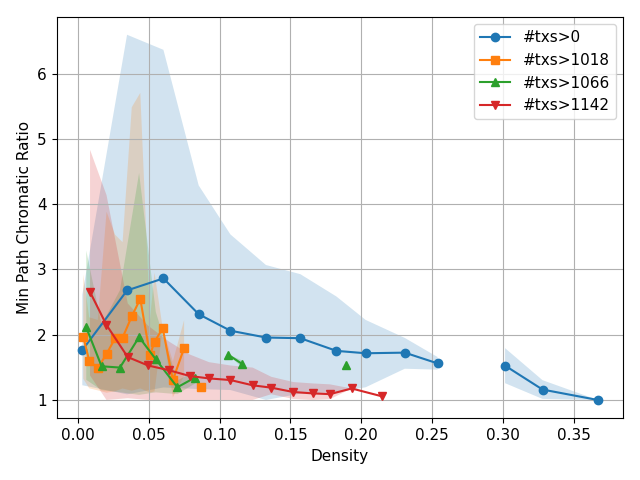}}
\caption{min-path-chromatic-ratio}
\label{fig:min-path-chromatic-ratio}
\end{figure}
\begin{figure}
\centerline{\includegraphics[width=\columnwidth]{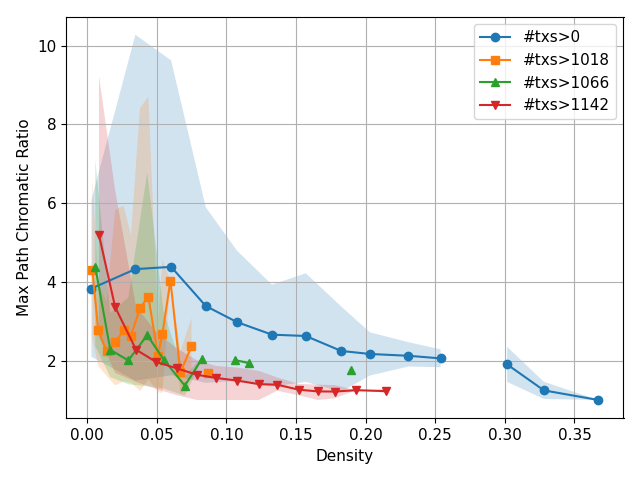}}
\caption{max-path-chromatic-ratio}
\label{fig:max-path-chromatic-ratio}
\end{figure}
\begin{figure}
\centerline{\includegraphics[width=\columnwidth]{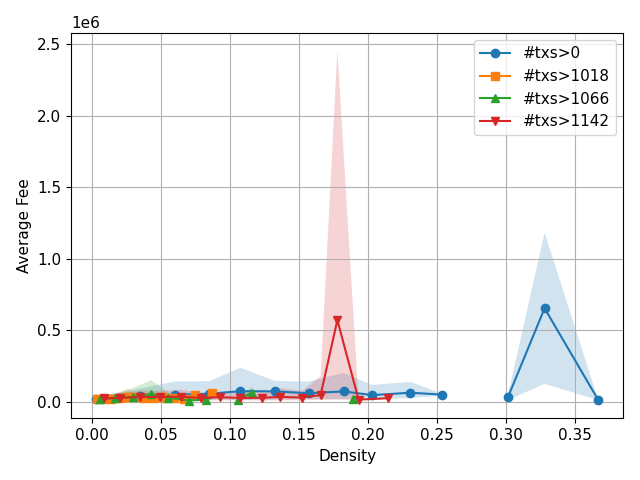}}
\caption{avg-fee}
\label{fig:avg-fee}
\end{figure}
\begin{figure}
\centerline{\includegraphics[width=\columnwidth]{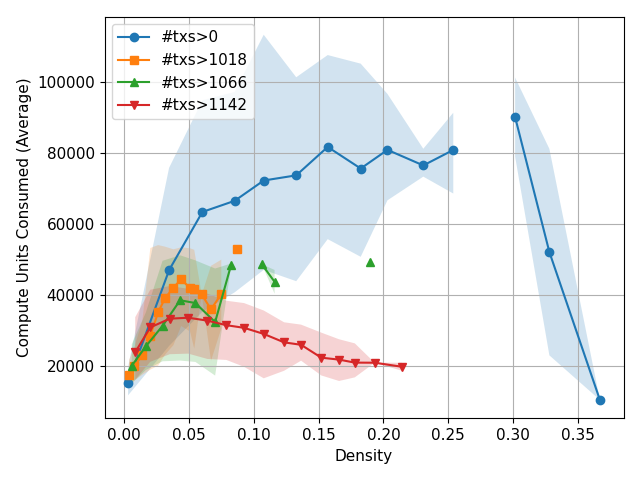}}
\caption{avg-computeUnitsConsumed}
\label{fig:avg-computeUnitsConsumed}
\end{figure}
\begin{figure}
\centerline{\includegraphics[width=\columnwidth]{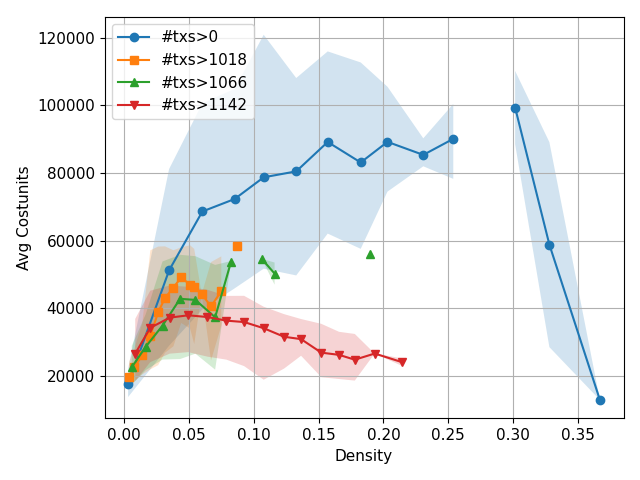}}
\caption{avg-costUnits}
\label{fig:avg-costUnits}
\end{figure}
\begin{figure}
\centerline{\includegraphics[width=\columnwidth]{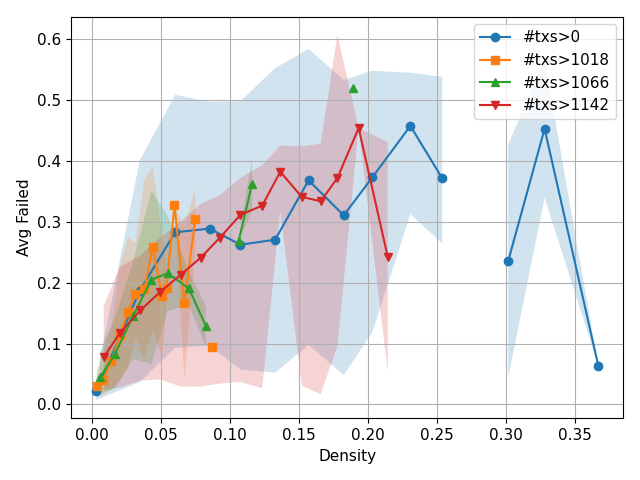}}
\caption{avg-failed}
\label{fig:avg-failed}
\end{figure}
\begin{figure}
\centerline{\includegraphics[width=\columnwidth]{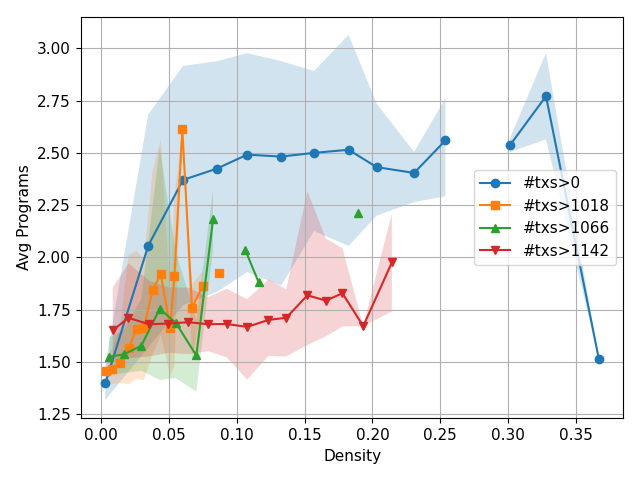}}
\caption{avg-programs}
\label{fig:avg-programs}
\end{figure}
\begin{figure}
\centerline{\includegraphics[width=\columnwidth]{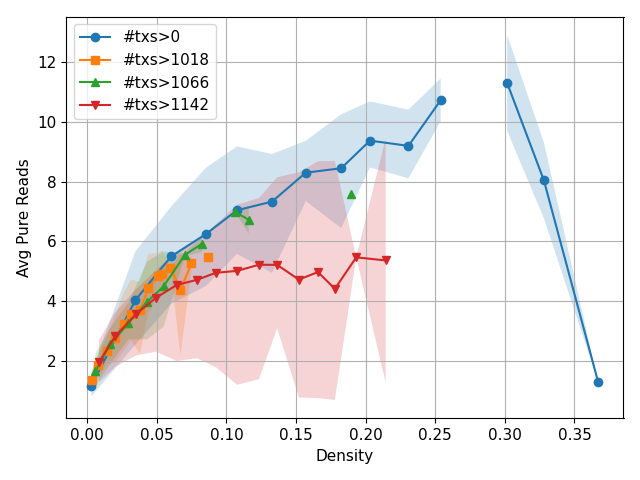}}
\caption{avg-pure-reads}
\label{fig:avg-pure-reads}
\end{figure}
\begin{figure}
\centerline{\includegraphics[width=\columnwidth]{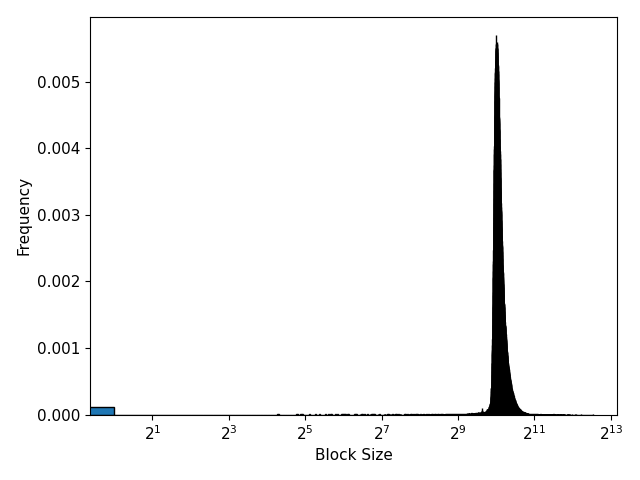}}
\caption{block-size-dist}
\label{fig:block-size-dist}
\end{figure}
\begin{figure}
\centerline{\includegraphics[width=\columnwidth]{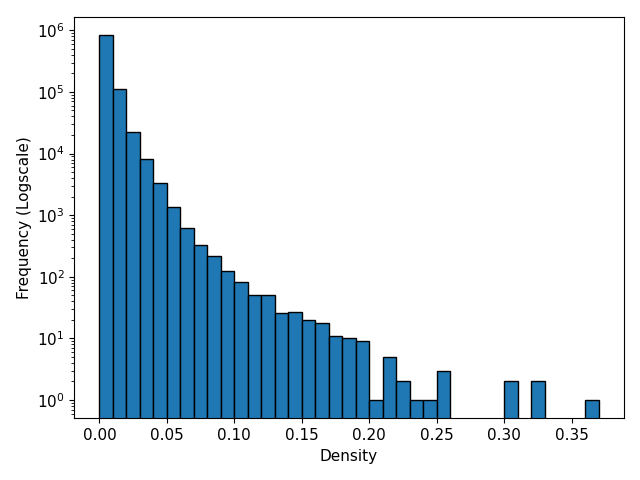}}
\caption{density-dist}
\label{fig:density-dist}
\end{figure}
\begin{figure}
\centerline{\includegraphics[width=\columnwidth]{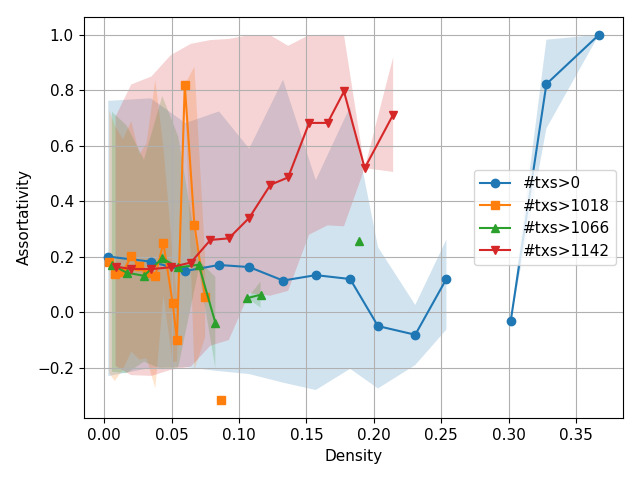}}
\caption{assortativity}
\label{fig:assortativity}
\end{figure}
\begin{figure}
\centerline{\includegraphics[width=\columnwidth]{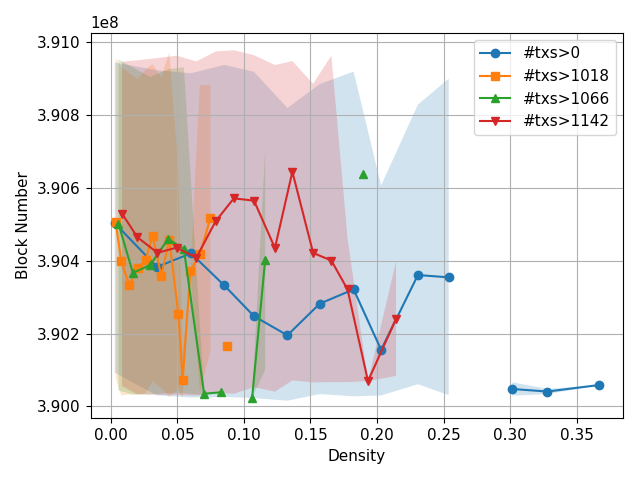}}
\caption{block-number}
\label{fig:block-number}
\end{figure}
\begin{figure}
\centerline{\includegraphics[width=\columnwidth]{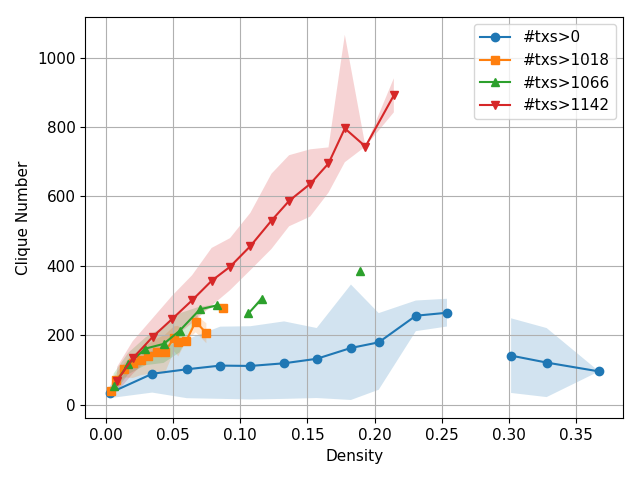}}
\caption{clique-number}
\label{fig:clique-number}
\end{figure}
\begin{figure}
\centerline{\includegraphics[width=\columnwidth]{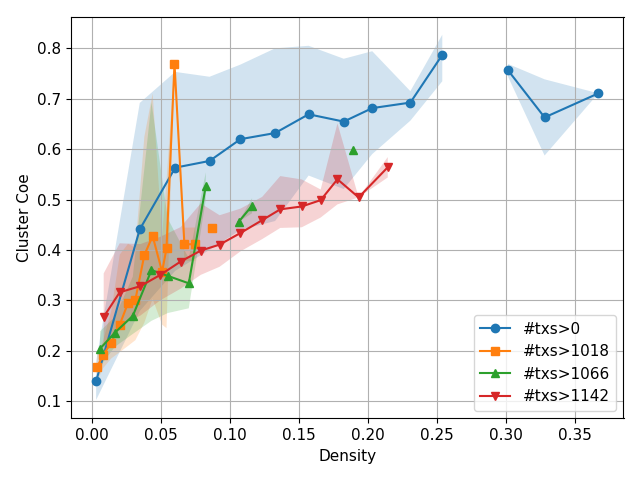}}
\caption{cluster-coe}
\label{fig:cluster-coe}
\end{figure}
\begin{figure}
\centerline{\includegraphics[width=\columnwidth]{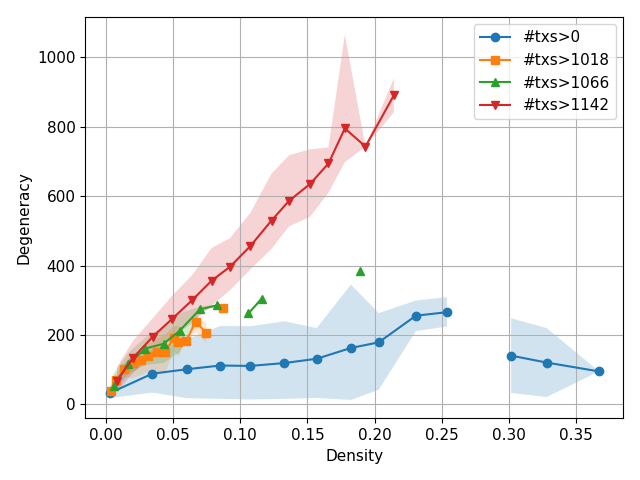}}
\caption{degeneracy}
\label{fig:degeneracy}
\end{figure}
\begin{figure}
\centerline{\includegraphics[width=\columnwidth]{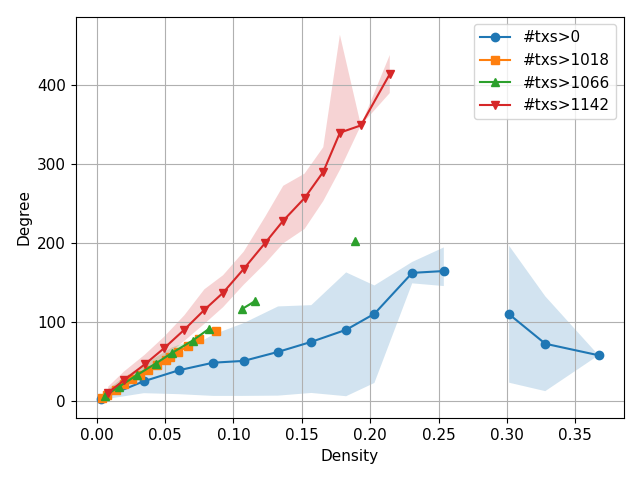}}
\caption{degree}
\label{fig:degree}
\end{figure}
\begin{figure}
\centerline{\includegraphics[width=\columnwidth]{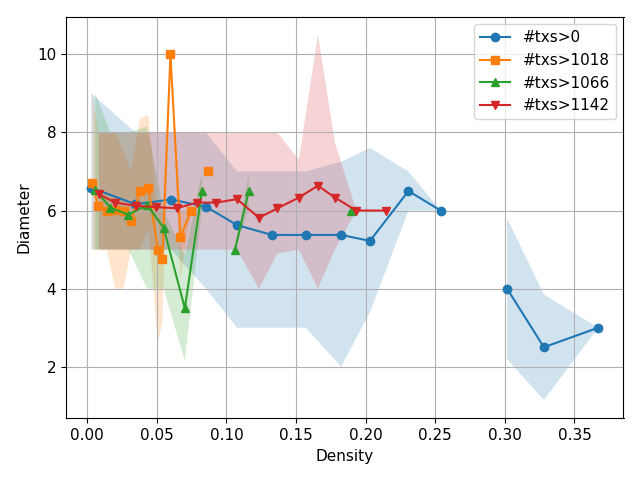}}
\caption{diameter}
\label{fig:diameter}
\end{figure}
\begin{figure}
\centerline{\includegraphics[width=\columnwidth]{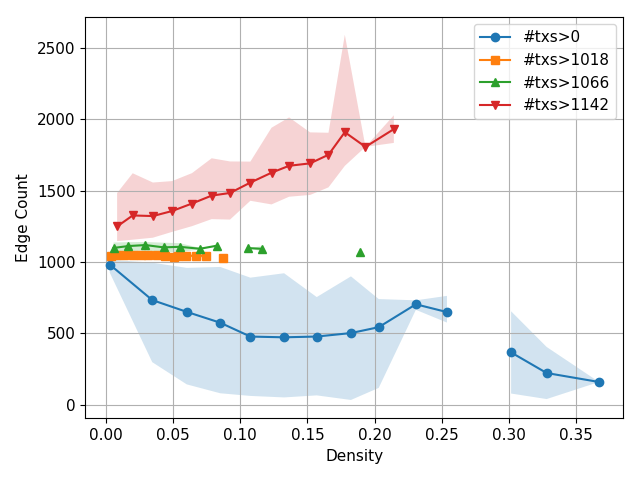}}
\caption{edge-count}
\label{fig:edge-count}
\end{figure}
\begin{figure}
\centerline{\includegraphics[width=\columnwidth]{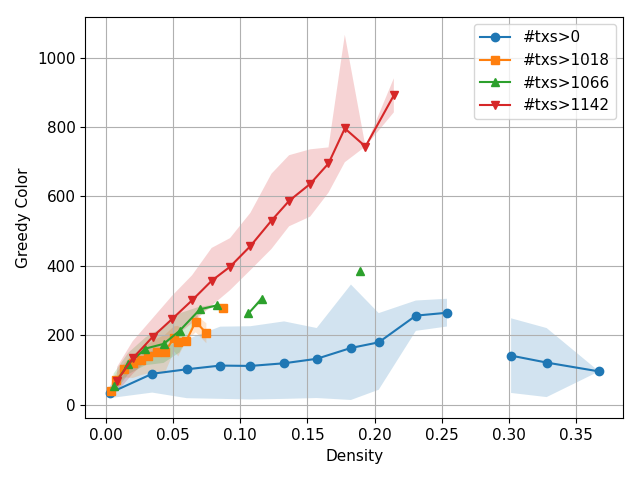}}
\caption{greedy-color}
\label{fig:greedy-color}
\end{figure}
\begin{figure}
\centerline{\includegraphics[width=\columnwidth]{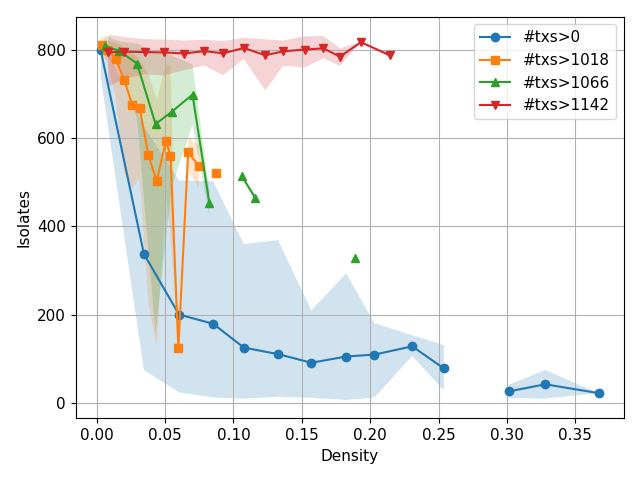}}
\caption{isolates}
\label{fig:isolates}
\end{figure}
\begin{figure}
\centerline{\includegraphics[width=\columnwidth]{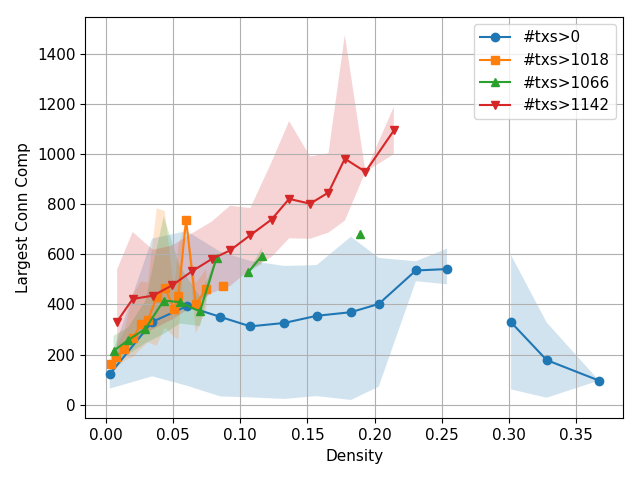}}
\caption{largest-conn-comp}
\label{fig:largest-conn-comp}
\end{figure}
\begin{figure}
\centerline{\includegraphics[width=\columnwidth]{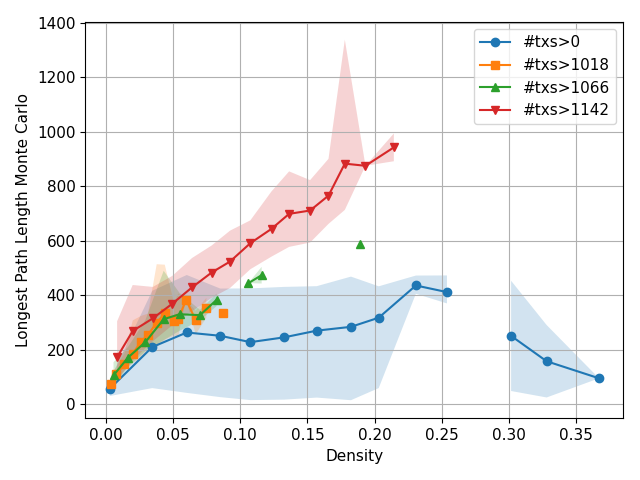}}
\caption{longest-path-length-monte-carlo}
\label{fig:longest-path-length-monte-carlo}
\end{figure}
\begin{figure}
\centerline{\includegraphics[width=\columnwidth]{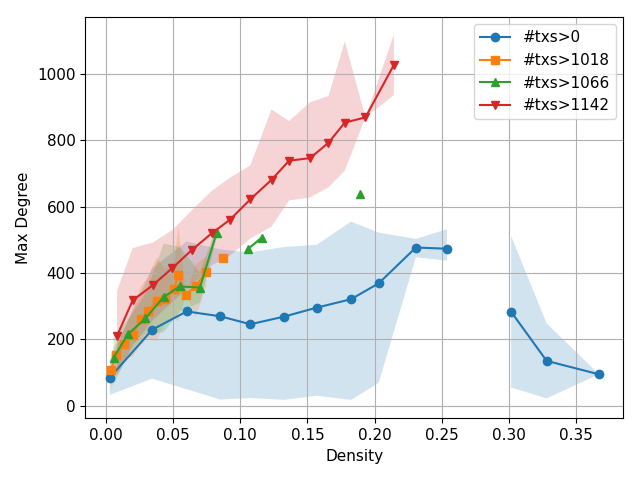}}
\caption{max-degree}
\label{fig:max-degree}
\end{figure}
\begin{figure}
\centerline{\includegraphics[width=\columnwidth]{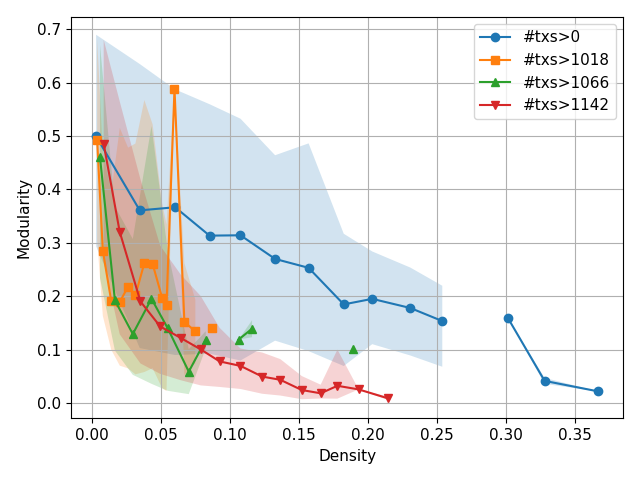}}
\caption{modularity}
\label{fig:modularity}
\end{figure}
\begin{figure}
\centerline{\includegraphics[width=\columnwidth]{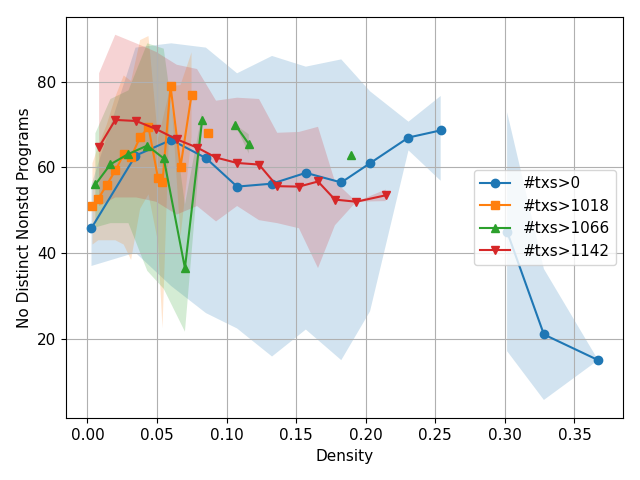}}
\caption{no-distinct-nonstd-programs}
\label{fig:no-distinct-nonstd-programs}
\end{figure}
\begin{figure}
\centerline{\includegraphics[width=\columnwidth]{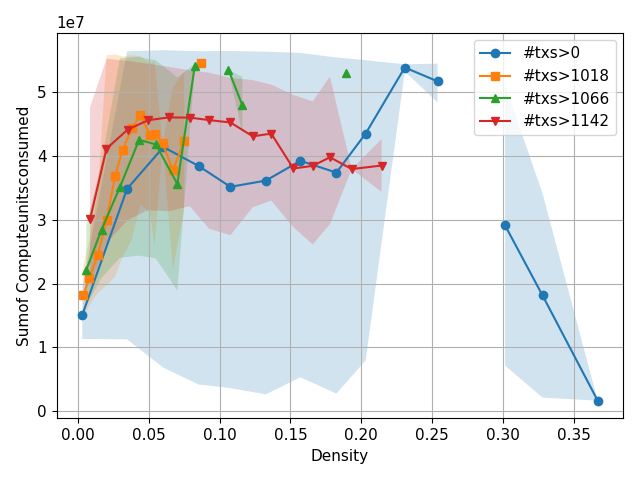}}
\caption{sumof-computeUnitsConsumed}
\label{fig:sumof-computeUnitsConsumed}
\end{figure}
\begin{figure}
\centerline{\includegraphics[width=\columnwidth]{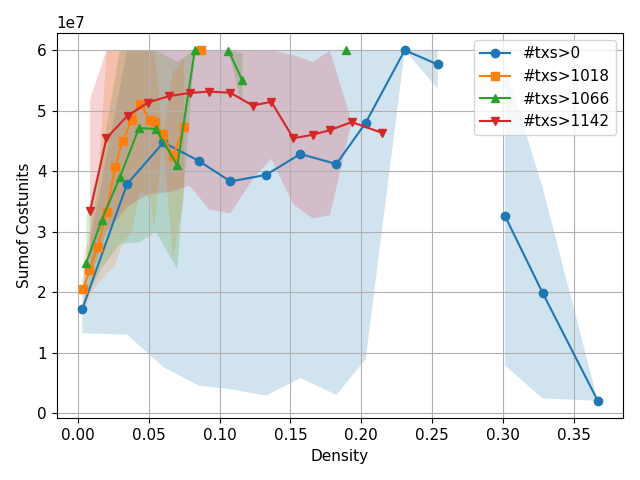}}
\caption{sumof-costUnits}
\label{fig:sumof-costUnits}
\end{figure}
\begin{figure}
\centerline{\includegraphics[width=\columnwidth]{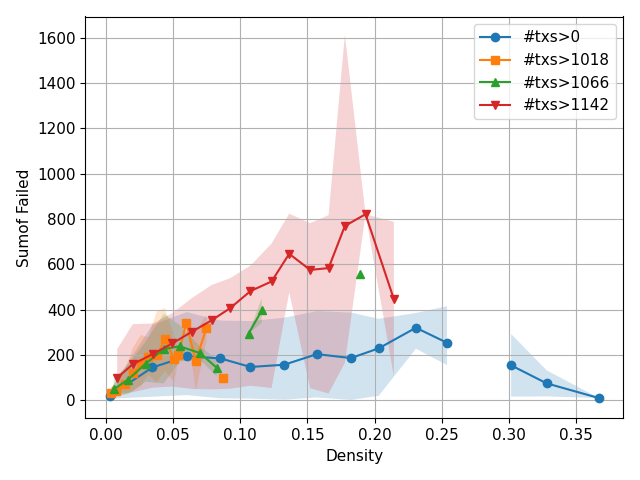}}
\caption{sumof-failed}
\label{fig:sumof-failed}
\end{figure}
\begin{figure}
\centerline{\includegraphics[width=\columnwidth]{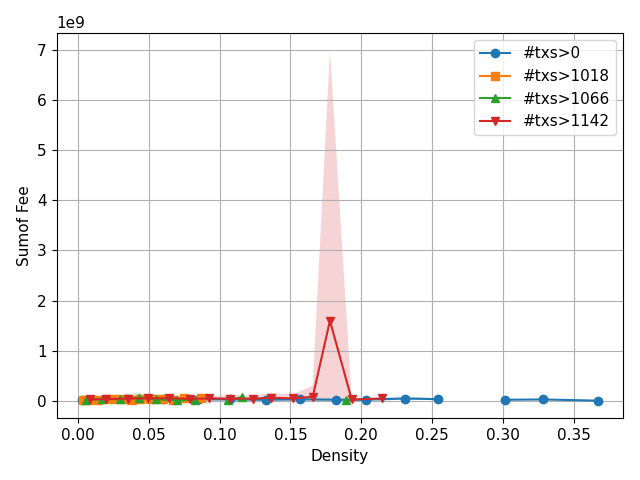}}
\caption{sumof-fee}
\label{fig:sumof-fee}
\end{figure}
\begin{figure}
\centerline{\includegraphics[width=\columnwidth]{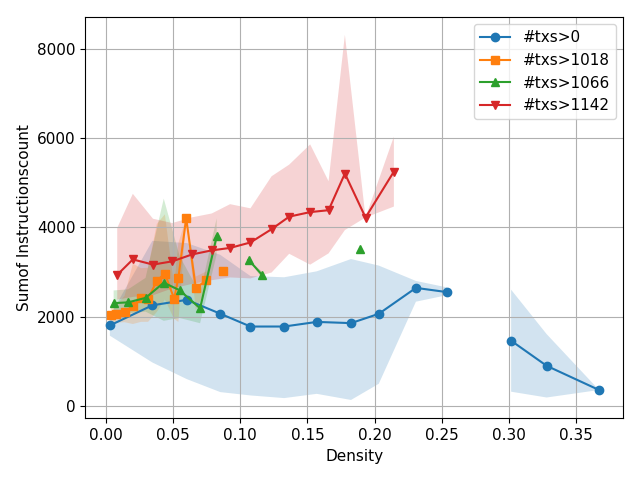}}
\caption{sumof-instructionsCount}
\label{fig:sumof-instructionsCount}
\end{figure}
\begin{figure}
\centerline{\includegraphics[width=\columnwidth]{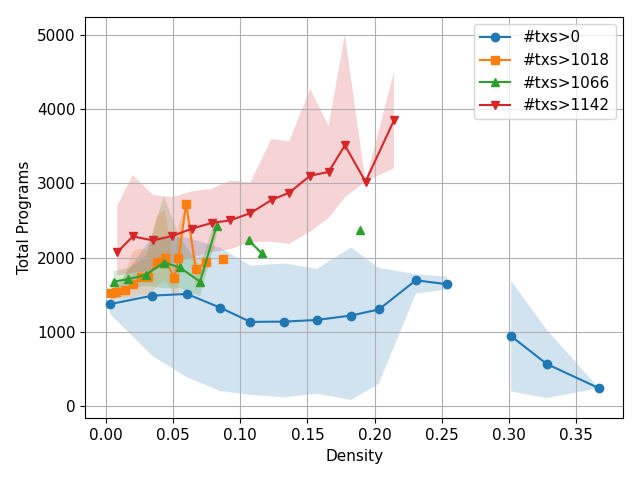}}
\caption{total-programs}
\label{fig:total-programs}
\end{figure}
\begin{figure}
\centerline{\includegraphics[width=\columnwidth]{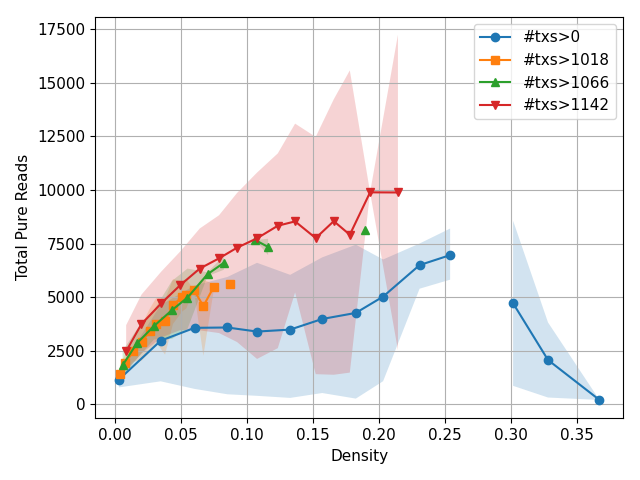}}
\caption{total-pure-reads}
\label{fig:total-pure-reads}
\end{figure}
\begin{figure}
\centerline{\includegraphics[width=\columnwidth]{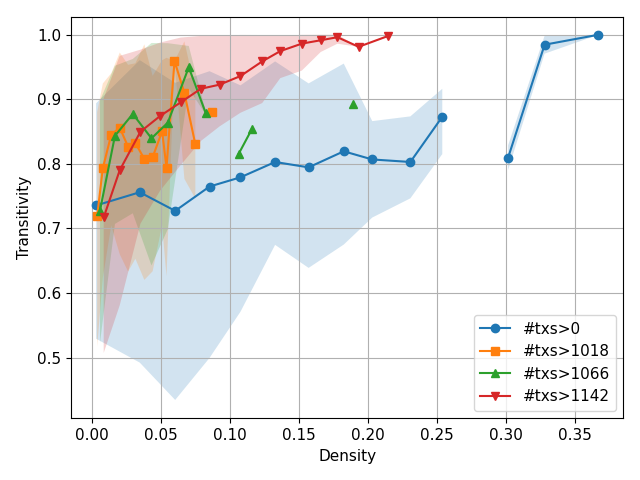}}
\caption{transitivity}
\label{fig:transitivity}
\end{figure}
\begin{figure}
\centerline{\includegraphics[width=\columnwidth]{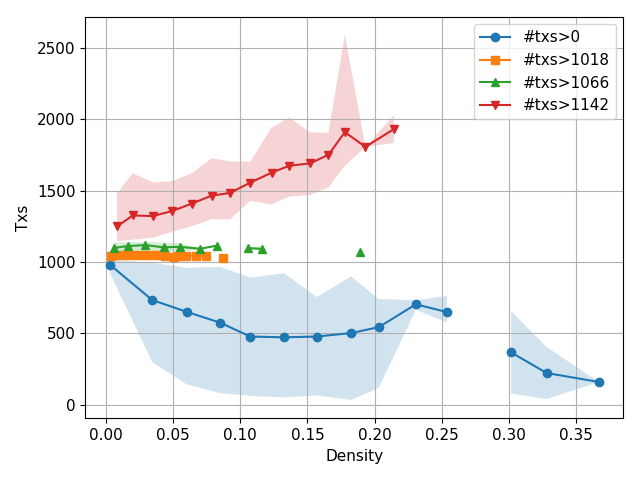}}
\caption{txs}
\label{fig:txs}
\end{figure}
\begin{figure}
\centerline{\includegraphics[width=\columnwidth]{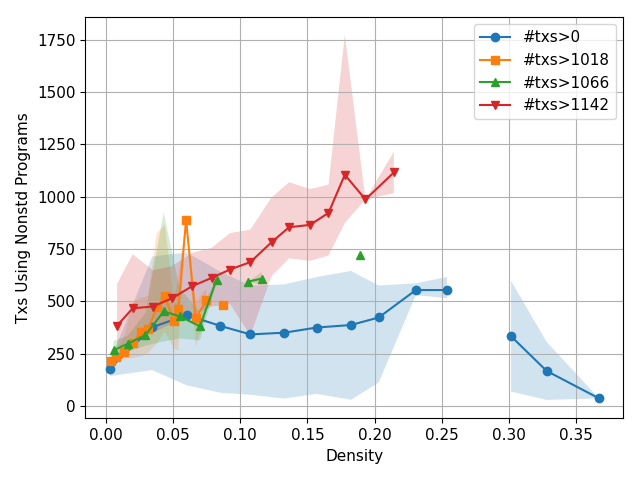}}
\caption{txs-using-nonstd-programs}
\label{fig:txs-using-nonstd-programs}
\end{figure}
\begin{figure}
\centerline{\includegraphics[width=\columnwidth]{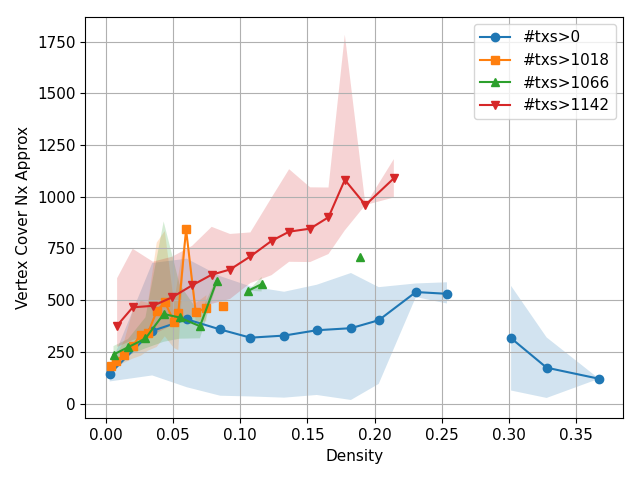}}
\caption{vertex-cover-nx-approx}
\label{fig:vertex-cover-nx-approx}
\end{figure}
\begin{figure}
\centerline{\includegraphics[width=\columnwidth]{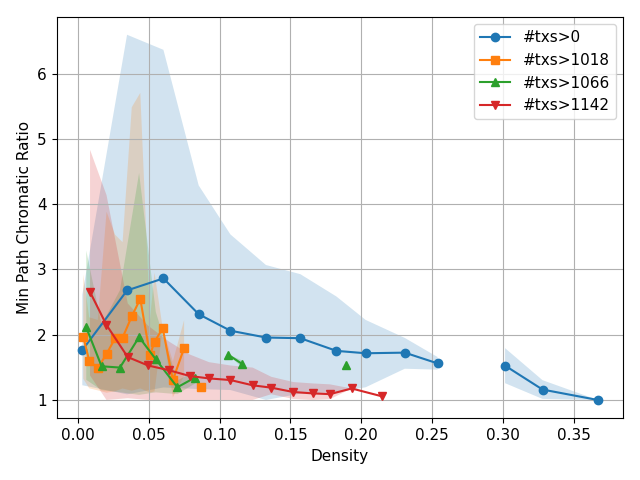}}
\caption{min-path-chromatic-ratio}
\label{fig:min-path-chromatic-ratio}
\end{figure}
\begin{figure}
\centerline{\includegraphics[width=\columnwidth]{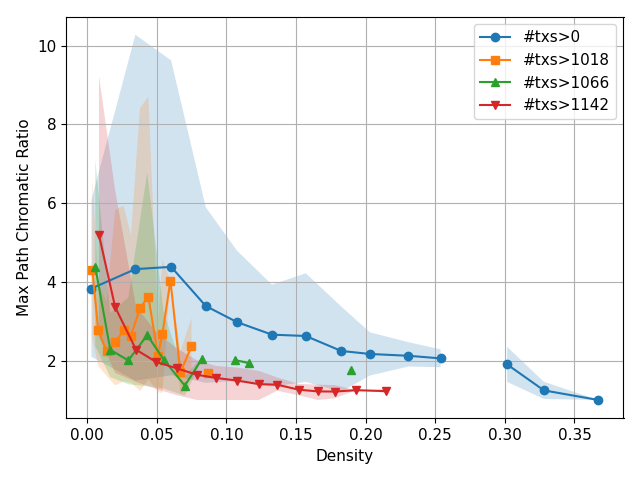}}
\caption{max-path-chromatic-ratio}
\label{fig:max-path-chromatic-ratio}
\end{figure}
\begin{figure}
\centerline{\includegraphics[width=\columnwidth]{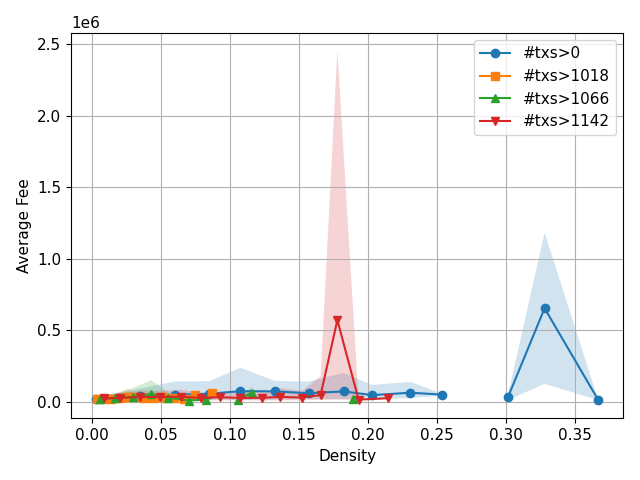}}
\caption{avg-fee}
\label{fig:avg-fee}
\end{figure}
\begin{figure}
\centerline{\includegraphics[width=\columnwidth]{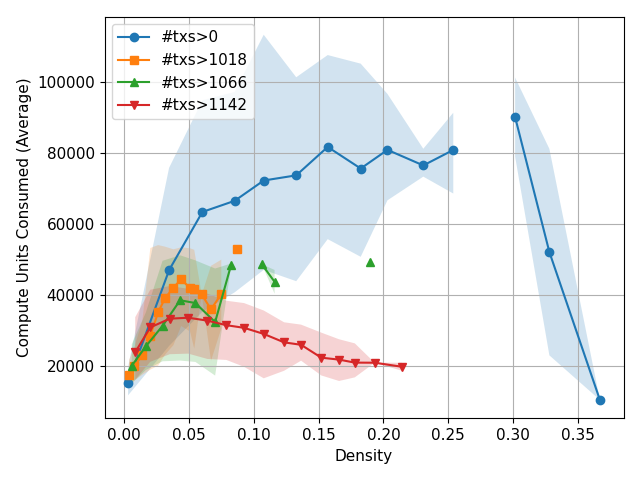}}
\caption{avg-computeUnitsConsumed}
\label{fig:avg-computeUnitsConsumed}
\end{figure}
\begin{figure}
\centerline{\includegraphics[width=\columnwidth]{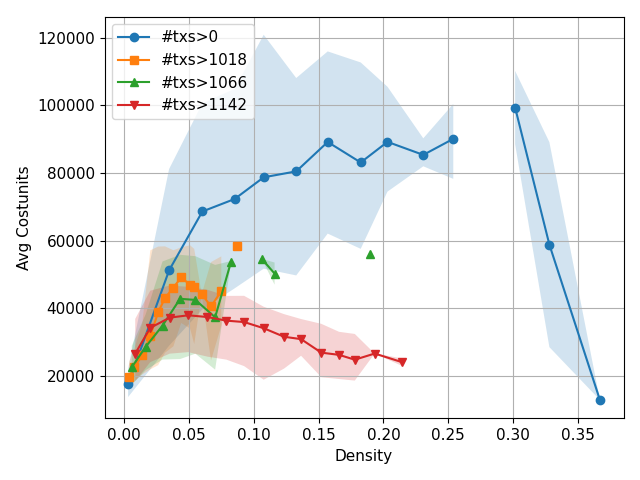}}
\caption{avg-costUnits}
\label{fig:avg-costUnits}
\end{figure}
\begin{figure}
\centerline{\includegraphics[width=\columnwidth]{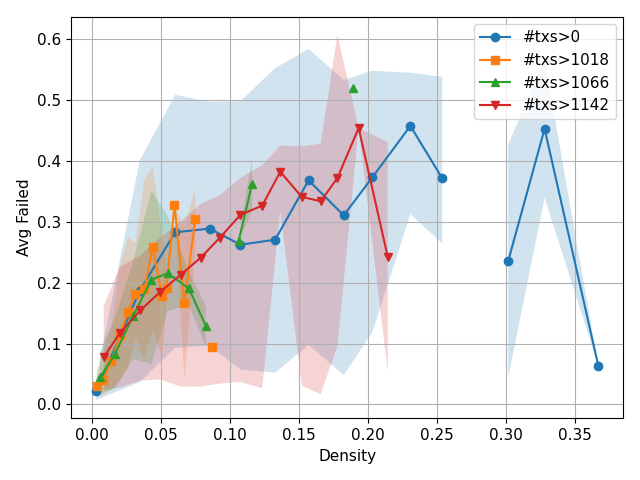}}
\caption{avg-failed}
\label{fig:avg-failed}
\end{figure}
\begin{figure}
\centerline{\includegraphics[width=\columnwidth]{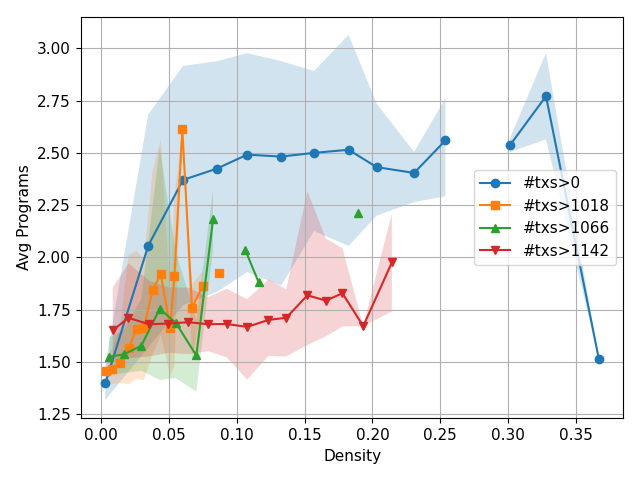}}
\caption{avg-programs}
\label{fig:avg-programs}
\end{figure}
\begin{figure}
\centerline{\includegraphics[width=\columnwidth]{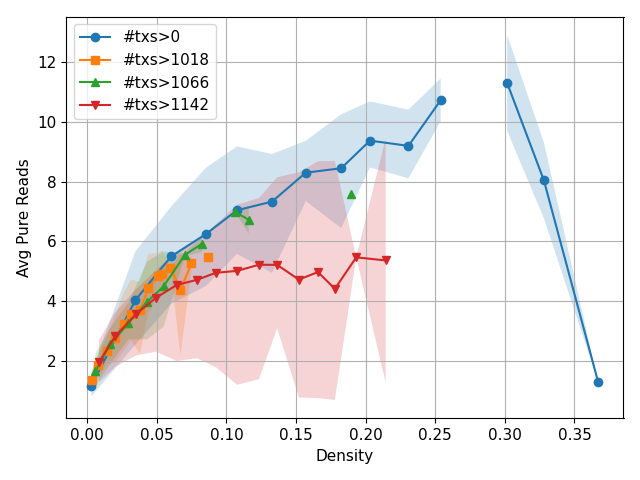}}
\caption{avg-pure-reads}
\label{fig:avg-pure-reads}
\end{figure}
\begin{figure}
\centerline{\includegraphics[width=\columnwidth]{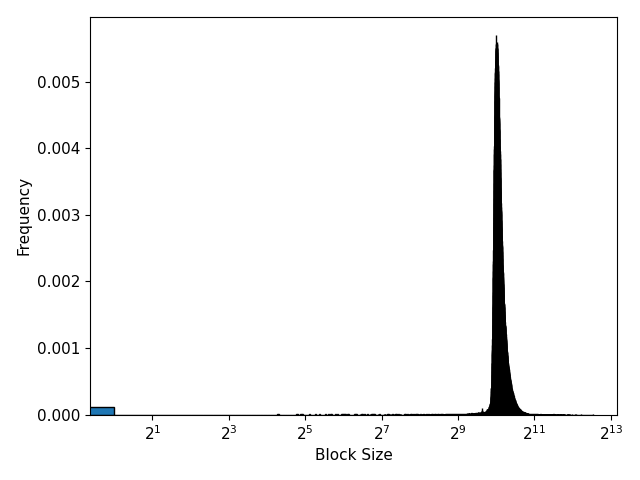}}
\caption{block-size-dist}
\label{fig:block-size-dist}
\end{figure}
\begin{figure}
\centerline{\includegraphics[width=\columnwidth]{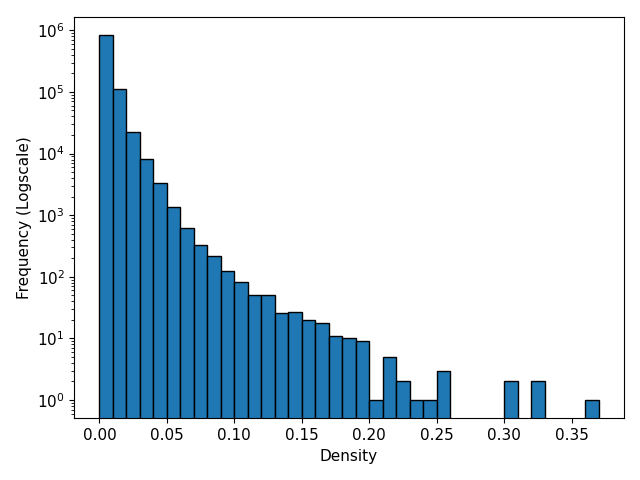}}
\caption{density-dist}
\label{fig:density-dist}
\end{figure}
\begin{figure}
\centerline{\includegraphics[width=\columnwidth]{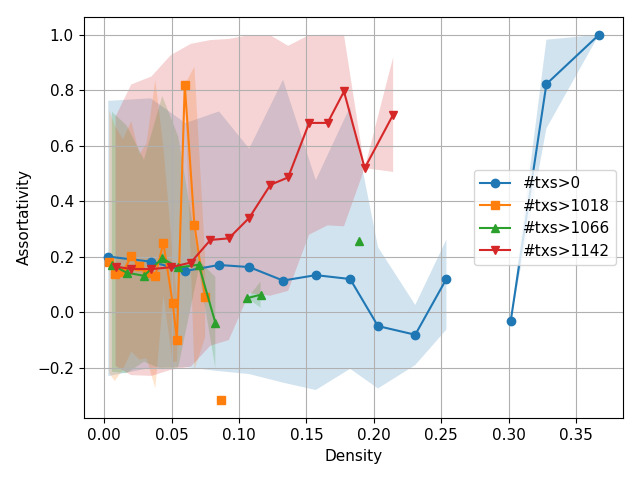}}
\caption{assortativity}
\label{fig:assortativity}
\end{figure}
\begin{figure}
\centerline{\includegraphics[width=\columnwidth]{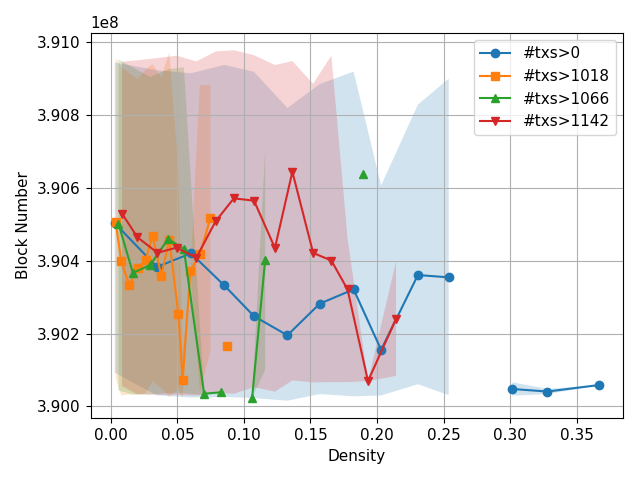}}
\caption{block-number}
\label{fig:block-number}
\end{figure}
\begin{figure}
\centerline{\includegraphics[width=\columnwidth]{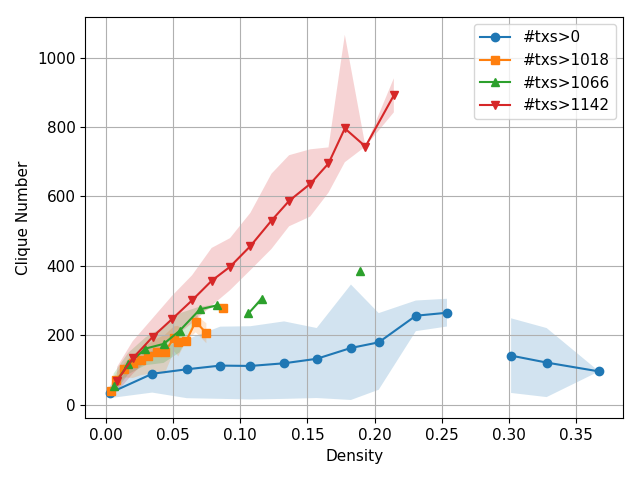}}
\caption{clique-number}
\label{fig:clique-number}
\end{figure}
\begin{figure}
\centerline{\includegraphics[width=\columnwidth]{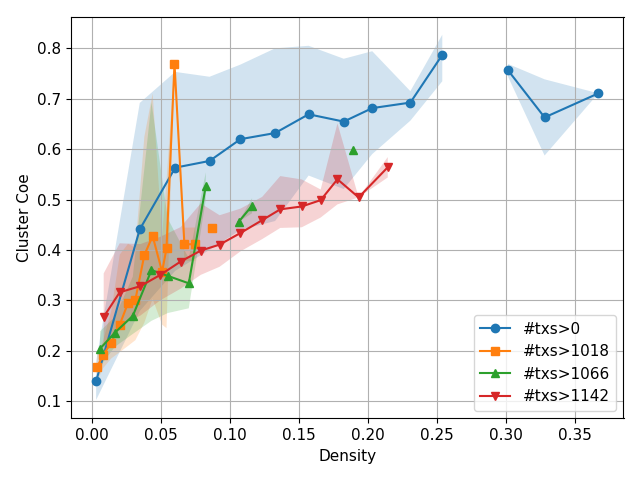}}
\caption{cluster-coe}
\label{fig:cluster-coe}
\end{figure}
\begin{figure}
\centerline{\includegraphics[width=\columnwidth]{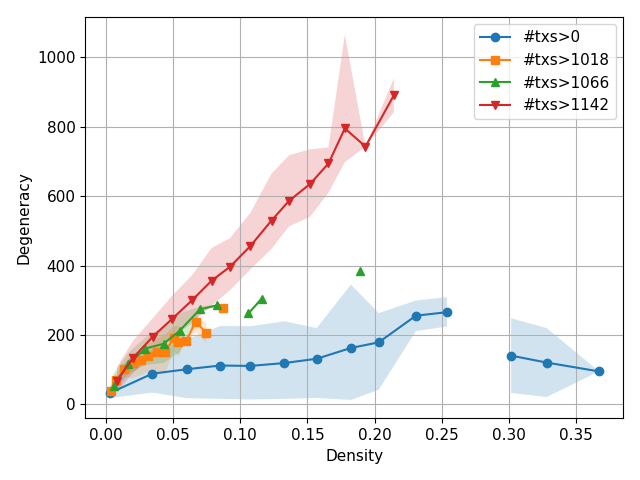}}
\caption{degeneracy}
\label{fig:degeneracy}
\end{figure}
\begin{figure}
\centerline{\includegraphics[width=\columnwidth]{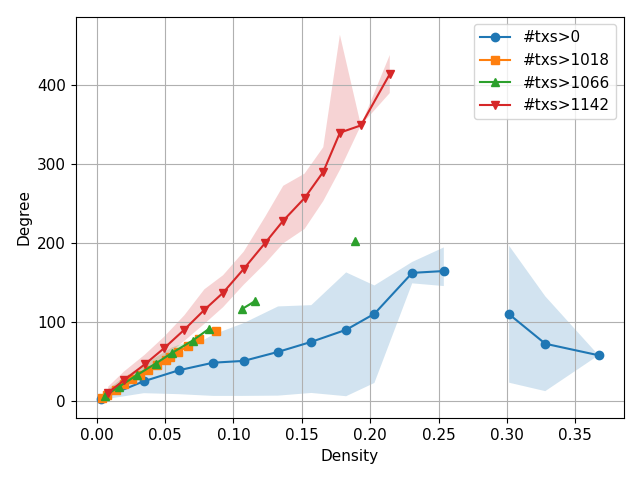}}
\caption{degree}
\label{fig:degree}
\end{figure}
\begin{figure}
\centerline{\includegraphics[width=\columnwidth]{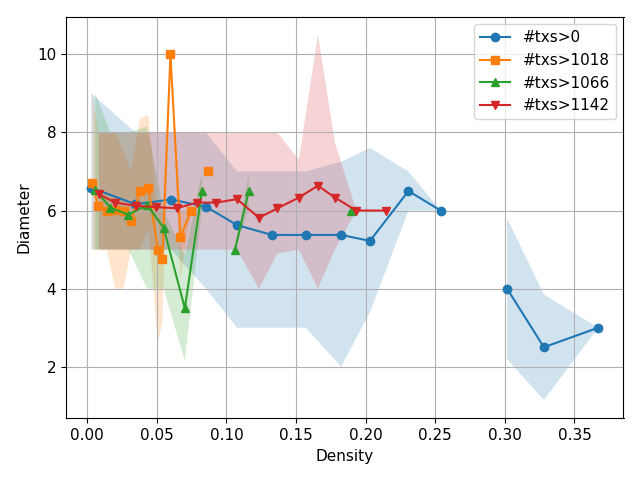}}
\caption{diameter}
\label{fig:diameter}
\end{figure}
\begin{figure}
\centerline{\includegraphics[width=\columnwidth]{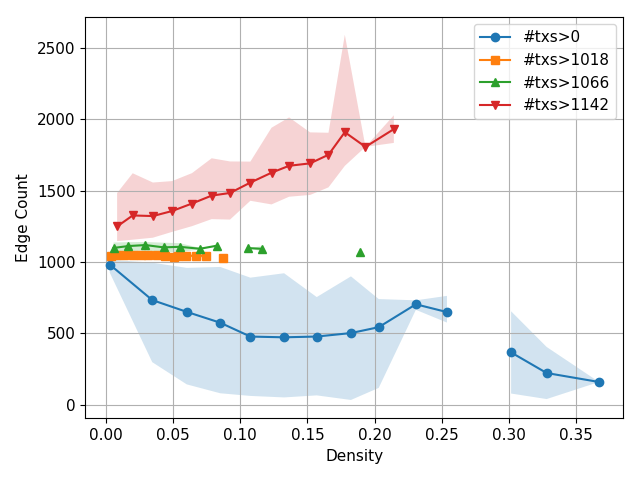}}
\caption{edge-count}
\label{fig:edge-count}
\end{figure}
\begin{figure}
\centerline{\includegraphics[width=\columnwidth]{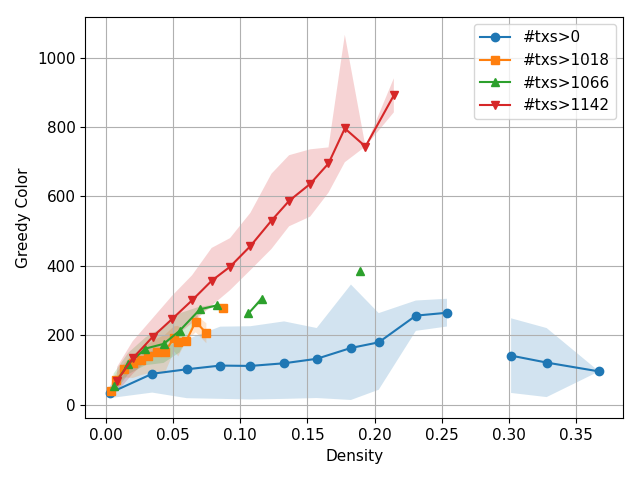}}
\caption{greedy-color}
\label{fig:greedy-color}
\end{figure}
\begin{figure}
\centerline{\includegraphics[width=\columnwidth]{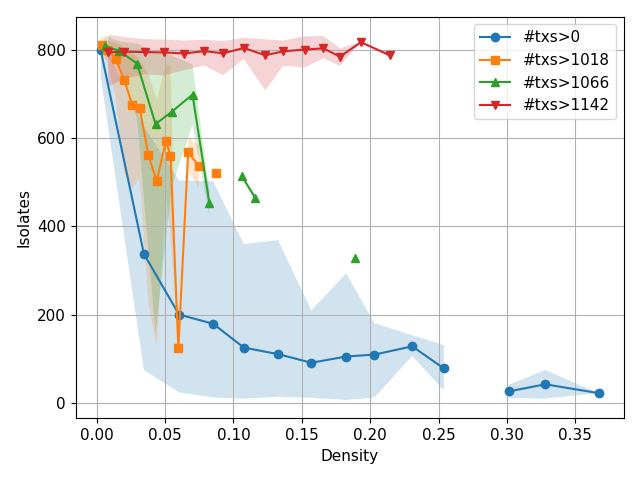}}
\caption{isolates}
\label{fig:isolates}
\end{figure}
\begin{figure}
\centerline{\includegraphics[width=\columnwidth]{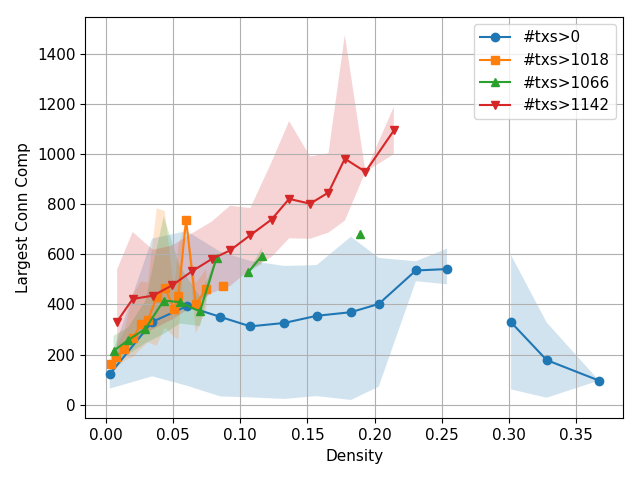}}
\caption{largest-conn-comp}
\label{fig:largest-conn-comp}
\end{figure}
\begin{figure}
\centerline{\includegraphics[width=\columnwidth]{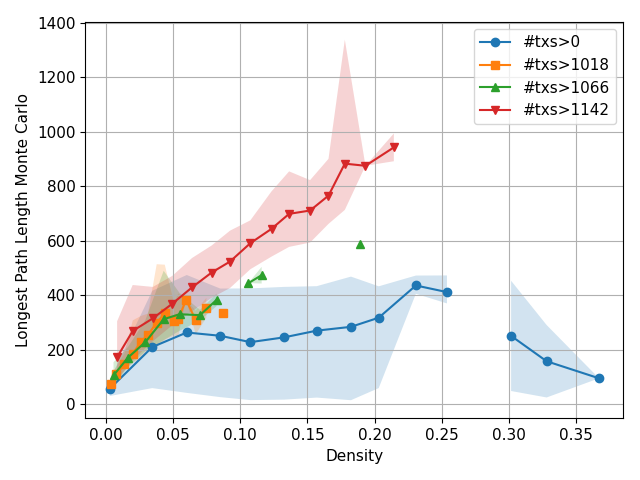}}
\caption{longest-path-length-monte-carlo}
\label{fig:longest-path-length-monte-carlo}
\end{figure}
\begin{figure}
\centerline{\includegraphics[width=\columnwidth]{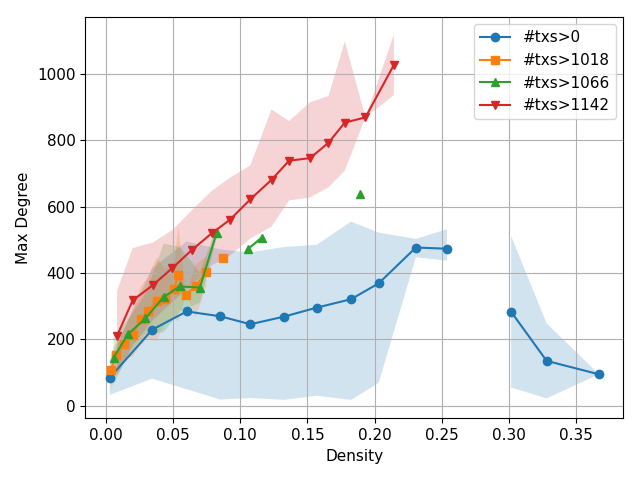}}
\caption{max-degree}
\label{fig:max-degree}
\end{figure}
\begin{figure}
\centerline{\includegraphics[width=\columnwidth]{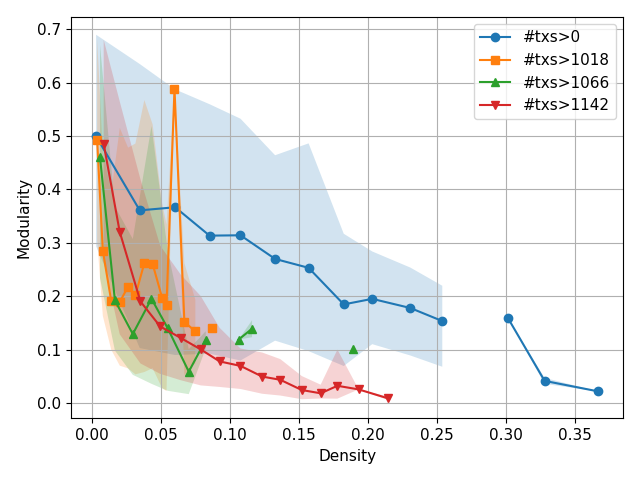}}
\caption{modularity}
\label{fig:modularity}
\end{figure}
\begin{figure}
\centerline{\includegraphics[width=\columnwidth]{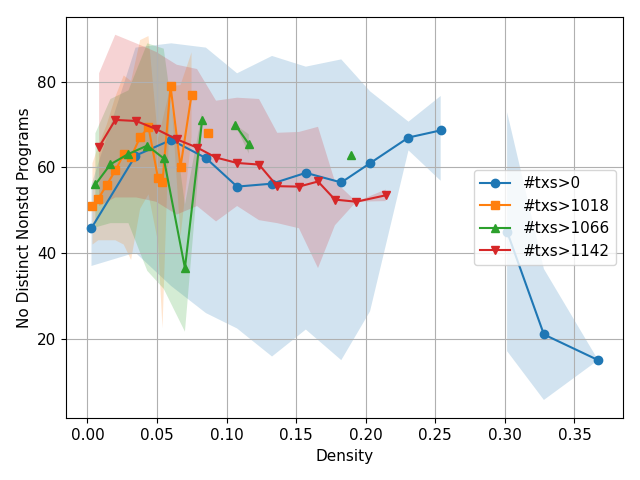}}
\caption{no-distinct-nonstd-programs}
\label{fig:no-distinct-nonstd-programs}
\end{figure}
\begin{figure}
\centerline{\includegraphics[width=\columnwidth]{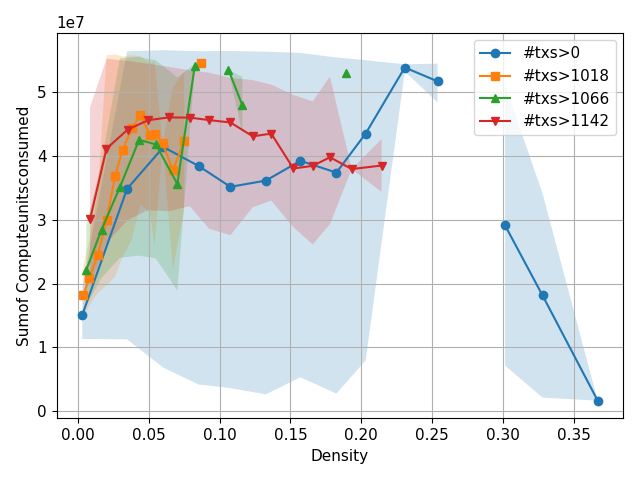}}
\caption{sumof-computeUnitsConsumed}
\label{fig:sumof-computeUnitsConsumed}
\end{figure}
\begin{figure}
\centerline{\includegraphics[width=\columnwidth]{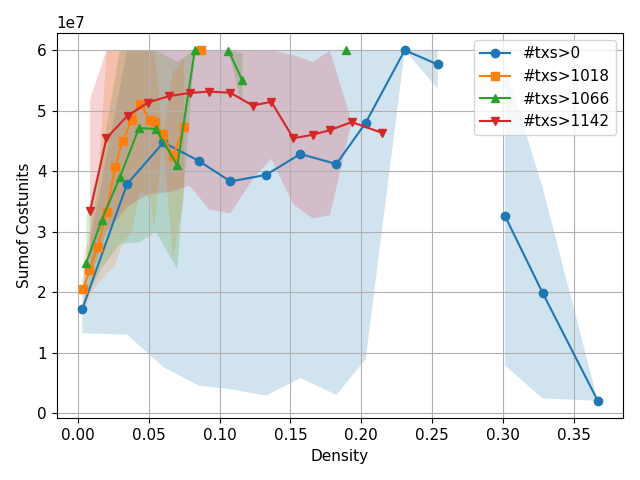}}
\caption{sumof-costUnits}
\label{fig:sumof-costUnits}
\end{figure}
\begin{figure}
\centerline{\includegraphics[width=\columnwidth]{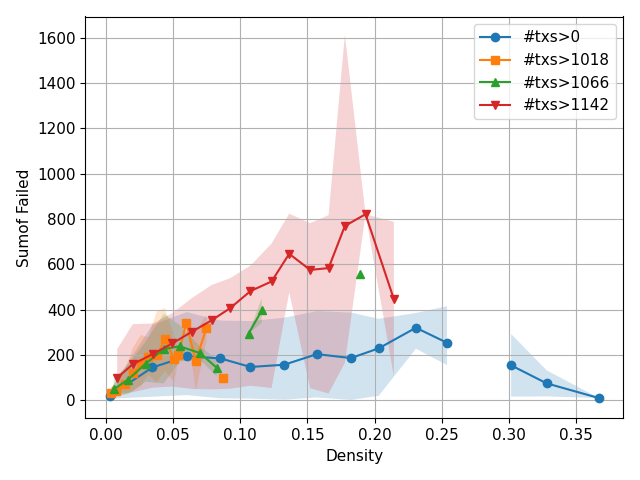}}
\caption{sumof-failed}
\label{fig:sumof-failed}
\end{figure}
\begin{figure}
\centerline{\includegraphics[width=\columnwidth]{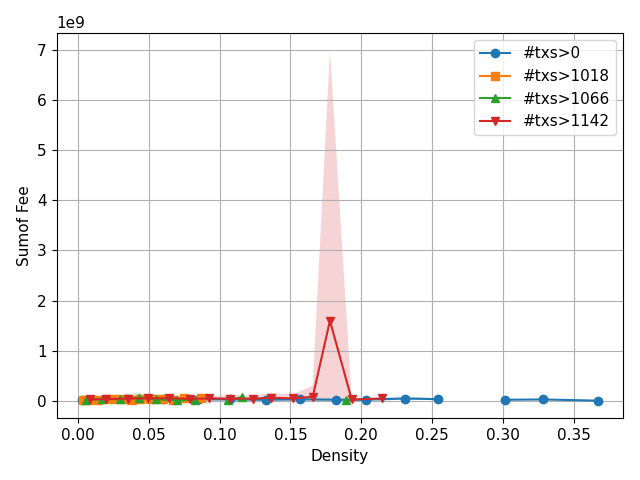}}
\caption{sumof-fee}
\label{fig:sumof-fee}
\end{figure}
\begin{figure}
\centerline{\includegraphics[width=\columnwidth]{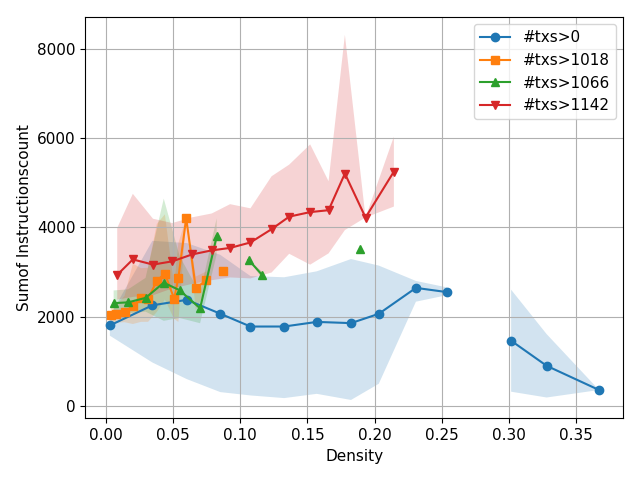}}
\caption{sumof-instructionsCount}
\label{fig:sumof-instructionsCount}
\end{figure}
\begin{figure}
\centerline{\includegraphics[width=\columnwidth]{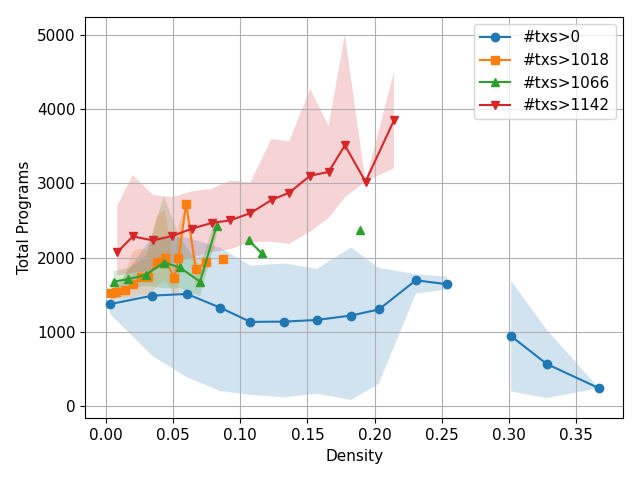}}
\caption{total-programs}
\label{fig:total-programs}
\end{figure}
\begin{figure}
\centerline{\includegraphics[width=\columnwidth]{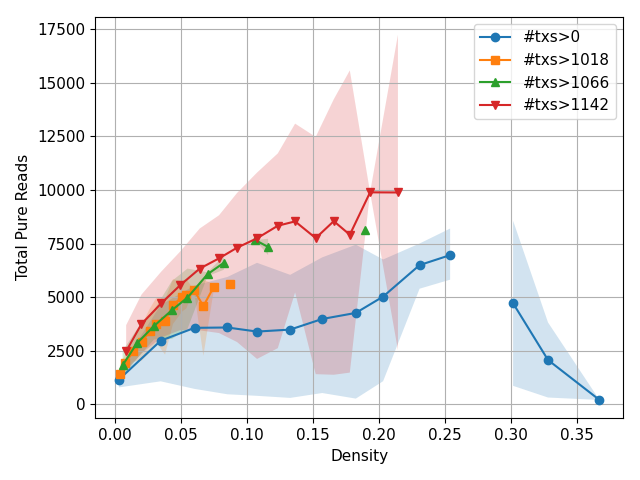}}
\caption{total-pure-reads}
\label{fig:total-pure-reads}
\end{figure}
\begin{figure}
\centerline{\includegraphics[width=\columnwidth]{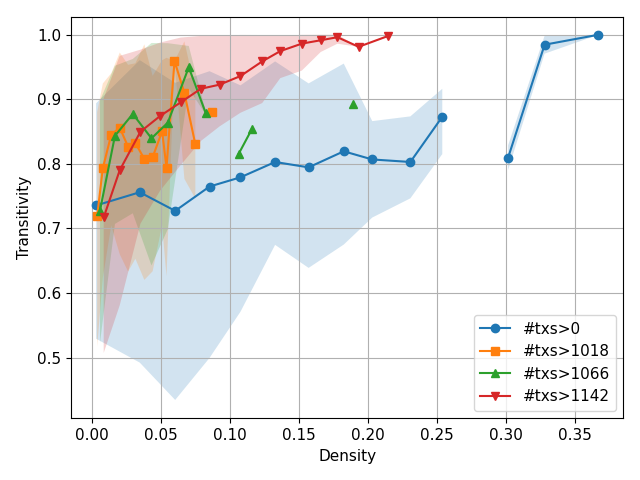}}
\caption{transitivity}
\label{fig:transitivity}
\end{figure}
\begin{figure}
\centerline{\includegraphics[width=\columnwidth]{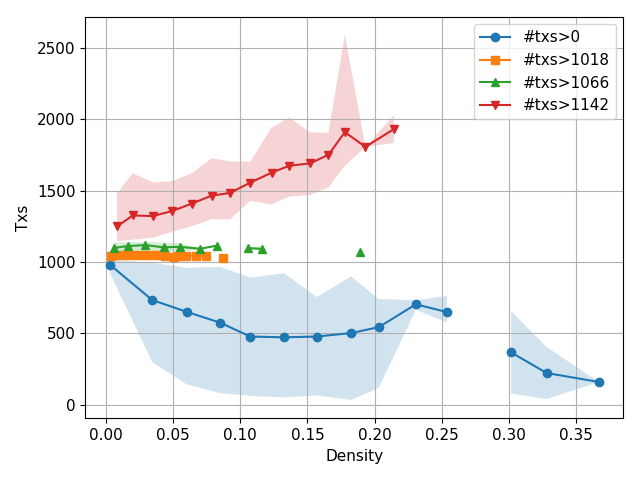}}
\caption{txs}
\label{fig:txs}
\end{figure}
\begin{figure}
\centerline{\includegraphics[width=\columnwidth]{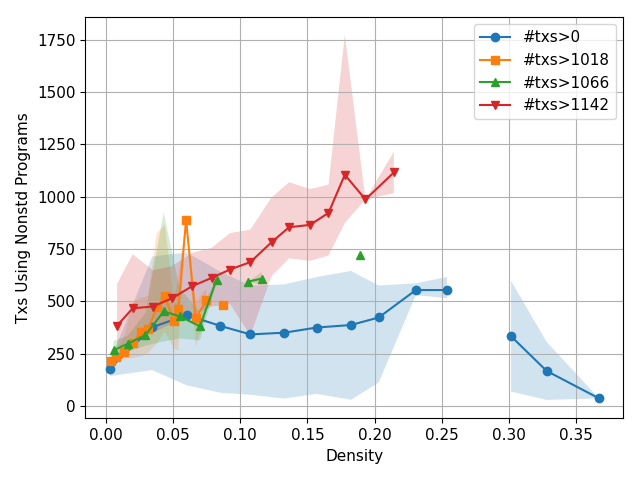}}
\caption{txs-using-nonstd-programs}
\label{fig:txs-using-nonstd-programs}
\end{figure}
\begin{figure}
\centerline{\includegraphics[width=\columnwidth]{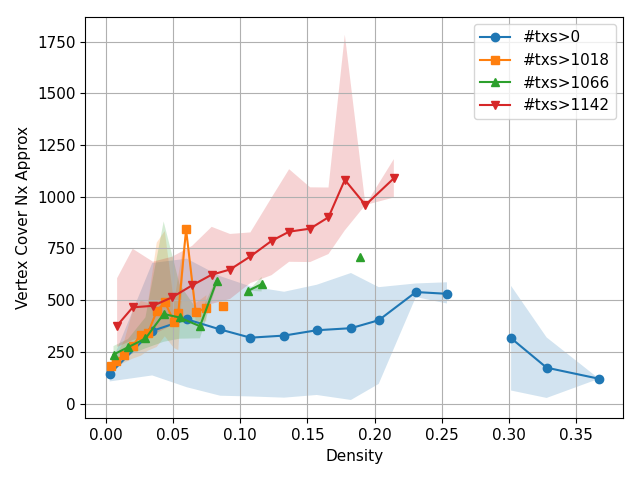}}
\caption{vertex-cover-nx-approx}
\label{fig:vertex-cover-nx-approx}
\end{figure}
\begin{figure}
\centerline{\includegraphics[width=\columnwidth]{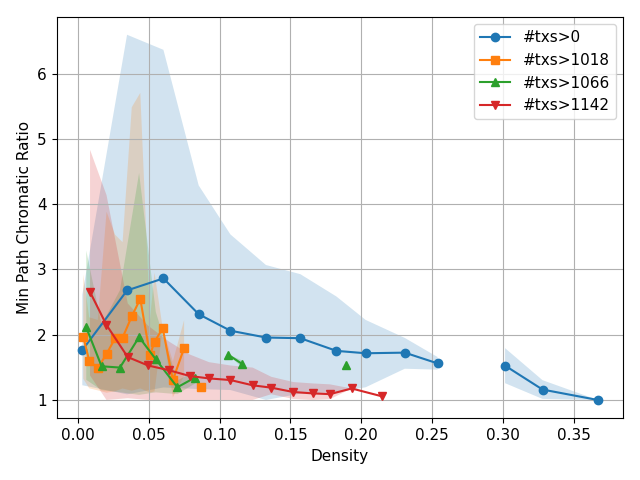}}
\caption{min-path-chromatic-ratio}
\label{fig:min-path-chromatic-ratio}
\end{figure}
\begin{figure}
\centerline{\includegraphics[width=\columnwidth]{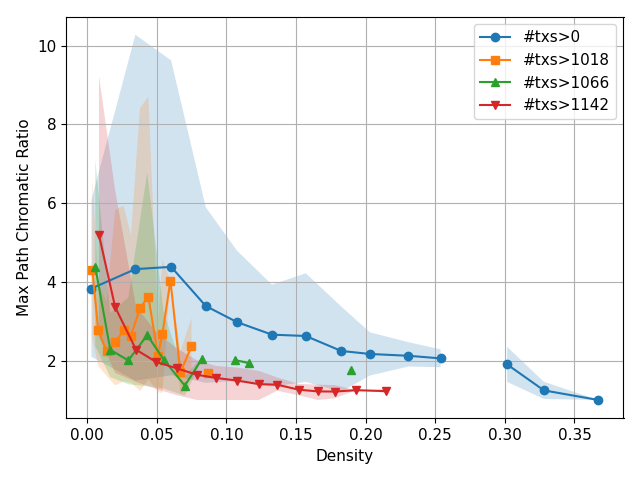}}
\caption{max-path-chromatic-ratio}
\label{fig:max-path-chromatic-ratio}
\end{figure}
\begin{figure}
\centerline{\includegraphics[width=\columnwidth]{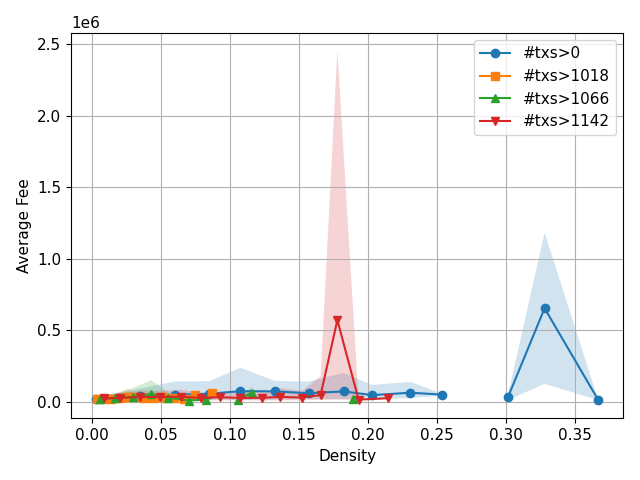}}
\caption{avg-fee}
\label{fig:avg-fee}
\end{figure}
\begin{figure}
\centerline{\includegraphics[width=\columnwidth]{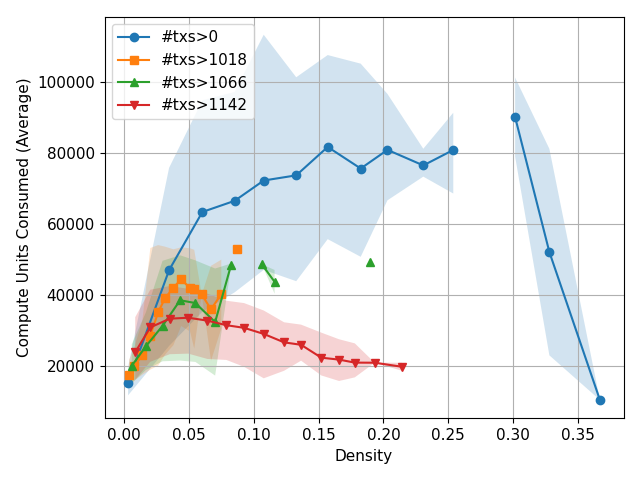}}
\caption{avg-computeUnitsConsumed}
\label{fig:avg-computeUnitsConsumed}
\end{figure}
\begin{figure}
\centerline{\includegraphics[width=\columnwidth]{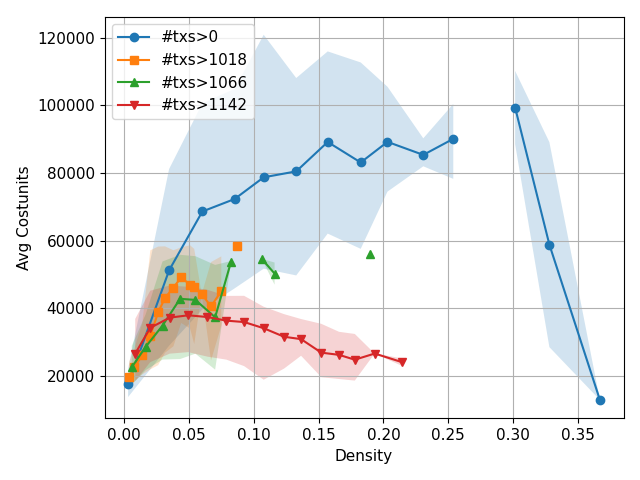}}
\caption{avg-costUnits}
\label{fig:avg-costUnits}
\end{figure}
\begin{figure}
\centerline{\includegraphics[width=\columnwidth]{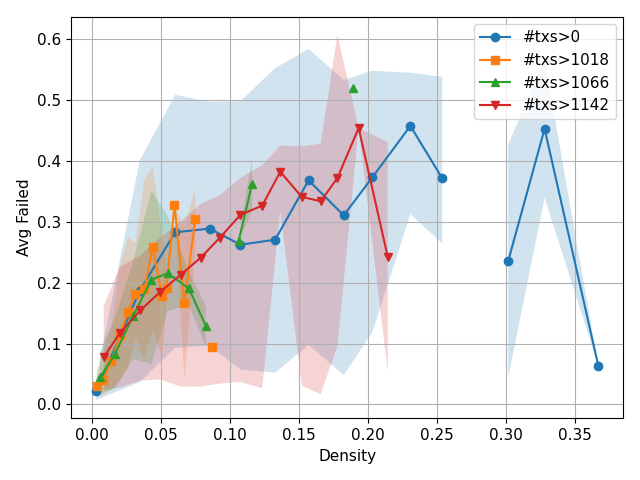}}
\caption{avg-failed}
\label{fig:avg-failed}
\end{figure}
\begin{figure}
\centerline{\includegraphics[width=\columnwidth]{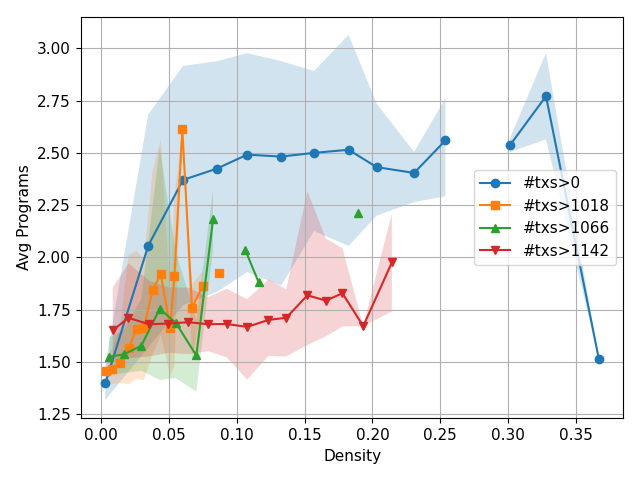}}
\caption{avg-programs}
\label{fig:avg-programs}
\end{figure}
\begin{figure}
\centerline{\includegraphics[width=\columnwidth]{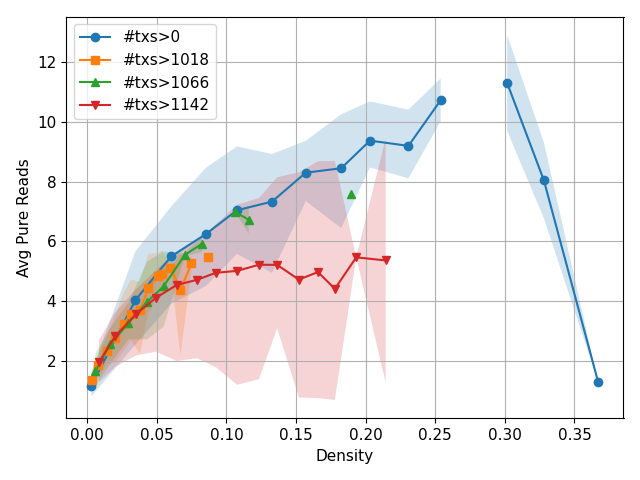}}
\caption{avg-pure-reads}
\label{fig:avg-pure-reads}
\end{figure}
\begin{figure}
\centerline{\includegraphics[width=\columnwidth]{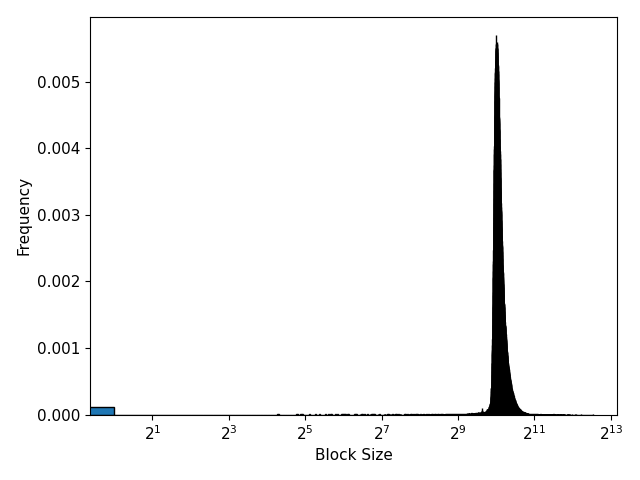}}
\caption{block-size-dist}
\label{fig:block-size-dist}
\end{figure}
\begin{figure}
\centerline{\includegraphics[width=\columnwidth]{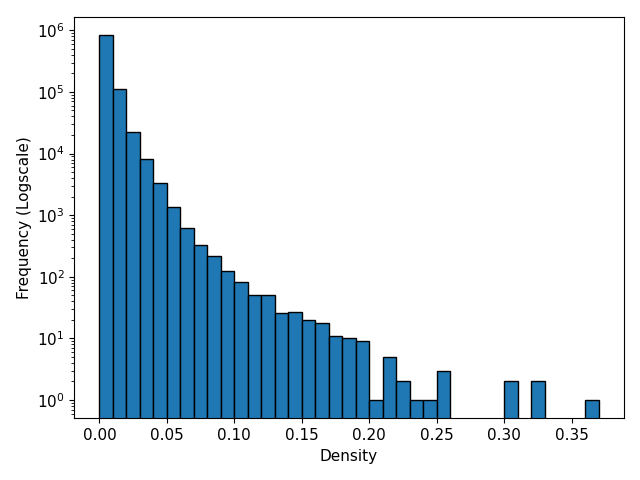}}
\caption{density-dist}
\label{fig:density-dist}
\end{figure}
\begin{figure}
\centerline{\includegraphics[width=\columnwidth]{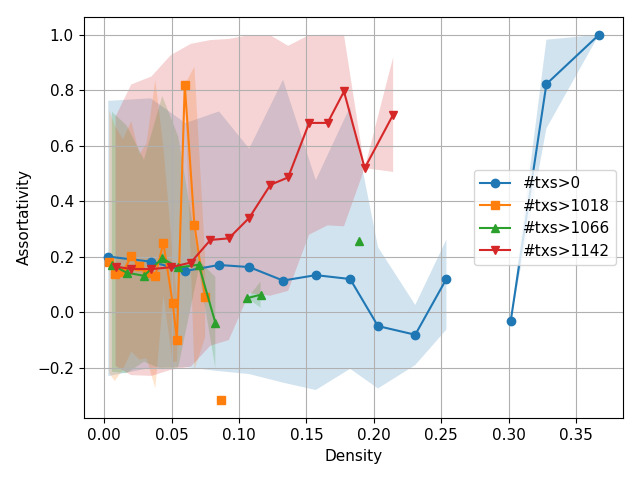}}
\caption{assortativity}
\label{fig:assortativity}
\end{figure}
\begin{figure}
\centerline{\includegraphics[width=\columnwidth]{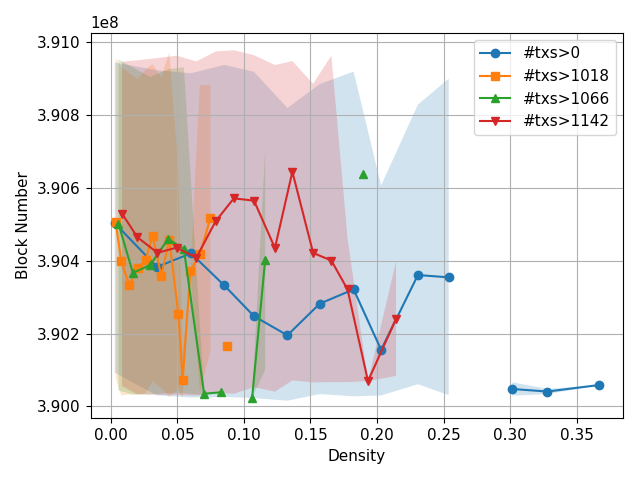}}
\caption{block-number}
\label{fig:block-number}
\end{figure}
\begin{figure}
\centerline{\includegraphics[width=\columnwidth]{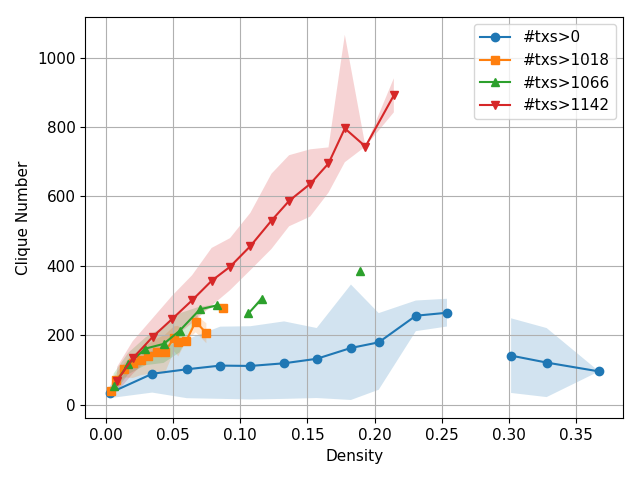}}
\caption{clique-number}
\label{fig:clique-number}
\end{figure}
\begin{figure}
\centerline{\includegraphics[width=\columnwidth]{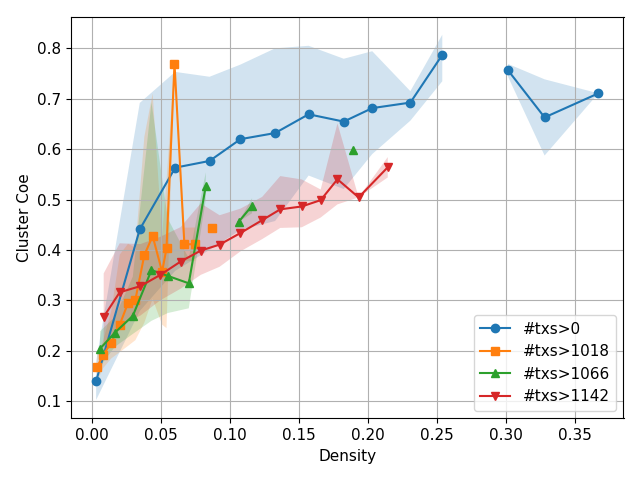}}
\caption{cluster-coe}
\label{fig:cluster-coe}
\end{figure}
\begin{figure}
\centerline{\includegraphics[width=\columnwidth]{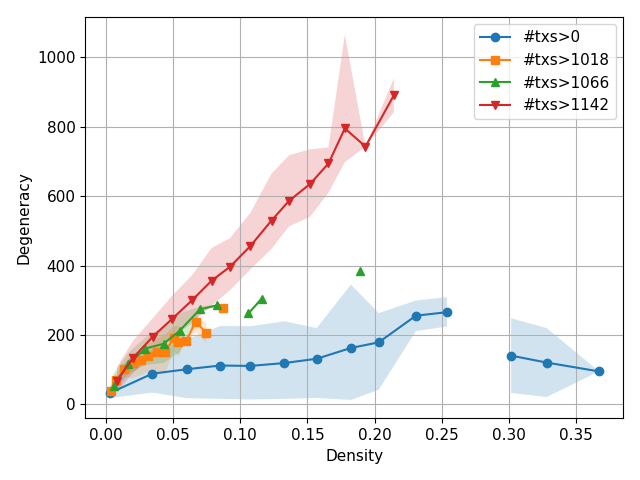}}
\caption{degeneracy}
\label{fig:degeneracy}
\end{figure}
\begin{figure}
\centerline{\includegraphics[width=\columnwidth]{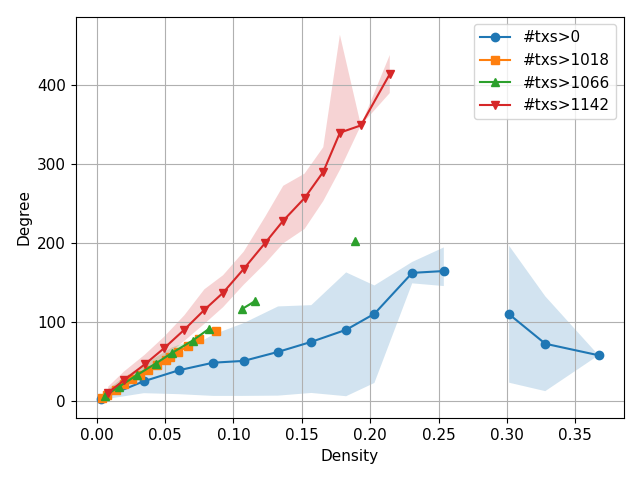}}
\caption{degree}
\label{fig:degree}
\end{figure}
\begin{figure}
\centerline{\includegraphics[width=\columnwidth]{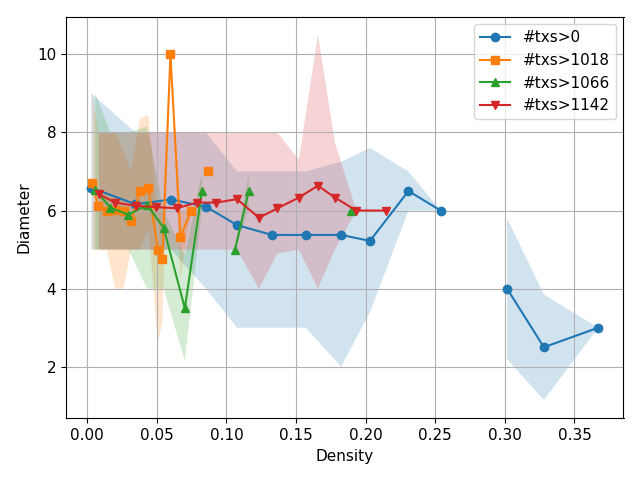}}
\caption{diameter}
\label{fig:diameter}
\end{figure}
\begin{figure}
\centerline{\includegraphics[width=\columnwidth]{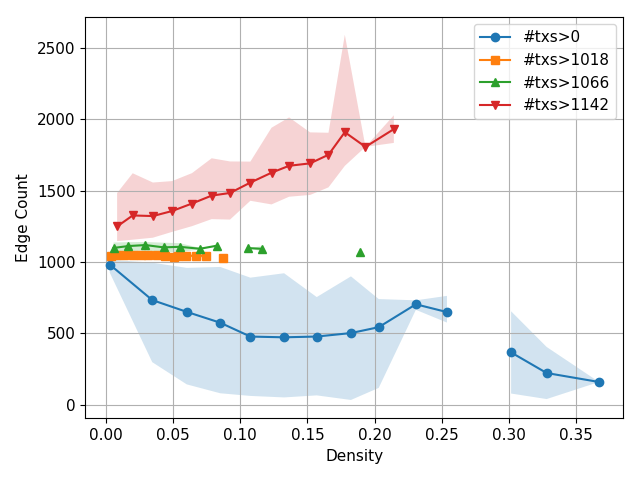}}
\caption{edge-count}
\label{fig:edge-count}
\end{figure}
\begin{figure}
\centerline{\includegraphics[width=\columnwidth]{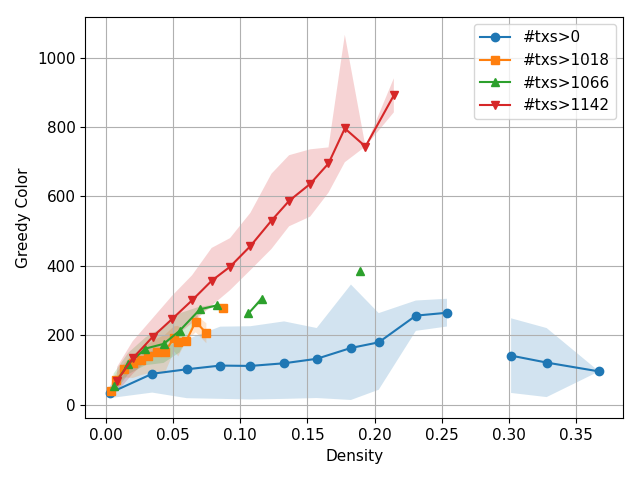}}
\caption{greedy-color}
\label{fig:greedy-color}
\end{figure}
\begin{figure}
\centerline{\includegraphics[width=\columnwidth]{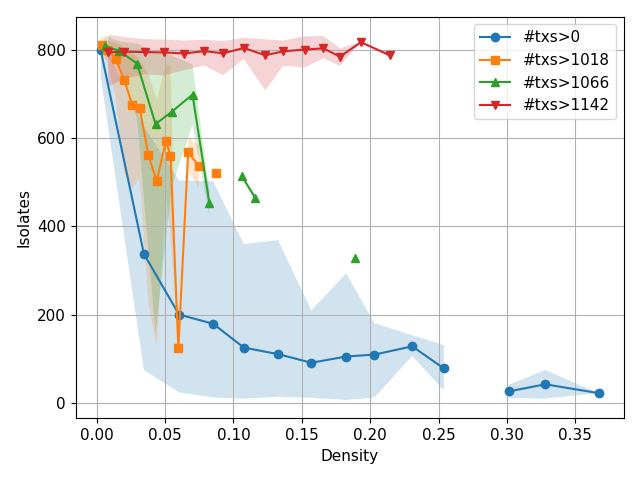}}
\caption{isolates}
\label{fig:isolates}
\end{figure}
\begin{figure}
\centerline{\includegraphics[width=\columnwidth]{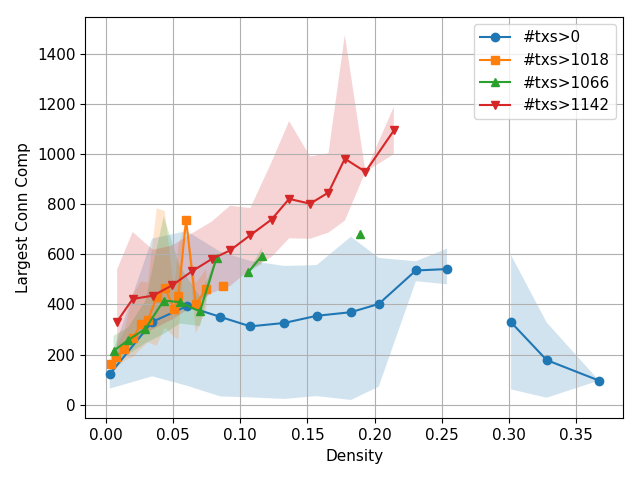}}
\caption{largest-conn-comp}
\label{fig:largest-conn-comp}
\end{figure}
\begin{figure}
\centerline{\includegraphics[width=\columnwidth]{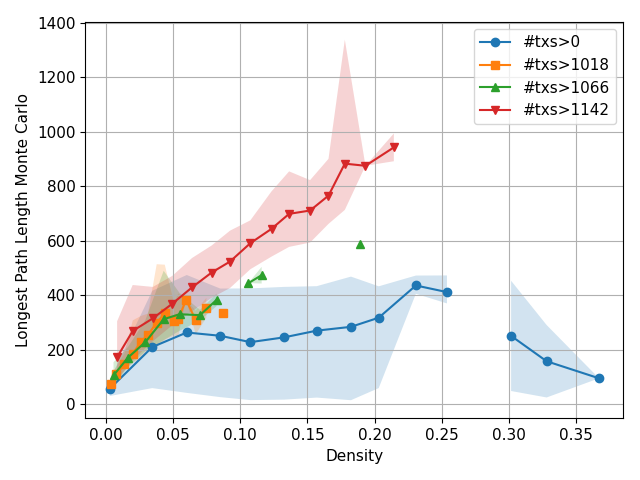}}
\caption{longest-path-length-monte-carlo}
\label{fig:longest-path-length-monte-carlo}
\end{figure}
\begin{figure}
\centerline{\includegraphics[width=\columnwidth]{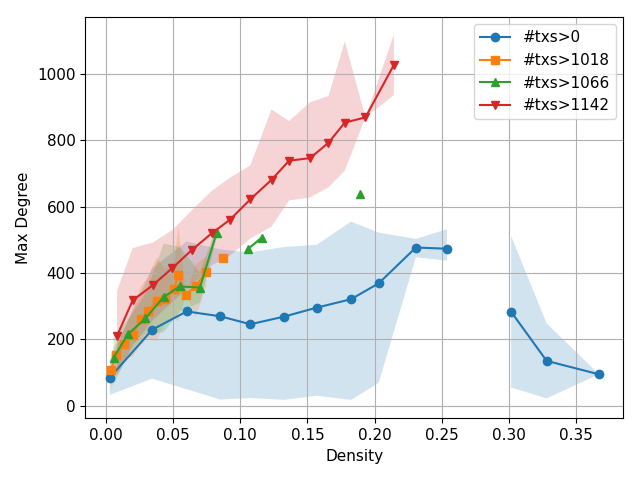}}
\caption{max-degree}
\label{fig:max-degree}
\end{figure}
\begin{figure}
\centerline{\includegraphics[width=\columnwidth]{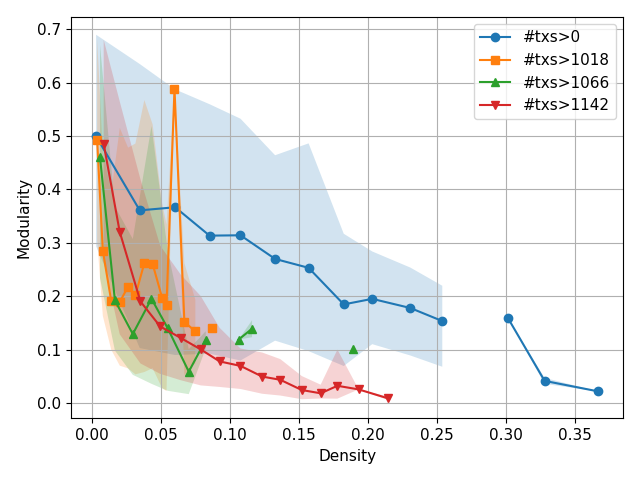}}
\caption{modularity}
\label{fig:modularity}
\end{figure}
\begin{figure}
\centerline{\includegraphics[width=\columnwidth]{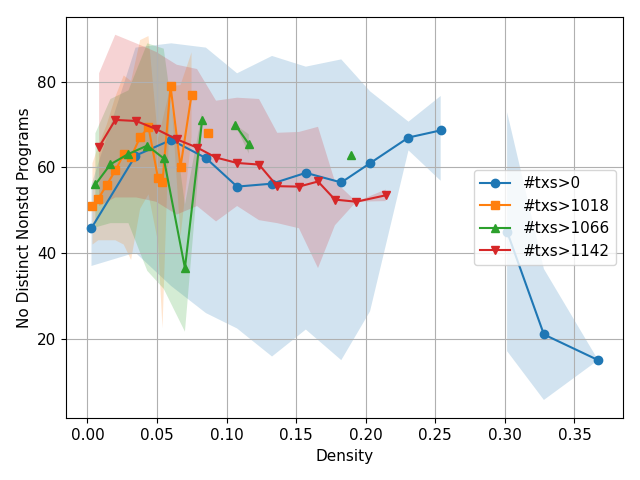}}
\caption{no-distinct-nonstd-programs}
\label{fig:no-distinct-nonstd-programs}
\end{figure}
\begin{figure}
\centerline{\includegraphics[width=\columnwidth]{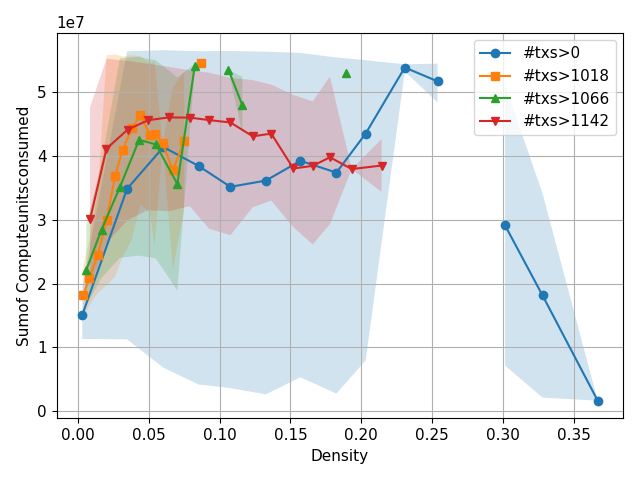}}
\caption{sumof-computeUnitsConsumed}
\label{fig:sumof-computeUnitsConsumed}
\end{figure}
\begin{figure}
\centerline{\includegraphics[width=\columnwidth]{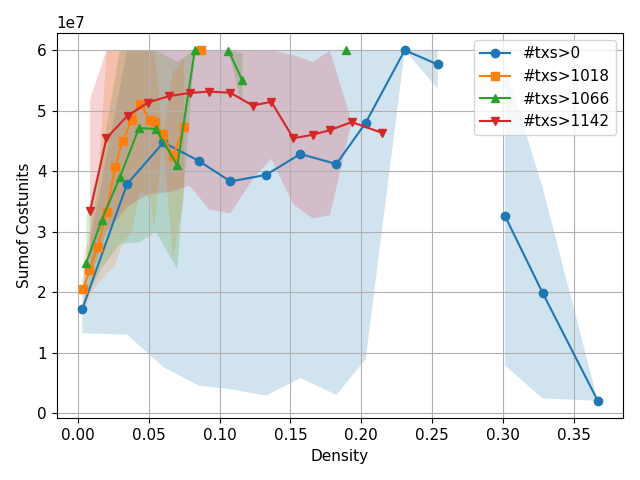}}
\caption{sumof-costUnits}
\label{fig:sumof-costUnits}
\end{figure}
\begin{figure}
\centerline{\includegraphics[width=\columnwidth]{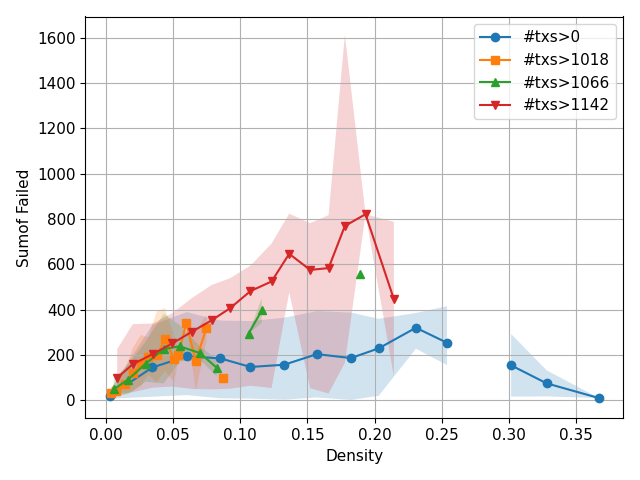}}
\caption{sumof-failed}
\label{fig:sumof-failed}
\end{figure}
\begin{figure}
\centerline{\includegraphics[width=\columnwidth]{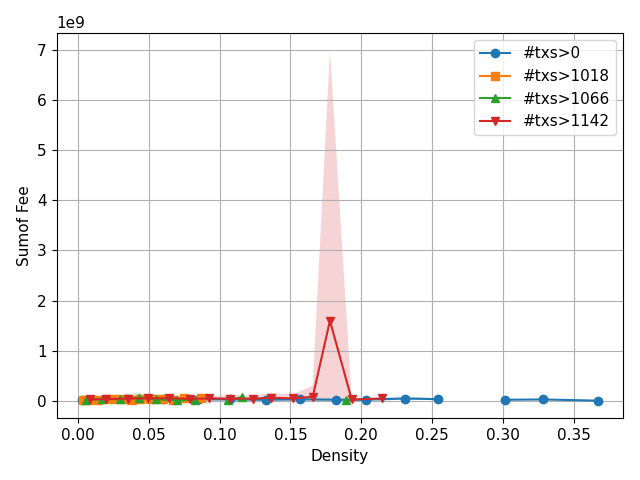}}
\caption{sumof-fee}
\label{fig:sumof-fee}
\end{figure}
\begin{figure}
\centerline{\includegraphics[width=\columnwidth]{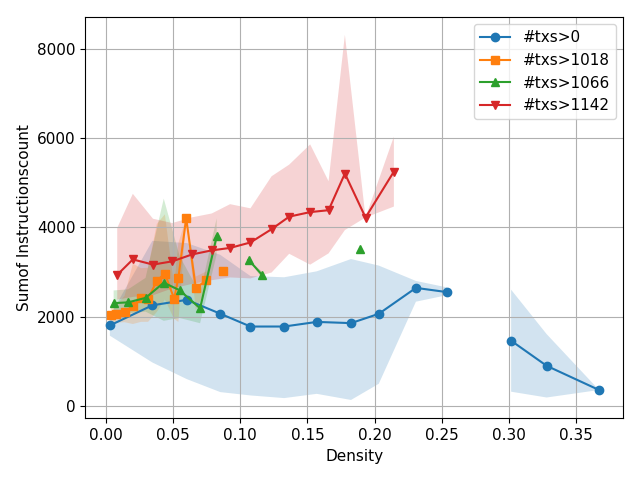}}
\caption{sumof-instructionsCount}
\label{fig:sumof-instructionsCount}
\end{figure}
\begin{figure}
\centerline{\includegraphics[width=\columnwidth]{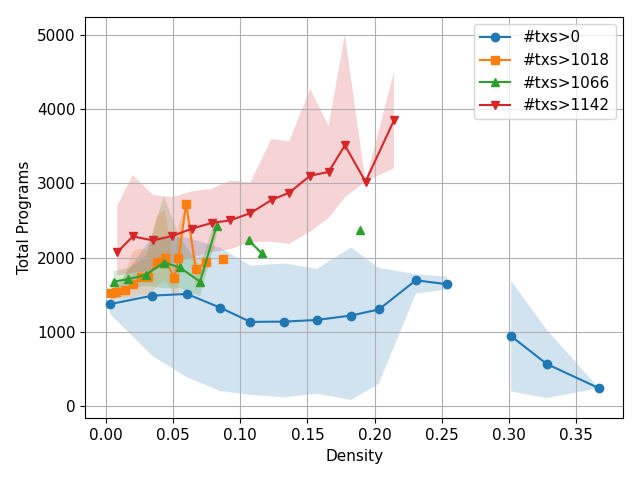}}
\caption{total-programs}
\label{fig:total-programs}
\end{figure}
\begin{figure}
\centerline{\includegraphics[width=\columnwidth]{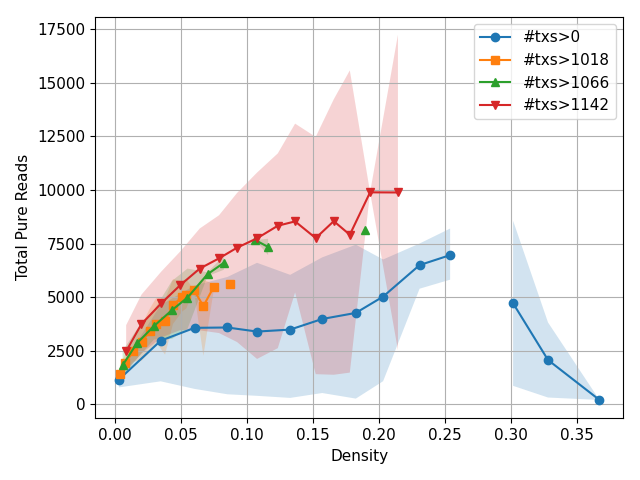}}
\caption{total-pure-reads}
\label{fig:total-pure-reads}
\end{figure}
\begin{figure}
\centerline{\includegraphics[width=\columnwidth]{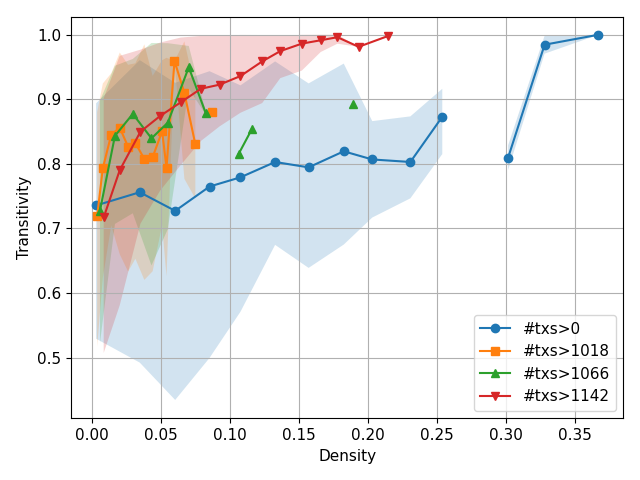}}
\caption{transitivity}
\label{fig:transitivity}
\end{figure}
\begin{figure}
\centerline{\includegraphics[width=\columnwidth]{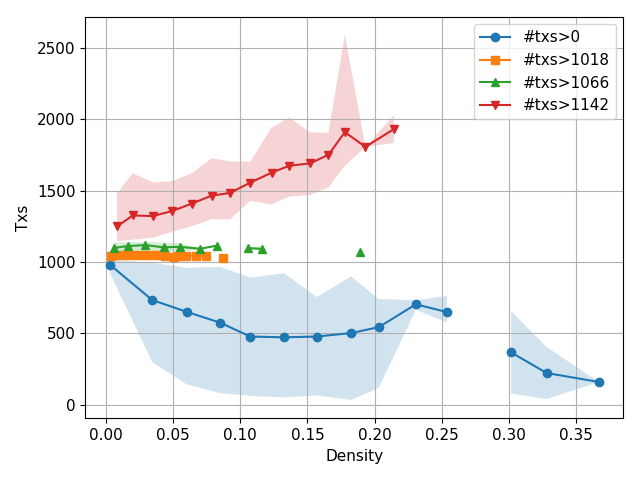}}
\caption{txs}
\label{fig:txs}
\end{figure}
\begin{figure}
\centerline{\includegraphics[width=\columnwidth]{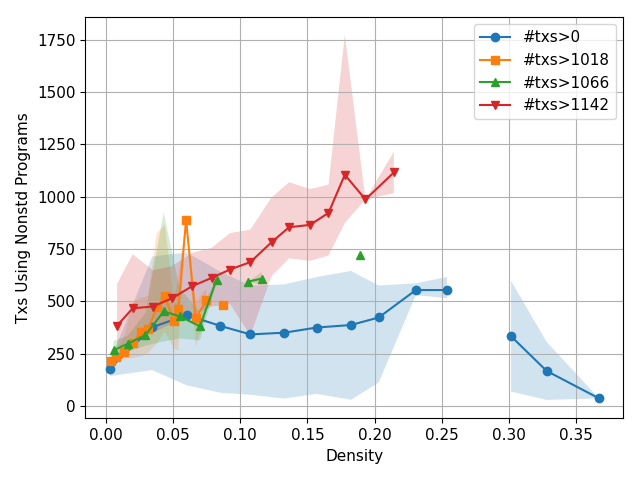}}
\caption{txs-using-nonstd-programs}
\label{fig:txs-using-nonstd-programs}
\end{figure}
\begin{figure}
\centerline{\includegraphics[width=\columnwidth]{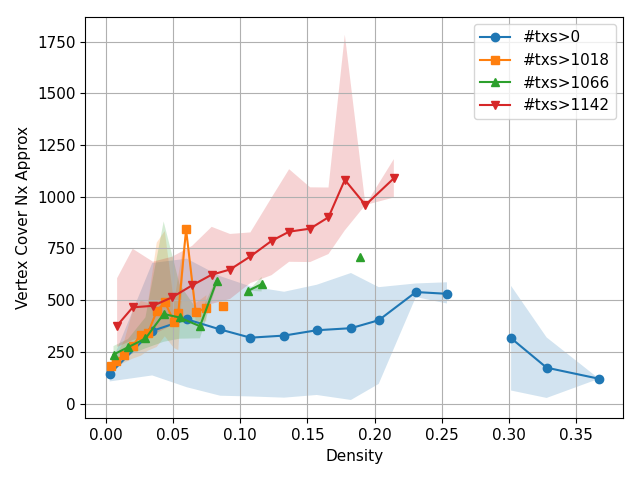}}
\caption{vertex-cover-nx-approx}
\label{fig:vertex-cover-nx-approx}
\end{figure}
\begin{figure}
\centerline{\includegraphics[width=\columnwidth]{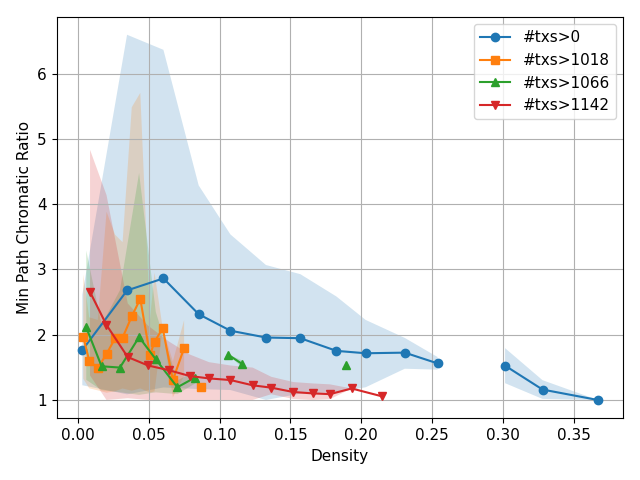}}
\caption{min-path-chromatic-ratio}
\label{fig:min-path-chromatic-ratio}
\end{figure}
\begin{figure}
\centerline{\includegraphics[width=\columnwidth]{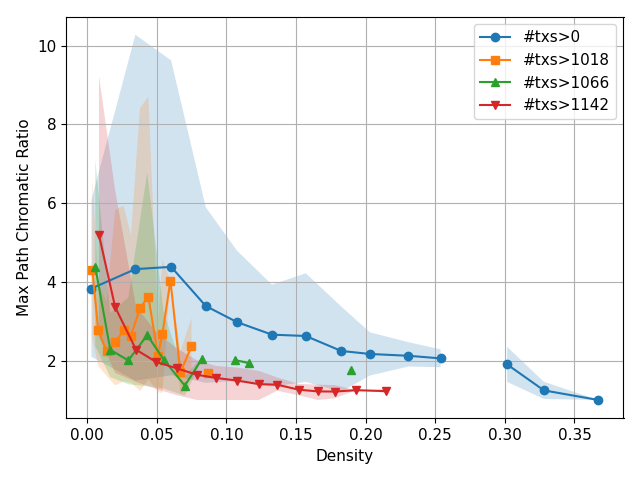}}
\caption{max-path-chromatic-ratio}
\label{fig:max-path-chromatic-ratio}
\end{figure}
\begin{figure}
\centerline{\includegraphics[width=\columnwidth]{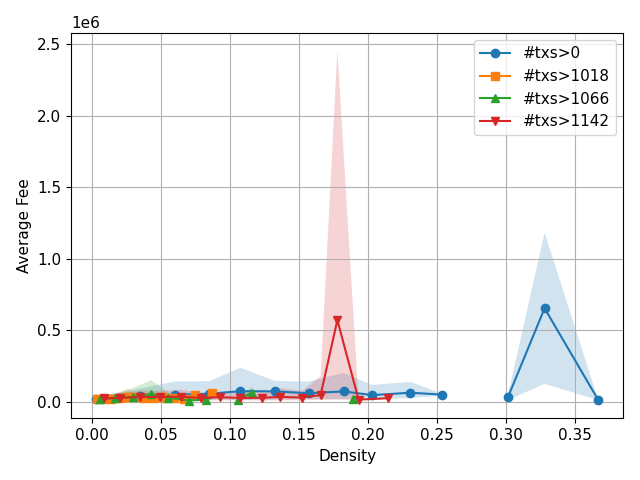}}
\caption{avg-fee}
\label{fig:avg-fee}
\end{figure}
\begin{figure}
\centerline{\includegraphics[width=\columnwidth]{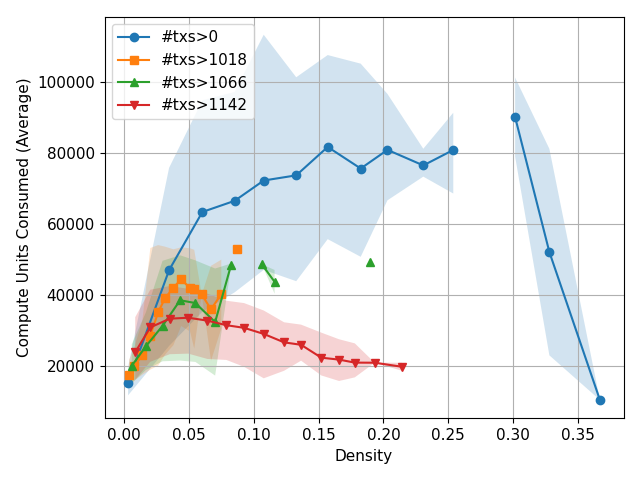}}
\caption{avg-computeUnitsConsumed}
\label{fig:avg-computeUnitsConsumed}
\end{figure}
\begin{figure}
\centerline{\includegraphics[width=\columnwidth]{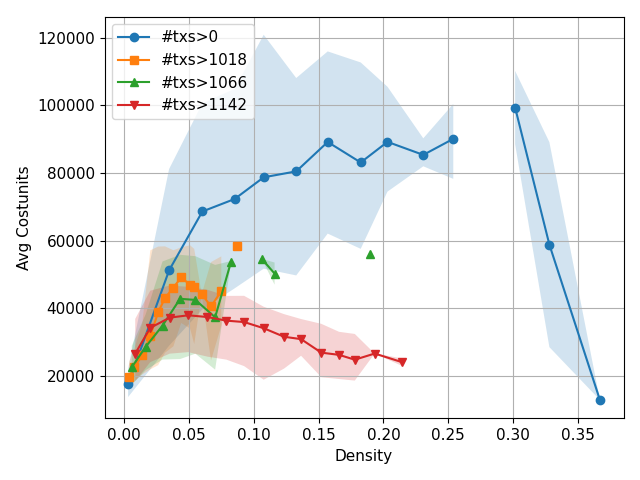}}
\caption{avg-costUnits}
\label{fig:avg-costUnits}
\end{figure}
\begin{figure}
\centerline{\includegraphics[width=\columnwidth]{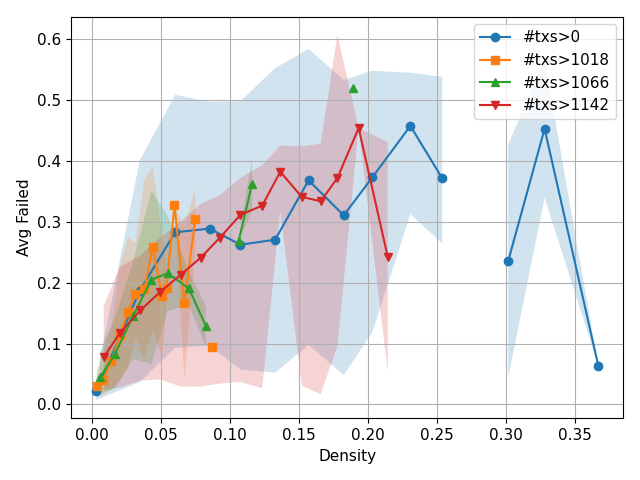}}
\caption{avg-failed}
\label{fig:avg-failed}
\end{figure}
\begin{figure}
\centerline{\includegraphics[width=\columnwidth]{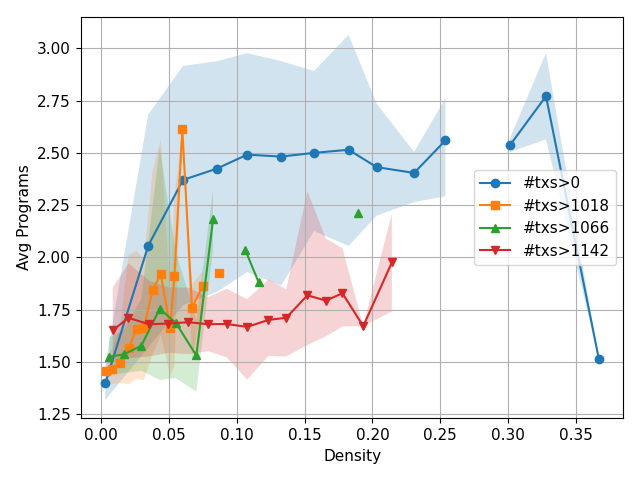}}
\caption{avg-programs}
\label{fig:avg-programs}
\end{figure}
\begin{figure}
\centerline{\includegraphics[width=\columnwidth]{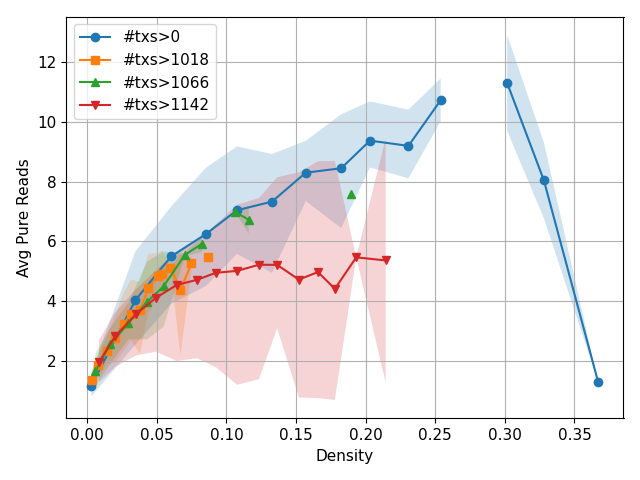}}
\caption{avg-pure-reads}
\label{fig:avg-pure-reads}
\end{figure}
\begin{figure}
\centerline{\includegraphics[width=\columnwidth]{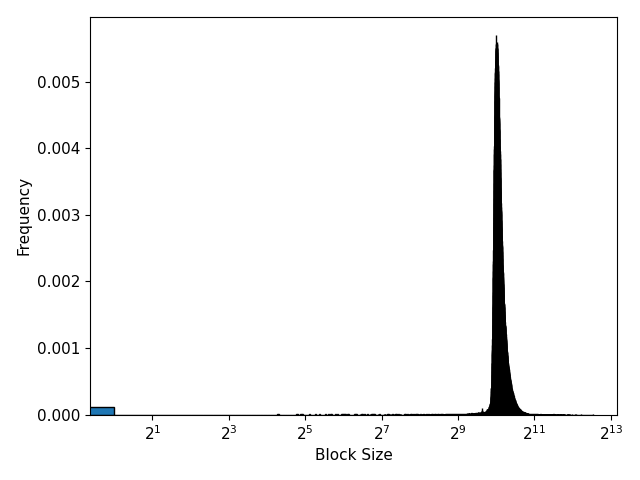}}
\caption{block-size-dist}
\label{fig:block-size-dist}
\end{figure}
\begin{figure}
\centerline{\includegraphics[width=\columnwidth]{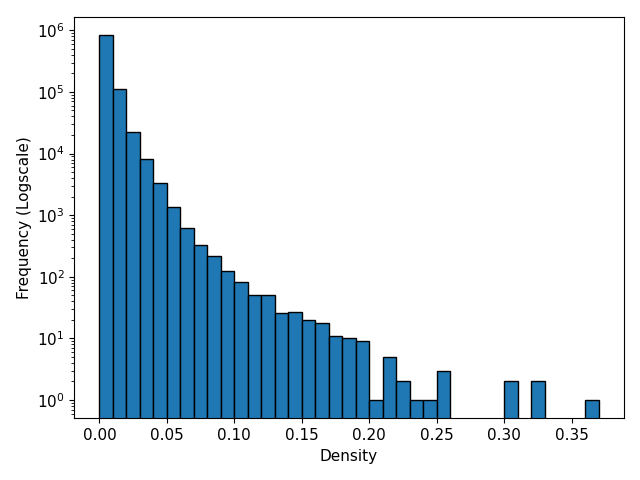}}
\caption{density-dist}
\label{fig:density-dist}
\end{figure}
\begin{figure}
\centerline{\includegraphics[width=\columnwidth]{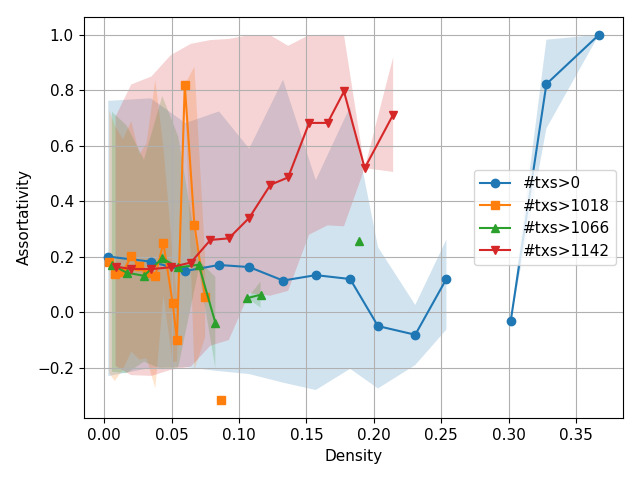}}
\caption{assortativity}
\label{fig:assortativity}
\end{figure}
\begin{figure}
\centerline{\includegraphics[width=\columnwidth]{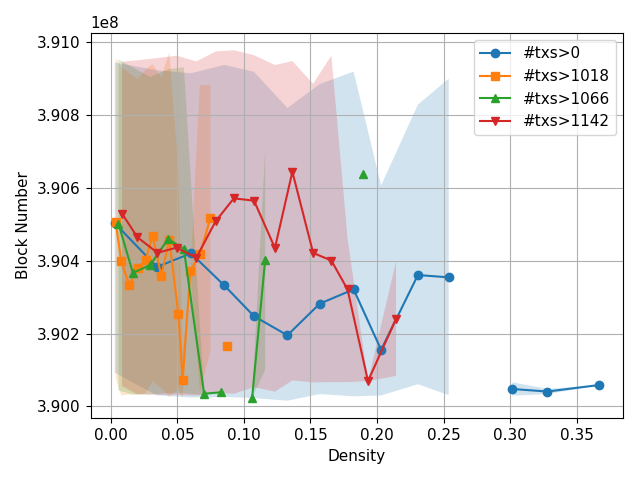}}
\caption{block-number}
\label{fig:block-number}
\end{figure}
\begin{figure}
\centerline{\includegraphics[width=\columnwidth]{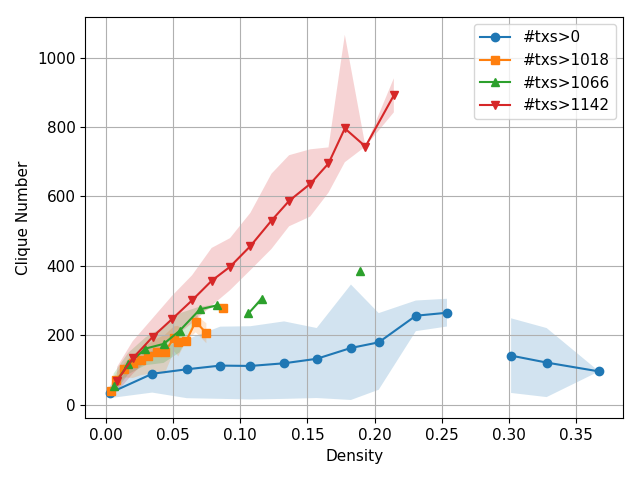}}
\caption{clique-number}
\label{fig:clique-number}
\end{figure}
\begin{figure}
\centerline{\includegraphics[width=\columnwidth]{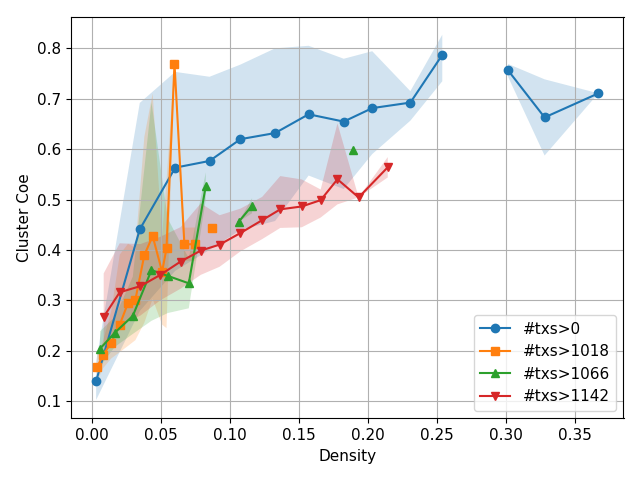}}
\caption{cluster-coe}
\label{fig:cluster-coe}
\end{figure}
\begin{figure}
\centerline{\includegraphics[width=\columnwidth]{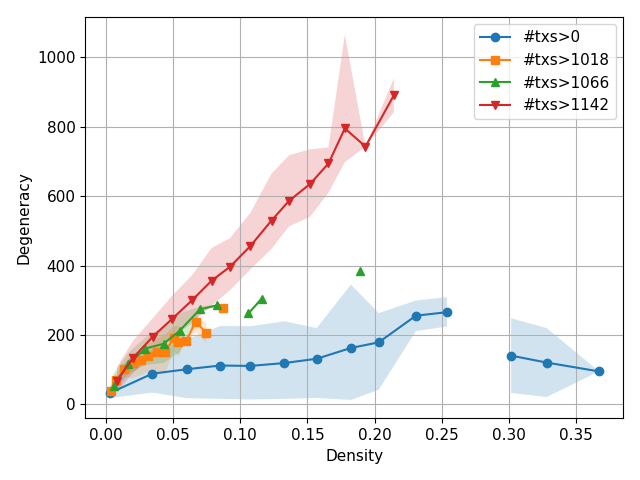}}
\caption{degeneracy}
\label{fig:degeneracy}
\end{figure}
\begin{figure}
\centerline{\includegraphics[width=\columnwidth]{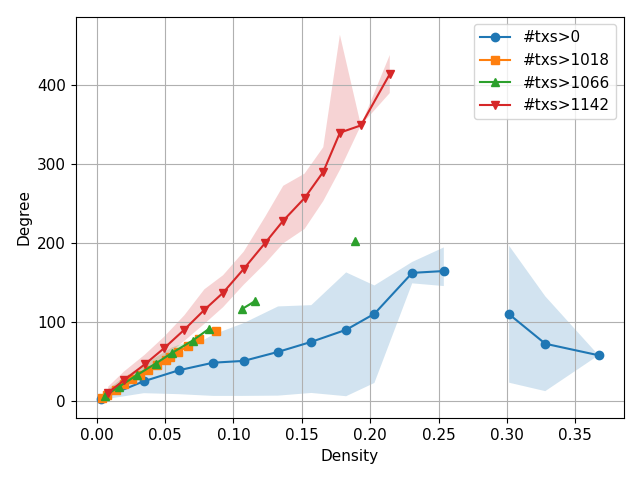}}
\caption{degree}
\label{fig:degree}
\end{figure}
\begin{figure}
\centerline{\includegraphics[width=\columnwidth]{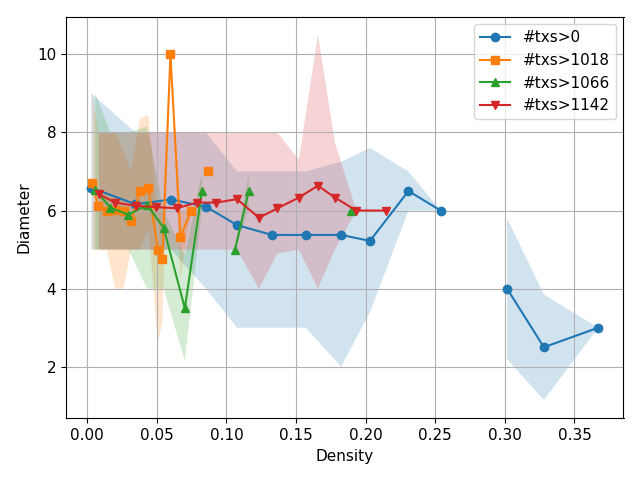}}
\caption{diameter}
\label{fig:diameter}
\end{figure}
\begin{figure}
\centerline{\includegraphics[width=\columnwidth]{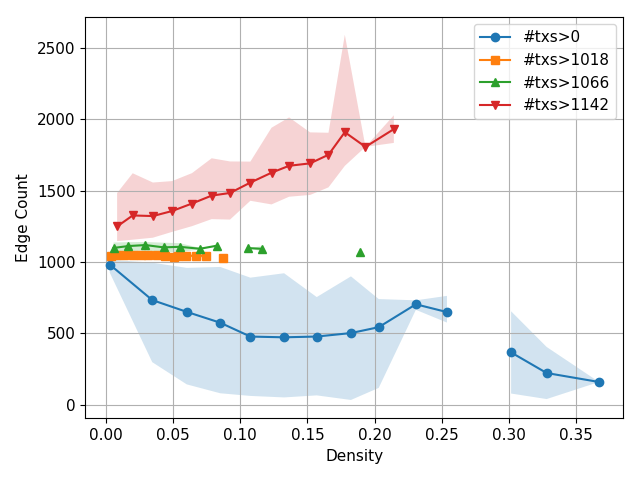}}
\caption{edge-count}
\label{fig:edge-count}
\end{figure}
\begin{figure}
\centerline{\includegraphics[width=\columnwidth]{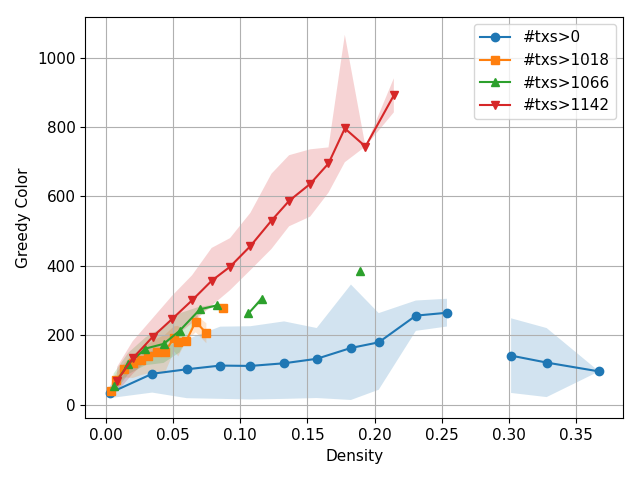}}
\caption{greedy-color}
\label{fig:greedy-color}
\end{figure}
\begin{figure}
\centerline{\includegraphics[width=\columnwidth]{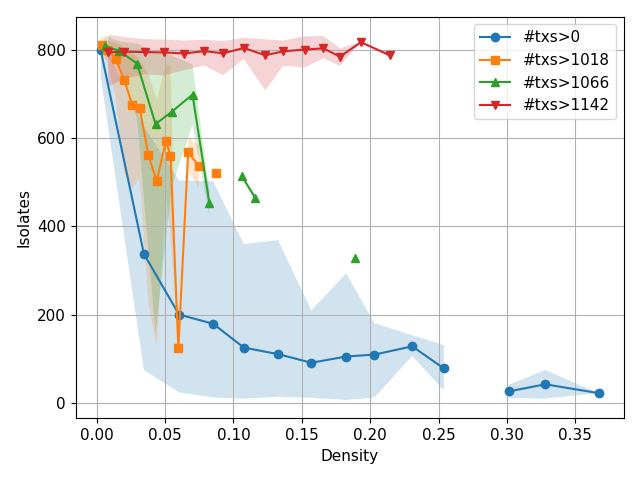}}
\caption{isolates}
\label{fig:isolates}
\end{figure}
\begin{figure}
\centerline{\includegraphics[width=\columnwidth]{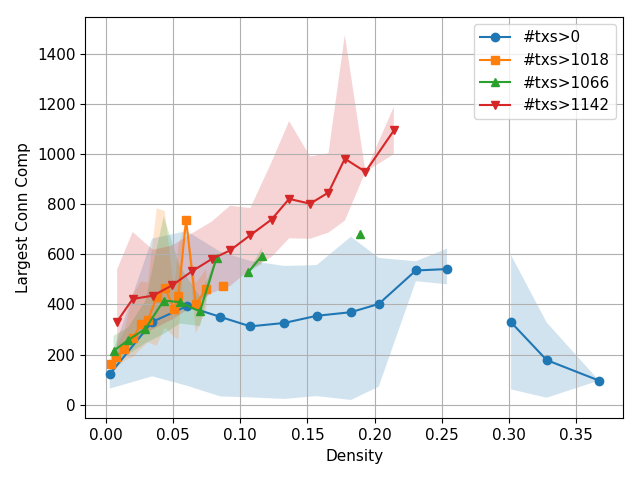}}
\caption{largest-conn-comp}
\label{fig:largest-conn-comp}
\end{figure}
\begin{figure}
\centerline{\includegraphics[width=\columnwidth]{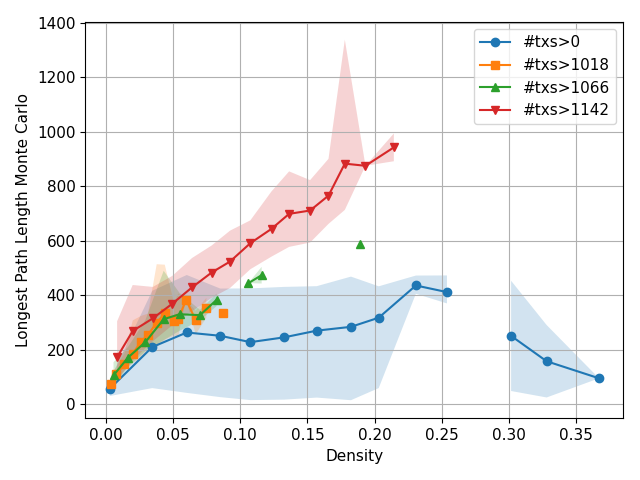}}
\caption{longest-path-length-monte-carlo}
\label{fig:longest-path-length-monte-carlo}
\end{figure}
\begin{figure}
\centerline{\includegraphics[width=\columnwidth]{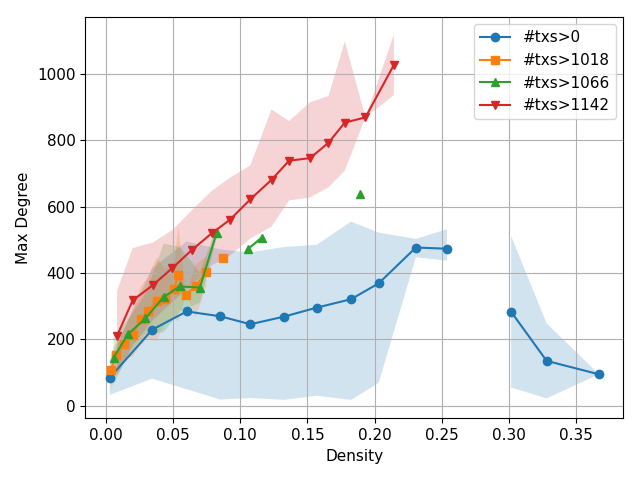}}
\caption{max-degree}
\label{fig:max-degree}
\end{figure}
\begin{figure}
\centerline{\includegraphics[width=\columnwidth]{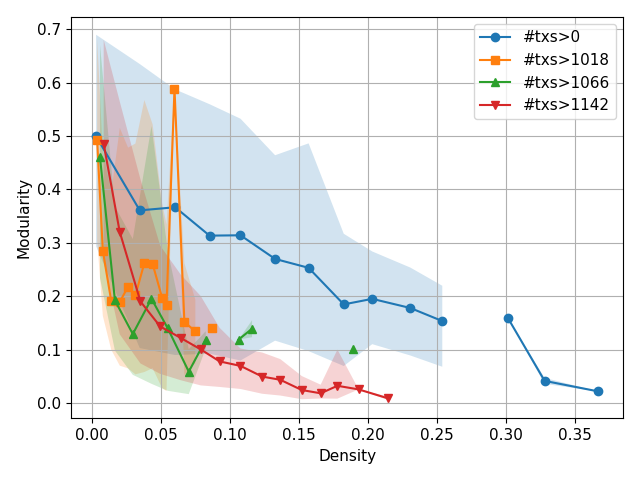}}
\caption{modularity}
\label{fig:modularity}
\end{figure}
\begin{figure}
\centerline{\includegraphics[width=\columnwidth]{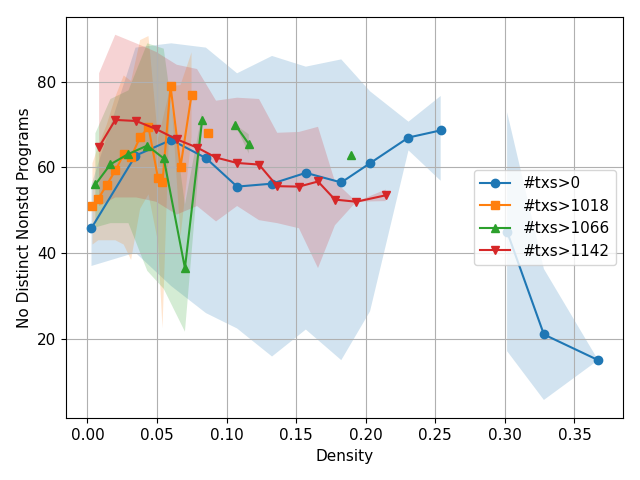}}
\caption{no-distinct-nonstd-programs}
\label{fig:no-distinct-nonstd-programs}
\end{figure}
\begin{figure}
\centerline{\includegraphics[width=\columnwidth]{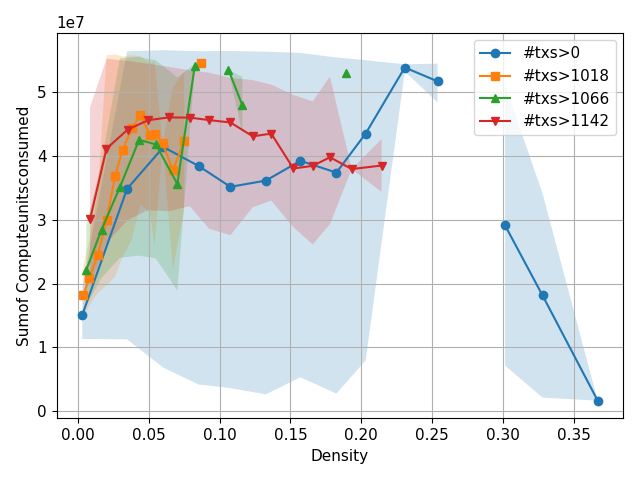}}
\caption{sumof-computeUnitsConsumed}
\label{fig:sumof-computeUnitsConsumed}
\end{figure}
\begin{figure}
\centerline{\includegraphics[width=\columnwidth]{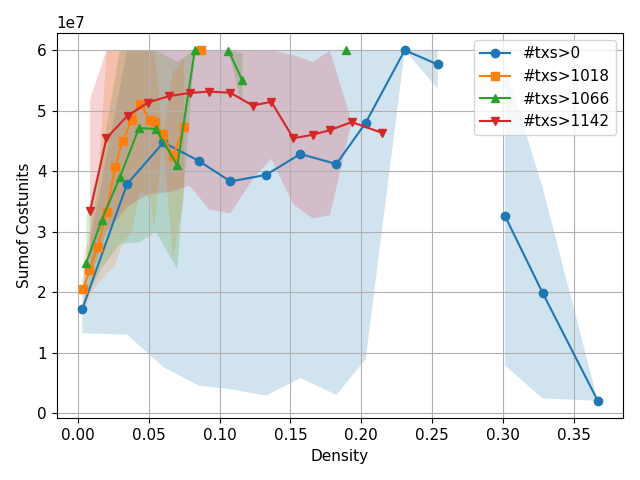}}
\caption{sumof-costUnits}
\label{fig:sumof-costUnits}
\end{figure}
\begin{figure}
\centerline{\includegraphics[width=\columnwidth]{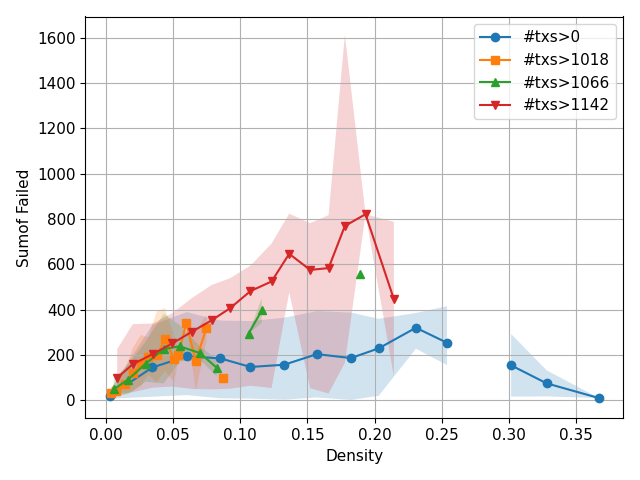}}
\caption{sumof-failed}
\label{fig:sumof-failed}
\end{figure}
\begin{figure}
\centerline{\includegraphics[width=\columnwidth]{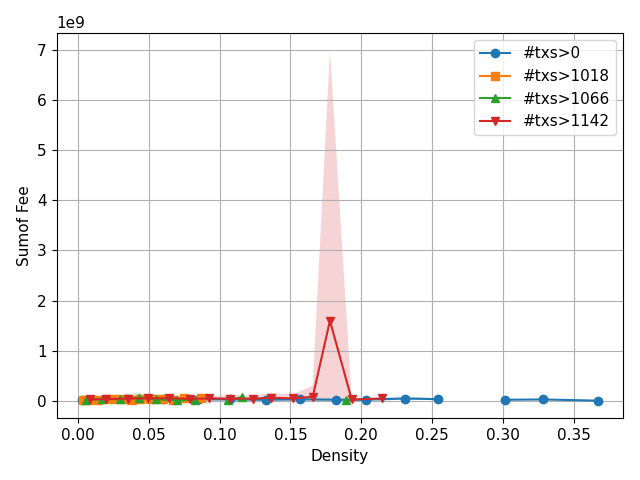}}
\caption{sumof-fee}
\label{fig:sumof-fee}
\end{figure}
\begin{figure}
\centerline{\includegraphics[width=\columnwidth]{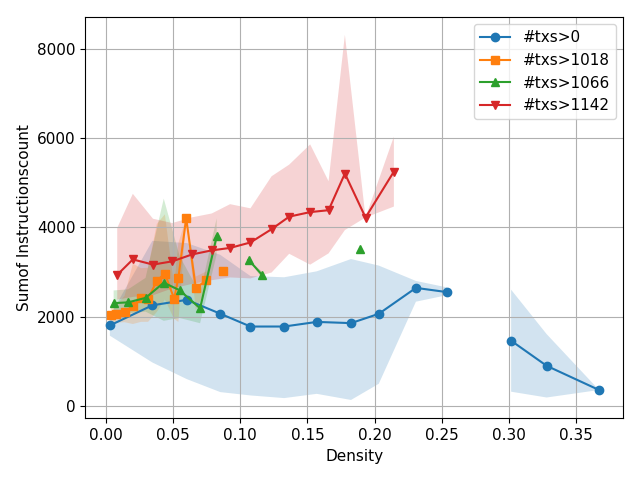}}
\caption{sumof-instructionsCount}
\label{fig:sumof-instructionsCount}
\end{figure}
\begin{figure}
\centerline{\includegraphics[width=\columnwidth]{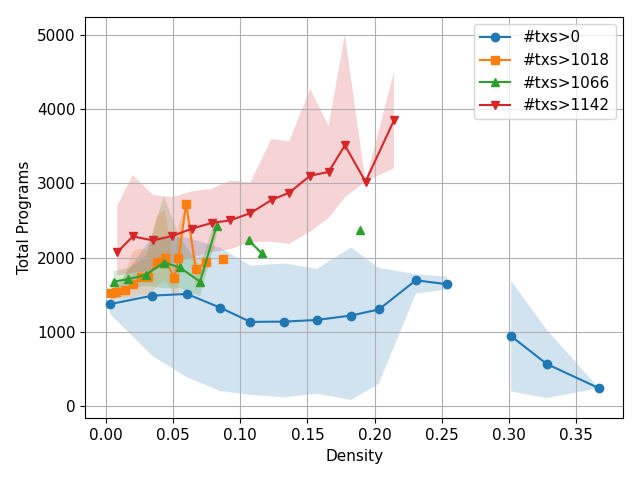}}
\caption{total-programs}
\label{fig:total-programs}
\end{figure}
\begin{figure}
\centerline{\includegraphics[width=\columnwidth]{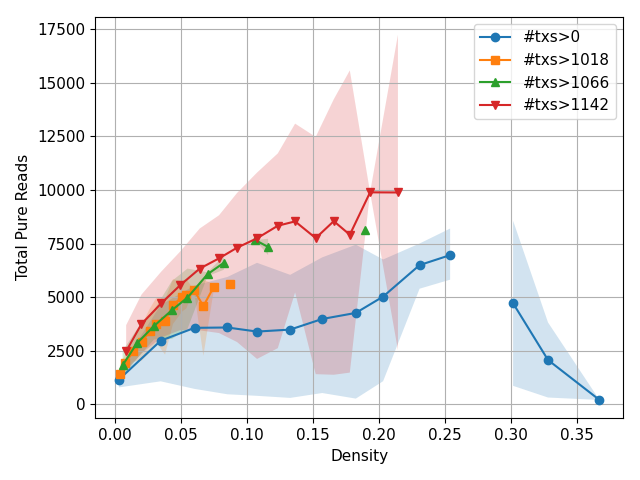}}
\caption{total-pure-reads}
\label{fig:total-pure-reads}
\end{figure}
\begin{figure}
\centerline{\includegraphics[width=\columnwidth]{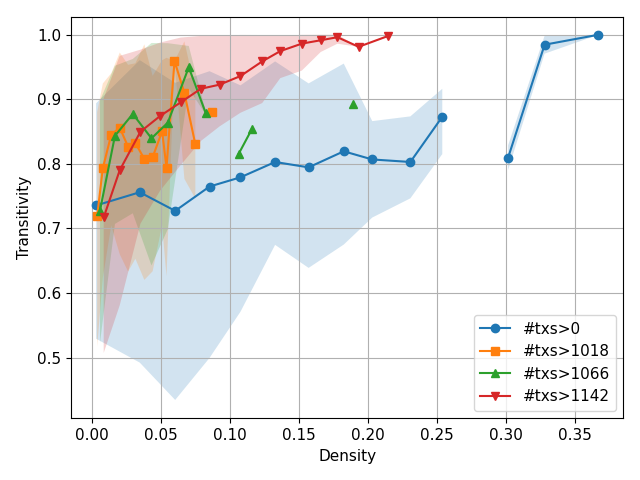}}
\caption{transitivity}
\label{fig:transitivity}
\end{figure}
\begin{figure}
\centerline{\includegraphics[width=\columnwidth]{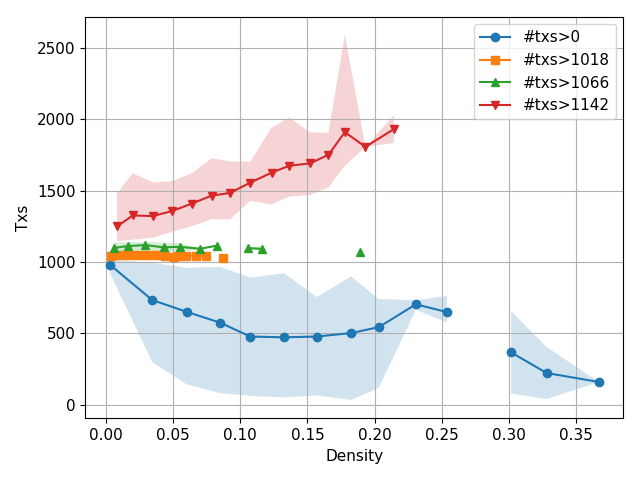}}
\caption{txs}
\label{fig:txs}
\end{figure}
\begin{figure}
\centerline{\includegraphics[width=\columnwidth]{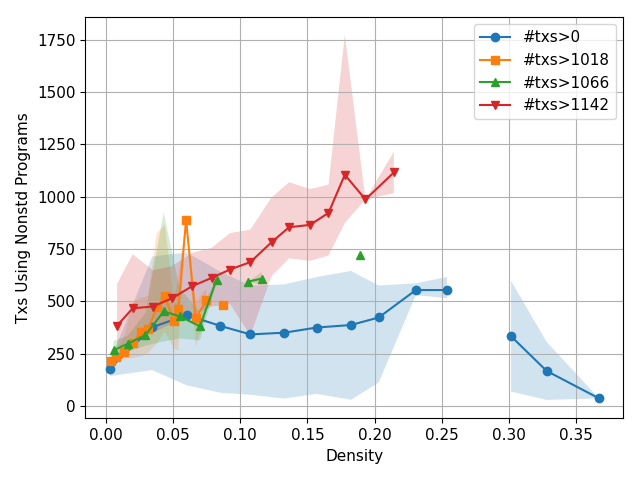}}
\caption{txs-using-nonstd-programs}
\label{fig:txs-using-nonstd-programs}
\end{figure}
\begin{figure}
\centerline{\includegraphics[width=\columnwidth]{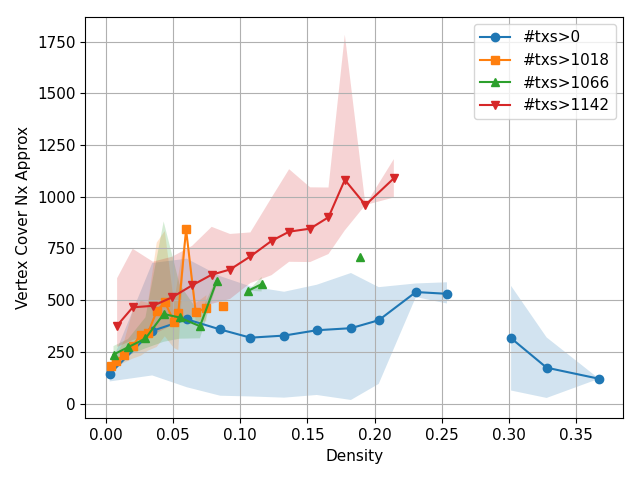}}
\caption{vertex-cover-nx-approx}
\label{fig:vertex-cover-nx-approx}
\end{figure}
\begin{figure}
\centerline{\includegraphics[width=\columnwidth]{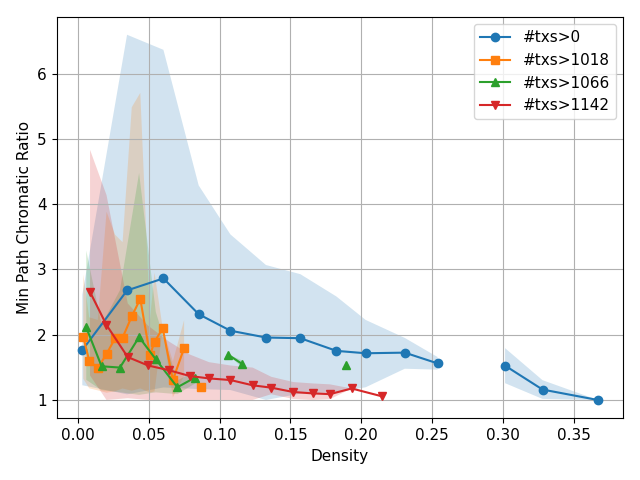}}
\caption{min-path-chromatic-ratio}
\label{fig:min-path-chromatic-ratio}
\end{figure}
\begin{figure}
\centerline{\includegraphics[width=\columnwidth]{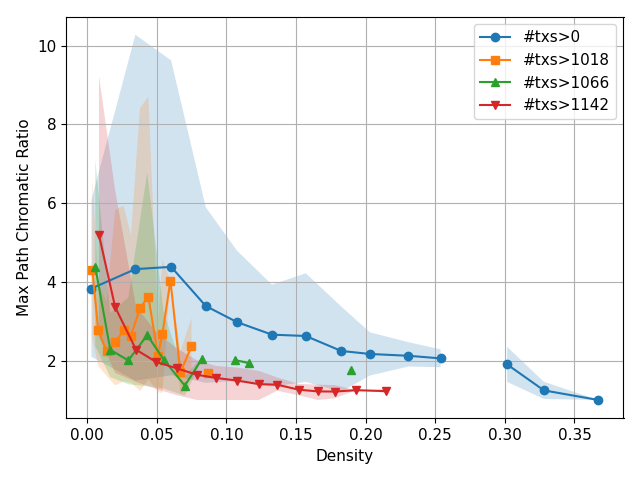}}
\caption{max-path-chromatic-ratio}
\label{fig:max-path-chromatic-ratio}
\end{figure}
\begin{figure}
\centerline{\includegraphics[width=\columnwidth]{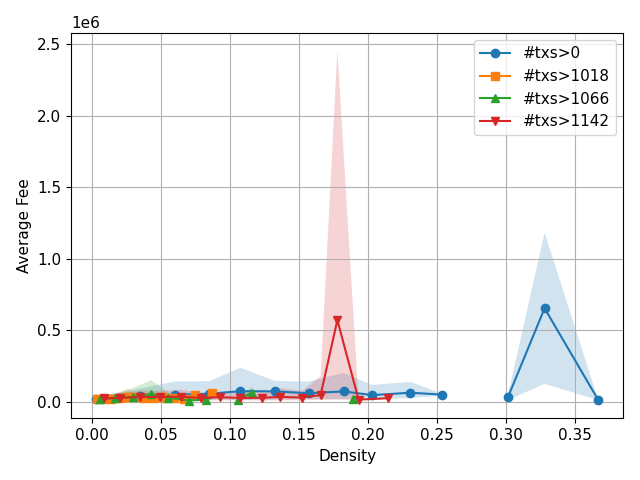}}
\caption{avg-fee}
\label{fig:avg-fee}
\end{figure}
\begin{figure}
\centerline{\includegraphics[width=\columnwidth]{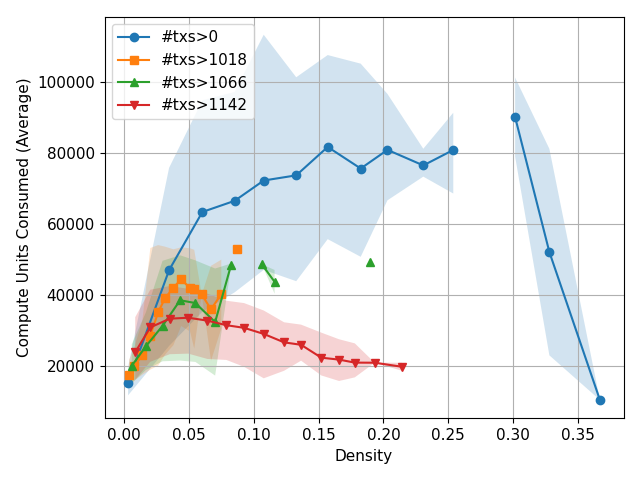}}
\caption{avg-computeUnitsConsumed}
\label{fig:avg-computeUnitsConsumed}
\end{figure}
\begin{figure}
\centerline{\includegraphics[width=\columnwidth]{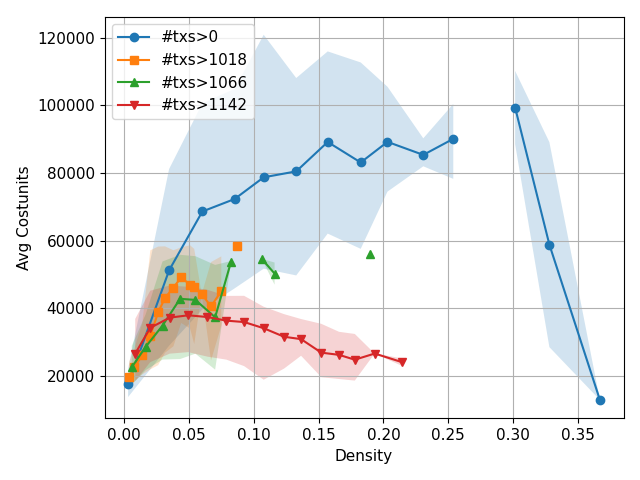}}
\caption{avg-costUnits}
\label{fig:avg-costUnits}
\end{figure}
\begin{figure}
\centerline{\includegraphics[width=\columnwidth]{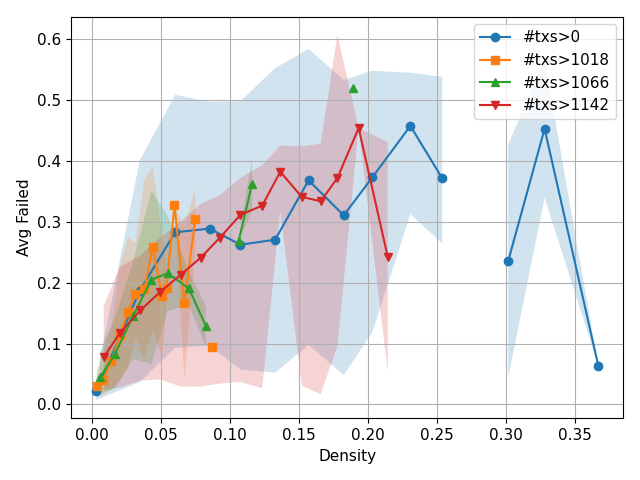}}
\caption{avg-failed}
\label{fig:avg-failed}
\end{figure}
\begin{figure}
\centerline{\includegraphics[width=\columnwidth]{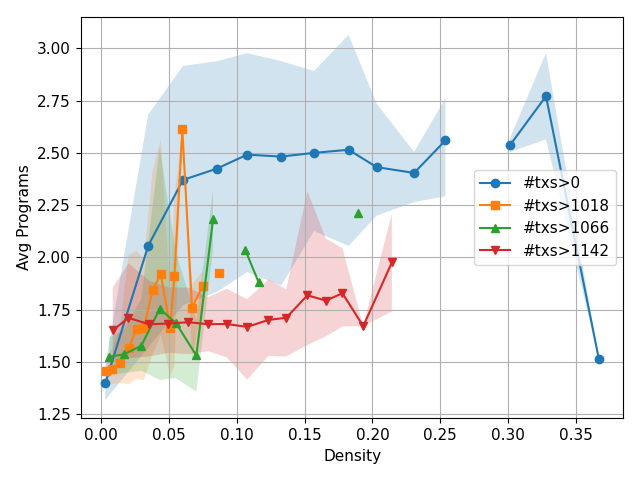}}
\caption{avg-programs}
\label{fig:avg-programs}
\end{figure}
\begin{figure}
\centerline{\includegraphics[width=\columnwidth]{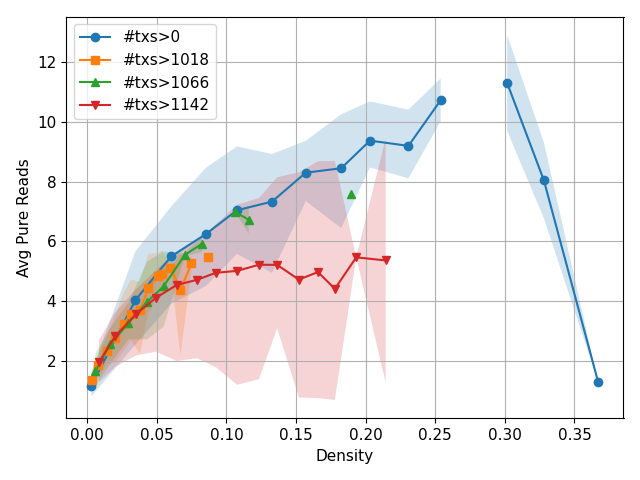}}
\caption{avg-pure-reads}
\label{fig:avg-pure-reads}
\end{figure}
\begin{figure}
\centerline{\includegraphics[width=\columnwidth]{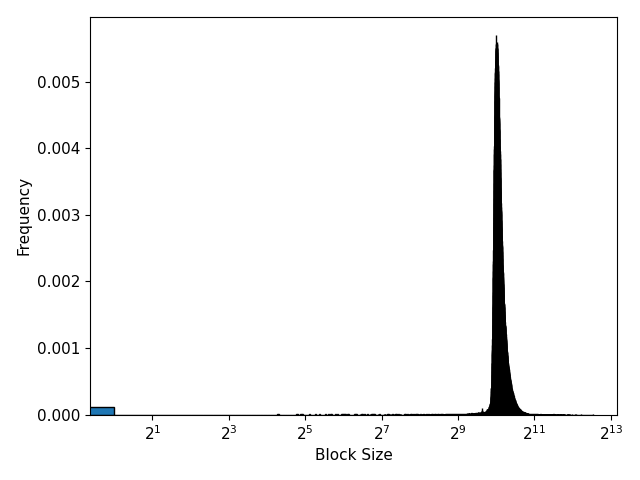}}
\caption{block-size-dist}
\label{fig:block-size-dist}
\end{figure}
\begin{figure}
\centerline{\includegraphics[width=\columnwidth]{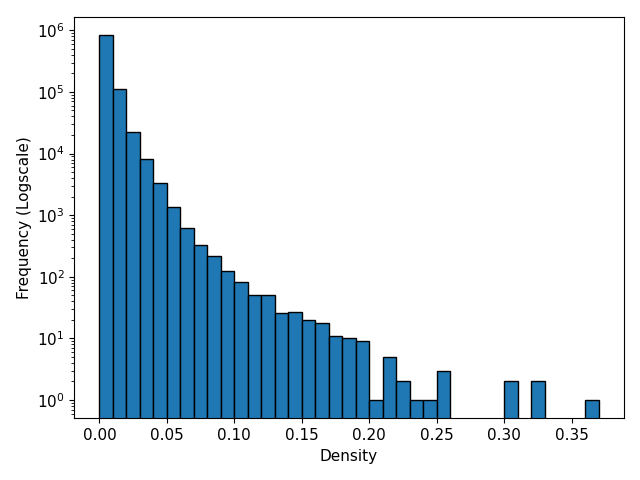}}
\caption{density-dist}
\label{fig:density-dist}
\end{figure}
\begin{figure}
\centerline{\includegraphics[width=\columnwidth]{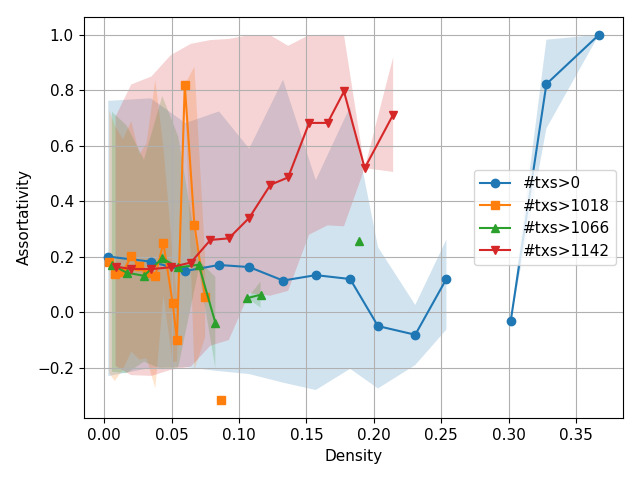}}
\caption{assortativity}
\label{fig:assortativity}
\end{figure}
\begin{figure}
\centerline{\includegraphics[width=\columnwidth]{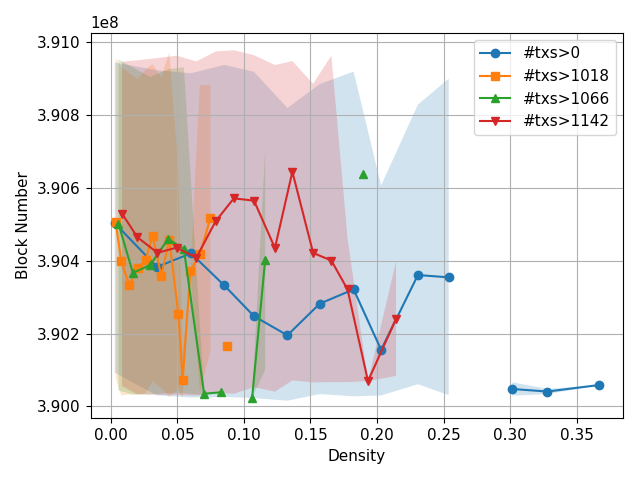}}
\caption{block-number}
\label{fig:block-number}
\end{figure}
\begin{figure}
\centerline{\includegraphics[width=\columnwidth]{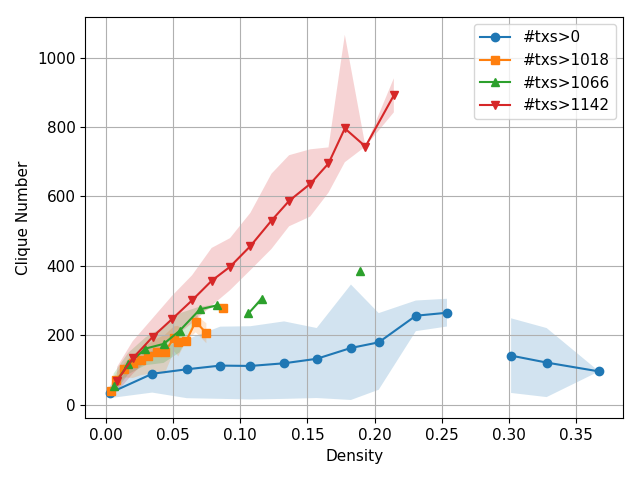}}
\caption{clique-number}
\label{fig:clique-number}
\end{figure}
\begin{figure}
\centerline{\includegraphics[width=\columnwidth]{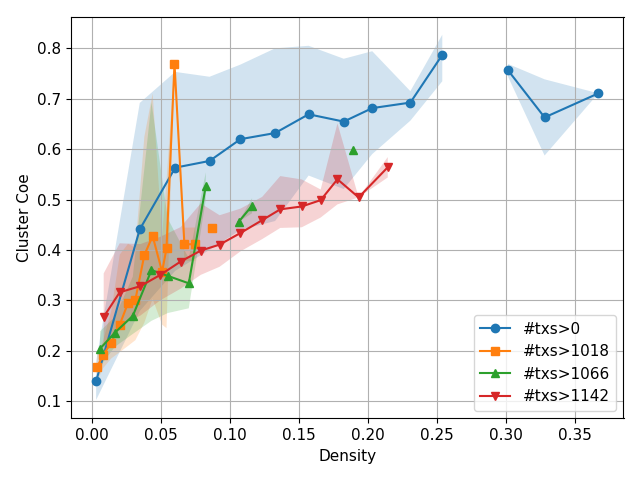}}
\caption{cluster-coe}
\label{fig:cluster-coe}
\end{figure}
\begin{figure}
\centerline{\includegraphics[width=\columnwidth]{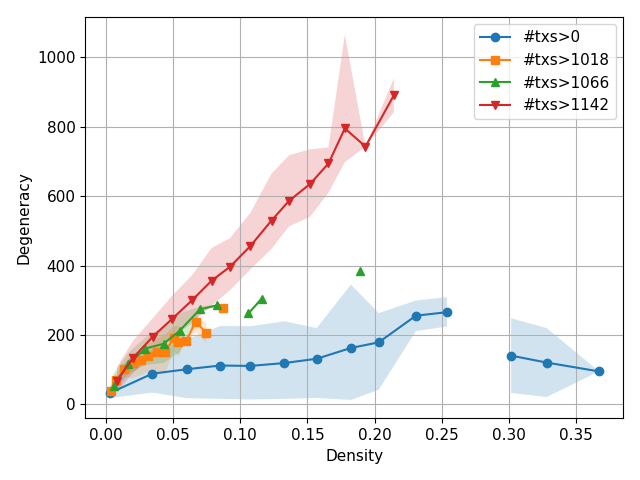}}
\caption{degeneracy}
\label{fig:degeneracy}
\end{figure}
\begin{figure}
\centerline{\includegraphics[width=\columnwidth]{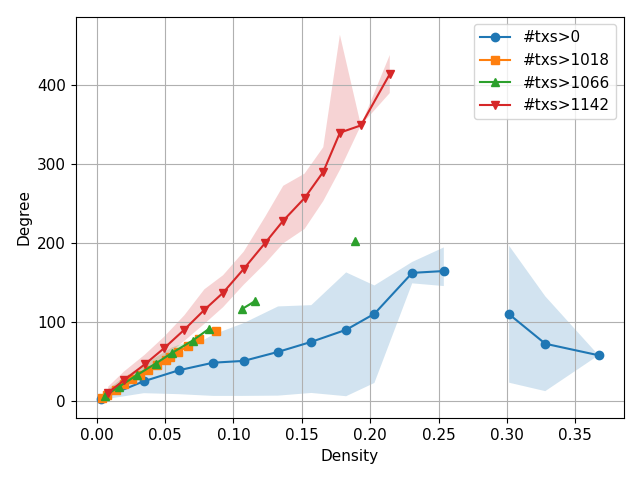}}
\caption{degree}
\label{fig:degree}
\end{figure}
\begin{figure}
\centerline{\includegraphics[width=\columnwidth]{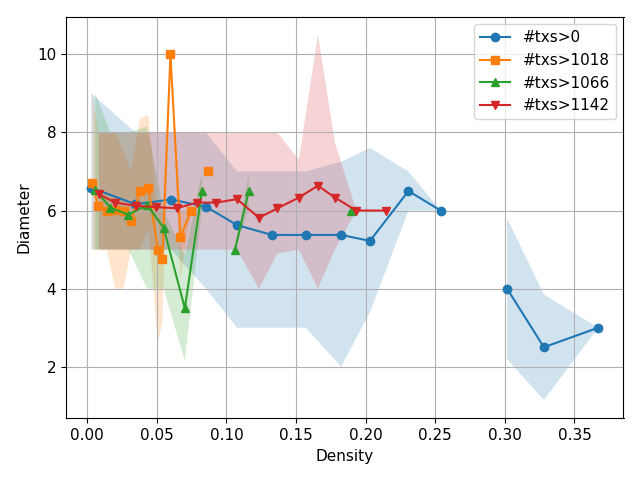}}
\caption{diameter}
\label{fig:diameter}
\end{figure}
\begin{figure}
\centerline{\includegraphics[width=\columnwidth]{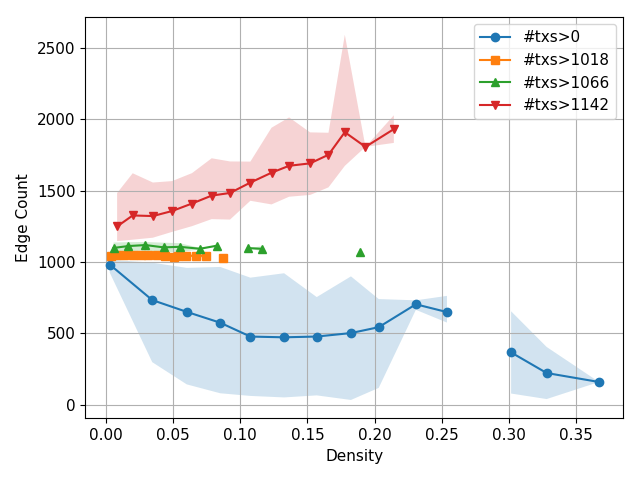}}
\caption{edge-count}
\label{fig:edge-count}
\end{figure}
\begin{figure}
\centerline{\includegraphics[width=\columnwidth]{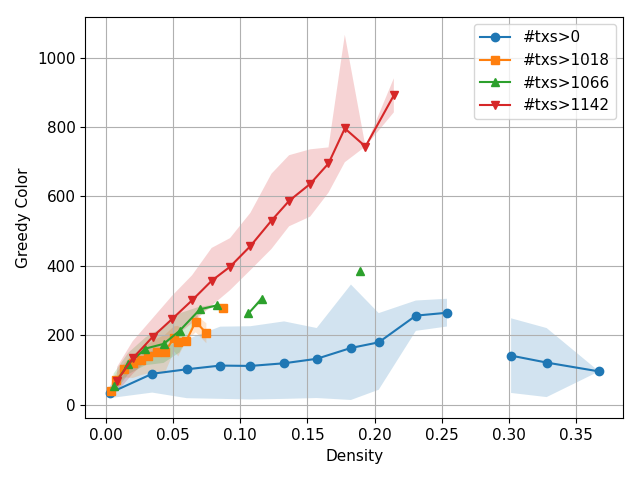}}
\caption{greedy-color}
\label{fig:greedy-color}
\end{figure}
\begin{figure}
\centerline{\includegraphics[width=\columnwidth]{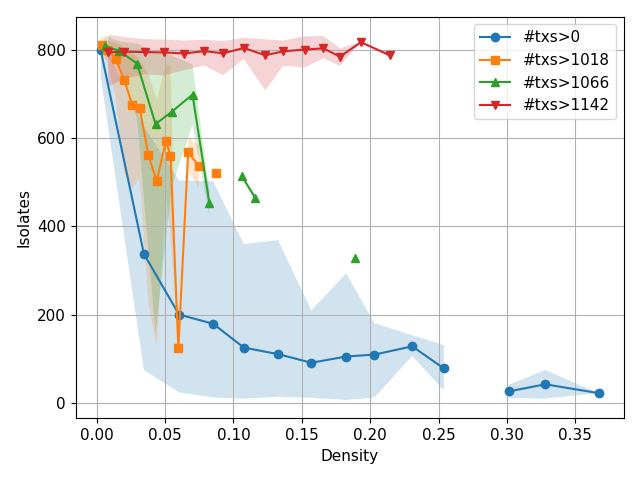}}
\caption{isolates}
\label{fig:isolates}
\end{figure}
\begin{figure}
\centerline{\includegraphics[width=\columnwidth]{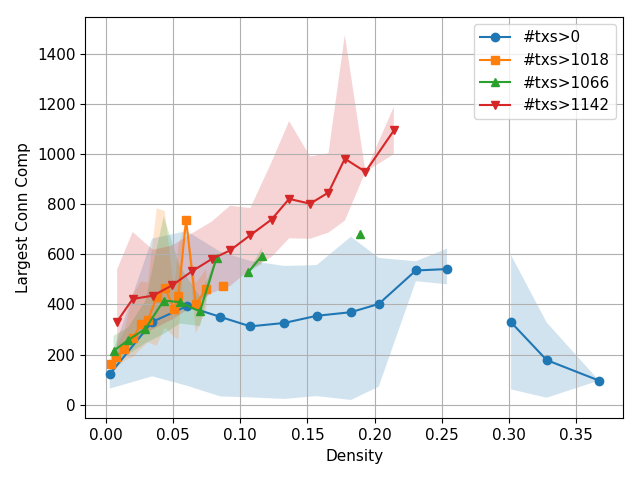}}
\caption{largest-conn-comp}
\label{fig:largest-conn-comp}
\end{figure}
\begin{figure}
\centerline{\includegraphics[width=\columnwidth]{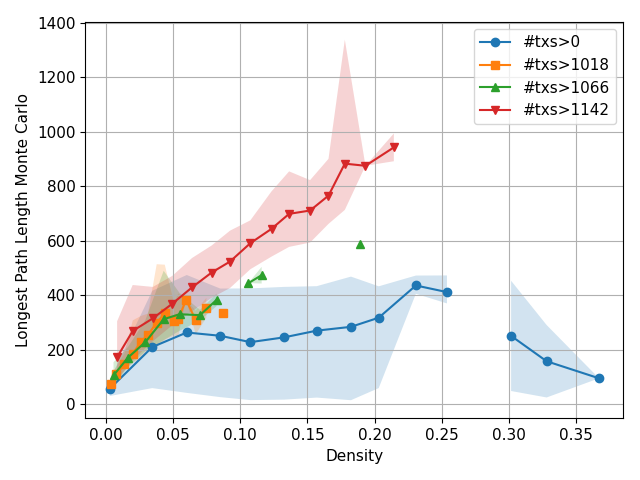}}
\caption{longest-path-length-monte-carlo}
\label{fig:longest-path-length-monte-carlo}
\end{figure}
\begin{figure}
\centerline{\includegraphics[width=\columnwidth]{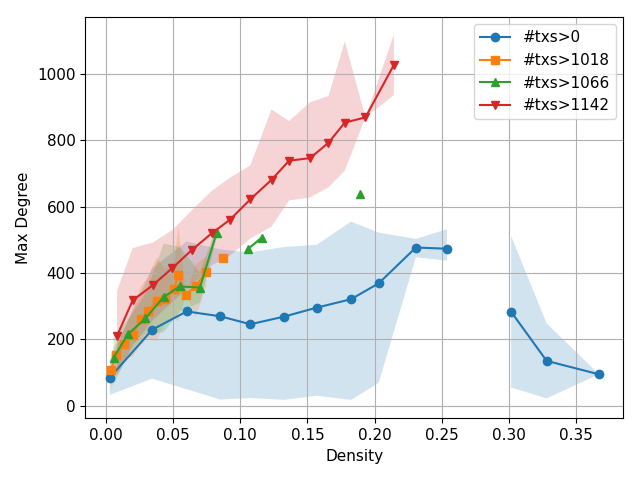}}
\caption{max-degree}
\label{fig:max-degree}
\end{figure}
\begin{figure}
\centerline{\includegraphics[width=\columnwidth]{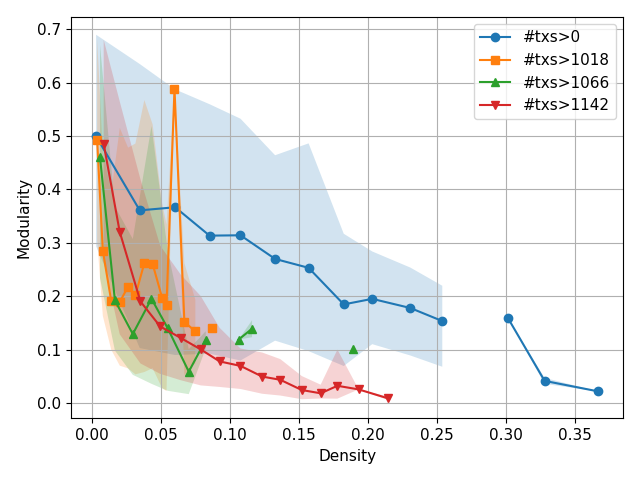}}
\caption{modularity}
\label{fig:modularity}
\end{figure}
\begin{figure}
\centerline{\includegraphics[width=\columnwidth]{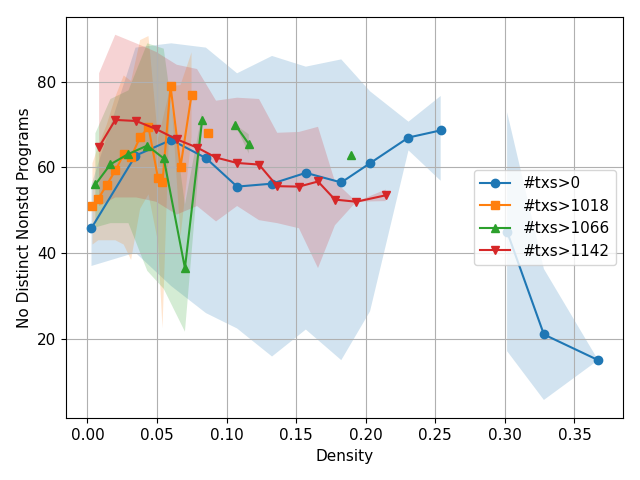}}
\caption{no-distinct-nonstd-programs}
\label{fig:no-distinct-nonstd-programs}
\end{figure}
\begin{figure}
\centerline{\includegraphics[width=\columnwidth]{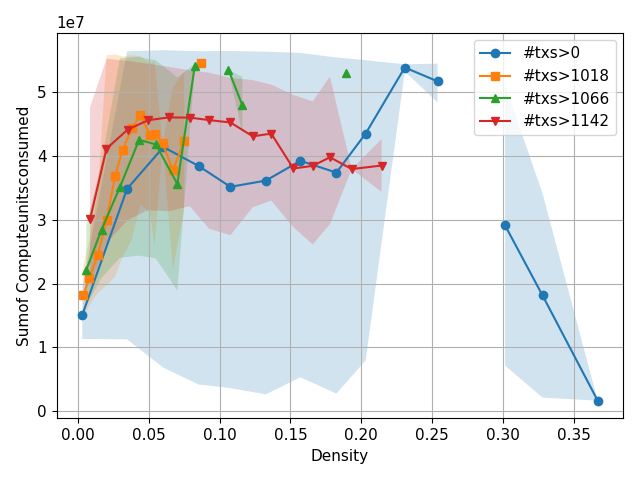}}
\caption{sumof-computeUnitsConsumed}
\label{fig:sumof-computeUnitsConsumed}
\end{figure}
\begin{figure}
\centerline{\includegraphics[width=\columnwidth]{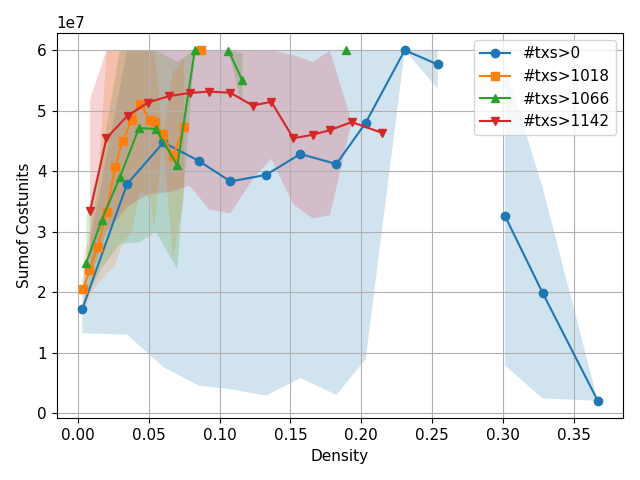}}
\caption{sumof-costUnits}
\label{fig:sumof-costUnits}
\end{figure}
\begin{figure}
\centerline{\includegraphics[width=\columnwidth]{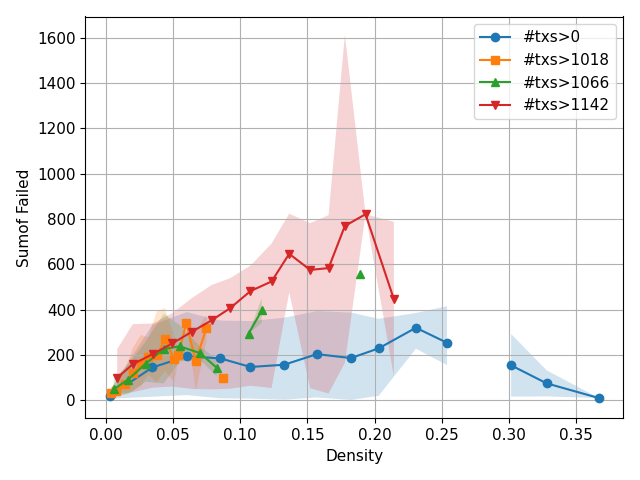}}
\caption{sumof-failed}
\label{fig:sumof-failed}
\end{figure}
\begin{figure}
\centerline{\includegraphics[width=\columnwidth]{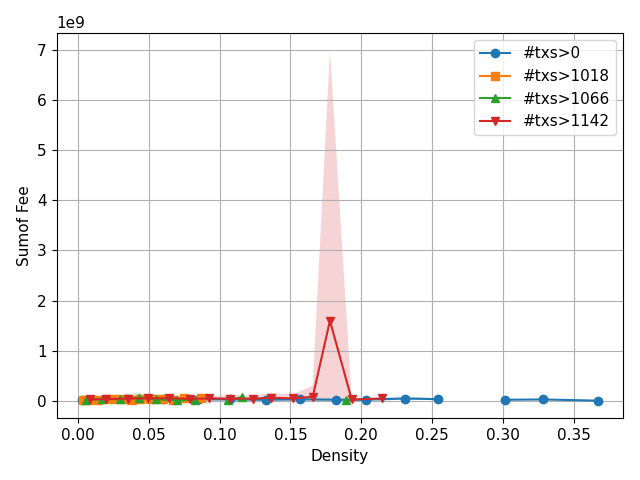}}
\caption{sumof-fee}
\label{fig:sumof-fee}
\end{figure}
\begin{figure}
\centerline{\includegraphics[width=\columnwidth]{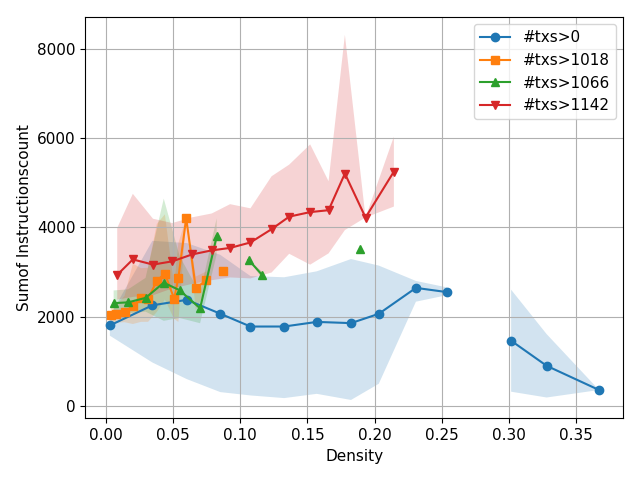}}
\caption{sumof-instructionsCount}
\label{fig:sumof-instructionsCount}
\end{figure}
\begin{figure}
\centerline{\includegraphics[width=\columnwidth]{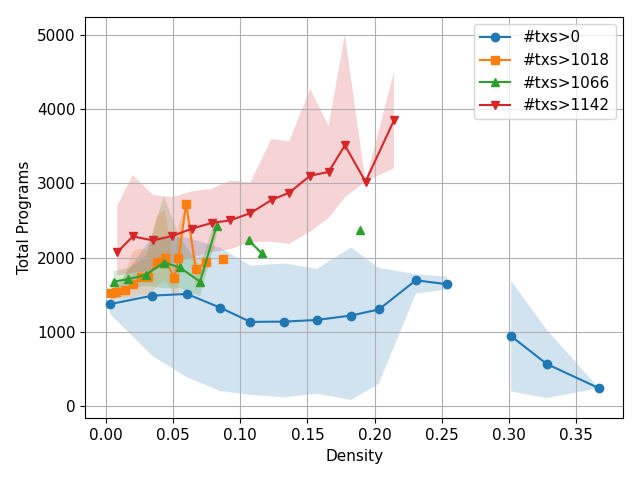}}
\caption{total-programs}
\label{fig:total-programs}
\end{figure}
\begin{figure}
\centerline{\includegraphics[width=\columnwidth]{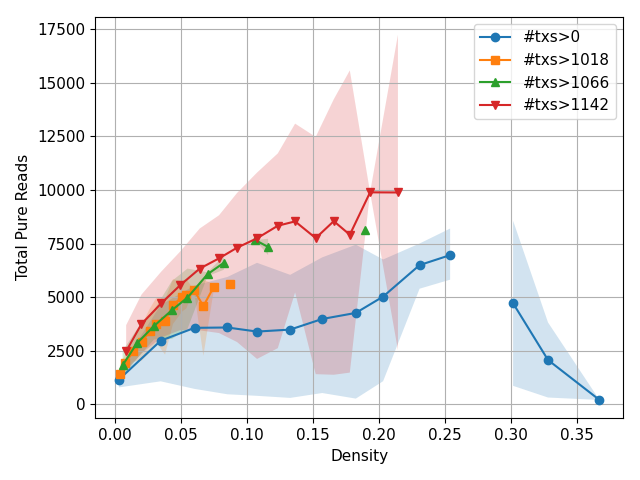}}
\caption{total-pure-reads}
\label{fig:total-pure-reads}
\end{figure}
\begin{figure}
\centerline{\includegraphics[width=\columnwidth]{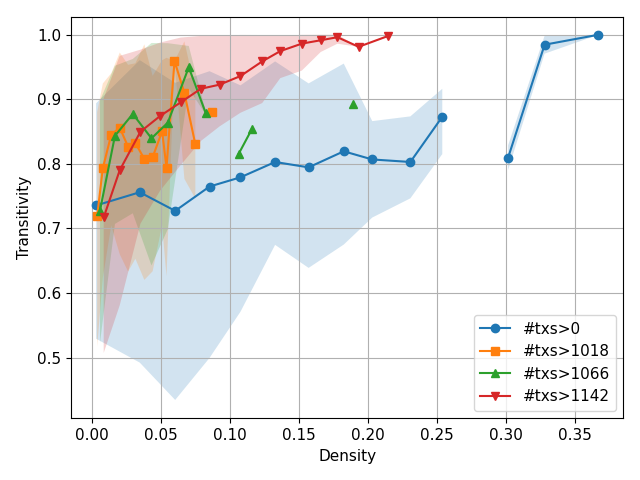}}
\caption{transitivity}
\label{fig:transitivity}
\end{figure}
\begin{figure}
\centerline{\includegraphics[width=\columnwidth]{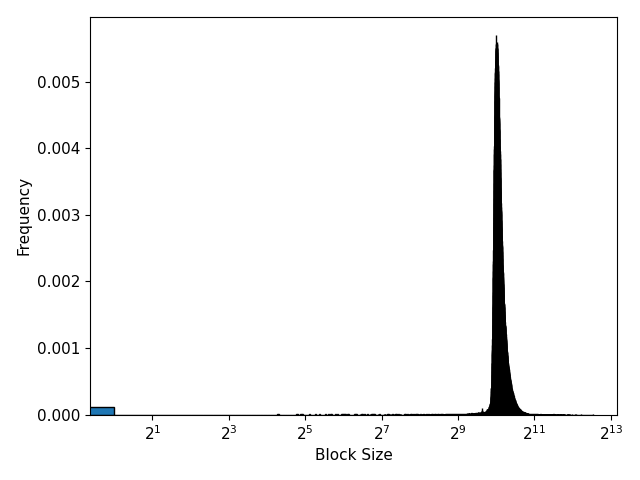}}
\caption{block-size-dist}
\label{fig:block-size-dist}
\end{figure}
\begin{figure}
\centerline{\includegraphics[width=\columnwidth]{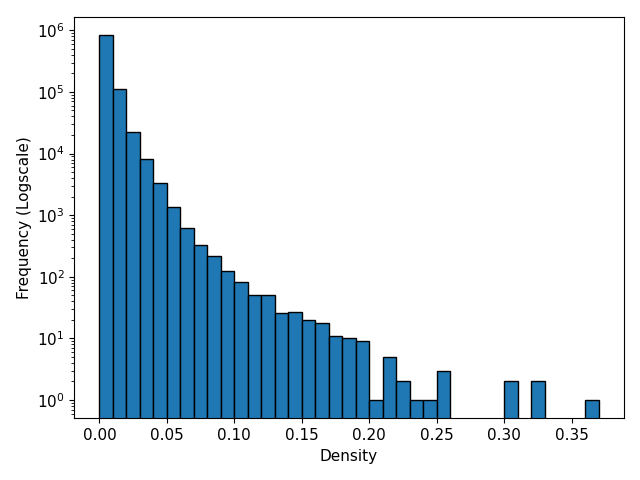}}
\caption{density-dist}
\label{fig:density-dist}
\end{figure}
\begin{figure}
\centerline{\includegraphics[width=\columnwidth]{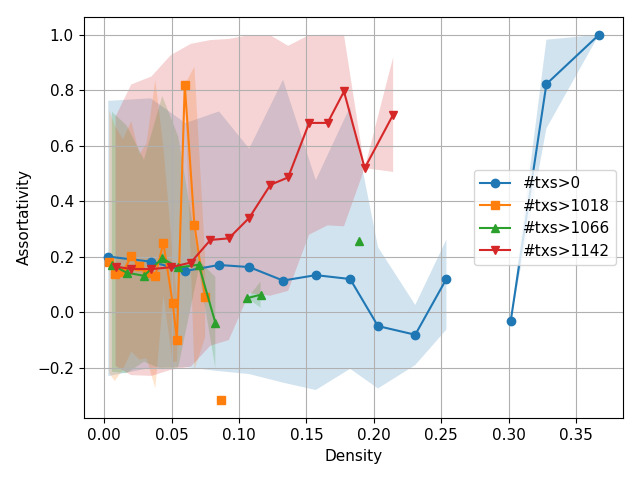}}
\caption{assortativity}
\label{fig:assortativity}
\end{figure}
\begin{figure}
\centerline{\includegraphics[width=\columnwidth]{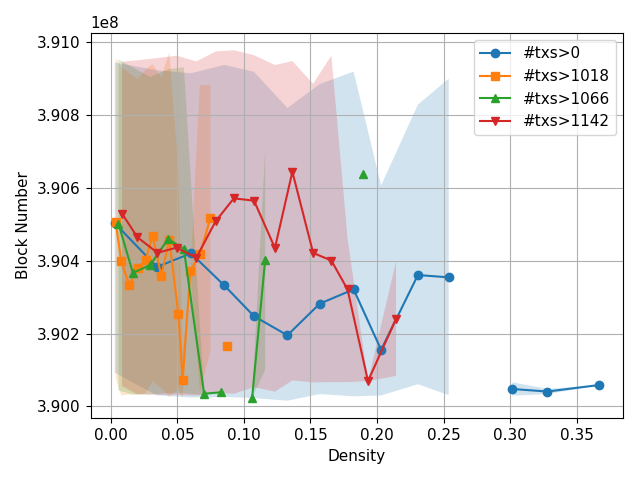}}
\caption{block-number}
\label{fig:block-number}
\end{figure}
\begin{figure}
\centerline{\includegraphics[width=\columnwidth]{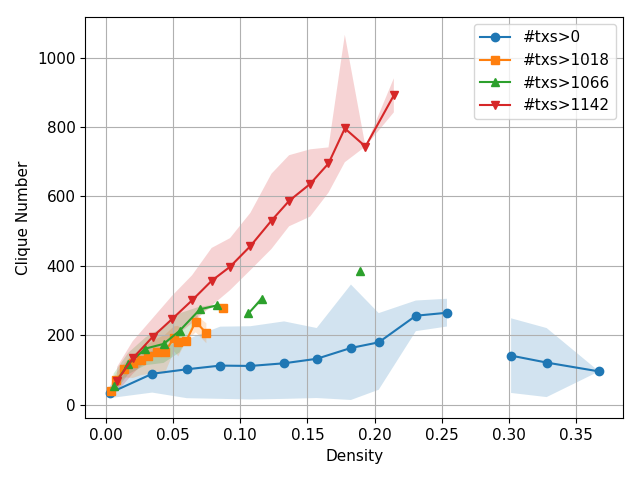}}
\caption{clique-number}
\label{fig:clique-number}
\end{figure}
\begin{figure}
\centerline{\includegraphics[width=\columnwidth]{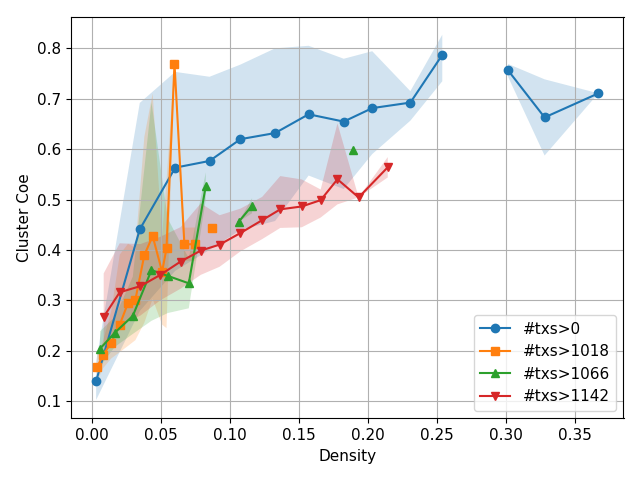}}
\caption{cluster-coe}
\label{fig:cluster-coe}
\end{figure}
\begin{figure}
\centerline{\includegraphics[width=\columnwidth]{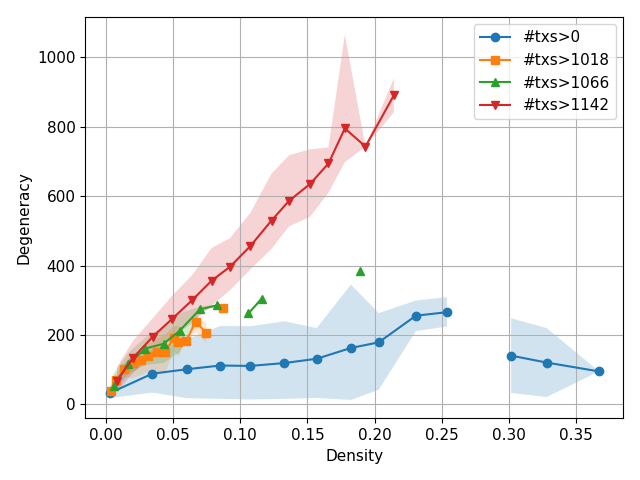}}
\caption{degeneracy}
\label{fig:degeneracy}
\end{figure}
\begin{figure}
\centerline{\includegraphics[width=\columnwidth]{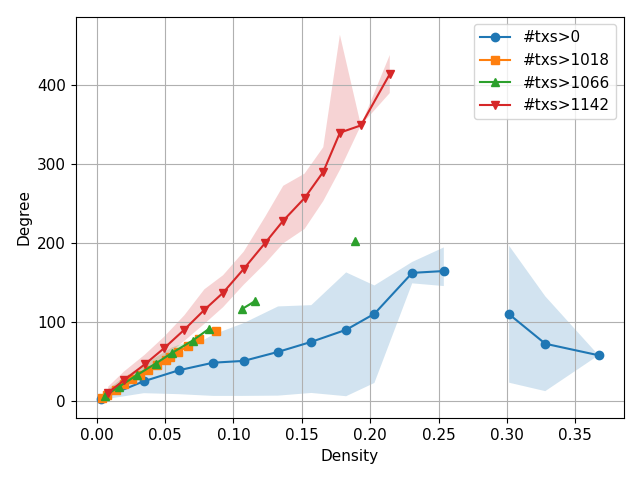}}
\caption{degree}
\label{fig:degree}
\end{figure}
\begin{figure}
\centerline{\includegraphics[width=\columnwidth]{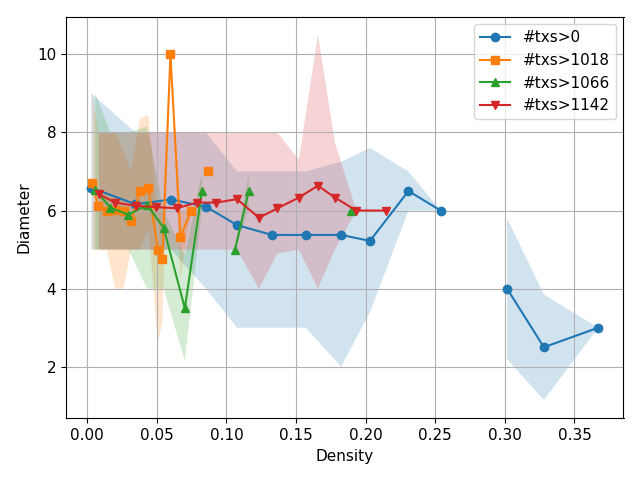}}
\caption{diameter}
\label{fig:diameter}
\end{figure}
\begin{figure}
\centerline{\includegraphics[width=\columnwidth]{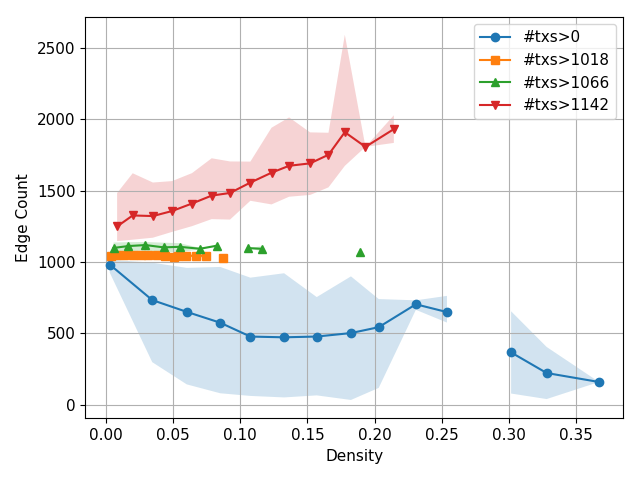}}
\caption{edge-count}
\label{fig:edge-count}
\end{figure}
\begin{figure}
\centerline{\includegraphics[width=\columnwidth]{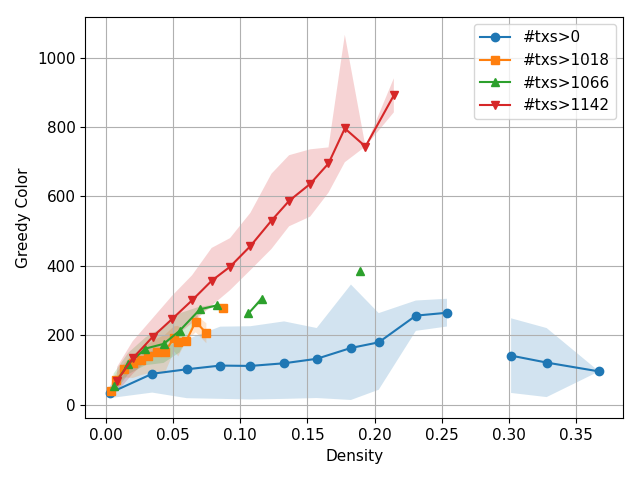}}
\caption{greedy-color}
\label{fig:greedy-color}
\end{figure}
\begin{figure}
\centerline{\includegraphics[width=\columnwidth]{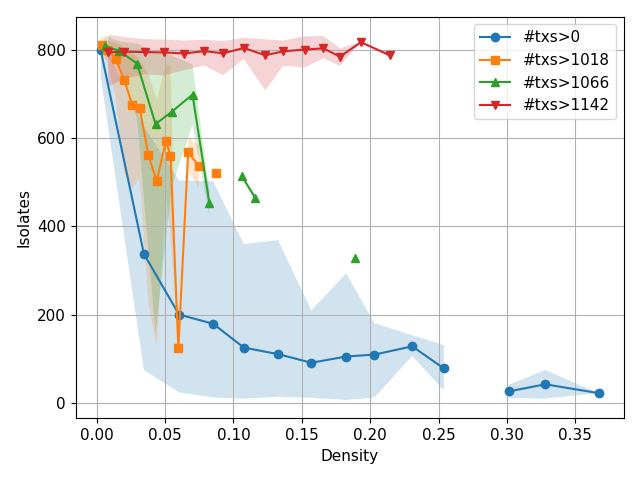}}
\caption{isolates}
\label{fig:isolates}
\end{figure}
\begin{figure}
\centerline{\includegraphics[width=\columnwidth]{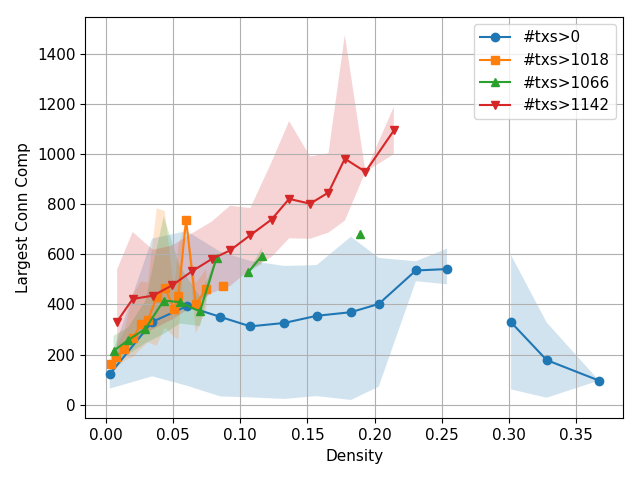}}
\caption{largest-conn-comp}
\label{fig:largest-conn-comp}
\end{figure}
\begin{figure}
\centerline{\includegraphics[width=\columnwidth]{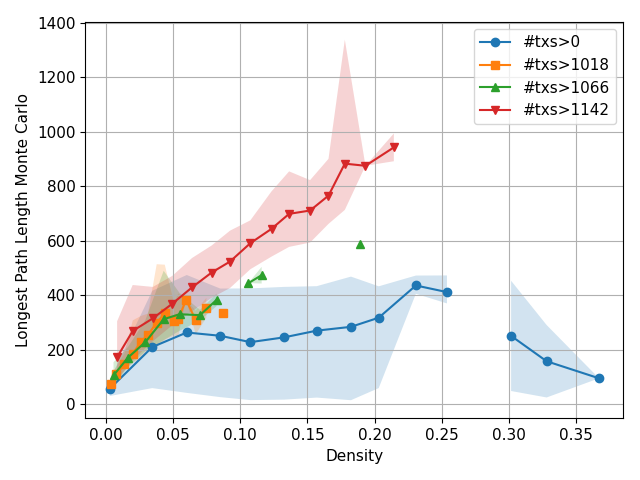}}
\caption{longest-path-length-monte-carlo}
\label{fig:longest-path-length-monte-carlo}
\end{figure}
\begin{figure}
\centerline{\includegraphics[width=\columnwidth]{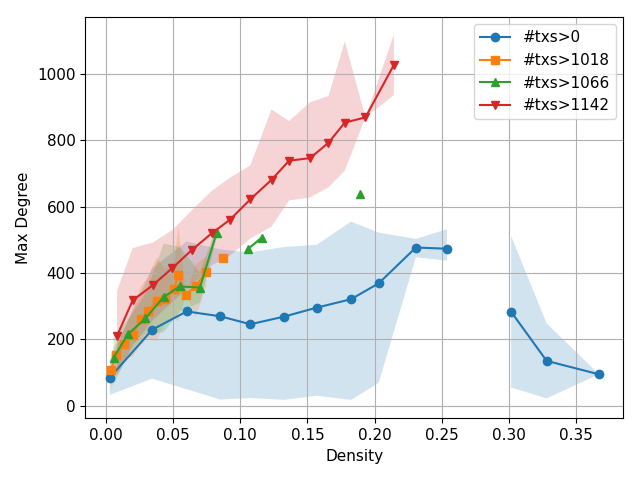}}
\caption{max-degree}
\label{fig:max-degree}
\end{figure}
\begin{figure}
\centerline{\includegraphics[width=\columnwidth]{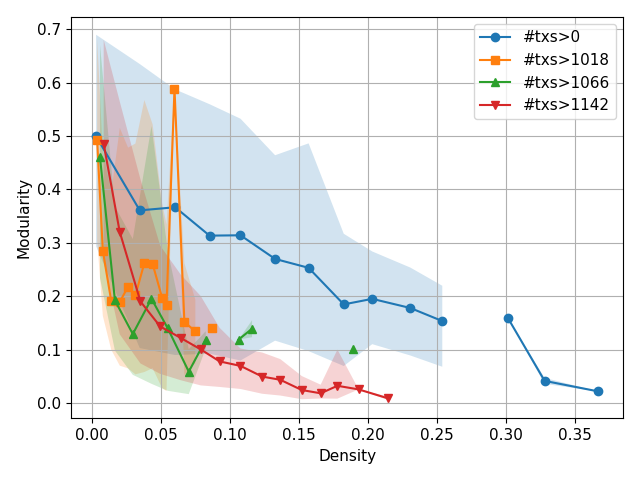}}
\caption{modularity}
\label{fig:modularity}
\end{figure}
\begin{figure}
\centerline{\includegraphics[width=\columnwidth]{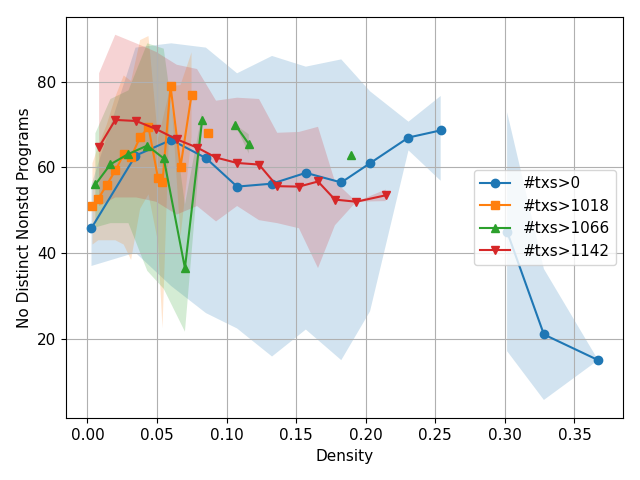}}
\caption{no-distinct-nonstd-programs}
\label{fig:no-distinct-nonstd-programs}
\end{figure}
\begin{figure}
\centerline{\includegraphics[width=\columnwidth]{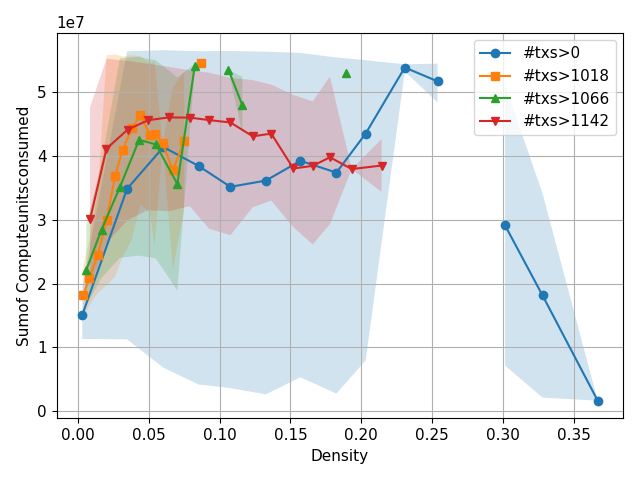}}
\caption{sumof-computeUnitsConsumed}
\label{fig:sumof-computeUnitsConsumed}
\end{figure}
\begin{figure}
\centerline{\includegraphics[width=\columnwidth]{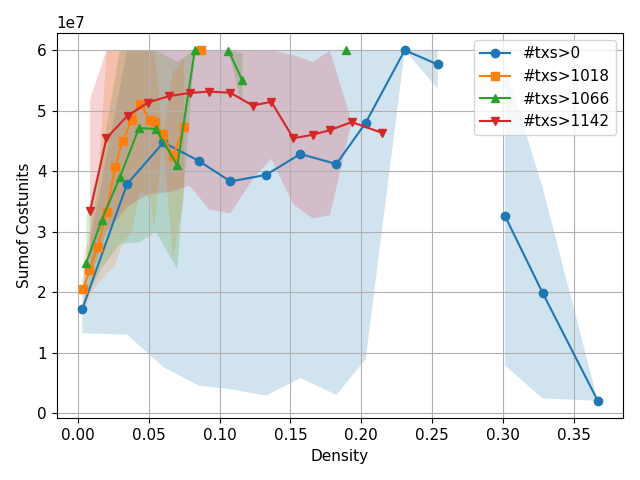}}
\caption{sumof-costUnits}
\label{fig:sumof-costUnits}
\end{figure}
\begin{figure}
\centerline{\includegraphics[width=\columnwidth]{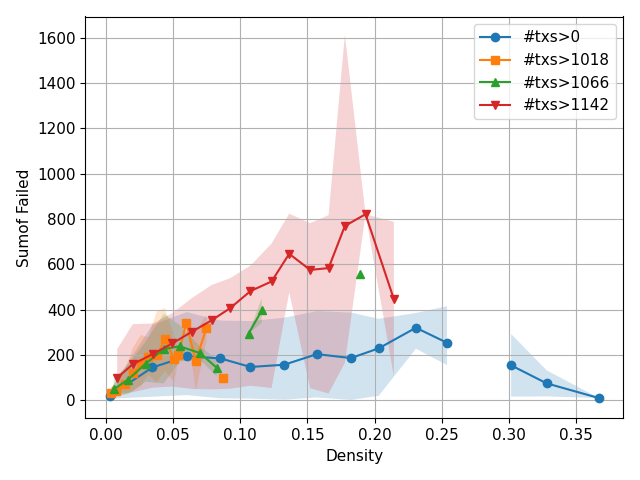}}
\caption{sumof-failed}
\label{fig:sumof-failed}
\end{figure}
\begin{figure}
\centerline{\includegraphics[width=\columnwidth]{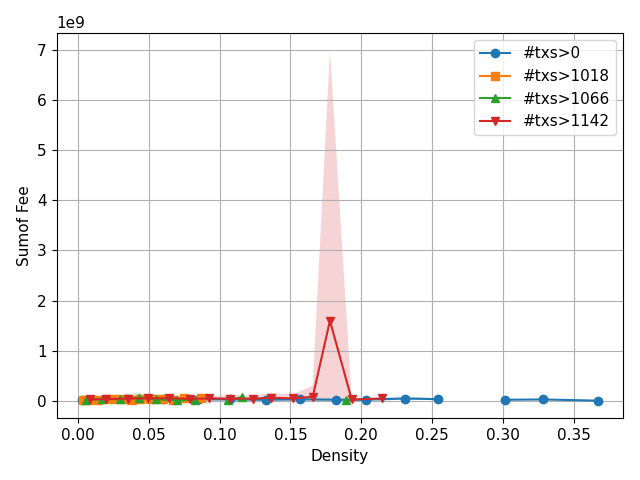}}
\caption{sumof-fee}
\label{fig:sumof-fee}
\end{figure}
\begin{figure}
\centerline{\includegraphics[width=\columnwidth]{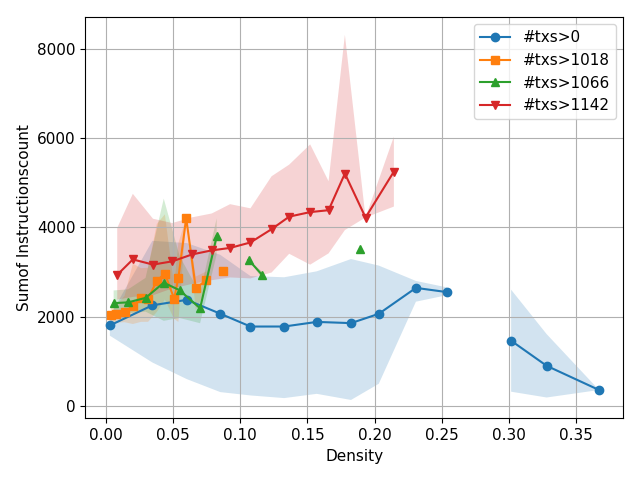}}
\caption{sumof-instructionsCount}
\label{fig:sumof-instructionsCount}
\end{figure}
\begin{figure}
\centerline{\includegraphics[width=\columnwidth]{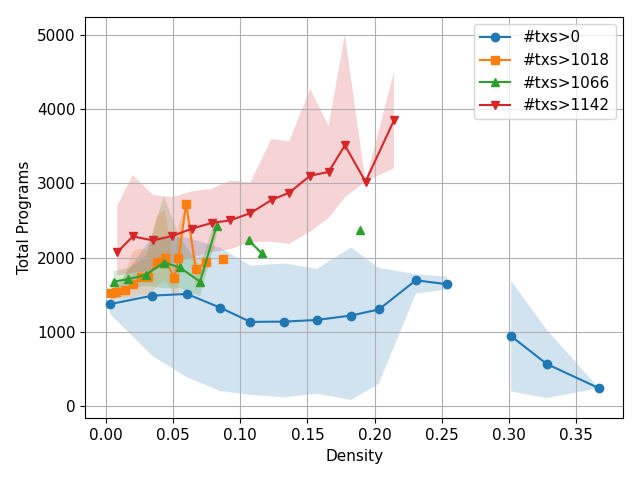}}
\caption{total-programs}
\label{fig:total-programs}
\end{figure}
\begin{figure}
\centerline{\includegraphics[width=\columnwidth]{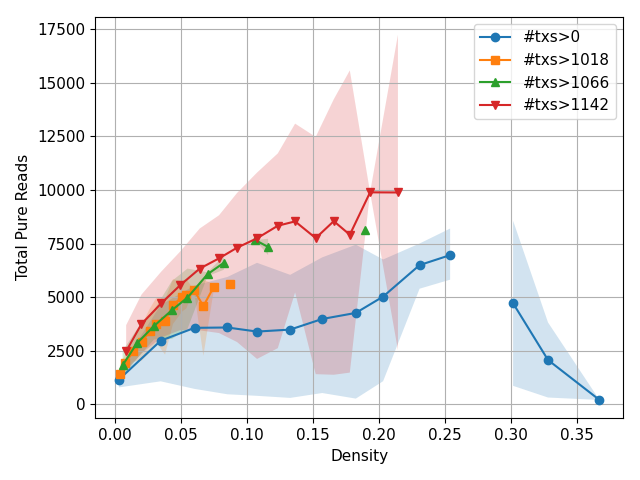}}
\caption{total-pure-reads}
\label{fig:total-pure-reads}
\end{figure}
\begin{figure}
\centerline{\includegraphics[width=\columnwidth]{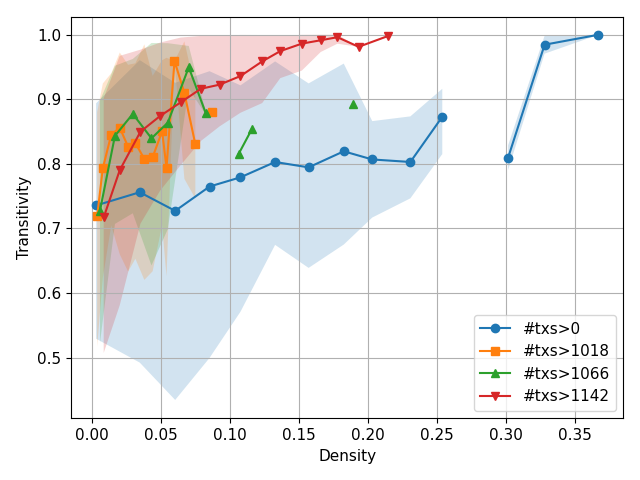}}
\caption{transitivity}
\label{fig:transitivity}
\end{figure}
\begin{figure}
\centerline{\includegraphics[width=\columnwidth]{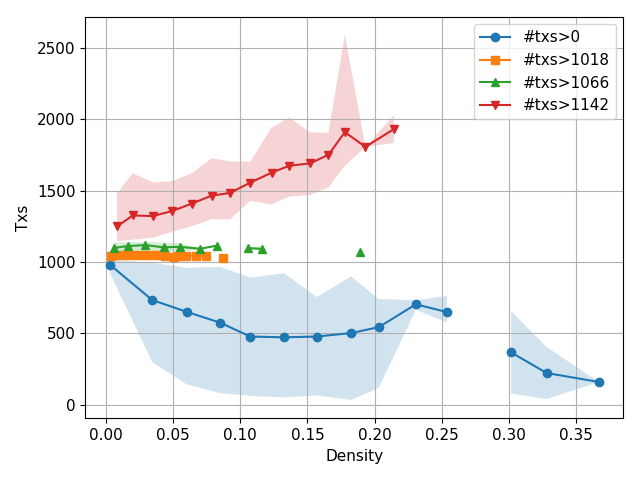}}
\caption{txs}
\label{fig:txs}
\end{figure}
\begin{figure}
\centerline{\includegraphics[width=\columnwidth]{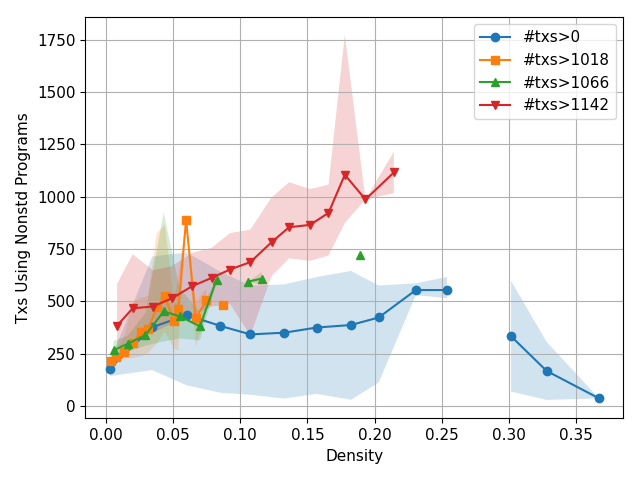}}
\caption{txs-using-nonstd-programs}
\label{fig:txs-using-nonstd-programs}
\end{figure}
\begin{figure}
\centerline{\includegraphics[width=\columnwidth]{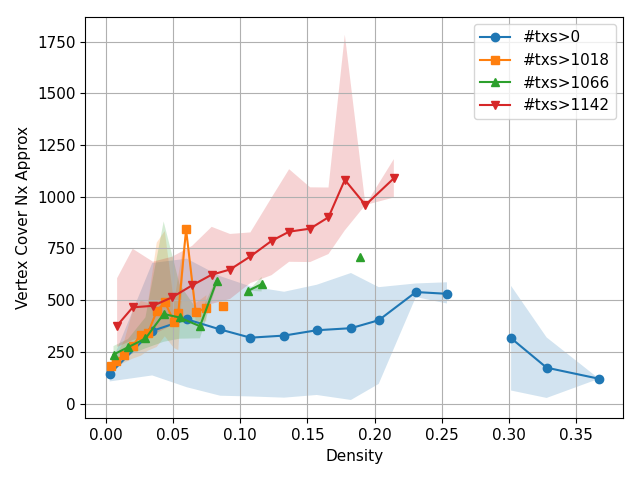}}
\caption{vertex-cover-nx-approx}
\label{fig:vertex-cover-nx-approx}
\end{figure}
\begin{figure}
\centerline{\includegraphics[width=\columnwidth]{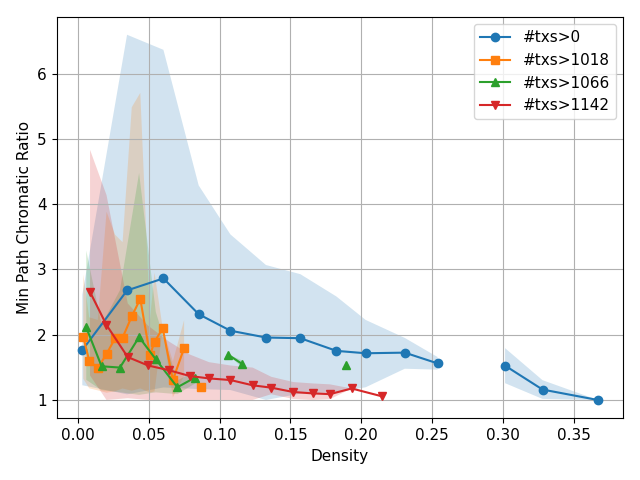}}
\caption{min-path-chromatic-ratio}
\label{fig:min-path-chromatic-ratio}
\end{figure}
\begin{figure}
\centerline{\includegraphics[width=\columnwidth]{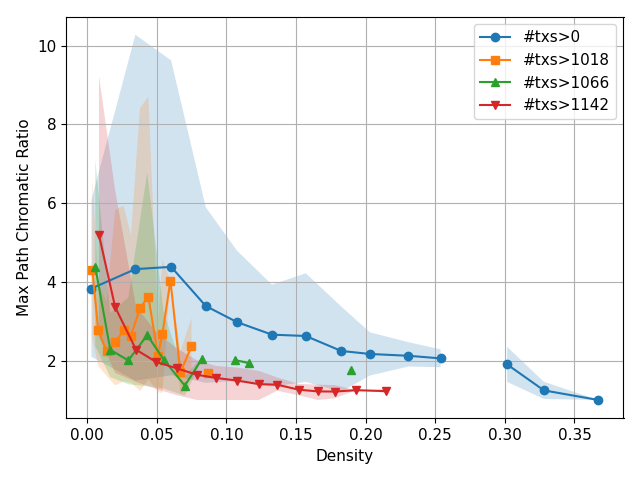}}
\caption{max-path-chromatic-ratio}
\label{fig:max-path-chromatic-ratio}
\end{figure}
\begin{figure}
\centerline{\includegraphics[width=\columnwidth]{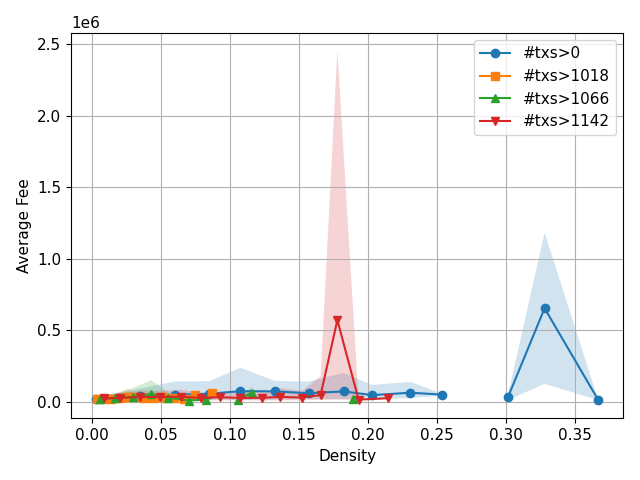}}
\caption{avg-fee}
\label{fig:avg-fee}
\end{figure}
\begin{figure}
\centerline{\includegraphics[width=\columnwidth]{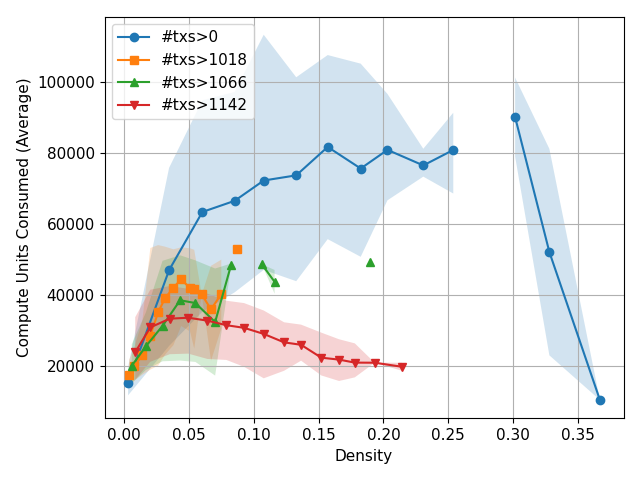}}
\caption{avg-computeUnitsConsumed}
\label{fig:avg-computeUnitsConsumed}
\end{figure}
\begin{figure}
\centerline{\includegraphics[width=\columnwidth]{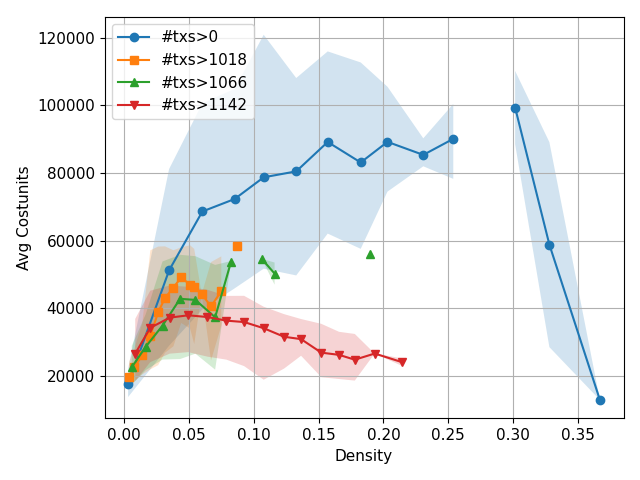}}
\caption{avg-costUnits}
\label{fig:avg-costUnits}
\end{figure}
\begin{figure}
\centerline{\includegraphics[width=\columnwidth]{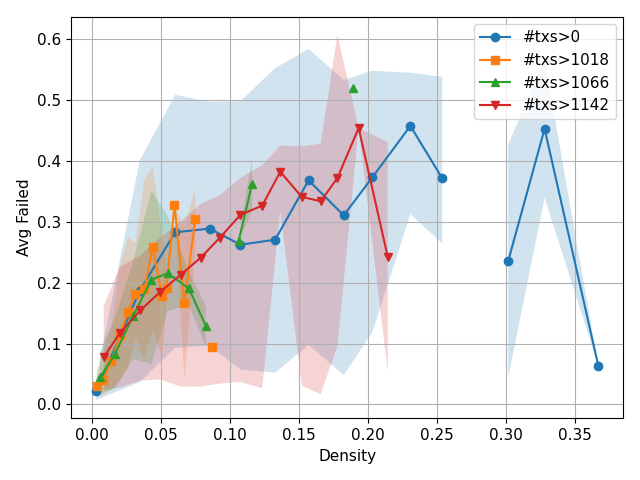}}
\caption{avg-failed}
\label{fig:avg-failed}
\end{figure}
\begin{figure}
\centerline{\includegraphics[width=\columnwidth]{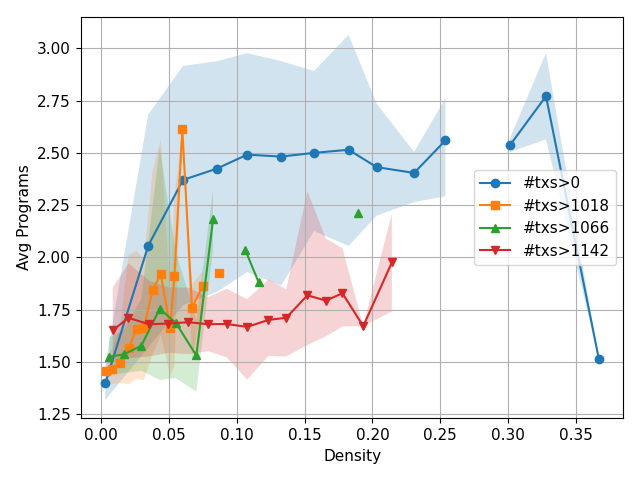}}
\caption{avg-programs}
\label{fig:avg-programs}
\end{figure}
\begin{figure}
\centerline{\includegraphics[width=\columnwidth]{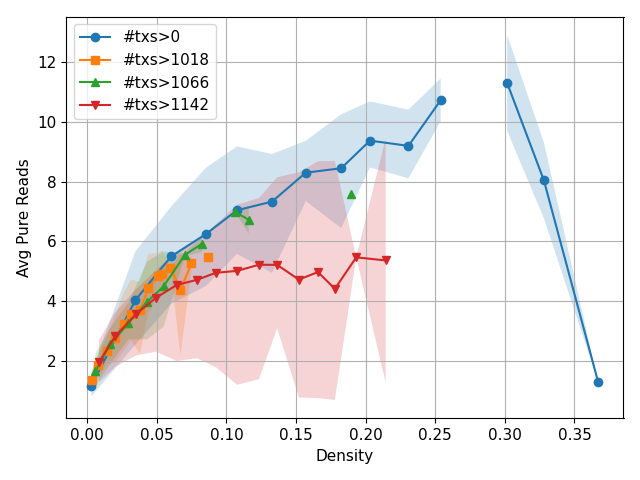}}
\caption{avg-pure-reads}
\label{fig:avg-pure-reads}
\end{figure}
\begin{figure}
\centerline{\includegraphics[width=\columnwidth]{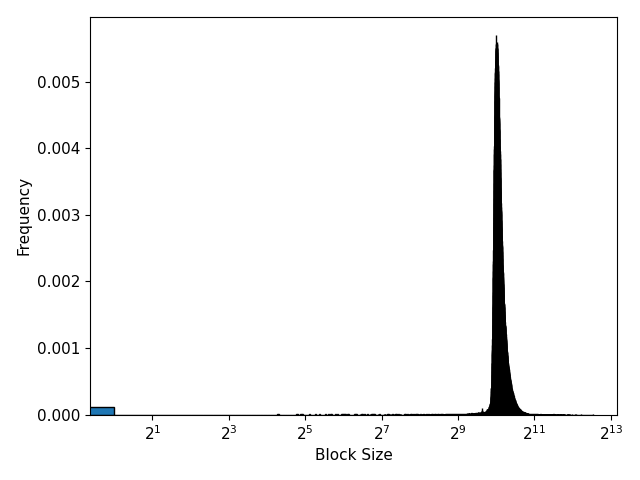}}
\caption{block-size-dist}
\label{fig:block-size-dist}
\end{figure}
\begin{figure}
\centerline{\includegraphics[width=\columnwidth]{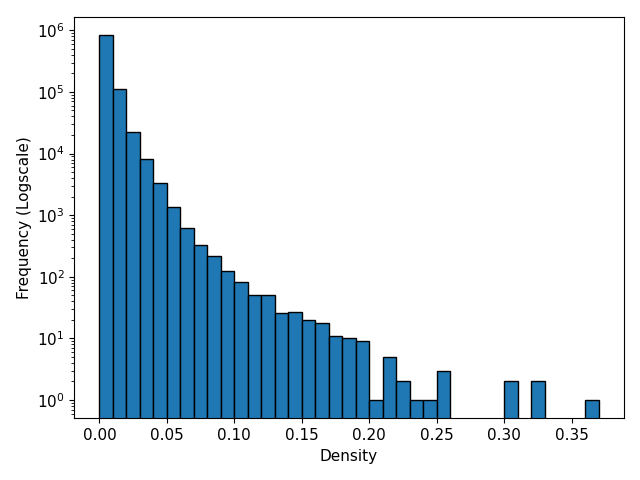}}
\caption{density-dist}
\label{fig:density-dist}
\end{figure}
\begin{figure}
\centerline{\includegraphics[width=\columnwidth]{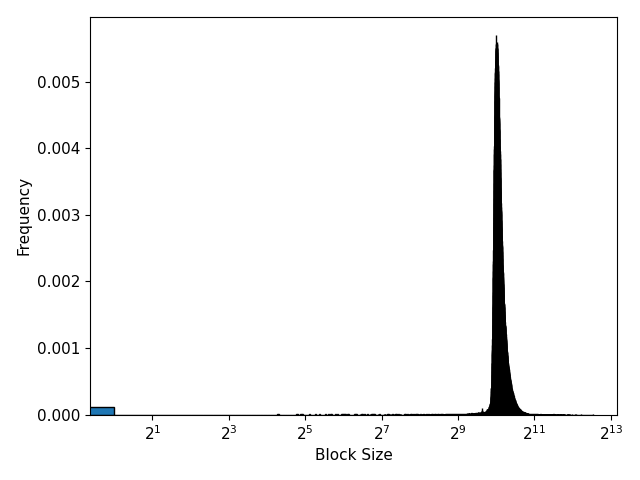}}
\caption{block-size-dist}
\label{fig:block-size-dist}
\end{figure}
\begin{figure}
\centerline{\includegraphics[width=\columnwidth]{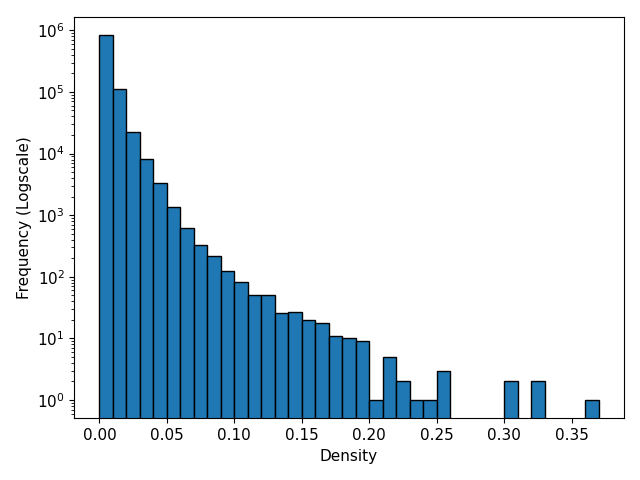}}
\caption{density-dist}
\label{fig:density-dist}
\end{figure}
\begin{figure}
\centerline{\includegraphics[width=\columnwidth]{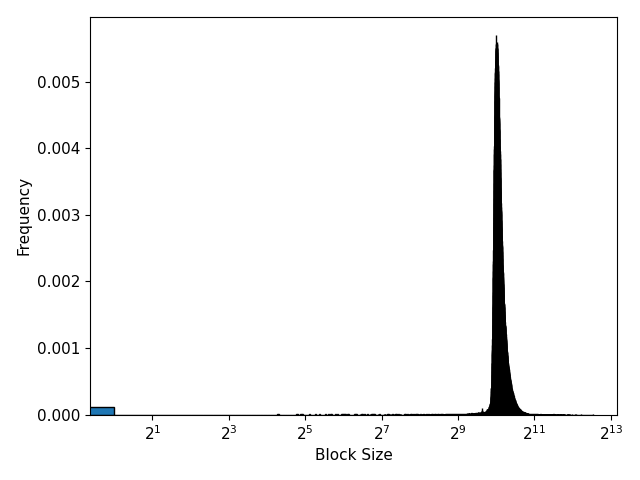}}
\caption{block-size-dist}
\label{fig:block-size-dist}
\end{figure}
\begin{figure}
\centerline{\includegraphics[width=\columnwidth]{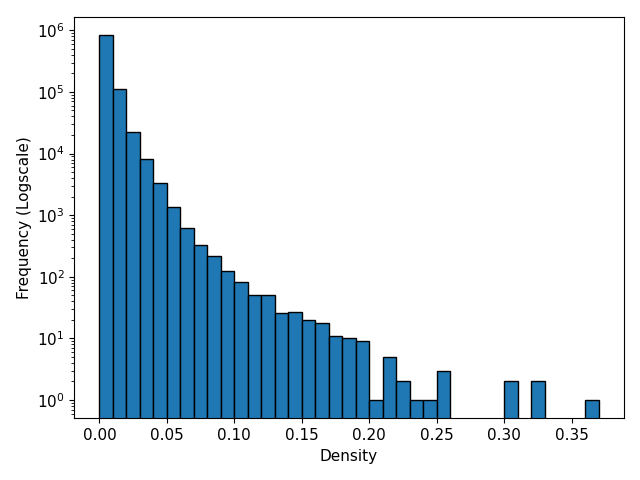}}
\caption{density-dist}
\label{fig:density-dist}
\end{figure}
\begin{figure}
\centerline{\includegraphics[width=\columnwidth]{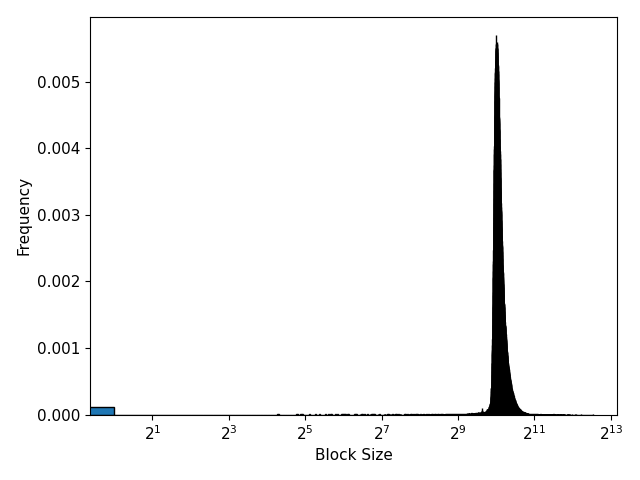}}
\caption{block-size-dist}
\label{fig:block-size-dist}
\end{figure}
\begin{figure}
\centerline{\includegraphics[width=\columnwidth]{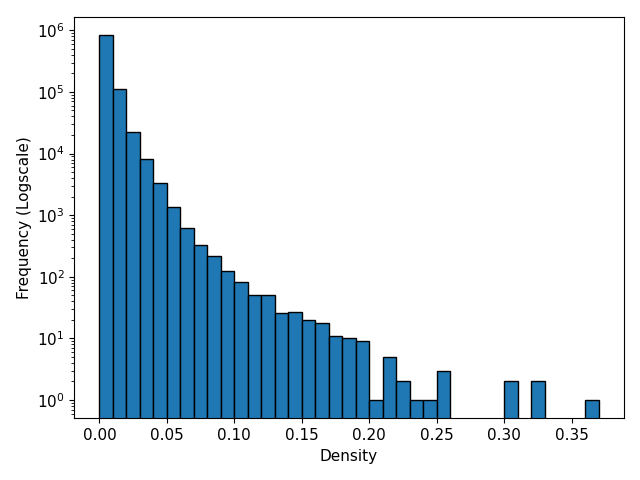}}
\caption{density-dist}
\label{fig:density-dist}
\end{figure}
\begin{figure}
\centerline{\includegraphics[width=\columnwidth]{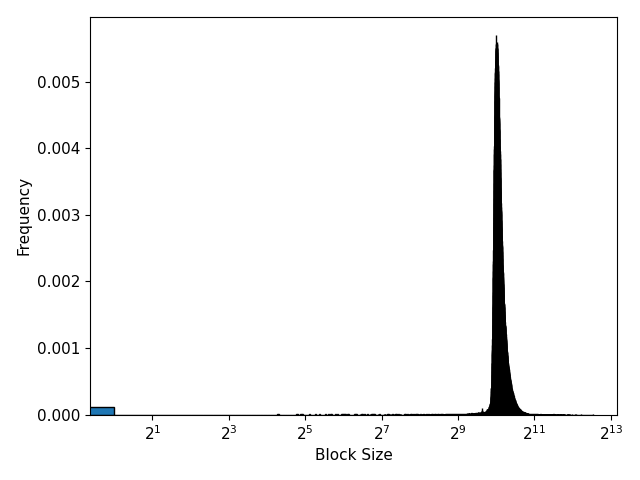}}
\caption{block-size-dist}
\label{fig:block-size-dist}
\end{figure}
\begin{figure}
\centerline{\includegraphics[width=\columnwidth]{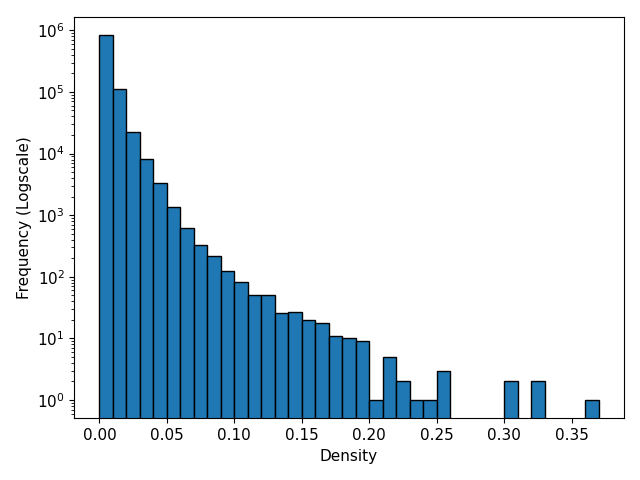}}
\caption{density-dist}
\label{fig:density-dist}
\end{figure}
\begin{figure}
\centerline{\includegraphics[width=\columnwidth]{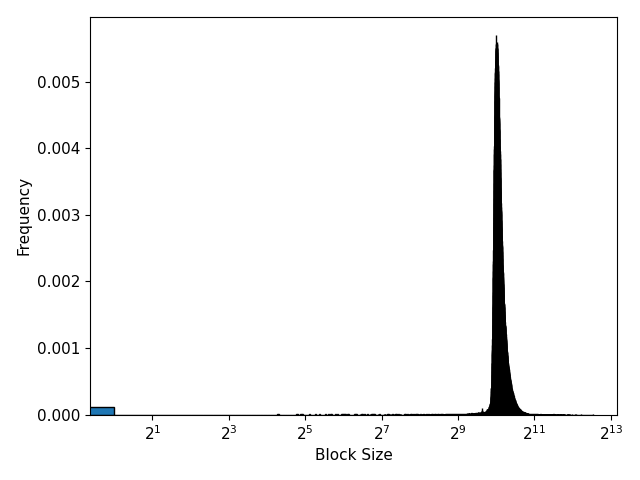}}
\caption{block-size-dist}
\label{fig:block-size-dist}
\end{figure}
\begin{figure}
\centerline{\includegraphics[width=\columnwidth]{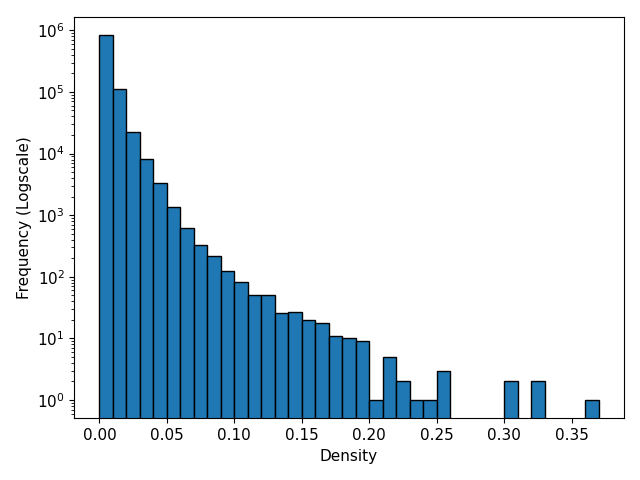}}
\caption{density-dist}
\label{fig:density-dist}
\end{figure}
\begin{figure}
\centerline{\includegraphics[width=\columnwidth]{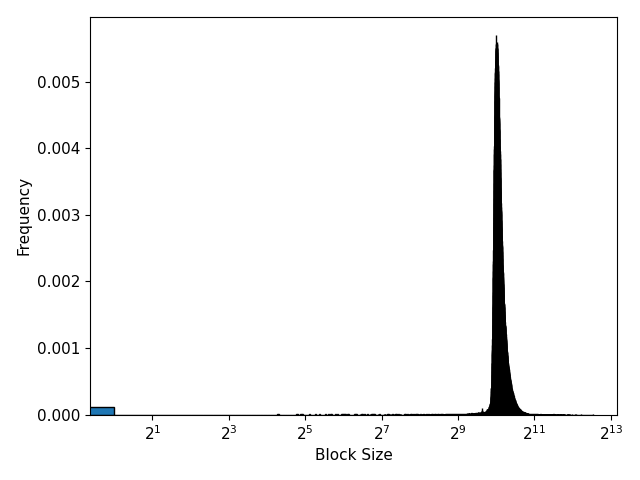}}
\caption{block-size-dist}
\label{fig:block-size-dist}
\end{figure}
\begin{figure}
\centerline{\includegraphics[width=\columnwidth]{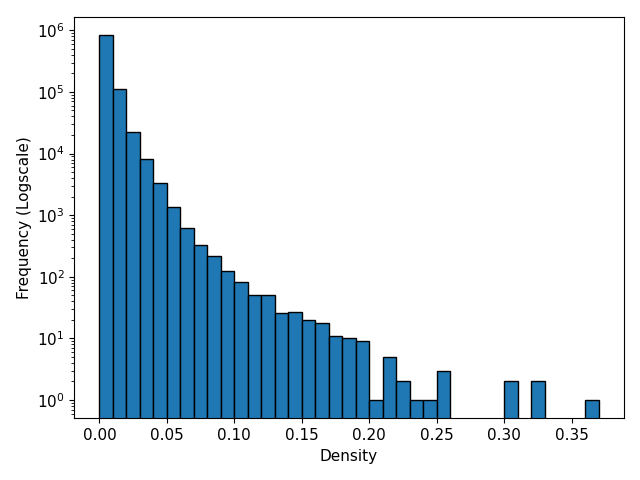}}
\caption{density-dist}
\label{fig:density-dist}
\end{figure}
\begin{figure}
\centerline{\includegraphics[width=\columnwidth]{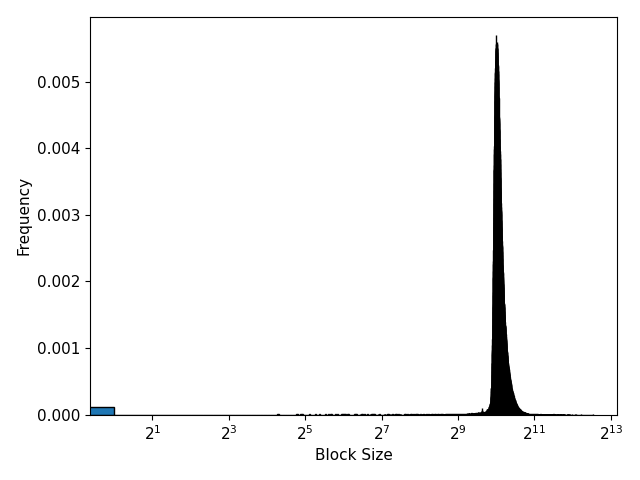}}
\caption{block-size-dist}
\label{fig:block-size-dist}
\end{figure}
\begin{figure}
\centerline{\includegraphics[width=\columnwidth]{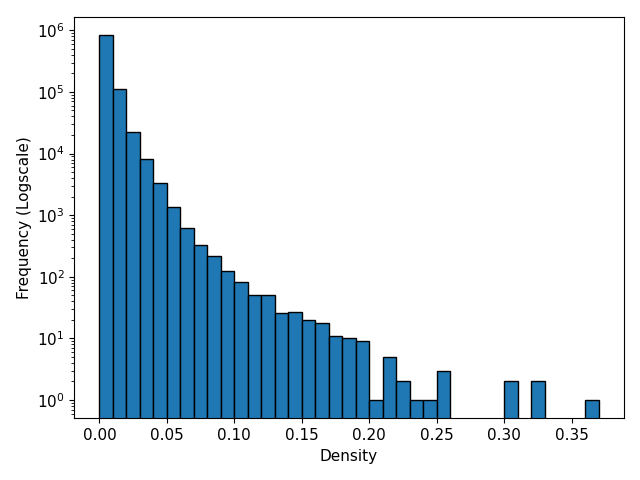}}
\caption{density-dist}
\label{fig:density-dist}
\end{figure}
\begin{figure}
\centerline{\includegraphics[width=\columnwidth]{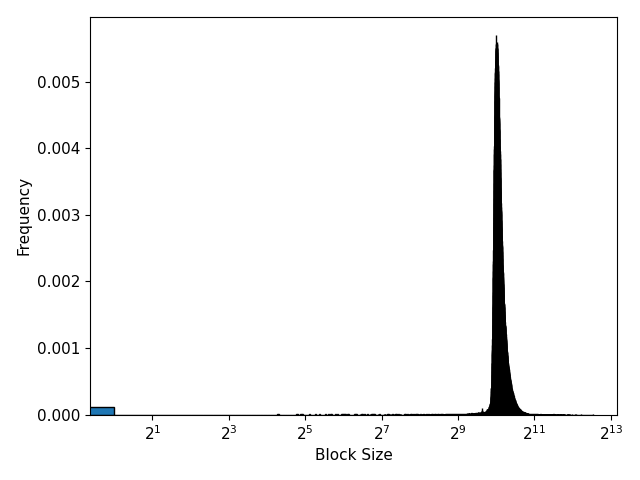}}
\caption{block-size-dist}
\label{fig:block-size-dist}
\end{figure}
\begin{figure}
\centerline{\includegraphics[width=\columnwidth]{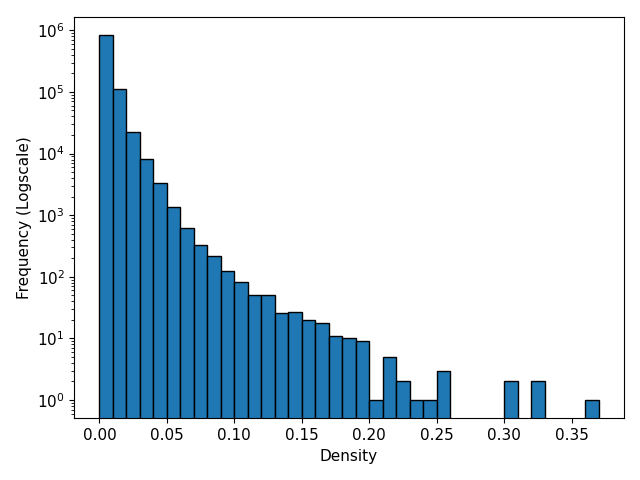}}
\caption{density-dist}
\label{fig:density-dist}
\end{figure}
\begin{figure}
\centerline{\includegraphics[width=\columnwidth]{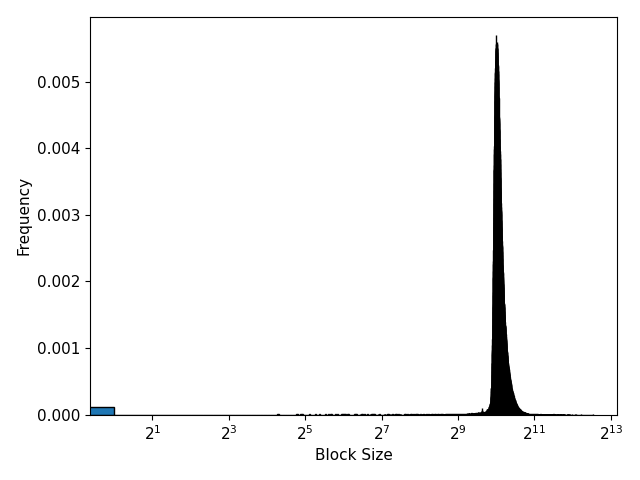}}
\caption{block-size-dist}
\label{fig:block-size-dist}
\end{figure}
\begin{figure}
\centerline{\includegraphics[width=\columnwidth]{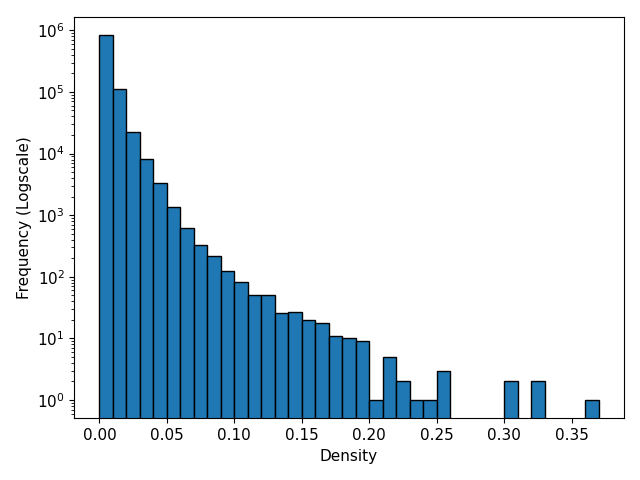}}
\caption{density-dist}
\label{fig:density-dist}
\end{figure}
\begin{figure}
\centerline{\includegraphics[width=\columnwidth]{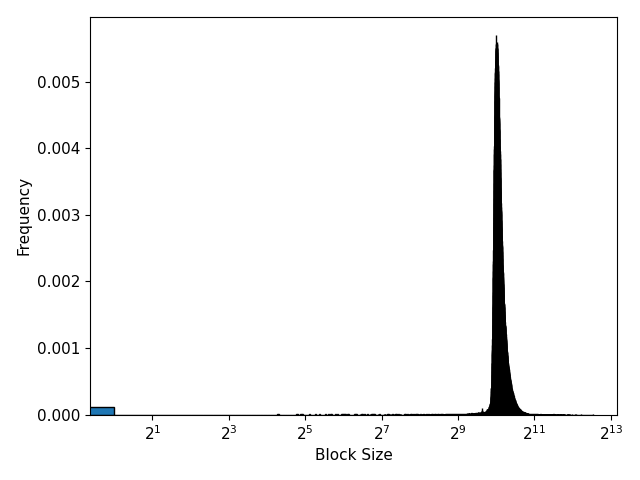}}
\caption{block-size-dist}
\label{fig:block-size-dist}
\end{figure}
\begin{figure}
\centerline{\includegraphics[width=\columnwidth]{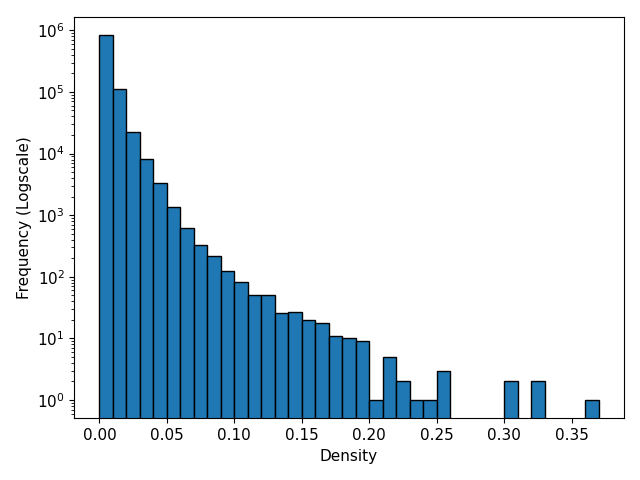}}
\caption{density-dist}
\label{fig:density-dist}
\end{figure}
\begin{figure}
\centerline{\includegraphics[width=\columnwidth]{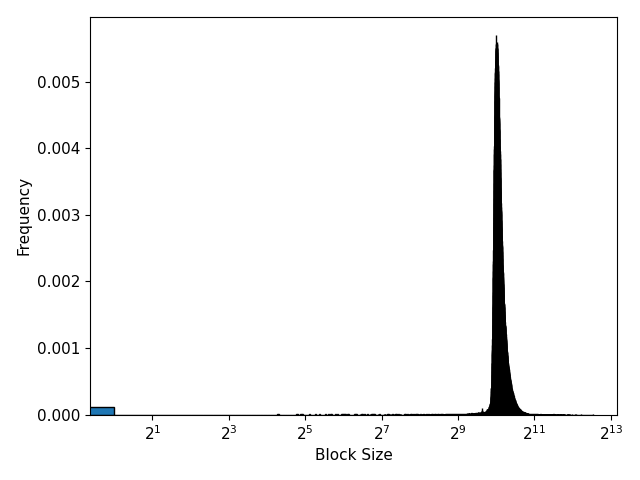}}
\caption{block-size-dist}
\label{fig:block-size-dist}
\end{figure}
\begin{figure}
\centerline{\includegraphics[width=\columnwidth]{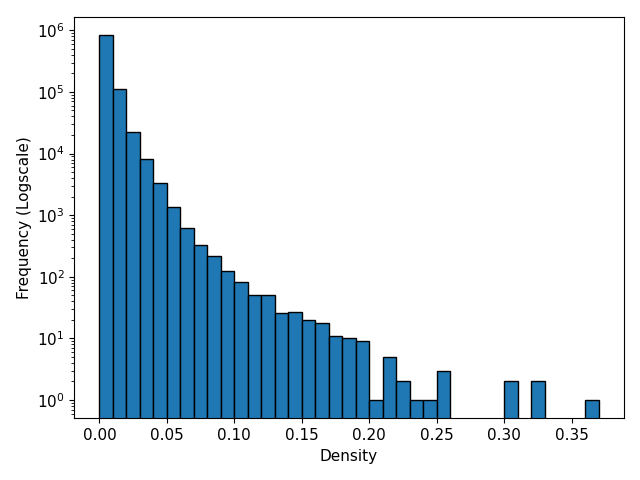}}
\caption{density-dist}
\label{fig:density-dist}
\end{figure}
\begin{figure}
\centerline{\includegraphics[width=\columnwidth]{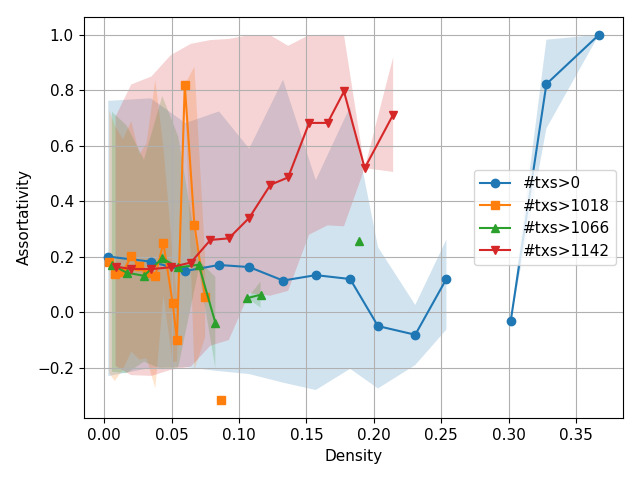}}
\caption{assortativity}
\label{fig:assortativity}
\end{figure}
\begin{figure}
\centerline{\includegraphics[width=\columnwidth]{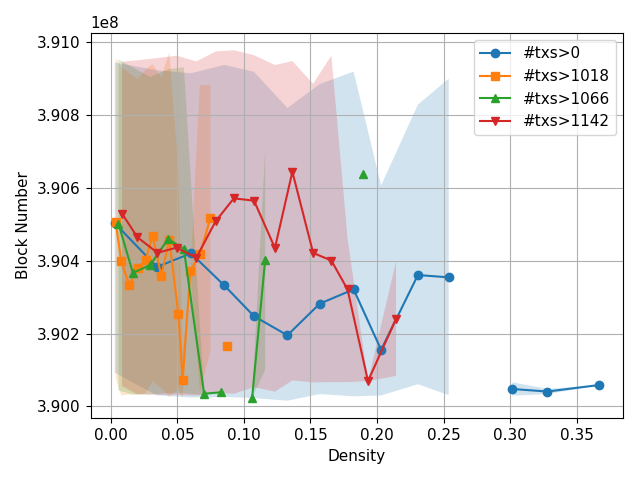}}
\caption{block-number}
\label{fig:block-number}
\end{figure}
\begin{figure}
\centerline{\includegraphics[width=\columnwidth]{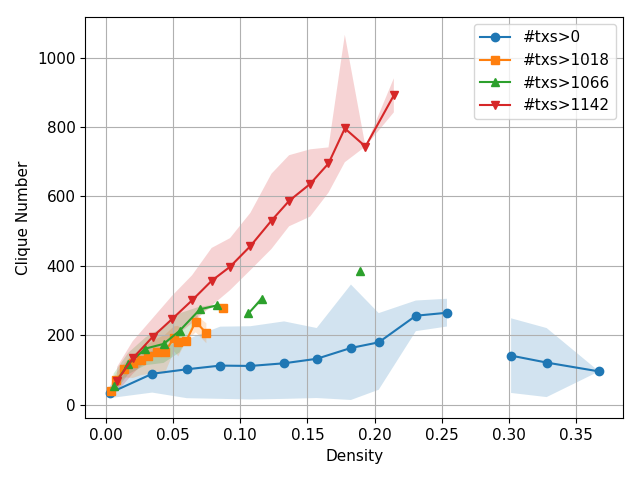}}
\caption{clique-number}
\label{fig:clique-number}
\end{figure}
\begin{figure}
\centerline{\includegraphics[width=\columnwidth]{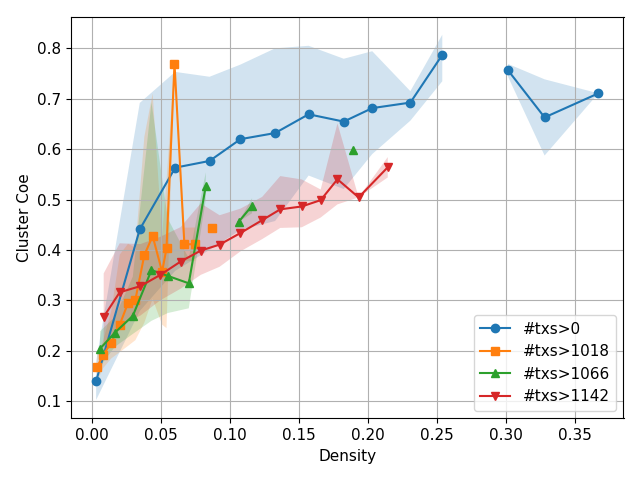}}
\caption{cluster-coe}
\label{fig:cluster-coe}
\end{figure}
\begin{figure}
\centerline{\includegraphics[width=\columnwidth]{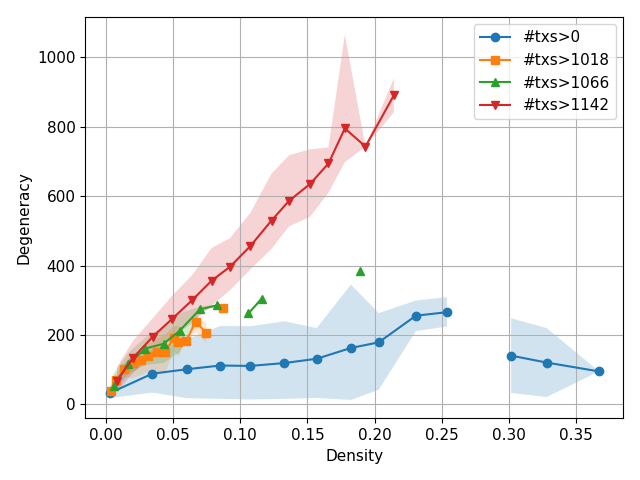}}
\caption{degeneracy}
\label{fig:degeneracy}
\end{figure}
\begin{figure}
\centerline{\includegraphics[width=\columnwidth]{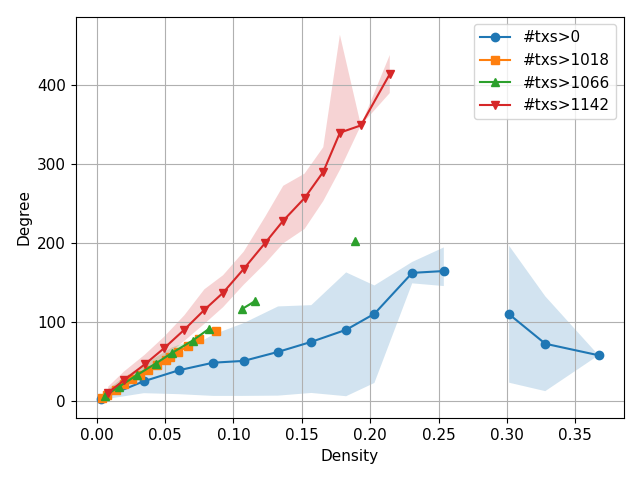}}
\caption{degree}
\label{fig:degree}
\end{figure}
\begin{figure}
\centerline{\includegraphics[width=\columnwidth]{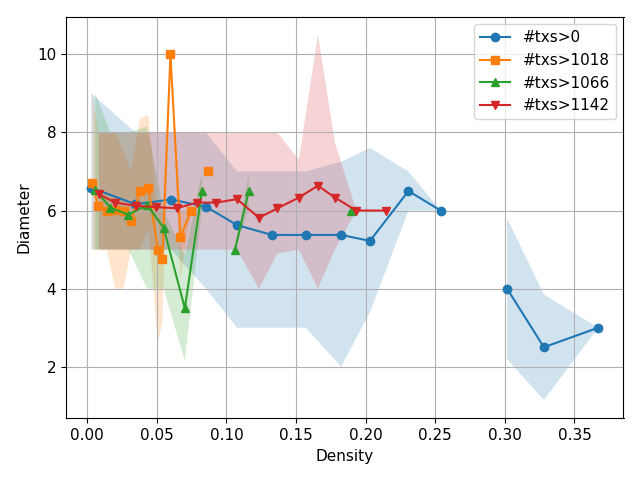}}
\caption{diameter}
\label{fig:diameter}
\end{figure}
\begin{figure}
\centerline{\includegraphics[width=\columnwidth]{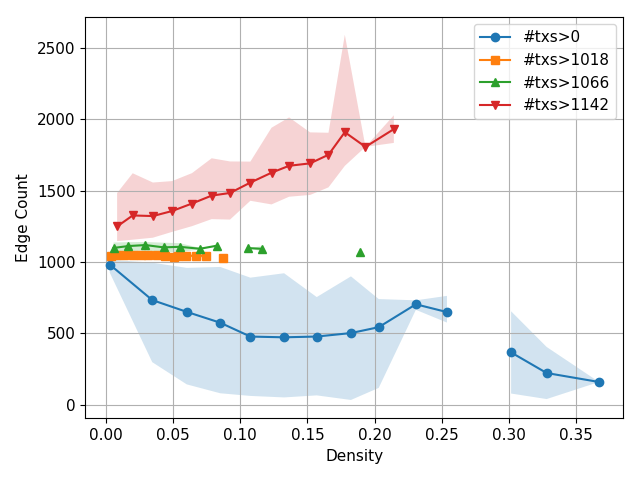}}
\caption{edge-count}
\label{fig:edge-count}
\end{figure}
\begin{figure}
\centerline{\includegraphics[width=\columnwidth]{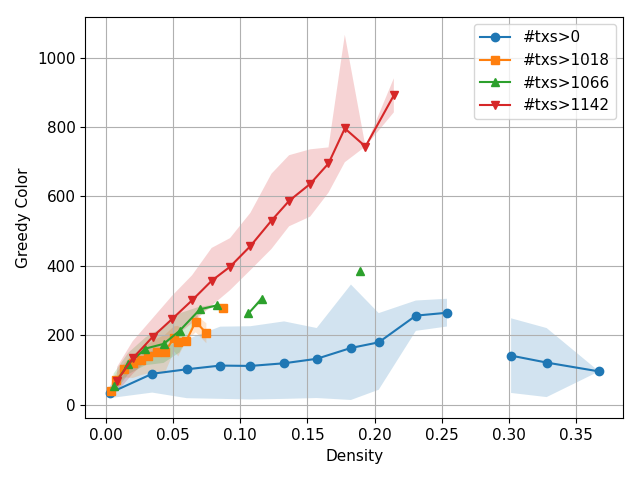}}
\caption{greedy-color}
\label{fig:greedy-color}
\end{figure}
\begin{figure}
\centerline{\includegraphics[width=\columnwidth]{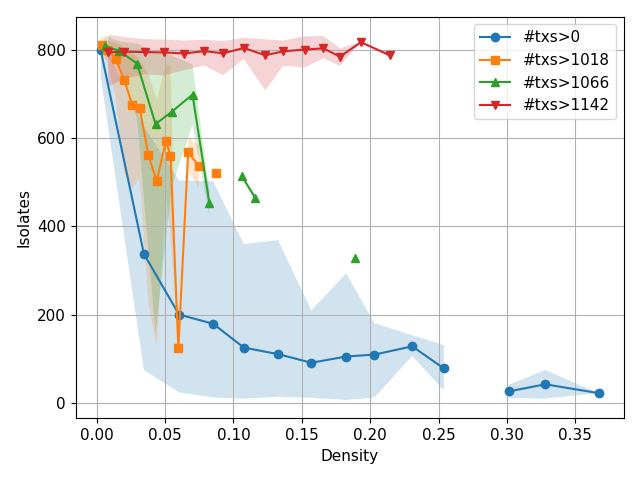}}
\caption{isolates}
\label{fig:isolates}
\end{figure}
\begin{figure}
\centerline{\includegraphics[width=\columnwidth]{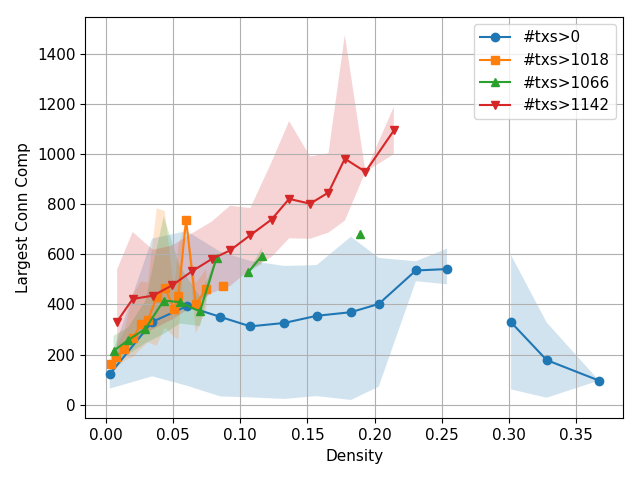}}
\caption{largest-conn-comp}
\label{fig:largest-conn-comp}
\end{figure}
\begin{figure}
\centerline{\includegraphics[width=\columnwidth]{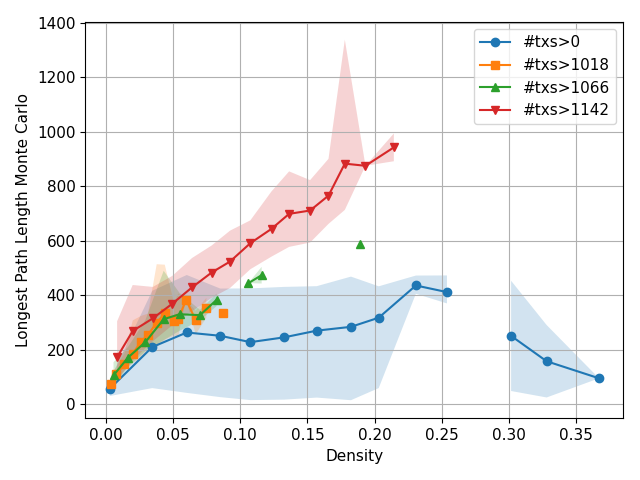}}
\caption{longest-path-length-monte-carlo}
\label{fig:longest-path-length-monte-carlo}
\end{figure}
\begin{figure}
\centerline{\includegraphics[width=\columnwidth]{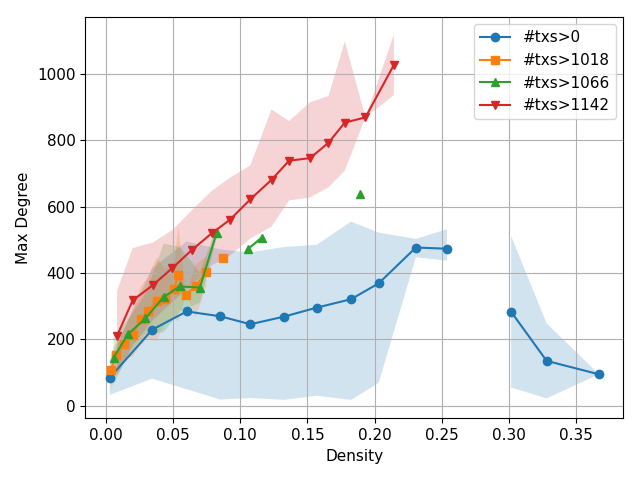}}
\caption{max-degree}
\label{fig:max-degree}
\end{figure}
\begin{figure}
\centerline{\includegraphics[width=\columnwidth]{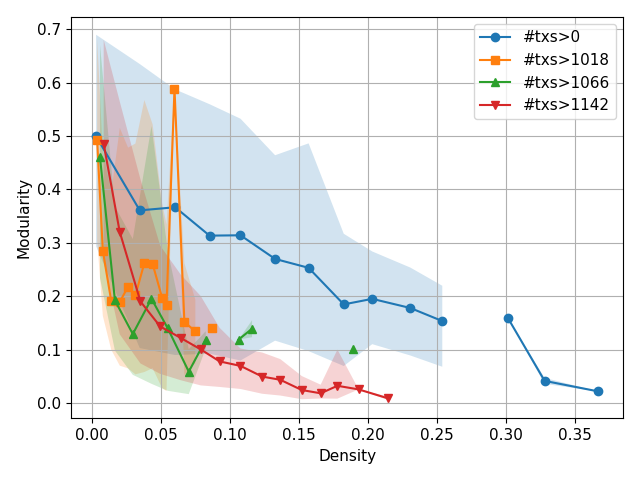}}
\caption{modularity}
\label{fig:modularity}
\end{figure}
\begin{figure}
\centerline{\includegraphics[width=\columnwidth]{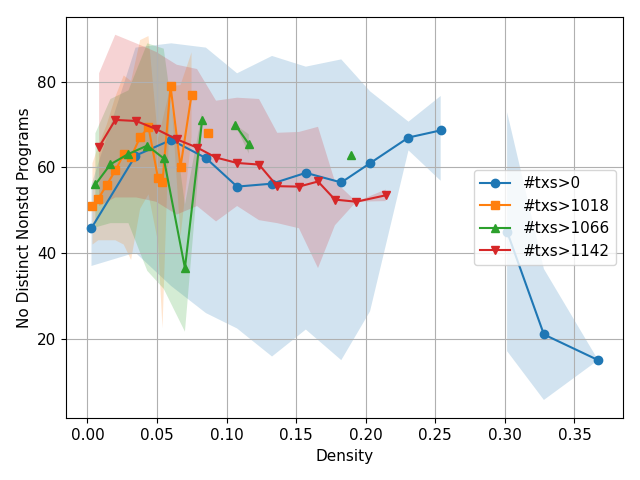}}
\caption{no-distinct-nonstd-programs}
\label{fig:no-distinct-nonstd-programs}
\end{figure}
\begin{figure}
\centerline{\includegraphics[width=\columnwidth]{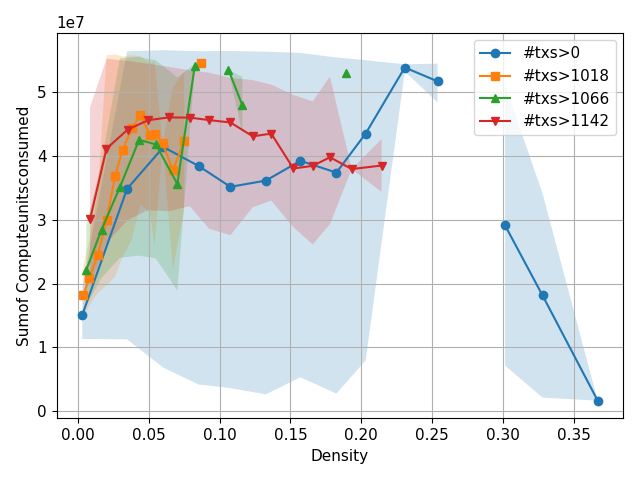}}
\caption{sumof-computeUnitsConsumed}
\label{fig:sumof-computeUnitsConsumed}
\end{figure}
\begin{figure}
\centerline{\includegraphics[width=\columnwidth]{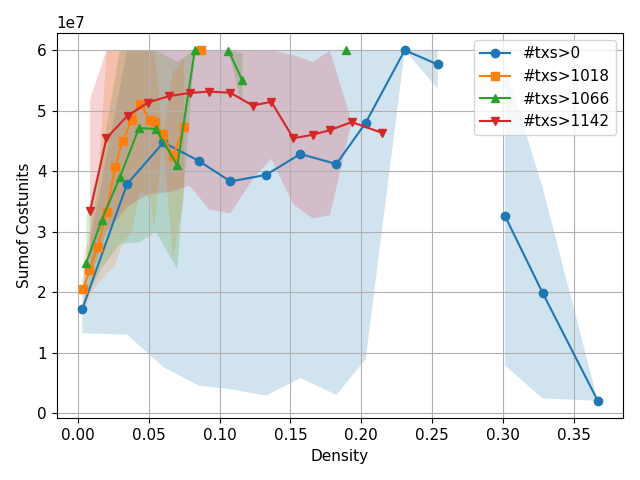}}
\caption{sumof-costUnits}
\label{fig:sumof-costUnits}
\end{figure}
\begin{figure}
\centerline{\includegraphics[width=\columnwidth]{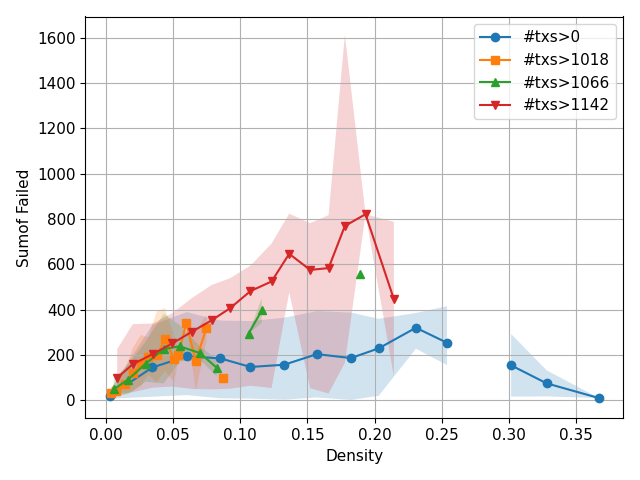}}
\caption{sumof-failed}
\label{fig:sumof-failed}
\end{figure}
\begin{figure}
\centerline{\includegraphics[width=\columnwidth]{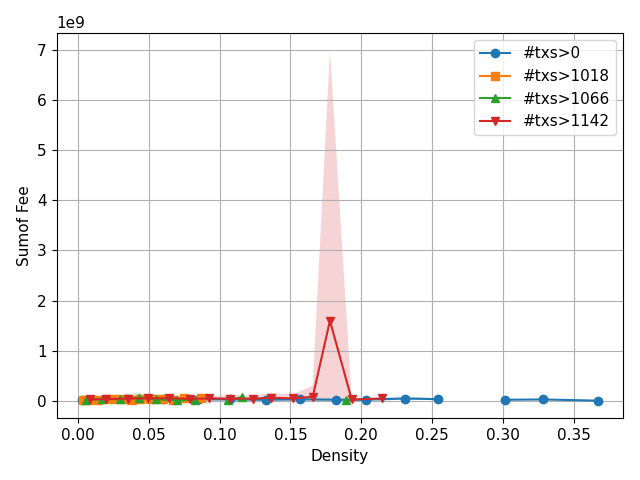}}
\caption{sumof-fee}
\label{fig:sumof-fee}
\end{figure}
\begin{figure}
\centerline{\includegraphics[width=\columnwidth]{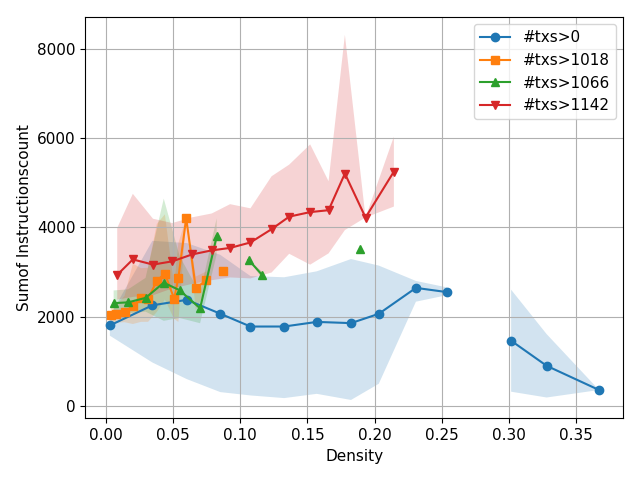}}
\caption{sumof-instructionsCount}
\label{fig:sumof-instructionsCount}
\end{figure}
\begin{figure}
\centerline{\includegraphics[width=\columnwidth]{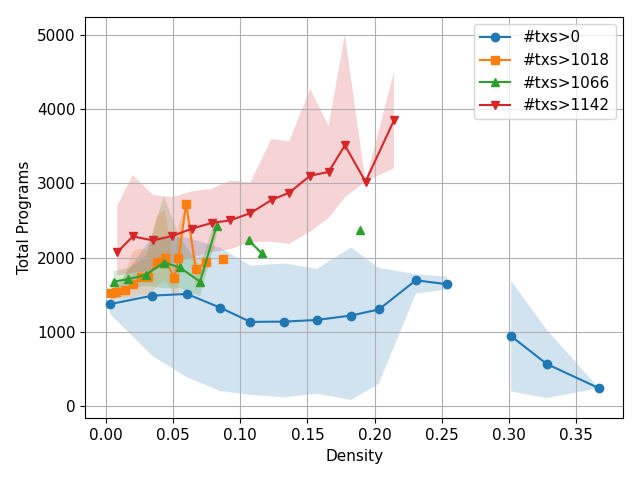}}
\caption{total-programs}
\label{fig:total-programs}
\end{figure}
\begin{figure}
\centerline{\includegraphics[width=\columnwidth]{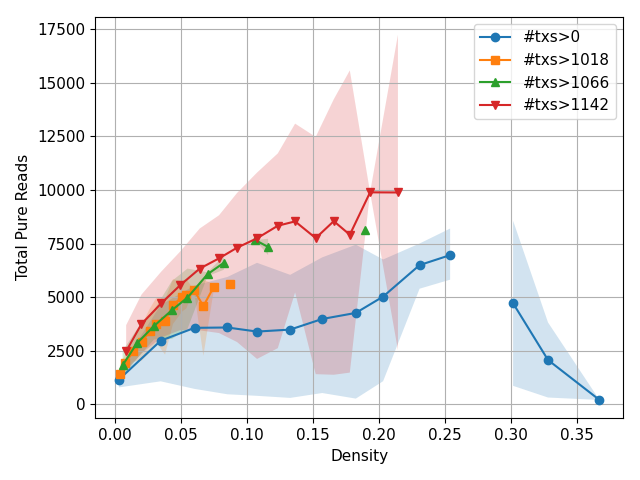}}
\caption{total-pure-reads}
\label{fig:total-pure-reads}
\end{figure}
\begin{figure}
\centerline{\includegraphics[width=\columnwidth]{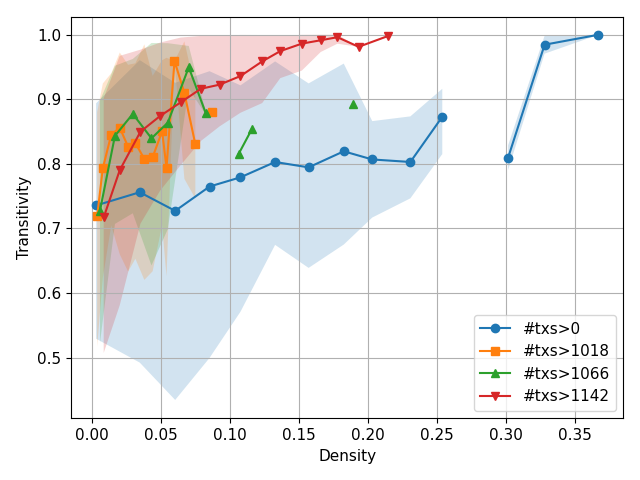}}
\caption{transitivity}
\label{fig:transitivity}
\end{figure}
\begin{figure}
\centerline{\includegraphics[width=\columnwidth]{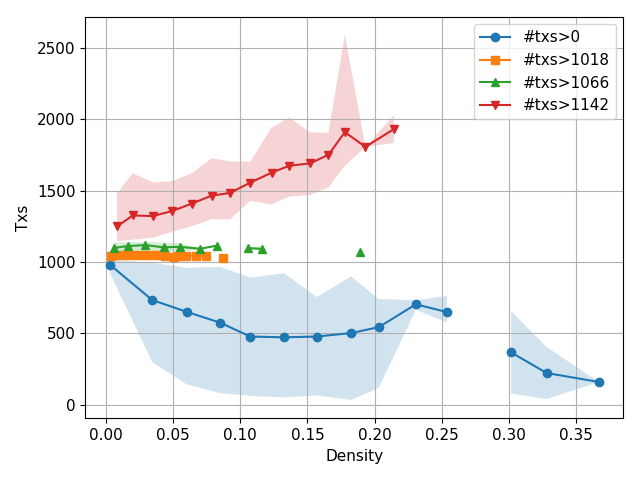}}
\caption{txs}
\label{fig:txs}
\end{figure}
\begin{figure}
\centerline{\includegraphics[width=\columnwidth]{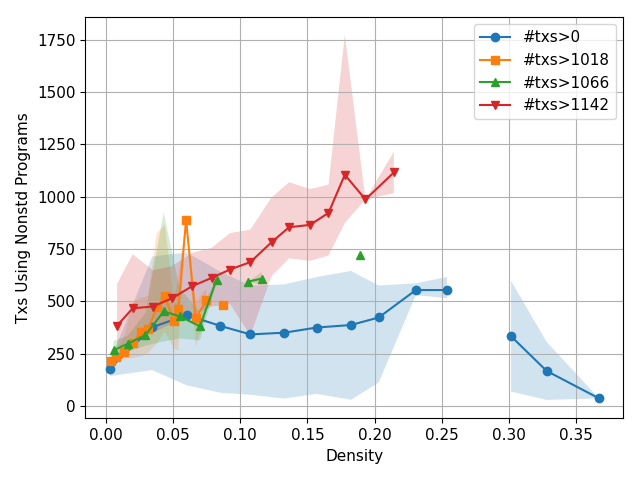}}
\caption{txs-using-nonstd-programs}
\label{fig:txs-using-nonstd-programs}
\end{figure}
\begin{figure}
\centerline{\includegraphics[width=\columnwidth]{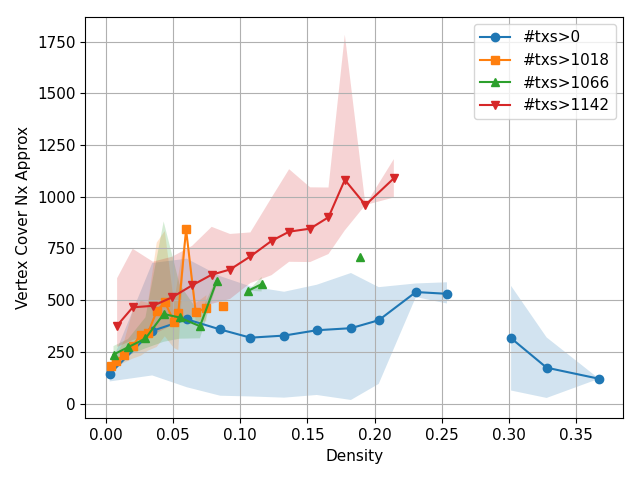}}
\caption{vertex-cover-nx-approx}
\label{fig:vertex-cover-nx-approx}
\end{figure}
\begin{figure}
\centerline{\includegraphics[width=\columnwidth]{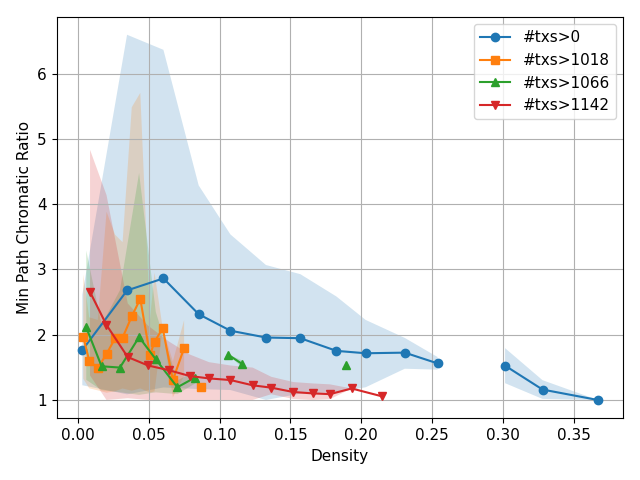}}
\caption{min-path-chromatic-ratio}
\label{fig:min-path-chromatic-ratio}
\end{figure}
\begin{figure}
\centerline{\includegraphics[width=\columnwidth]{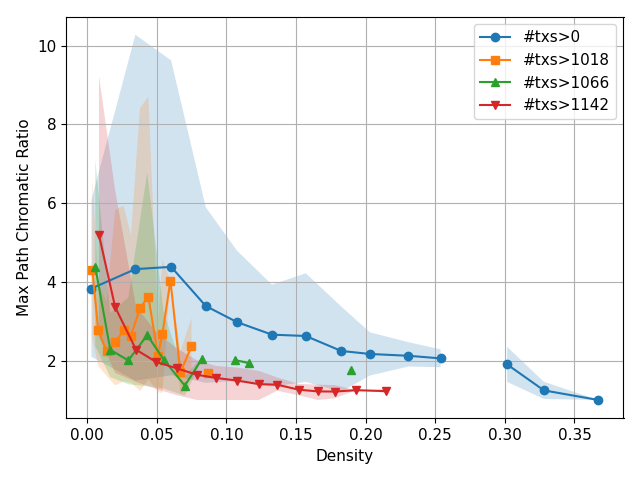}}
\caption{max-path-chromatic-ratio}
\label{fig:max-path-chromatic-ratio}
\end{figure}
\begin{figure}
\centerline{\includegraphics[width=\columnwidth]{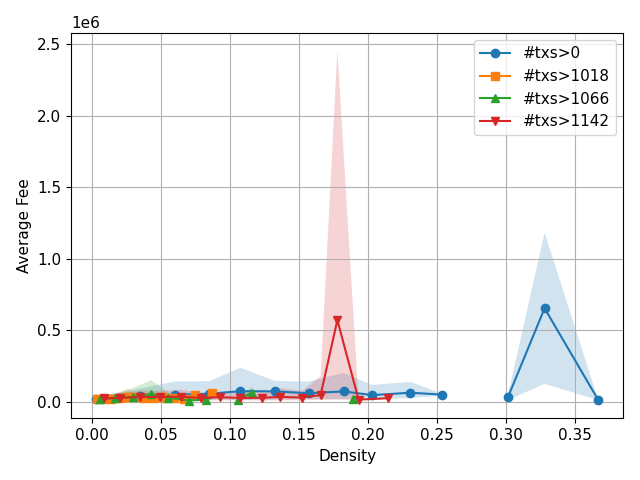}}
\caption{avg-fee}
\label{fig:avg-fee}
\end{figure}
\begin{figure}
\centerline{\includegraphics[width=\columnwidth]{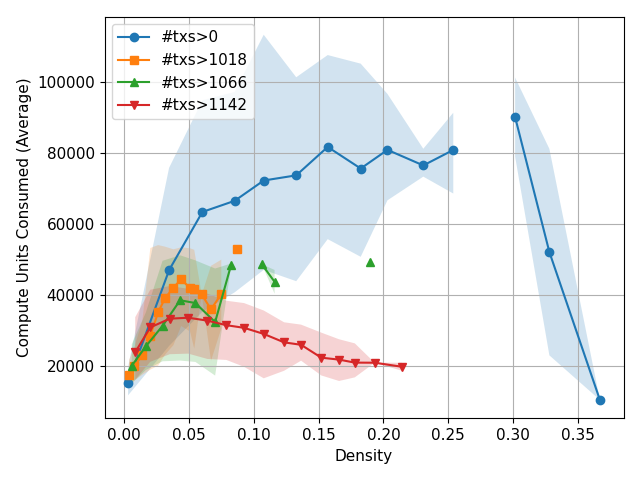}}
\caption{avg-computeUnitsConsumed}
\label{fig:avg-computeUnitsConsumed}
\end{figure}
\begin{figure}
\centerline{\includegraphics[width=\columnwidth]{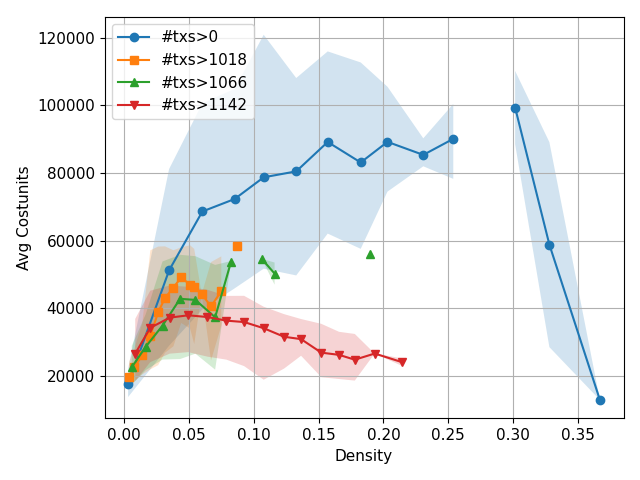}}
\caption{avg-costUnits}
\label{fig:avg-costUnits}
\end{figure}
\begin{figure}
\centerline{\includegraphics[width=\columnwidth]{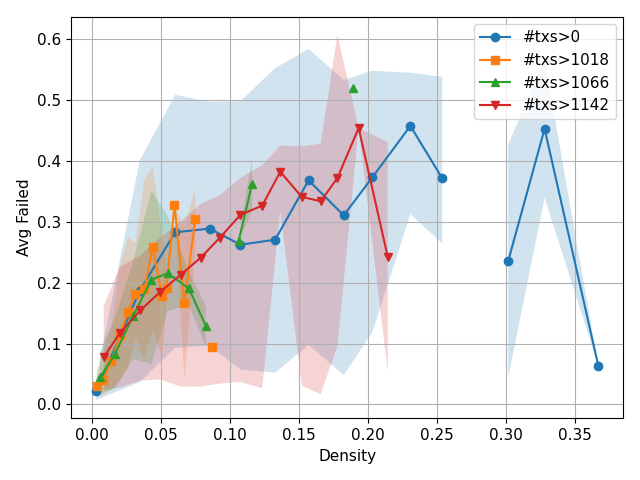}}
\caption{avg-failed}
\label{fig:avg-failed}
\end{figure}
\begin{figure}
\centerline{\includegraphics[width=\columnwidth]{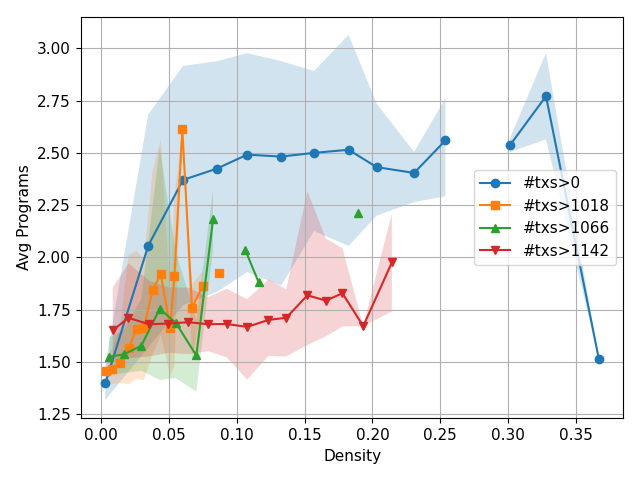}}
\caption{avg-programs}
\label{fig:avg-programs}
\end{figure}
\begin{figure}
\centerline{\includegraphics[width=\columnwidth]{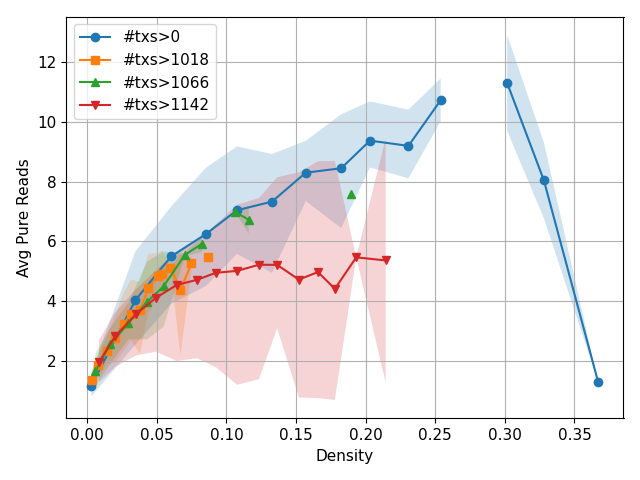}}
\caption{avg-pure-reads}
\label{fig:avg-pure-reads}
\end{figure}
\begin{figure}
\centerline{\includegraphics[width=\columnwidth]{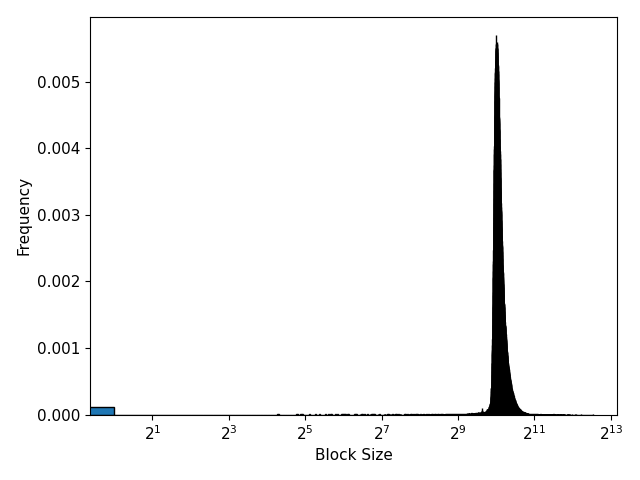}}
\caption{block-size-dist}
\label{fig:block-size-dist}
\end{figure}
\begin{figure}
\centerline{\includegraphics[width=\columnwidth]{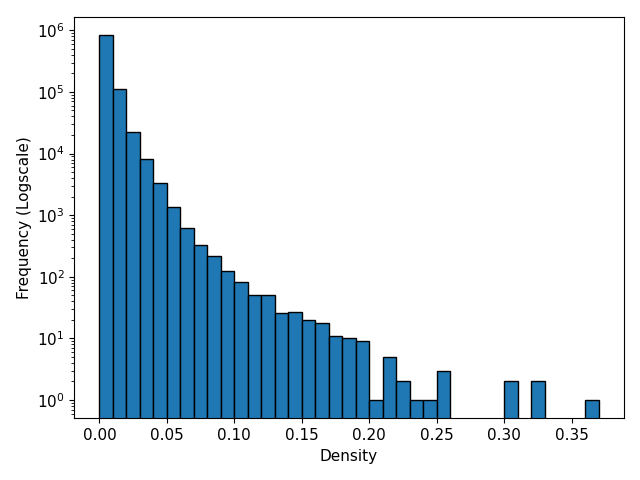}}
\caption{density-dist}
\label{fig:density-dist}
\end{figure}
\begin{figure}
\centerline{\includegraphics[width=\columnwidth]{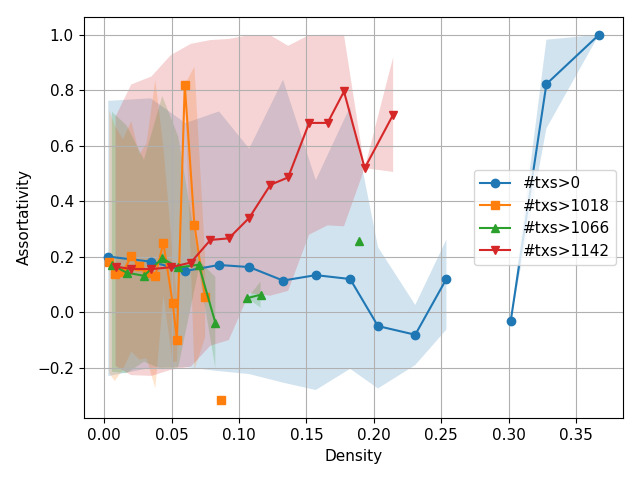}}
\caption{assortativity}
\label{fig:assortativity}
\end{figure}
\begin{figure}
\centerline{\includegraphics[width=\columnwidth]{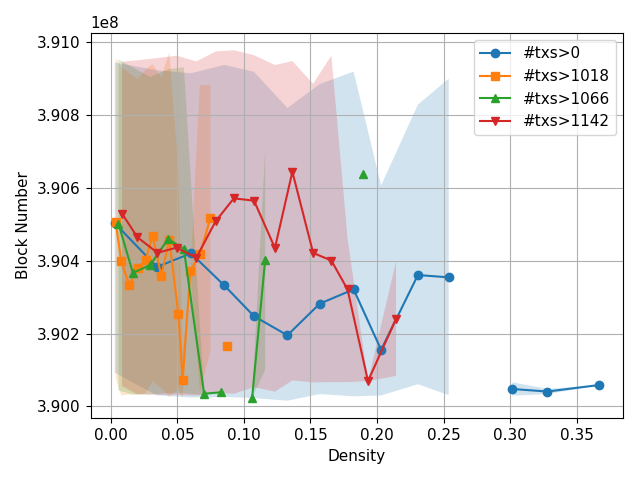}}
\caption{block-number}
\label{fig:block-number}
\end{figure}
\begin{figure}
\centerline{\includegraphics[width=\columnwidth]{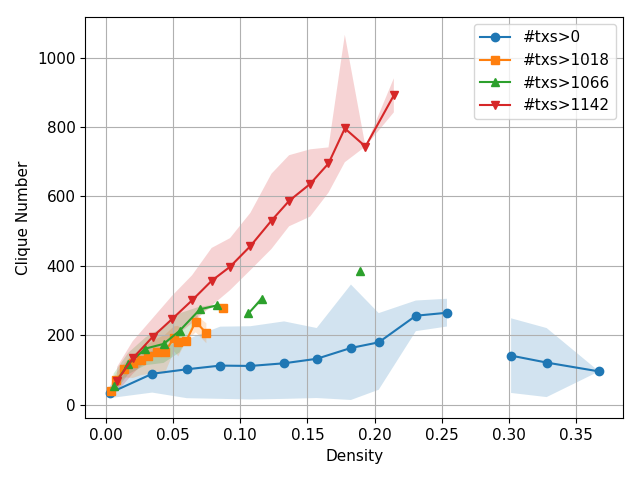}}
\caption{clique-number}
\label{fig:clique-number}
\end{figure}
\begin{figure}
\centerline{\includegraphics[width=\columnwidth]{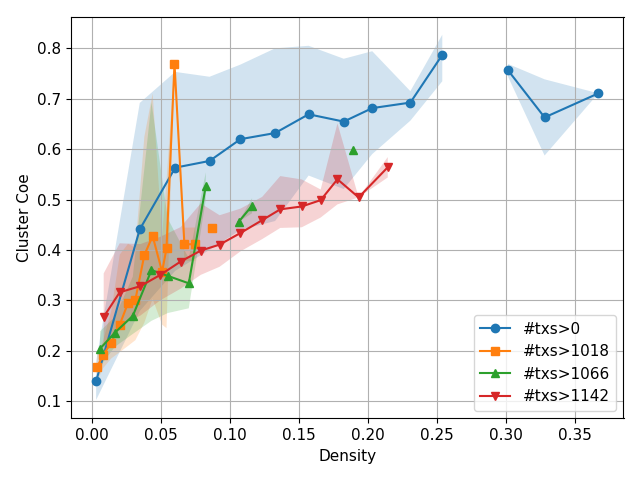}}
\caption{cluster-coe}
\label{fig:cluster-coe}
\end{figure}
\begin{figure}
\centerline{\includegraphics[width=\columnwidth]{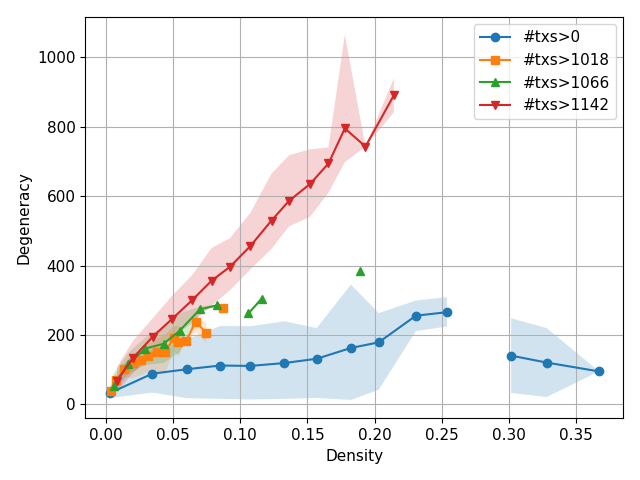}}
\caption{degeneracy}
\label{fig:degeneracy}
\end{figure}
\begin{figure}
\centerline{\includegraphics[width=\columnwidth]{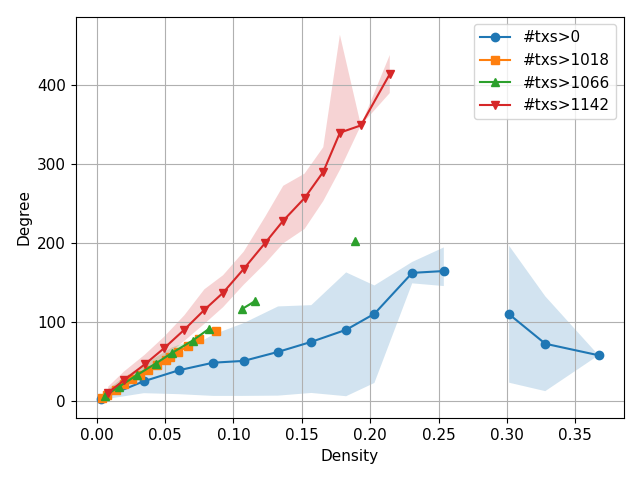}}
\caption{degree}
\label{fig:degree}
\end{figure}
\begin{figure}
\centerline{\includegraphics[width=\columnwidth]{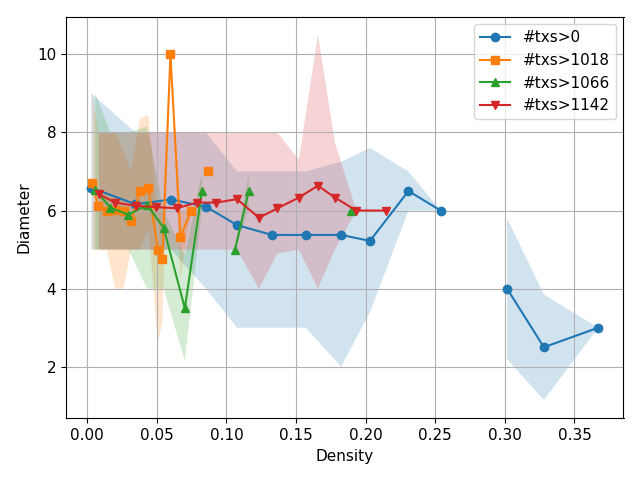}}
\caption{diameter}
\label{fig:diameter}
\end{figure}
\begin{figure}
\centerline{\includegraphics[width=\columnwidth]{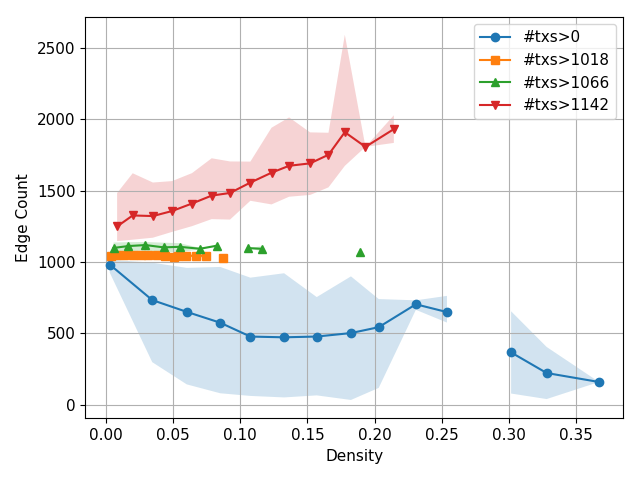}}
\caption{edge-count}
\label{fig:edge-count}
\end{figure}
\begin{figure}
\centerline{\includegraphics[width=\columnwidth]{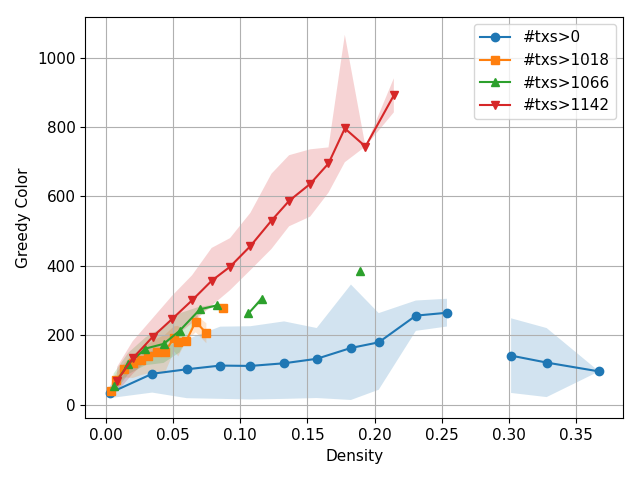}}
\caption{greedy-color}
\label{fig:greedy-color}
\end{figure}
\begin{figure}
\centerline{\includegraphics[width=\columnwidth]{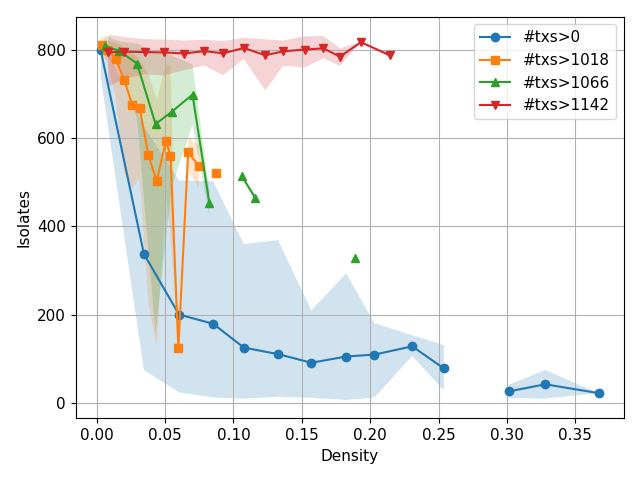}}
\caption{isolates}
\label{fig:isolates}
\end{figure}
\begin{figure}
\centerline{\includegraphics[width=\columnwidth]{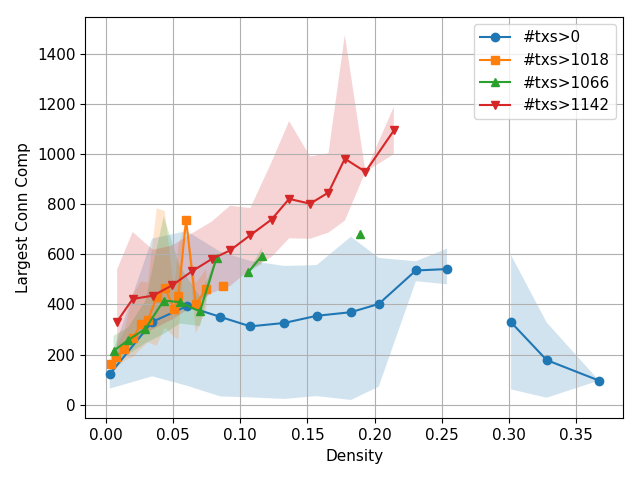}}
\caption{largest-conn-comp}
\label{fig:largest-conn-comp}
\end{figure}
\begin{figure}
\centerline{\includegraphics[width=\columnwidth]{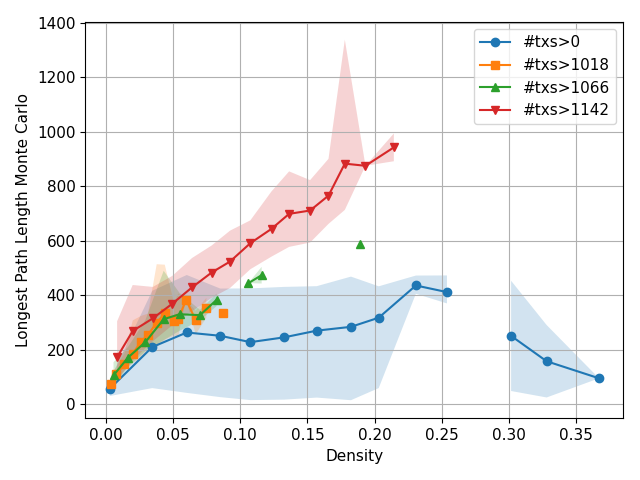}}
\caption{longest-path-length-monte-carlo}
\label{fig:longest-path-length-monte-carlo}
\end{figure}
\begin{figure}
\centerline{\includegraphics[width=\columnwidth]{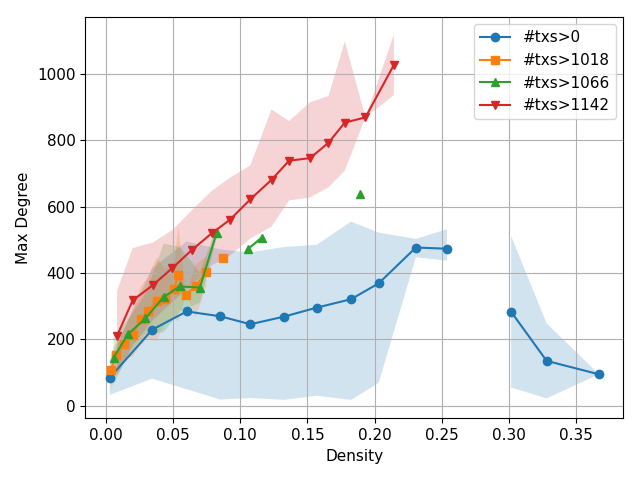}}
\caption{max-degree}
\label{fig:max-degree}
\end{figure}
\begin{figure}
\centerline{\includegraphics[width=\columnwidth]{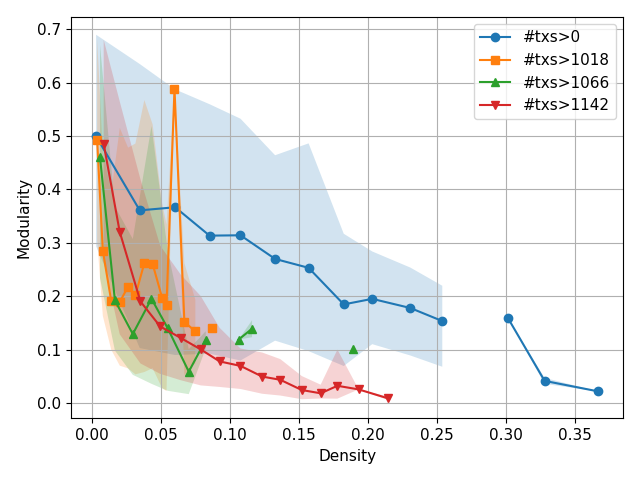}}
\caption{modularity}
\label{fig:modularity}
\end{figure}
\begin{figure}
\centerline{\includegraphics[width=\columnwidth]{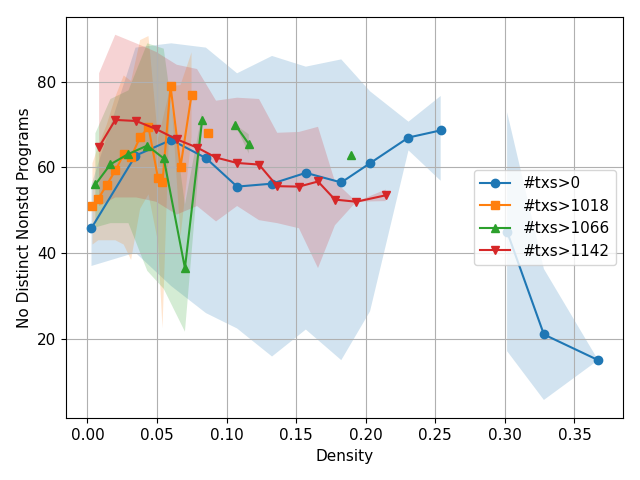}}
\caption{no-distinct-nonstd-programs}
\label{fig:no-distinct-nonstd-programs}
\end{figure}
\begin{figure}
\centerline{\includegraphics[width=\columnwidth]{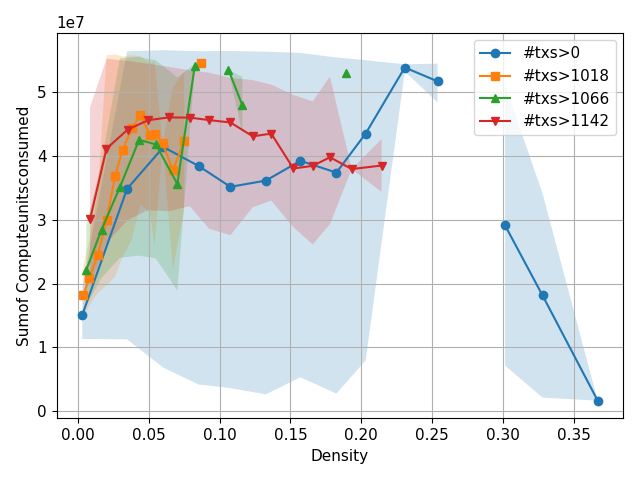}}
\caption{sumof-computeUnitsConsumed}
\label{fig:sumof-computeUnitsConsumed}
\end{figure}
\begin{figure}
\centerline{\includegraphics[width=\columnwidth]{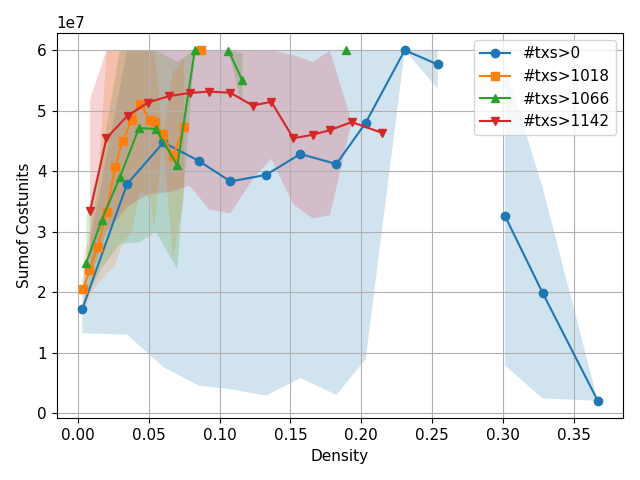}}
\caption{sumof-costUnits}
\label{fig:sumof-costUnits}
\end{figure}
\begin{figure}
\centerline{\includegraphics[width=\columnwidth]{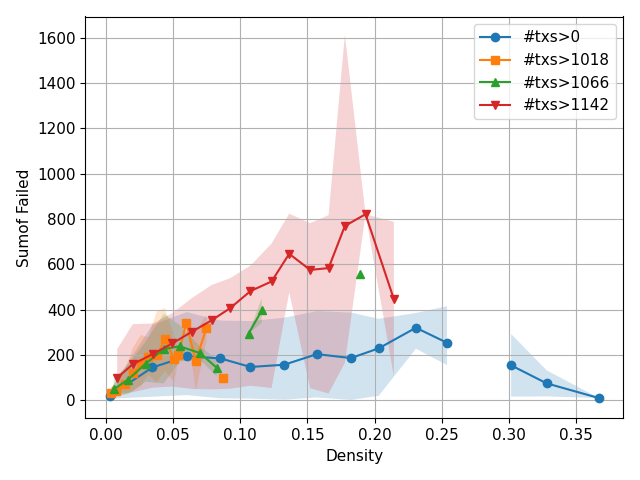}}
\caption{sumof-failed}
\label{fig:sumof-failed}
\end{figure}
\begin{figure}
\centerline{\includegraphics[width=\columnwidth]{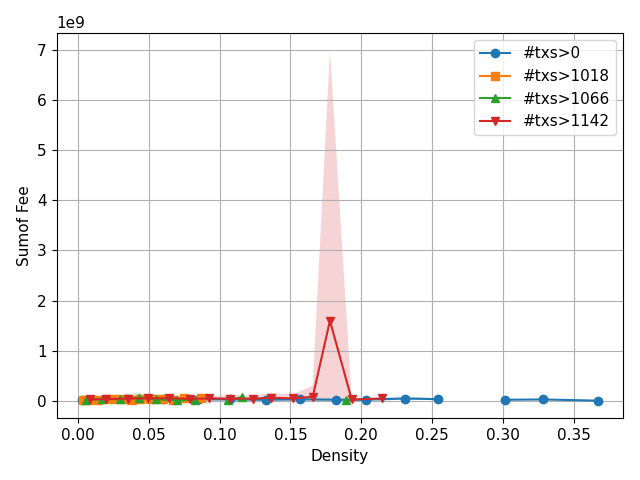}}
\caption{sumof-fee}
\label{fig:sumof-fee}
\end{figure}
\begin{figure}
\centerline{\includegraphics[width=\columnwidth]{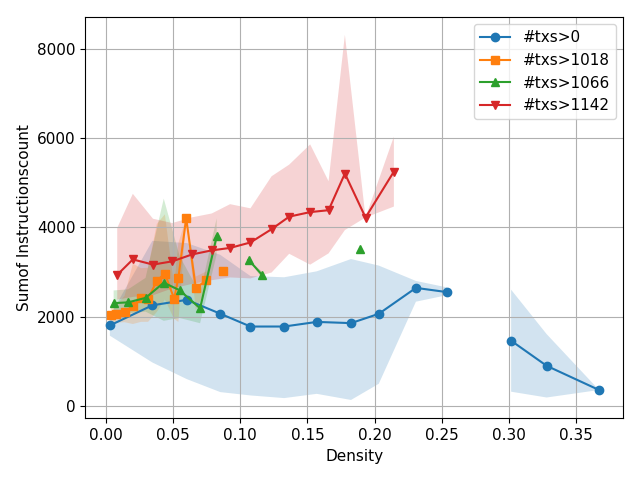}}
\caption{sumof-instructionsCount}
\label{fig:sumof-instructionsCount}
\end{figure}
\begin{figure}
\centerline{\includegraphics[width=\columnwidth]{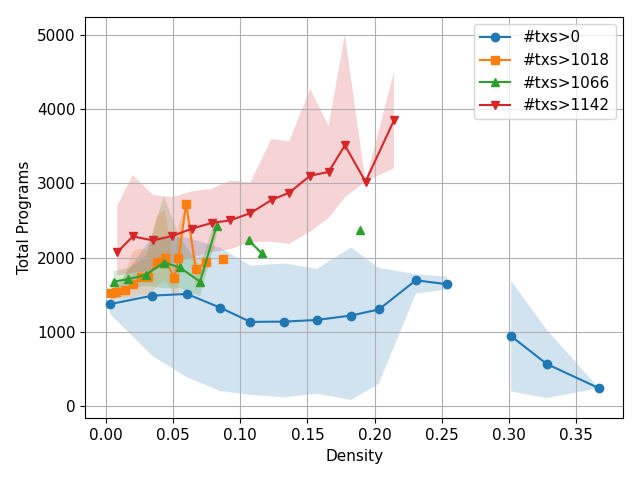}}
\caption{total-programs}
\label{fig:total-programs}
\end{figure}
\begin{figure}
\centerline{\includegraphics[width=\columnwidth]{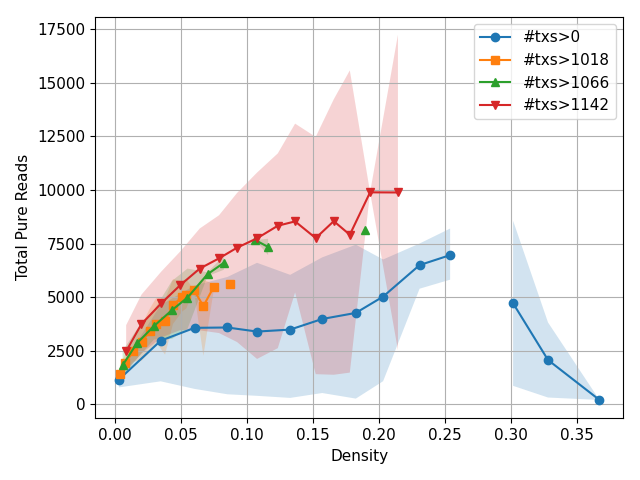}}
\caption{total-pure-reads}
\label{fig:total-pure-reads}
\end{figure}
\begin{figure}
\centerline{\includegraphics[width=\columnwidth]{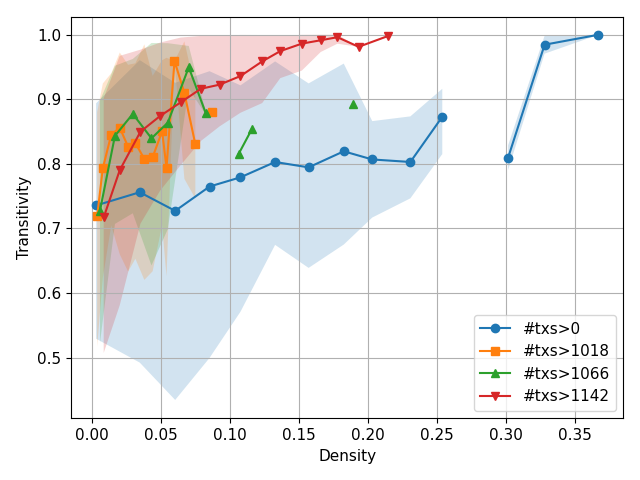}}
\caption{transitivity}
\label{fig:transitivity}
\end{figure}
\begin{figure}
\centerline{\includegraphics[width=\columnwidth]{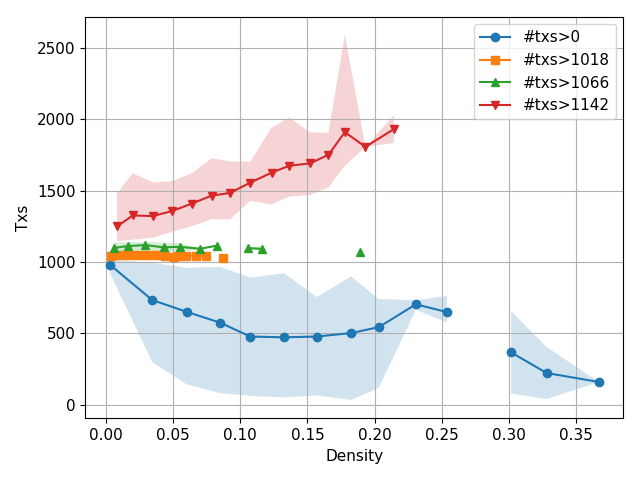}}
\caption{txs}
\label{fig:txs}
\end{figure}
\begin{figure}
\centerline{\includegraphics[width=\columnwidth]{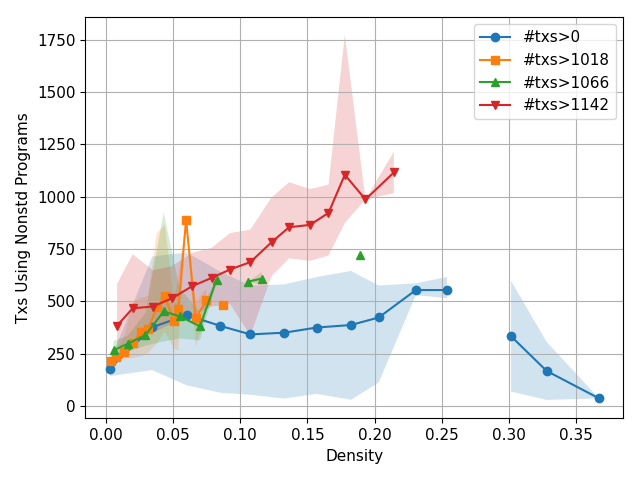}}
\caption{txs-using-nonstd-programs}
\label{fig:txs-using-nonstd-programs}
\end{figure}
\begin{figure}
\centerline{\includegraphics[width=\columnwidth]{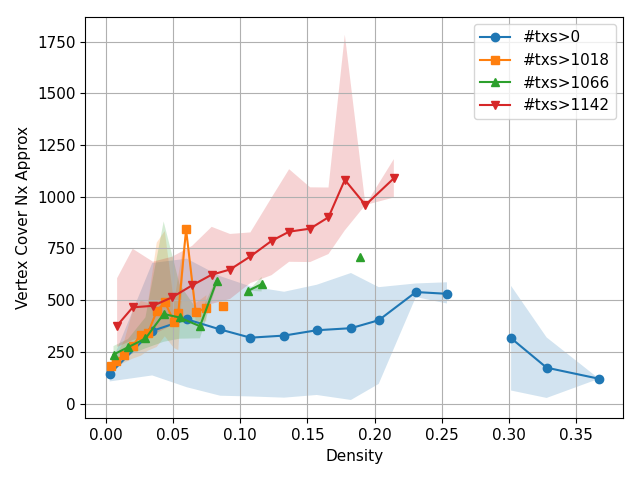}}
\caption{vertex-cover-nx-approx}
\label{fig:vertex-cover-nx-approx}
\end{figure}
\begin{figure}
\centerline{\includegraphics[width=\columnwidth]{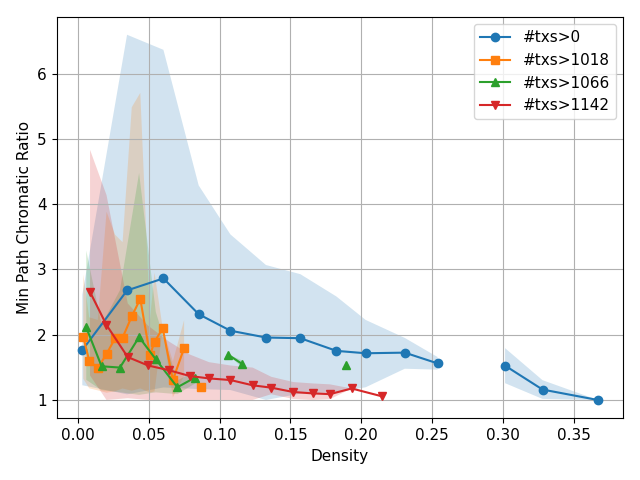}}
\caption{min-path-chromatic-ratio}
\label{fig:min-path-chromatic-ratio}
\end{figure}
\begin{figure}
\centerline{\includegraphics[width=\columnwidth]{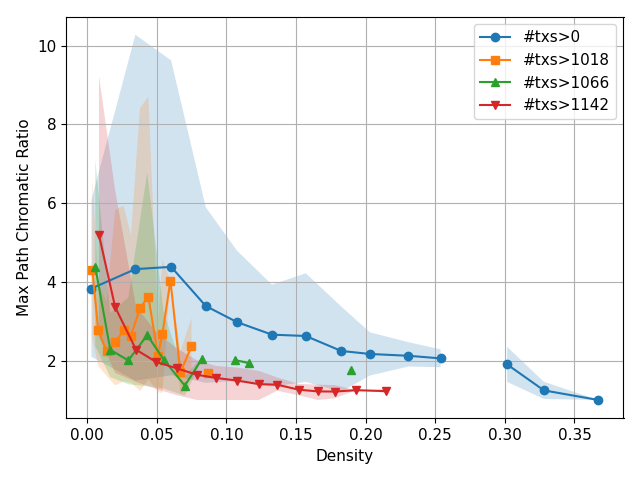}}
\caption{max-path-chromatic-ratio}
\label{fig:max-path-chromatic-ratio}
\end{figure}
\begin{figure}
\centerline{\includegraphics[width=\columnwidth]{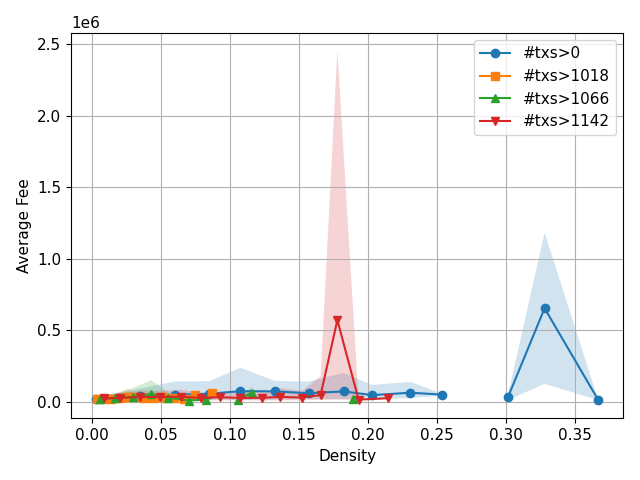}}
\caption{avg-fee}
\label{fig:avg-fee}
\end{figure}
\begin{figure}
\centerline{\includegraphics[width=\columnwidth]{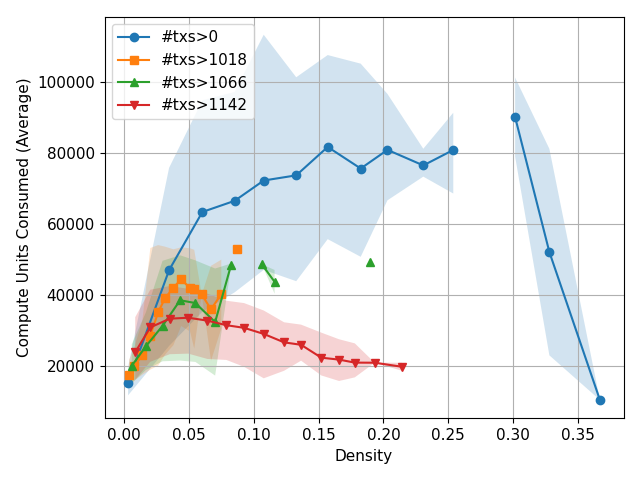}}
\caption{avg-computeUnitsConsumed}
\label{fig:avg-computeUnitsConsumed}
\end{figure}
\begin{figure}
\centerline{\includegraphics[width=\columnwidth]{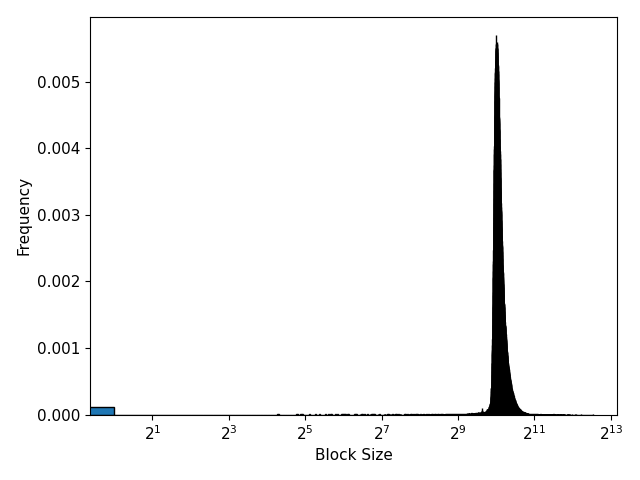}}
\caption{block-size-dist}
\label{fig:block-size-dist}
\end{figure}
\begin{figure}
\centerline{\includegraphics[width=\columnwidth]{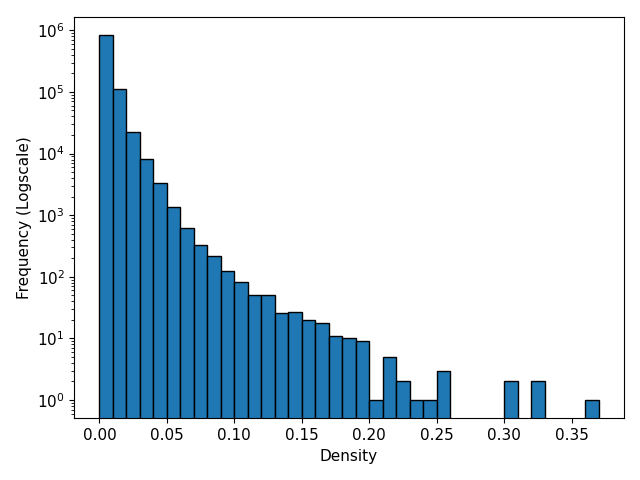}}
\caption{density-dist}
\label{fig:density-dist}
\end{figure}
\begin{figure}
\centerline{\includegraphics[width=\columnwidth]{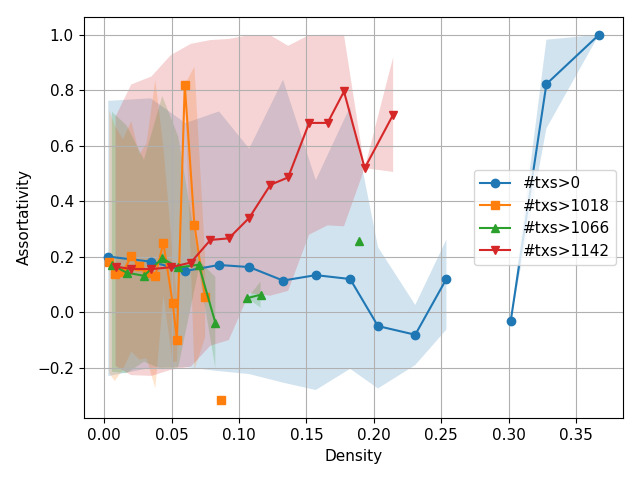}}
\caption{assortativity}
\label{fig:assortativity}
\end{figure}
\begin{figure}
\centerline{\includegraphics[width=\columnwidth]{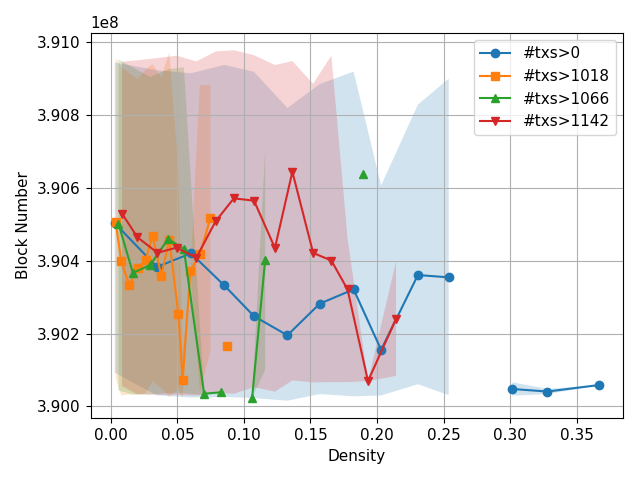}}
\caption{block-number}
\label{fig:block-number}
\end{figure}
\begin{figure}
\centerline{\includegraphics[width=\columnwidth]{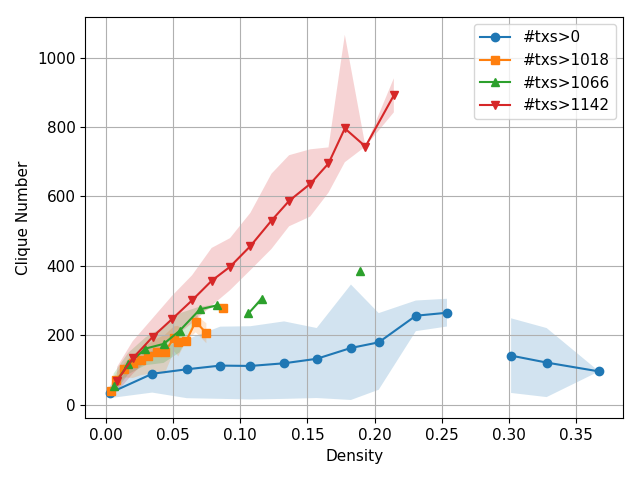}}
\caption{clique-number}
\label{fig:clique-number}
\end{figure}
\begin{figure}
\centerline{\includegraphics[width=\columnwidth]{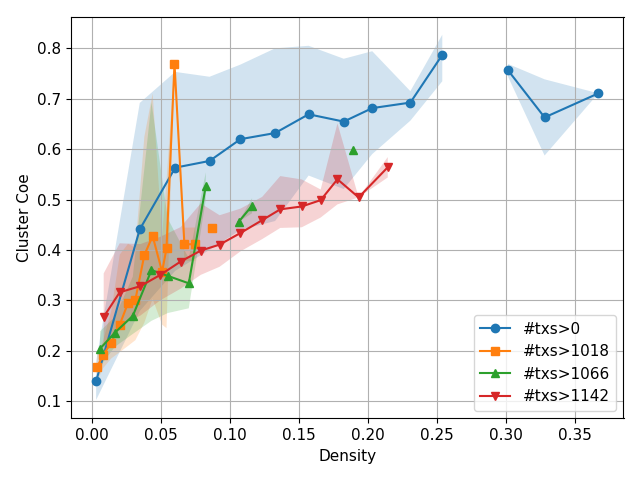}}
\caption{cluster-coe}
\label{fig:cluster-coe}
\end{figure}
\begin{figure}
\centerline{\includegraphics[width=\columnwidth]{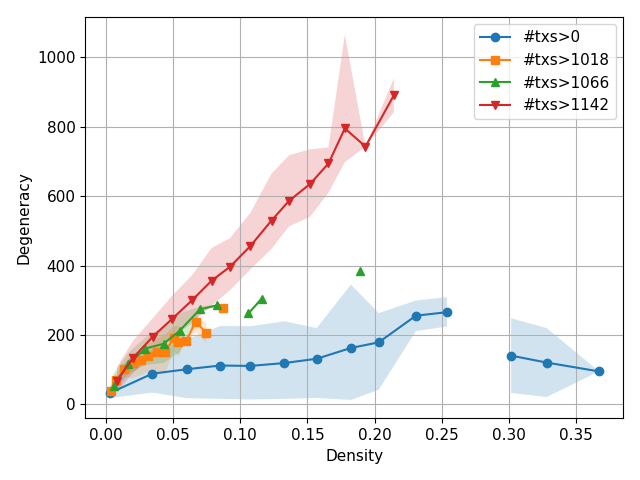}}
\caption{degeneracy}
\label{fig:degeneracy}
\end{figure}
\begin{figure}
\centerline{\includegraphics[width=\columnwidth]{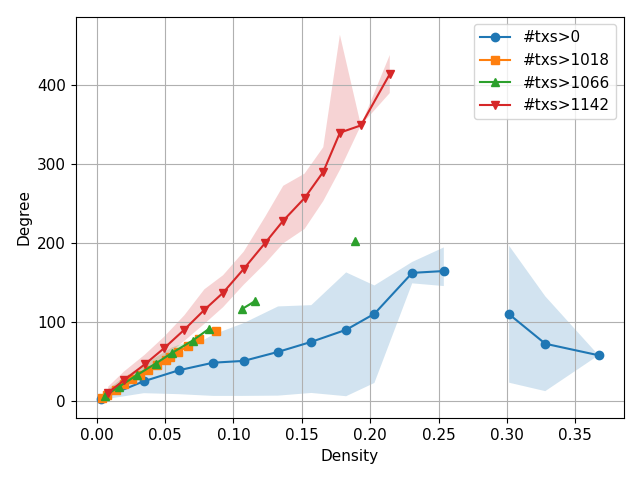}}
\caption{degree}
\label{fig:degree}
\end{figure}
\begin{figure}
\centerline{\includegraphics[width=\columnwidth]{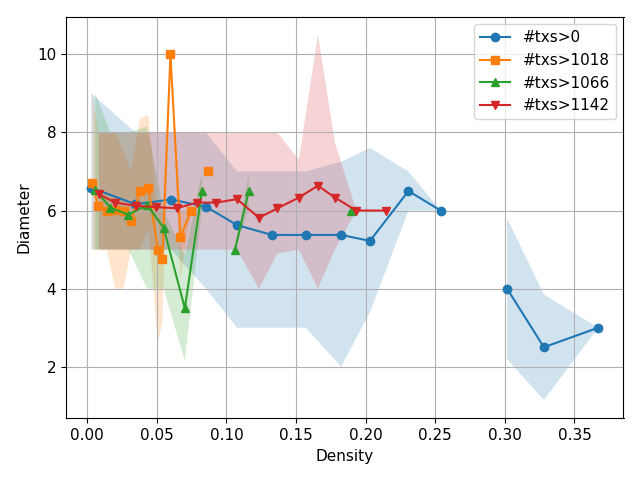}}
\caption{diameter}
\label{fig:diameter}
\end{figure}
\begin{figure}
\centerline{\includegraphics[width=\columnwidth]{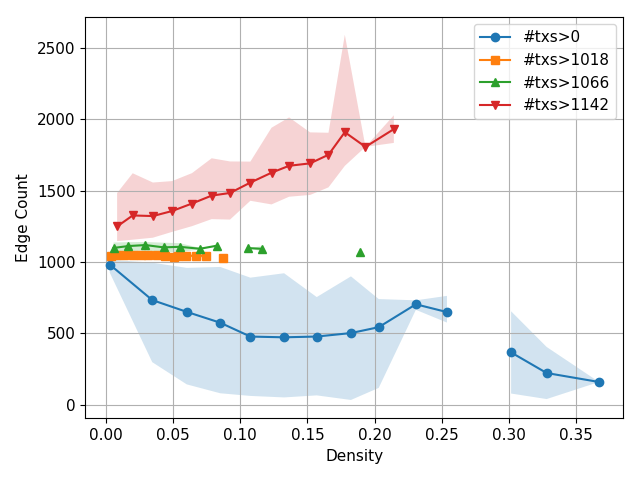}}
\caption{edge-count}
\label{fig:edge-count}
\end{figure}
\begin{figure}
\centerline{\includegraphics[width=\columnwidth]{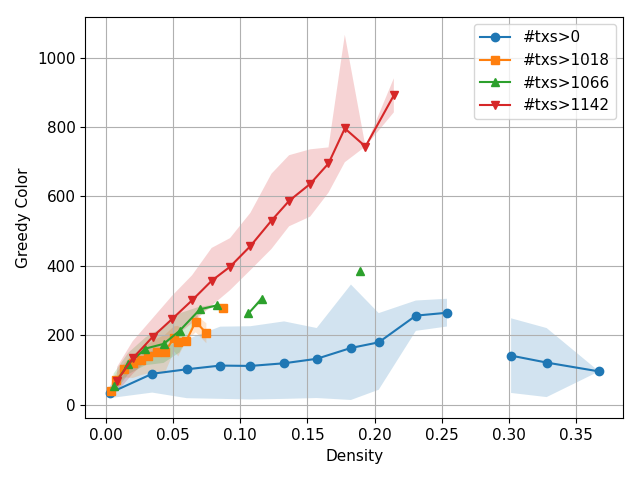}}
\caption{greedy-color}
\label{fig:greedy-color}
\end{figure}
\begin{figure}
\centerline{\includegraphics[width=\columnwidth]{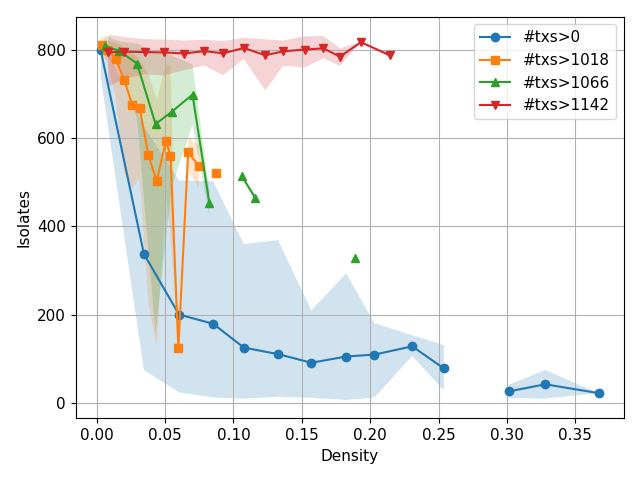}}
\caption{isolates}
\label{fig:isolates}
\end{figure}
\begin{figure}
\centerline{\includegraphics[width=\columnwidth]{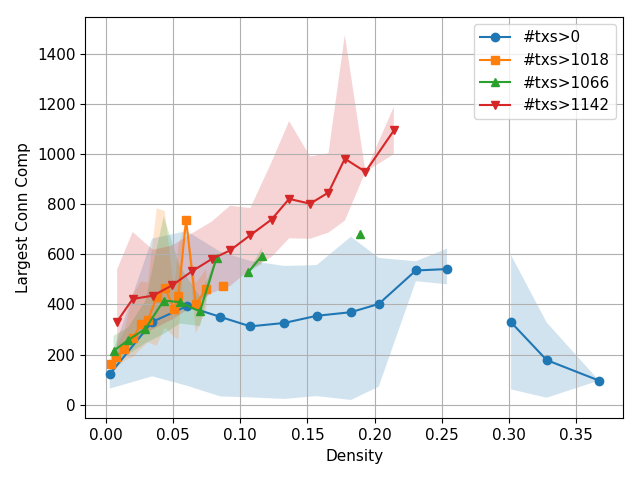}}
\caption{largest-conn-comp}
\label{fig:largest-conn-comp}
\end{figure}
\begin{figure}
\centerline{\includegraphics[width=\columnwidth]{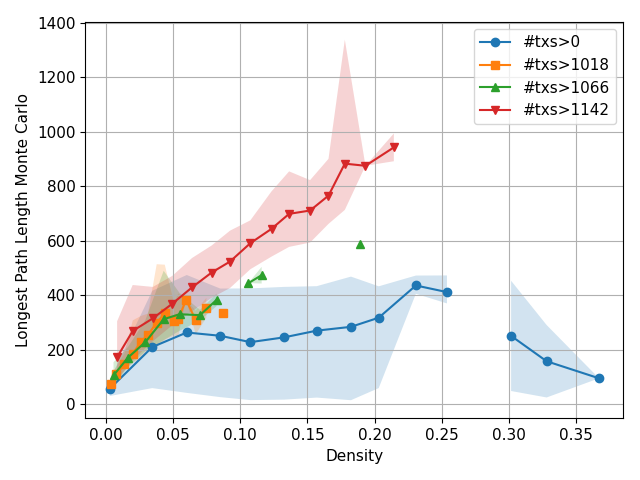}}
\caption{longest-path-length-monte-carlo}
\label{fig:longest-path-length-monte-carlo}
\end{figure}
\begin{figure}
\centerline{\includegraphics[width=\columnwidth]{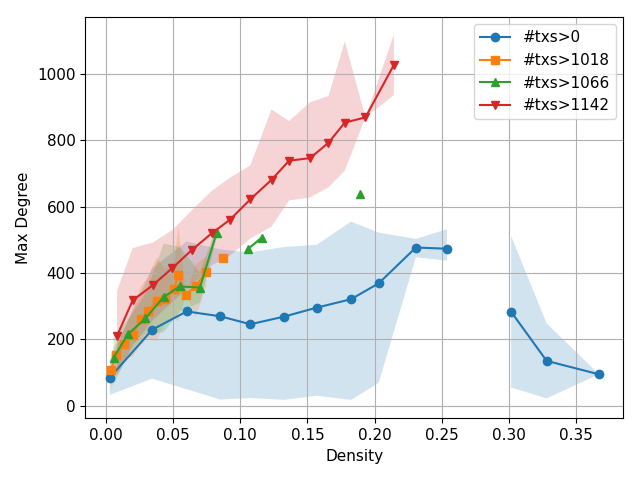}}
\caption{max-degree}
\label{fig:max-degree}
\end{figure}
\begin{figure}
\centerline{\includegraphics[width=\columnwidth]{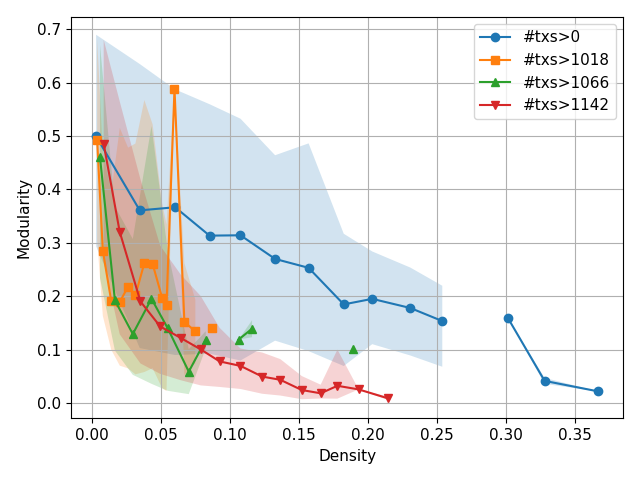}}
\caption{modularity}
\label{fig:modularity}
\end{figure}
\begin{figure}
\centerline{\includegraphics[width=\columnwidth]{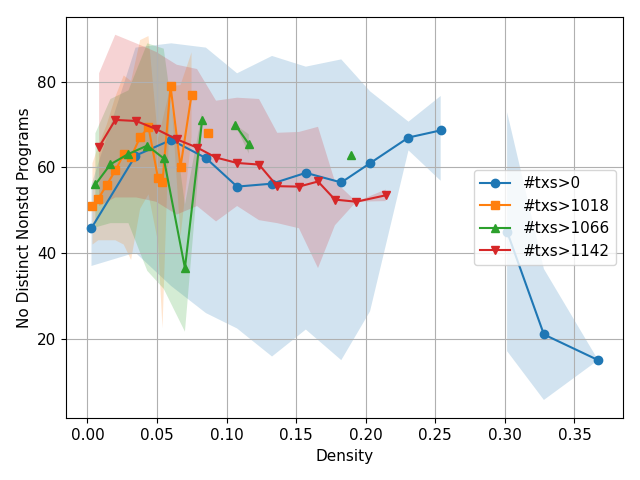}}
\caption{no-distinct-nonstd-programs}
\label{fig:no-distinct-nonstd-programs}
\end{figure}
\begin{figure}
\centerline{\includegraphics[width=\columnwidth]{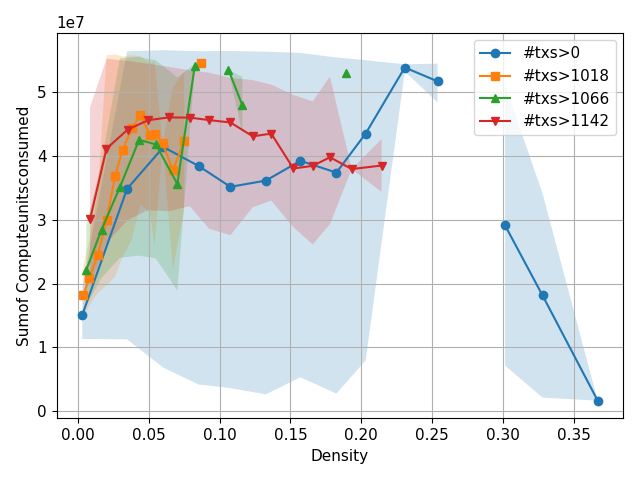}}
\caption{sumof-computeUnitsConsumed}
\label{fig:sumof-computeUnitsConsumed}
\end{figure}
\begin{figure}
\centerline{\includegraphics[width=\columnwidth]{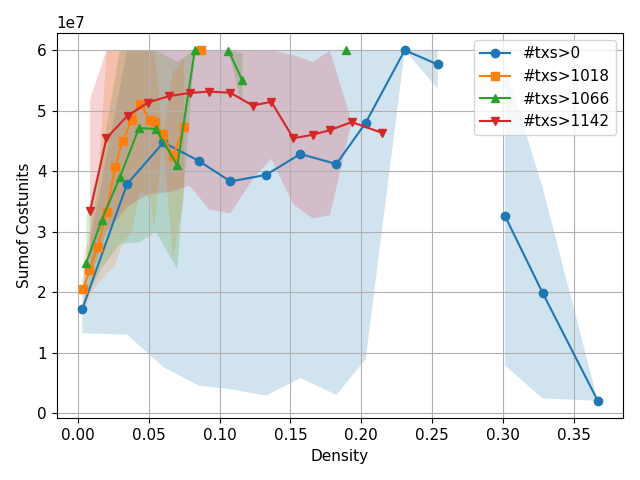}}
\caption{sumof-costUnits}
\label{fig:sumof-costUnits}
\end{figure}
\begin{figure}
\centerline{\includegraphics[width=\columnwidth]{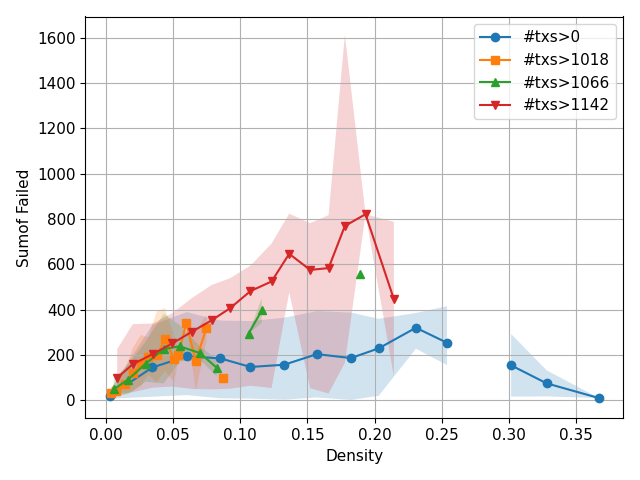}}
\caption{sumof-failed}
\label{fig:sumof-failed}
\end{figure}
\begin{figure}
\centerline{\includegraphics[width=\columnwidth]{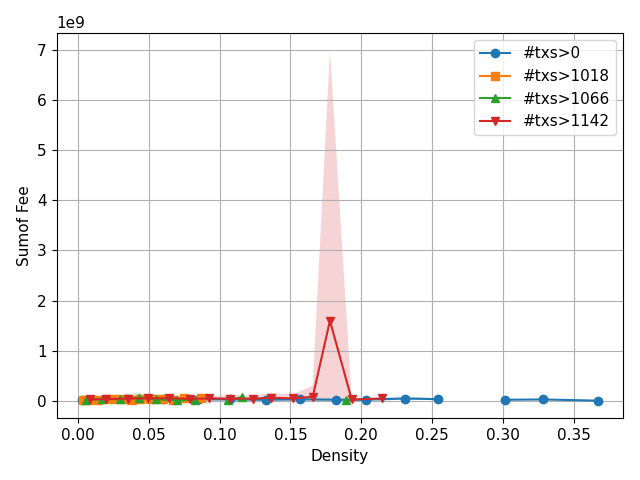}}
\caption{sumof-fee}
\label{fig:sumof-fee}
\end{figure}
\begin{figure}
\centerline{\includegraphics[width=\columnwidth]{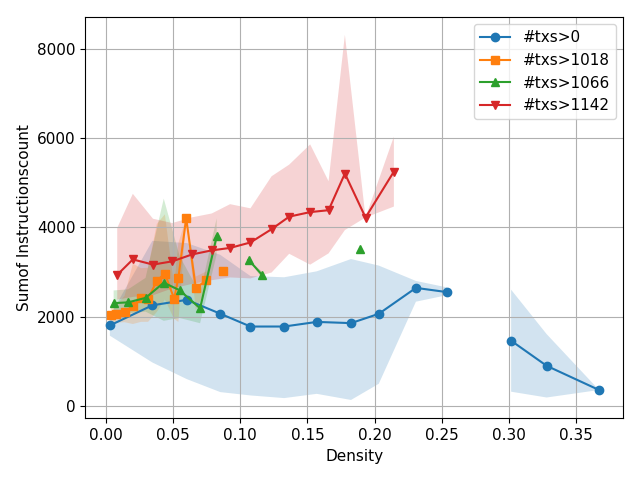}}
\caption{sumof-instructionsCount}
\label{fig:sumof-instructionsCount}
\end{figure}
\begin{figure}
\centerline{\includegraphics[width=\columnwidth]{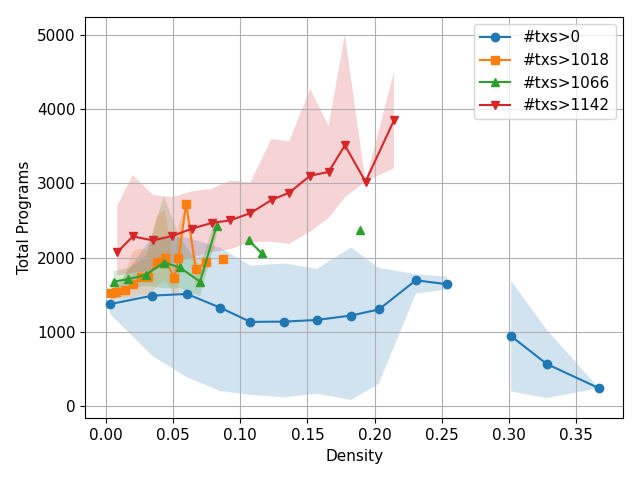}}
\caption{total-programs}
\label{fig:total-programs}
\end{figure}
\begin{figure}
\centerline{\includegraphics[width=\columnwidth]{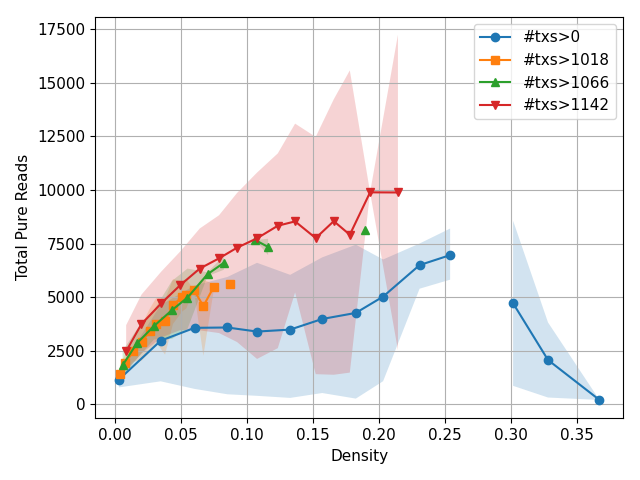}}
\caption{total-pure-reads}
\label{fig:total-pure-reads}
\end{figure}
\begin{figure}
\centerline{\includegraphics[width=\columnwidth]{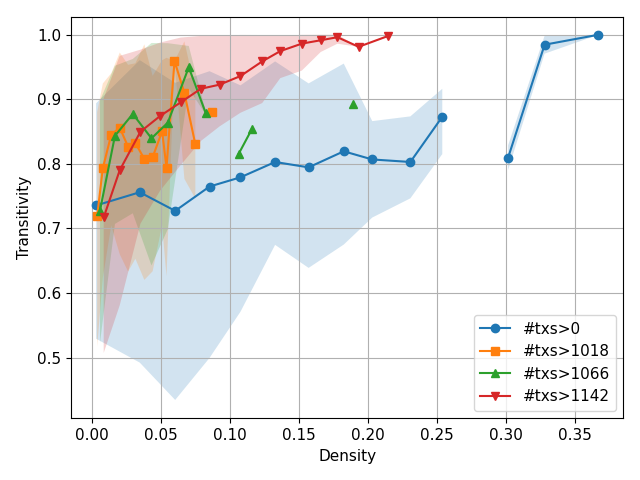}}
\caption{transitivity}
\label{fig:transitivity}
\end{figure}
\begin{figure}
\centerline{\includegraphics[width=\columnwidth]{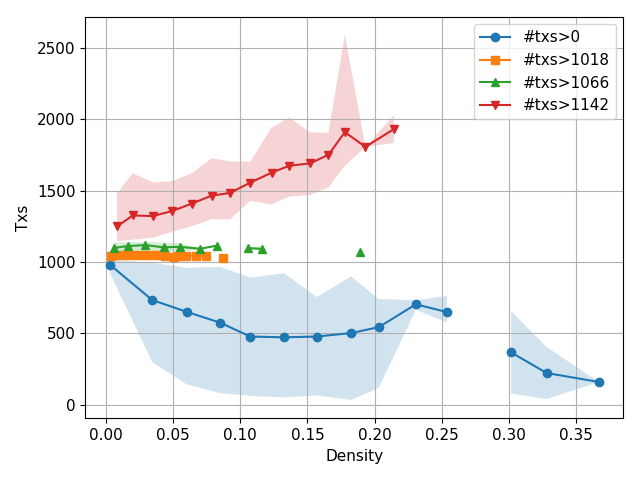}}
\caption{txs}
\label{fig:txs}
\end{figure}
\begin{figure}
\centerline{\includegraphics[width=\columnwidth]{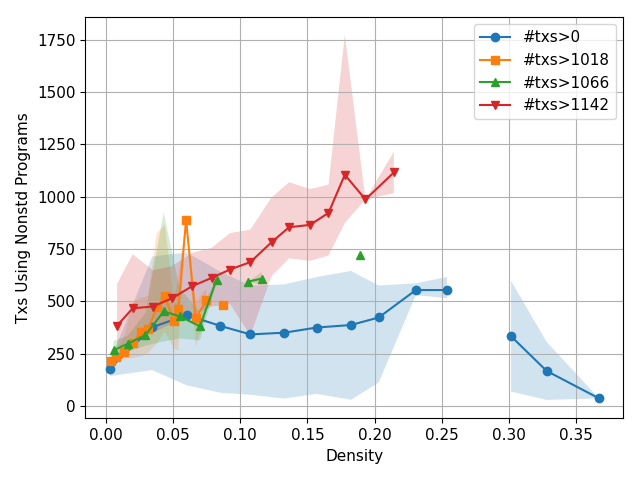}}
\caption{txs-using-nonstd-programs}
\label{fig:txs-using-nonstd-programs}
\end{figure}
\begin{figure}
\centerline{\includegraphics[width=\columnwidth]{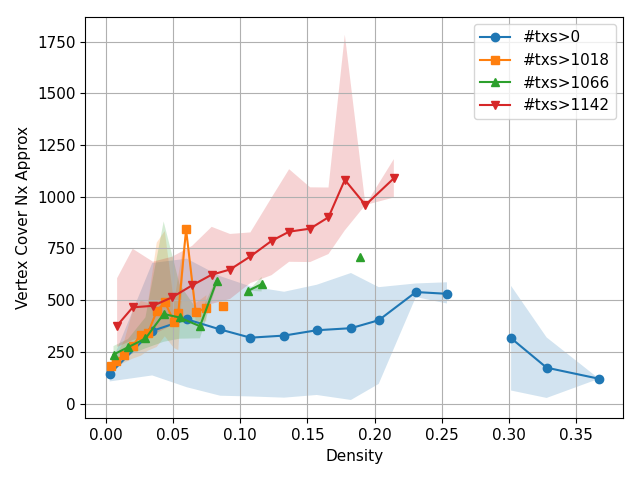}}
\caption{vertex-cover-nx-approx}
\label{fig:vertex-cover-nx-approx}
\end{figure}
\begin{figure}
\centerline{\includegraphics[width=\columnwidth]{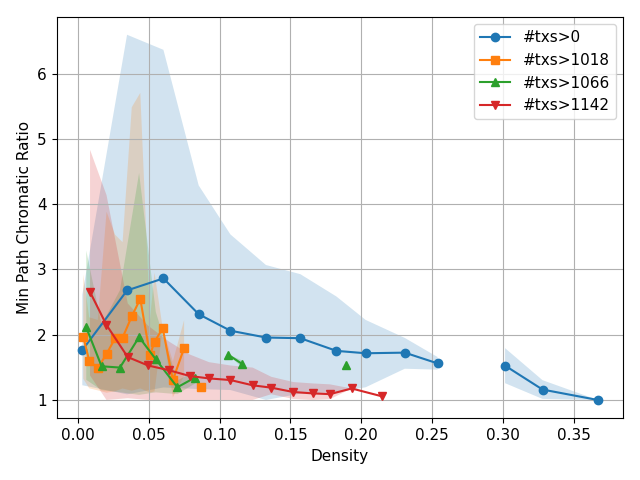}}
\caption{min-path-chromatic-ratio}
\label{fig:min-path-chromatic-ratio}
\end{figure}
\begin{figure}
\centerline{\includegraphics[width=\columnwidth]{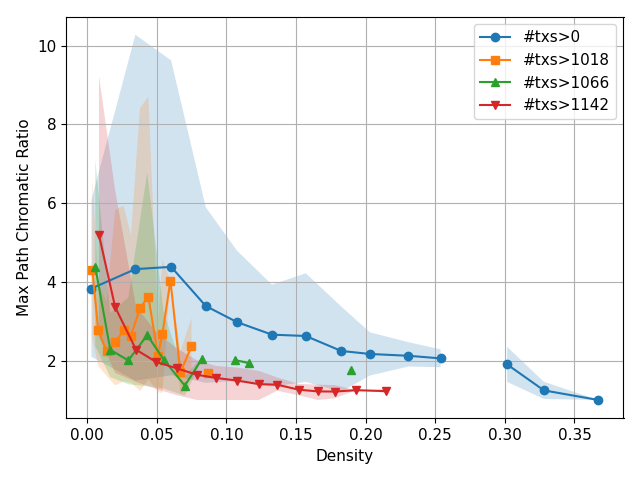}}
\caption{max-path-chromatic-ratio}
\label{fig:max-path-chromatic-ratio}
\end{figure}
\begin{figure}
\centerline{\includegraphics[width=\columnwidth]{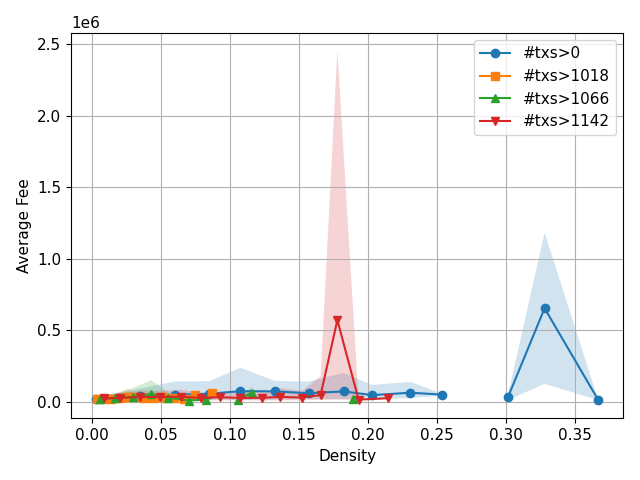}}
\caption{avg-fee}
\label{fig:avg-fee}
\end{figure}
\begin{figure}
\centerline{\includegraphics[width=\columnwidth]{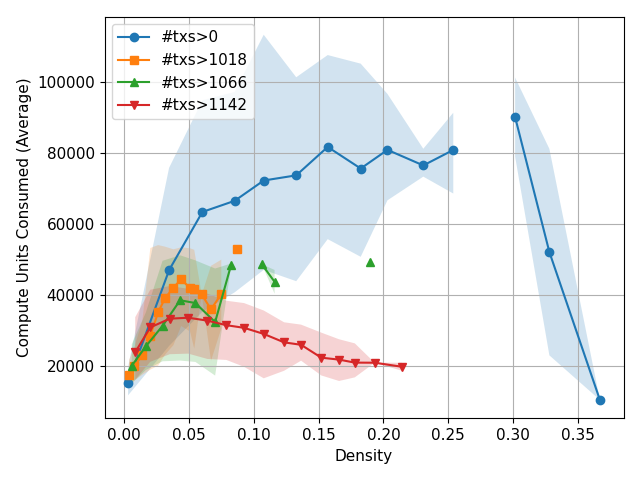}}
\caption{avg-computeUnitsConsumed}
\label{fig:avg-computeUnitsConsumed}
\end{figure}
\begin{figure}
\centerline{\includegraphics[width=\columnwidth]{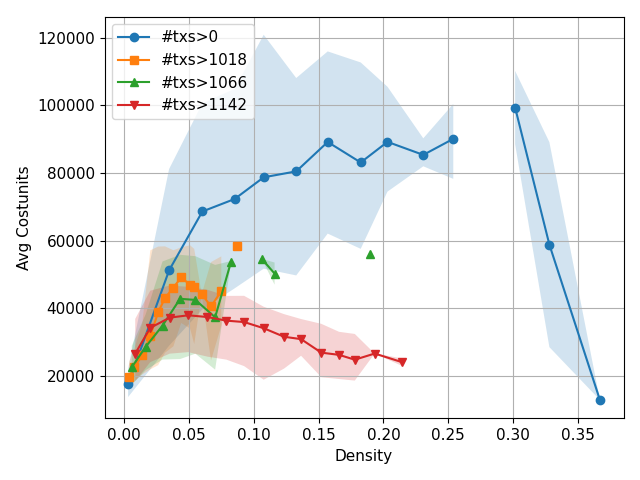}}
\caption{avg-costUnits}
\label{fig:avg-costUnits}
\end{figure}
\begin{figure}
\centerline{\includegraphics[width=\columnwidth]{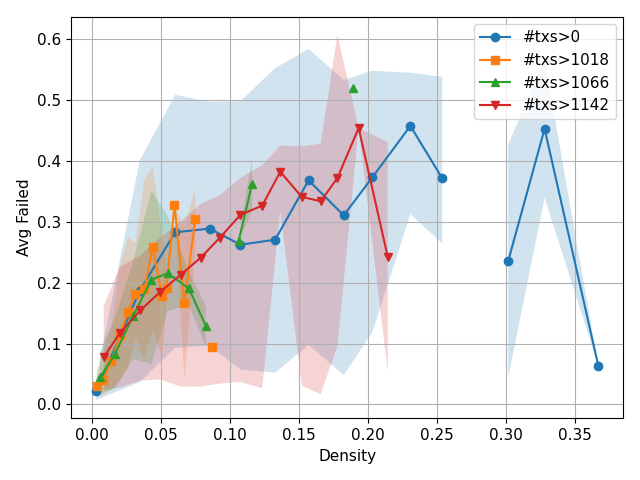}}
\caption{avg-failed}
\label{fig:avg-failed}
\end{figure}
\begin{figure}
\centerline{\includegraphics[width=\columnwidth]{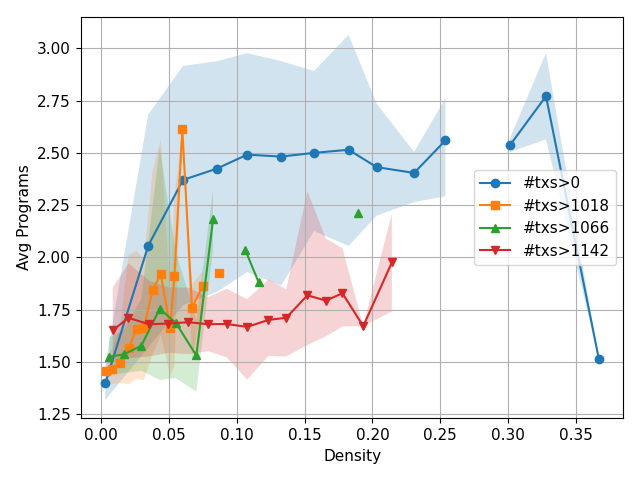}}
\caption{avg-programs}
\label{fig:avg-programs}
\end{figure}
\begin{figure}
\centerline{\includegraphics[width=\columnwidth]{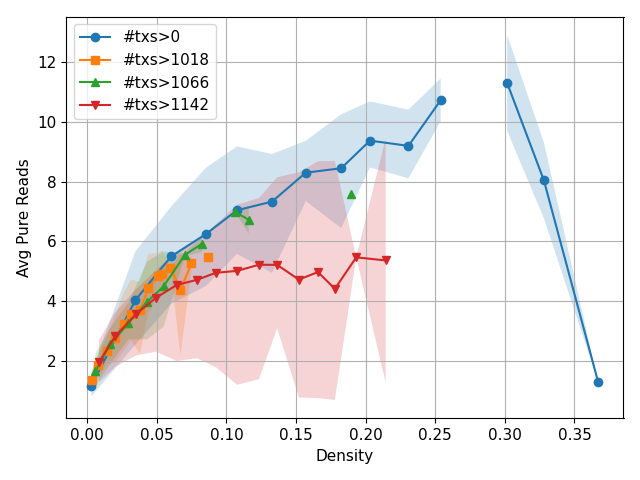}}
\caption{avg-pure-reads}
\label{fig:avg-pure-reads}
\end{figure}
\begin{figure}
\centerline{\includegraphics[width=\columnwidth]{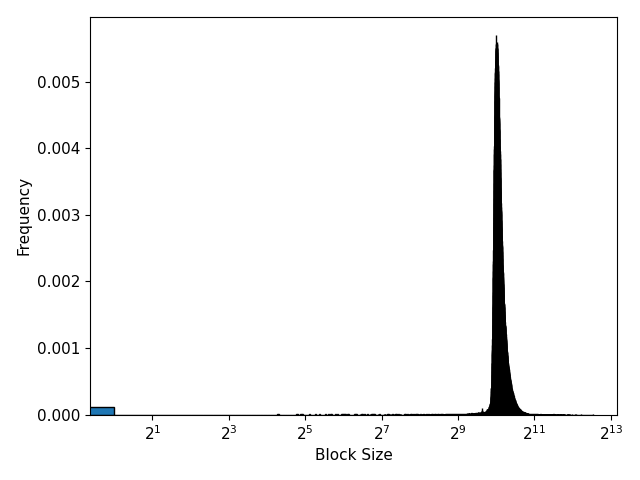}}
\caption{block-size-dist}
\label{fig:block-size-dist}
\end{figure}
\begin{figure}
\centerline{\includegraphics[width=\columnwidth]{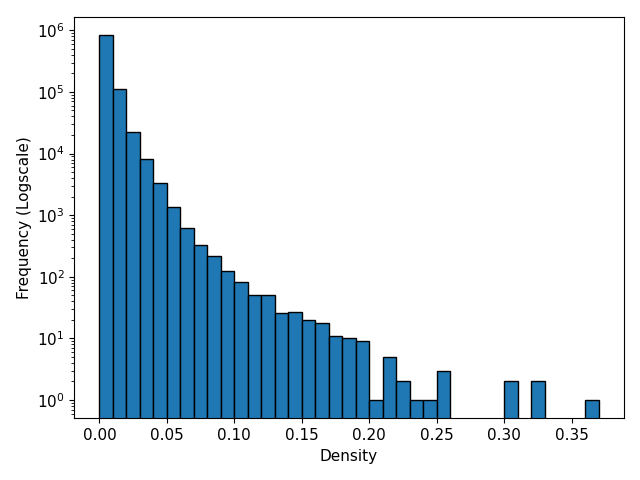}}
\caption{density-dist}
\label{fig:density-dist}
\end{figure}
\begin{figure}
\centerline{\includegraphics[width=\columnwidth]{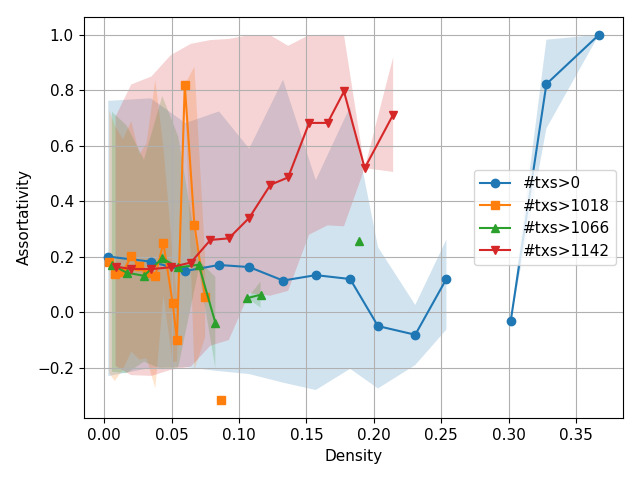}}
\caption{assortativity}
\label{fig:assortativity}
\end{figure}
\begin{figure}
\centerline{\includegraphics[width=\columnwidth]{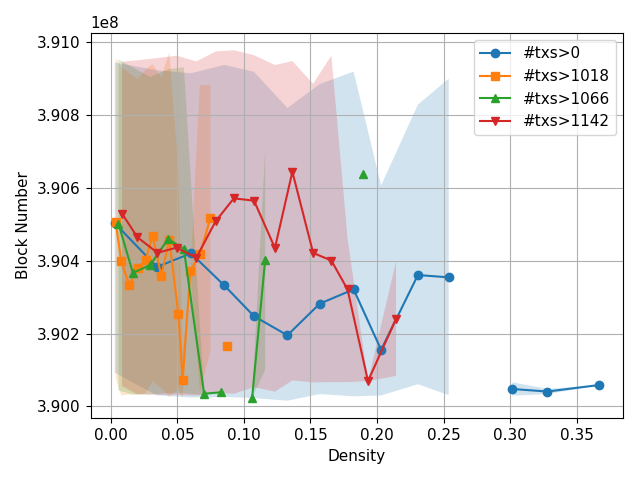}}
\caption{block-number}
\label{fig:block-number}
\end{figure}
\begin{figure}
\centerline{\includegraphics[width=\columnwidth]{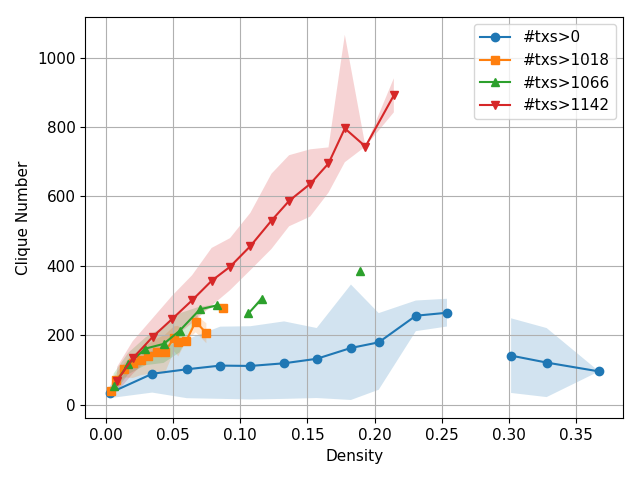}}
\caption{clique-number}
\label{fig:clique-number}
\end{figure}
\begin{figure}
\centerline{\includegraphics[width=\columnwidth]{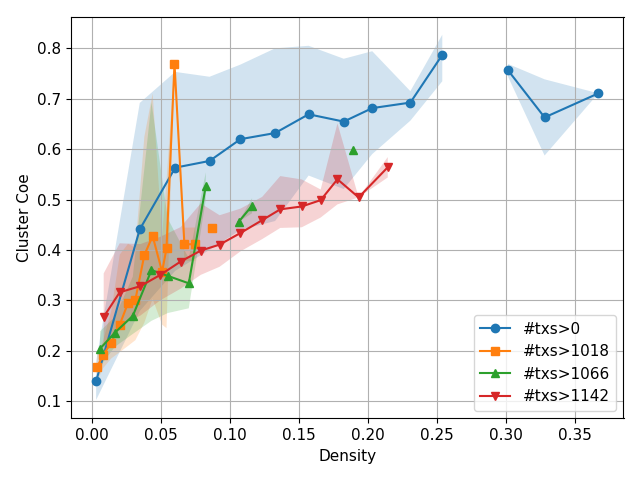}}
\caption{cluster-coe}
\label{fig:cluster-coe}
\end{figure}
\begin{figure}
\centerline{\includegraphics[width=\columnwidth]{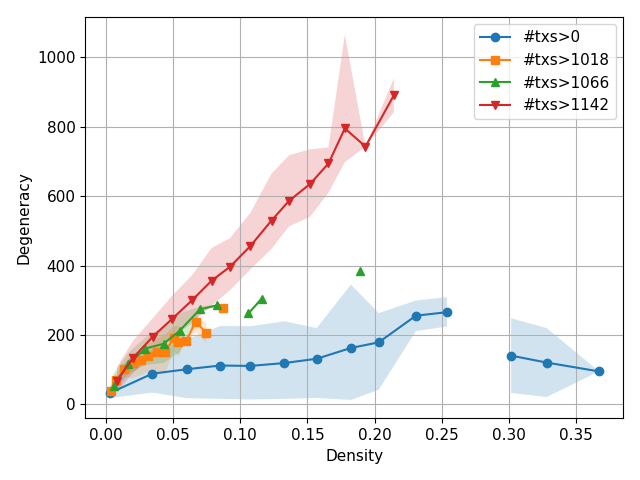}}
\caption{degeneracy}
\label{fig:degeneracy}
\end{figure}
\begin{figure}
\centerline{\includegraphics[width=\columnwidth]{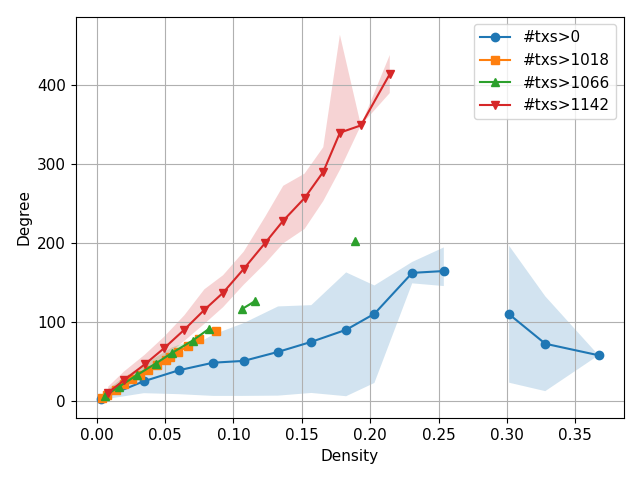}}
\caption{degree}
\label{fig:degree}
\end{figure}
\begin{figure}
\centerline{\includegraphics[width=\columnwidth]{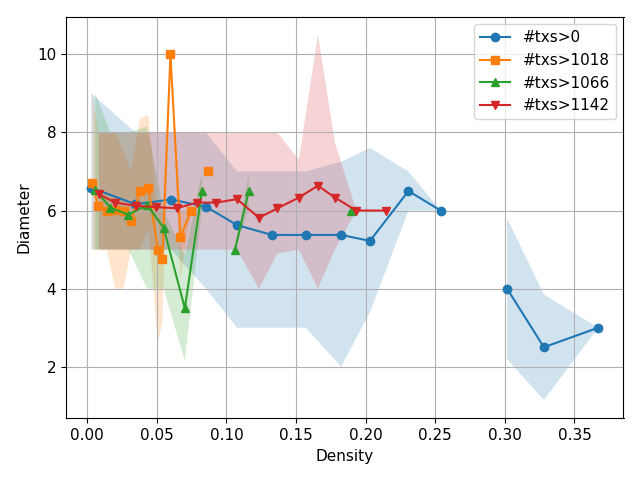}}
\caption{diameter}
\label{fig:diameter}
\end{figure}
\begin{figure}
\centerline{\includegraphics[width=\columnwidth]{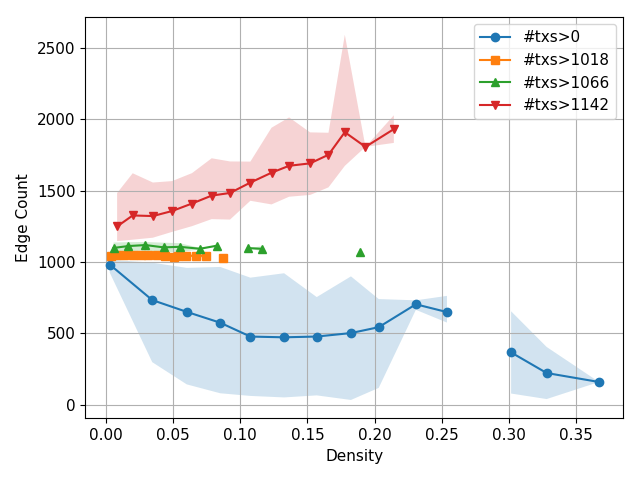}}
\caption{edge-count}
\label{fig:edge-count}
\end{figure}
\begin{figure}
\centerline{\includegraphics[width=\columnwidth]{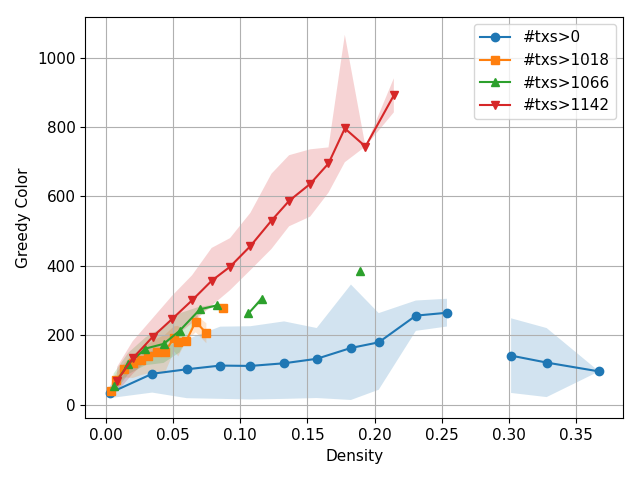}}
\caption{greedy-color}
\label{fig:greedy-color}
\end{figure}
\begin{figure}
\centerline{\includegraphics[width=\columnwidth]{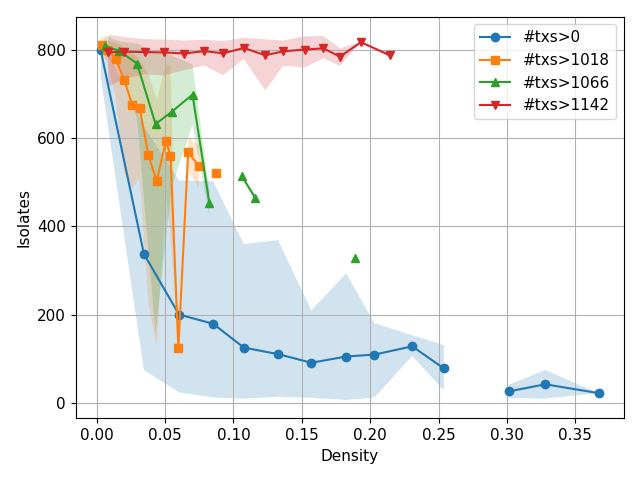}}
\caption{isolates}
\label{fig:isolates}
\end{figure}
\begin{figure}
\centerline{\includegraphics[width=\columnwidth]{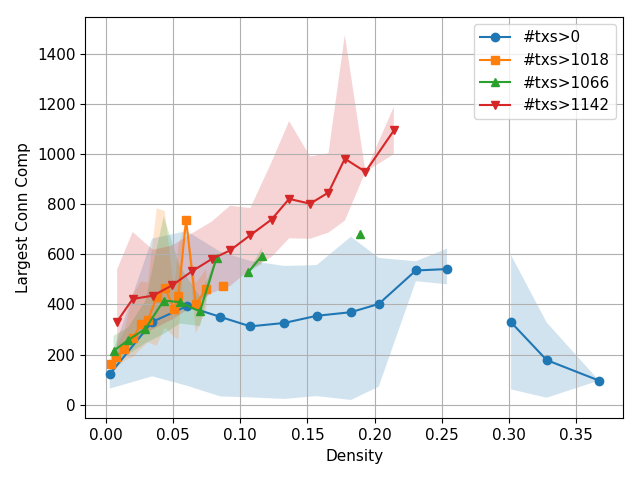}}
\caption{largest-conn-comp}
\label{fig:largest-conn-comp}
\end{figure}
\begin{figure}
\centerline{\includegraphics[width=\columnwidth]{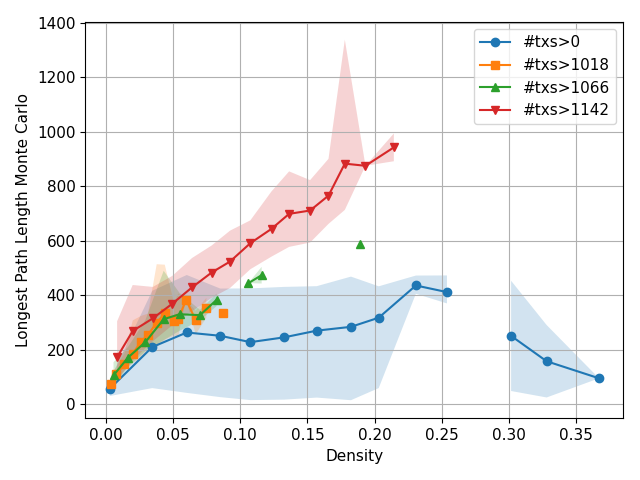}}
\caption{longest-path-length-monte-carlo}
\label{fig:longest-path-length-monte-carlo}
\end{figure}
\begin{figure}
\centerline{\includegraphics[width=\columnwidth]{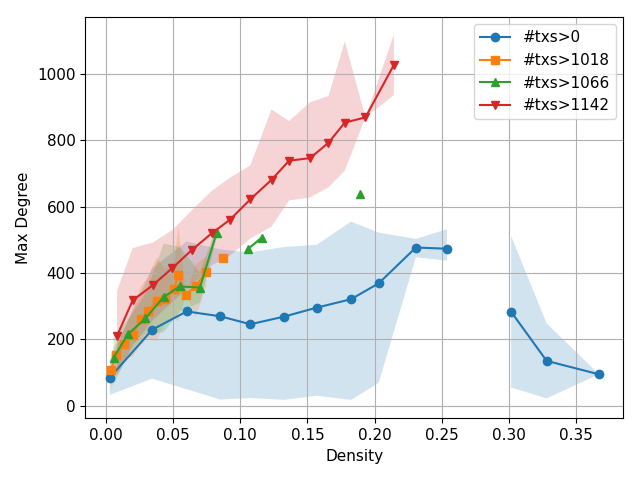}}
\caption{max-degree}
\label{fig:max-degree}
\end{figure}
\begin{figure}
\centerline{\includegraphics[width=\columnwidth]{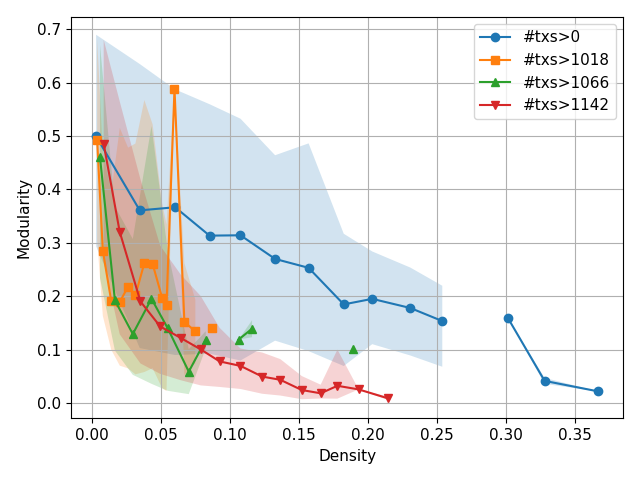}}
\caption{modularity}
\label{fig:modularity}
\end{figure}
\begin{figure}
\centerline{\includegraphics[width=\columnwidth]{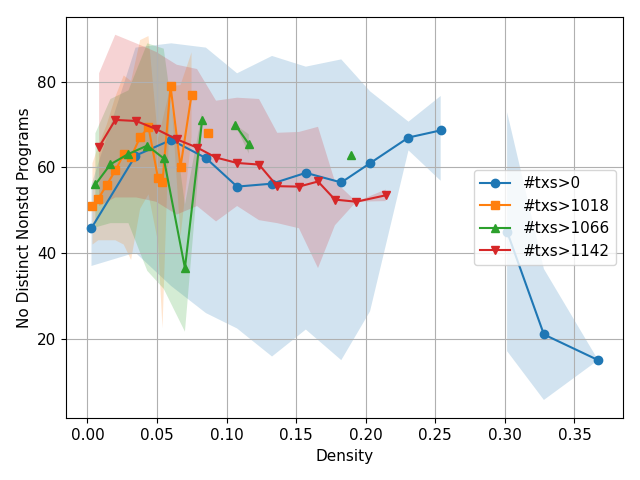}}
\caption{no-distinct-nonstd-programs}
\label{fig:no-distinct-nonstd-programs}
\end{figure}
\begin{figure}
\centerline{\includegraphics[width=\columnwidth]{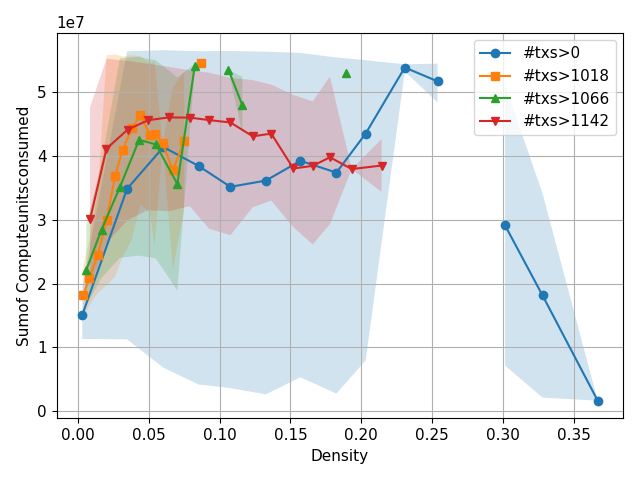}}
\caption{sumof-computeUnitsConsumed}
\label{fig:sumof-computeUnitsConsumed}
\end{figure}
\begin{figure}
\centerline{\includegraphics[width=\columnwidth]{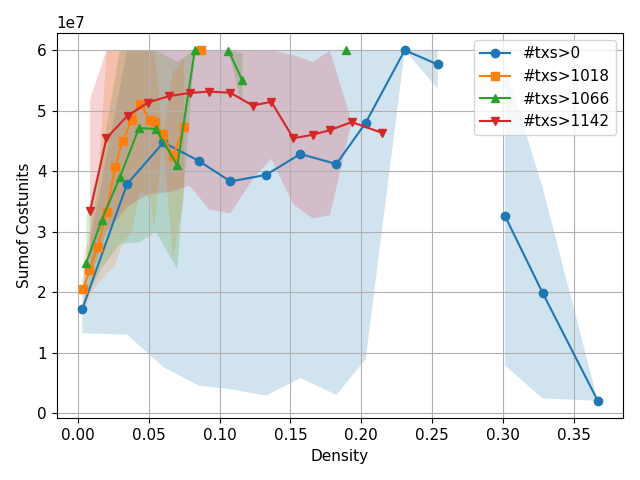}}
\caption{sumof-costUnits}
\label{fig:sumof-costUnits}
\end{figure}
\begin{figure}
\centerline{\includegraphics[width=\columnwidth]{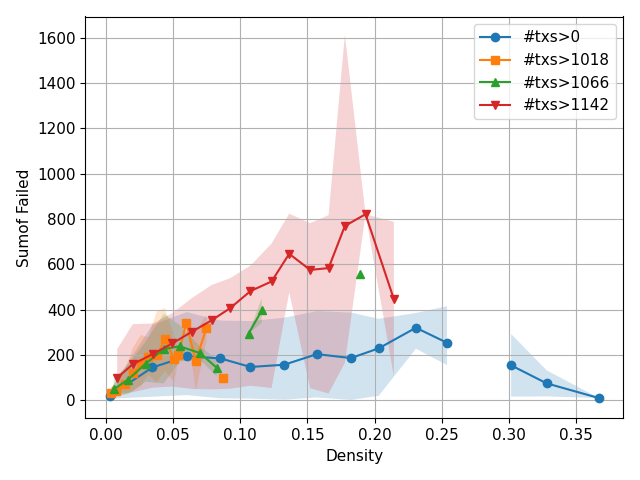}}
\caption{sumof-failed}
\label{fig:sumof-failed}
\end{figure}
\begin{figure}
\centerline{\includegraphics[width=\columnwidth]{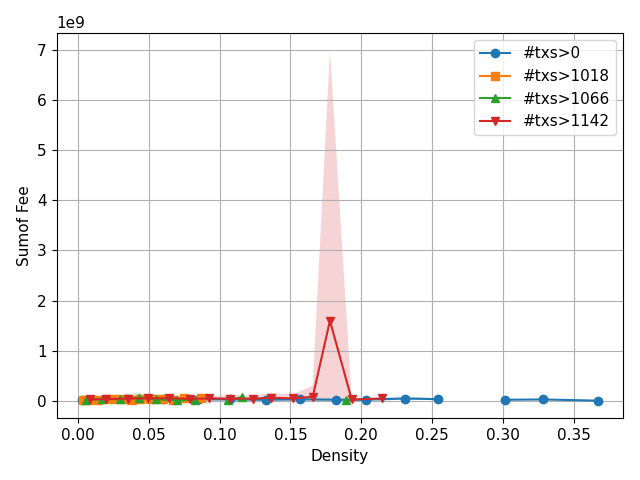}}
\caption{sumof-fee}
\label{fig:sumof-fee}
\end{figure}
\begin{figure}
\centerline{\includegraphics[width=\columnwidth]{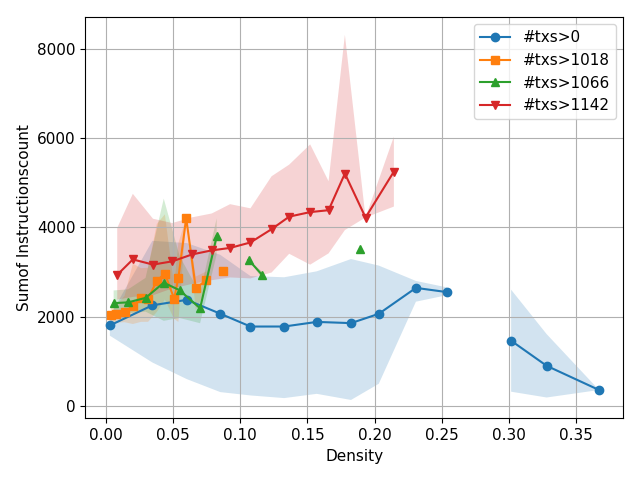}}
\caption{sumof-instructionsCount}
\label{fig:sumof-instructionsCount}
\end{figure}
\begin{figure}
\centerline{\includegraphics[width=\columnwidth]{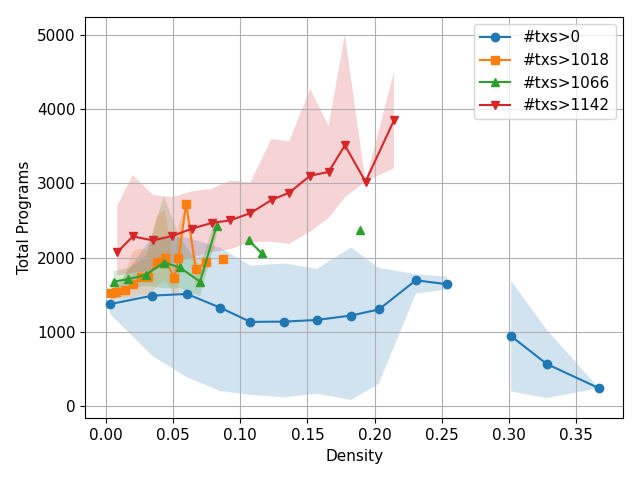}}
\caption{total-programs}
\label{fig:total-programs}
\end{figure}
\begin{figure}
\centerline{\includegraphics[width=\columnwidth]{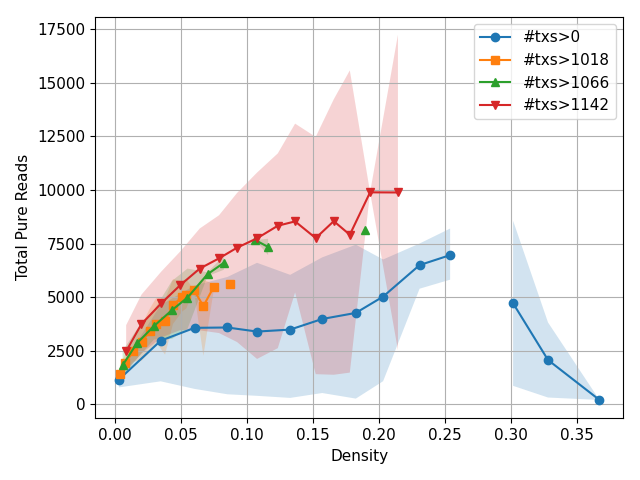}}
\caption{total-pure-reads}
\label{fig:total-pure-reads}
\end{figure}
\begin{figure}
\centerline{\includegraphics[width=\columnwidth]{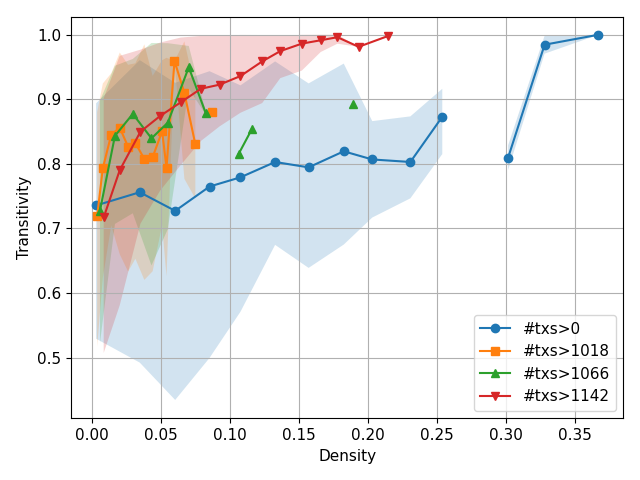}}
\caption{transitivity}
\label{fig:transitivity}
\end{figure}
\begin{figure}
\centerline{\includegraphics[width=\columnwidth]{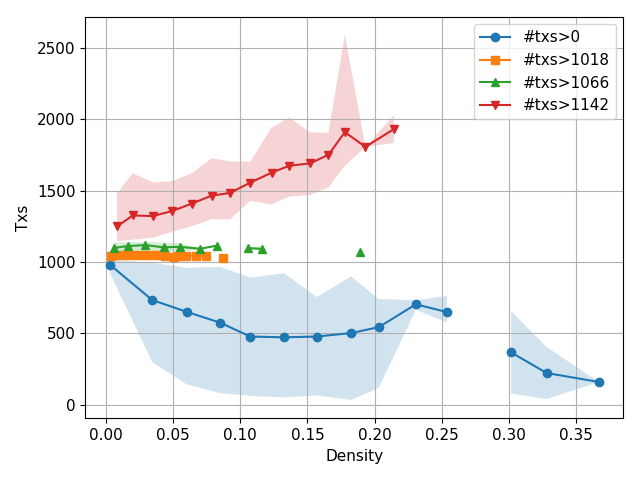}}
\caption{txs}
\label{fig:txs}
\end{figure}
\begin{figure}
\centerline{\includegraphics[width=\columnwidth]{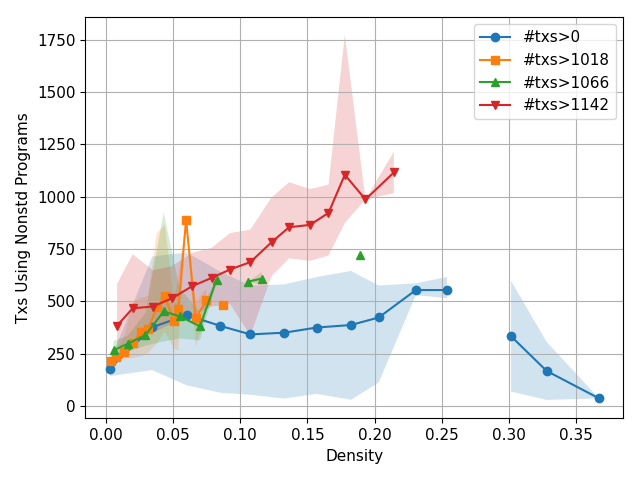}}
\caption{txs-using-nonstd-programs}
\label{fig:txs-using-nonstd-programs}
\end{figure}
\begin{figure}
\centerline{\includegraphics[width=\columnwidth]{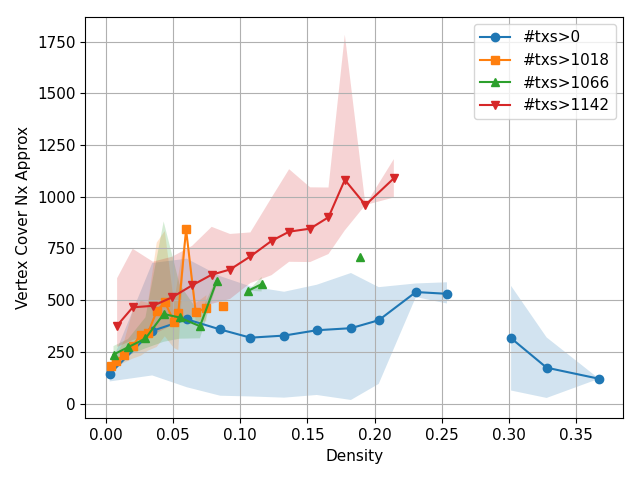}}
\caption{vertex-cover-nx-approx}
\label{fig:vertex-cover-nx-approx}
\end{figure}
\begin{figure}
\centerline{\includegraphics[width=\columnwidth]{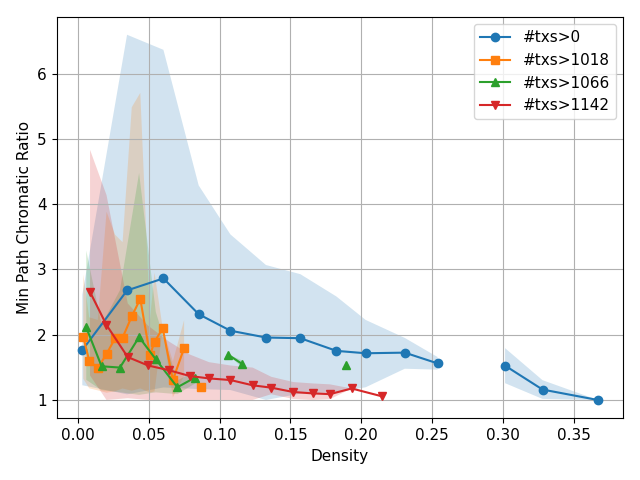}}
\caption{min-path-chromatic-ratio}
\label{fig:min-path-chromatic-ratio}
\end{figure}
\begin{figure}
\centerline{\includegraphics[width=\columnwidth]{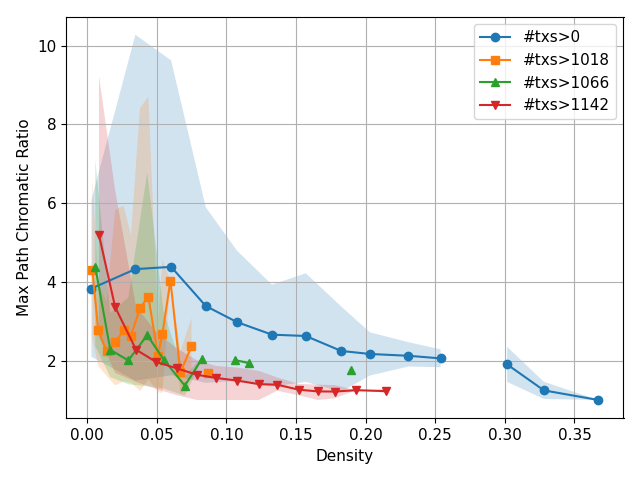}}
\caption{max-path-chromatic-ratio}
\label{fig:max-path-chromatic-ratio}
\end{figure}
\begin{figure}
\centerline{\includegraphics[width=\columnwidth]{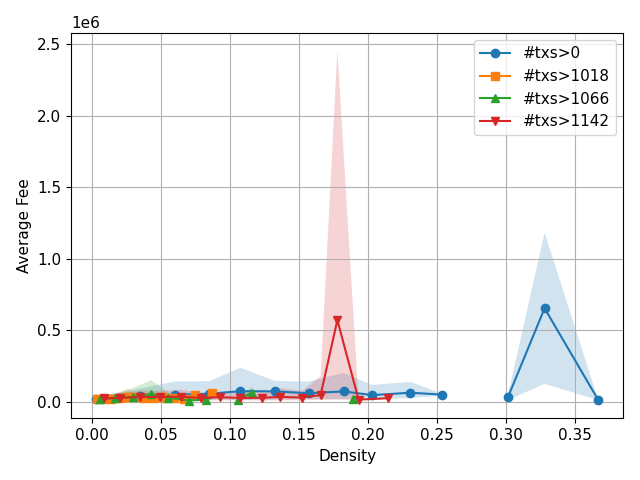}}
\caption{avg-fee}
\label{fig:avg-fee}
\end{figure}
\begin{figure}
\centerline{\includegraphics[width=\columnwidth]{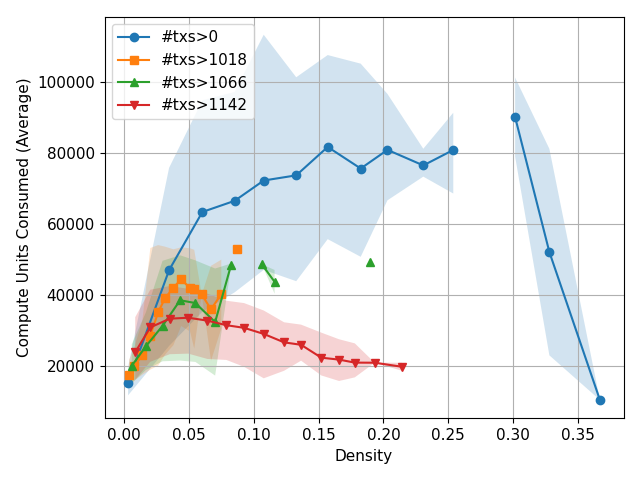}}
\caption{avg-computeUnitsConsumed}
\label{fig:avg-computeUnitsConsumed}
\end{figure}
\begin{figure}
\centerline{\includegraphics[width=\columnwidth]{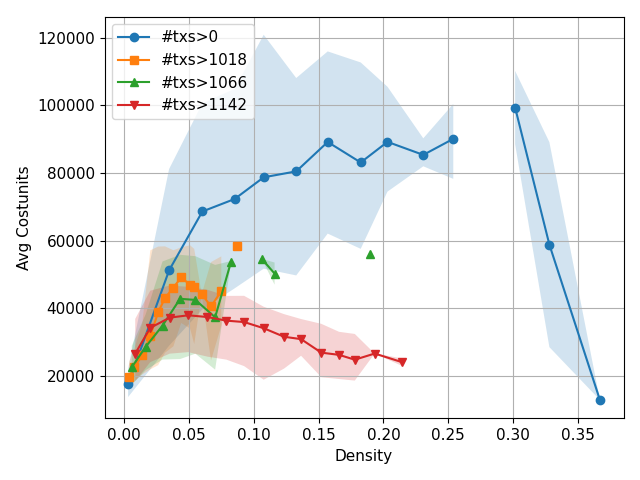}}
\caption{avg-costUnits}
\label{fig:avg-costUnits}
\end{figure}
\begin{figure}
\centerline{\includegraphics[width=\columnwidth]{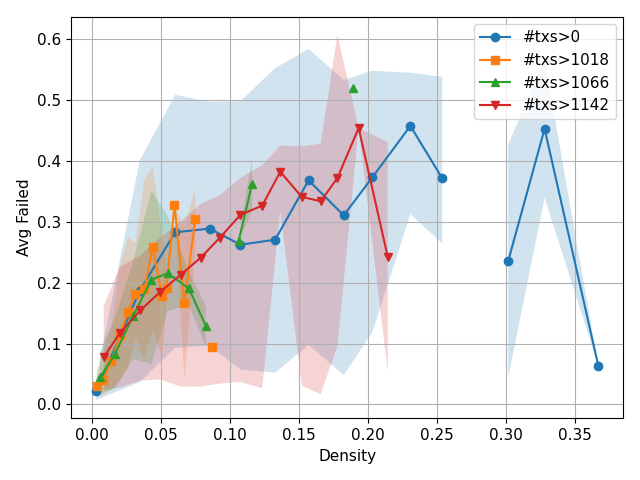}}
\caption{avg-failed}
\label{fig:avg-failed}
\end{figure}
\begin{figure}
\centerline{\includegraphics[width=\columnwidth]{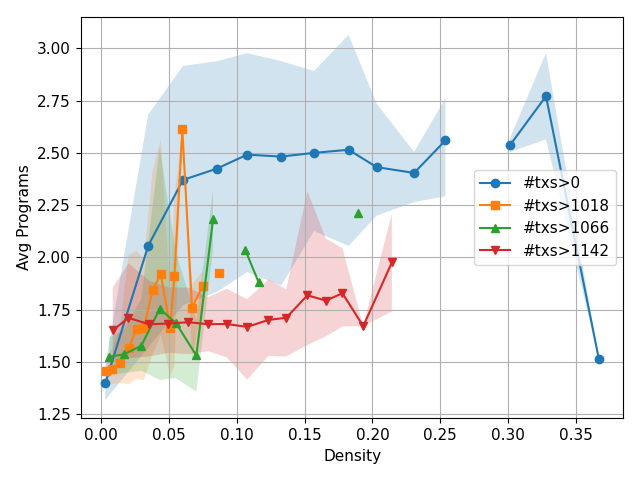}}
\caption{avg-programs}
\label{fig:avg-programs}
\end{figure}
\begin{figure}
\centerline{\includegraphics[width=\columnwidth]{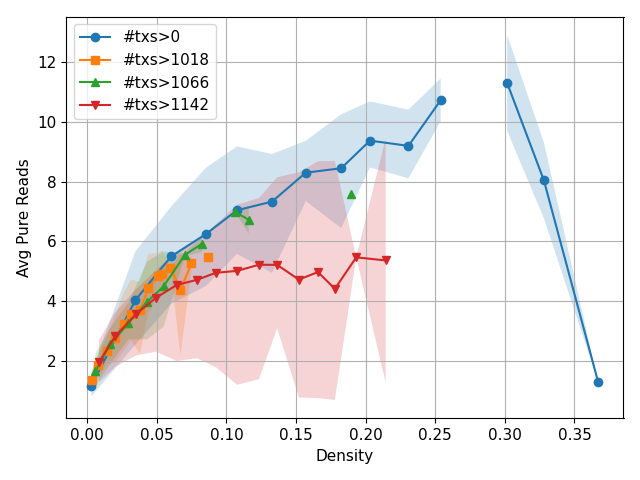}}
\caption{avg-pure-reads}
\label{fig:avg-pure-reads}
\end{figure}
[]
{}
\begin{figure}
\centerline{\includegraphics[width=\columnwidth]{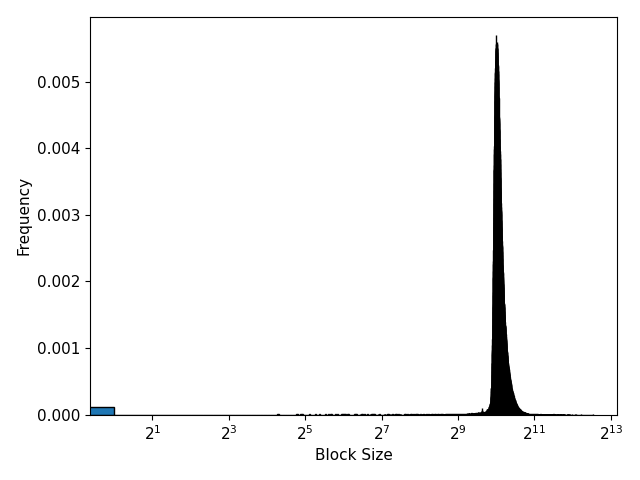}}
\caption{block-size-dist}
\label{fig:block-size-dist}
\end{figure}
\begin{figure}
\centerline{\includegraphics[width=\columnwidth]{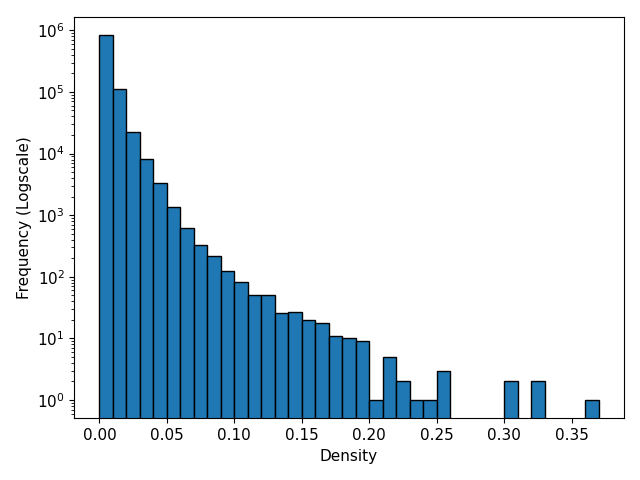}}
\caption{density-dist}
\label{fig:density-dist}
\end{figure}
\begin{figure}
\centerline{\includegraphics[width=\columnwidth]{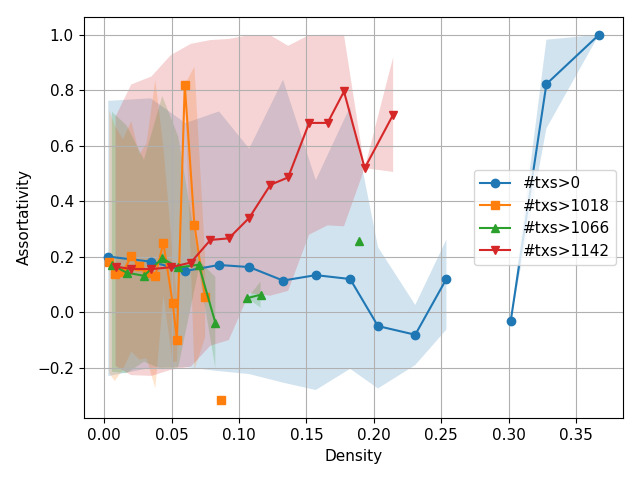}}
\caption{assortativity}
\label{fig:assortativity}
\end{figure}
\begin{figure}
\centerline{\includegraphics[width=\columnwidth]{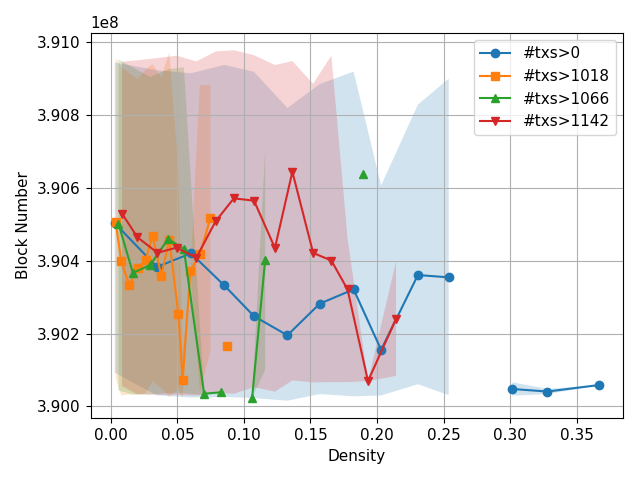}}
\caption{block-number}
\label{fig:block-number}
\end{figure}
\begin{figure}
\centerline{\includegraphics[width=\columnwidth]{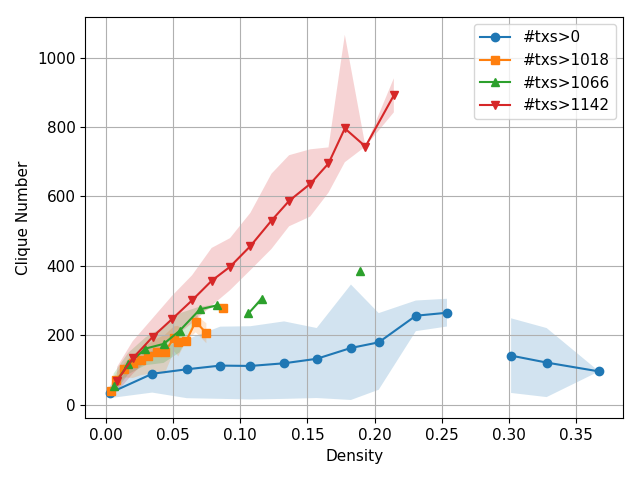}}
\caption{clique-number}
\label{fig:clique-number}
\end{figure}
\begin{figure}
\centerline{\includegraphics[width=\columnwidth]{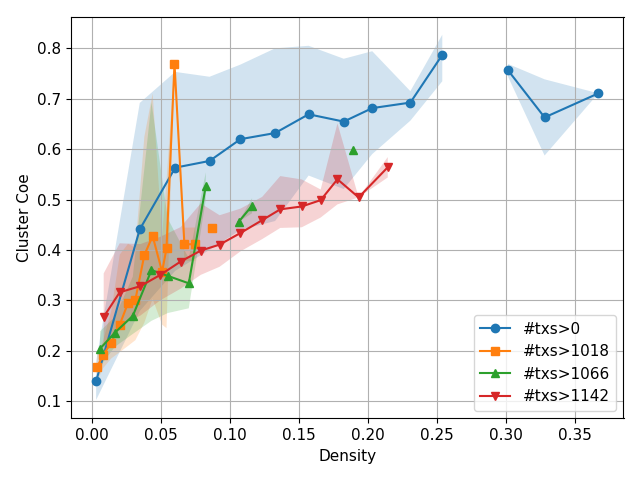}}
\caption{cluster-coe}
\label{fig:cluster-coe}
\end{figure}
\begin{figure}
\centerline{\includegraphics[width=\columnwidth]{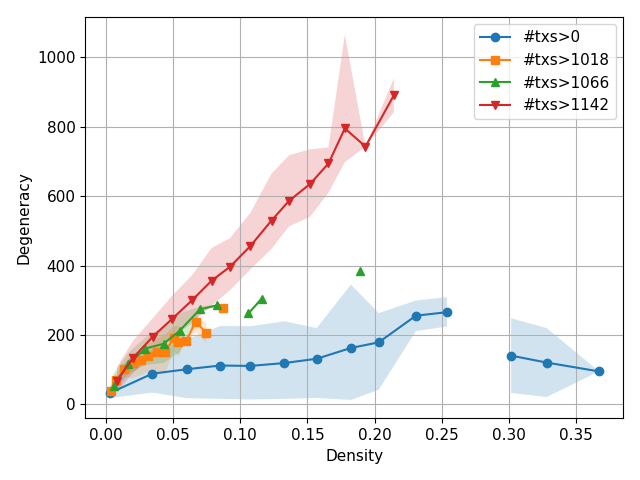}}
\caption{degeneracy}
\label{fig:degeneracy}
\end{figure}
\begin{figure}
\centerline{\includegraphics[width=\columnwidth]{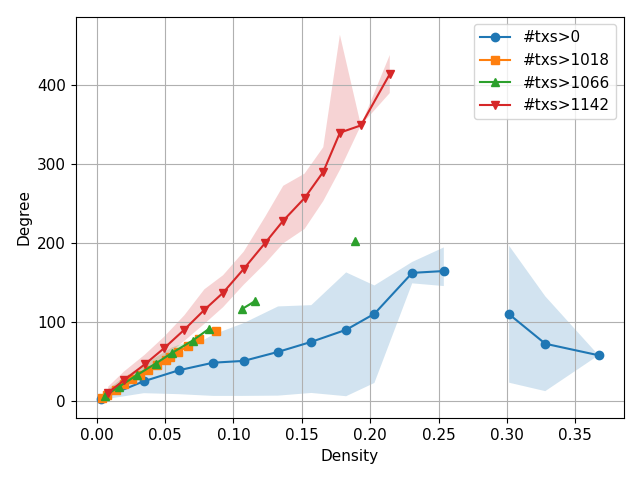}}
\caption{degree}
\label{fig:degree}
\end{figure}
\begin{figure}
\centerline{\includegraphics[width=\columnwidth]{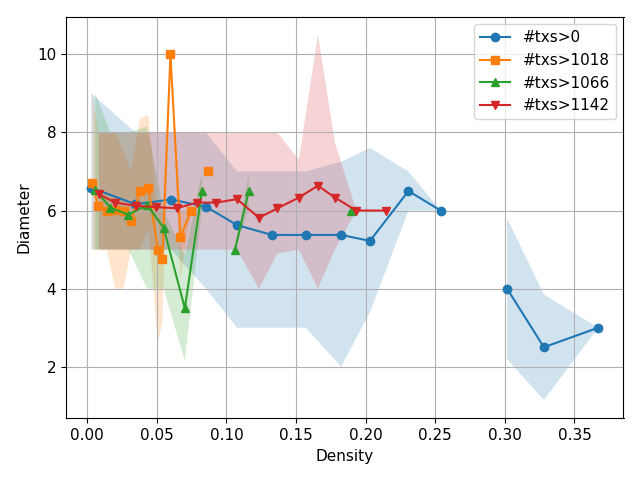}}
\caption{diameter}
\label{fig:diameter}
\end{figure}
\begin{figure}
\centerline{\includegraphics[width=\columnwidth]{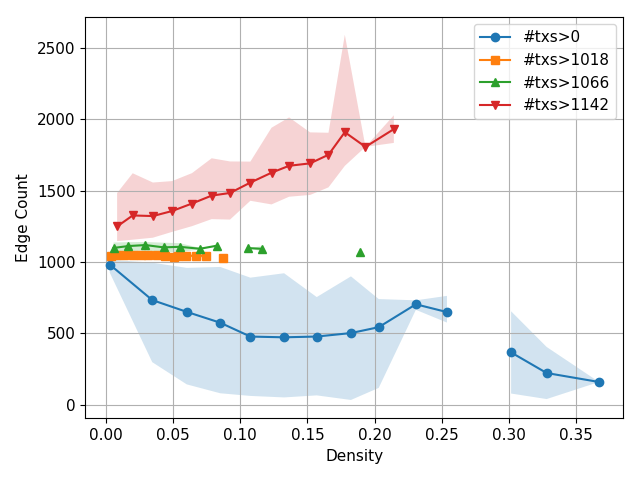}}
\caption{edge-count}
\label{fig:edge-count}
\end{figure}
\begin{figure}
\centerline{\includegraphics[width=\columnwidth]{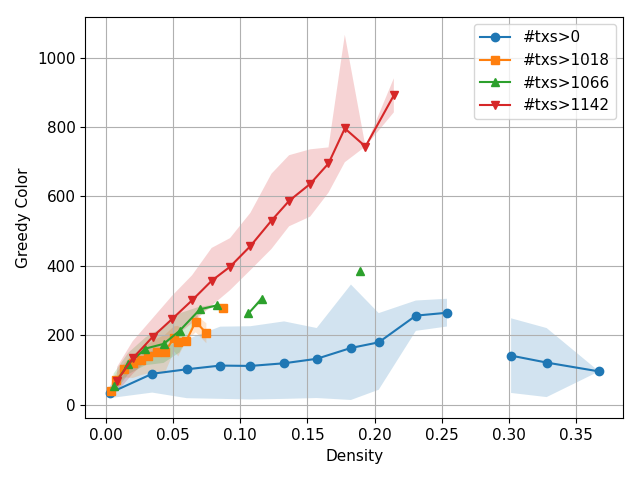}}
\caption{greedy-color}
\label{fig:greedy-color}
\end{figure}
\begin{figure}
\centerline{\includegraphics[width=\columnwidth]{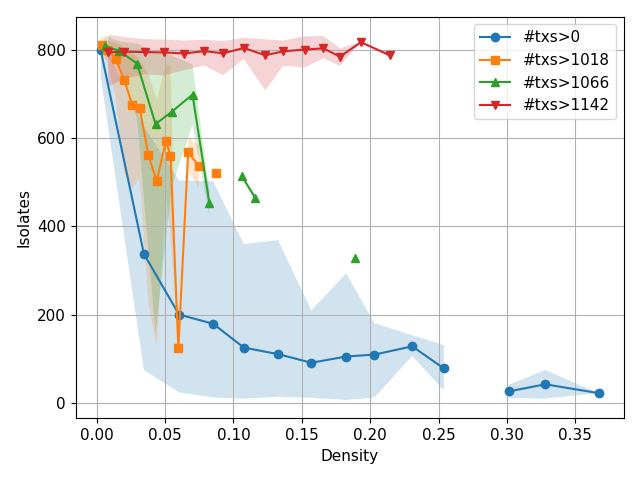}}
\caption{isolates}
\label{fig:isolates}
\end{figure}
\begin{figure}
\centerline{\includegraphics[width=\columnwidth]{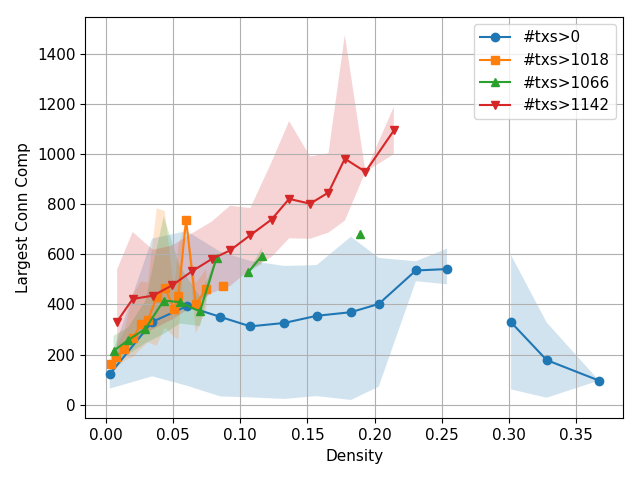}}
\caption{largest-conn-comp}
\label{fig:largest-conn-comp}
\end{figure}
\begin{figure}
\centerline{\includegraphics[width=\columnwidth]{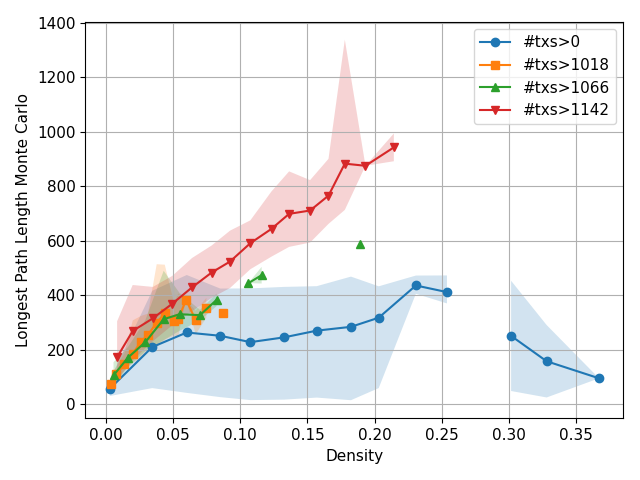}}
\caption{longest-path-length-monte-carlo}
\label{fig:longest-path-length-monte-carlo}
\end{figure}
\begin{figure}
\centerline{\includegraphics[width=\columnwidth]{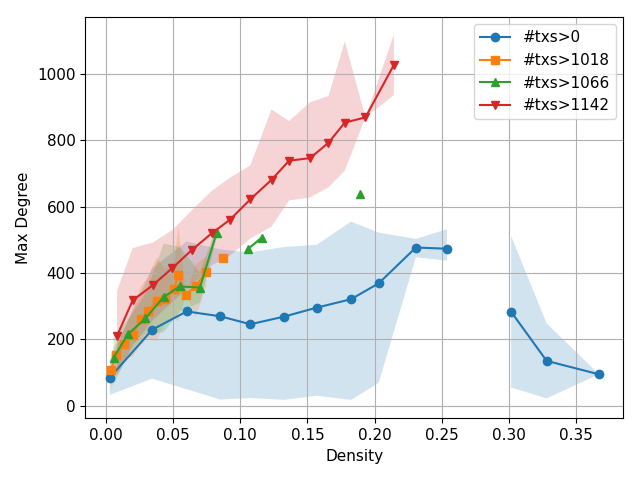}}
\caption{max-degree}
\label{fig:max-degree}
\end{figure}
\begin{figure}
\centerline{\includegraphics[width=\columnwidth]{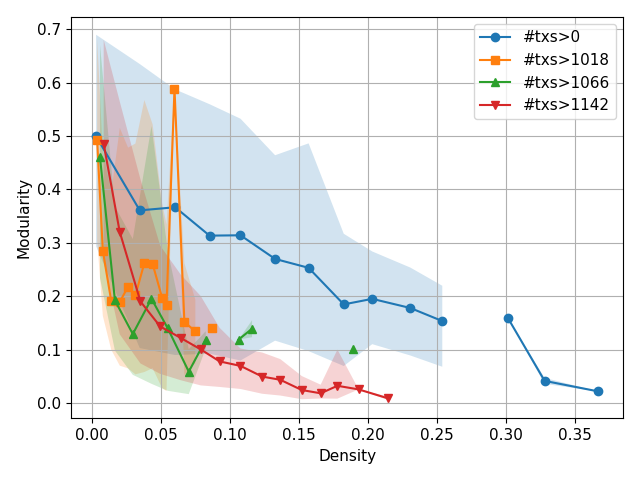}}
\caption{modularity}
\label{fig:modularity}
\end{figure}
\begin{figure}
\centerline{\includegraphics[width=\columnwidth]{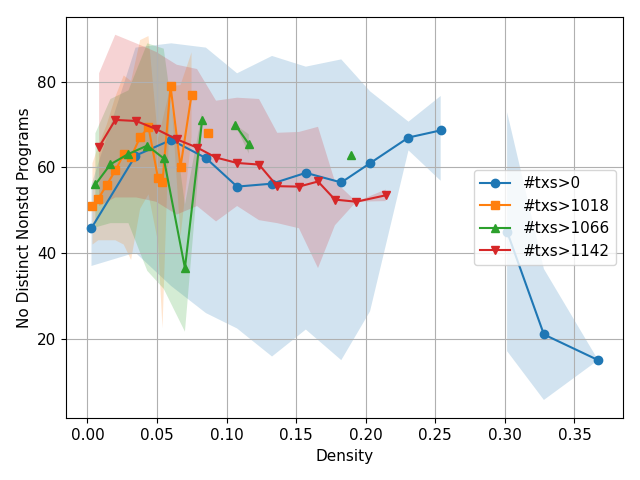}}
\caption{no-distinct-nonstd-programs}
\label{fig:no-distinct-nonstd-programs}
\end{figure}
\begin{figure}
\centerline{\includegraphics[width=\columnwidth]{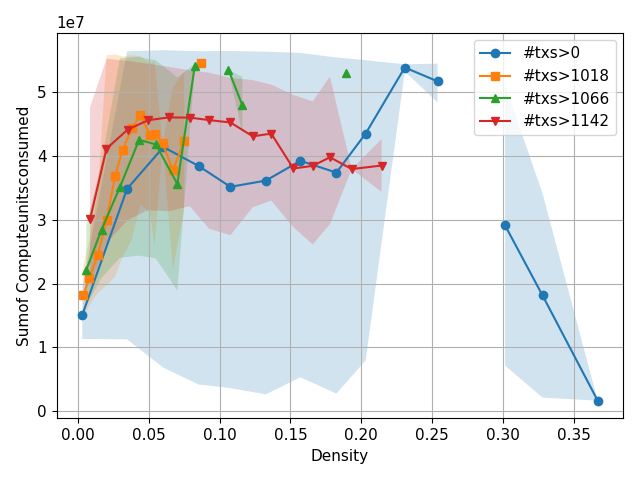}}
\caption{sumof-computeUnitsConsumed}
\label{fig:sumof-computeUnitsConsumed}
\end{figure}
\begin{figure}
\centerline{\includegraphics[width=\columnwidth]{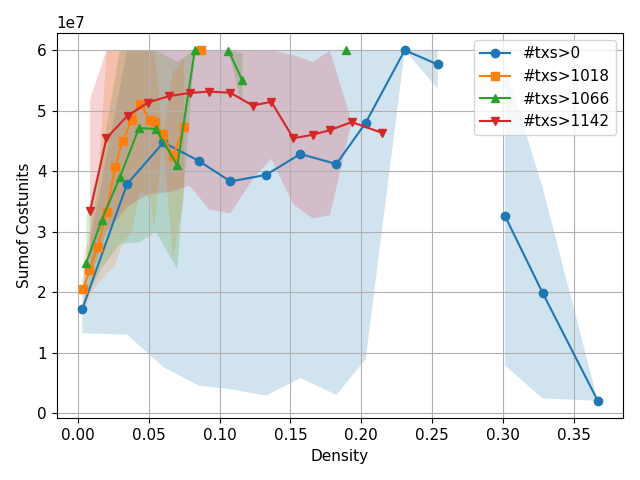}}
\caption{sumof-costUnits}
\label{fig:sumof-costUnits}
\end{figure}
\begin{figure}
\centerline{\includegraphics[width=\columnwidth]{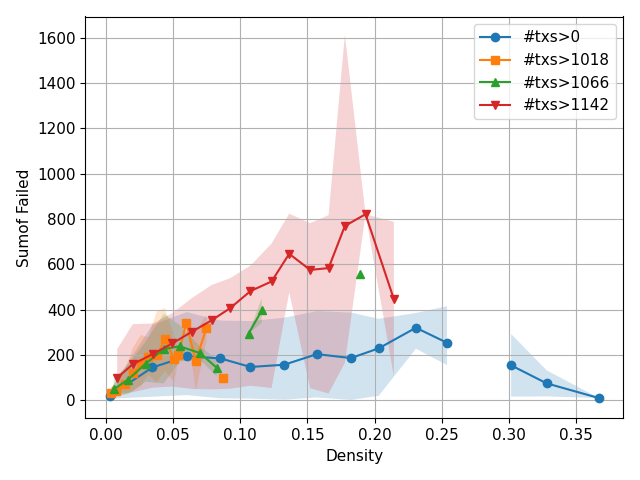}}
\caption{sumof-failed}
\label{fig:sumof-failed}
\end{figure}
\begin{figure}
\centerline{\includegraphics[width=\columnwidth]{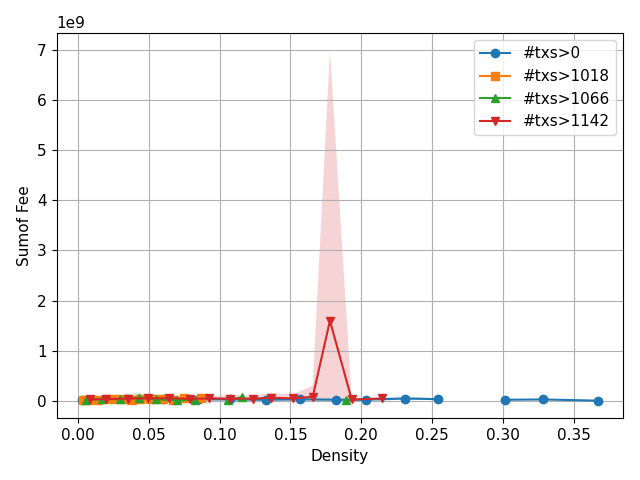}}
\caption{sumof-fee}
\label{fig:sumof-fee}
\end{figure}
\begin{figure}
\centerline{\includegraphics[width=\columnwidth]{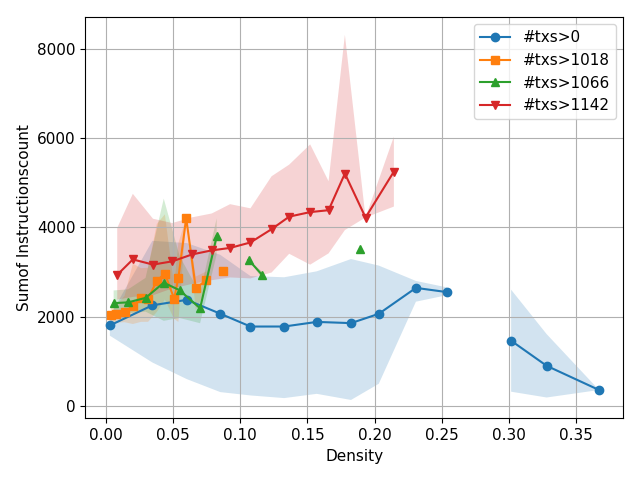}}
\caption{sumof-instructionsCount}
\label{fig:sumof-instructionsCount}
\end{figure}
\begin{figure}
\centerline{\includegraphics[width=\columnwidth]{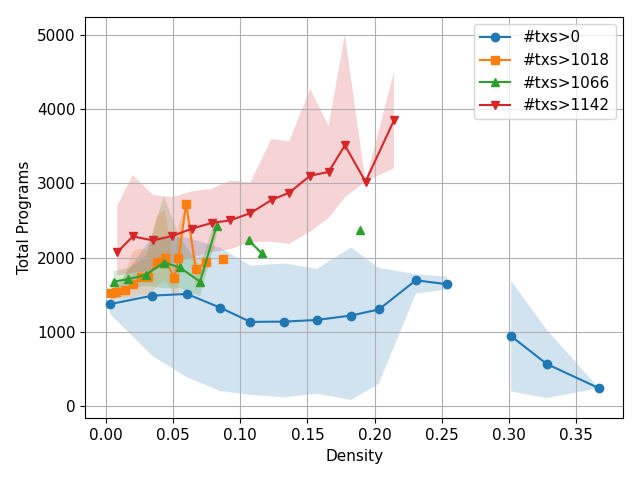}}
\caption{total-programs}
\label{fig:total-programs}
\end{figure}
\begin{figure}
\centerline{\includegraphics[width=\columnwidth]{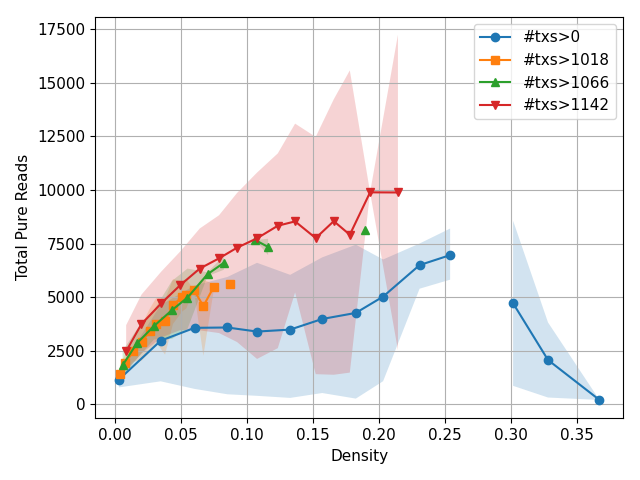}}
\caption{total-pure-reads}
\label{fig:total-pure-reads}
\end{figure}
\begin{figure}
\centerline{\includegraphics[width=\columnwidth]{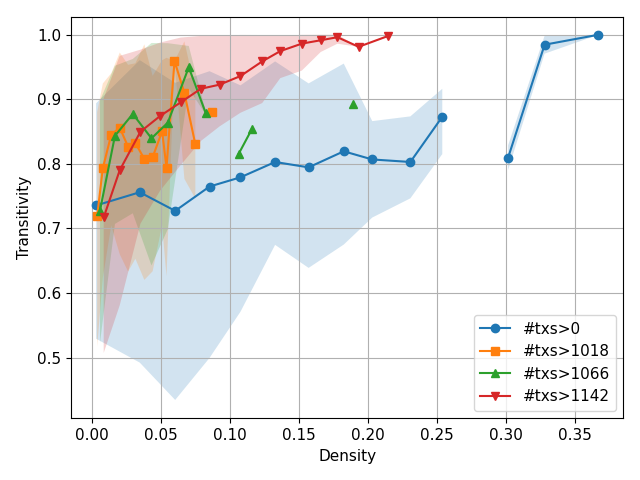}}
\caption{transitivity}
\label{fig:transitivity}
\end{figure}
\begin{figure}
\centerline{\includegraphics[width=\columnwidth]{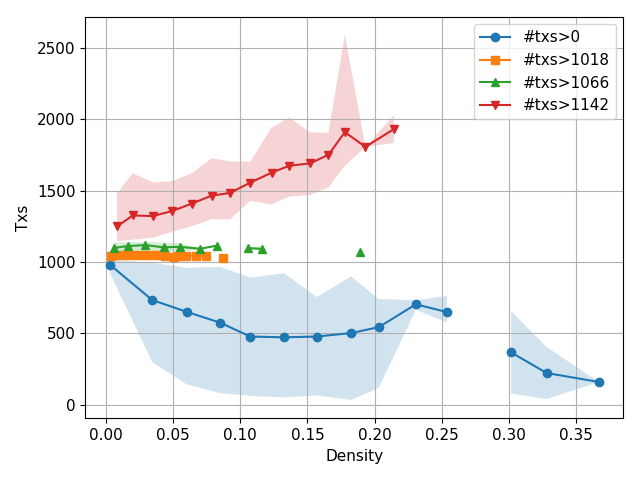}}
\caption{txs}
\label{fig:txs}
\end{figure}
\begin{figure}
\centerline{\includegraphics[width=\columnwidth]{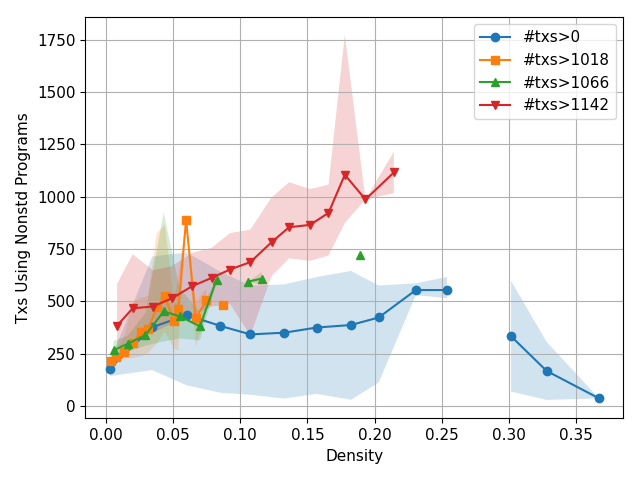}}
\caption{txs-using-nonstd-programs}
\label{fig:txs-using-nonstd-programs}
\end{figure}
\begin{figure}
\centerline{\includegraphics[width=\columnwidth]{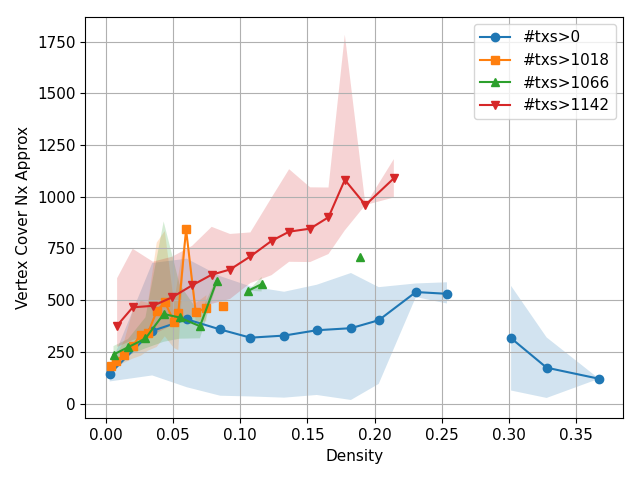}}
\caption{vertex-cover-nx-approx}
\label{fig:vertex-cover-nx-approx}
\end{figure}
\begin{figure}
\centerline{\includegraphics[width=\columnwidth]{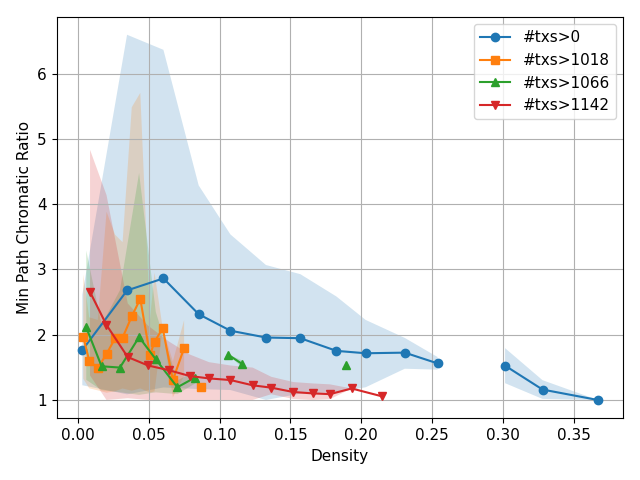}}
\caption{min-path-chromatic-ratio}
\label{fig:min-path-chromatic-ratio}
\end{figure}
\begin{figure}
\centerline{\includegraphics[width=\columnwidth]{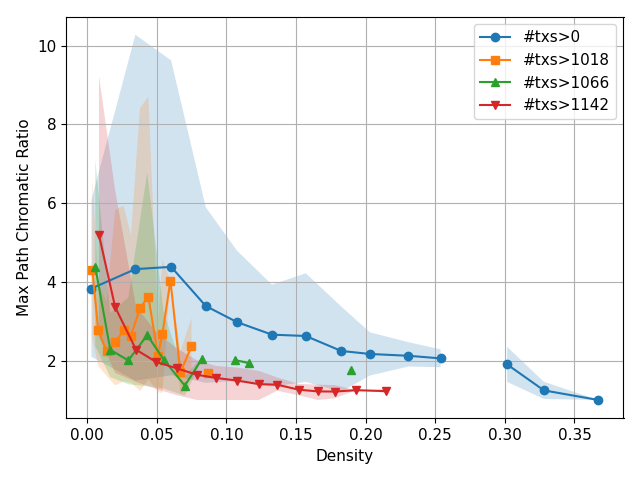}}
\caption{max-path-chromatic-ratio}
\label{fig:max-path-chromatic-ratio}
\end{figure}
\begin{figure}
\centerline{\includegraphics[width=\columnwidth]{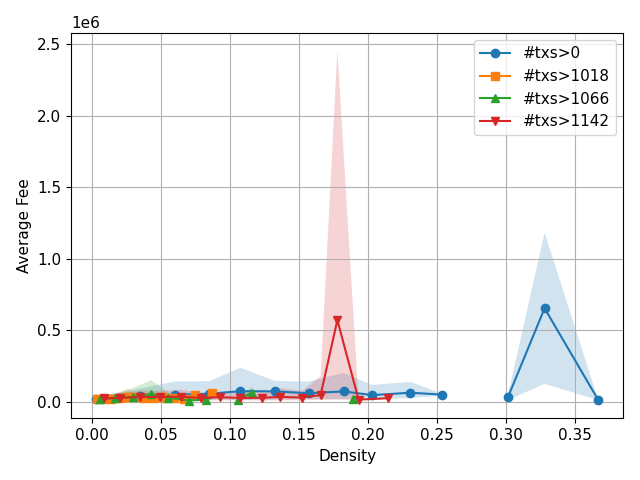}}
\caption{avg-fee}
\label{fig:avg-fee}
\end{figure}
\begin{figure}
\centerline{\includegraphics[width=\columnwidth]{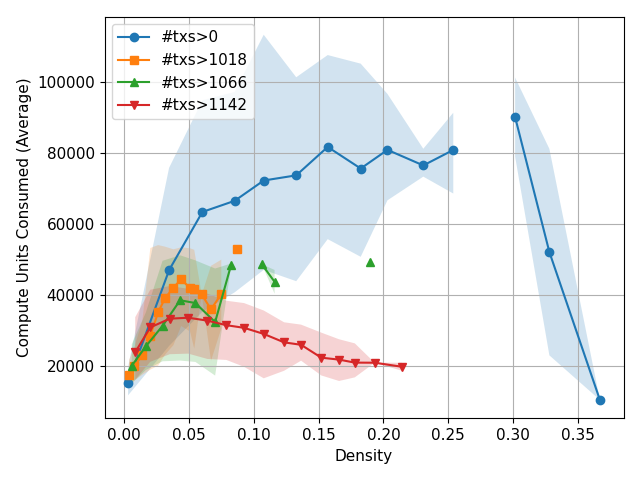}}
\caption{avg-computeUnitsConsumed}
\label{fig:avg-computeUnitsConsumed}
\end{figure}
\begin{figure}
\centerline{\includegraphics[width=\columnwidth]{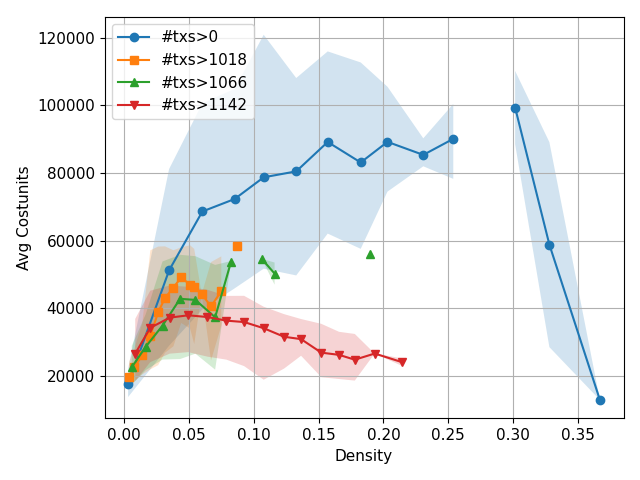}}
\caption{avg-costUnits}
\label{fig:avg-costUnits}
\end{figure}
\begin{figure}
\centerline{\includegraphics[width=\columnwidth]{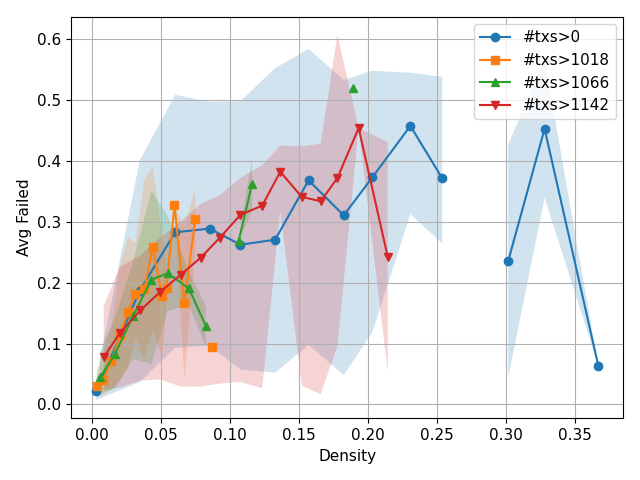}}
\caption{avg-failed}
\label{fig:avg-failed}
\end{figure}
\begin{figure}
\centerline{\includegraphics[width=\columnwidth]{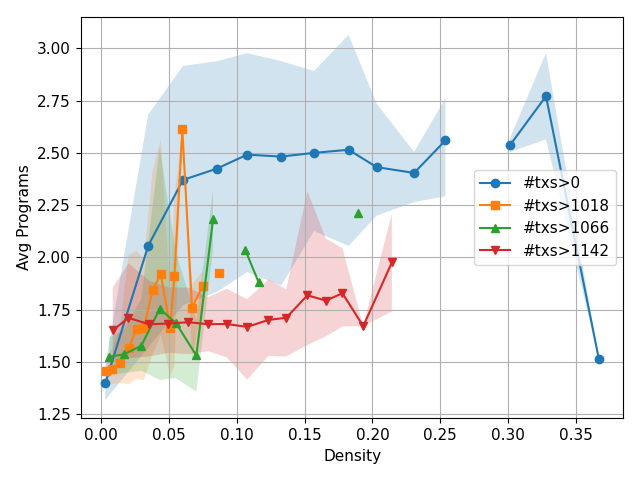}}
\caption{avg-programs}
\label{fig:avg-programs}
\end{figure}
\begin{figure}
\centerline{\includegraphics[width=\columnwidth]{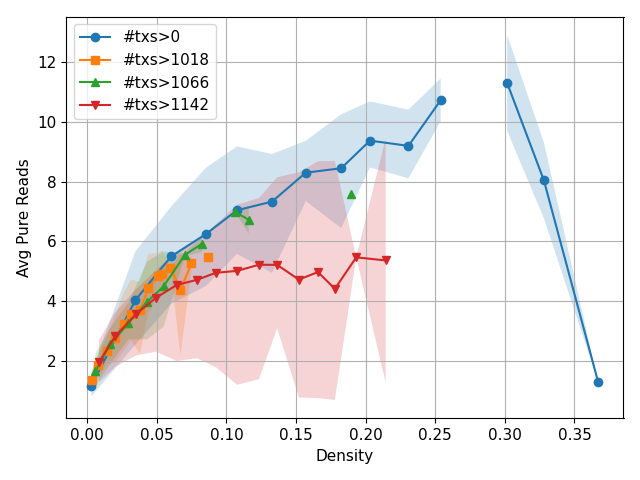}}
\caption{avg-pure-reads}
\label{fig:avg-pure-reads}
\end{figure}
\begin{figure}
\centerline{\includegraphics[width=\columnwidth]{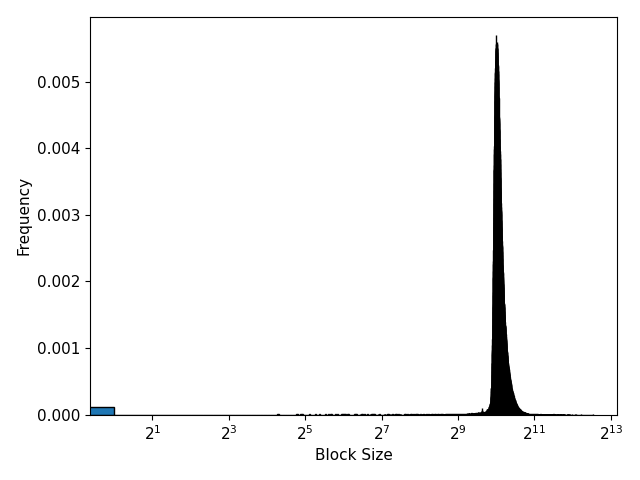}}
\caption{block-size-dist}
\label{fig:block-size-dist}
\end{figure}
\begin{figure}
\centerline{\includegraphics[width=\columnwidth]{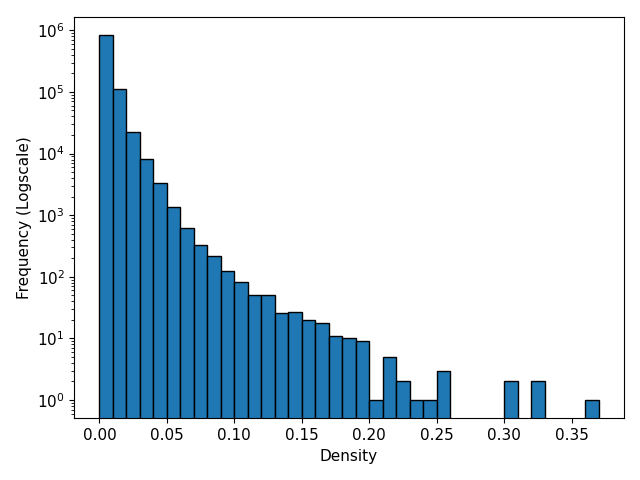}}
\caption{density-dist}
\label{fig:density-dist}
\end{figure}
\begin{figure}
\centerline{\includegraphics[width=\columnwidth]{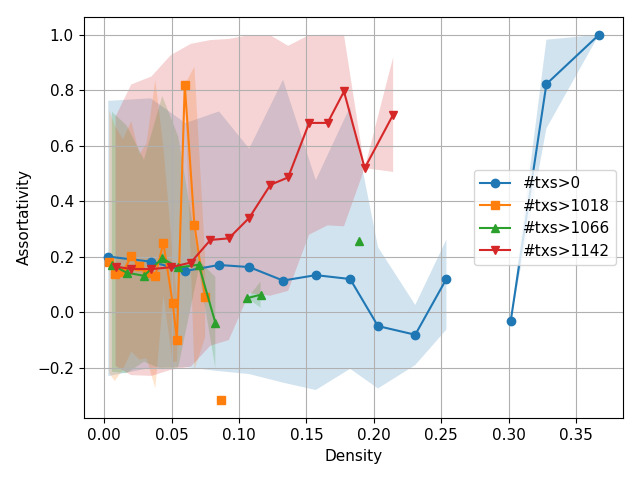}}
\caption{assortativity}
\label{fig:assortativity}
\end{figure}
\begin{figure}
\centerline{\includegraphics[width=\columnwidth]{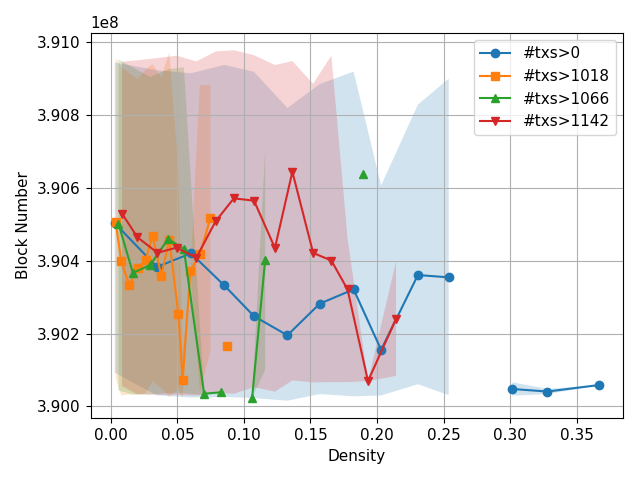}}
\caption{block-number}
\label{fig:block-number}
\end{figure}
\begin{figure}
\centerline{\includegraphics[width=\columnwidth]{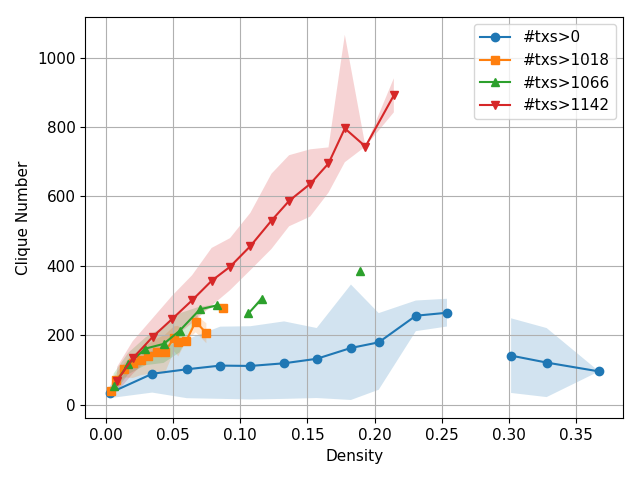}}
\caption{clique-number}
\label{fig:clique-number}
\end{figure}
\begin{figure}
\centerline{\includegraphics[width=\columnwidth]{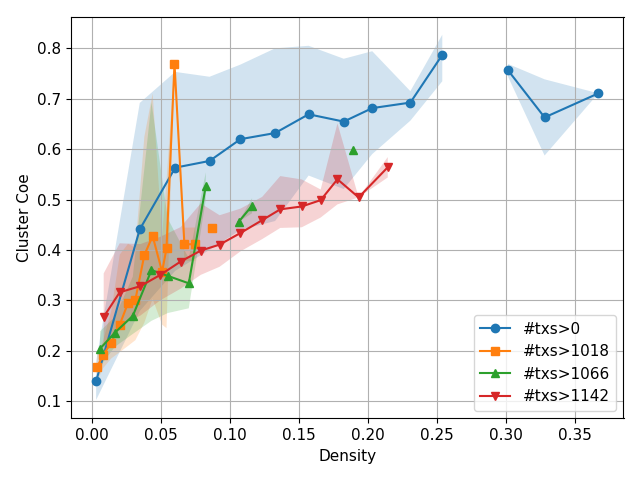}}
\caption{cluster-coe}
\label{fig:cluster-coe}
\end{figure}
\begin{figure}
\centerline{\includegraphics[width=\columnwidth]{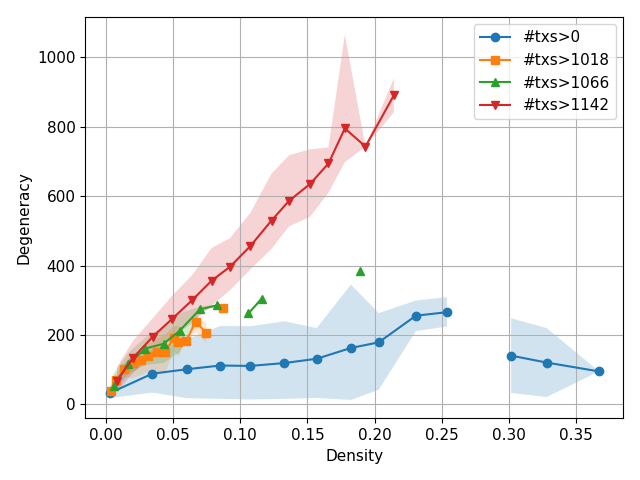}}
\caption{degeneracy}
\label{fig:degeneracy}
\end{figure}
\begin{figure}
\centerline{\includegraphics[width=\columnwidth]{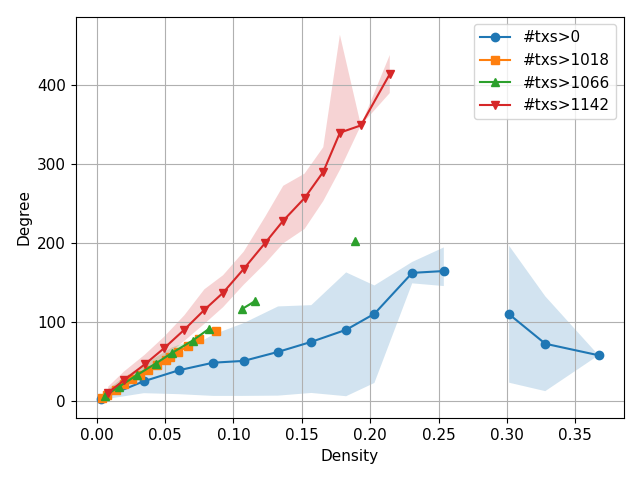}}
\caption{degree}
\label{fig:degree}
\end{figure}
\begin{figure}
\centerline{\includegraphics[width=\columnwidth]{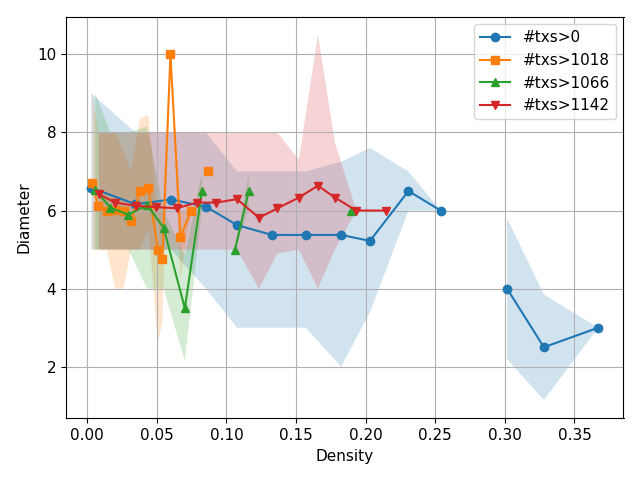}}
\caption{diameter}
\label{fig:diameter}
\end{figure}
\begin{figure}
\centerline{\includegraphics[width=\columnwidth]{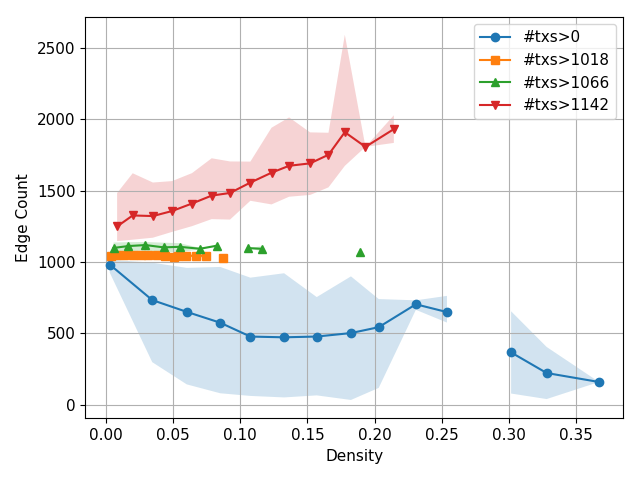}}
\caption{edge-count}
\label{fig:edge-count}
\end{figure}
\begin{figure}
\centerline{\includegraphics[width=\columnwidth]{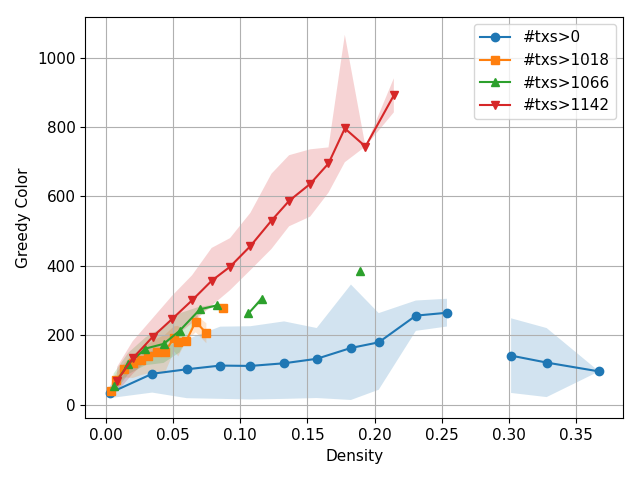}}
\caption{greedy-color}
\label{fig:greedy-color}
\end{figure}
\begin{figure}
\centerline{\includegraphics[width=\columnwidth]{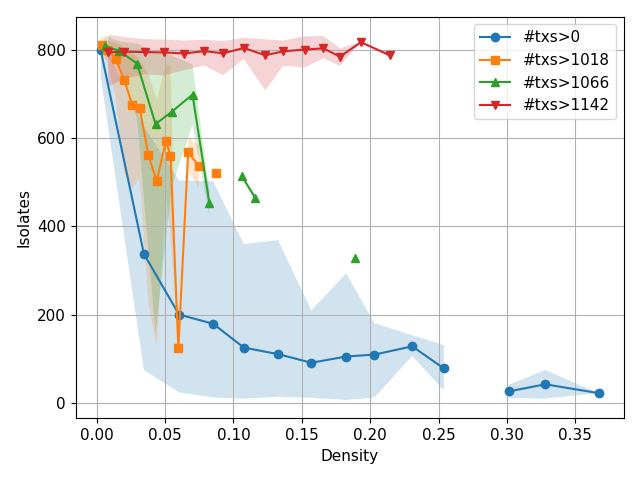}}
\caption{isolates}
\label{fig:isolates}
\end{figure}
\begin{figure}
\centerline{\includegraphics[width=\columnwidth]{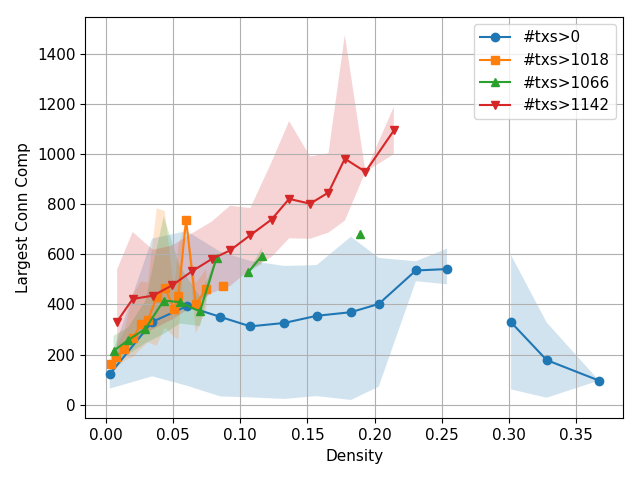}}
\caption{largest-conn-comp}
\label{fig:largest-conn-comp}
\end{figure}
\begin{figure}
\centerline{\includegraphics[width=\columnwidth]{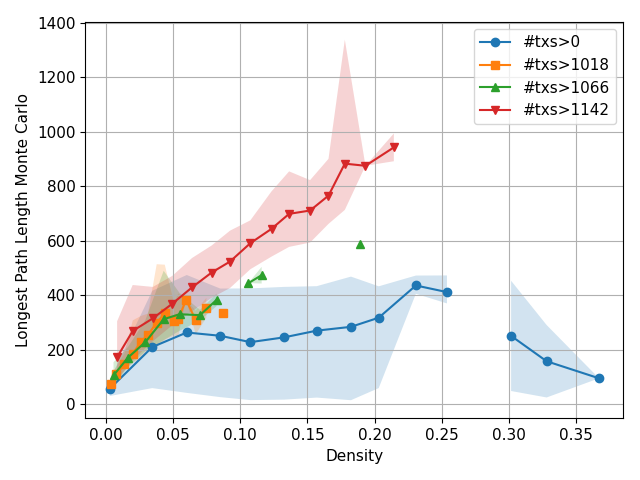}}
\caption{longest-path-length-monte-carlo}
\label{fig:longest-path-length-monte-carlo}
\end{figure}
\begin{figure}
\centerline{\includegraphics[width=\columnwidth]{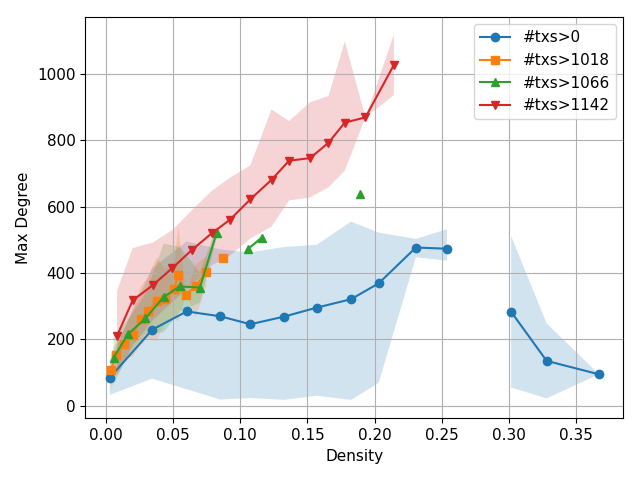}}
\caption{max-degree}
\label{fig:max-degree}
\end{figure}
\begin{figure}
\centerline{\includegraphics[width=\columnwidth]{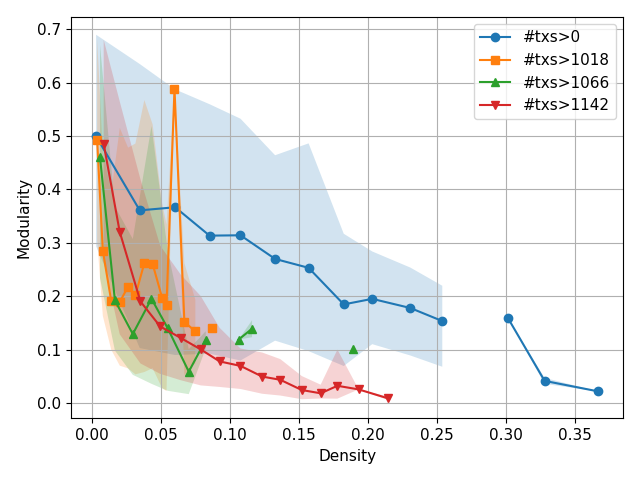}}
\caption{modularity}
\label{fig:modularity}
\end{figure}
\begin{figure}
\centerline{\includegraphics[width=\columnwidth]{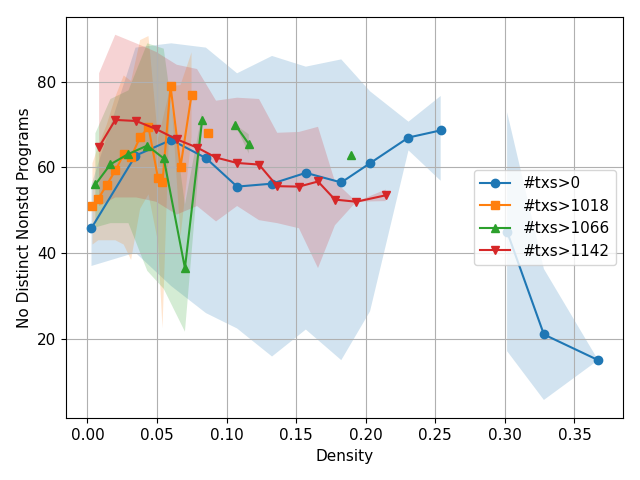}}
\caption{no-distinct-nonstd-programs}
\label{fig:no-distinct-nonstd-programs}
\end{figure}
\begin{figure}
\centerline{\includegraphics[width=\columnwidth]{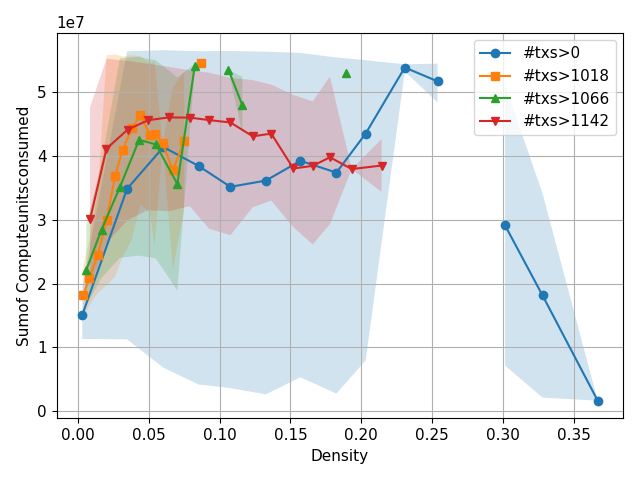}}
\caption{sumof-computeUnitsConsumed}
\label{fig:sumof-computeUnitsConsumed}
\end{figure}
\begin{figure}
\centerline{\includegraphics[width=\columnwidth]{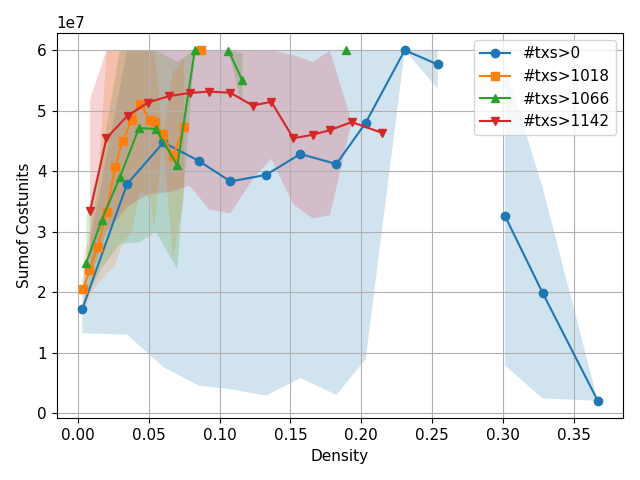}}
\caption{sumof-costUnits}
\label{fig:sumof-costUnits}
\end{figure}
\begin{figure}
\centerline{\includegraphics[width=\columnwidth]{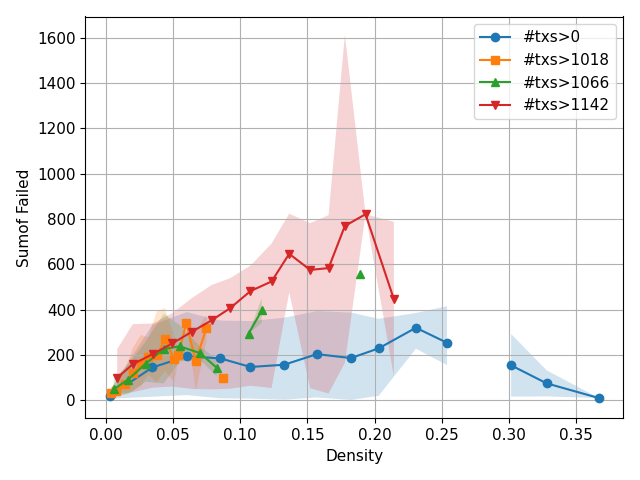}}
\caption{sumof-failed}
\label{fig:sumof-failed}
\end{figure}
\begin{figure}
\centerline{\includegraphics[width=\columnwidth]{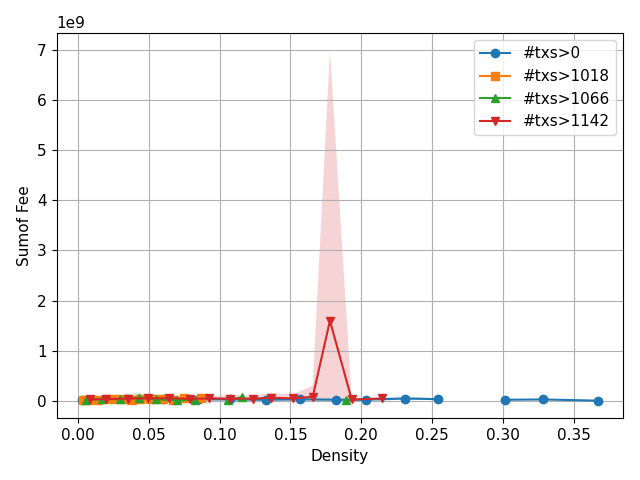}}
\caption{sumof-fee}
\label{fig:sumof-fee}
\end{figure}
\begin{figure}
\centerline{\includegraphics[width=\columnwidth]{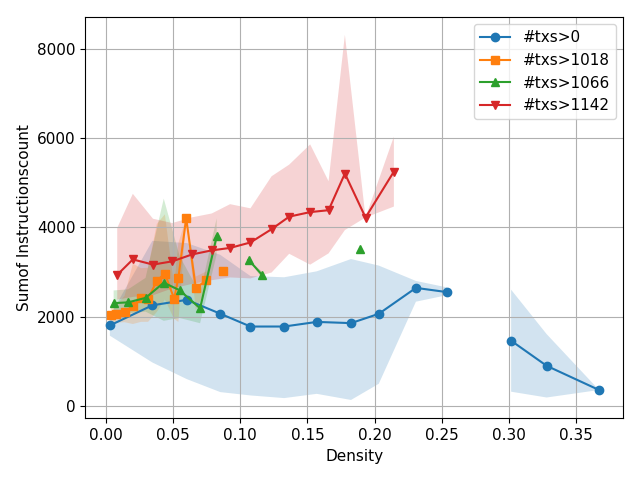}}
\caption{sumof-instructionsCount}
\label{fig:sumof-instructionsCount}
\end{figure}
\begin{figure}
\centerline{\includegraphics[width=\columnwidth]{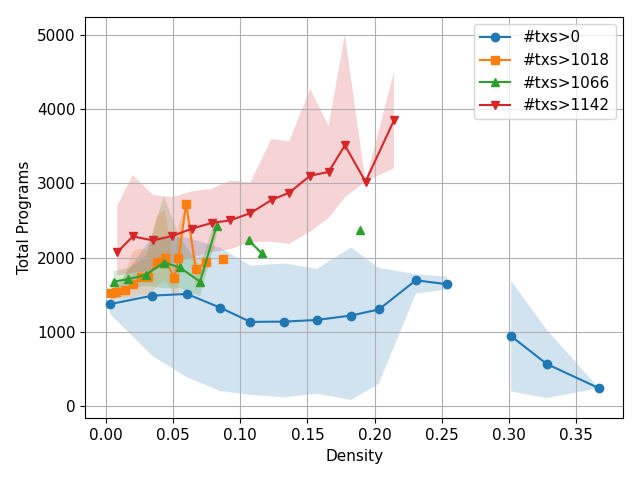}}
\caption{total-programs}
\label{fig:total-programs}
\end{figure}
\begin{figure}
\centerline{\includegraphics[width=\columnwidth]{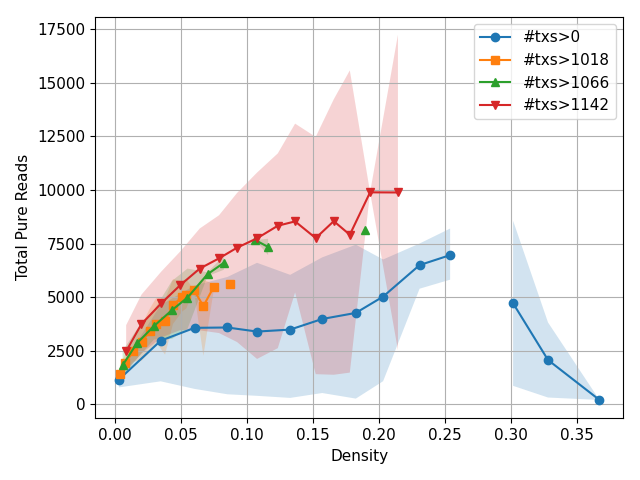}}
\caption{total-pure-reads}
\label{fig:total-pure-reads}
\end{figure}
\begin{figure}
\centerline{\includegraphics[width=\columnwidth]{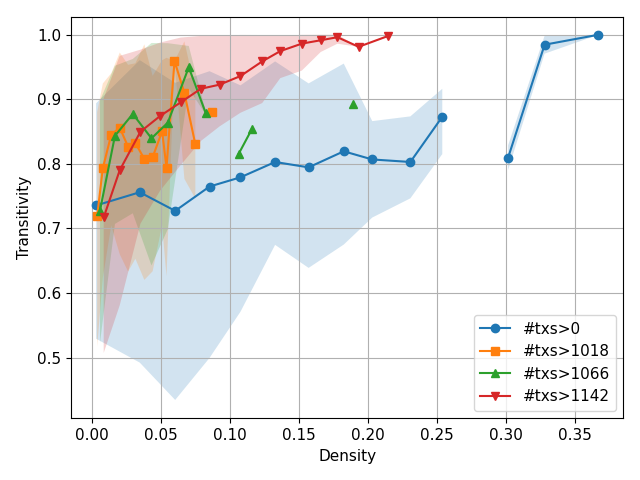}}
\caption{transitivity}
\label{fig:transitivity}
\end{figure}
\begin{figure}
\centerline{\includegraphics[width=\columnwidth]{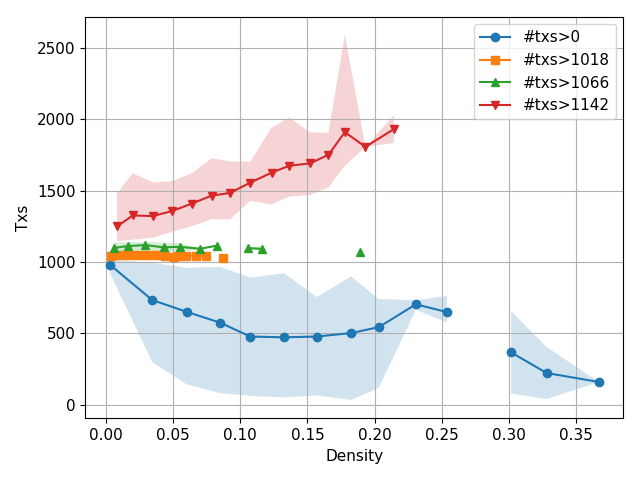}}
\caption{txs}
\label{fig:txs}
\end{figure}
\begin{figure}
\centerline{\includegraphics[width=\columnwidth]{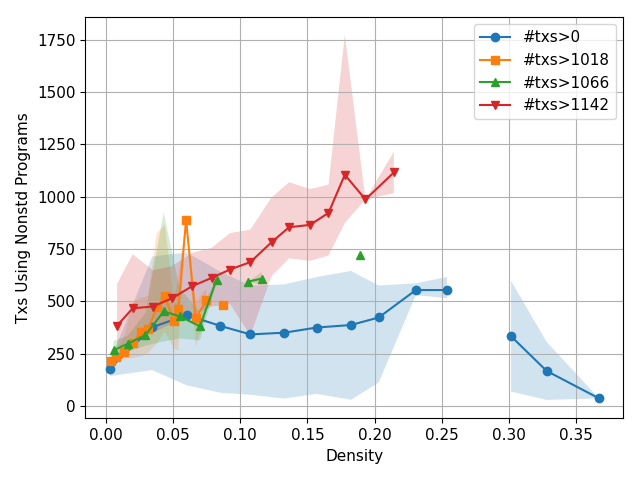}}
\caption{txs-using-nonstd-programs}
\label{fig:txs-using-nonstd-programs}
\end{figure}
\begin{figure}
\centerline{\includegraphics[width=\columnwidth]{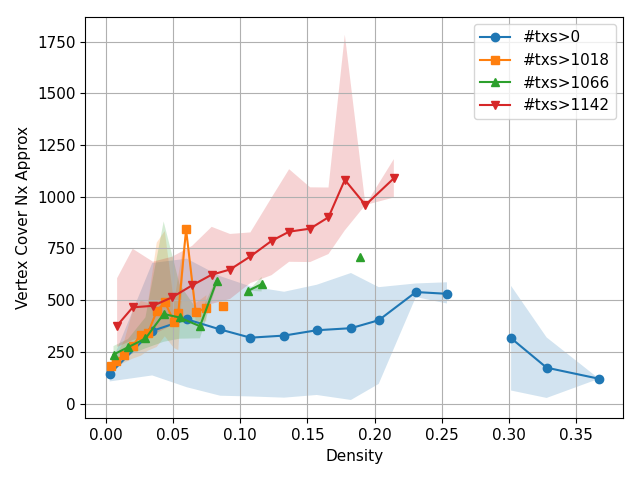}}
\caption{vertex-cover-nx-approx}
\label{fig:vertex-cover-nx-approx}
\end{figure}
\begin{figure}
\centerline{\includegraphics[width=\columnwidth]{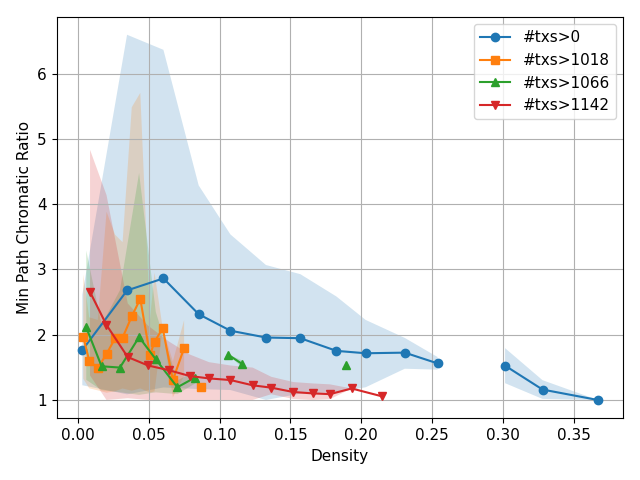}}
\caption{min-path-chromatic-ratio}
\label{fig:min-path-chromatic-ratio}
\end{figure}
\begin{figure}
\centerline{\includegraphics[width=\columnwidth]{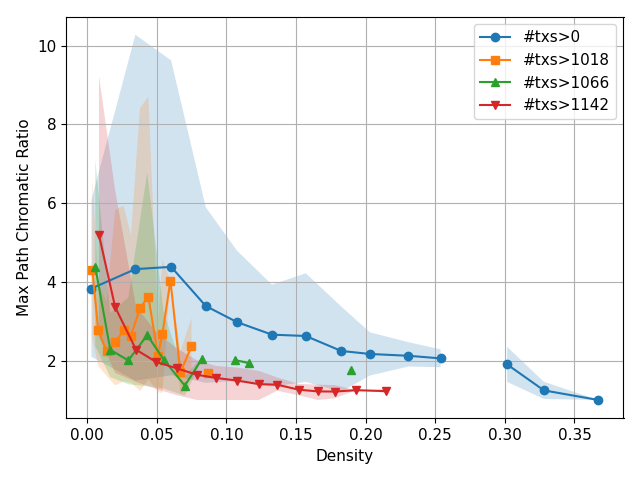}}
\caption{max-path-chromatic-ratio}
\label{fig:max-path-chromatic-ratio}
\end{figure}
\begin{figure}
\centerline{\includegraphics[width=\columnwidth]{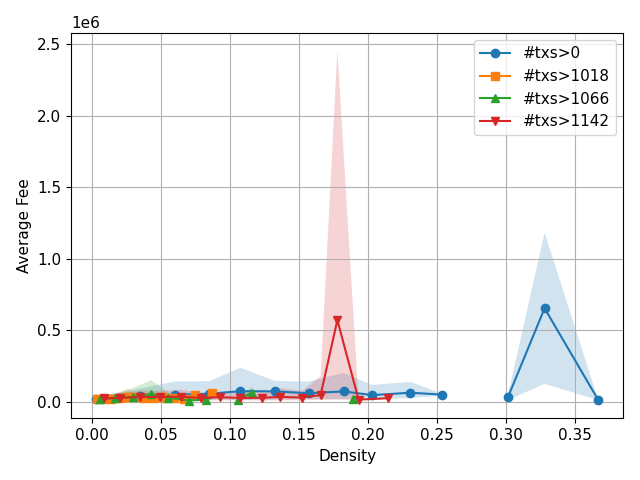}}
\caption{avg-fee}
\label{fig:avg-fee}
\end{figure}
\begin{figure}
\centerline{\includegraphics[width=\columnwidth]{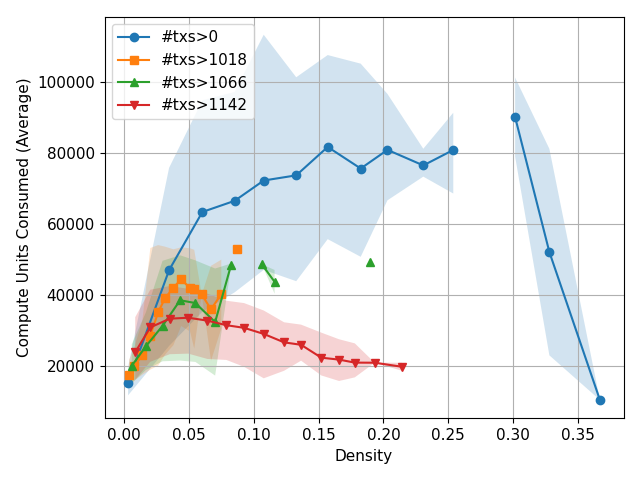}}
\caption{avg-computeUnitsConsumed}
\label{fig:avg-computeUnitsConsumed}
\end{figure}
\begin{figure}
\centerline{\includegraphics[width=\columnwidth]{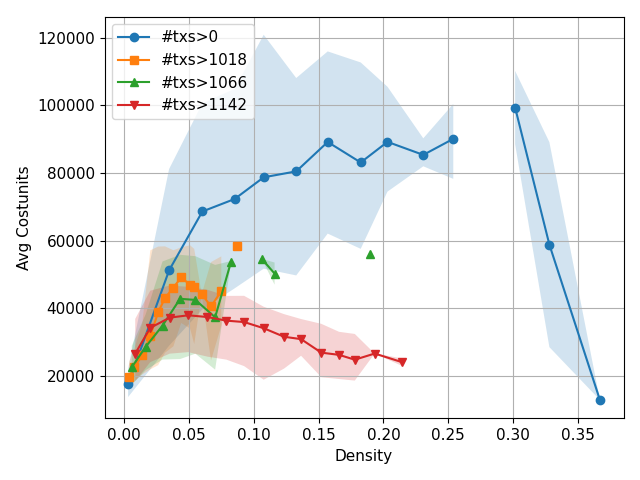}}
\caption{avg-costUnits}
\label{fig:avg-costUnits}
\end{figure}
\begin{figure}
\centerline{\includegraphics[width=\columnwidth]{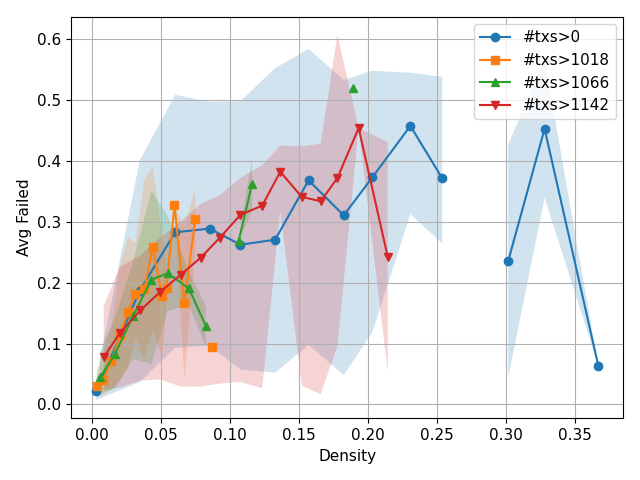}}
\caption{avg-failed}
\label{fig:avg-failed}
\end{figure}
\begin{figure}
\centerline{\includegraphics[width=\columnwidth]{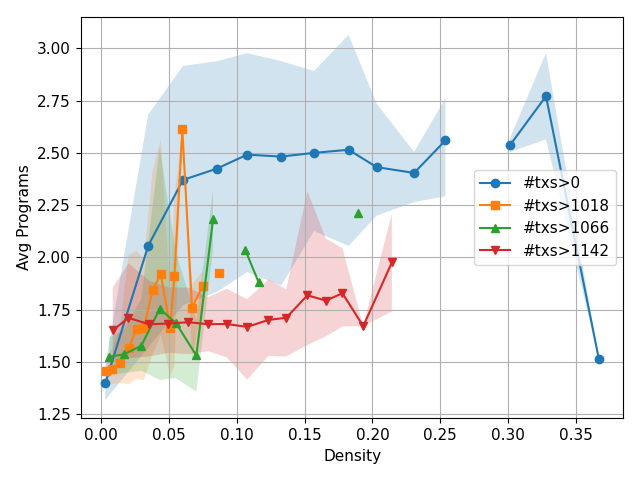}}
\caption{avg-programs}
\label{fig:avg-programs}
\end{figure}
\begin{figure}
\centerline{\includegraphics[width=\columnwidth]{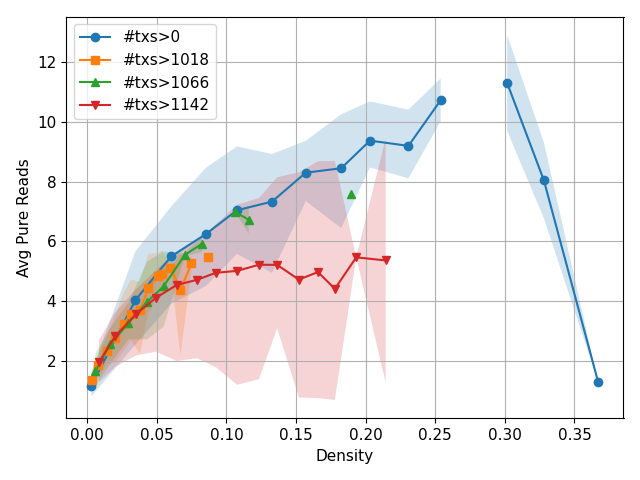}}
\caption{avg-pure-reads}
\label{fig:avg-pure-reads}
\end{figure}
\begin{figure}
\centerline{\includegraphics[width=\columnwidth]{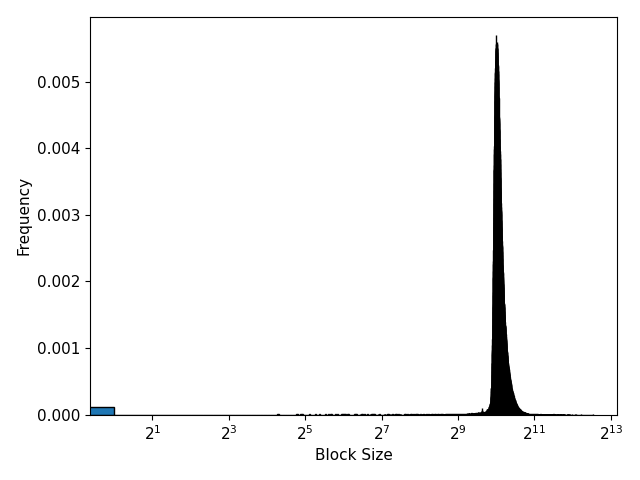}}
\caption{block-size-dist}
\label{fig:block-size-dist}
\end{figure}
\begin{figure}
\centerline{\includegraphics[width=\columnwidth]{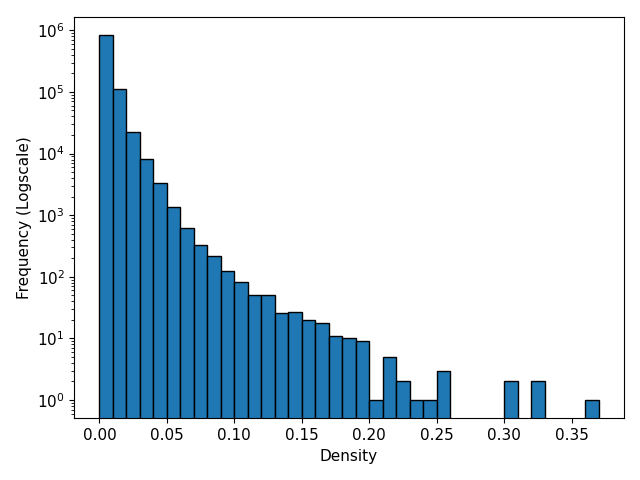}}
\caption{density-dist}
\label{fig:density-dist}
\end{figure}
\begin{figure}
\centerline{\includegraphics[width=\columnwidth]{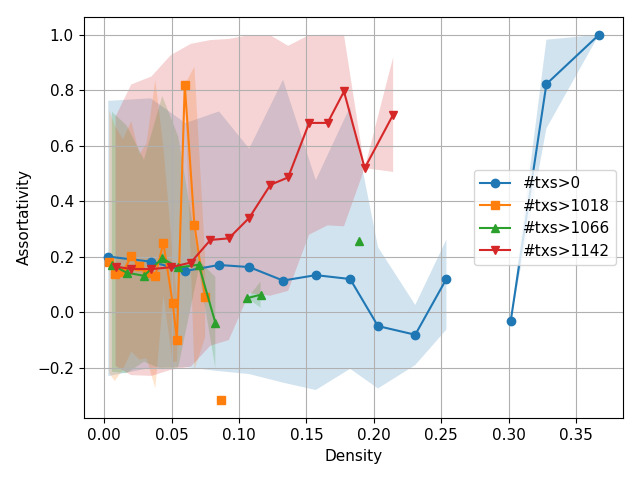}}
\caption{assortativity}
\label{fig:assortativity}
\end{figure}
\begin{figure}
\centerline{\includegraphics[width=\columnwidth]{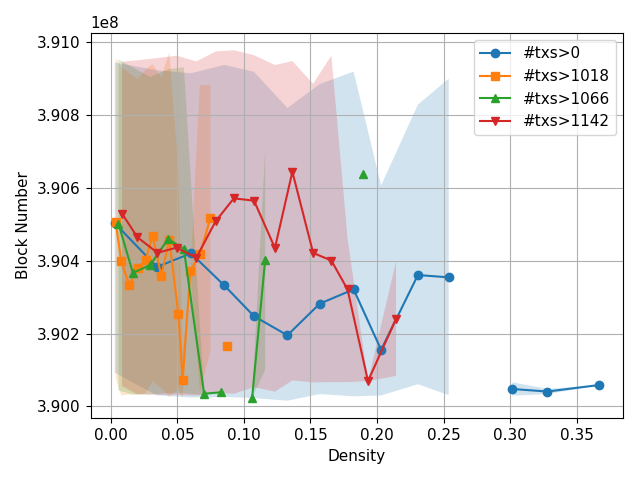}}
\caption{block-number}
\label{fig:block-number}
\end{figure}
\begin{figure}
\centerline{\includegraphics[width=\columnwidth]{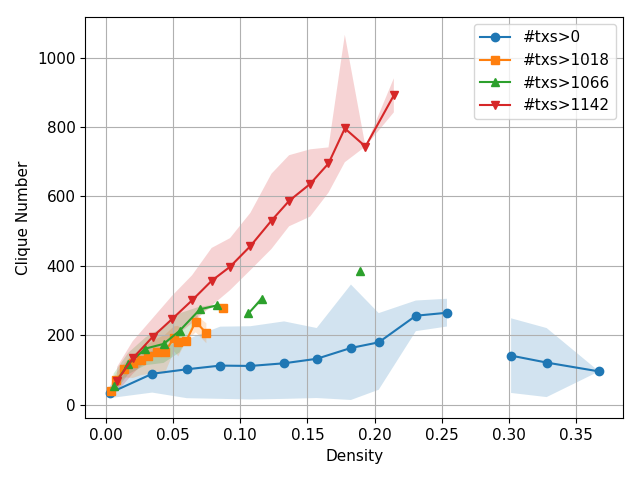}}
\caption{clique-number}
\label{fig:clique-number}
\end{figure}
\begin{figure}
\centerline{\includegraphics[width=\columnwidth]{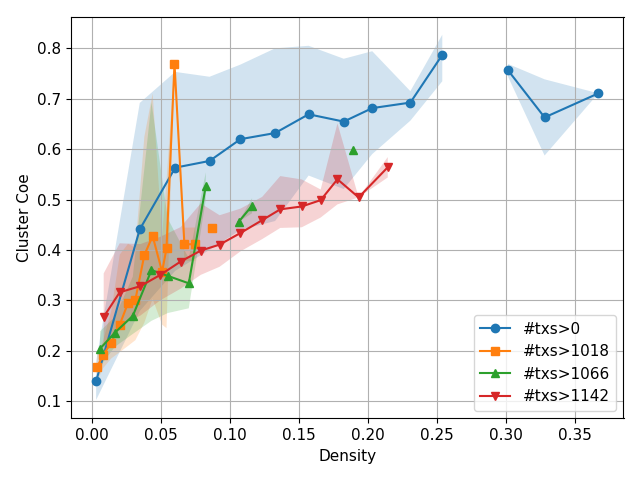}}
\caption{cluster-coe}
\label{fig:cluster-coe}
\end{figure}
\begin{figure}
\centerline{\includegraphics[width=\columnwidth]{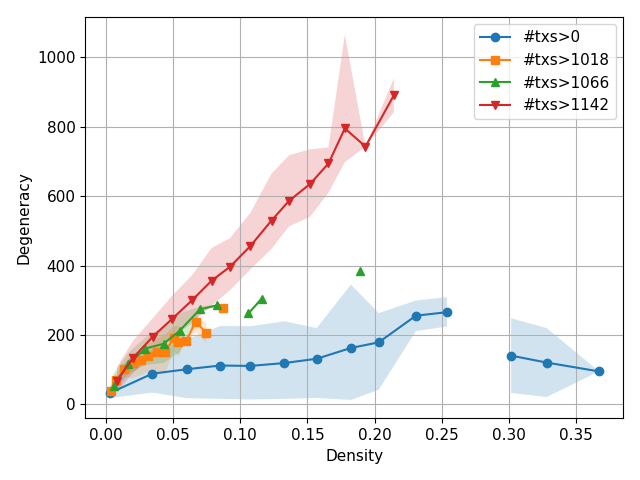}}
\caption{degeneracy}
\label{fig:degeneracy}
\end{figure}
\begin{figure}
\centerline{\includegraphics[width=\columnwidth]{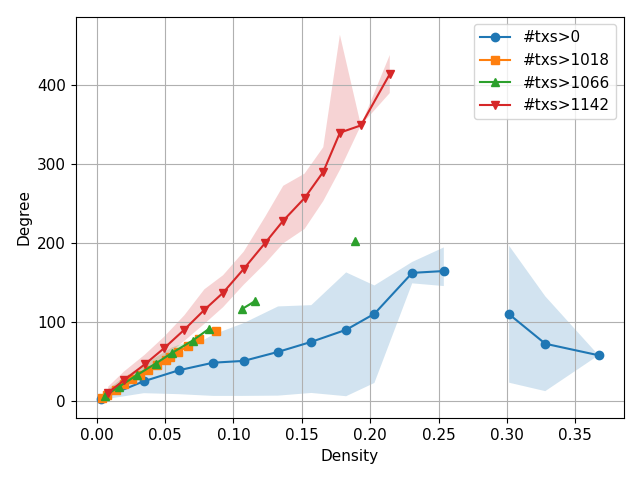}}
\caption{degree}
\label{fig:degree}
\end{figure}
\begin{figure}
\centerline{\includegraphics[width=\columnwidth]{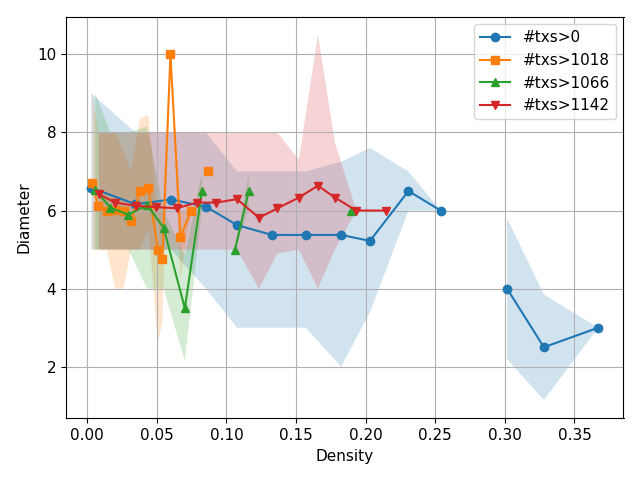}}
\caption{diameter}
\label{fig:diameter}
\end{figure}
\begin{figure}
\centerline{\includegraphics[width=\columnwidth]{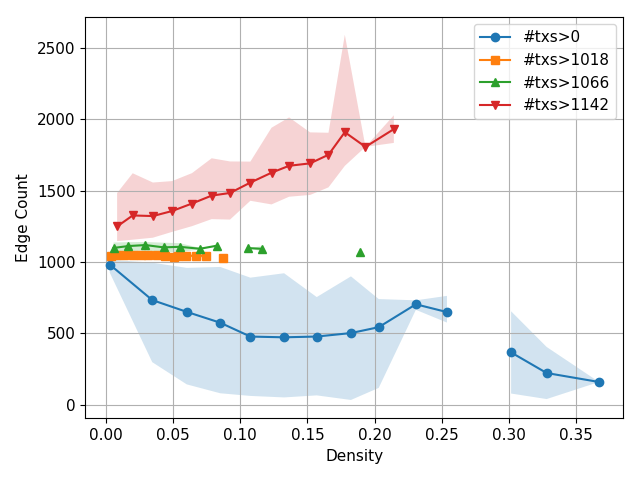}}
\caption{edge-count}
\label{fig:edge-count}
\end{figure}
\begin{figure}
\centerline{\includegraphics[width=\columnwidth]{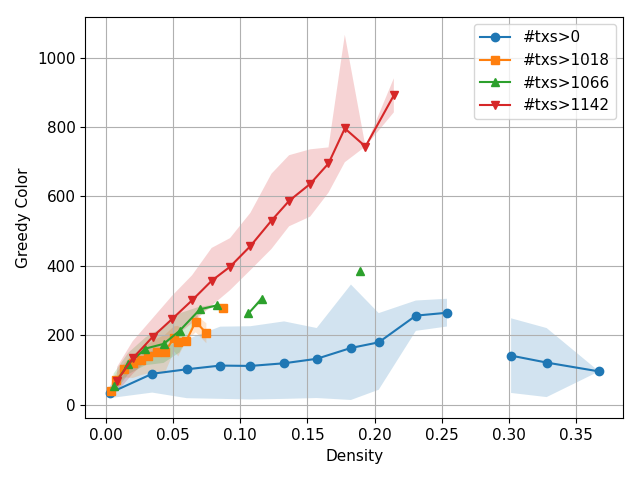}}
\caption{greedy-color}
\label{fig:greedy-color}
\end{figure}
\begin{figure}
\centerline{\includegraphics[width=\columnwidth]{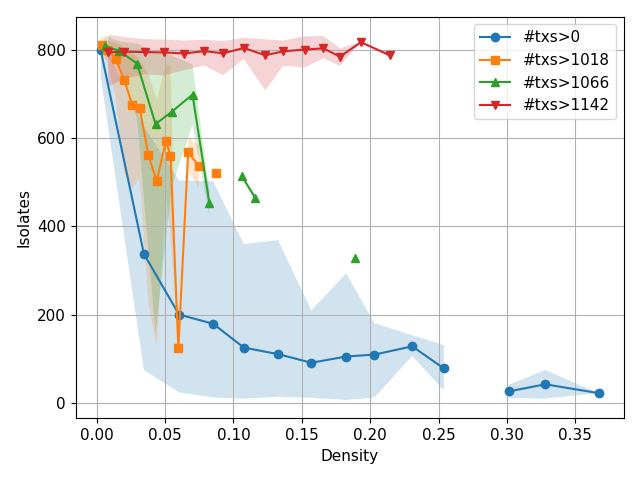}}
\caption{isolates}
\label{fig:isolates}
\end{figure}
\begin{figure}
\centerline{\includegraphics[width=\columnwidth]{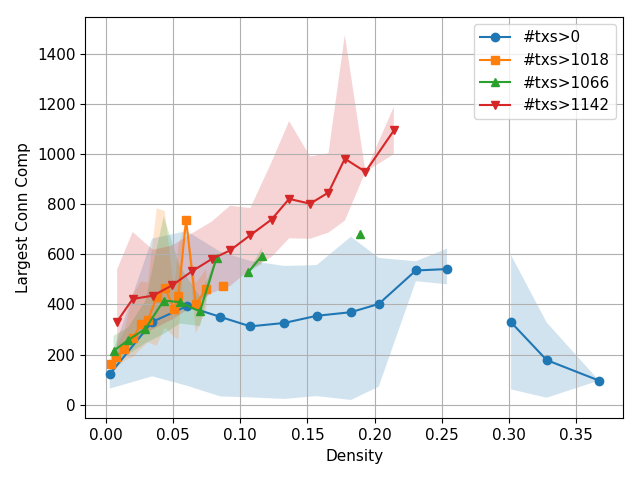}}
\caption{largest-conn-comp}
\label{fig:largest-conn-comp}
\end{figure}
\begin{figure}
\centerline{\includegraphics[width=\columnwidth]{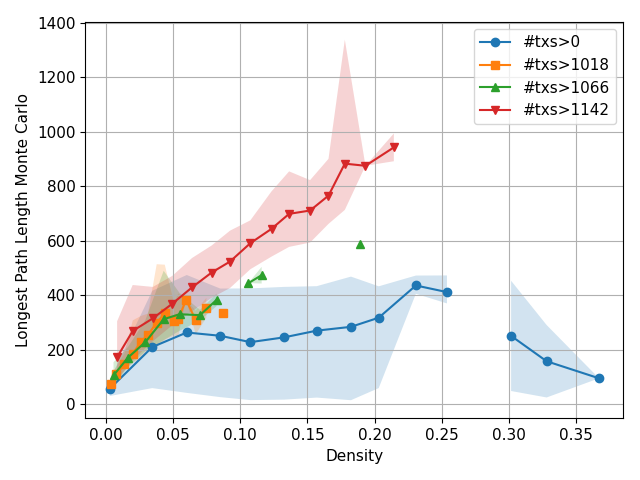}}
\caption{longest-path-length-monte-carlo}
\label{fig:longest-path-length-monte-carlo}
\end{figure}
\begin{figure}
\centerline{\includegraphics[width=\columnwidth]{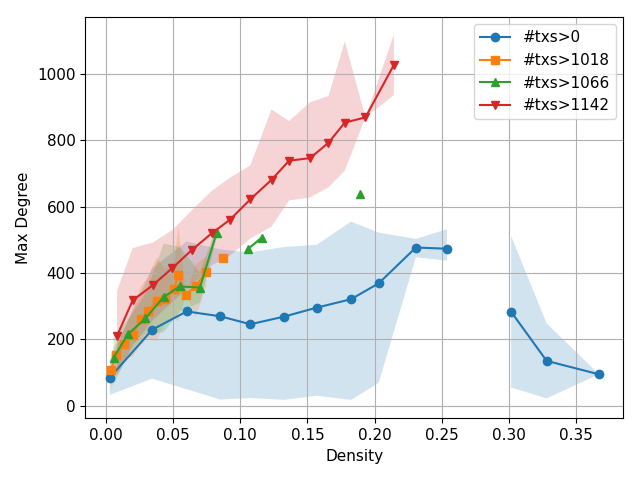}}
\caption{max-degree}
\label{fig:max-degree}
\end{figure}
\begin{figure}
\centerline{\includegraphics[width=\columnwidth]{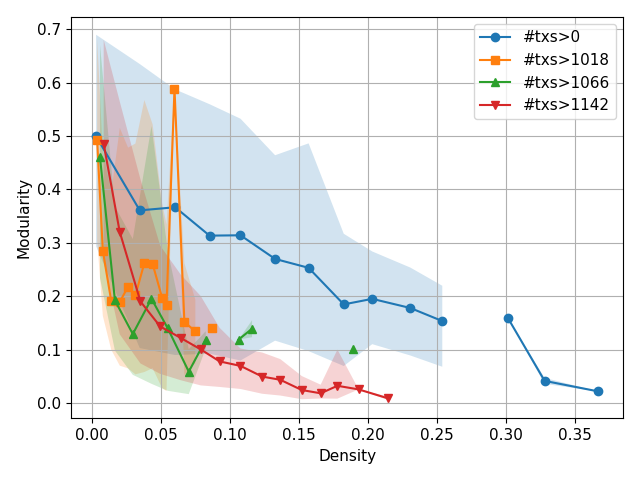}}
\caption{modularity}
\label{fig:modularity}
\end{figure}
\begin{figure}
\centerline{\includegraphics[width=\columnwidth]{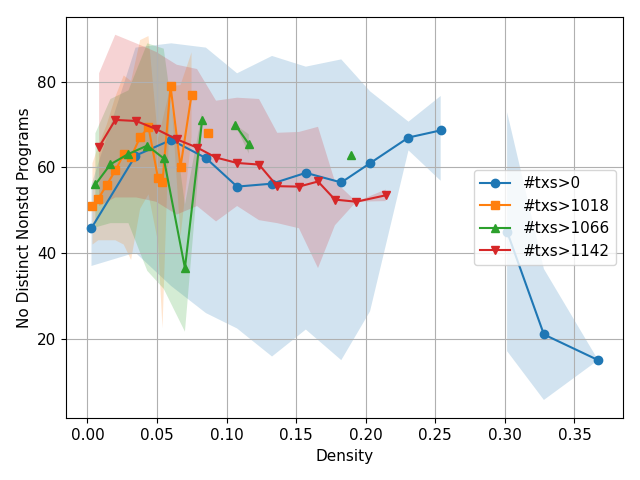}}
\caption{no-distinct-nonstd-programs}
\label{fig:no-distinct-nonstd-programs}
\end{figure}
\begin{figure}
\centerline{\includegraphics[width=\columnwidth]{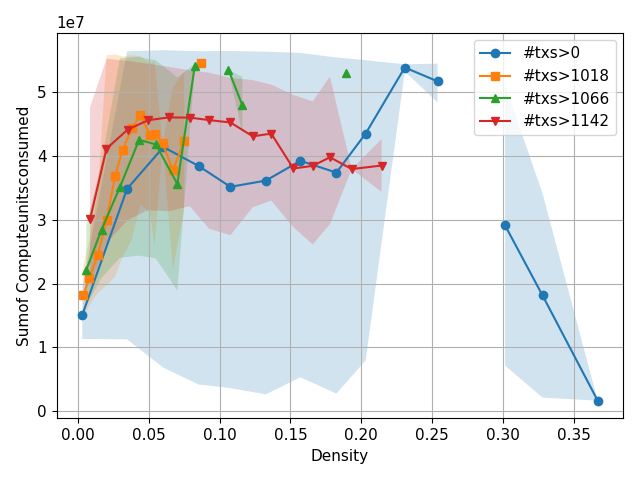}}
\caption{sumof-computeUnitsConsumed}
\label{fig:sumof-computeUnitsConsumed}
\end{figure}
\begin{figure}
\centerline{\includegraphics[width=\columnwidth]{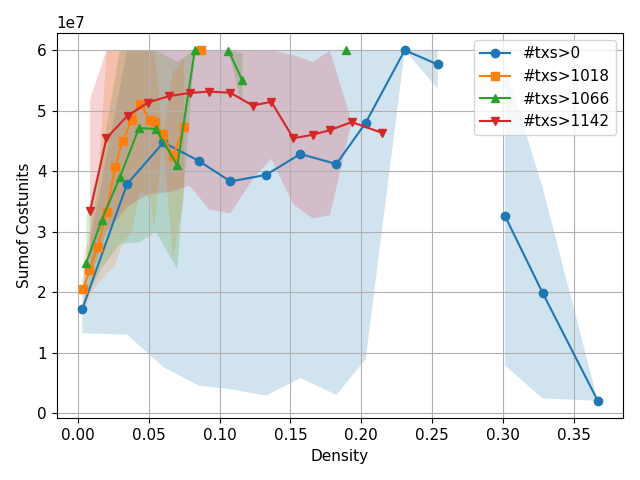}}
\caption{sumof-costUnits}
\label{fig:sumof-costUnits}
\end{figure}
\begin{figure}
\centerline{\includegraphics[width=\columnwidth]{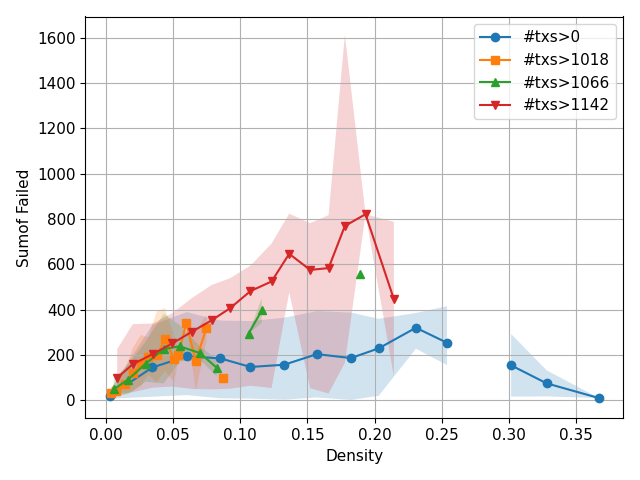}}
\caption{sumof-failed}
\label{fig:sumof-failed}
\end{figure}
\begin{figure}
\centerline{\includegraphics[width=\columnwidth]{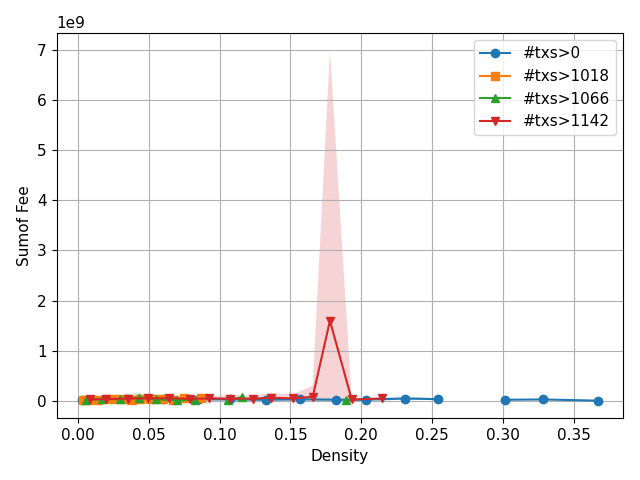}}
\caption{sumof-fee}
\label{fig:sumof-fee}
\end{figure}
\begin{figure}
\centerline{\includegraphics[width=\columnwidth]{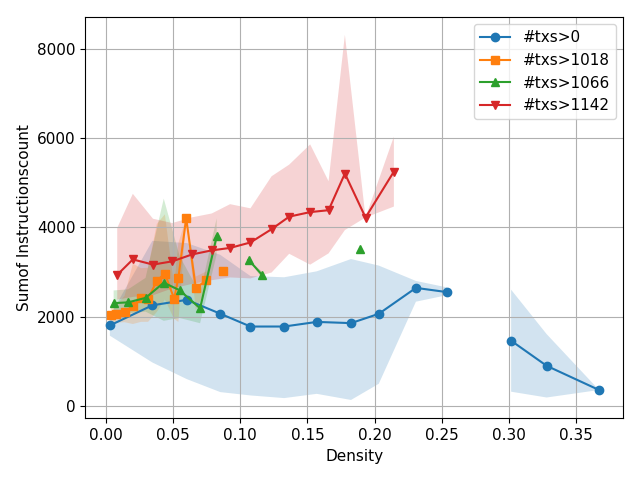}}
\caption{sumof-instructionsCount}
\label{fig:sumof-instructionsCount}
\end{figure}
\begin{figure}
\centerline{\includegraphics[width=\columnwidth]{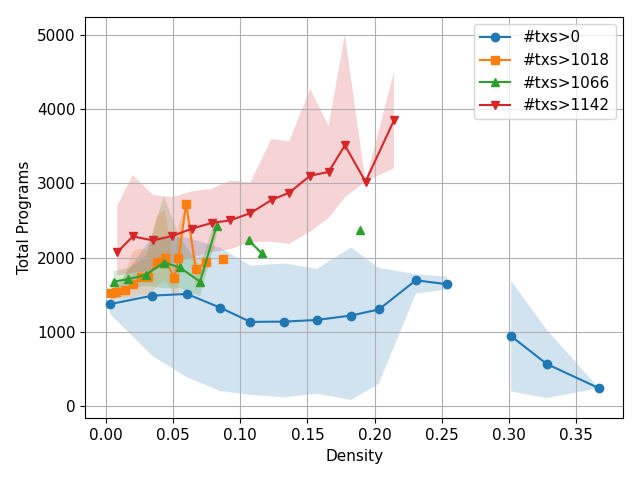}}
\caption{total-programs}
\label{fig:total-programs}
\end{figure}
\begin{figure}
\centerline{\includegraphics[width=\columnwidth]{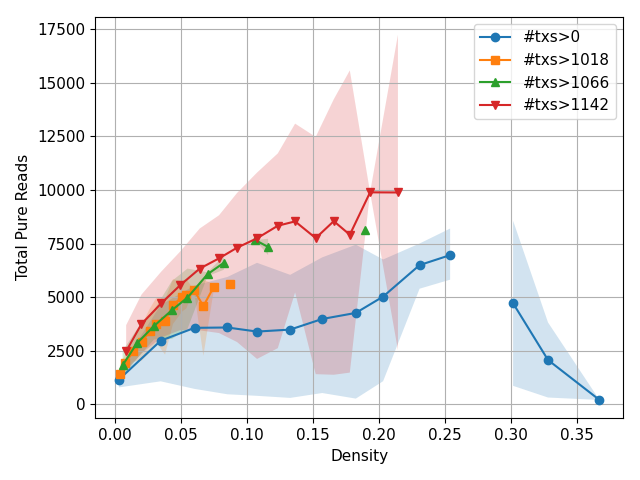}}
\caption{total-pure-reads}
\label{fig:total-pure-reads}
\end{figure}
\begin{figure}
\centerline{\includegraphics[width=\columnwidth]{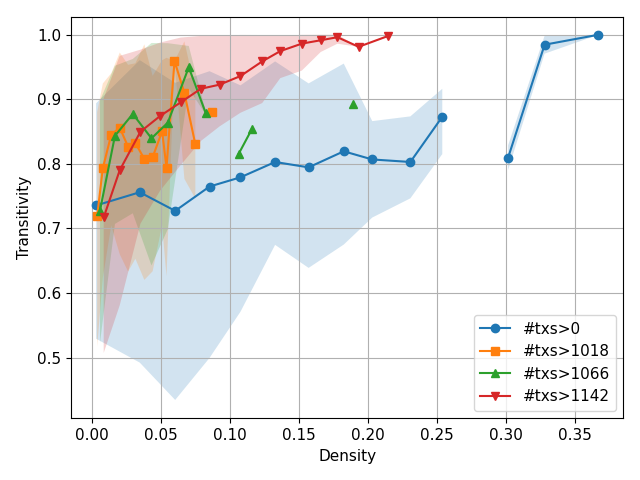}}
\caption{transitivity}
\label{fig:transitivity}
\end{figure}
\begin{figure}
\centerline{\includegraphics[width=\columnwidth]{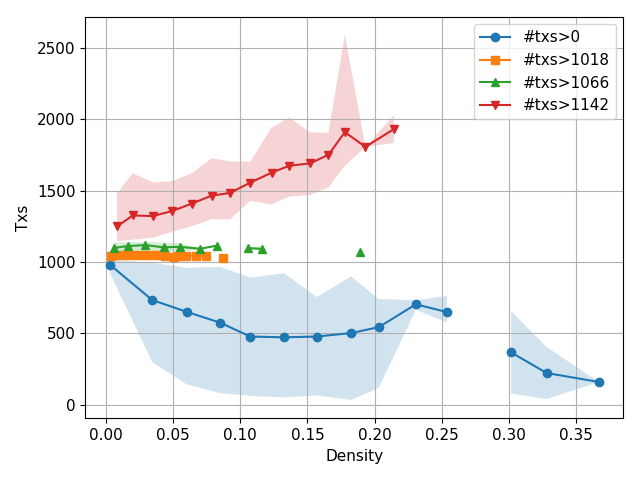}}
\caption{txs}
\label{fig:txs}
\end{figure}
\begin{figure}
\centerline{\includegraphics[width=\columnwidth]{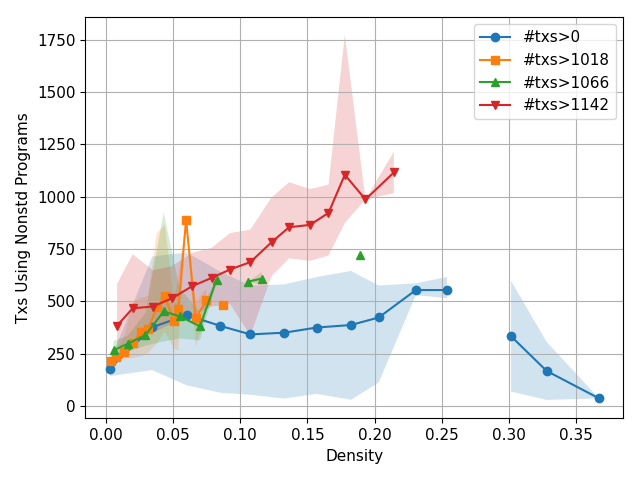}}
\caption{txs-using-nonstd-programs}
\label{fig:txs-using-nonstd-programs}
\end{figure}
\begin{figure}
\centerline{\includegraphics[width=\columnwidth]{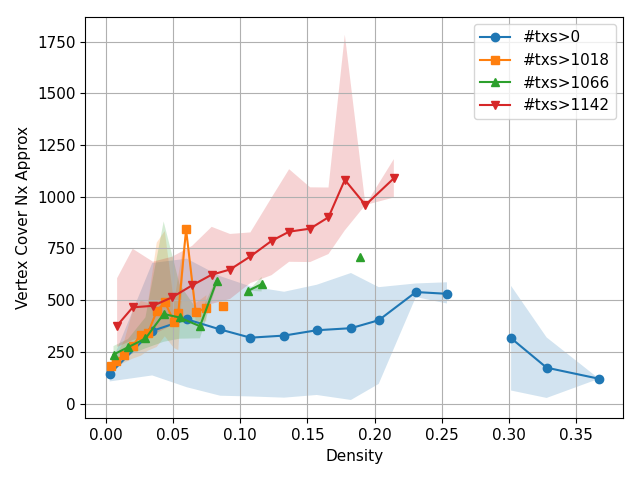}}
\caption{vertex-cover-nx-approx}
\label{fig:vertex-cover-nx-approx}
\end{figure}
\begin{figure}
\centerline{\includegraphics[width=\columnwidth]{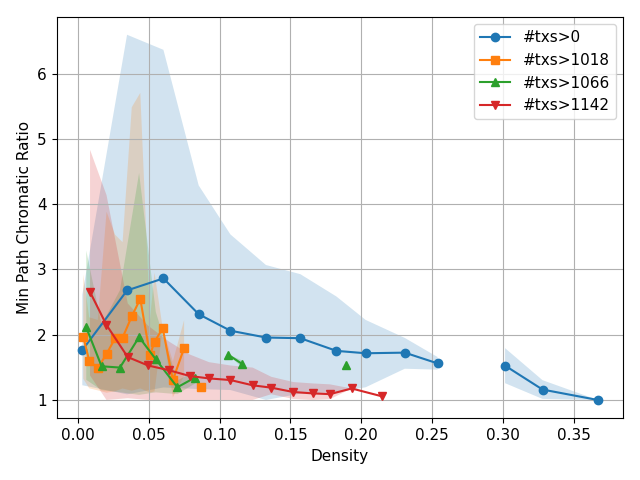}}
\caption{min-path-chromatic-ratio}
\label{fig:min-path-chromatic-ratio}
\end{figure}
\begin{figure}
\centerline{\includegraphics[width=\columnwidth]{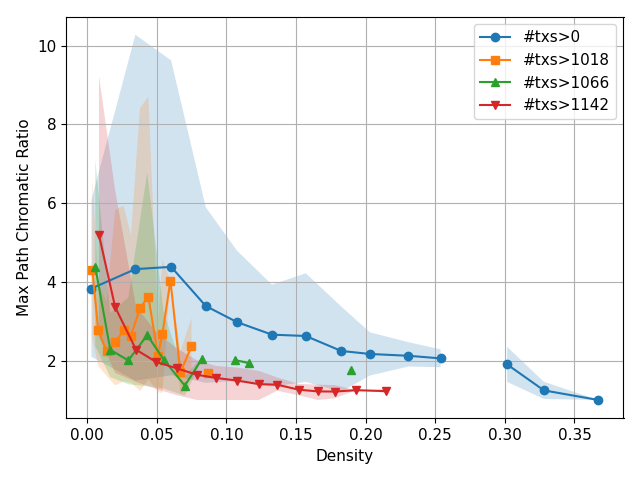}}
\caption{max-path-chromatic-ratio}
\label{fig:max-path-chromatic-ratio}
\end{figure}
\begin{figure}
\centerline{\includegraphics[width=\columnwidth]{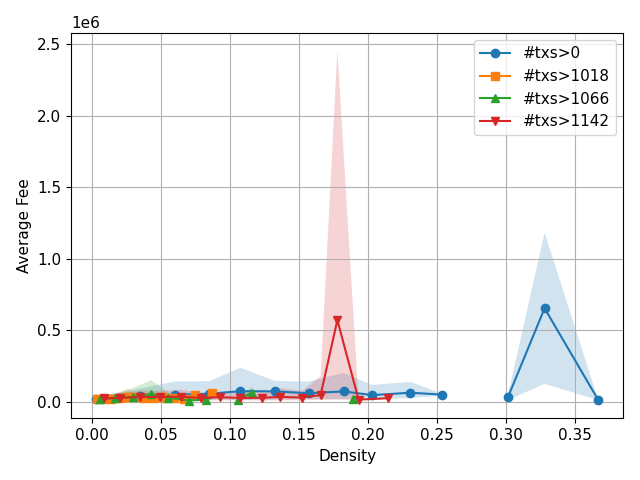}}
\caption{avg-fee}
\label{fig:avg-fee}
\end{figure}
\begin{figure}
\centerline{\includegraphics[width=\columnwidth]{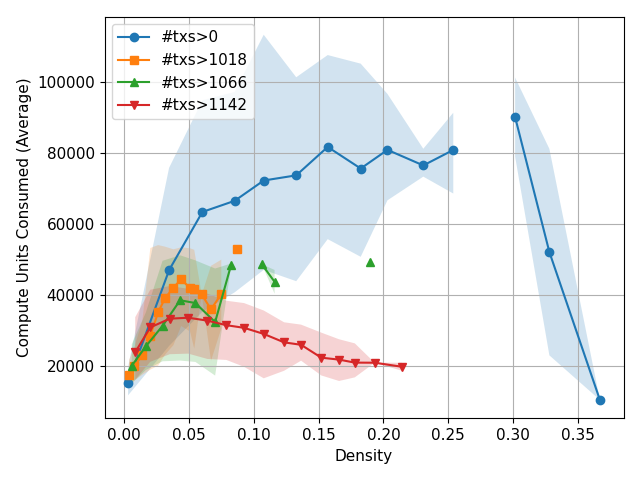}}
\caption{avg-computeUnitsConsumed}
\label{fig:avg-computeUnitsConsumed}
\end{figure}
\begin{figure}
\centerline{\includegraphics[width=\columnwidth]{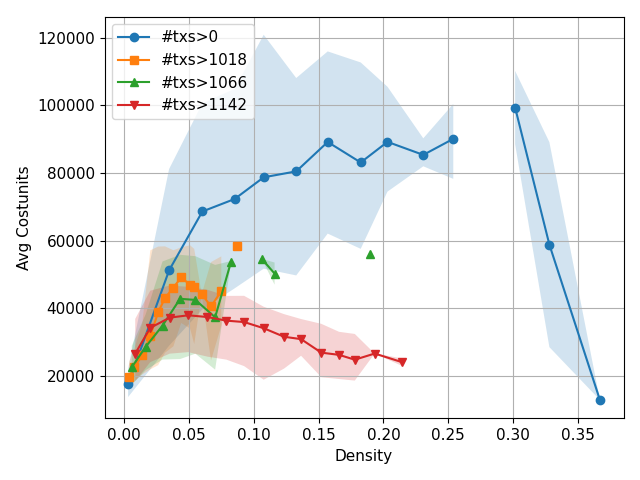}}
\caption{avg-costUnits}
\label{fig:avg-costUnits}
\end{figure}
\begin{figure}
\centerline{\includegraphics[width=\columnwidth]{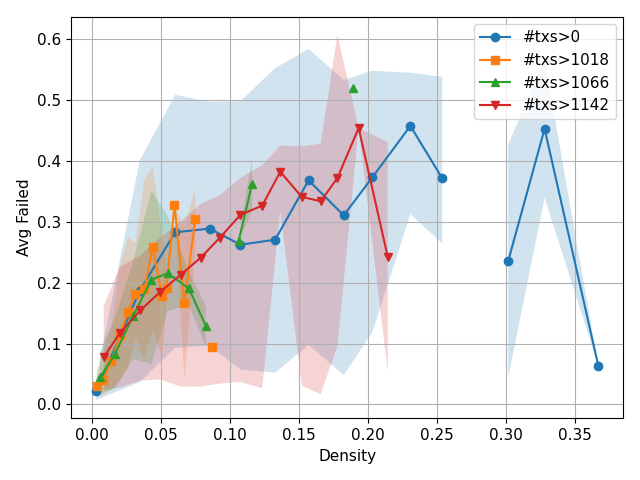}}
\caption{avg-failed}
\label{fig:avg-failed}
\end{figure}
\begin{figure}
\centerline{\includegraphics[width=\columnwidth]{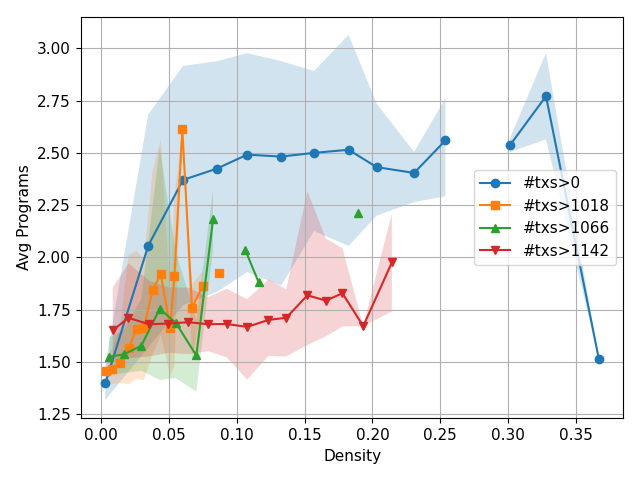}}
\caption{avg-programs}
\label{fig:avg-programs}
\end{figure}
\begin{figure}
\centerline{\includegraphics[width=\columnwidth]{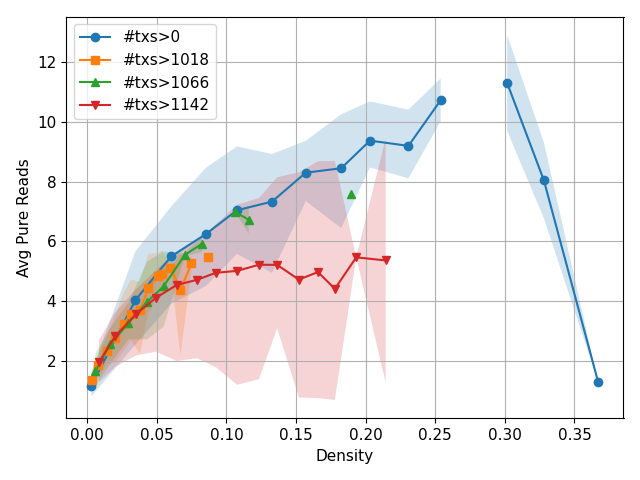}}
\caption{avg-pure-reads}
\label{fig:avg-pure-reads}
\end{figure}
\begin{figure}
\centerline{\includegraphics[width=\columnwidth]{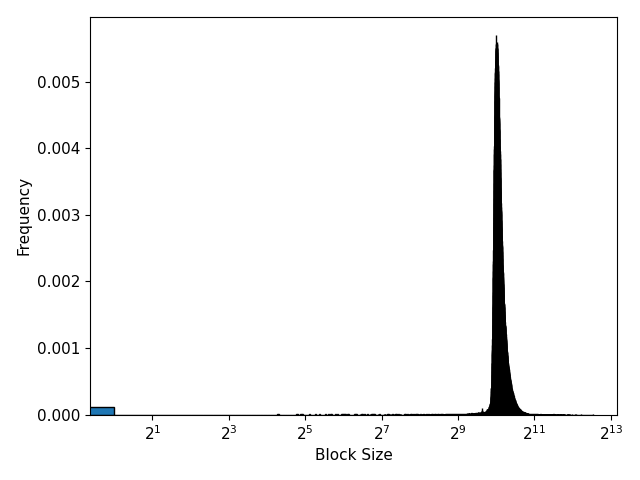}}
\caption{block-size-dist}
\label{fig:block-size-dist}
\end{figure}
\begin{figure}
\centerline{\includegraphics[width=\columnwidth]{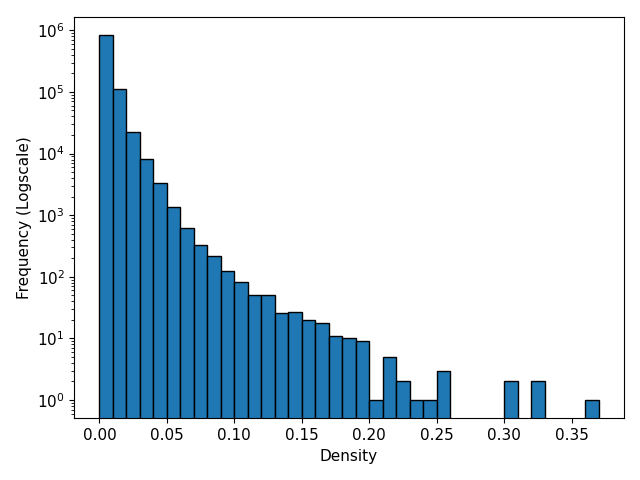}}
\caption{density-dist}
\label{fig:density-dist}
\end{figure}
\begin{figure}
\centerline{\includegraphics[width=\columnwidth]{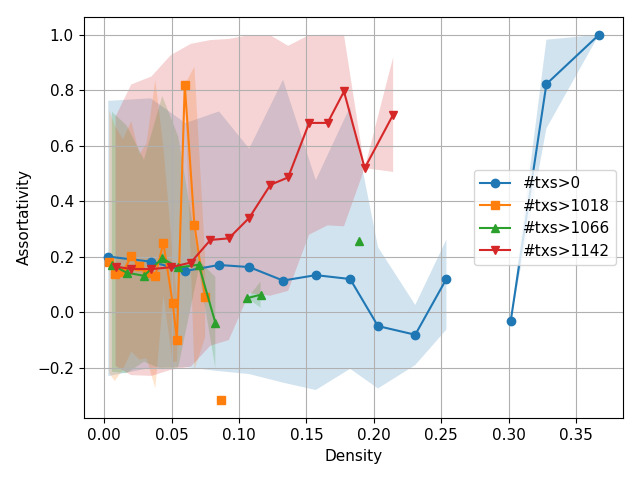}}
\caption{assortativity}
\label{fig:assortativity}
\end{figure}
\begin{figure}
\centerline{\includegraphics[width=\columnwidth]{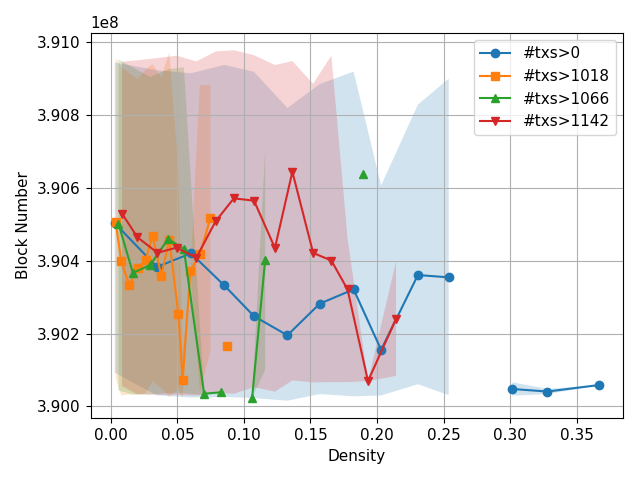}}
\caption{block-number}
\label{fig:block-number}
\end{figure}
\begin{figure}
\centerline{\includegraphics[width=\columnwidth]{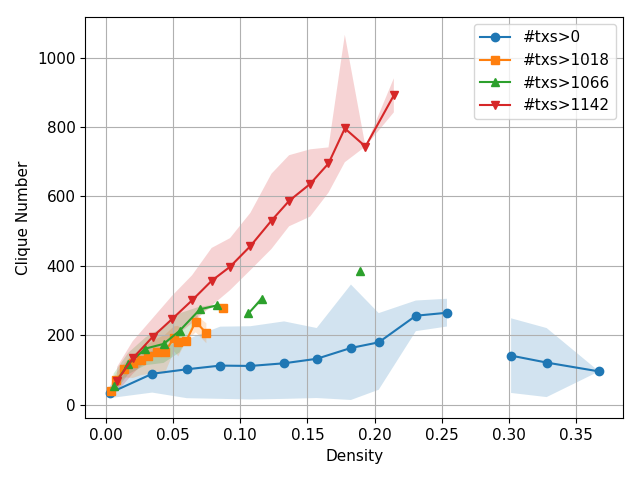}}
\caption{clique-number}
\label{fig:clique-number}
\end{figure}
\begin{figure}
\centerline{\includegraphics[width=\columnwidth]{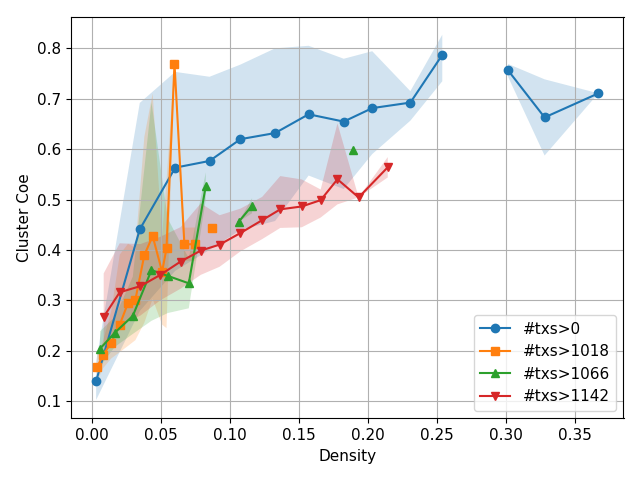}}
\caption{cluster-coe}
\label{fig:cluster-coe}
\end{figure}
\begin{figure}
\centerline{\includegraphics[width=\columnwidth]{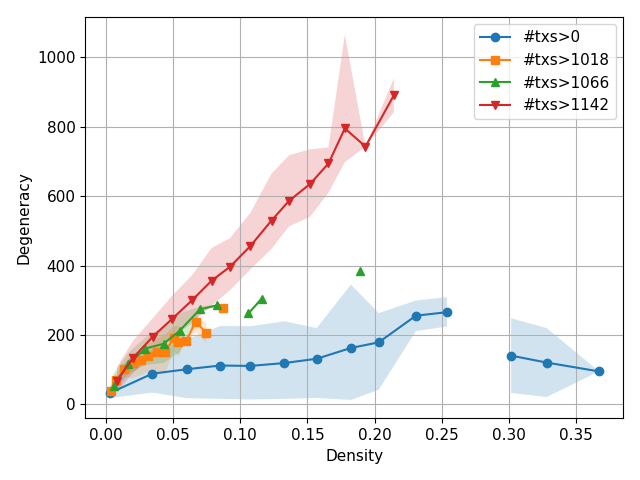}}
\caption{degeneracy}
\label{fig:degeneracy}
\end{figure}
\begin{figure}
\centerline{\includegraphics[width=\columnwidth]{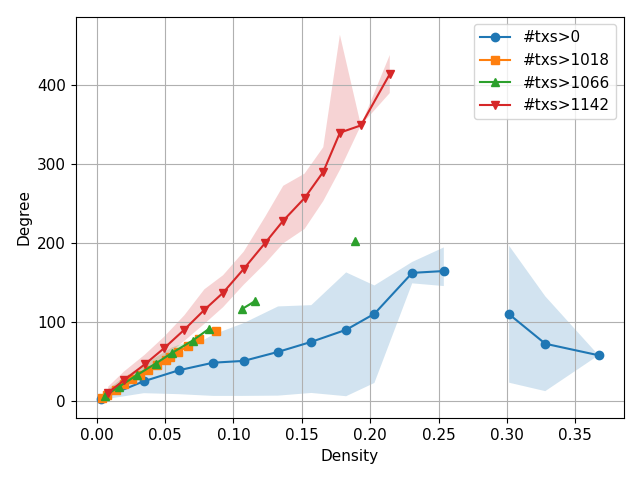}}
\caption{degree}
\label{fig:degree}
\end{figure}
\begin{figure}
\centerline{\includegraphics[width=\columnwidth]{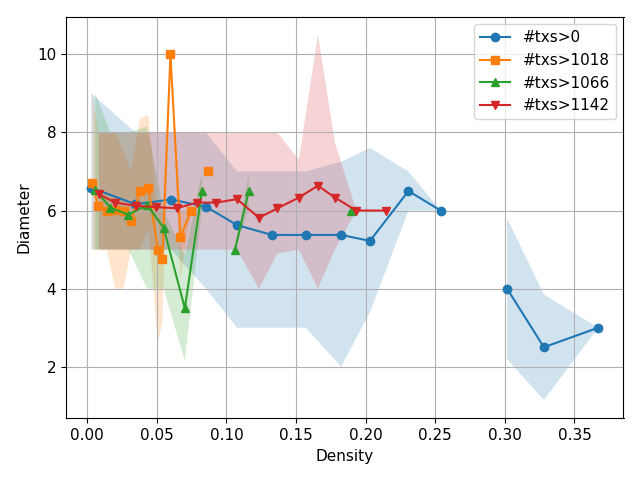}}
\caption{diameter}
\label{fig:diameter}
\end{figure}
\begin{figure}
\centerline{\includegraphics[width=\columnwidth]{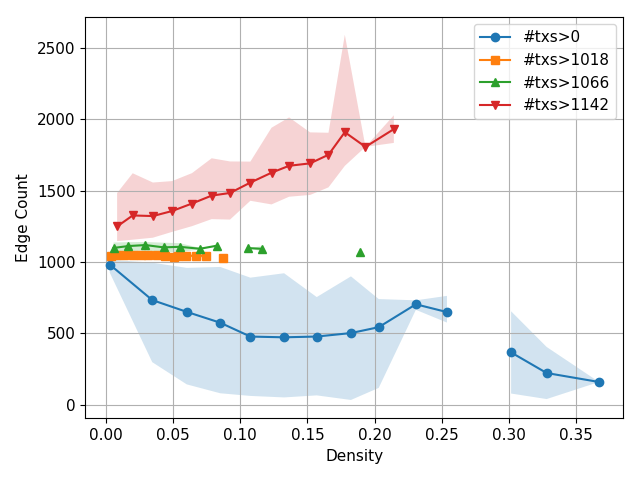}}
\caption{edge-count}
\label{fig:edge-count}
\end{figure}
\begin{figure}
\centerline{\includegraphics[width=\columnwidth]{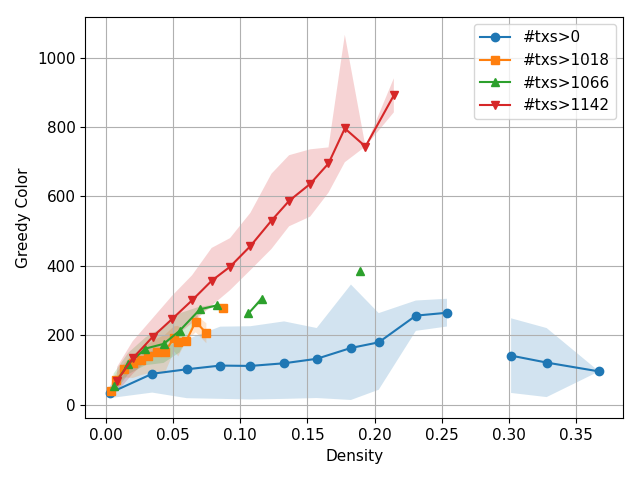}}
\caption{greedy-color}
\label{fig:greedy-color}
\end{figure}
\begin{figure}
\centerline{\includegraphics[width=\columnwidth]{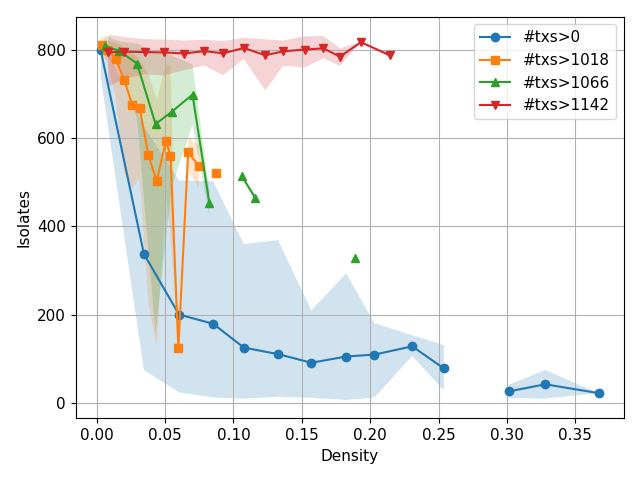}}
\caption{isolates}
\label{fig:isolates}
\end{figure}
\begin{figure}
\centerline{\includegraphics[width=\columnwidth]{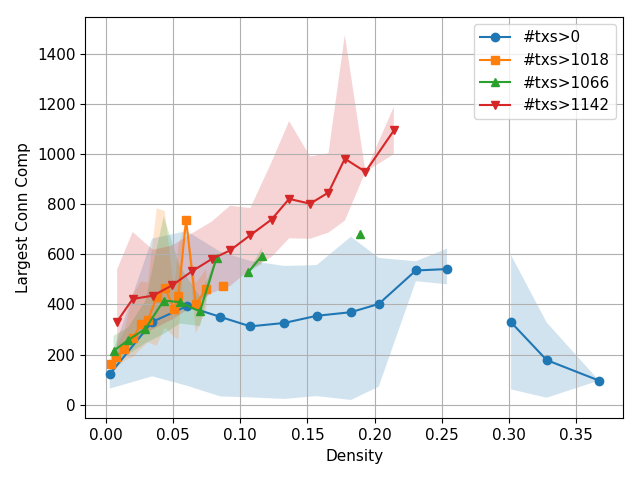}}
\caption{largest-conn-comp}
\label{fig:largest-conn-comp}
\end{figure}
\begin{figure}
\centerline{\includegraphics[width=\columnwidth]{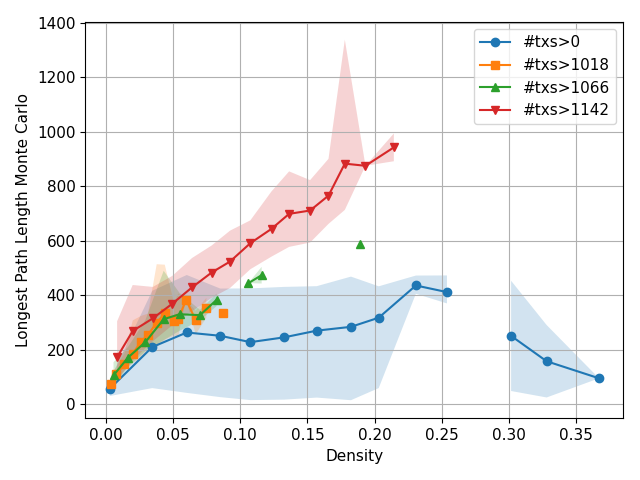}}
\caption{longest-path-length-monte-carlo}
\label{fig:longest-path-length-monte-carlo}
\end{figure}
\begin{figure}
\centerline{\includegraphics[width=\columnwidth]{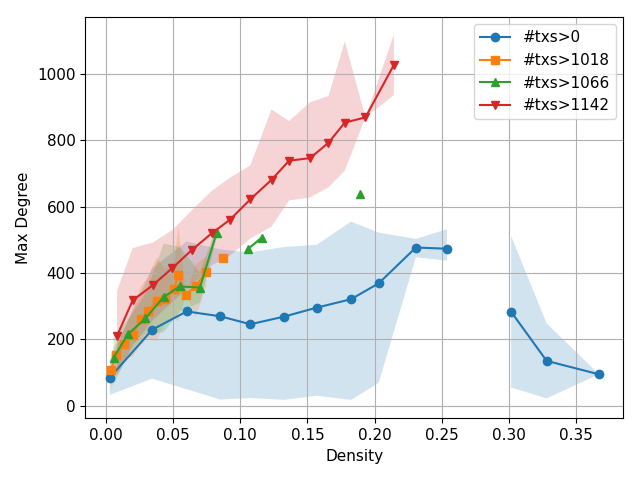}}
\caption{max-degree}
\label{fig:max-degree}
\end{figure}
\begin{figure}
\centerline{\includegraphics[width=\columnwidth]{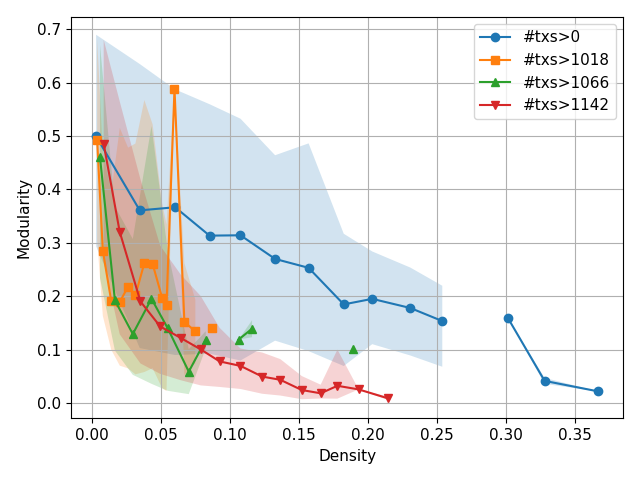}}
\caption{modularity}
\label{fig:modularity}
\end{figure}
\begin{figure}
\centerline{\includegraphics[width=\columnwidth]{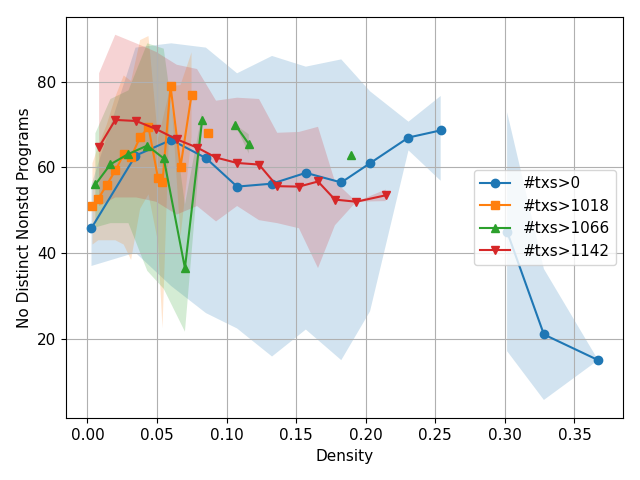}}
\caption{no-distinct-nonstd-programs}
\label{fig:no-distinct-nonstd-programs}
\end{figure}
\begin{figure}
\centerline{\includegraphics[width=\columnwidth]{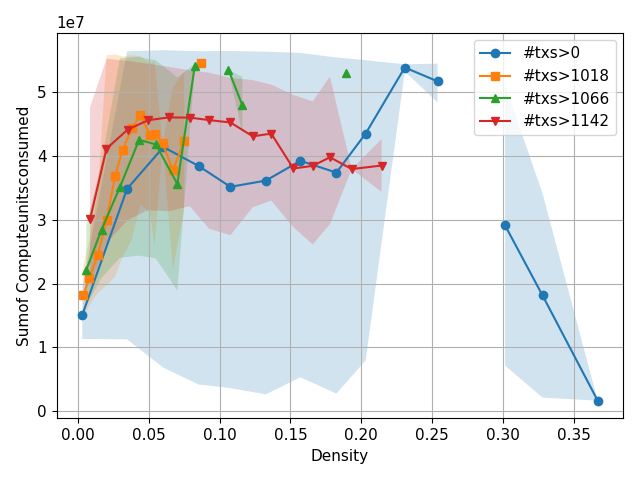}}
\caption{sumof-computeUnitsConsumed}
\label{fig:sumof-computeUnitsConsumed}
\end{figure}
\begin{figure}
\centerline{\includegraphics[width=\columnwidth]{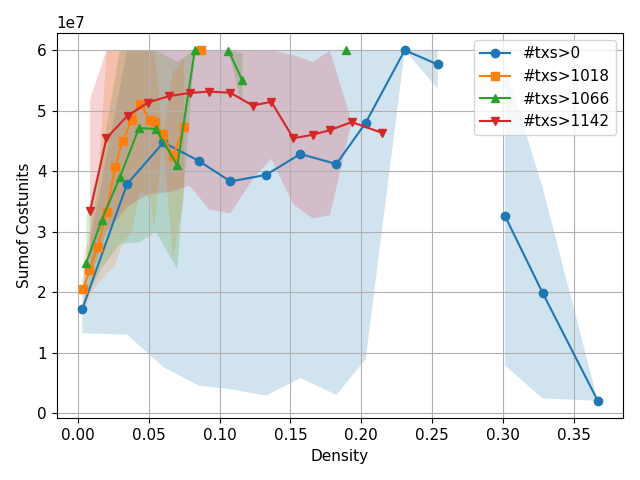}}
\caption{sumof-costUnits}
\label{fig:sumof-costUnits}
\end{figure}
\begin{figure}
\centerline{\includegraphics[width=\columnwidth]{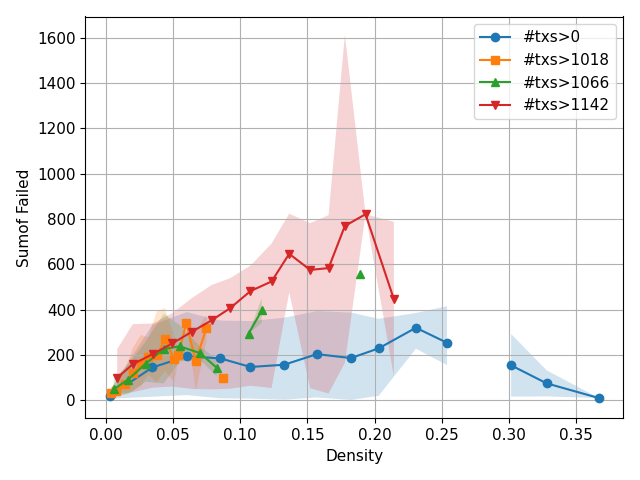}}
\caption{sumof-failed}
\label{fig:sumof-failed}
\end{figure}
\begin{figure}
\centerline{\includegraphics[width=\columnwidth]{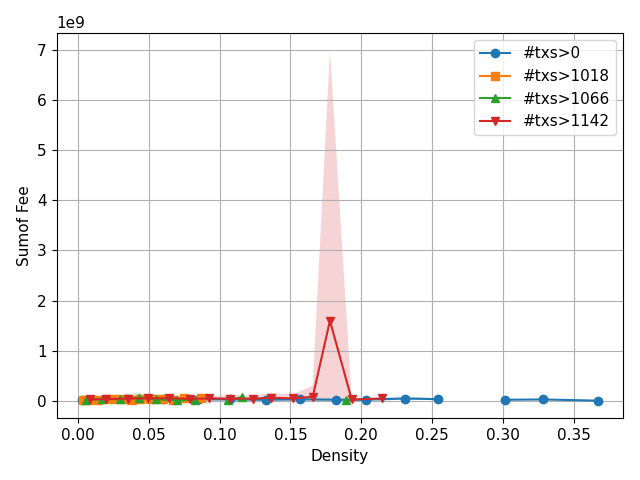}}
\caption{sumof-fee}
\label{fig:sumof-fee}
\end{figure}
\begin{figure}
\centerline{\includegraphics[width=\columnwidth]{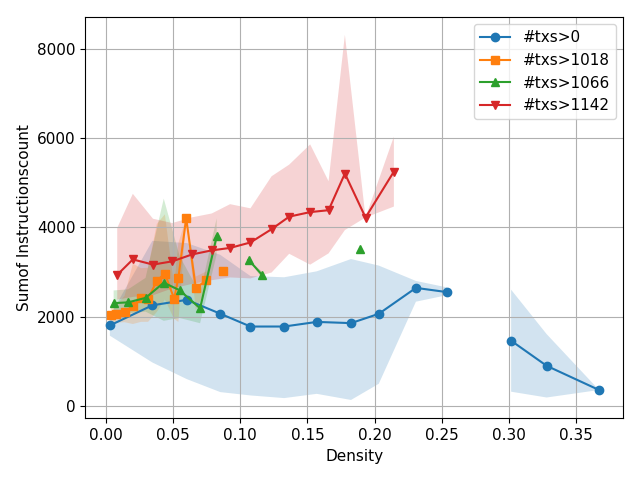}}
\caption{sumof-instructionsCount}
\label{fig:sumof-instructionsCount}
\end{figure}
\begin{figure}
\centerline{\includegraphics[width=\columnwidth]{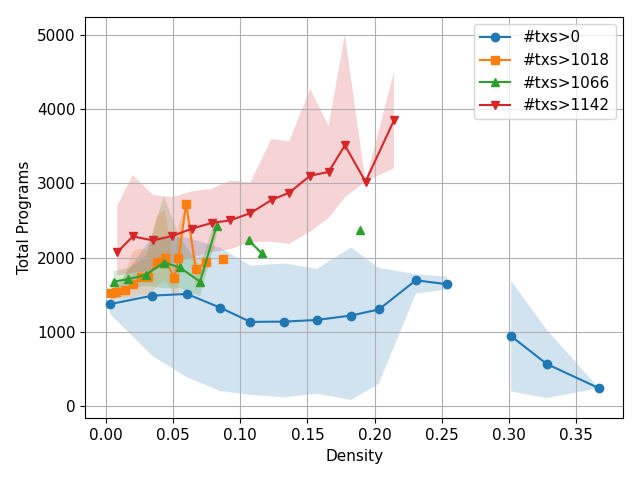}}
\caption{total-programs}
\label{fig:total-programs}
\end{figure}
\begin{figure}
\centerline{\includegraphics[width=\columnwidth]{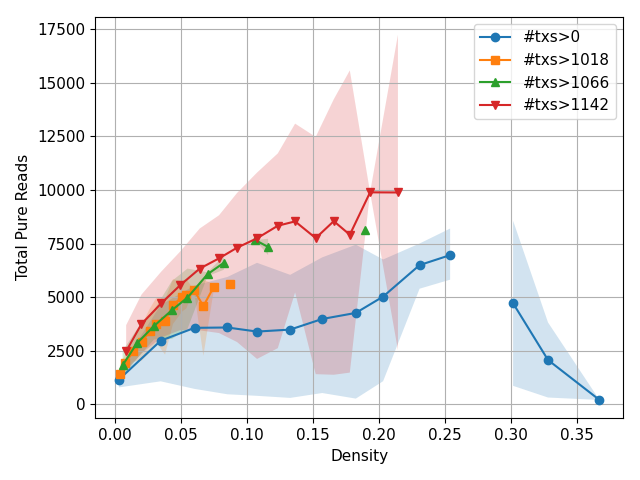}}
\caption{total-pure-reads}
\label{fig:total-pure-reads}
\end{figure}
\begin{figure}
\centerline{\includegraphics[width=\columnwidth]{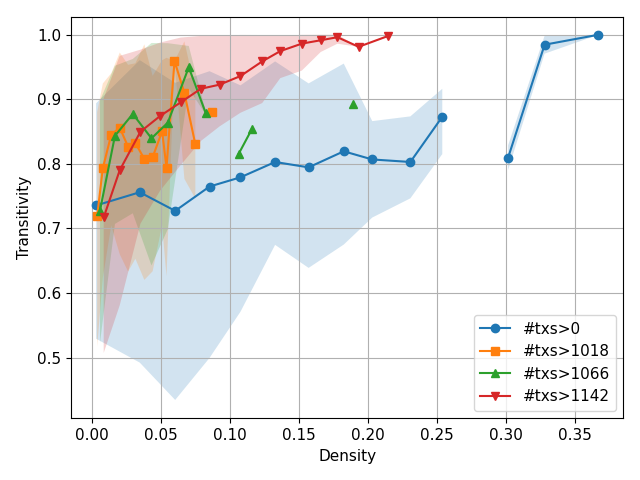}}
\caption{transitivity}
\label{fig:transitivity}
\end{figure}
\begin{figure}
\centerline{\includegraphics[width=\columnwidth]{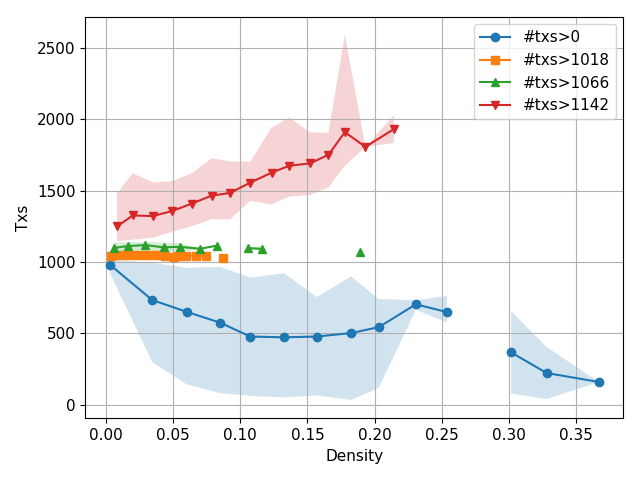}}
\caption{txs}
\label{fig:txs}
\end{figure}
\begin{figure}
\centerline{\includegraphics[width=\columnwidth]{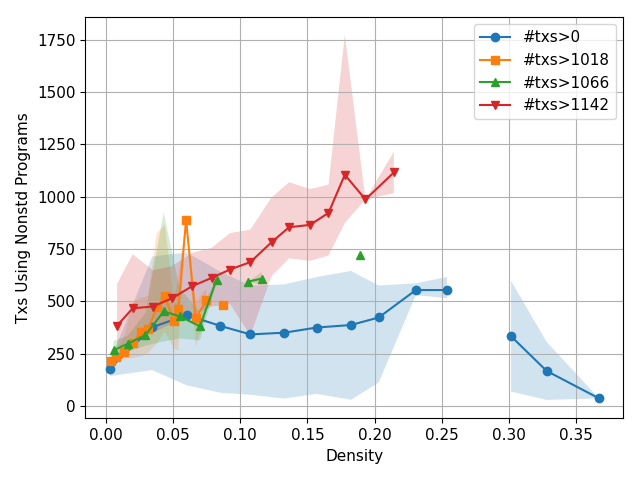}}
\caption{txs-using-nonstd-programs}
\label{fig:txs-using-nonstd-programs}
\end{figure}
\begin{figure}
\centerline{\includegraphics[width=\columnwidth]{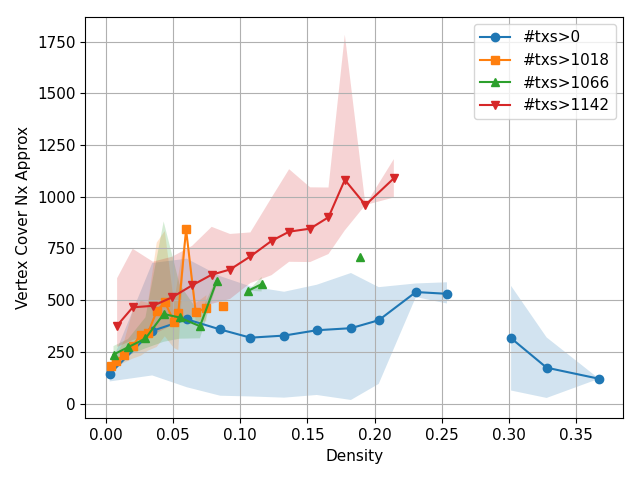}}
\caption{vertex-cover-nx-approx}
\label{fig:vertex-cover-nx-approx}
\end{figure}
\begin{figure}
\centerline{\includegraphics[width=\columnwidth]{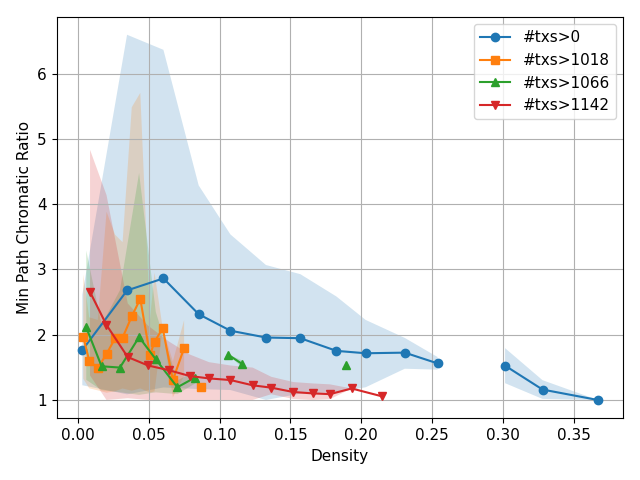}}
\caption{min-path-chromatic-ratio}
\label{fig:min-path-chromatic-ratio}
\end{figure}
\begin{figure}
\centerline{\includegraphics[width=\columnwidth]{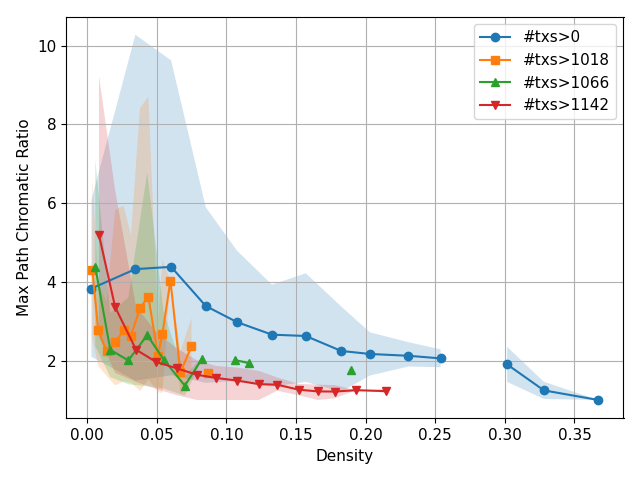}}
\caption{max-path-chromatic-ratio}
\label{fig:max-path-chromatic-ratio}
\end{figure}
\begin{figure}
\centerline{\includegraphics[width=\columnwidth]{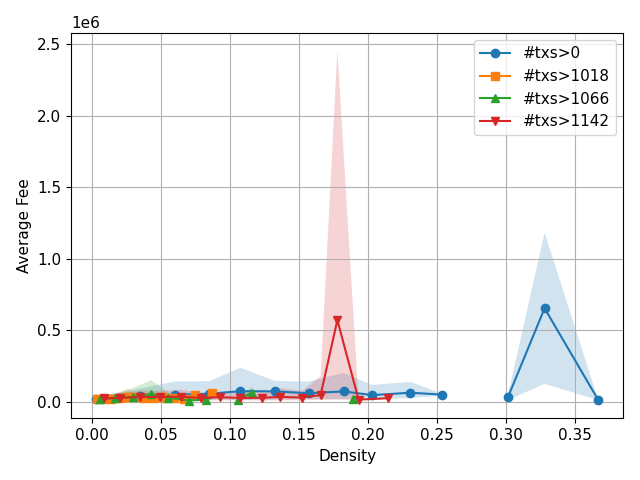}}
\caption{avg-fee}
\label{fig:avg-fee}
\end{figure}
\begin{figure}
\centerline{\includegraphics[width=\columnwidth]{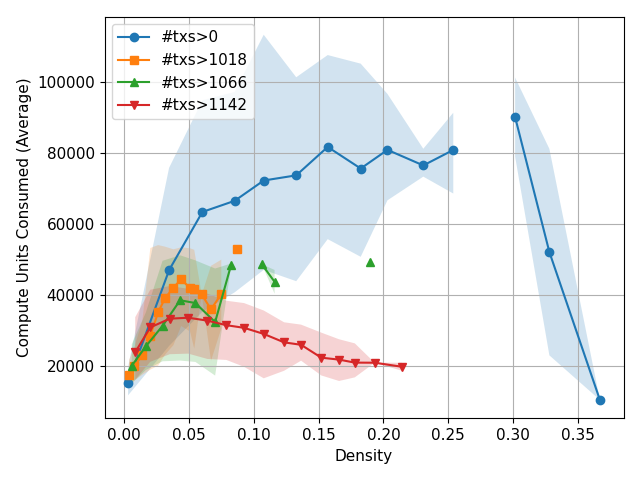}}
\caption{avg-computeUnitsConsumed}
\label{fig:avg-computeUnitsConsumed}
\end{figure}
\begin{figure}
\centerline{\includegraphics[width=\columnwidth]{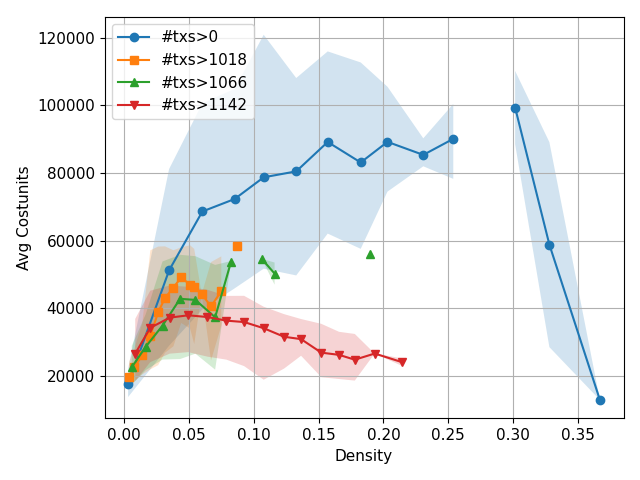}}
\caption{avg-costUnits}
\label{fig:avg-costUnits}
\end{figure}
\begin{figure}
\centerline{\includegraphics[width=\columnwidth]{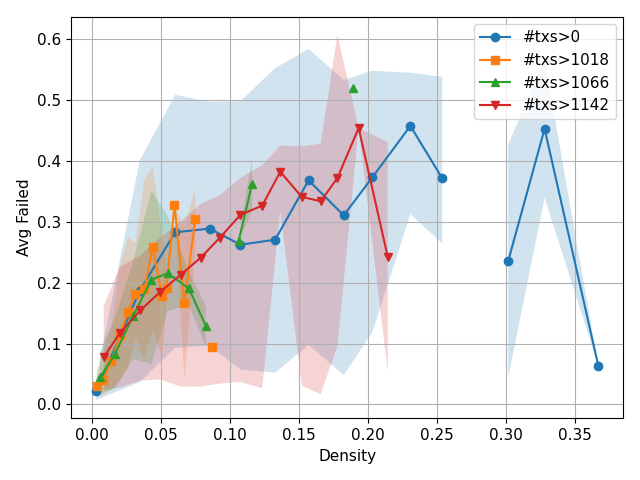}}
\caption{avg-failed}
\label{fig:avg-failed}
\end{figure}
\begin{figure}
\centerline{\includegraphics[width=\columnwidth]{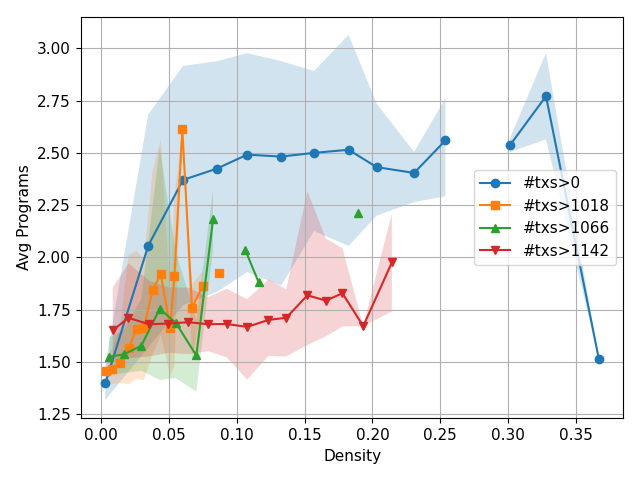}}
\caption{avg-programs}
\label{fig:avg-programs}
\end{figure}
\begin{figure}
\centerline{\includegraphics[width=\columnwidth]{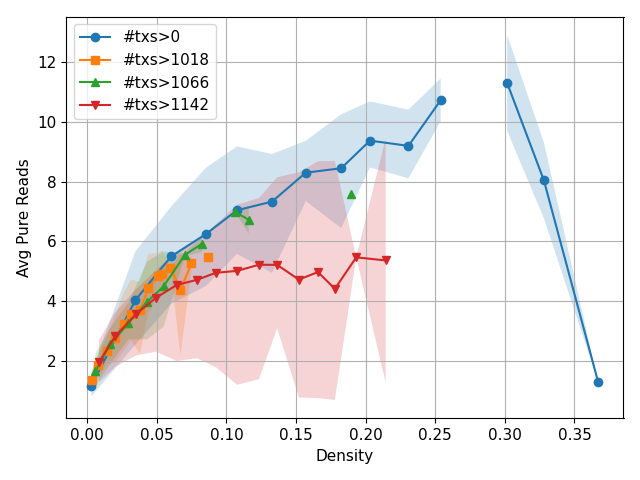}}
\caption{avg-pure-reads}
\label{fig:avg-pure-reads}
\end{figure}
\begin{figure}
\centerline{\includegraphics[width=\columnwidth]{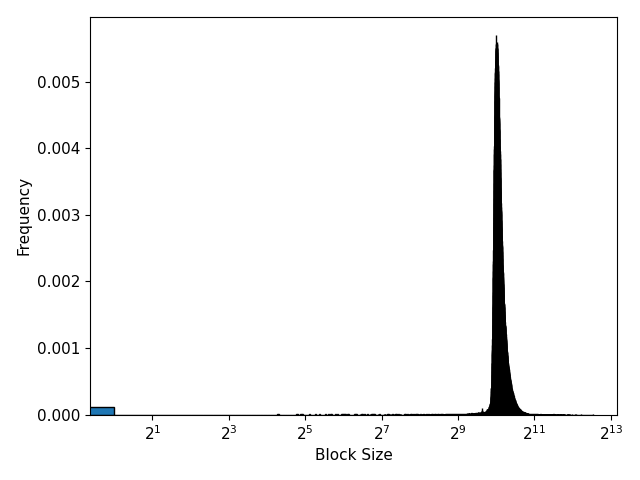}}
\caption{block-size-dist}
\label{fig:block-size-dist}
\end{figure}
\begin{figure}
\centerline{\includegraphics[width=\columnwidth]{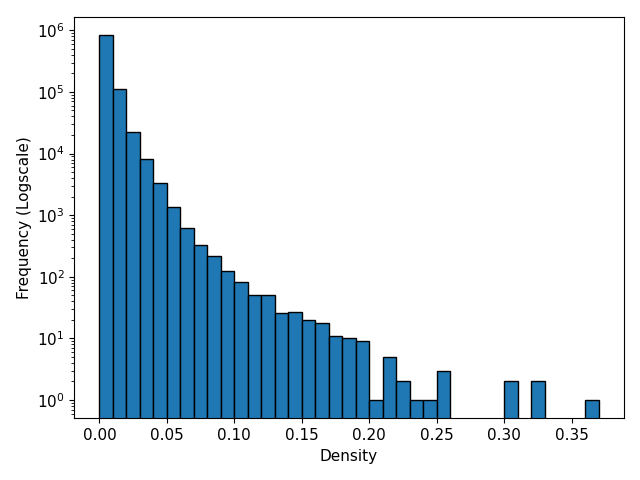}}
\caption{density-dist}
\label{fig:density-dist}
\end{figure}
\begin{figure}
\centerline{\includegraphics[width=\columnwidth]{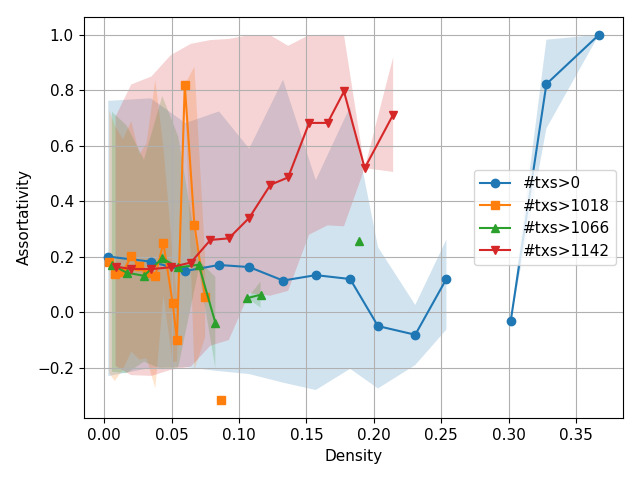}}
\caption{assortativity}
\label{fig:assortativity}
\end{figure}
\begin{figure}
\centerline{\includegraphics[width=\columnwidth]{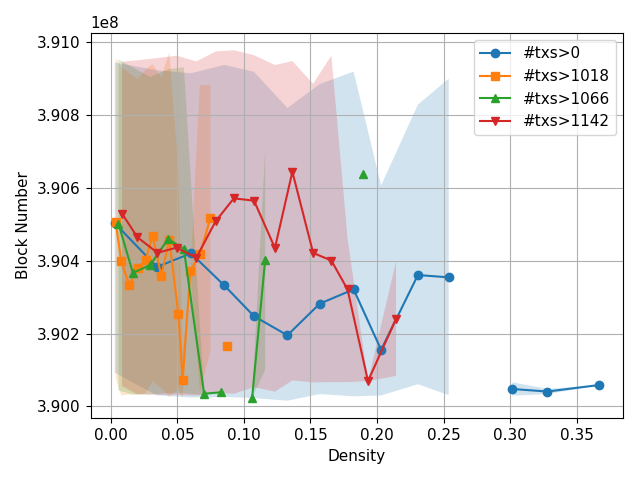}}
\caption{block-number}
\label{fig:block-number}
\end{figure}
\begin{figure}
\centerline{\includegraphics[width=\columnwidth]{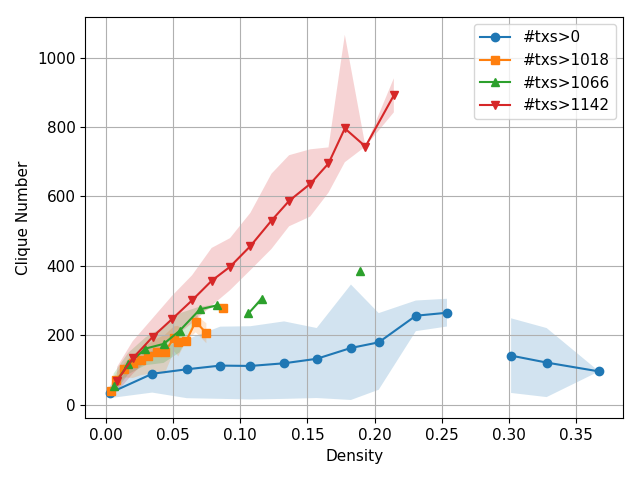}}
\caption{clique-number}
\label{fig:clique-number}
\end{figure}
\begin{figure}
\centerline{\includegraphics[width=\columnwidth]{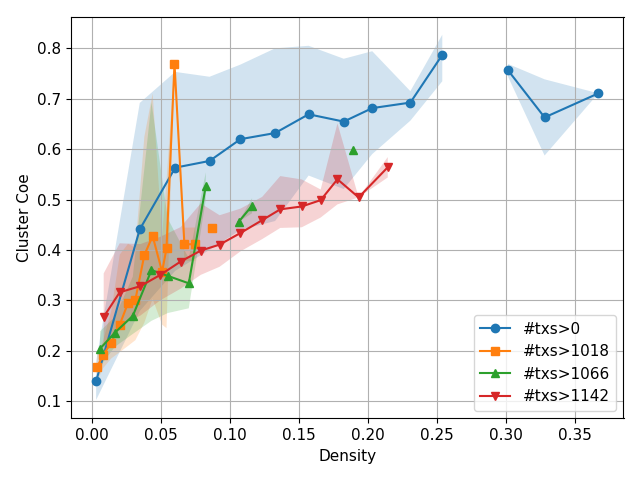}}
\caption{cluster-coe}
\label{fig:cluster-coe}
\end{figure}
\begin{figure}
\centerline{\includegraphics[width=\columnwidth]{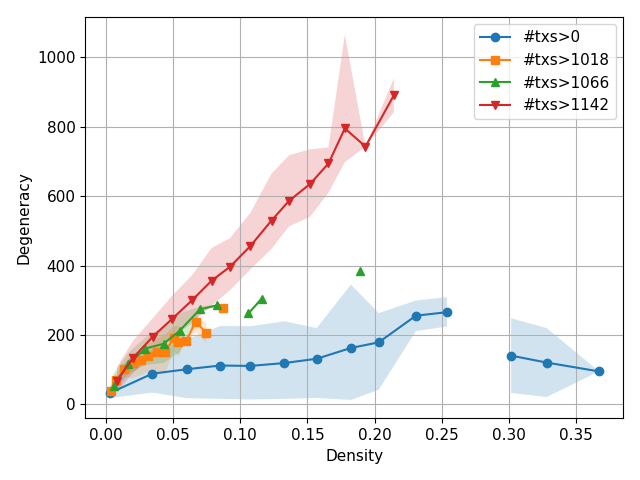}}
\caption{degeneracy}
\label{fig:degeneracy}
\end{figure}
\begin{figure}
\centerline{\includegraphics[width=\columnwidth]{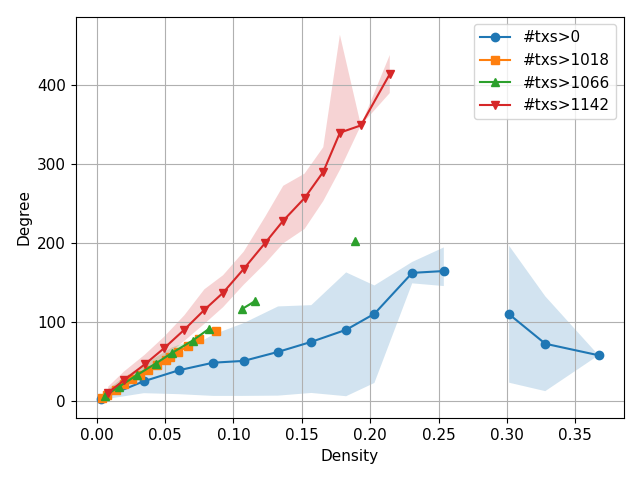}}
\caption{degree}
\label{fig:degree}
\end{figure}
\begin{figure}
\centerline{\includegraphics[width=\columnwidth]{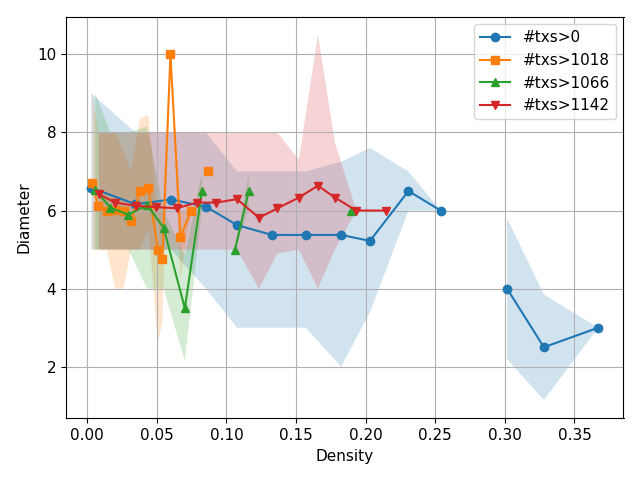}}
\caption{diameter}
\label{fig:diameter}
\end{figure}
\begin{figure}
\centerline{\includegraphics[width=\columnwidth]{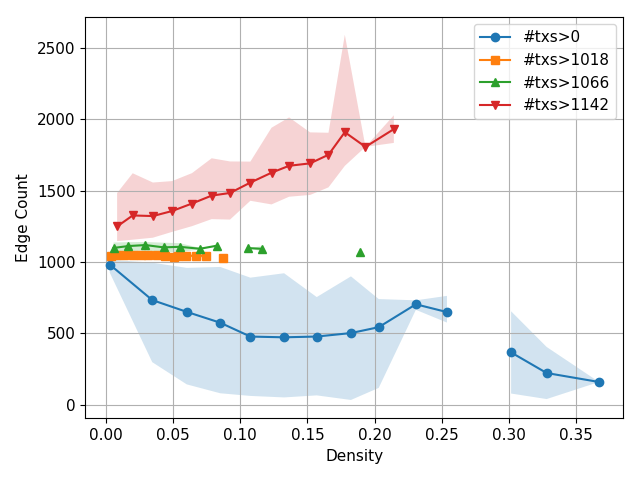}}
\caption{edge-count}
\label{fig:edge-count}
\end{figure}
\begin{figure}
\centerline{\includegraphics[width=\columnwidth]{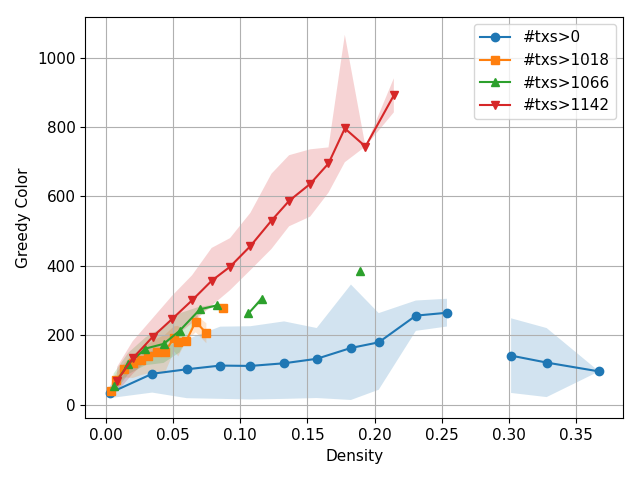}}
\caption{greedy-color}
\label{fig:greedy-color}
\end{figure}
\begin{figure}
\centerline{\includegraphics[width=\columnwidth]{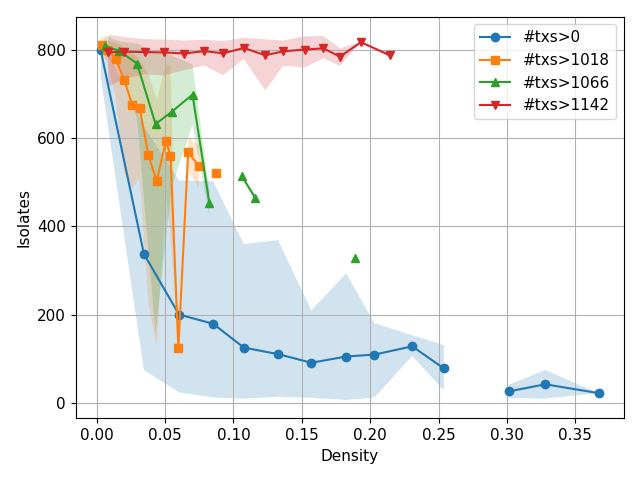}}
\caption{isolates}
\label{fig:isolates}
\end{figure}
\begin{figure}
\centerline{\includegraphics[width=\columnwidth]{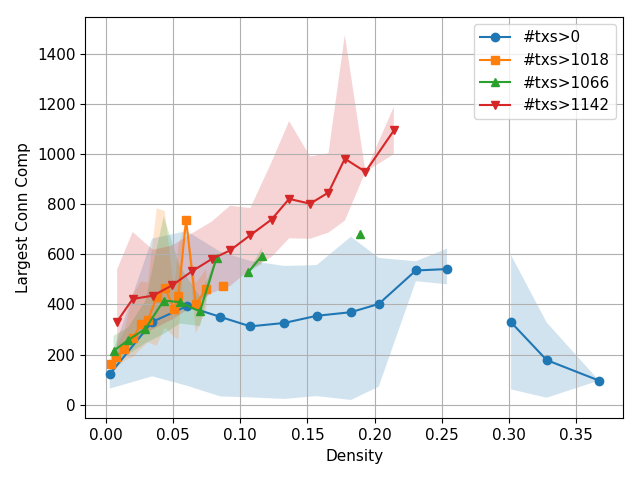}}
\caption{largest-conn-comp}
\label{fig:largest-conn-comp}
\end{figure}
\begin{figure}
\centerline{\includegraphics[width=\columnwidth]{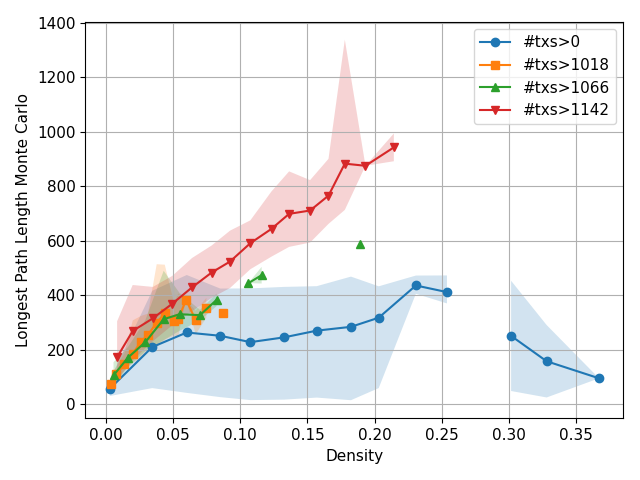}}
\caption{longest-path-length-monte-carlo}
\label{fig:longest-path-length-monte-carlo}
\end{figure}
\begin{figure}
\centerline{\includegraphics[width=\columnwidth]{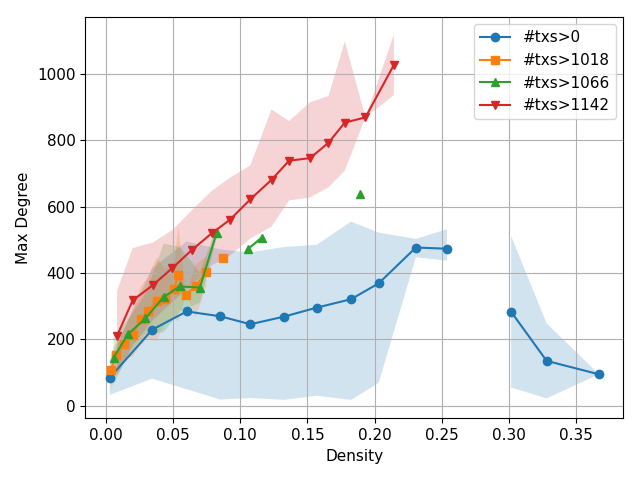}}
\caption{max-degree}
\label{fig:max-degree}
\end{figure}
\begin{figure}
\centerline{\includegraphics[width=\columnwidth]{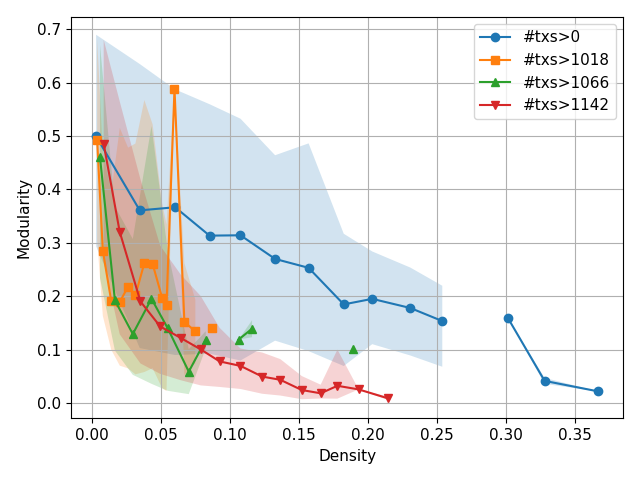}}
\caption{modularity}
\label{fig:modularity}
\end{figure}
\begin{figure}
\centerline{\includegraphics[width=\columnwidth]{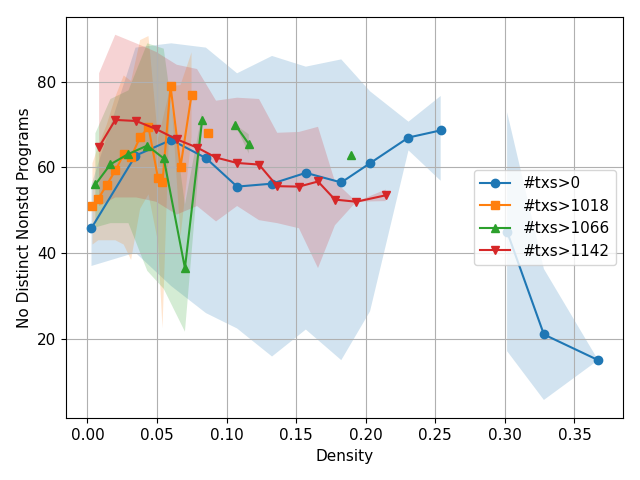}}
\caption{no-distinct-nonstd-programs}
\label{fig:no-distinct-nonstd-programs}
\end{figure}
\begin{figure}
\centerline{\includegraphics[width=\columnwidth]{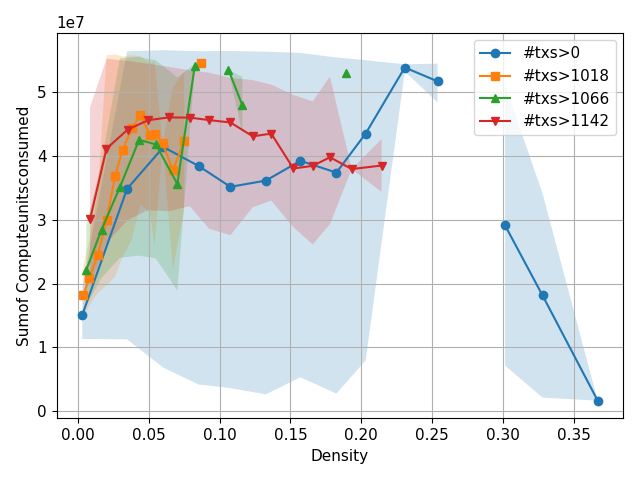}}
\caption{sumof-computeUnitsConsumed}
\label{fig:sumof-computeUnitsConsumed}
\end{figure}
\begin{figure}
\centerline{\includegraphics[width=\columnwidth]{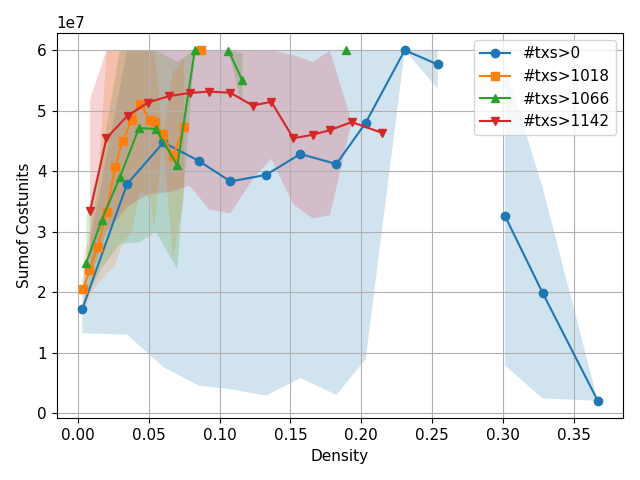}}
\caption{sumof-costUnits}
\label{fig:sumof-costUnits}
\end{figure}
\begin{figure}
\centerline{\includegraphics[width=\columnwidth]{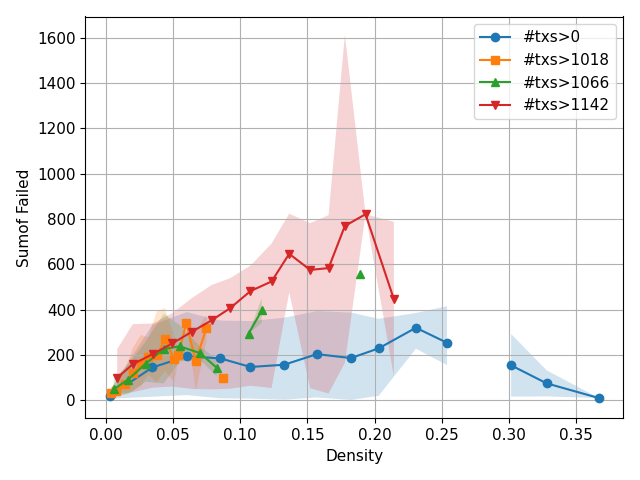}}
\caption{sumof-failed}
\label{fig:sumof-failed}
\end{figure}
\begin{figure}
\centerline{\includegraphics[width=\columnwidth]{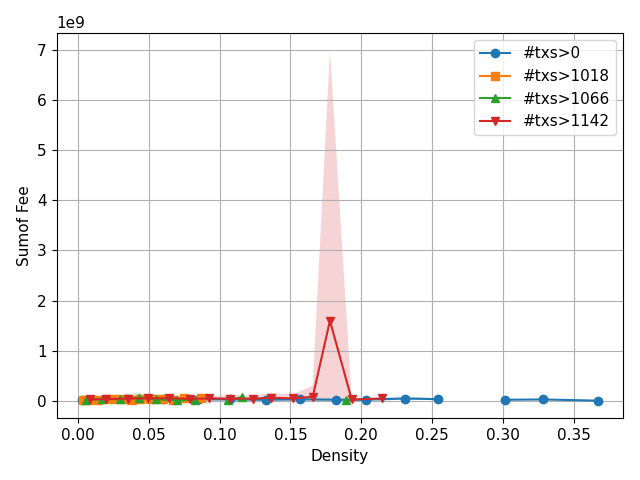}}
\caption{sumof-fee}
\label{fig:sumof-fee}
\end{figure}
\begin{figure}
\centerline{\includegraphics[width=\columnwidth]{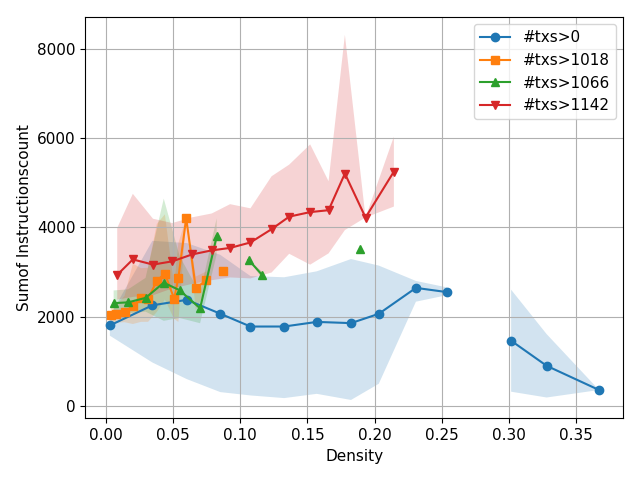}}
\caption{sumof-instructionsCount}
\label{fig:sumof-instructionsCount}
\end{figure}
\begin{figure}
\centerline{\includegraphics[width=\columnwidth]{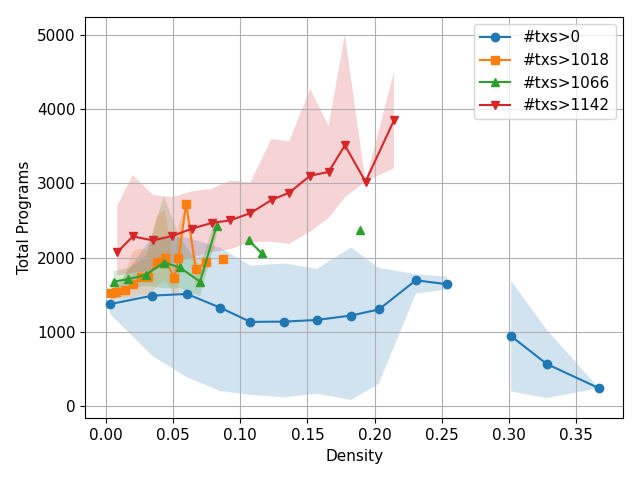}}
\caption{total-programs}
\label{fig:total-programs}
\end{figure}
\begin{figure}
\centerline{\includegraphics[width=\columnwidth]{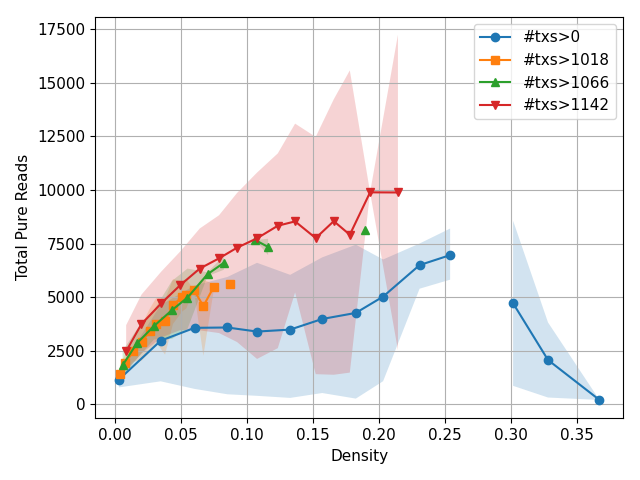}}
\caption{total-pure-reads}
\label{fig:total-pure-reads}
\end{figure}
\begin{figure}
\centerline{\includegraphics[width=\columnwidth]{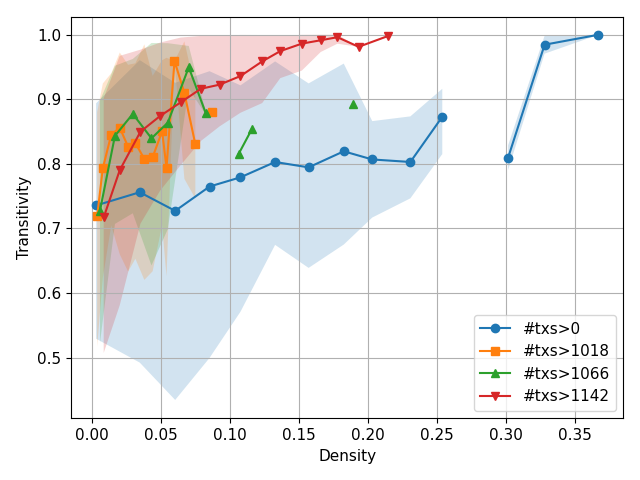}}
\caption{transitivity}
\label{fig:transitivity}
\end{figure}
\begin{figure}
\centerline{\includegraphics[width=\columnwidth]{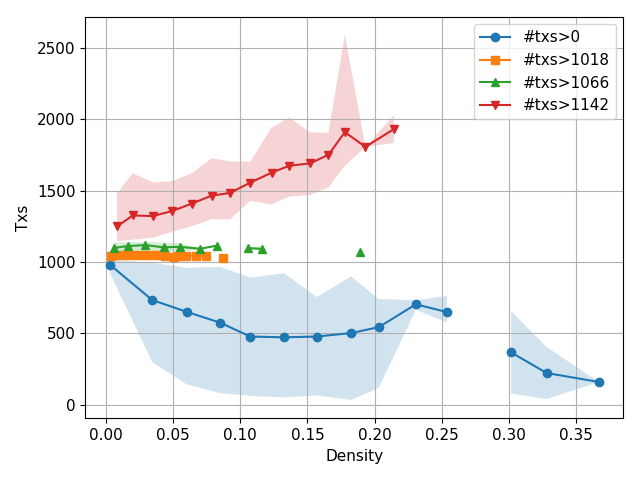}}
\caption{txs}
\label{fig:txs}
\end{figure}
\begin{figure}
\centerline{\includegraphics[width=\columnwidth]{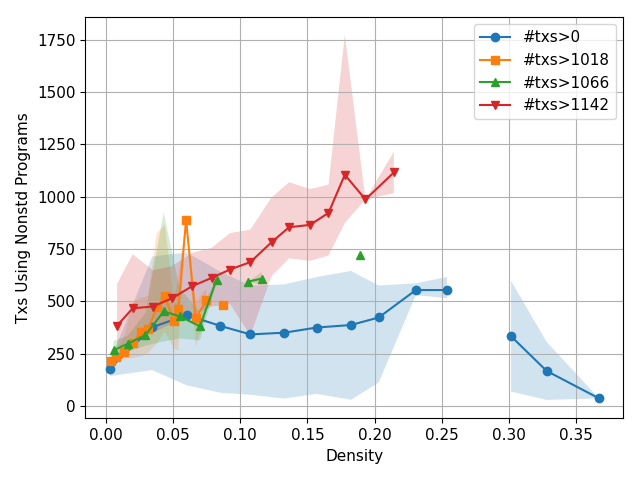}}
\caption{txs-using-nonstd-programs}
\label{fig:txs-using-nonstd-programs}
\end{figure}
\begin{figure}
\centerline{\includegraphics[width=\columnwidth]{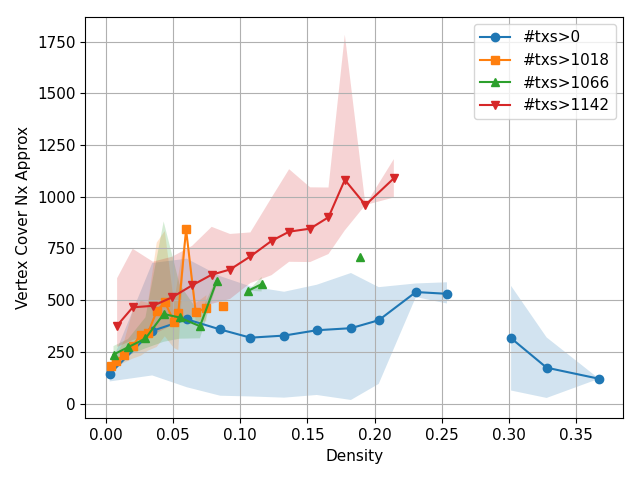}}
\caption{vertex-cover-nx-approx}
\label{fig:vertex-cover-nx-approx}
\end{figure}
\begin{figure}
\centerline{\includegraphics[width=\columnwidth]{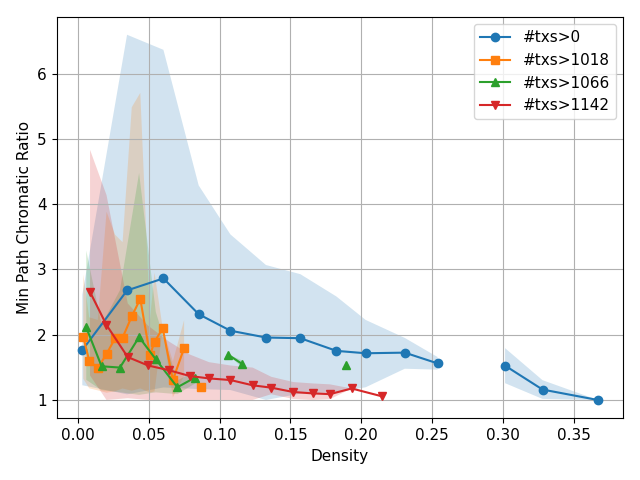}}
\caption{min-path-chromatic-ratio}
\label{fig:min-path-chromatic-ratio}
\end{figure}
\begin{figure}
\centerline{\includegraphics[width=\columnwidth]{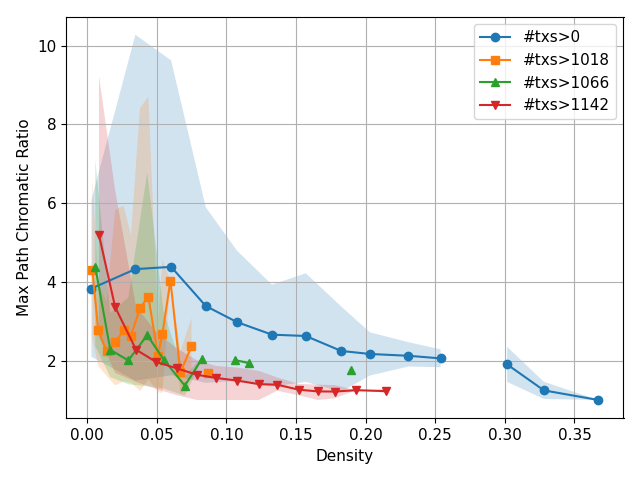}}
\caption{max-path-chromatic-ratio}
\label{fig:max-path-chromatic-ratio}
\end{figure}
\begin{figure}
\centerline{\includegraphics[width=\columnwidth]{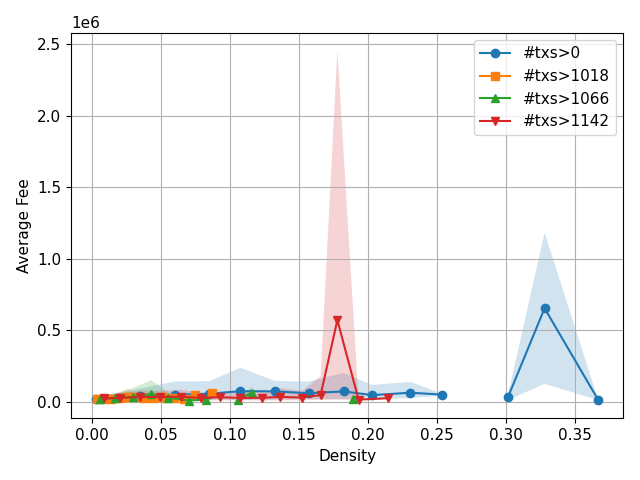}}
\caption{avg-fee}
\label{fig:avg-fee}
\end{figure}
\begin{figure}
\centerline{\includegraphics[width=\columnwidth]{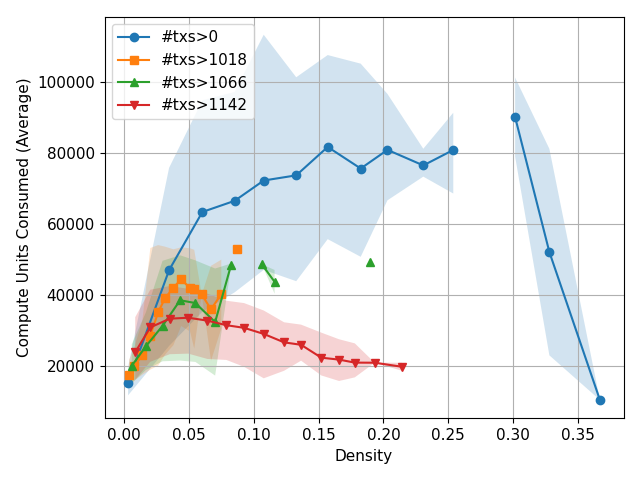}}
\caption{avg-computeUnitsConsumed}
\label{fig:avg-computeUnitsConsumed}
\end{figure}
\begin{figure}
\centerline{\includegraphics[width=\columnwidth]{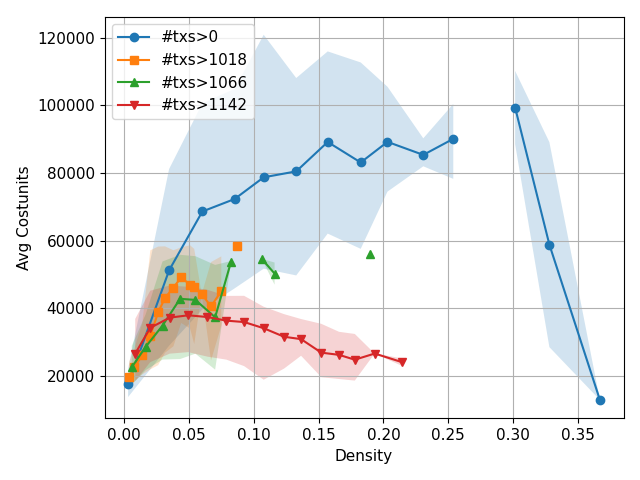}}
\caption{avg-costUnits}
\label{fig:avg-costUnits}
\end{figure}
\begin{figure}
\centerline{\includegraphics[width=\columnwidth]{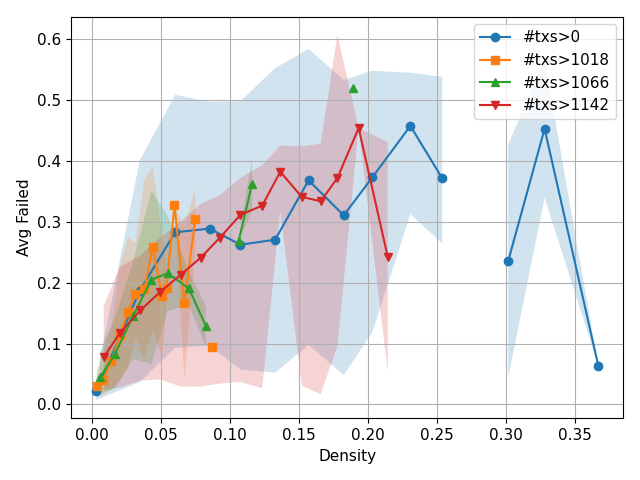}}
\caption{avg-failed}
\label{fig:avg-failed}
\end{figure}
\begin{figure}
\centerline{\includegraphics[width=\columnwidth]{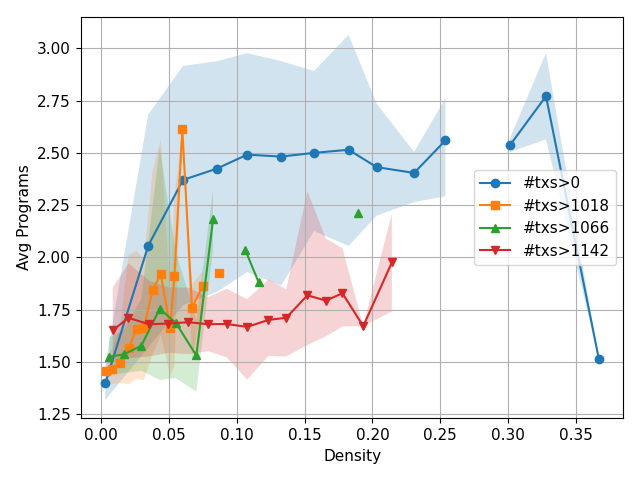}}
\caption{avg-programs}
\label{fig:avg-programs}
\end{figure}
\begin{figure}
\centerline{\includegraphics[width=\columnwidth]{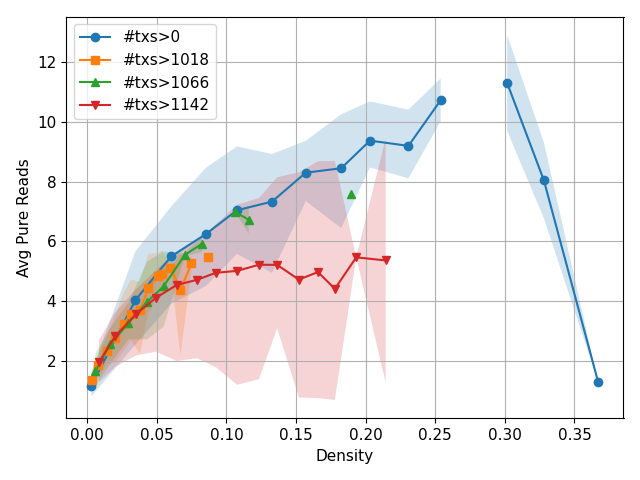}}
\caption{avg-pure-reads}
\label{fig:avg-pure-reads}
\end{figure}
\begin{figure}
\centerline{\includegraphics[width=\columnwidth]{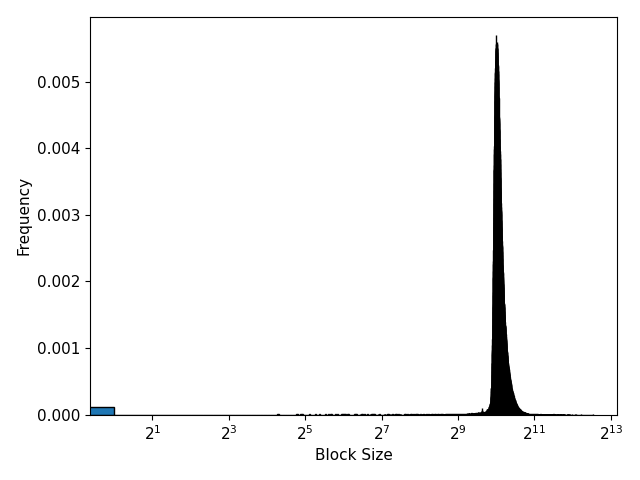}}
\caption{block-size-dist}
\label{fig:block-size-dist}
\end{figure}
\begin{figure}
\centerline{\includegraphics[width=\columnwidth]{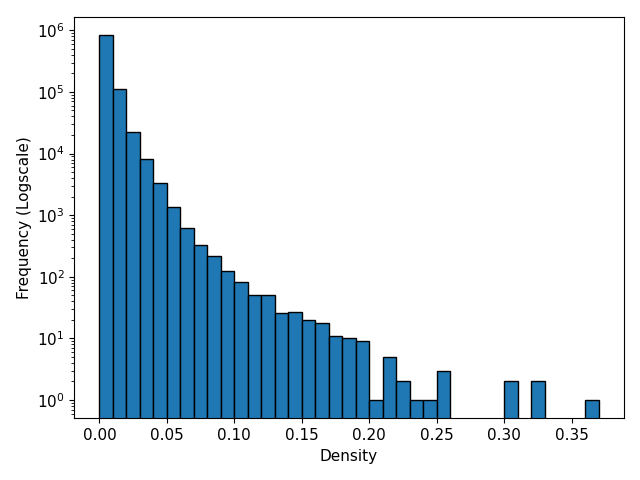}}
\caption{density-dist}
\label{fig:density-dist}
\end{figure}
\begin{figure}
\centerline{\includegraphics[width=\columnwidth]{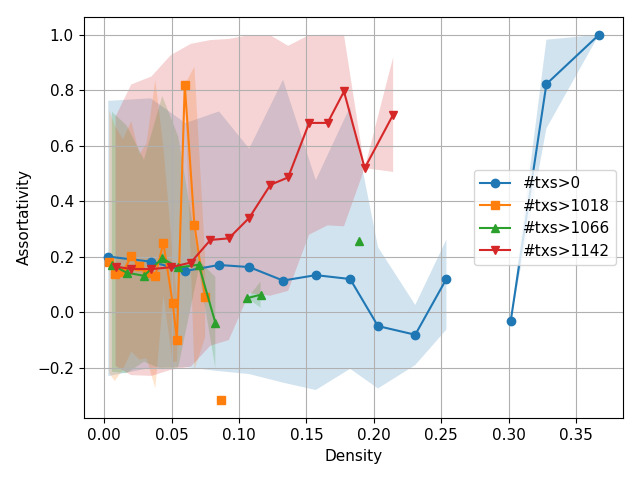}}
\caption{assortativity}
\label{fig:assortativity}
\end{figure}
\begin{figure}
\centerline{\includegraphics[width=\columnwidth]{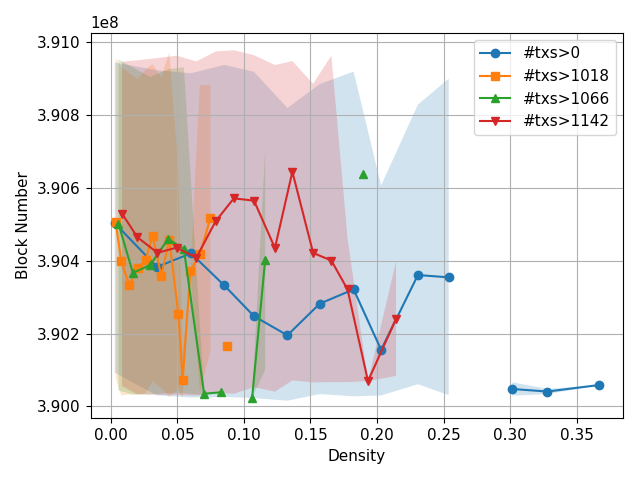}}
\caption{block-number}
\label{fig:block-number}
\end{figure}
\begin{figure}
\centerline{\includegraphics[width=\columnwidth]{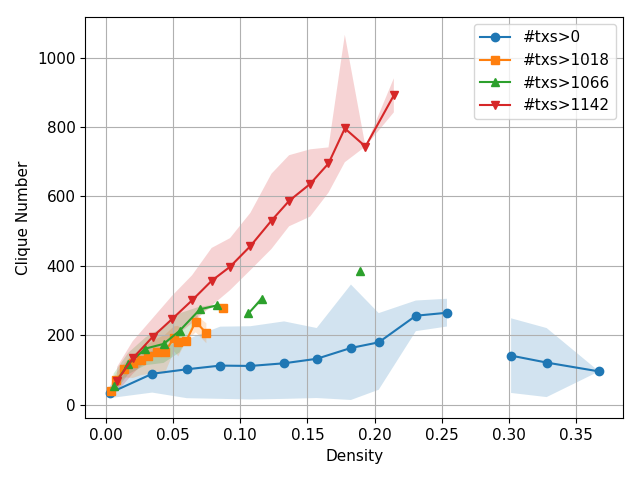}}
\caption{clique-number}
\label{fig:clique-number}
\end{figure}
\begin{figure}
\centerline{\includegraphics[width=\columnwidth]{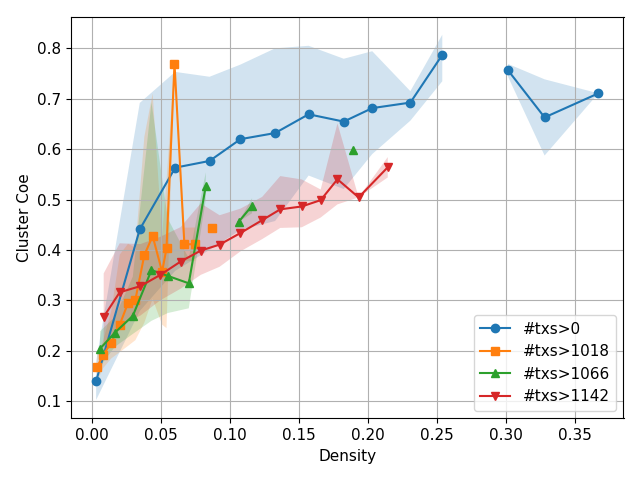}}
\caption{cluster-coe}
\label{fig:cluster-coe}
\end{figure}
\begin{figure}
\centerline{\includegraphics[width=\columnwidth]{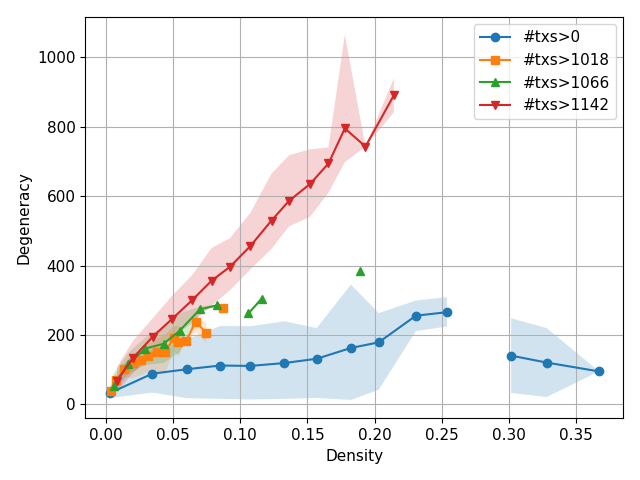}}
\caption{degeneracy}
\label{fig:degeneracy}
\end{figure}
\begin{figure}
\centerline{\includegraphics[width=\columnwidth]{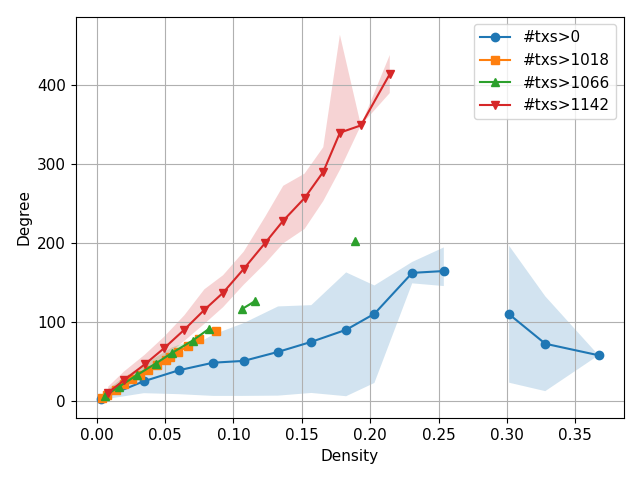}}
\caption{degree}
\label{fig:degree}
\end{figure}
\begin{figure}
\centerline{\includegraphics[width=\columnwidth]{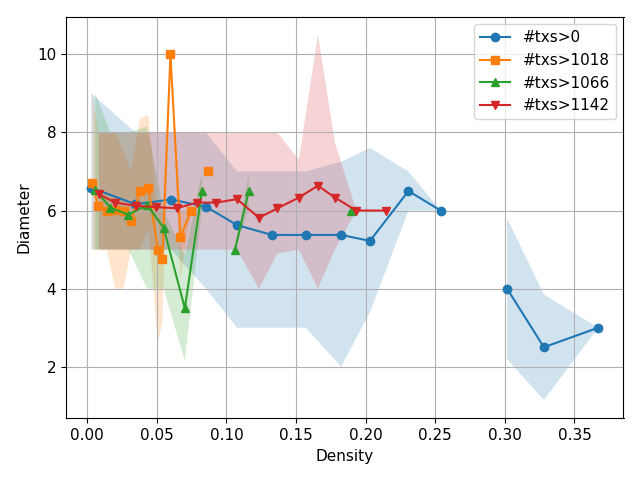}}
\caption{diameter}
\label{fig:diameter}
\end{figure}
\begin{figure}
\centerline{\includegraphics[width=\columnwidth]{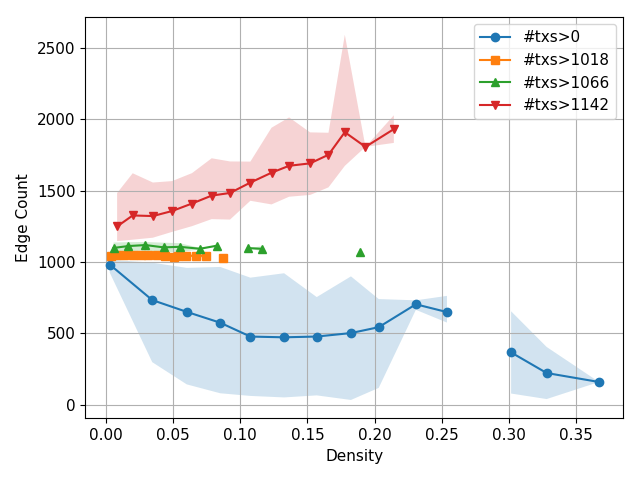}}
\caption{edge-count}
\label{fig:edge-count}
\end{figure}
\begin{figure}
\centerline{\includegraphics[width=\columnwidth]{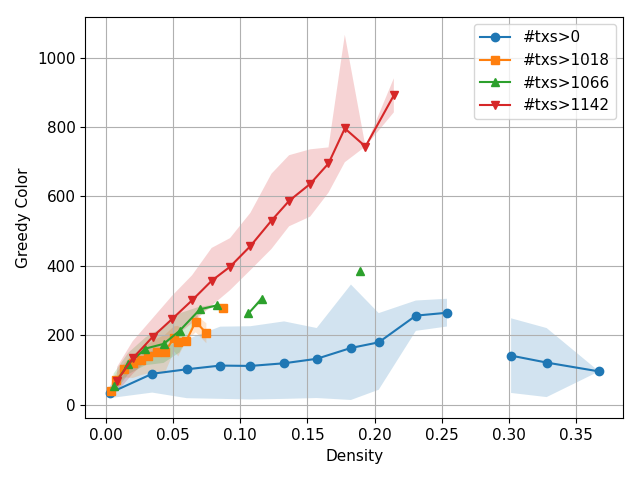}}
\caption{greedy-color}
\label{fig:greedy-color}
\end{figure}
\begin{figure}
\centerline{\includegraphics[width=\columnwidth]{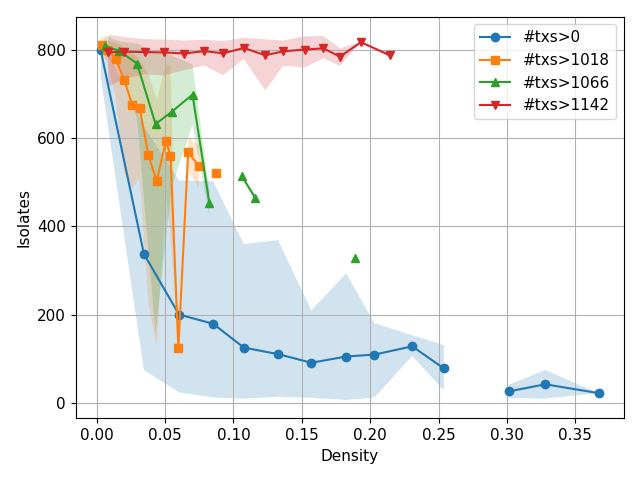}}
\caption{isolates}
\label{fig:isolates}
\end{figure}
\begin{figure}
\centerline{\includegraphics[width=\columnwidth]{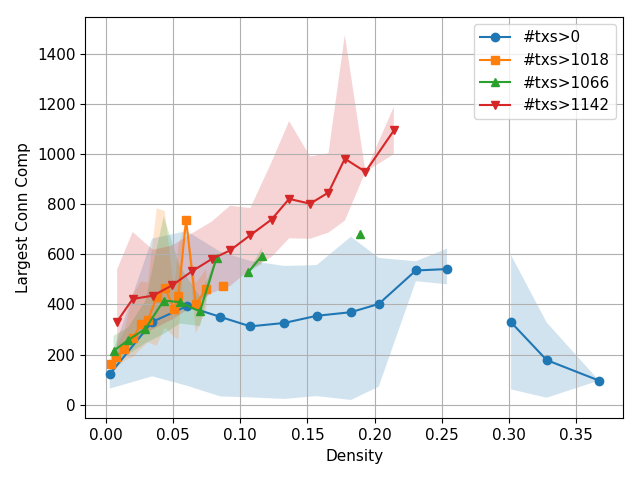}}
\caption{largest-conn-comp}
\label{fig:largest-conn-comp}
\end{figure}
\begin{figure}
\centerline{\includegraphics[width=\columnwidth]{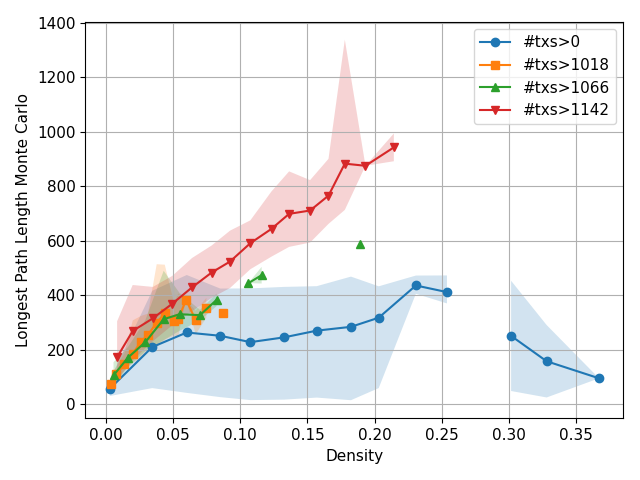}}
\caption{longest-path-length-monte-carlo}
\label{fig:longest-path-length-monte-carlo}
\end{figure}
\begin{figure}
\centerline{\includegraphics[width=\columnwidth]{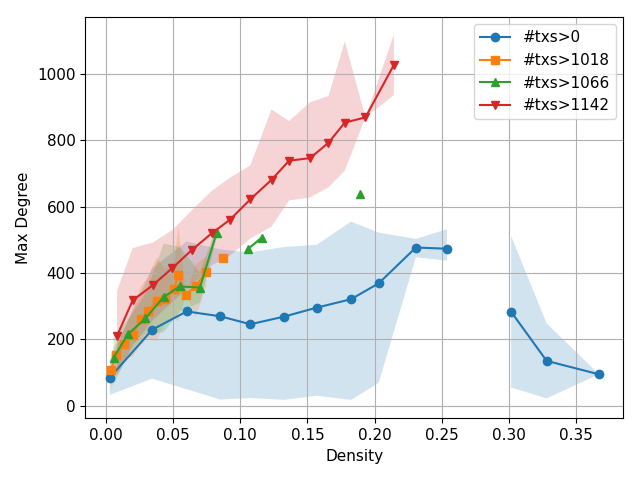}}
\caption{max-degree}
\label{fig:max-degree}
\end{figure}
\begin{figure}
\centerline{\includegraphics[width=\columnwidth]{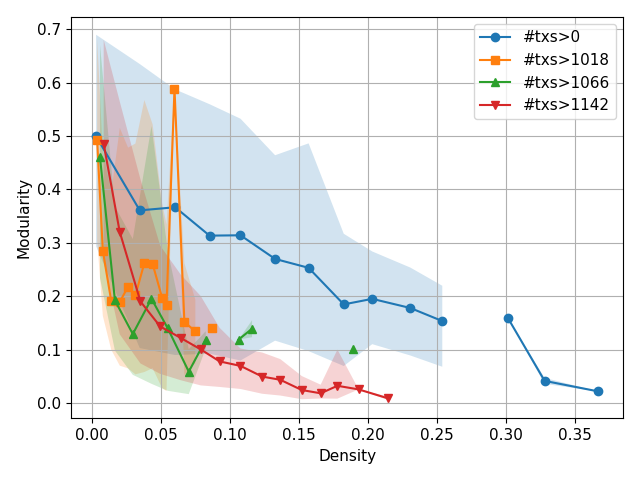}}
\caption{modularity}
\label{fig:modularity}
\end{figure}
\begin{figure}
\centerline{\includegraphics[width=\columnwidth]{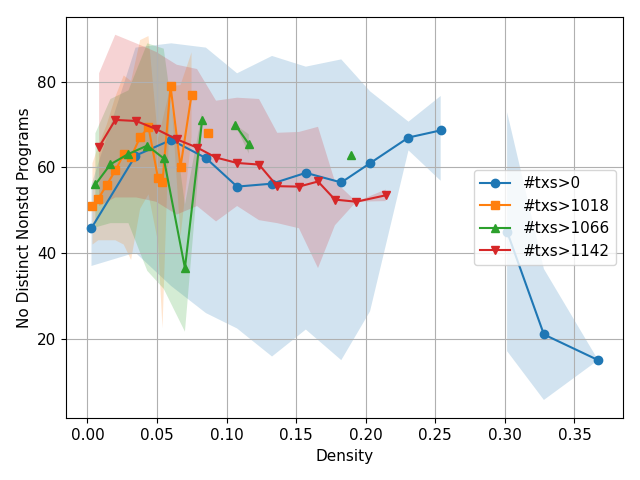}}
\caption{no-distinct-nonstd-programs}
\label{fig:no-distinct-nonstd-programs}
\end{figure}
\begin{figure}
\centerline{\includegraphics[width=\columnwidth]{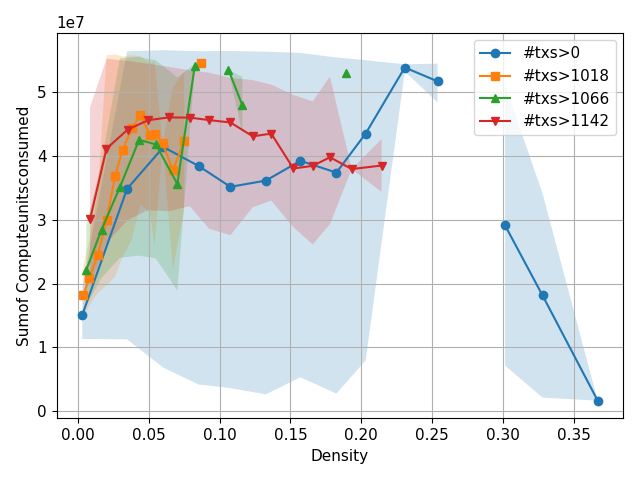}}
\caption{sumof-computeUnitsConsumed}
\label{fig:sumof-computeUnitsConsumed}
\end{figure}
\begin{figure}
\centerline{\includegraphics[width=\columnwidth]{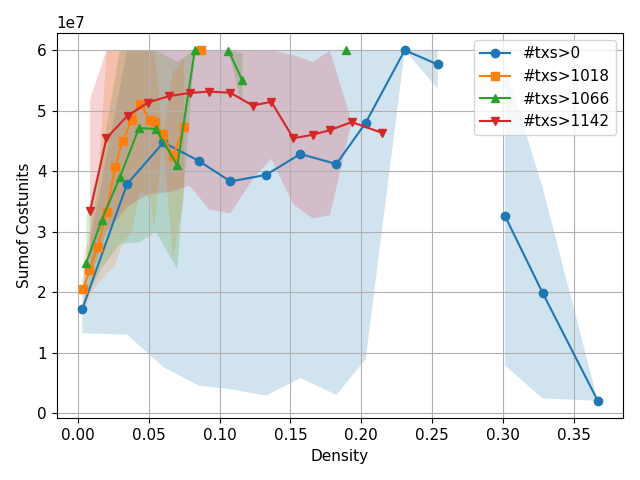}}
\caption{sumof-costUnits}
\label{fig:sumof-costUnits}
\end{figure}
\begin{figure}
\centerline{\includegraphics[width=\columnwidth]{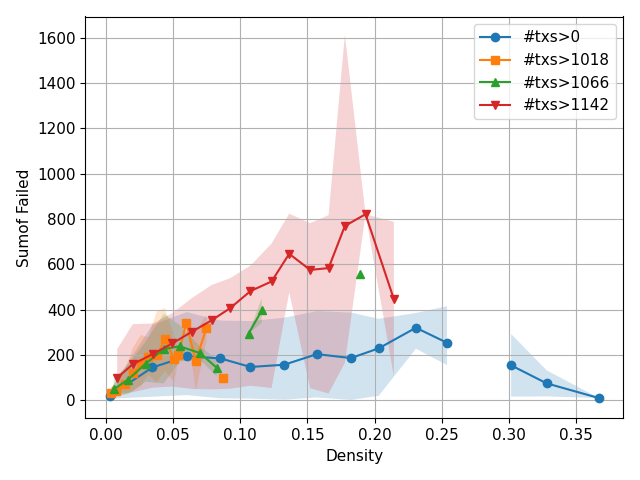}}
\caption{sumof-failed}
\label{fig:sumof-failed}
\end{figure}
\begin{figure}
\centerline{\includegraphics[width=\columnwidth]{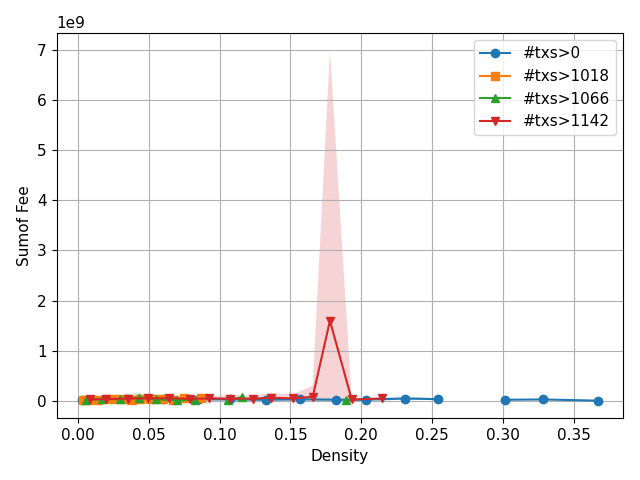}}
\caption{sumof-fee}
\label{fig:sumof-fee}
\end{figure}
\begin{figure}
\centerline{\includegraphics[width=\columnwidth]{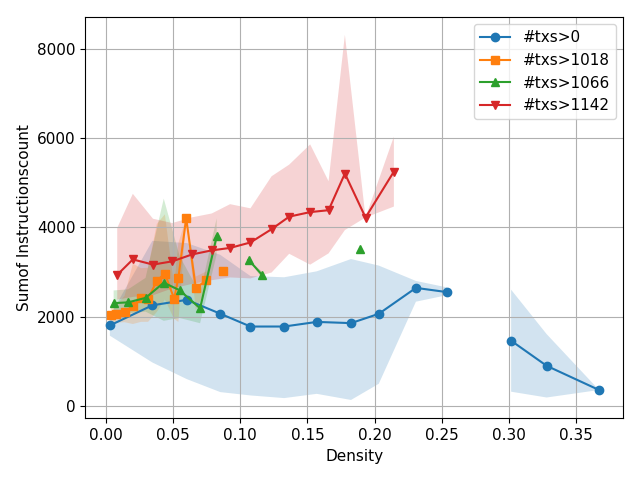}}
\caption{sumof-instructionsCount}
\label{fig:sumof-instructionsCount}
\end{figure}
\begin{figure}
\centerline{\includegraphics[width=\columnwidth]{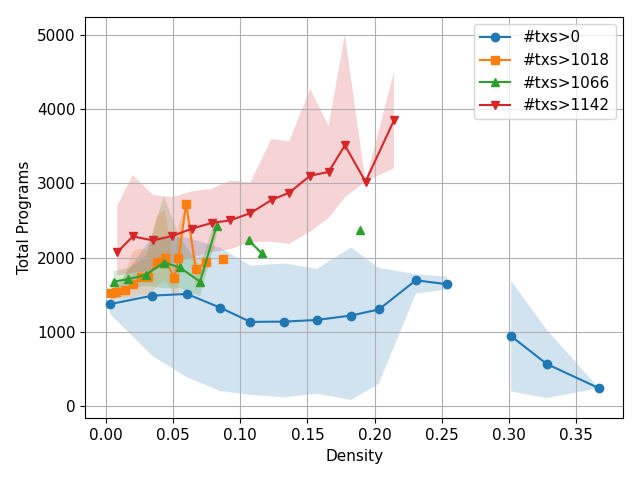}}
\caption{total-programs}
\label{fig:total-programs}
\end{figure}
\begin{figure}
\centerline{\includegraphics[width=\columnwidth]{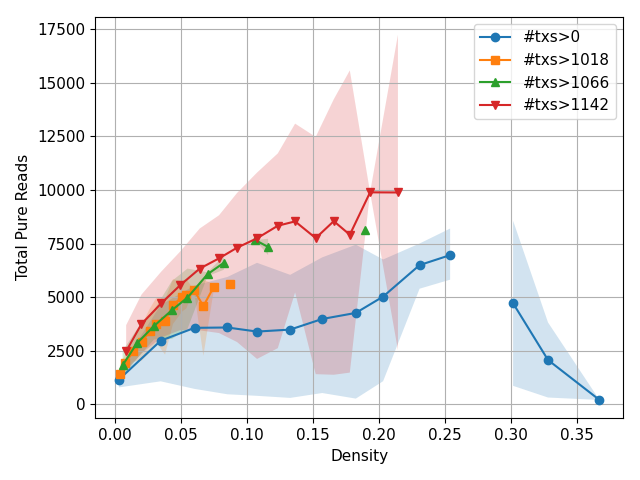}}
\caption{total-pure-reads}
\label{fig:total-pure-reads}
\end{figure}
\begin{figure}
\centerline{\includegraphics[width=\columnwidth]{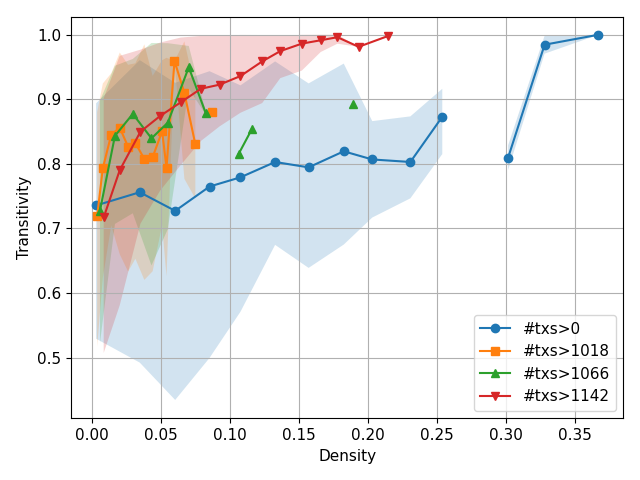}}
\caption{transitivity}
\label{fig:transitivity}
\end{figure}
\begin{figure}
\centerline{\includegraphics[width=\columnwidth]{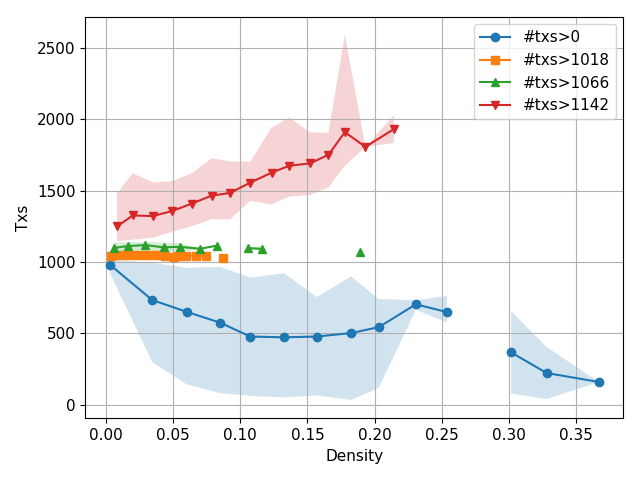}}
\caption{txs}
\label{fig:txs}
\end{figure}
\begin{figure}
\centerline{\includegraphics[width=\columnwidth]{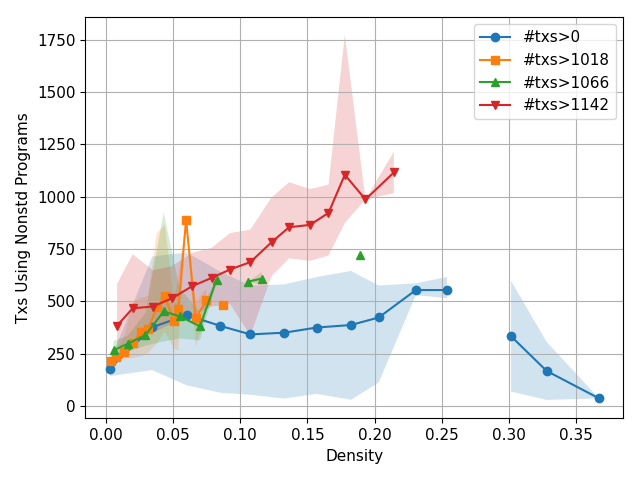}}
\caption{txs-using-nonstd-programs}
\label{fig:txs-using-nonstd-programs}
\end{figure}
\begin{figure}
\centerline{\includegraphics[width=\columnwidth]{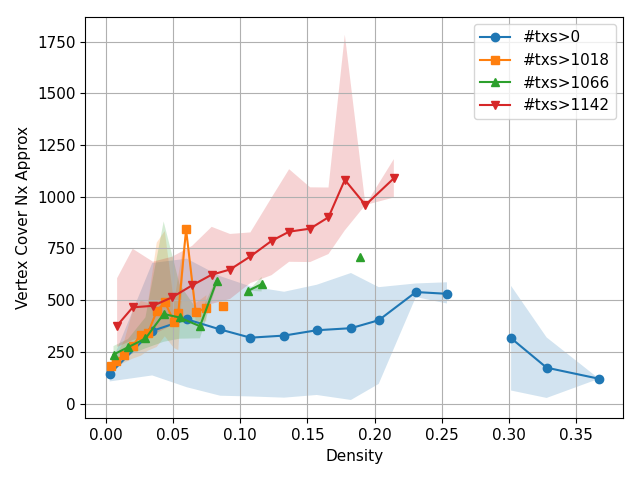}}
\caption{vertex-cover-nx-approx}
\label{fig:vertex-cover-nx-approx}
\end{figure}
\begin{figure}
\centerline{\includegraphics[width=\columnwidth]{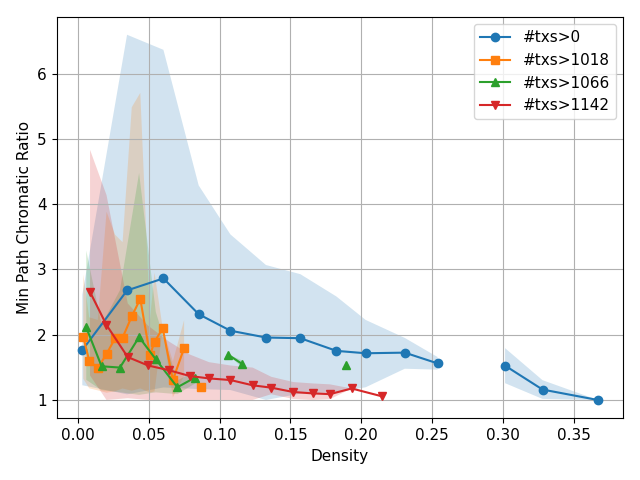}}
\caption{min-path-chromatic-ratio}
\label{fig:min-path-chromatic-ratio}
\end{figure}
\begin{figure}
\centerline{\includegraphics[width=\columnwidth]{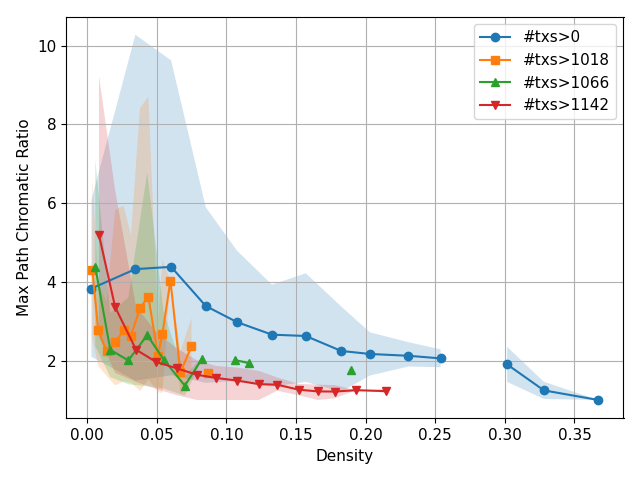}}
\caption{max-path-chromatic-ratio}
\label{fig:max-path-chromatic-ratio}
\end{figure}
\begin{figure}
\centerline{\includegraphics[width=\columnwidth]{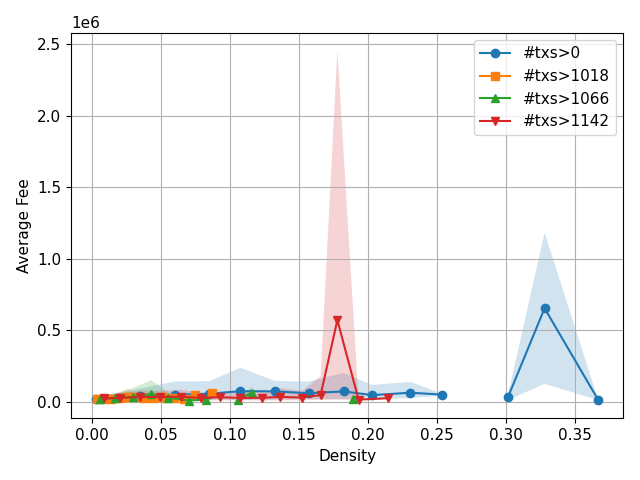}}
\caption{avg-fee}
\label{fig:avg-fee}
\end{figure}
\begin{figure}
\centerline{\includegraphics[width=\columnwidth]{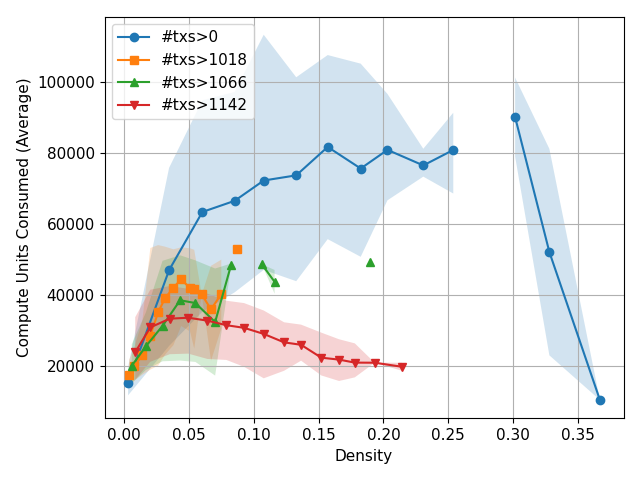}}
\caption{avg-computeUnitsConsumed}
\label{fig:avg-computeUnitsConsumed}
\end{figure}
\begin{figure}
\centerline{\includegraphics[width=\columnwidth]{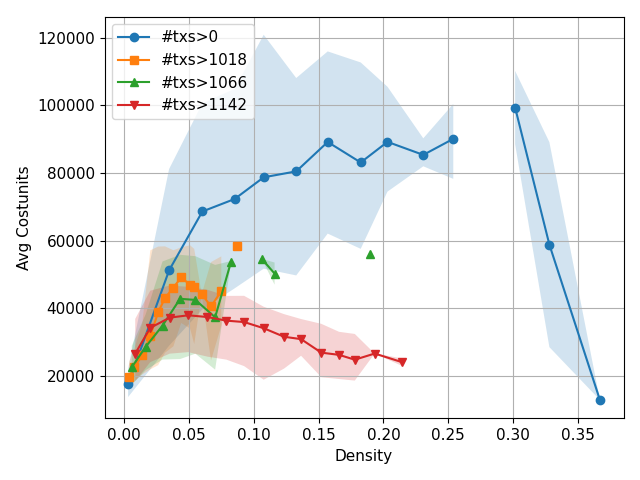}}
\caption{avg-costUnits}
\label{fig:avg-costUnits}
\end{figure}
\begin{figure}
\centerline{\includegraphics[width=\columnwidth]{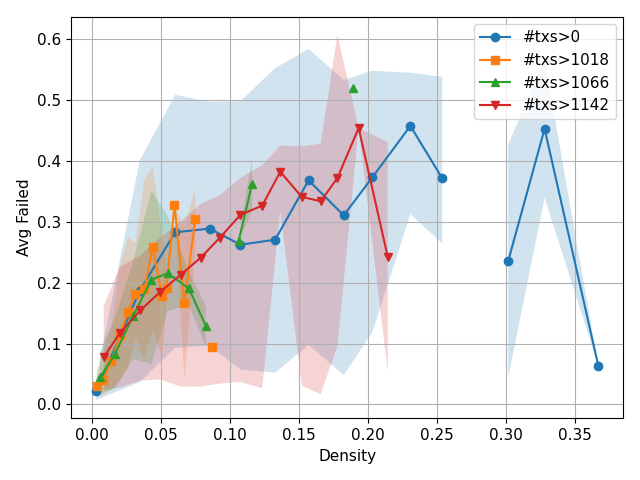}}
\caption{avg-failed}
\label{fig:avg-failed}
\end{figure}
\begin{figure}
\centerline{\includegraphics[width=\columnwidth]{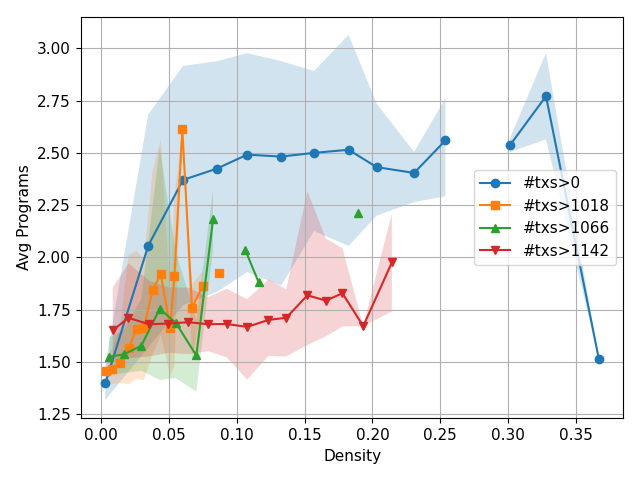}}
\caption{avg-programs}
\label{fig:avg-programs}
\end{figure}
\begin{figure}
\centerline{\includegraphics[width=\columnwidth]{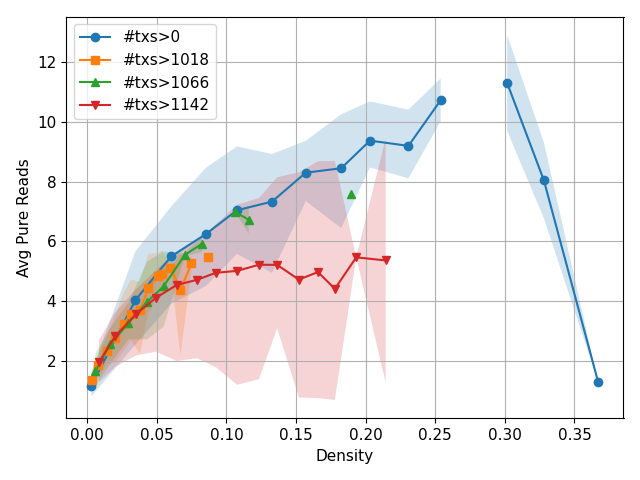}}
\caption{avg-pure-reads}
\label{fig:avg-pure-reads}
\end{figure}
\begin{figure}
\centerline{\includegraphics[width=\columnwidth]{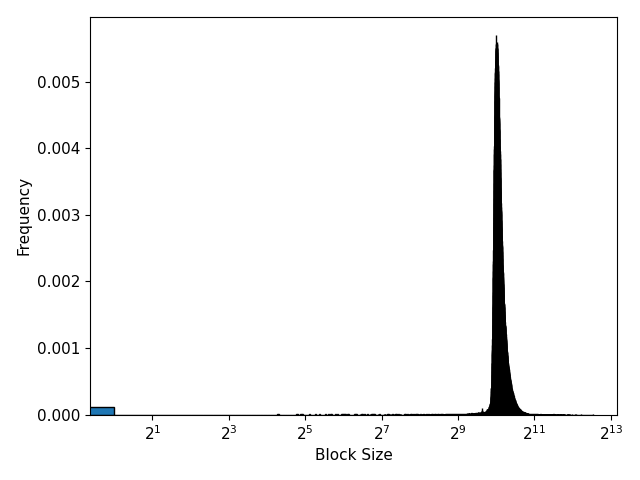}}
\caption{block-size-dist}
\label{fig:block-size-dist}
\end{figure}
\begin{figure}
\centerline{\includegraphics[width=\columnwidth]{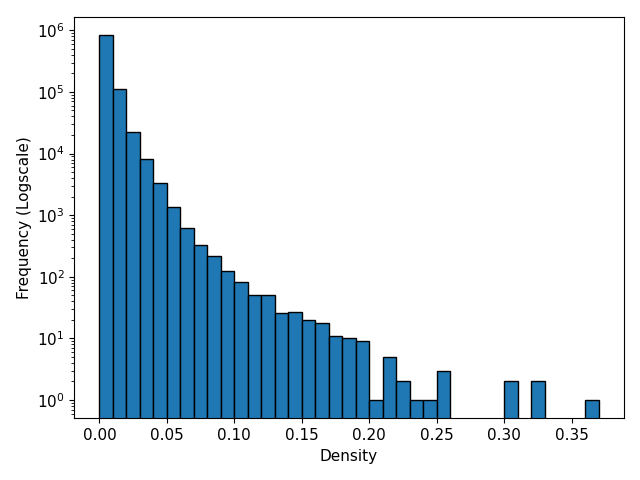}}
\caption{density-dist}
\label{fig:density-dist}
\end{figure}
\begin{figure}
\centerline{\includegraphics[width=\columnwidth]{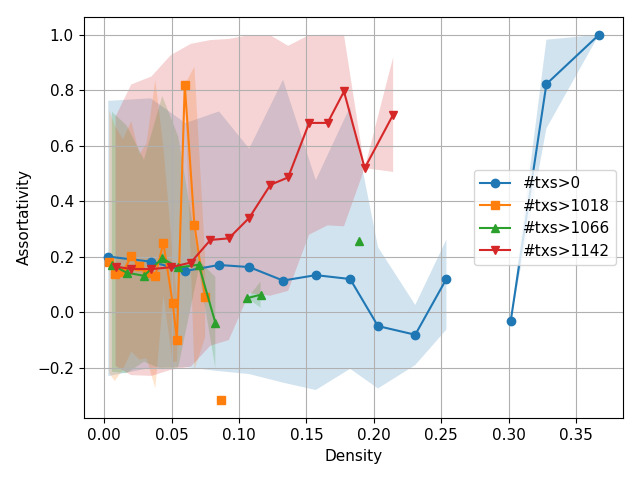}}
\caption{assortativity}
\label{fig:assortativity}
\end{figure}
\begin{figure}
\centerline{\includegraphics[width=\columnwidth]{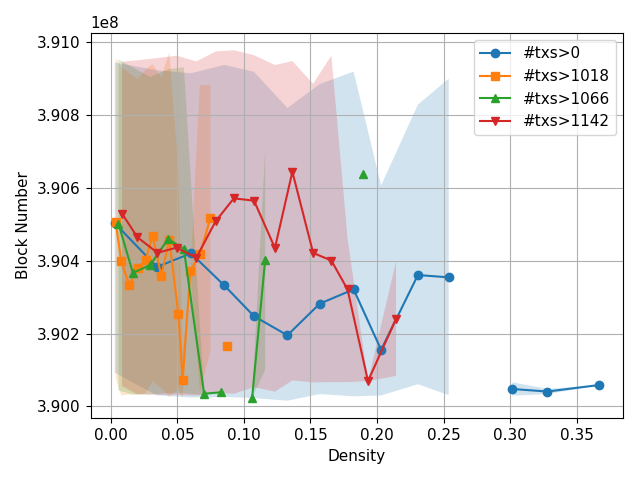}}
\caption{block-number}
\label{fig:block-number}
\end{figure}
\begin{figure}
\centerline{\includegraphics[width=\columnwidth]{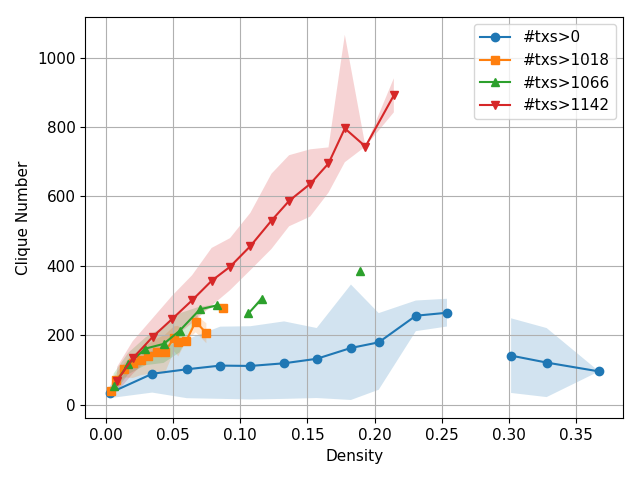}}
\caption{clique-number}
\label{fig:clique-number}
\end{figure}
\begin{figure}
\centerline{\includegraphics[width=\columnwidth]{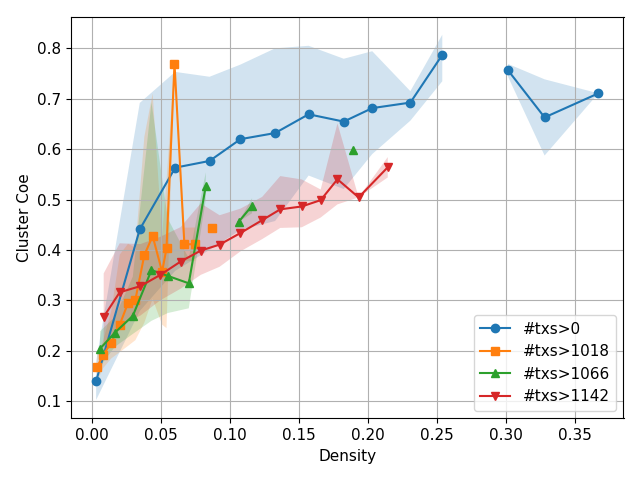}}
\caption{cluster-coe}
\label{fig:cluster-coe}
\end{figure}
\begin{figure}
\centerline{\includegraphics[width=\columnwidth]{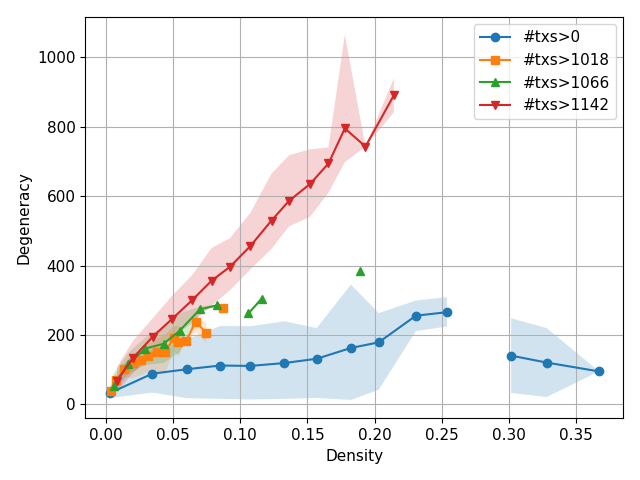}}
\caption{degeneracy}
\label{fig:degeneracy}
\end{figure}
\begin{figure}
\centerline{\includegraphics[width=\columnwidth]{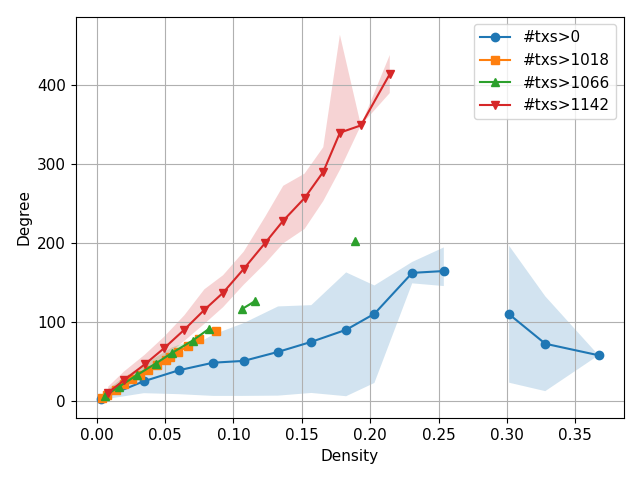}}
\caption{degree}
\label{fig:degree}
\end{figure}
\begin{figure}
\centerline{\includegraphics[width=\columnwidth]{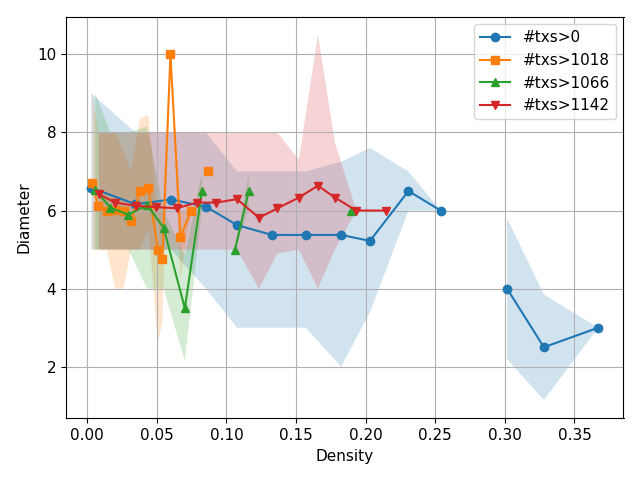}}
\caption{diameter}
\label{fig:diameter}
\end{figure}
\begin{figure}
\centerline{\includegraphics[width=\columnwidth]{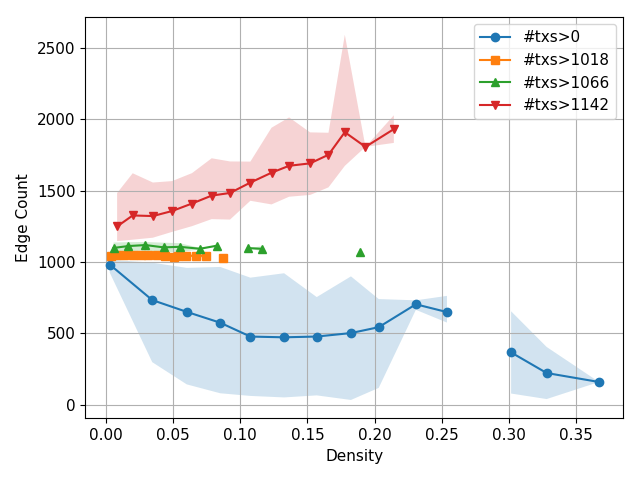}}
\caption{edge-count}
\label{fig:edge-count}
\end{figure}
\begin{figure}
\centerline{\includegraphics[width=\columnwidth]{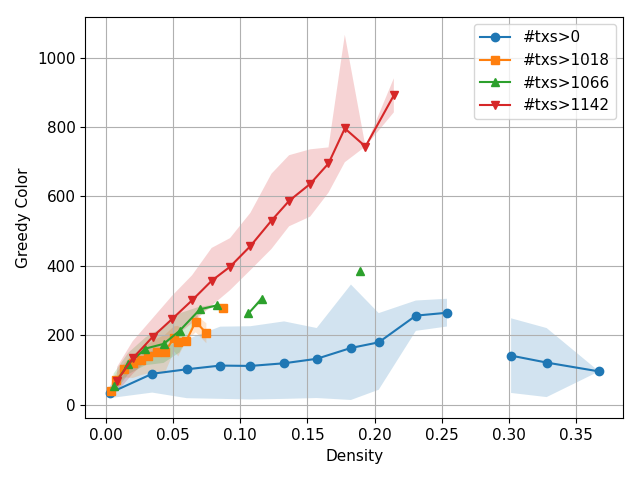}}
\caption{greedy-color}
\label{fig:greedy-color}
\end{figure}
\begin{figure}
\centerline{\includegraphics[width=\columnwidth]{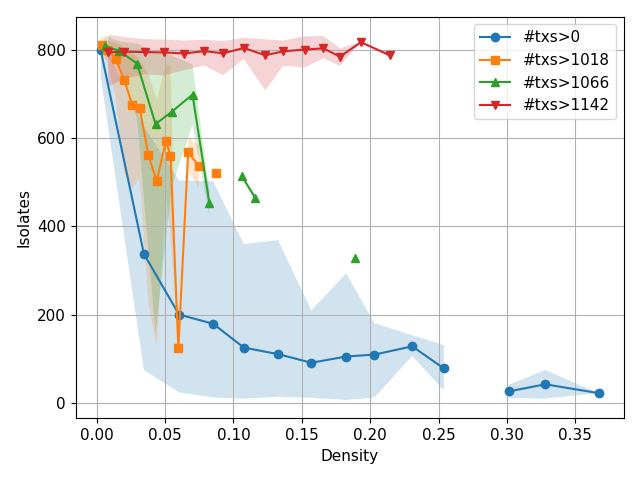}}
\caption{isolates}
\label{fig:isolates}
\end{figure}
\begin{figure}
\centerline{\includegraphics[width=\columnwidth]{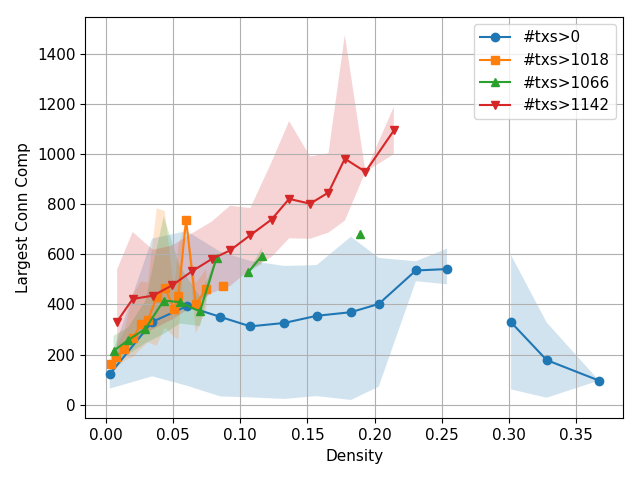}}
\caption{largest-conn-comp}
\label{fig:largest-conn-comp}
\end{figure}
\begin{figure}
\centerline{\includegraphics[width=\columnwidth]{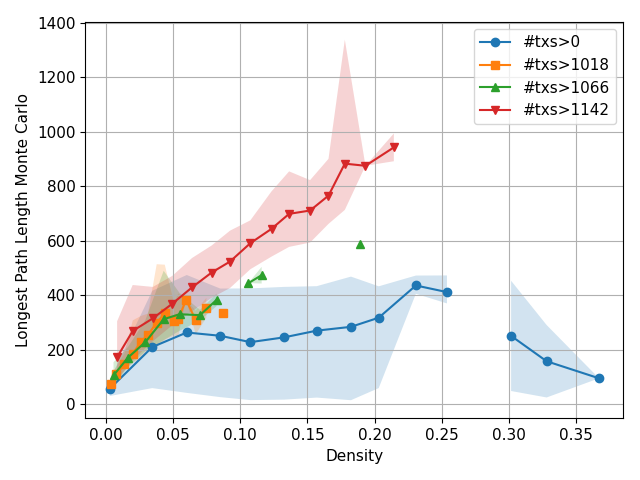}}
\caption{longest-path-length-monte-carlo}
\label{fig:longest-path-length-monte-carlo}
\end{figure}
\begin{figure}
\centerline{\includegraphics[width=\columnwidth]{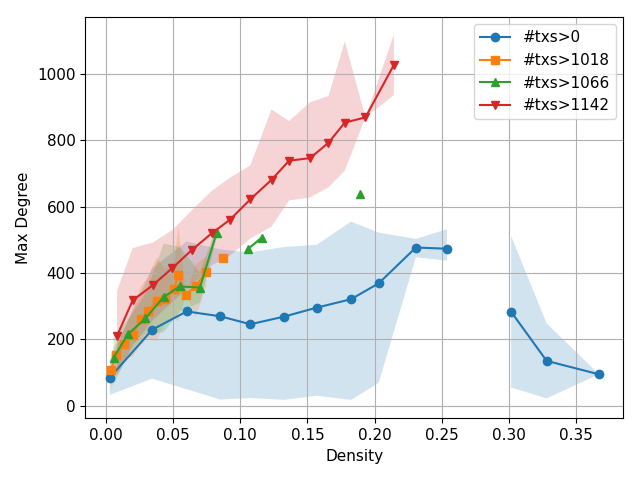}}
\caption{max-degree}
\label{fig:max-degree}
\end{figure}
\begin{figure}
\centerline{\includegraphics[width=\columnwidth]{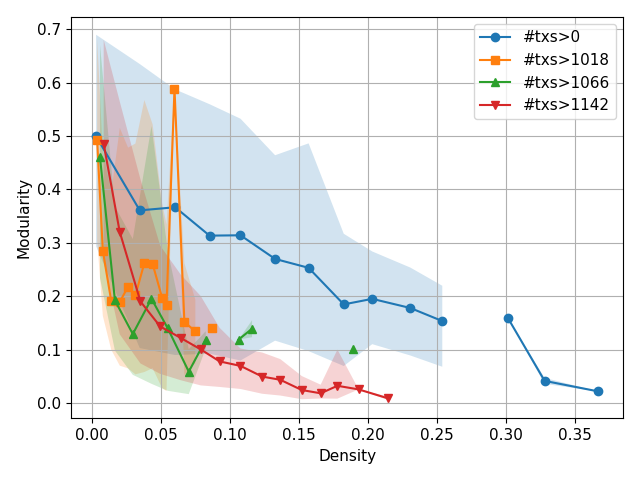}}
\caption{modularity}
\label{fig:modularity}
\end{figure}
\begin{figure}
\centerline{\includegraphics[width=\columnwidth]{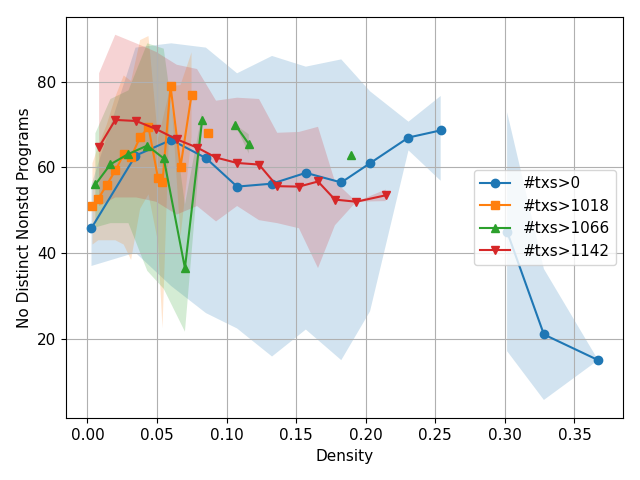}}
\caption{no-distinct-nonstd-programs}
\label{fig:no-distinct-nonstd-programs}
\end{figure}
\begin{figure}
\centerline{\includegraphics[width=\columnwidth]{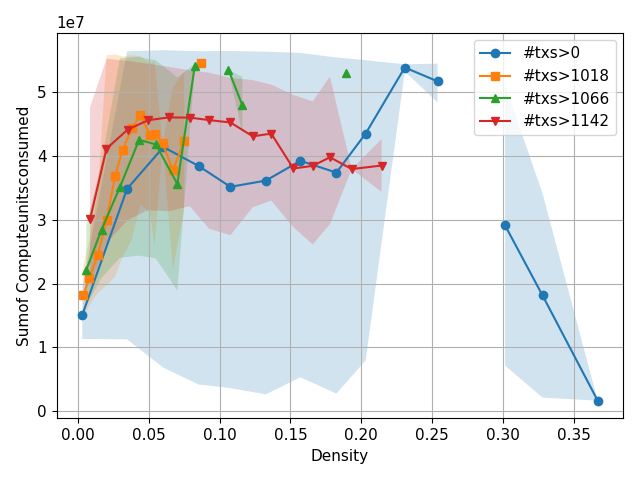}}
\caption{sumof-computeUnitsConsumed}
\label{fig:sumof-computeUnitsConsumed}
\end{figure}
\begin{figure}
\centerline{\includegraphics[width=\columnwidth]{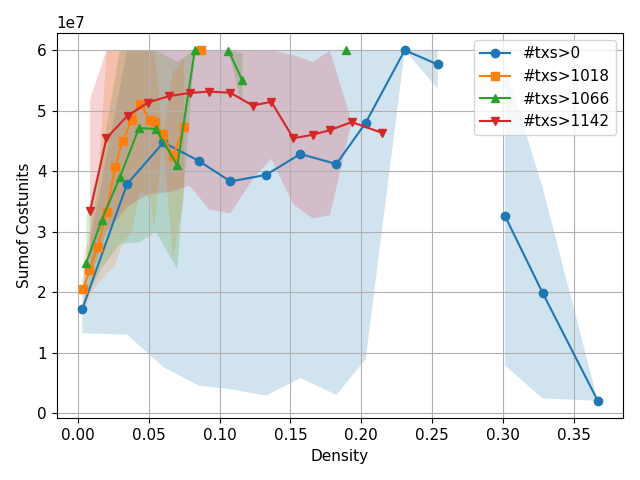}}
\caption{sumof-costUnits}
\label{fig:sumof-costUnits}
\end{figure}
\begin{figure}
\centerline{\includegraphics[width=\columnwidth]{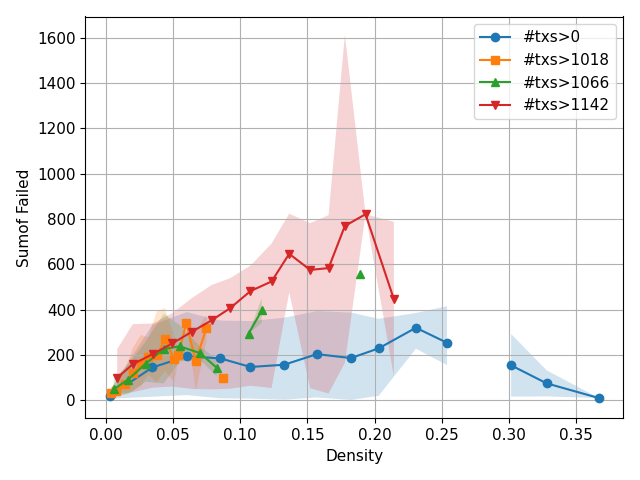}}
\caption{sumof-failed}
\label{fig:sumof-failed}
\end{figure}
\begin{figure}
\centerline{\includegraphics[width=\columnwidth]{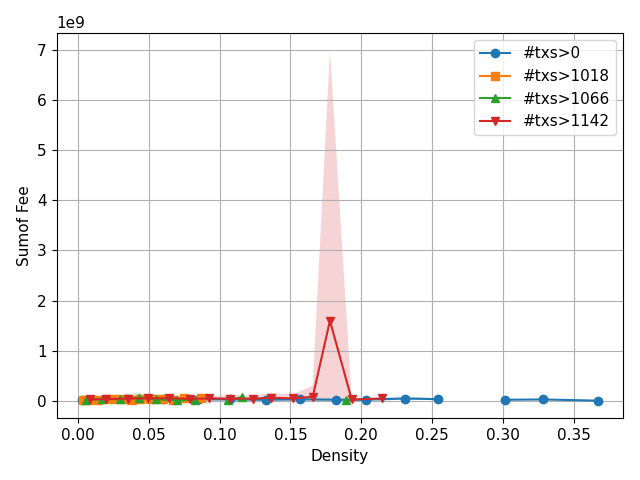}}
\caption{sumof-fee}
\label{fig:sumof-fee}
\end{figure}
\begin{figure}
\centerline{\includegraphics[width=\columnwidth]{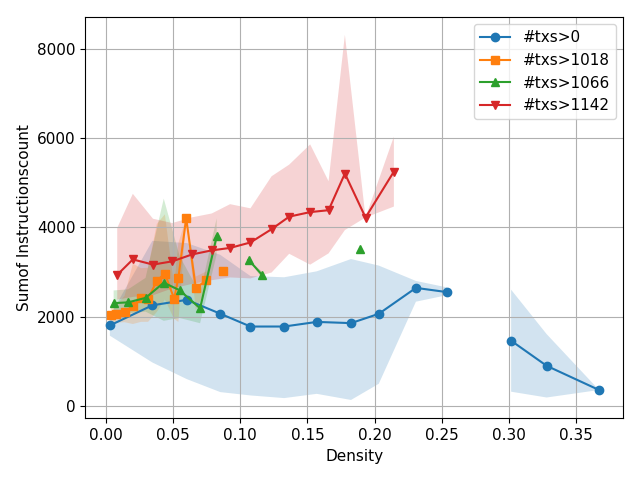}}
\caption{sumof-instructionsCount}
\label{fig:sumof-instructionsCount}
\end{figure}
\begin{figure}
\centerline{\includegraphics[width=\columnwidth]{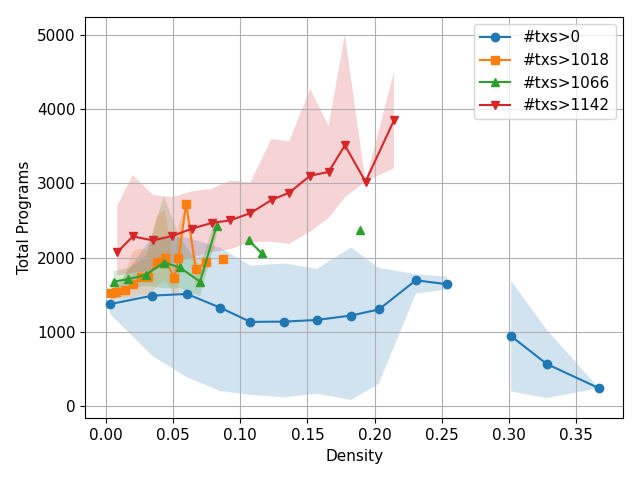}}
\caption{total-programs}
\label{fig:total-programs}
\end{figure}
\begin{figure}
\centerline{\includegraphics[width=\columnwidth]{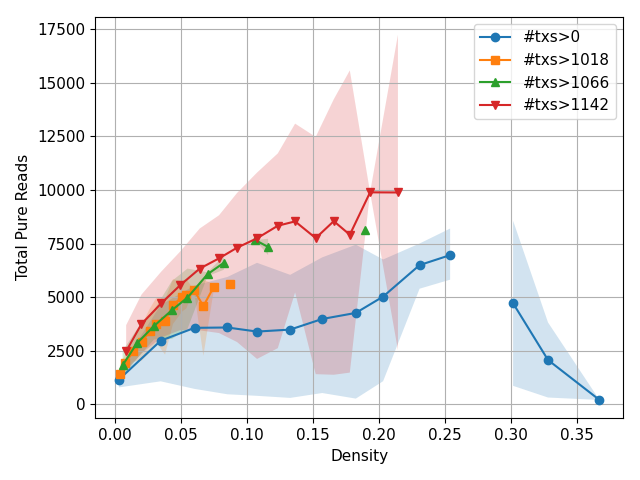}}
\caption{total-pure-reads}
\label{fig:total-pure-reads}
\end{figure}
\begin{figure}
\centerline{\includegraphics[width=\columnwidth]{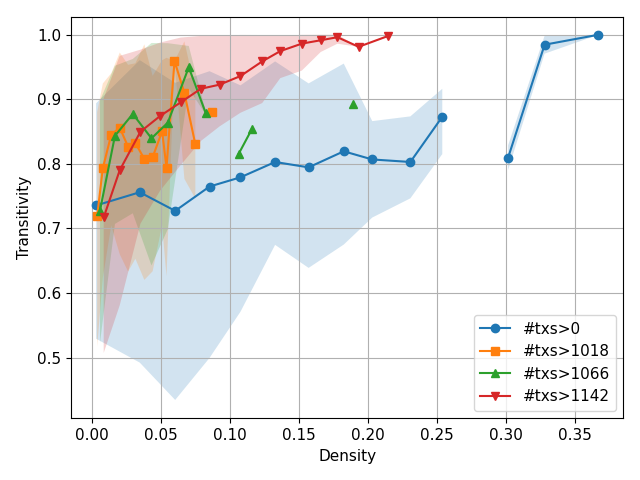}}
\caption{transitivity}
\label{fig:transitivity}
\end{figure}
\begin{figure}
\centerline{\includegraphics[width=\columnwidth]{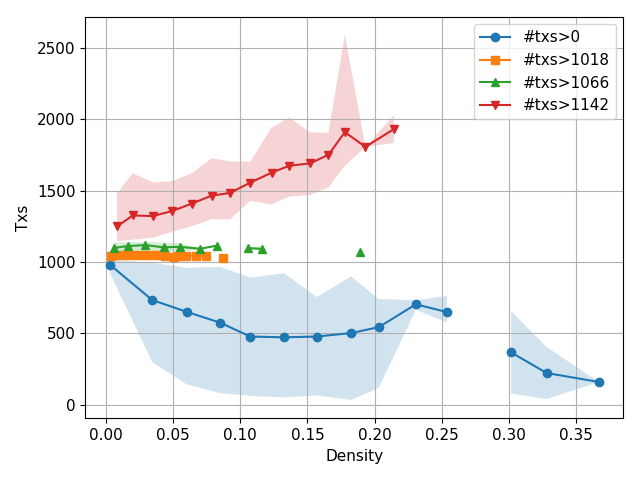}}
\caption{txs}
\label{fig:txs}
\end{figure}
\begin{figure}
\centerline{\includegraphics[width=\columnwidth]{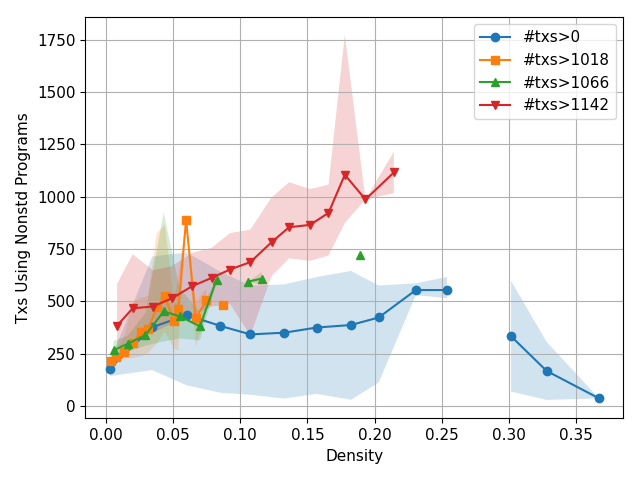}}
\caption{txs-using-nonstd-programs}
\label{fig:txs-using-nonstd-programs}
\end{figure}
\begin{figure}
\centerline{\includegraphics[width=\columnwidth]{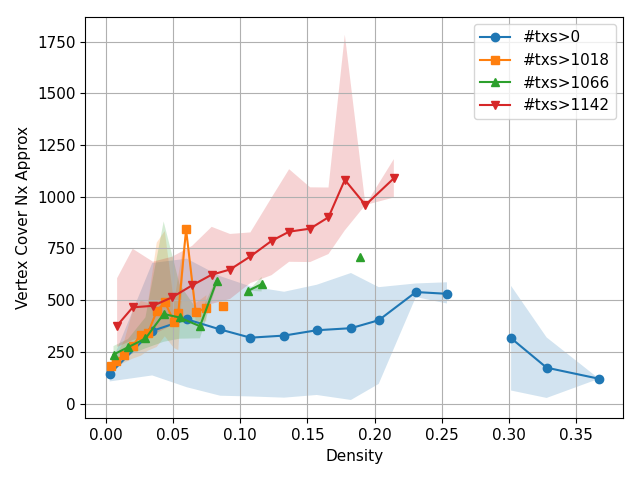}}
\caption{vertex-cover-nx-approx}
\label{fig:vertex-cover-nx-approx}
\end{figure}
\begin{figure}
\centerline{\includegraphics[width=\columnwidth]{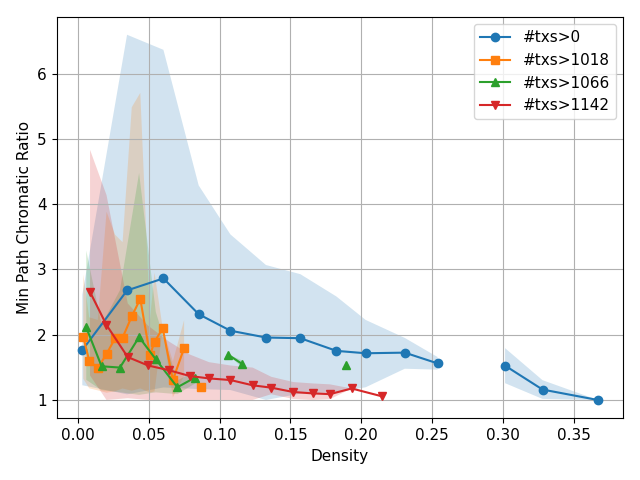}}
\caption{min-path-chromatic-ratio}
\label{fig:min-path-chromatic-ratio}
\end{figure}
\begin{figure}
\centerline{\includegraphics[width=\columnwidth]{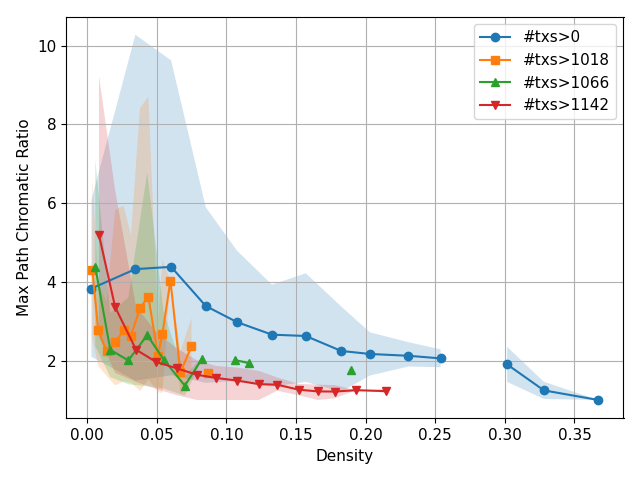}}
\caption{max-path-chromatic-ratio}
\label{fig:max-path-chromatic-ratio}
\end{figure}
\begin{figure}
\centerline{\includegraphics[width=\columnwidth]{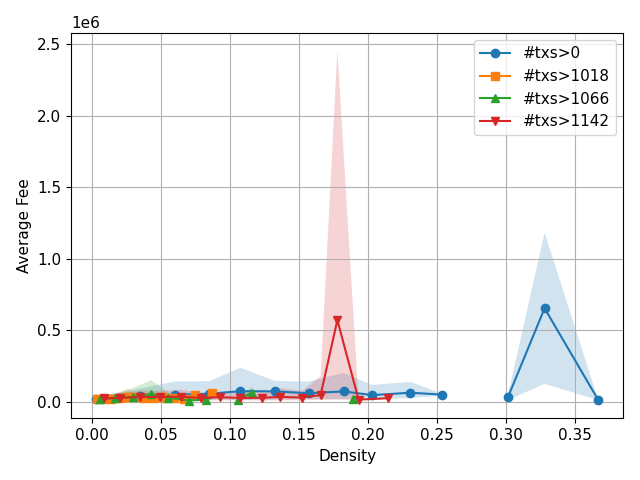}}
\caption{avg-fee}
\label{fig:avg-fee}
\end{figure}
\begin{figure}
\centerline{\includegraphics[width=\columnwidth]{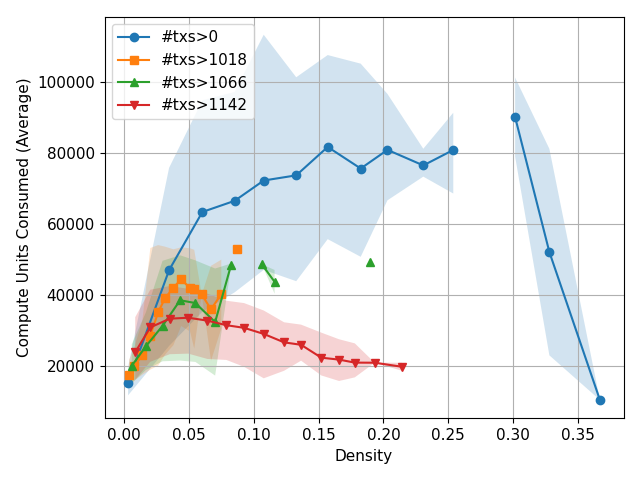}}
\caption{avg-computeUnitsConsumed}
\label{fig:avg-computeUnitsConsumed}
\end{figure}
\begin{figure}
\centerline{\includegraphics[width=\columnwidth]{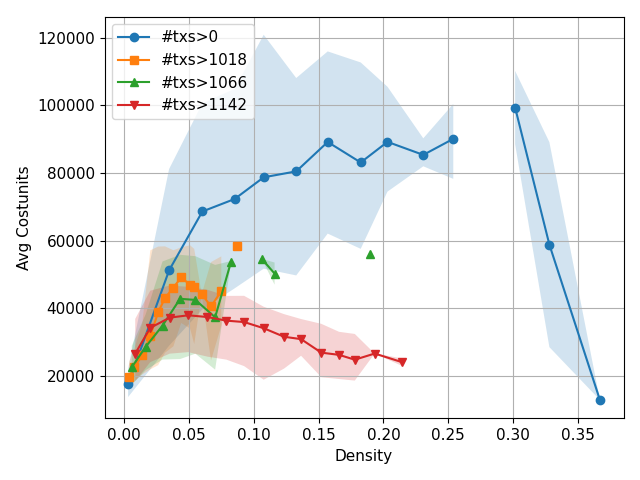}}
\caption{avg-costUnits}
\label{fig:avg-costUnits}
\end{figure}
\begin{figure}
\centerline{\includegraphics[width=\columnwidth]{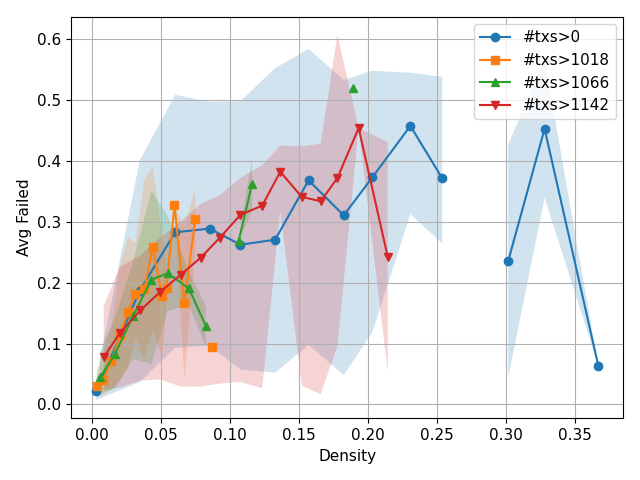}}
\caption{avg-failed}
\label{fig:avg-failed}
\end{figure}
\begin{figure}
\centerline{\includegraphics[width=\columnwidth]{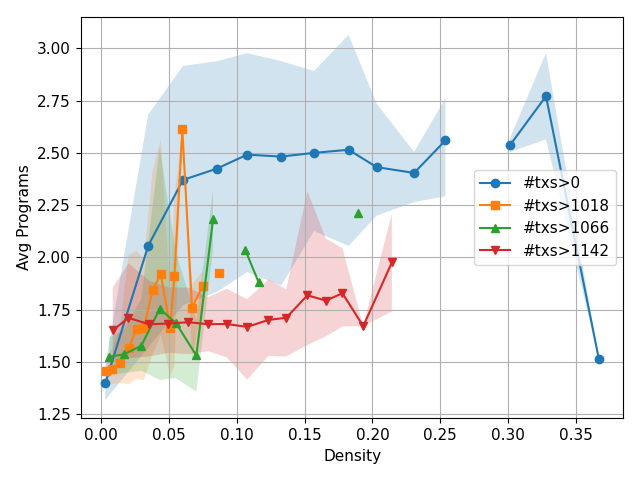}}
\caption{avg-programs}
\label{fig:avg-programs}
\end{figure}
\begin{figure}
\centerline{\includegraphics[width=\columnwidth]{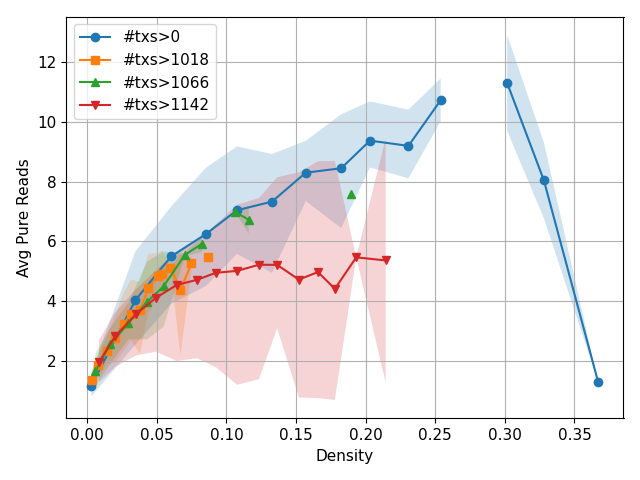}}
\caption{avg-pure-reads}
\label{fig:avg-pure-reads}
\end{figure}
\begin{figure}
\centerline{\includegraphics[width=\columnwidth]{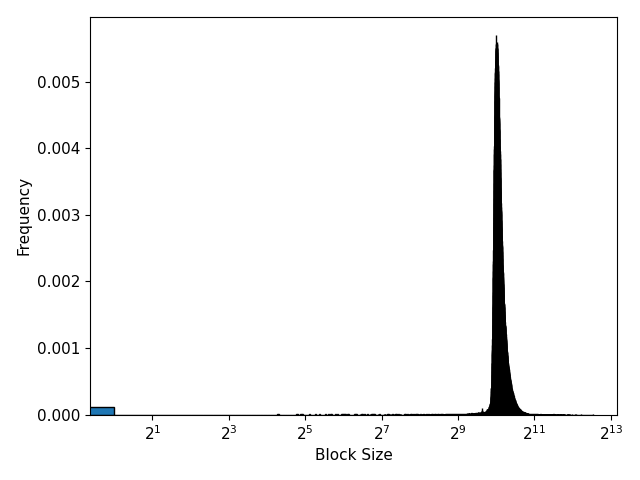}}
\caption{block-size-dist}
\label{fig:block-size-dist}
\end{figure}
\begin{figure}
\centerline{\includegraphics[width=\columnwidth]{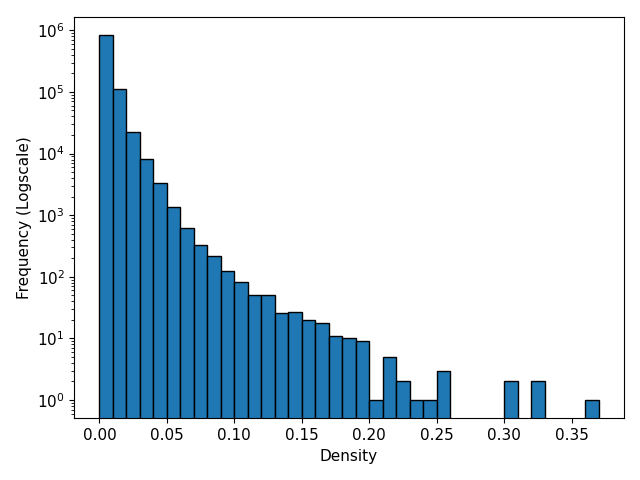}}
\caption{density-dist}
\label{fig:density-dist}
\end{figure}
\begin{figure}
\centerline{\includegraphics[width=\columnwidth]{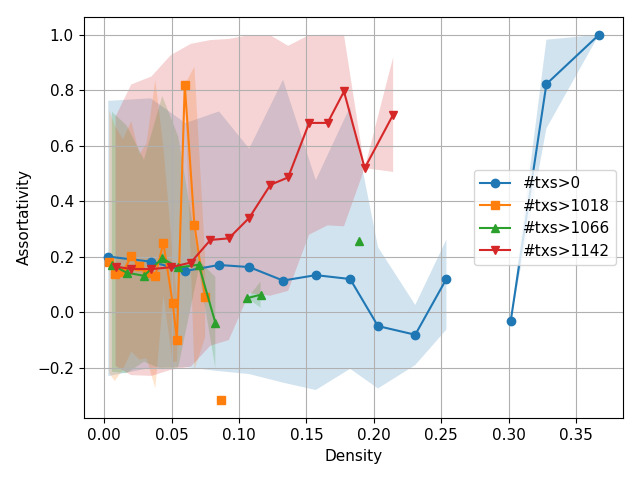}}
\caption{assortativity}
\label{fig:assortativity}
\end{figure}
\begin{figure}
\centerline{\includegraphics[width=\columnwidth]{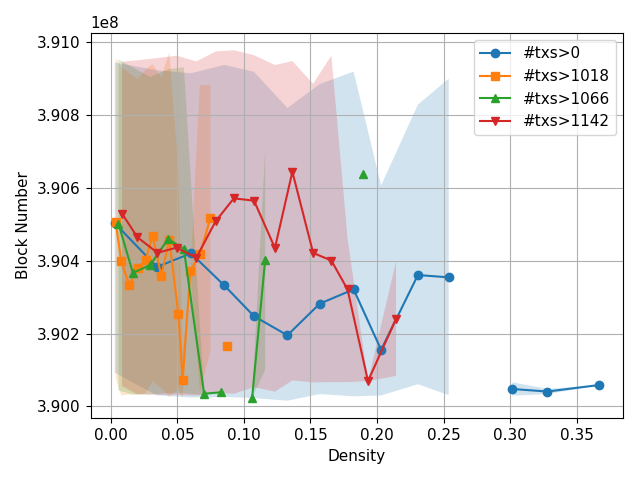}}
\caption{block-number}
\label{fig:block-number}
\end{figure}
\begin{figure}
\centerline{\includegraphics[width=\columnwidth]{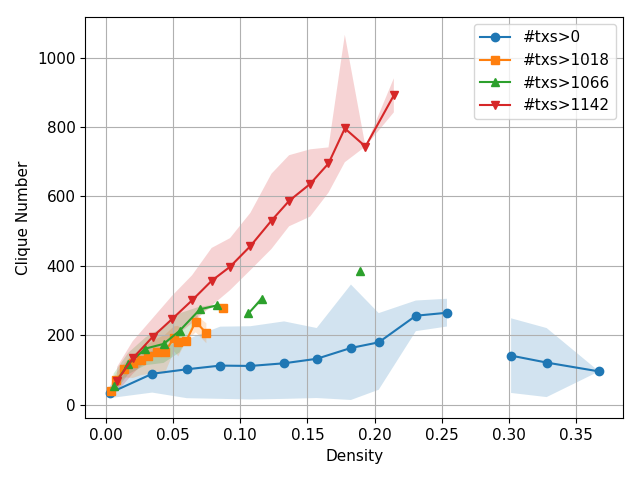}}
\caption{clique-number}
\label{fig:clique-number}
\end{figure}
\begin{figure}
\centerline{\includegraphics[width=\columnwidth]{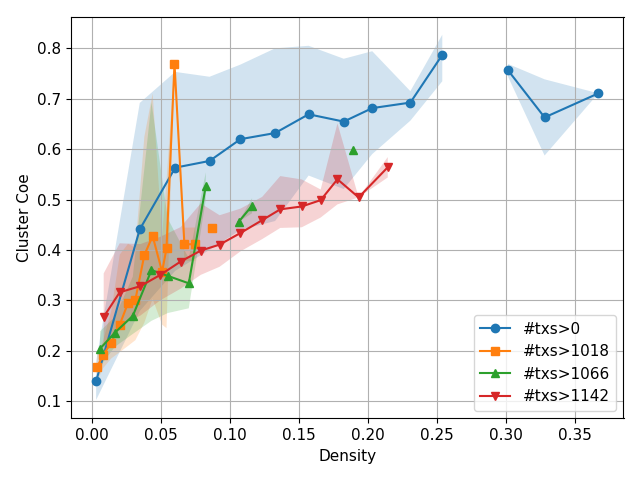}}
\caption{cluster-coe}
\label{fig:cluster-coe}
\end{figure}
\begin{figure}
\centerline{\includegraphics[width=\columnwidth]{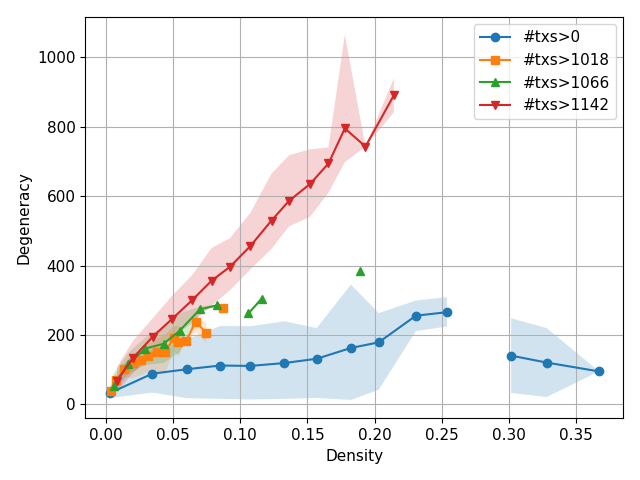}}
\caption{degeneracy}
\label{fig:degeneracy}
\end{figure}
\begin{figure}
\centerline{\includegraphics[width=\columnwidth]{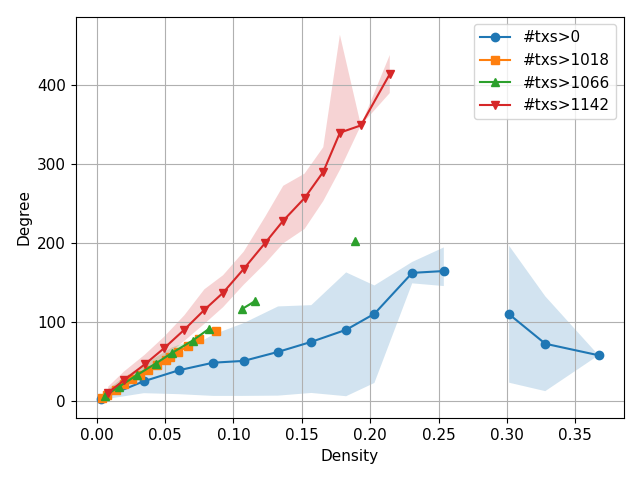}}
\caption{degree}
\label{fig:degree}
\end{figure}
\begin{figure}
\centerline{\includegraphics[width=\columnwidth]{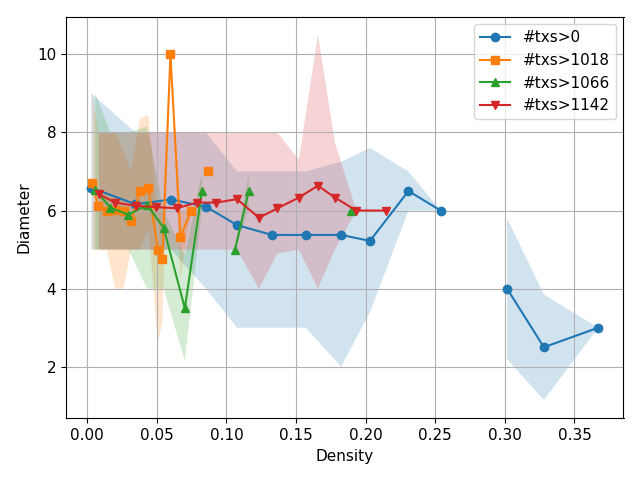}}
\caption{diameter}
\label{fig:diameter}
\end{figure}
\begin{figure}
\centerline{\includegraphics[width=\columnwidth]{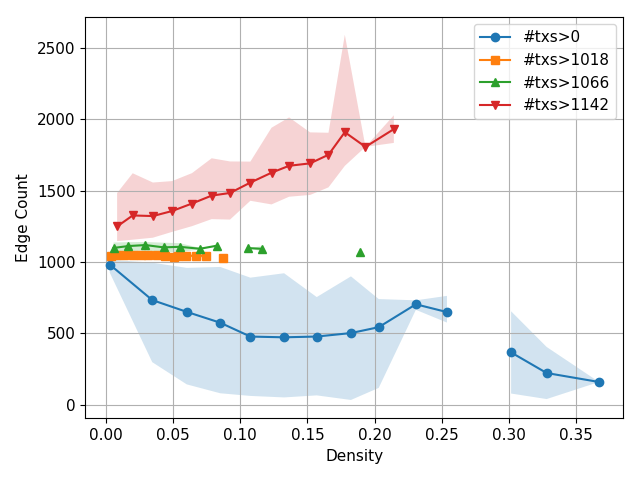}}
\caption{edge-count}
\label{fig:edge-count}
\end{figure}
\begin{figure}
\centerline{\includegraphics[width=\columnwidth]{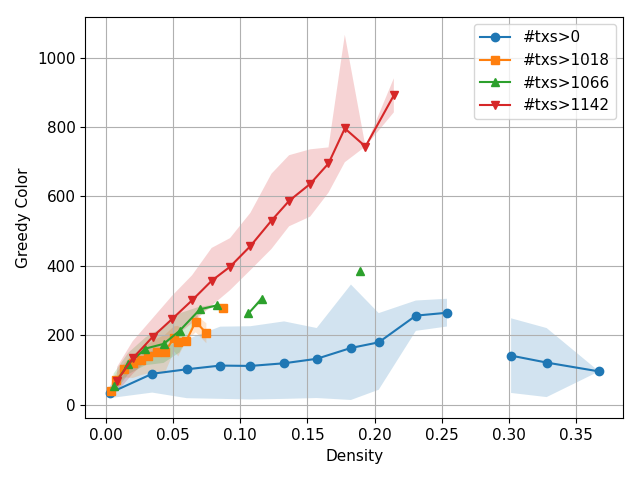}}
\caption{greedy-color}
\label{fig:greedy-color}
\end{figure}
\begin{figure}
\centerline{\includegraphics[width=\columnwidth]{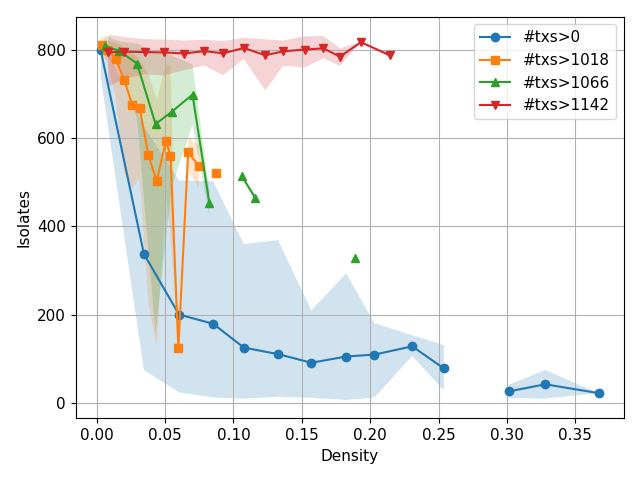}}
\caption{isolates}
\label{fig:isolates}
\end{figure}
\begin{figure}
\centerline{\includegraphics[width=\columnwidth]{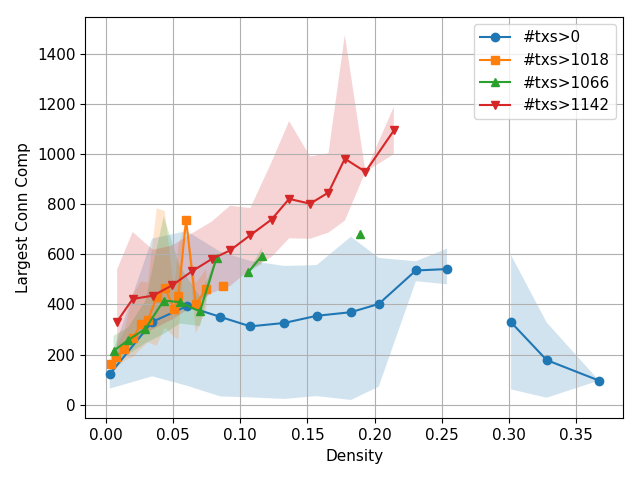}}
\caption{largest-conn-comp}
\label{fig:largest-conn-comp}
\end{figure}
\begin{figure}
\centerline{\includegraphics[width=\columnwidth]{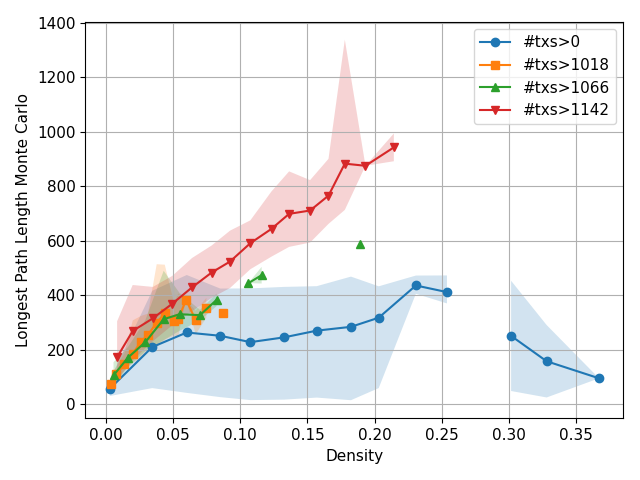}}
\caption{longest-path-length-monte-carlo}
\label{fig:longest-path-length-monte-carlo}
\end{figure}
\begin{figure}
\centerline{\includegraphics[width=\columnwidth]{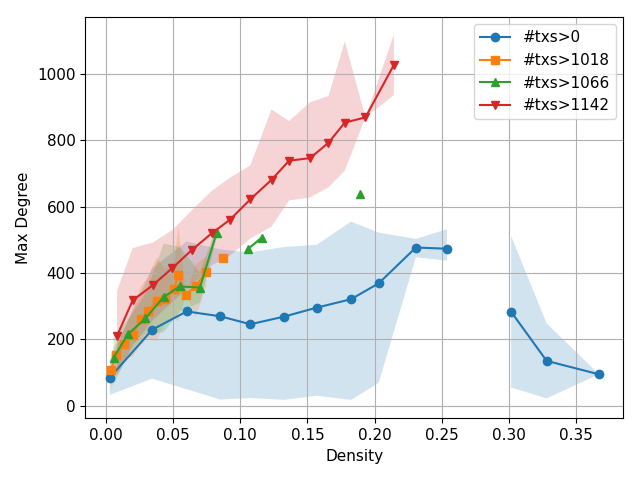}}
\caption{max-degree}
\label{fig:max-degree}
\end{figure}
\begin{figure}
\centerline{\includegraphics[width=\columnwidth]{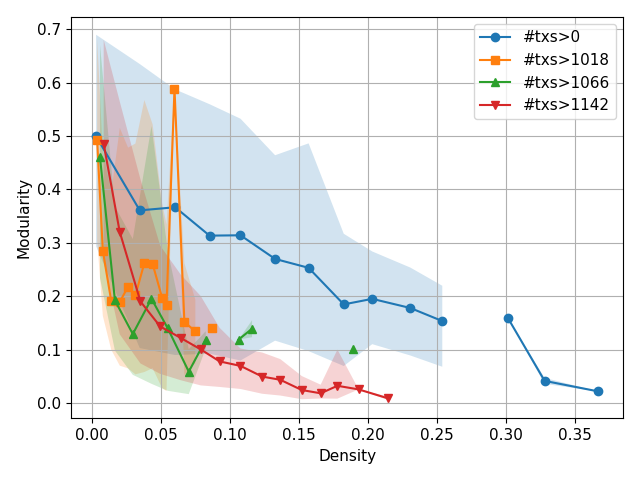}}
\caption{modularity}
\label{fig:modularity}
\end{figure}
\begin{figure}
\centerline{\includegraphics[width=\columnwidth]{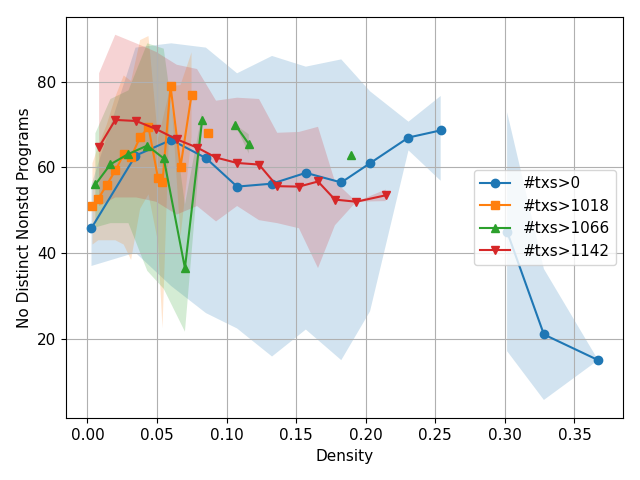}}
\caption{no-distinct-nonstd-programs}
\label{fig:no-distinct-nonstd-programs}
\end{figure}
\begin{figure}
\centerline{\includegraphics[width=\columnwidth]{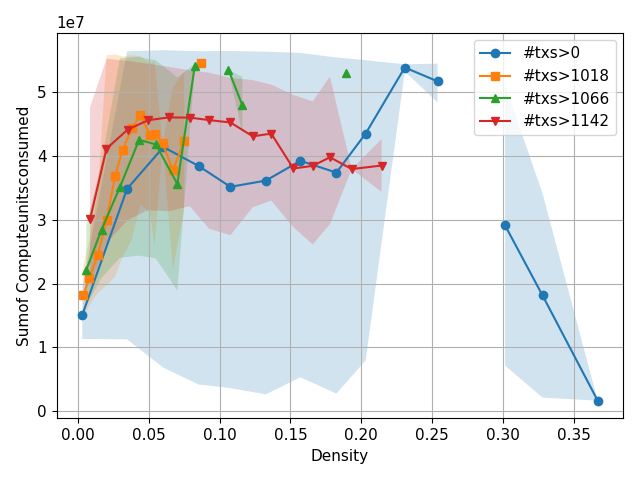}}
\caption{sumof-computeUnitsConsumed}
\label{fig:sumof-computeUnitsConsumed}
\end{figure}
\begin{figure}
\centerline{\includegraphics[width=\columnwidth]{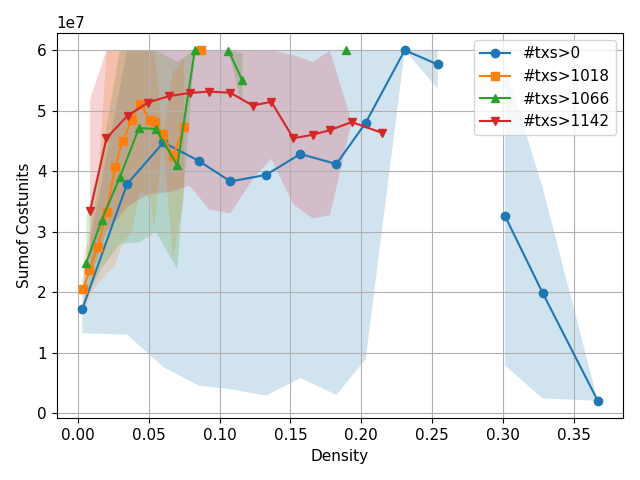}}
\caption{sumof-costUnits}
\label{fig:sumof-costUnits}
\end{figure}
\begin{figure}
\centerline{\includegraphics[width=\columnwidth]{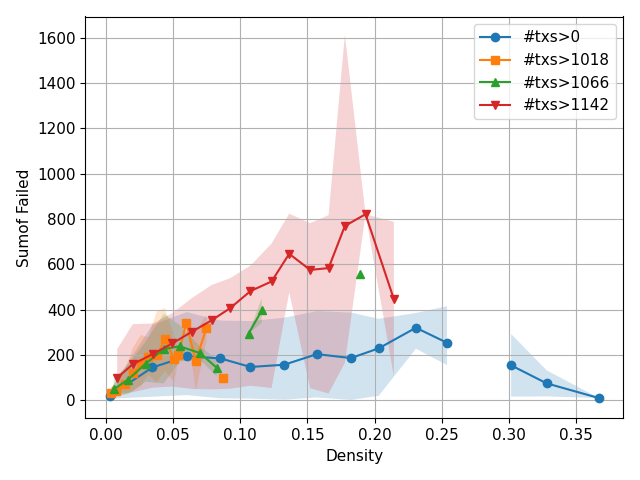}}
\caption{sumof-failed}
\label{fig:sumof-failed}
\end{figure}
\begin{figure}
\centerline{\includegraphics[width=\columnwidth]{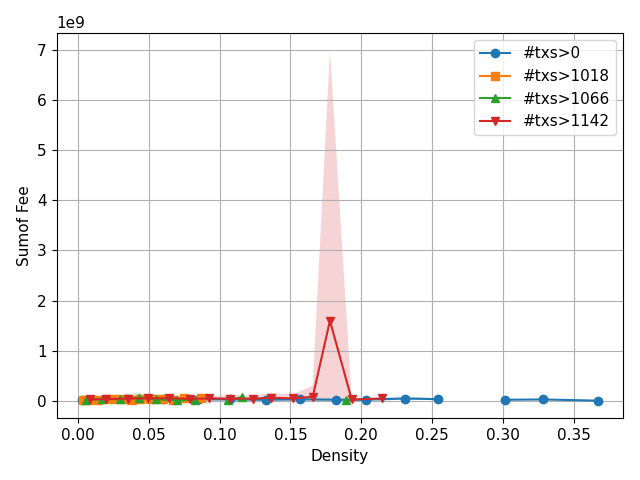}}
\caption{sumof-fee}
\label{fig:sumof-fee}
\end{figure}
\begin{figure}
\centerline{\includegraphics[width=\columnwidth]{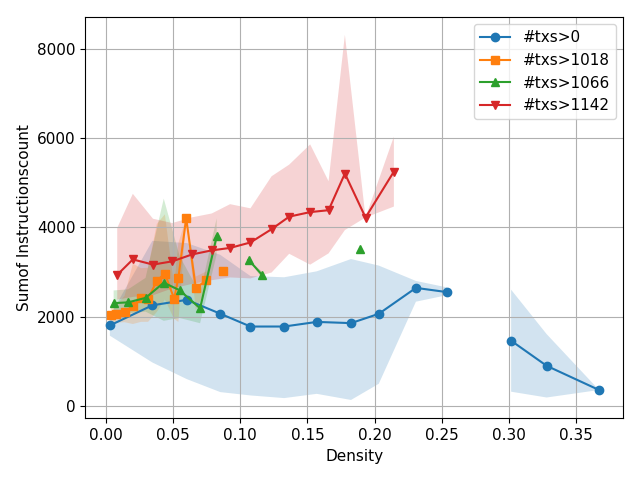}}
\caption{sumof-instructionsCount}
\label{fig:sumof-instructionsCount}
\end{figure}
\begin{figure}
\centerline{\includegraphics[width=\columnwidth]{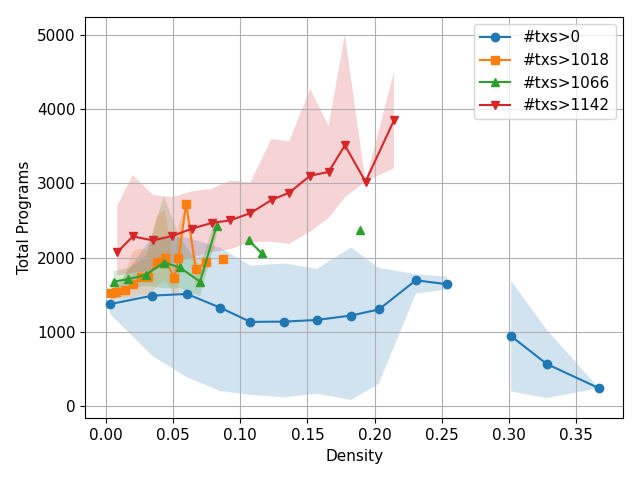}}
\caption{total-programs}
\label{fig:total-programs}
\end{figure}
\begin{figure}
\centerline{\includegraphics[width=\columnwidth]{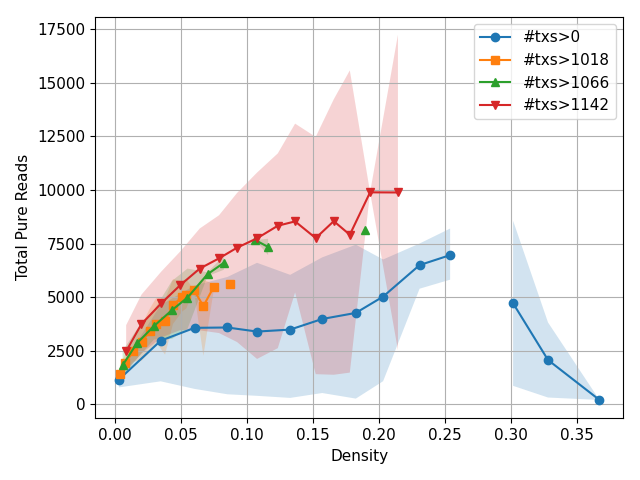}}
\caption{total-pure-reads}
\label{fig:total-pure-reads}
\end{figure}
\begin{figure}
\centerline{\includegraphics[width=\columnwidth]{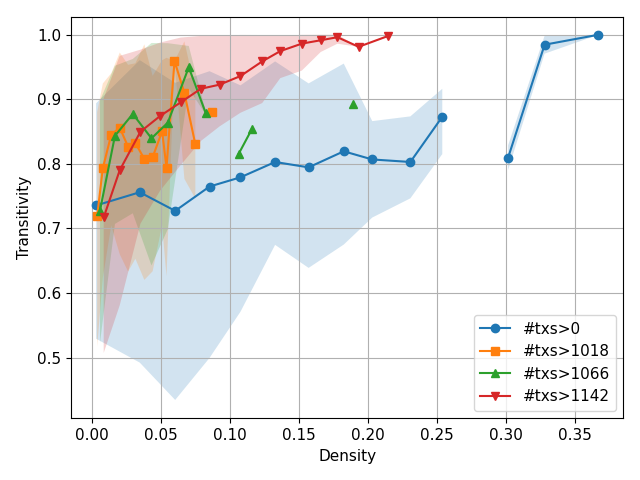}}
\caption{transitivity}
\label{fig:transitivity}
\end{figure}
\begin{figure}
\centerline{\includegraphics[width=\columnwidth]{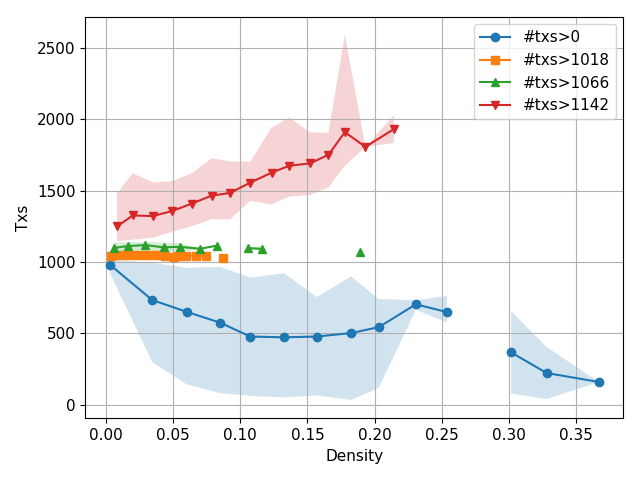}}
\caption{txs}
\label{fig:txs}
\end{figure}
\begin{figure}
\centerline{\includegraphics[width=\columnwidth]{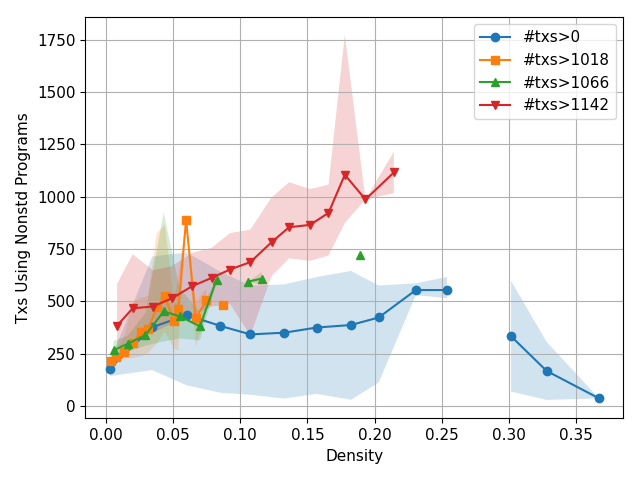}}
\caption{txs-using-nonstd-programs}
\label{fig:txs-using-nonstd-programs}
\end{figure}
\begin{figure}
\centerline{\includegraphics[width=\columnwidth]{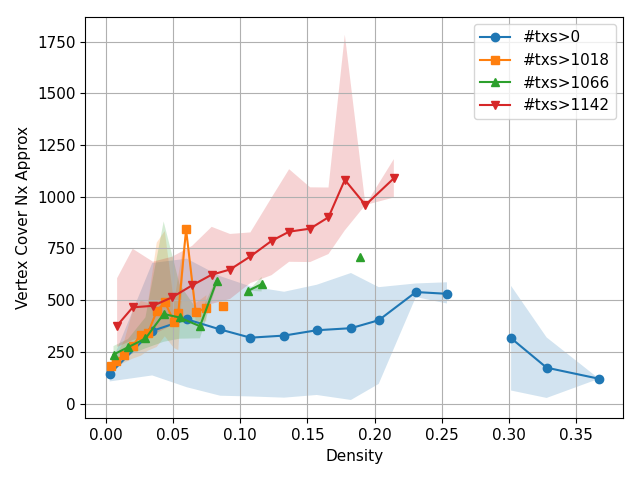}}
\caption{vertex-cover-nx-approx}
\label{fig:vertex-cover-nx-approx}
\end{figure}
\begin{figure}
\centerline{\includegraphics[width=\columnwidth]{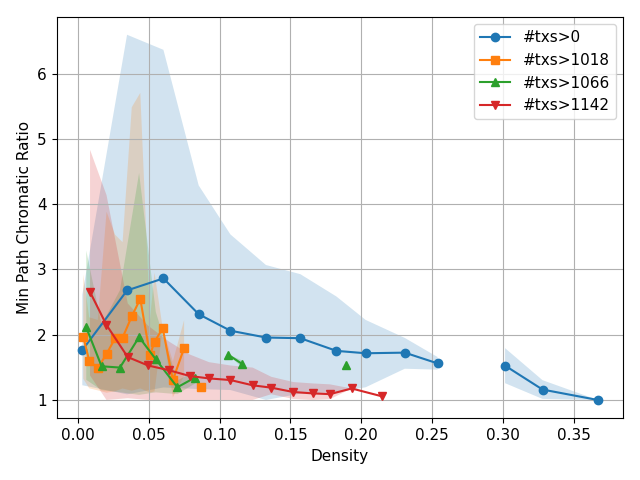}}
\caption{min-path-chromatic-ratio}
\label{fig:min-path-chromatic-ratio}
\end{figure}
\begin{figure}
\centerline{\includegraphics[width=\columnwidth]{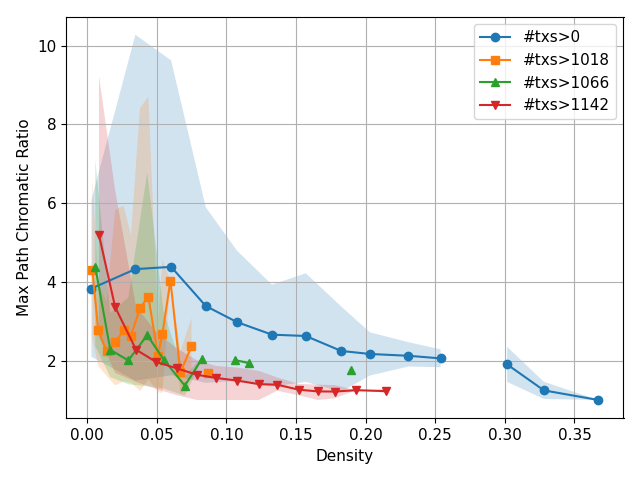}}
\caption{max-path-chromatic-ratio}
\label{fig:max-path-chromatic-ratio}
\end{figure}
\begin{figure}
\centerline{\includegraphics[width=\columnwidth]{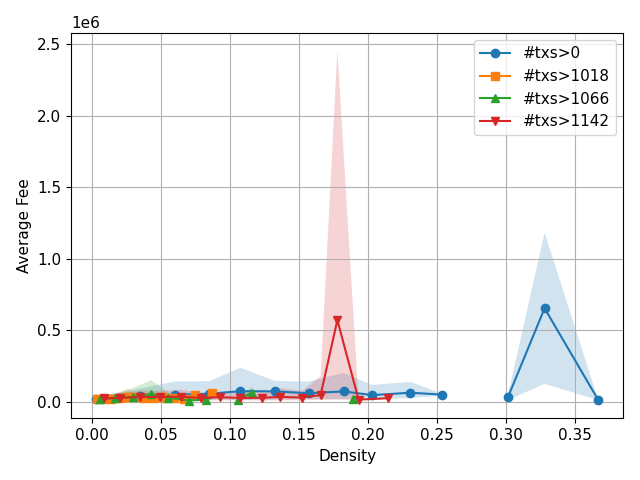}}
\caption{avg-fee}
\label{fig:avg-fee}
\end{figure}
\begin{figure}
\centerline{\includegraphics[width=\columnwidth]{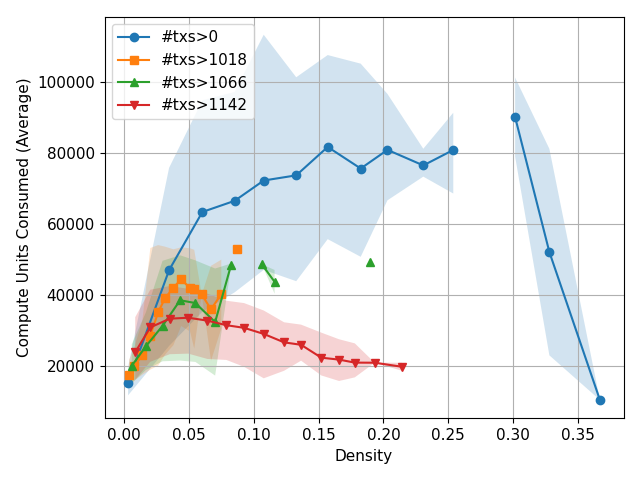}}
\caption{avg-computeUnitsConsumed}
\label{fig:avg-computeUnitsConsumed}
\end{figure}
\begin{figure}
\centerline{\includegraphics[width=\columnwidth]{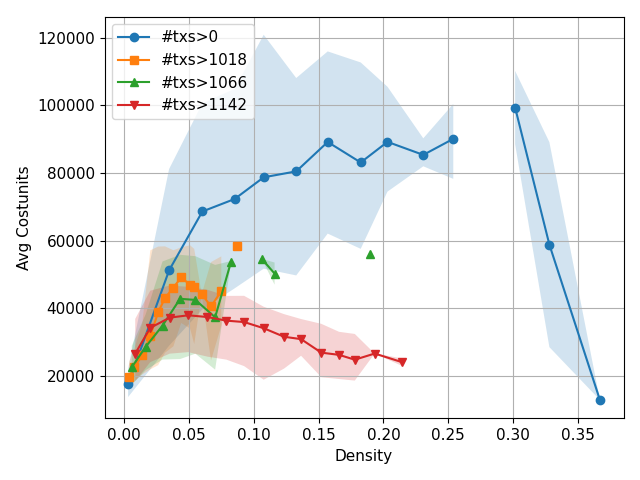}}
\caption{avg-costUnits}
\label{fig:avg-costUnits}
\end{figure}
\begin{figure}
\centerline{\includegraphics[width=\columnwidth]{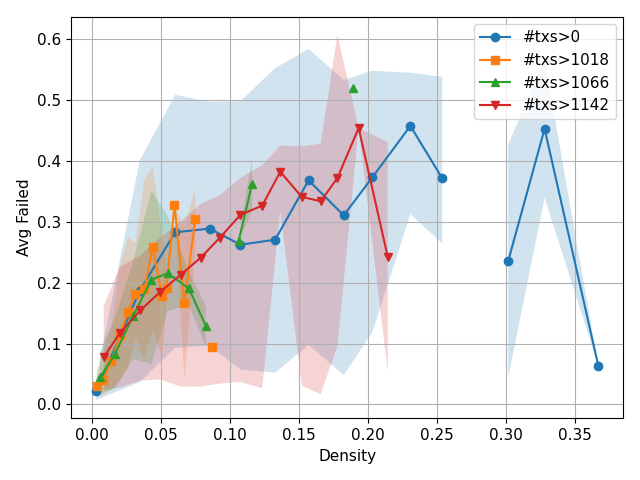}}
\caption{avg-failed}
\label{fig:avg-failed}
\end{figure}
\begin{figure}
\centerline{\includegraphics[width=\columnwidth]{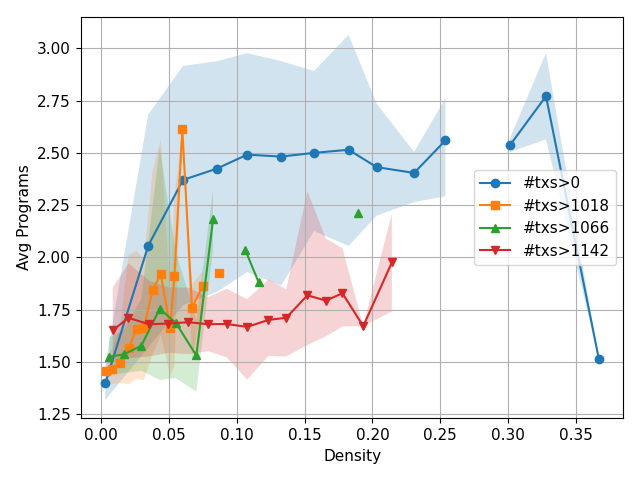}}
\caption{avg-programs}
\label{fig:avg-programs}
\end{figure}
\begin{figure}
\centerline{\includegraphics[width=\columnwidth]{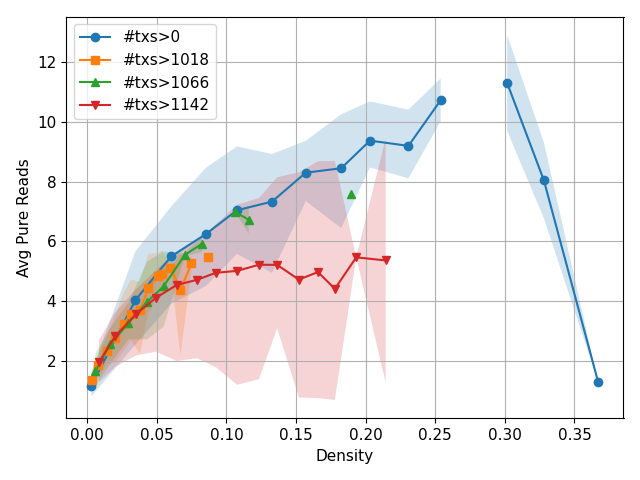}}
\caption{avg-pure-reads}
\label{fig:avg-pure-reads}
\end{figure}
\begin{figure}
\centerline{\includegraphics[width=\columnwidth]{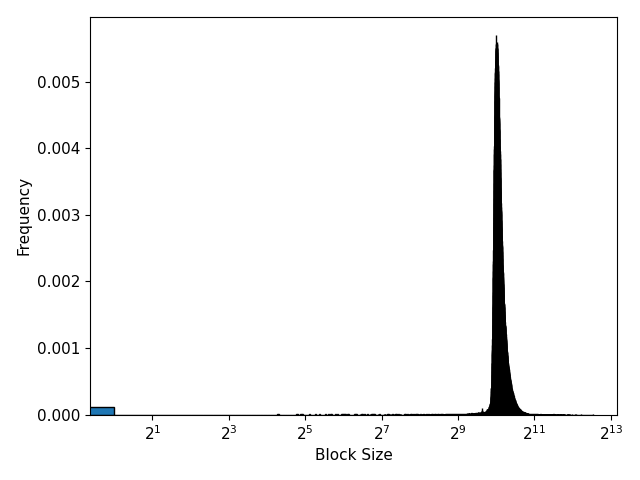}}
\caption{block-size-dist}
\label{fig:block-size-dist}
\end{figure}
\begin{figure}
\centerline{\includegraphics[width=\columnwidth]{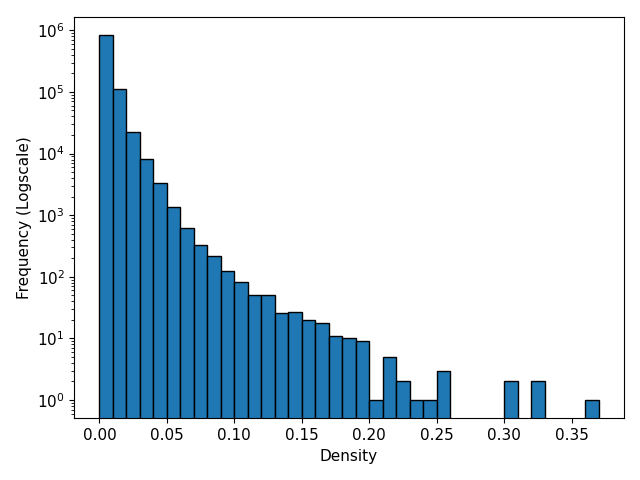}}
\caption{density-dist}
\label{fig:density-dist}
\end{figure}
\begin{figure}
\centerline{\includegraphics[width=\columnwidth]{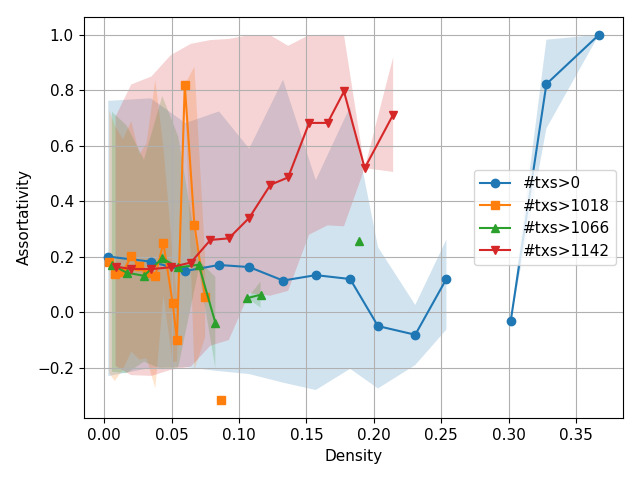}}
\caption{assortativity}
\label{fig:assortativity}
\end{figure}
\begin{figure}
\centerline{\includegraphics[width=\columnwidth]{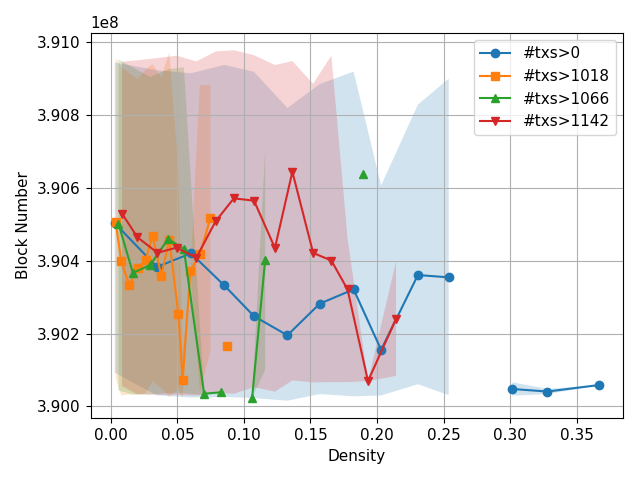}}
\caption{block-number}
\label{fig:block-number}
\end{figure}
\begin{figure}
\centerline{\includegraphics[width=\columnwidth]{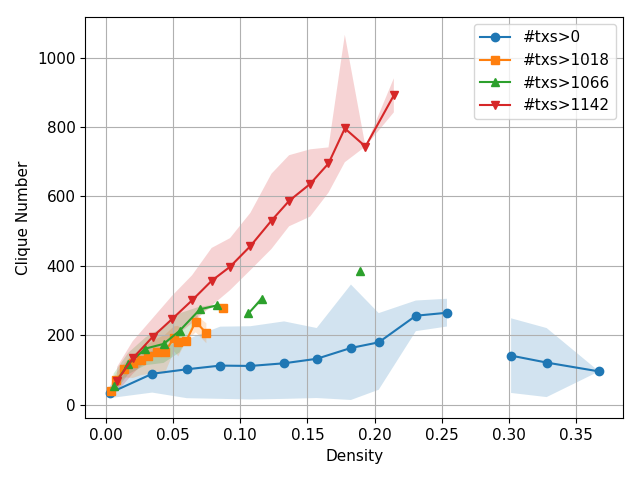}}
\caption{clique-number}
\label{fig:clique-number}
\end{figure}
\begin{figure}
\centerline{\includegraphics[width=\columnwidth]{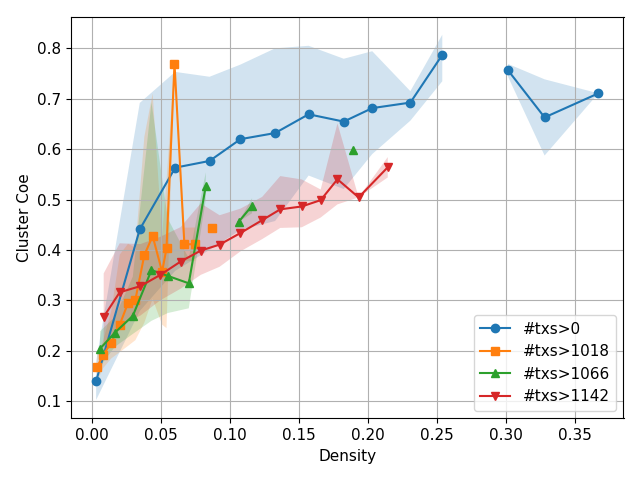}}
\caption{cluster-coe}
\label{fig:cluster-coe}
\end{figure}
\begin{figure}
\centerline{\includegraphics[width=\columnwidth]{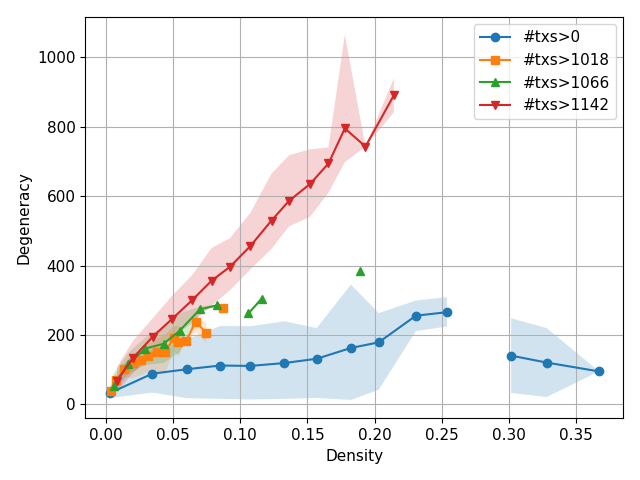}}
\caption{degeneracy}
\label{fig:degeneracy}
\end{figure}
\begin{figure}
\centerline{\includegraphics[width=\columnwidth]{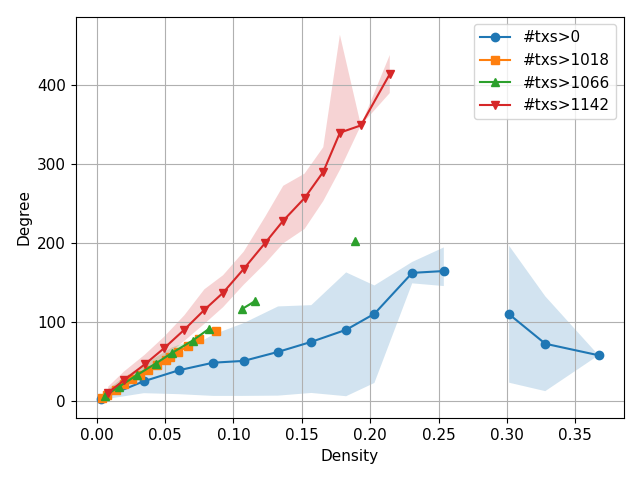}}
\caption{degree}
\label{fig:degree}
\end{figure}
\begin{figure}
\centerline{\includegraphics[width=\columnwidth]{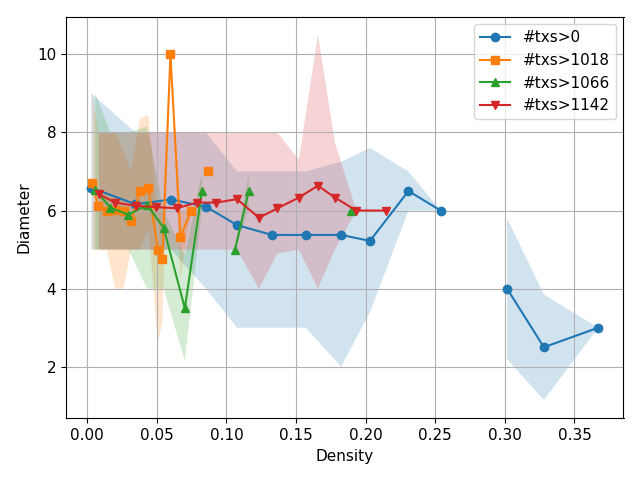}}
\caption{diameter}
\label{fig:diameter}
\end{figure}
\begin{figure}
\centerline{\includegraphics[width=\columnwidth]{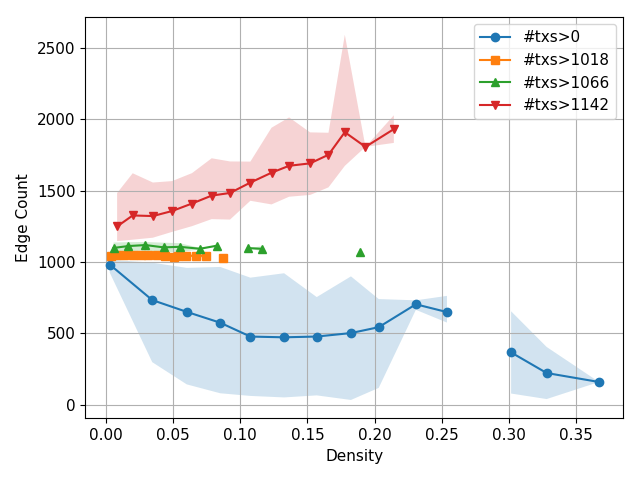}}
\caption{edge-count}
\label{fig:edge-count}
\end{figure}
\begin{figure}
\centerline{\includegraphics[width=\columnwidth]{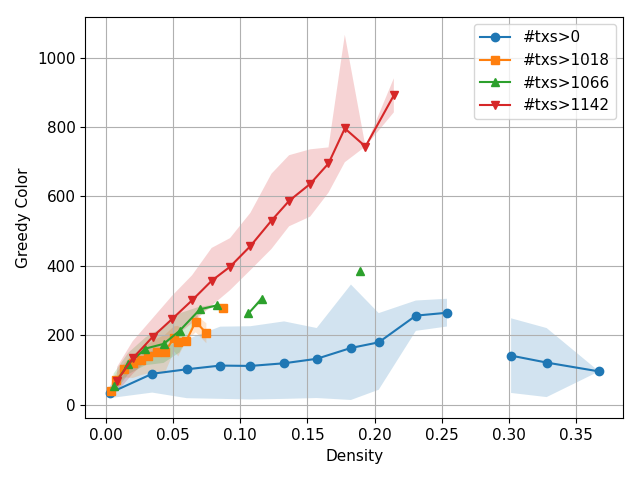}}
\caption{greedy-color}
\label{fig:greedy-color}
\end{figure}
\begin{figure}
\centerline{\includegraphics[width=\columnwidth]{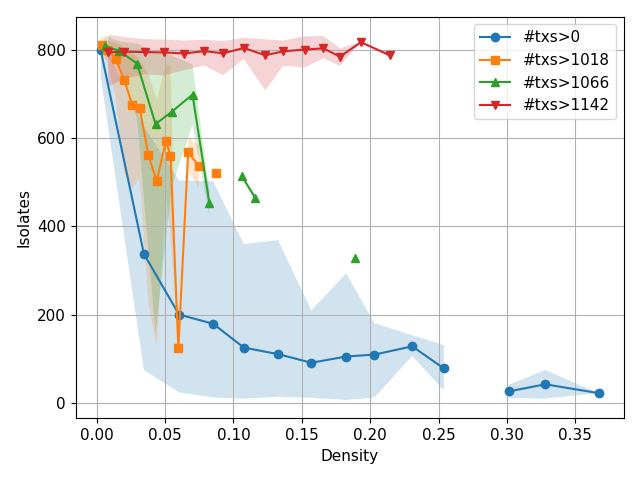}}
\caption{isolates}
\label{fig:isolates}
\end{figure}
\begin{figure}
\centerline{\includegraphics[width=\columnwidth]{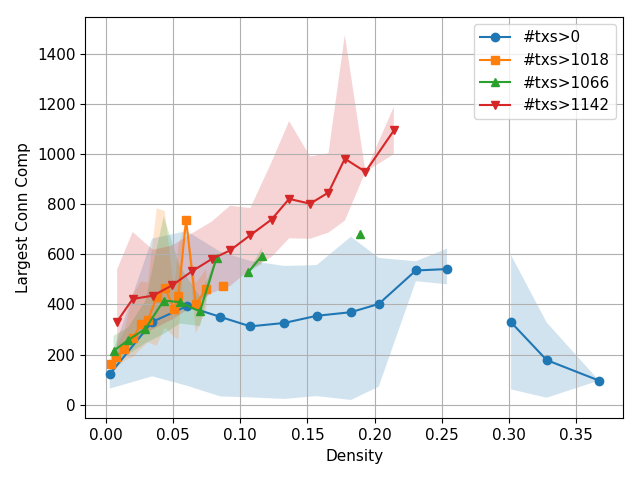}}
\caption{largest-conn-comp}
\label{fig:largest-conn-comp}
\end{figure}
\begin{figure}
\centerline{\includegraphics[width=\columnwidth]{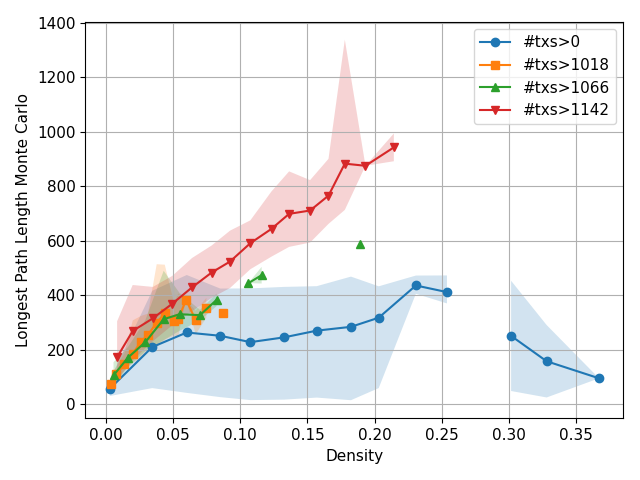}}
\caption{longest-path-length-monte-carlo}
\label{fig:longest-path-length-monte-carlo}
\end{figure}
\begin{figure}
\centerline{\includegraphics[width=\columnwidth]{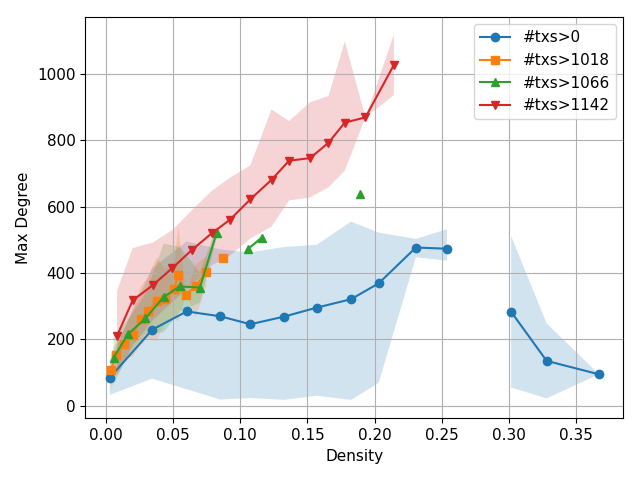}}
\caption{max-degree}
\label{fig:max-degree}
\end{figure}
\begin{figure}
\centerline{\includegraphics[width=\columnwidth]{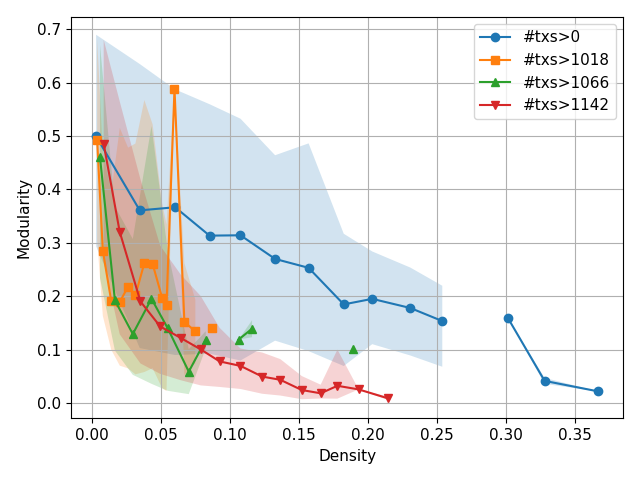}}
\caption{modularity}
\label{fig:modularity}
\end{figure}
\begin{figure}
\centerline{\includegraphics[width=\columnwidth]{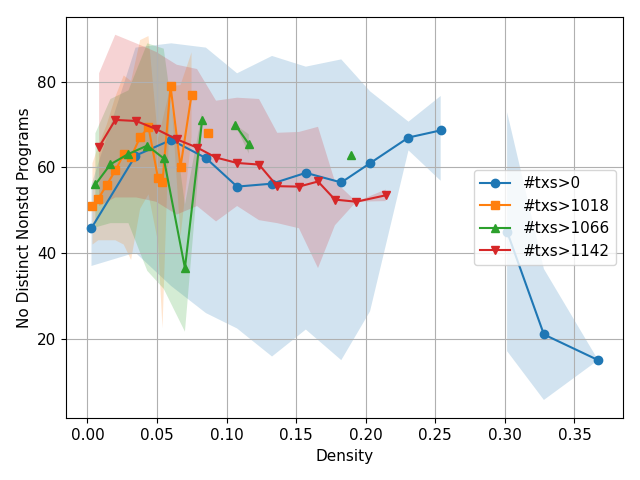}}
\caption{no-distinct-nonstd-programs}
\label{fig:no-distinct-nonstd-programs}
\end{figure}
\begin{figure}
\centerline{\includegraphics[width=\columnwidth]{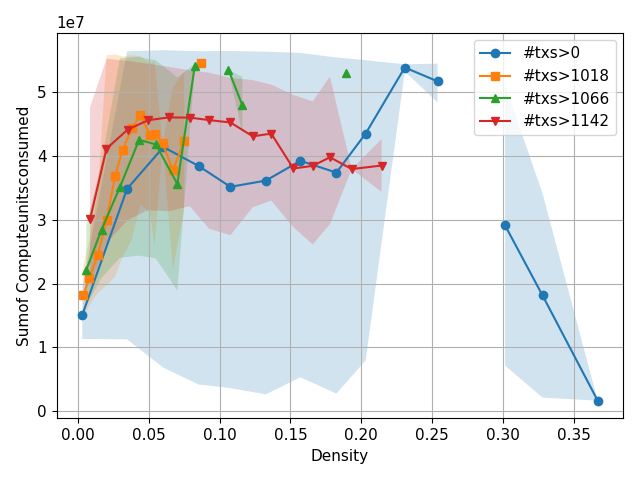}}
\caption{sumof-computeUnitsConsumed}
\label{fig:sumof-computeUnitsConsumed}
\end{figure}
\begin{figure}
\centerline{\includegraphics[width=\columnwidth]{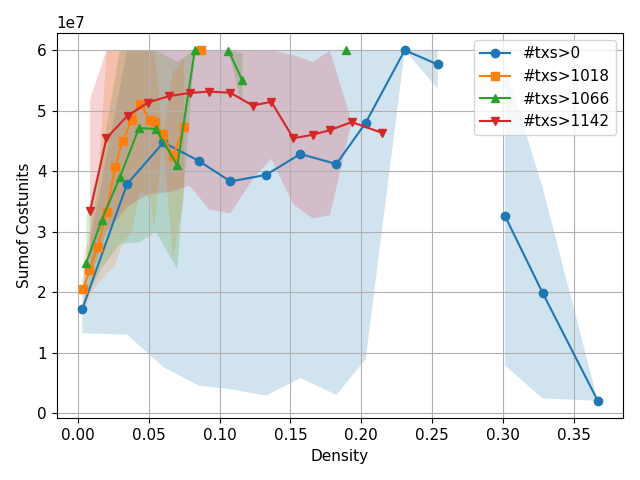}}
\caption{sumof-costUnits}
\label{fig:sumof-costUnits}
\end{figure}
\begin{figure}
\centerline{\includegraphics[width=\columnwidth]{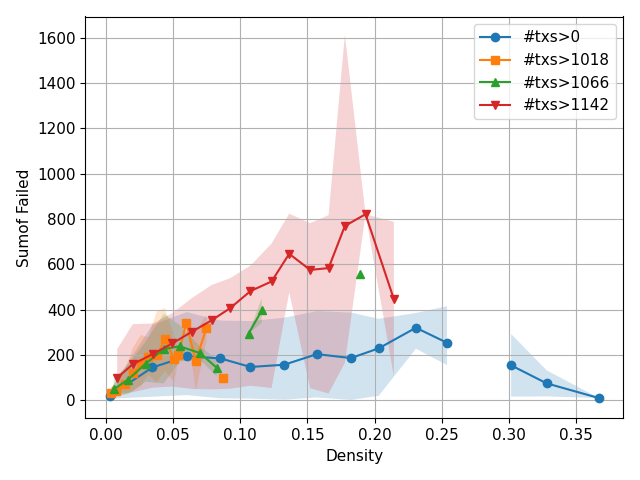}}
\caption{sumof-failed}
\label{fig:sumof-failed}
\end{figure}
\begin{figure}
\centerline{\includegraphics[width=\columnwidth]{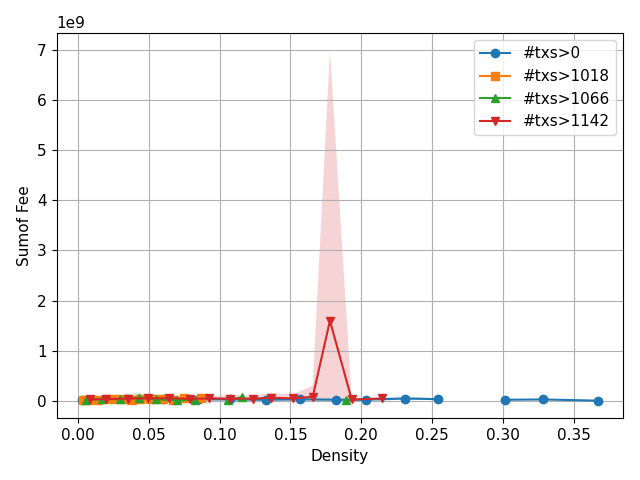}}
\caption{sumof-fee}
\label{fig:sumof-fee}
\end{figure}
\begin{figure}
\centerline{\includegraphics[width=\columnwidth]{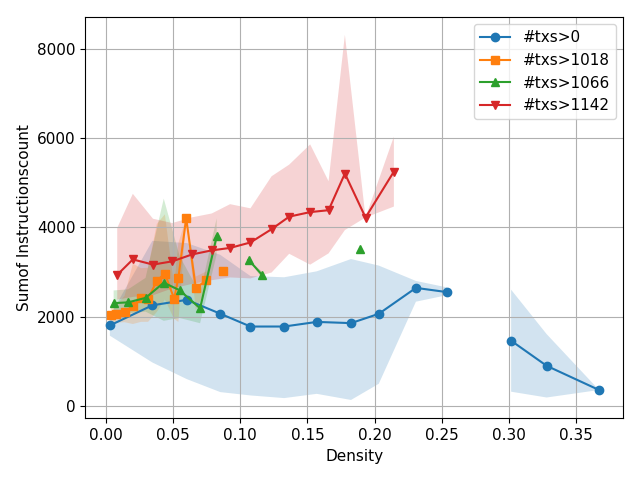}}
\caption{sumof-instructionsCount}
\label{fig:sumof-instructionsCount}
\end{figure}
\begin{figure}
\centerline{\includegraphics[width=\columnwidth]{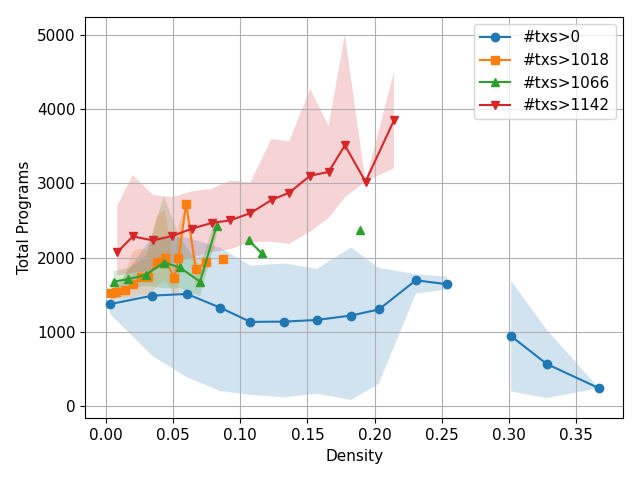}}
\caption{total-programs}
\label{fig:total-programs}
\end{figure}
\begin{figure}
\centerline{\includegraphics[width=\columnwidth]{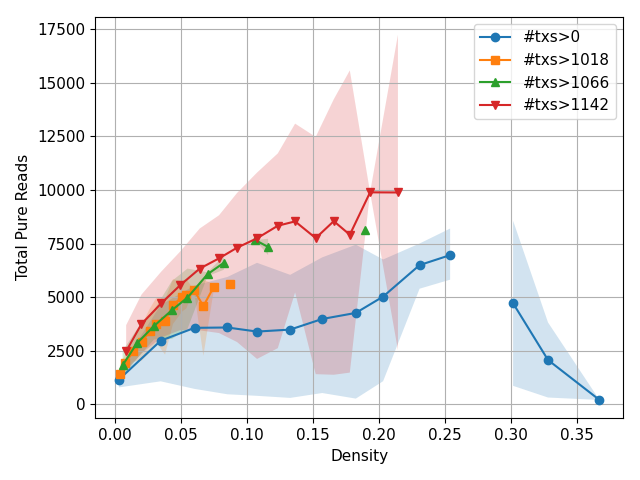}}
\caption{total-pure-reads}
\label{fig:total-pure-reads}
\end{figure}
\begin{figure}
\centerline{\includegraphics[width=\columnwidth]{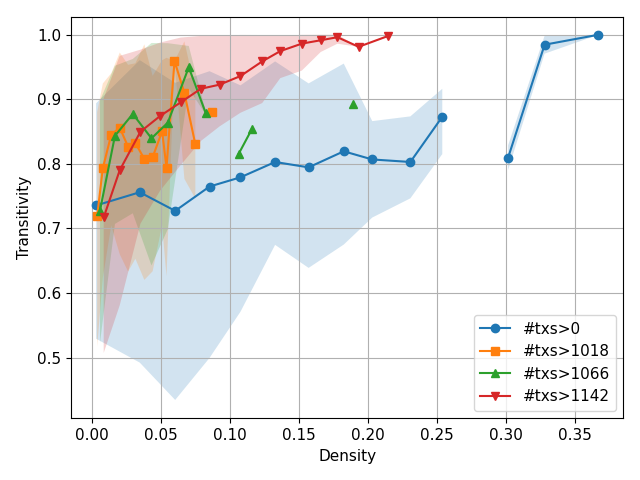}}
\caption{transitivity}
\label{fig:transitivity}
\end{figure}
\begin{figure}
\centerline{\includegraphics[width=\columnwidth]{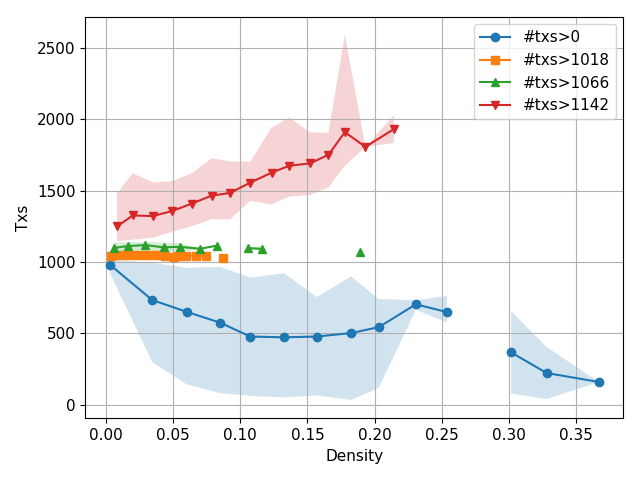}}
\caption{txs}
\label{fig:txs}
\end{figure}
\begin{figure}
\centerline{\includegraphics[width=\columnwidth]{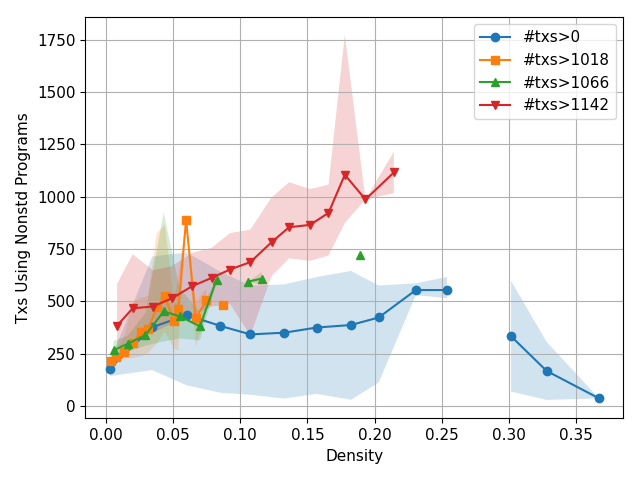}}
\caption{txs-using-nonstd-programs}
\label{fig:txs-using-nonstd-programs}
\end{figure}
\begin{figure}
\centerline{\includegraphics[width=\columnwidth]{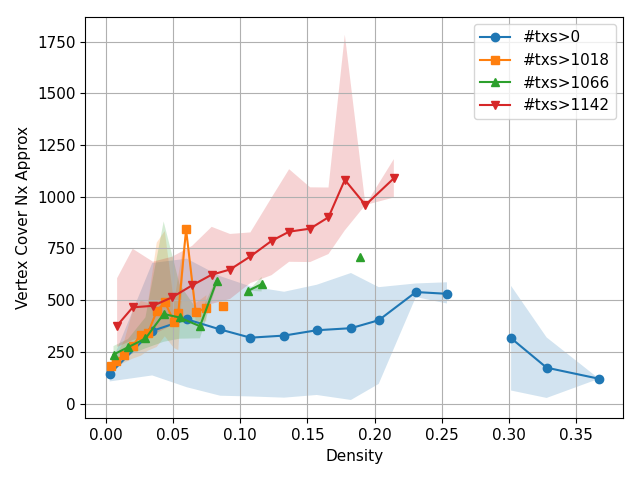}}
\caption{vertex-cover-nx-approx}
\label{fig:vertex-cover-nx-approx}
\end{figure}
\begin{figure}
\centerline{\includegraphics[width=\columnwidth]{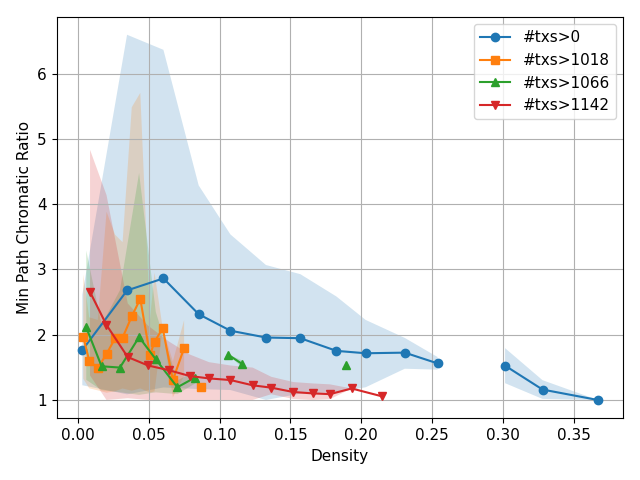}}
\caption{min-path-chromatic-ratio}
\label{fig:min-path-chromatic-ratio}
\end{figure}
\begin{figure}
\centerline{\includegraphics[width=\columnwidth]{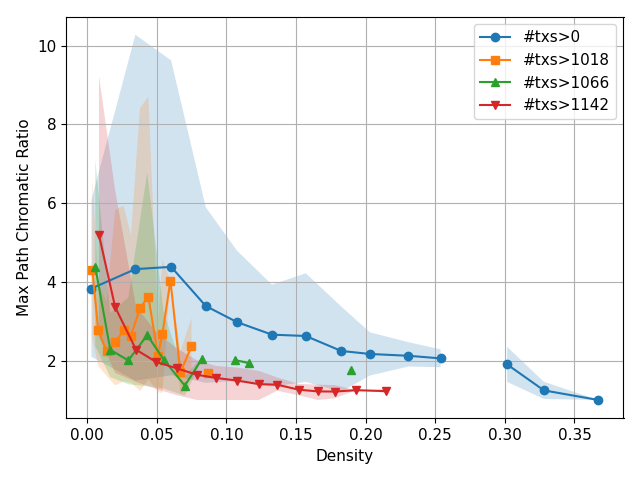}}
\caption{max-path-chromatic-ratio}
\label{fig:max-path-chromatic-ratio}
\end{figure}
\begin{figure}
\centerline{\includegraphics[width=\columnwidth]{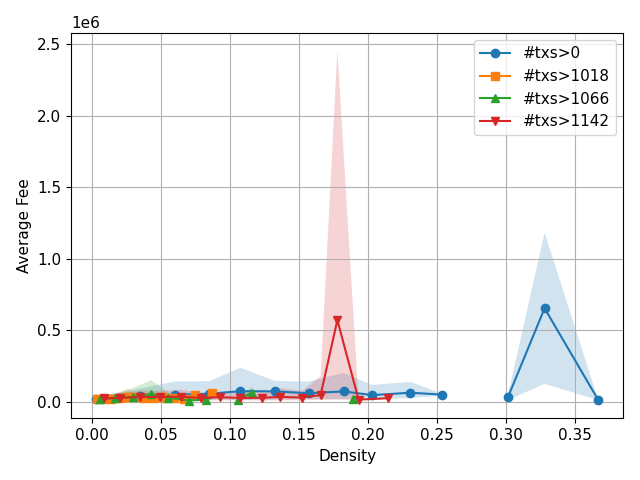}}
\caption{avg-fee}
\label{fig:avg-fee}
\end{figure}
\begin{figure}
\centerline{\includegraphics[width=\columnwidth]{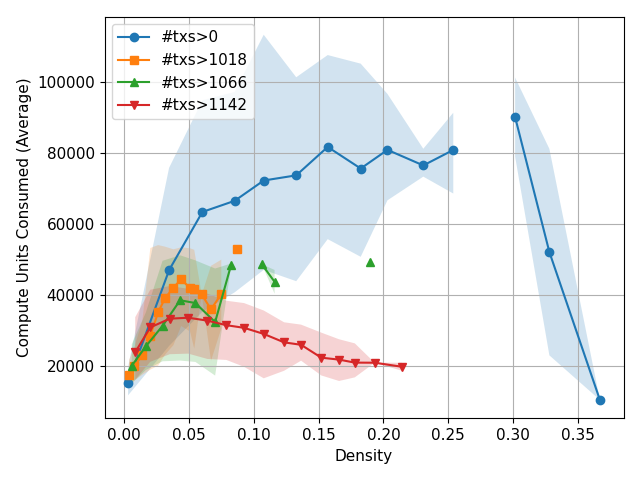}}
\caption{avg-computeUnitsConsumed}
\label{fig:avg-computeUnitsConsumed}
\end{figure}
\begin{figure}
\centerline{\includegraphics[width=\columnwidth]{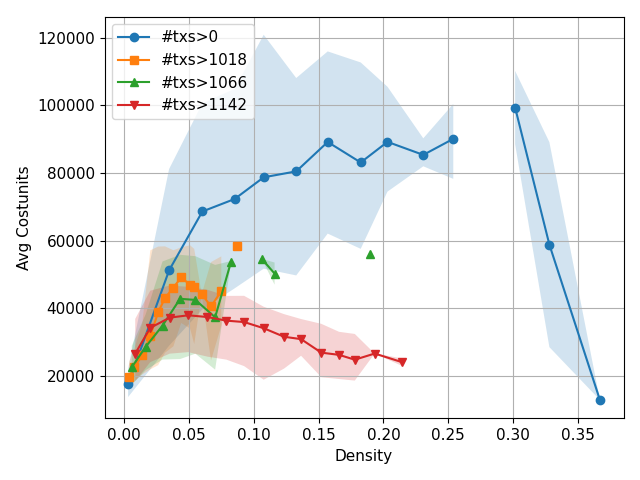}}
\caption{avg-costUnits}
\label{fig:avg-costUnits}
\end{figure}
\begin{figure}
\centerline{\includegraphics[width=\columnwidth]{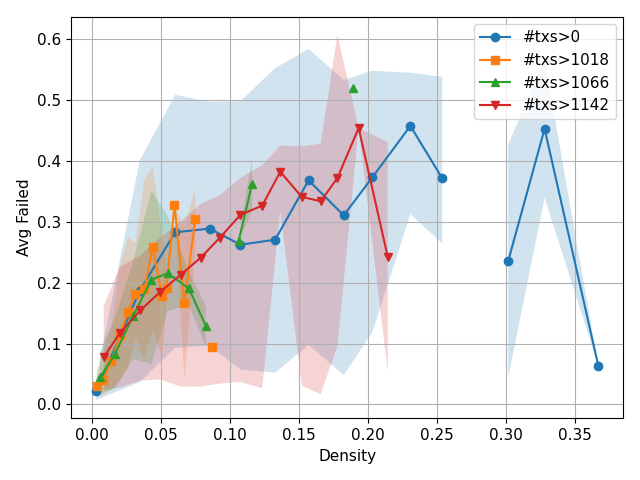}}
\caption{avg-failed}
\label{fig:avg-failed}
\end{figure}
\begin{figure}
\centerline{\includegraphics[width=\columnwidth]{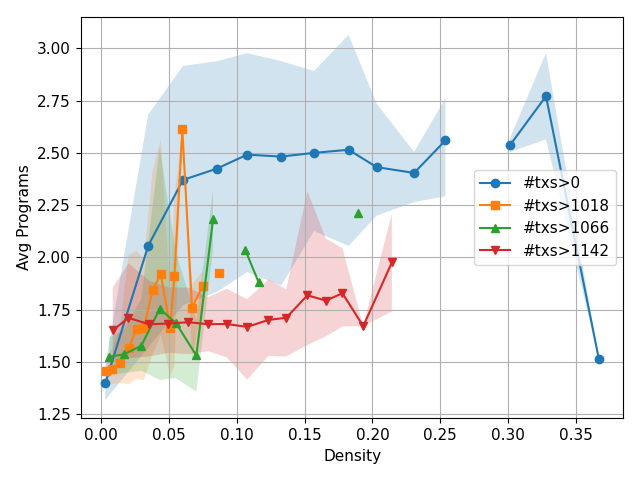}}
\caption{avg-programs}
\label{fig:avg-programs}
\end{figure}
\begin{figure}
\centerline{\includegraphics[width=\columnwidth]{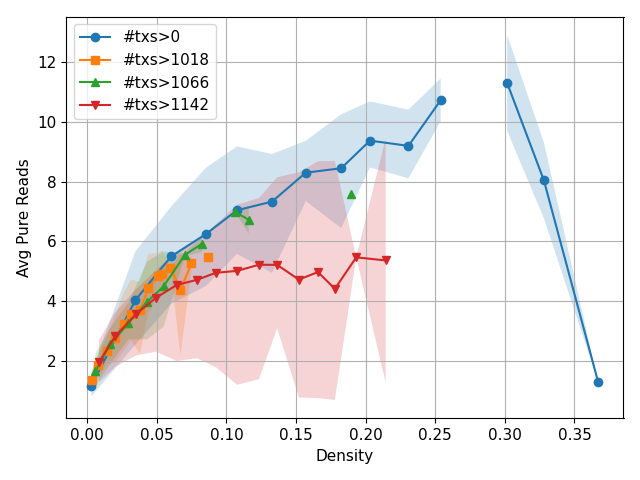}}
\caption{avg-pure-reads}
\label{fig:avg-pure-reads}
\end{figure}
\begin{figure}
\centerline{\includegraphics[width=\columnwidth]{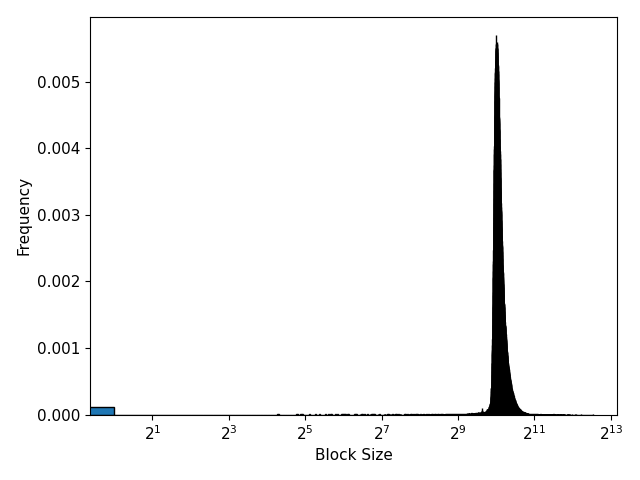}}
\caption{block-size-dist}
\label{fig:block-size-dist}
\end{figure}
\begin{figure}
\centerline{\includegraphics[width=\columnwidth]{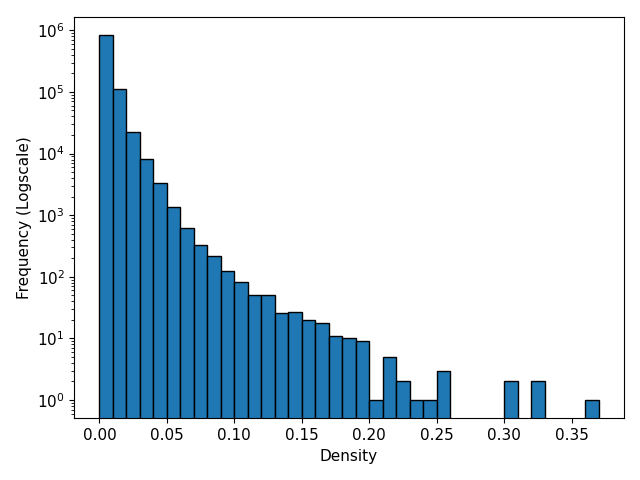}}
\caption{density-dist}
\label{fig:density-dist}
\end{figure}
\begin{figure}
\centerline{\includegraphics[width=\columnwidth]{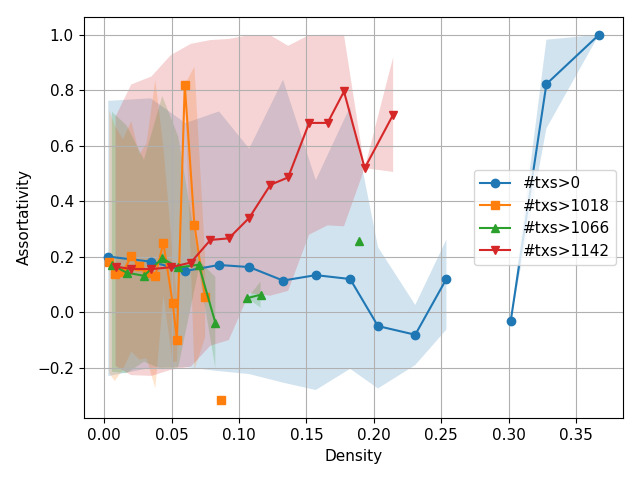}}
\caption{assortativity}
\label{fig:assortativity}
\end{figure}
\begin{figure}
\centerline{\includegraphics[width=\columnwidth]{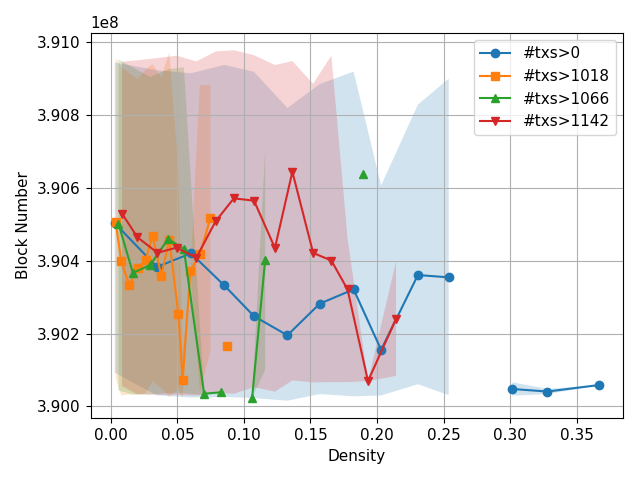}}
\caption{block-number}
\label{fig:block-number}
\end{figure}
\begin{figure}
\centerline{\includegraphics[width=\columnwidth]{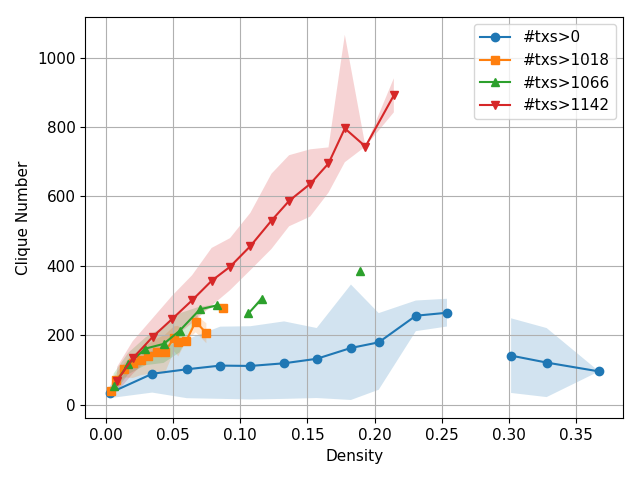}}
\caption{clique-number}
\label{fig:clique-number}
\end{figure}
\begin{figure}
\centerline{\includegraphics[width=\columnwidth]{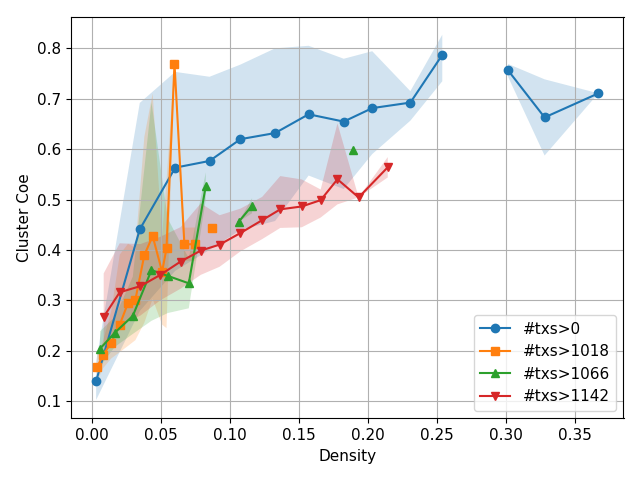}}
\caption{cluster-coe}
\label{fig:cluster-coe}
\end{figure}
\begin{figure}
\centerline{\includegraphics[width=\columnwidth]{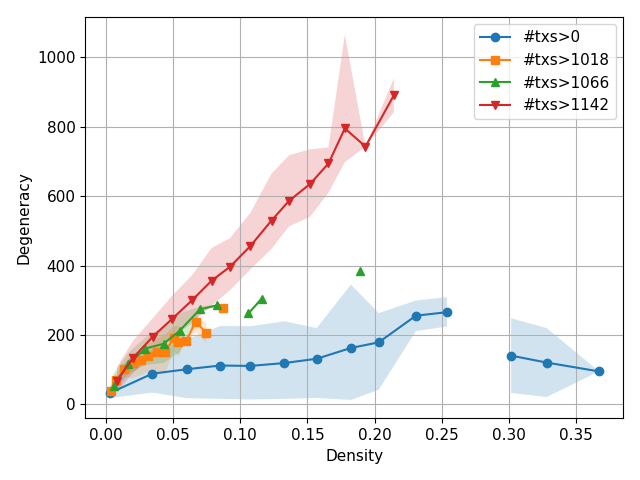}}
\caption{degeneracy}
\label{fig:degeneracy}
\end{figure}
\begin{figure}
\centerline{\includegraphics[width=\columnwidth]{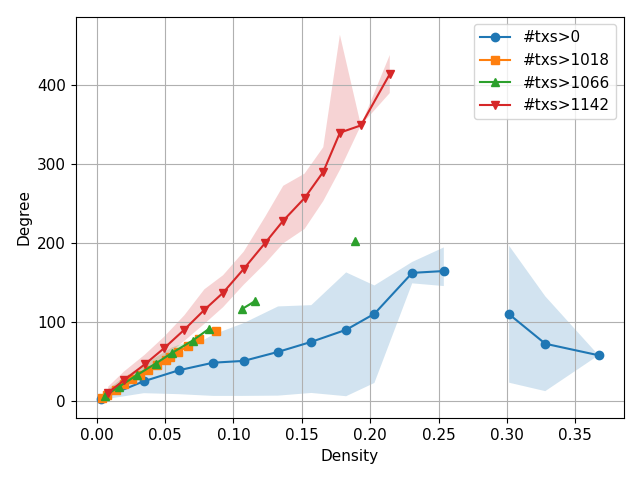}}
\caption{degree}
\label{fig:degree}
\end{figure}
\begin{figure}
\centerline{\includegraphics[width=\columnwidth]{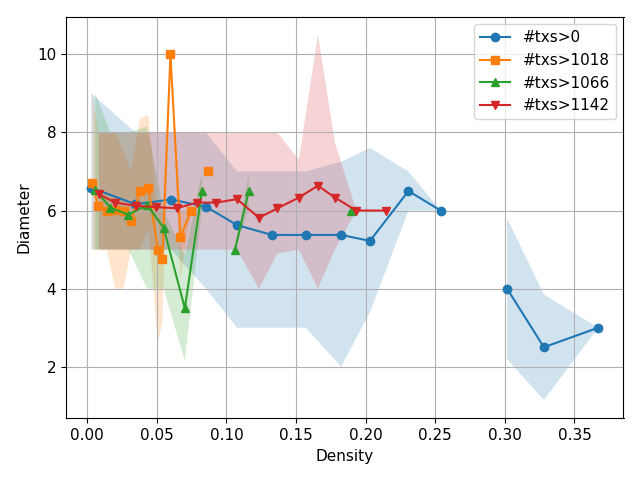}}
\caption{diameter}
\label{fig:diameter}
\end{figure}
\begin{figure}
\centerline{\includegraphics[width=\columnwidth]{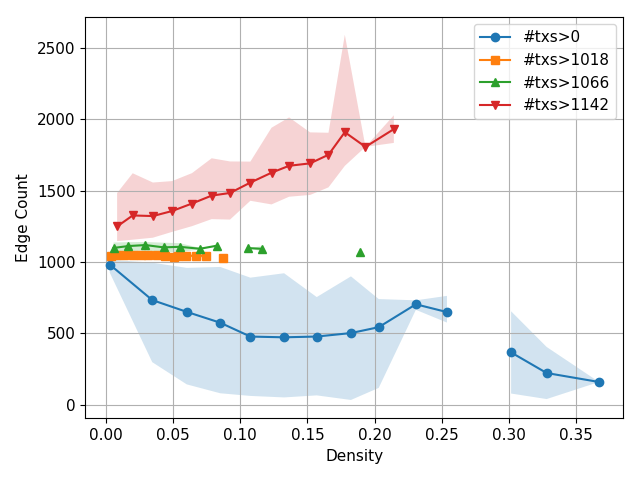}}
\caption{edge-count}
\label{fig:edge-count}
\end{figure}
\begin{figure}
\centerline{\includegraphics[width=\columnwidth]{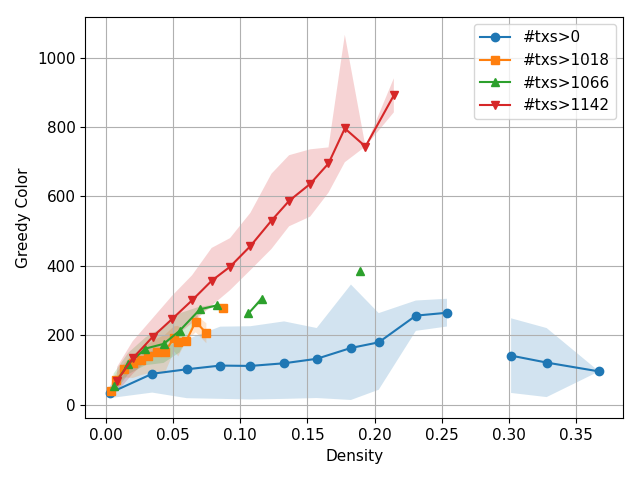}}
\caption{greedy-color}
\label{fig:greedy-color}
\end{figure}
\begin{figure}
\centerline{\includegraphics[width=\columnwidth]{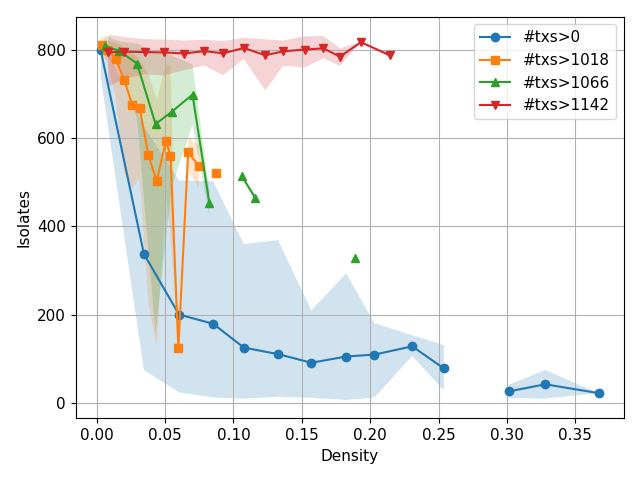}}
\caption{isolates}
\label{fig:isolates}
\end{figure}
\begin{figure}
\centerline{\includegraphics[width=\columnwidth]{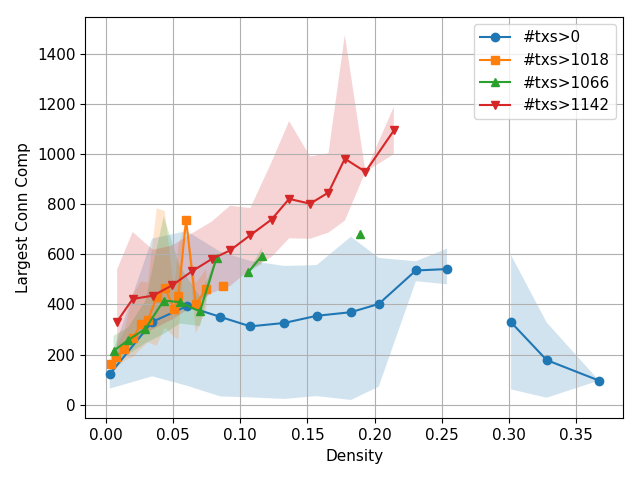}}
\caption{largest-conn-comp}
\label{fig:largest-conn-comp}
\end{figure}
\begin{figure}
\centerline{\includegraphics[width=\columnwidth]{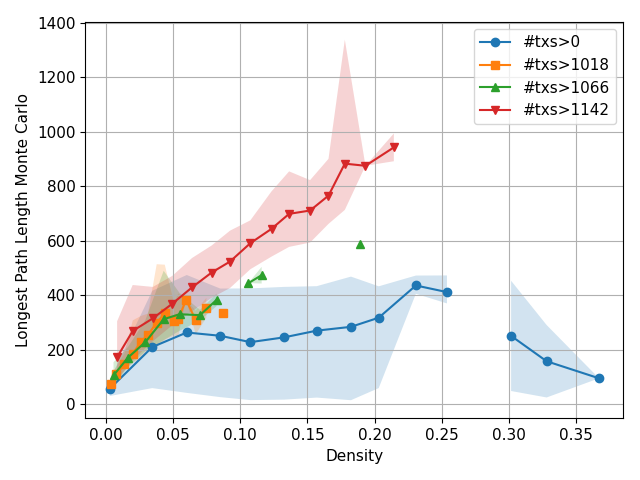}}
\caption{longest-path-length-monte-carlo}
\label{fig:longest-path-length-monte-carlo}
\end{figure}
\begin{figure}
\centerline{\includegraphics[width=\columnwidth]{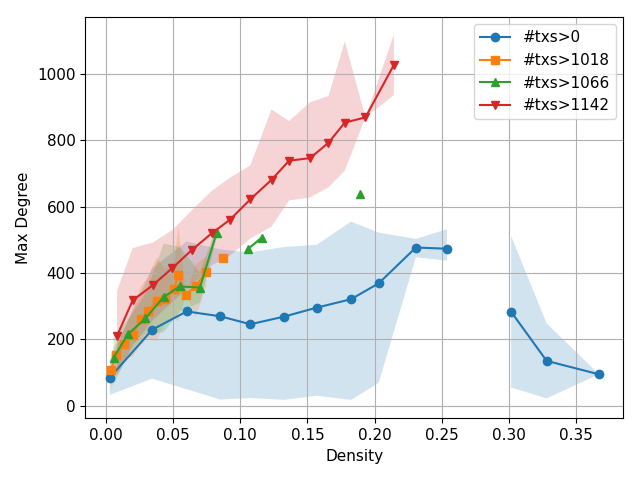}}
\caption{max-degree}
\label{fig:max-degree}
\end{figure}
\begin{figure}
\centerline{\includegraphics[width=\columnwidth]{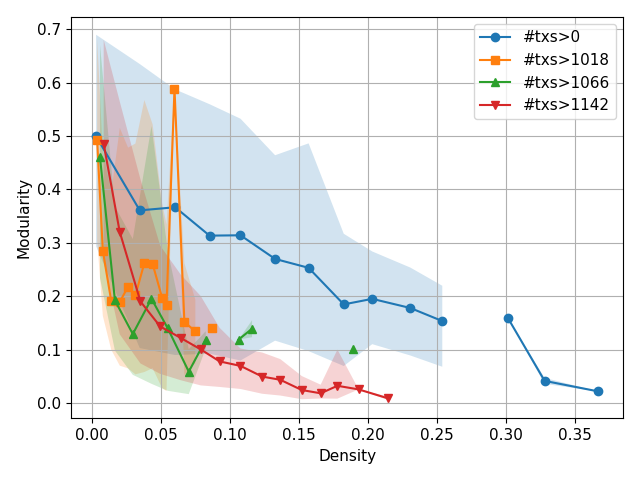}}
\caption{modularity}
\label{fig:modularity}
\end{figure}
\begin{figure}
\centerline{\includegraphics[width=\columnwidth]{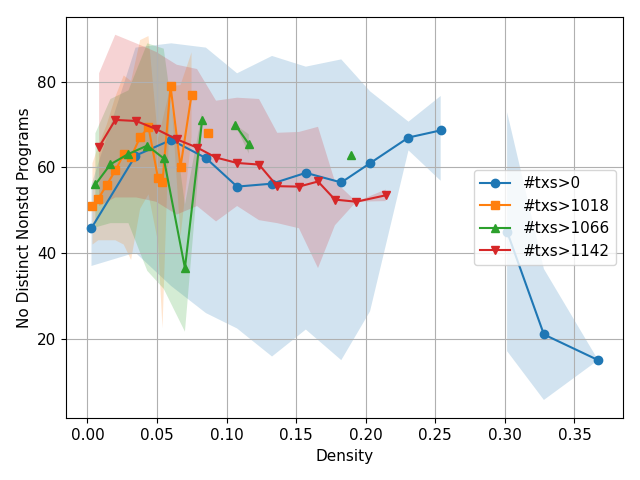}}
\caption{no-distinct-nonstd-programs}
\label{fig:no-distinct-nonstd-programs}
\end{figure}
\begin{figure}
\centerline{\includegraphics[width=\columnwidth]{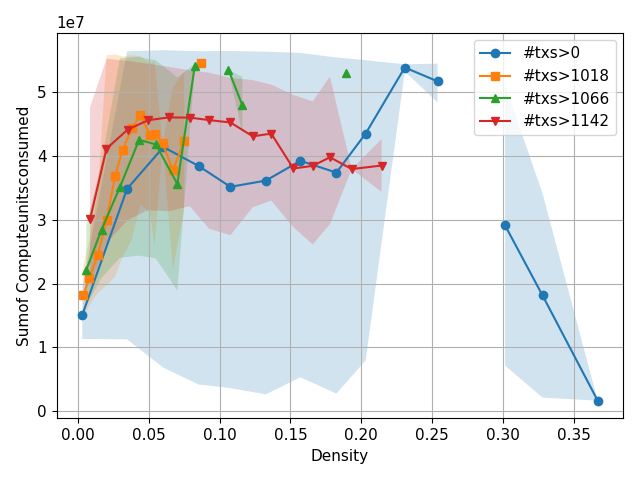}}
\caption{sumof-computeUnitsConsumed}
\label{fig:sumof-computeUnitsConsumed}
\end{figure}
\begin{figure}
\centerline{\includegraphics[width=\columnwidth]{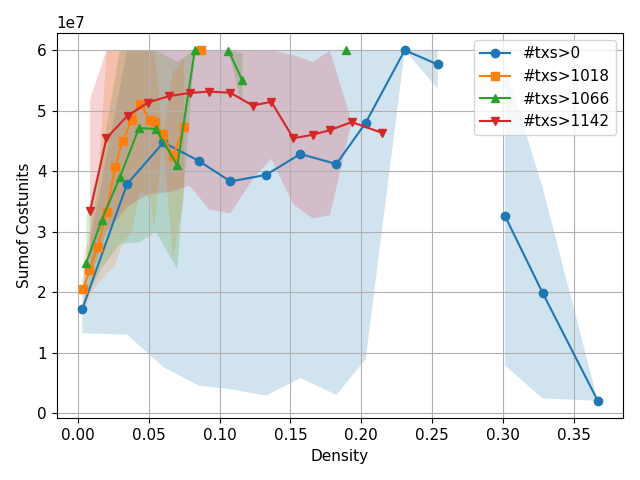}}
\caption{sumof-costUnits}
\label{fig:sumof-costUnits}
\end{figure}
\begin{figure}
\centerline{\includegraphics[width=\columnwidth]{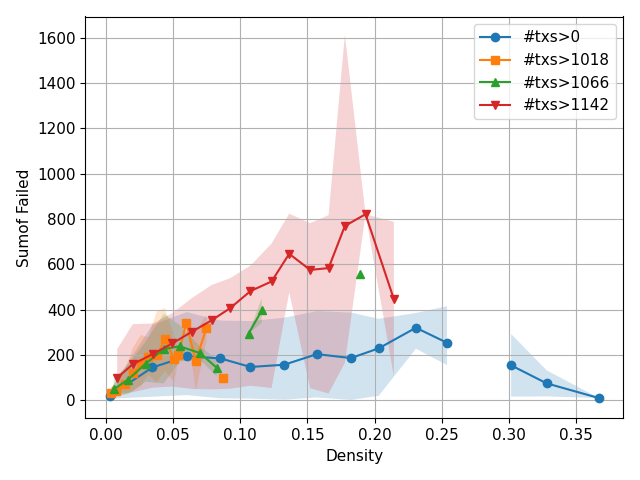}}
\caption{sumof-failed}
\label{fig:sumof-failed}
\end{figure}
\begin{figure}
\centerline{\includegraphics[width=\columnwidth]{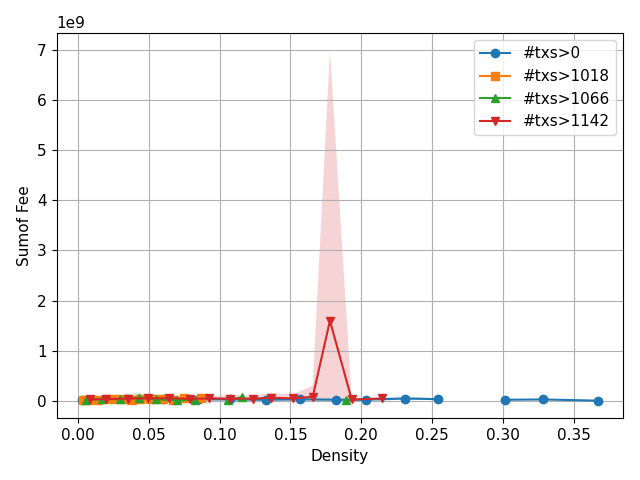}}
\caption{sumof-fee}
\label{fig:sumof-fee}
\end{figure}
\begin{figure}
\centerline{\includegraphics[width=\columnwidth]{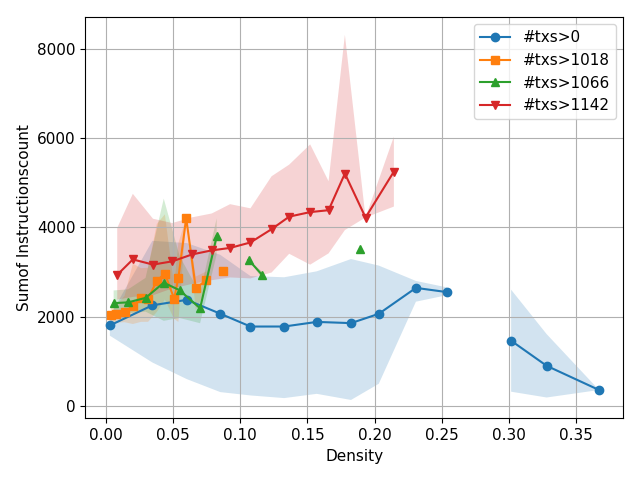}}
\caption{sumof-instructionsCount}
\label{fig:sumof-instructionsCount}
\end{figure}
\begin{figure}
\centerline{\includegraphics[width=\columnwidth]{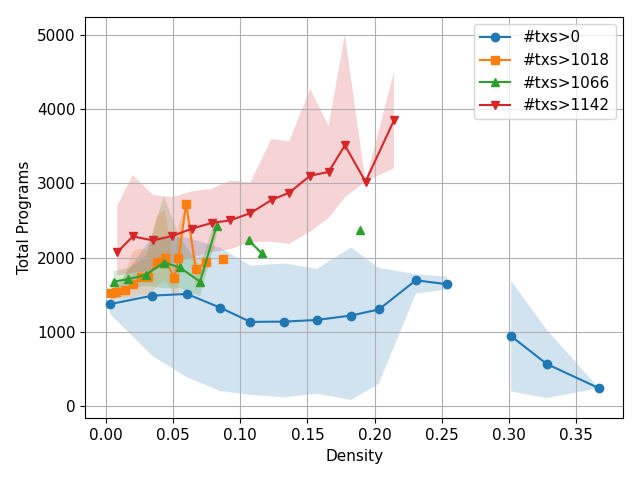}}
\caption{total-programs}
\label{fig:total-programs}
\end{figure}
\begin{figure}
\centerline{\includegraphics[width=\columnwidth]{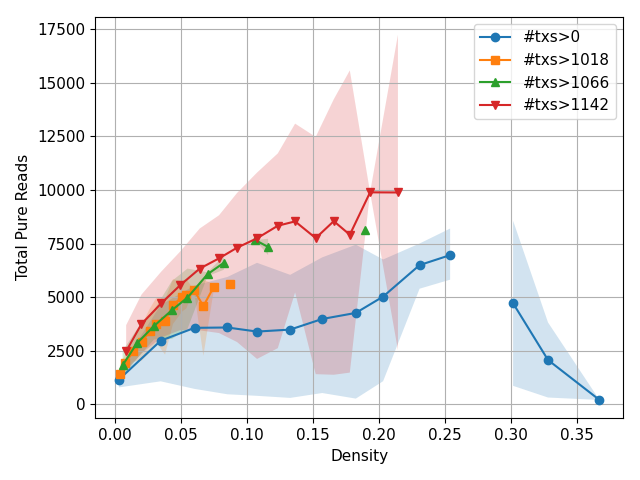}}
\caption{total-pure-reads}
\label{fig:total-pure-reads}
\end{figure}
\begin{figure}
\centerline{\includegraphics[width=\columnwidth]{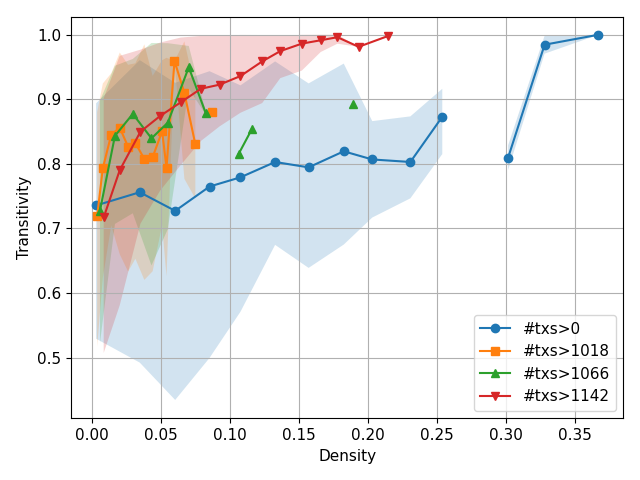}}
\caption{transitivity}
\label{fig:transitivity}
\end{figure}
\begin{figure}
\centerline{\includegraphics[width=\columnwidth]{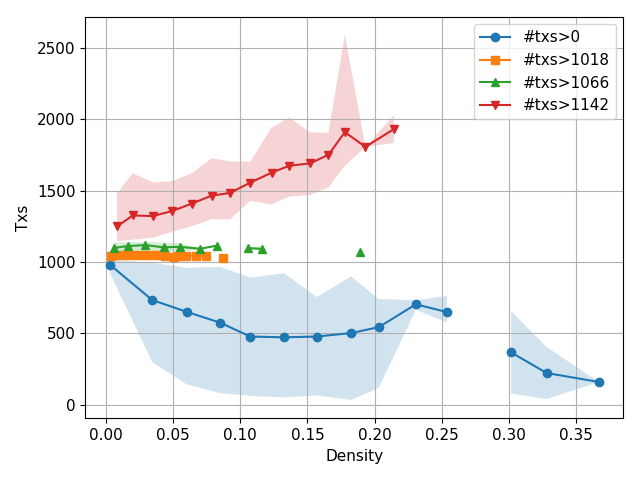}}
\caption{txs}
\label{fig:txs}
\end{figure}
\begin{figure}
\centerline{\includegraphics[width=\columnwidth]{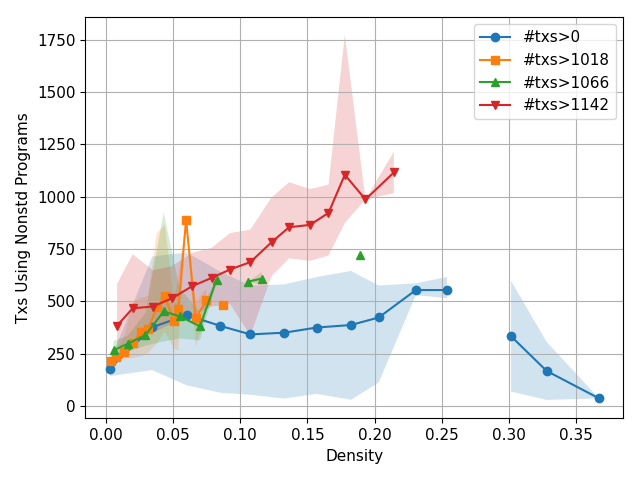}}
\caption{txs-using-nonstd-programs}
\label{fig:txs-using-nonstd-programs}
\end{figure}
\begin{figure}
\centerline{\includegraphics[width=\columnwidth]{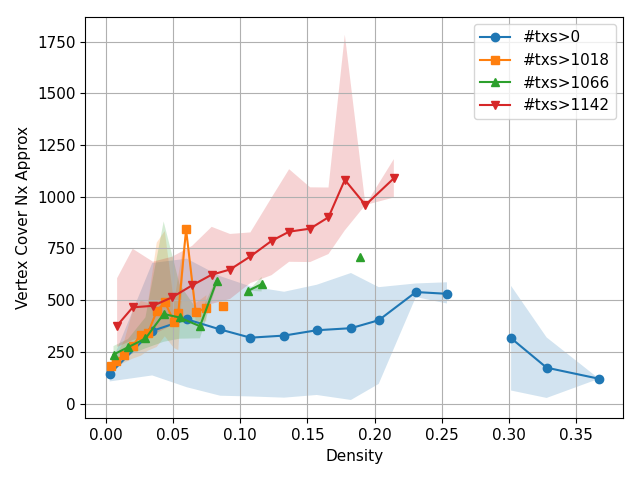}}
\caption{vertex-cover-nx-approx}
\label{fig:vertex-cover-nx-approx}
\end{figure}
\begin{figure}
\centerline{\includegraphics[width=\columnwidth]{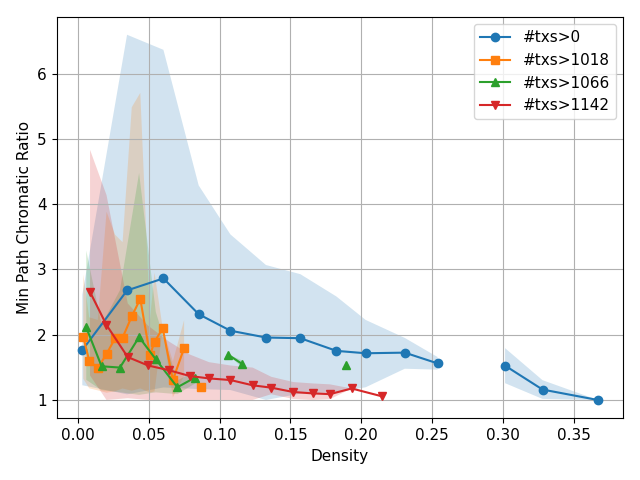}}
\caption{min-path-chromatic-ratio}
\label{fig:min-path-chromatic-ratio}
\end{figure}
\begin{figure}
\centerline{\includegraphics[width=\columnwidth]{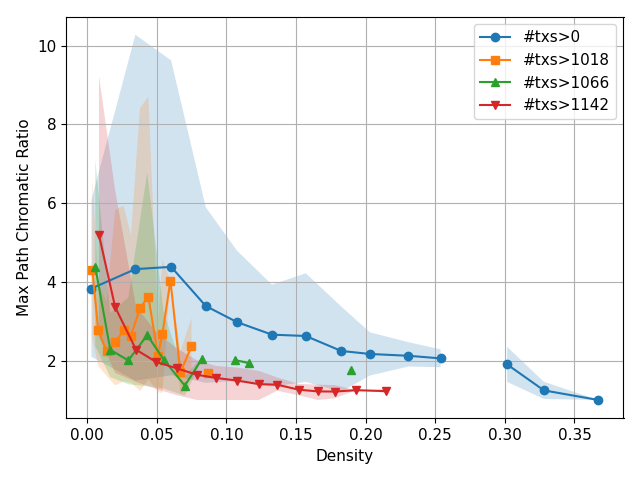}}
\caption{max-path-chromatic-ratio}
\label{fig:max-path-chromatic-ratio}
\end{figure}
\begin{figure}
\centerline{\includegraphics[width=\columnwidth]{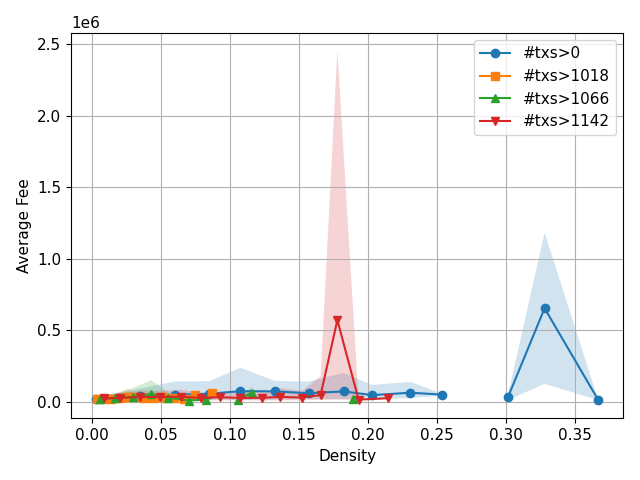}}
\caption{avg-fee}
\label{fig:avg-fee}
\end{figure}
\begin{figure}
\centerline{\includegraphics[width=\columnwidth]{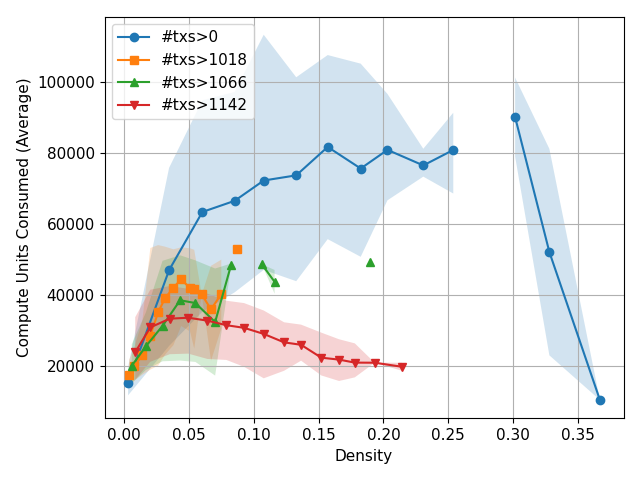}}
\caption{avg-computeUnitsConsumed}
\label{fig:avg-computeUnitsConsumed}
\end{figure}
\begin{figure}
\centerline{\includegraphics[width=\columnwidth]{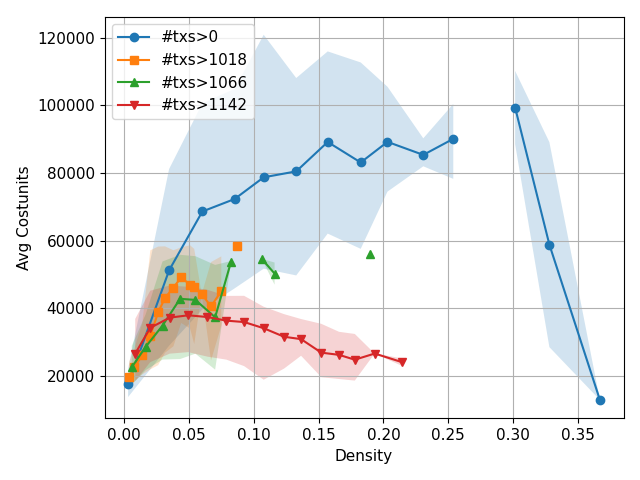}}
\caption{avg-costUnits}
\label{fig:avg-costUnits}
\end{figure}
\begin{figure}
\centerline{\includegraphics[width=\columnwidth]{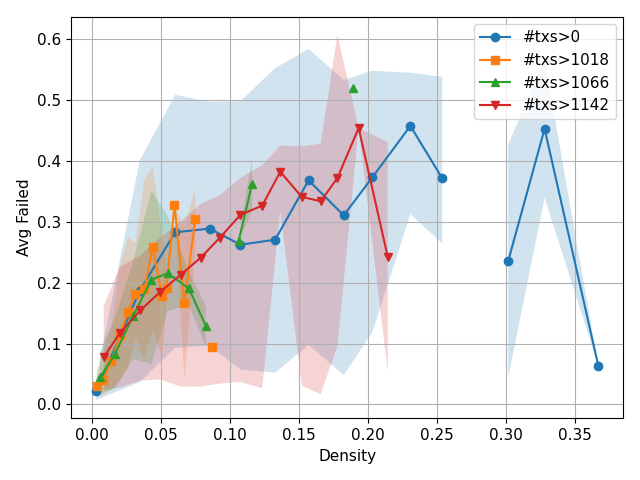}}
\caption{avg-failed}
\label{fig:avg-failed}
\end{figure}
\begin{figure}
\centerline{\includegraphics[width=\columnwidth]{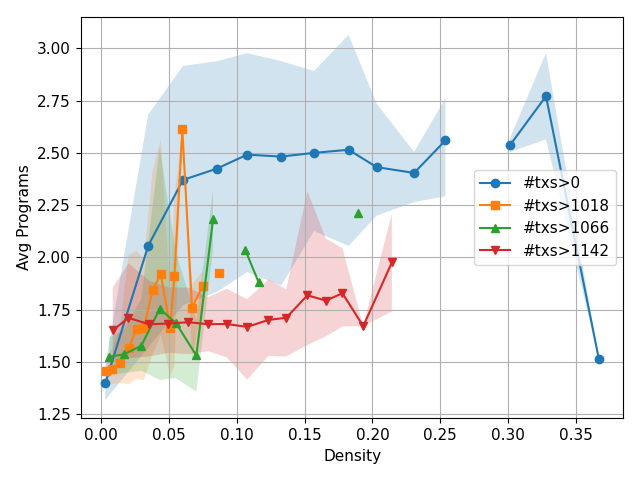}}
\caption{avg-programs}
\label{fig:avg-programs}
\end{figure}
\begin{figure}
\centerline{\includegraphics[width=\columnwidth]{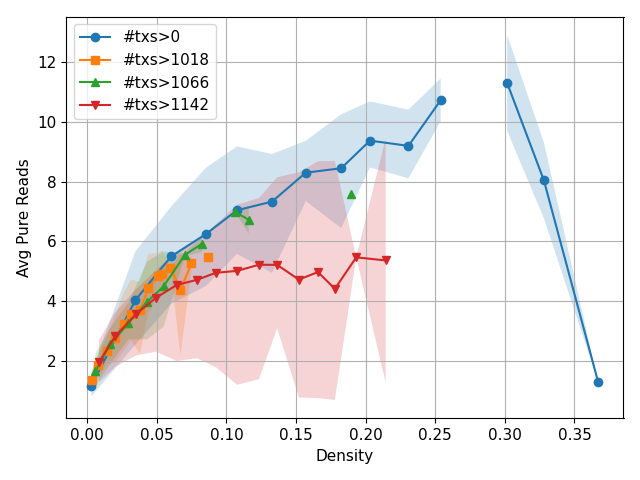}}
\caption{avg-pure-reads}
\label{fig:avg-pure-reads}
\end{figure}
\begin{figure}
\centerline{\includegraphics[width=\columnwidth]{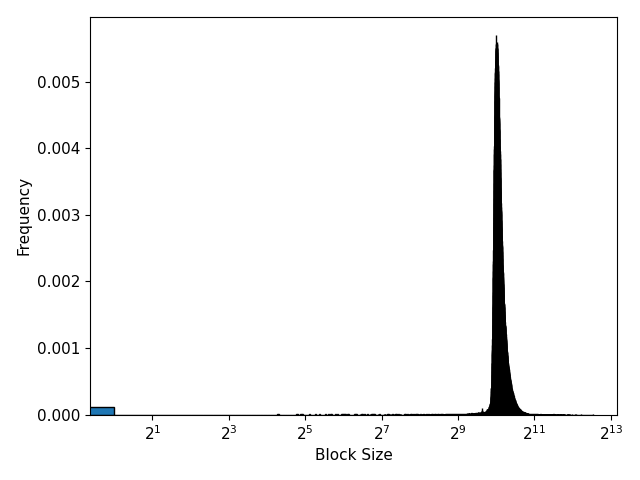}}
\caption{block-size-dist}
\label{fig:block-size-dist}
\end{figure}
\begin{figure}
\centerline{\includegraphics[width=\columnwidth]{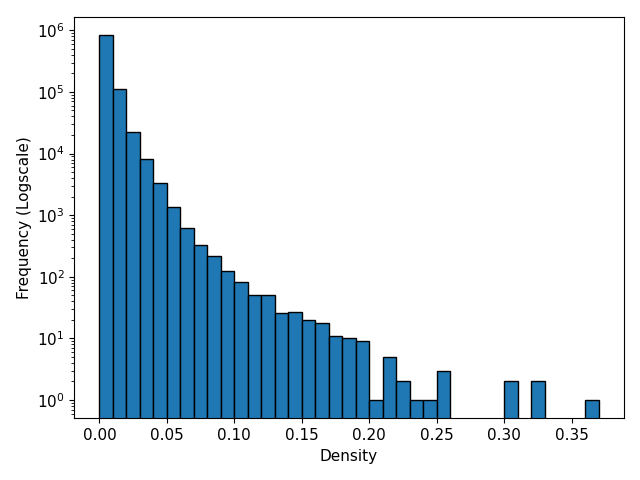}}
\caption{density-dist}
\label{fig:density-dist}
\end{figure}
\begin{figure}
\centerline{\includegraphics[width=\columnwidth]{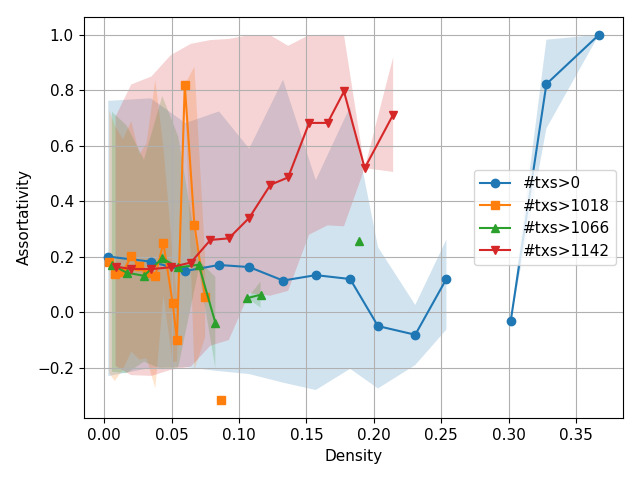}}
\caption{assortativity}
\label{fig:assortativity}
\end{figure}
\begin{figure}
\centerline{\includegraphics[width=\columnwidth]{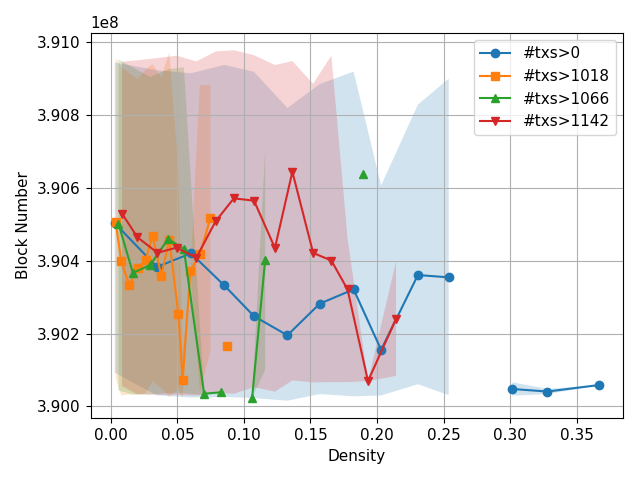}}
\caption{block-number}
\label{fig:block-number}
\end{figure}
\begin{figure}
\centerline{\includegraphics[width=\columnwidth]{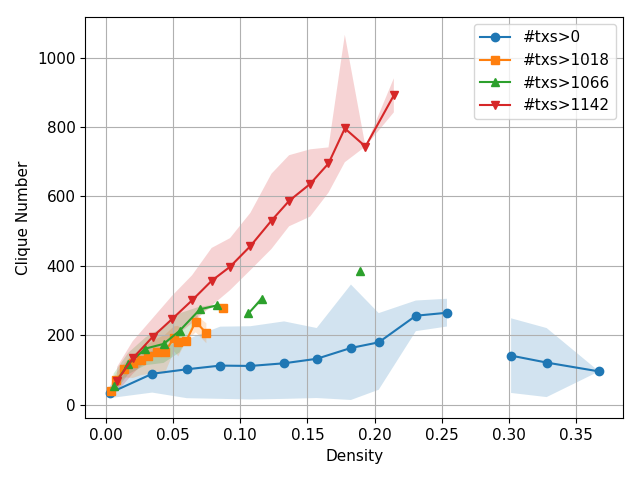}}
\caption{clique-number}
\label{fig:clique-number}
\end{figure}
\begin{figure}
\centerline{\includegraphics[width=\columnwidth]{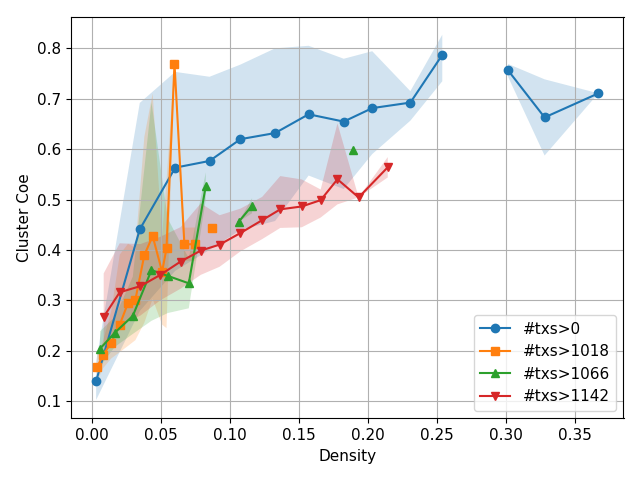}}
\caption{cluster-coe}
\label{fig:cluster-coe}
\end{figure}
\begin{figure}
\centerline{\includegraphics[width=\columnwidth]{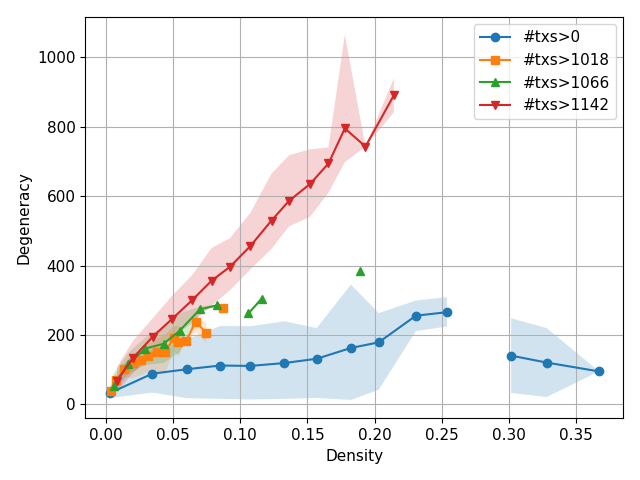}}
\caption{degeneracy}
\label{fig:degeneracy}
\end{figure}
\begin{figure}
\centerline{\includegraphics[width=\columnwidth]{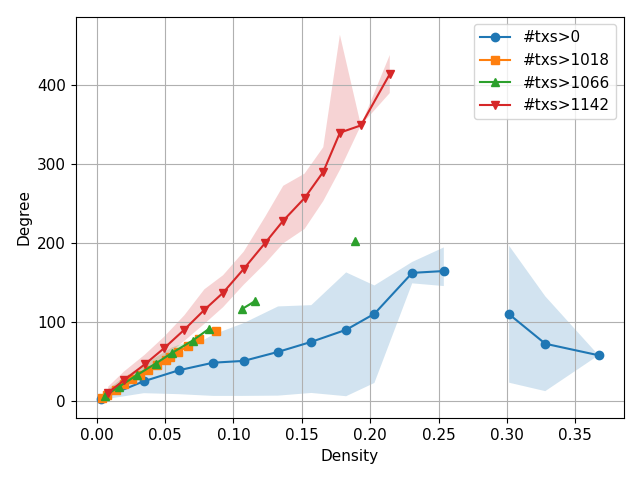}}
\caption{degree}
\label{fig:degree}
\end{figure}
\begin{figure}
\centerline{\includegraphics[width=\columnwidth]{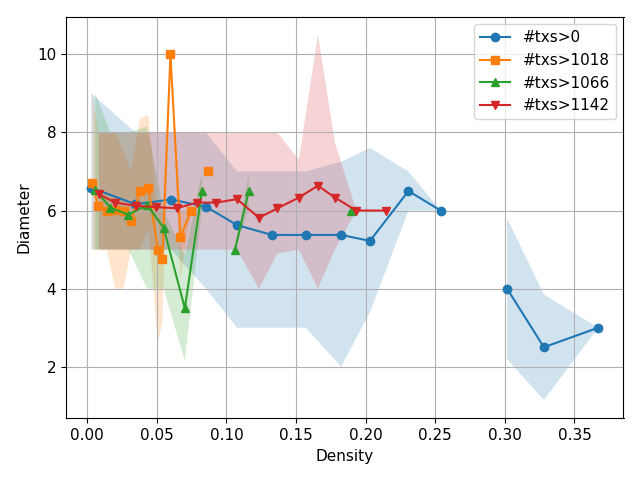}}
\caption{diameter}
\label{fig:diameter}
\end{figure}
\begin{figure}
\centerline{\includegraphics[width=\columnwidth]{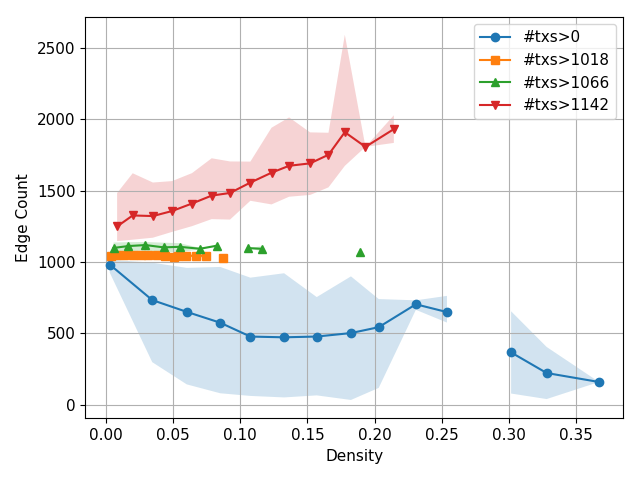}}
\caption{edge-count}
\label{fig:edge-count}
\end{figure}
\begin{figure}
\centerline{\includegraphics[width=\columnwidth]{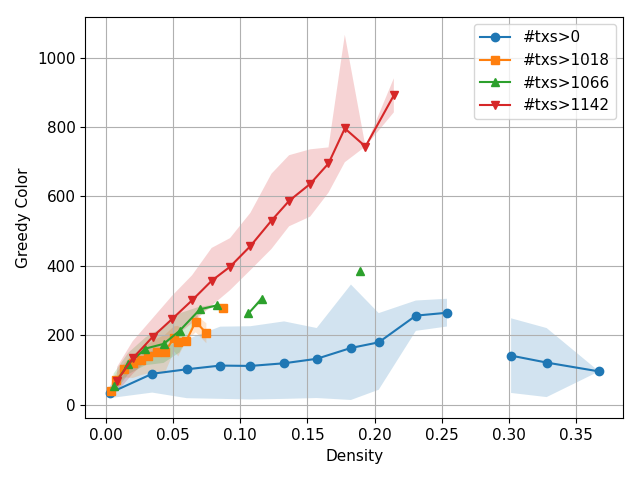}}
\caption{greedy-color}
\label{fig:greedy-color}
\end{figure}
\begin{figure}
\centerline{\includegraphics[width=\columnwidth]{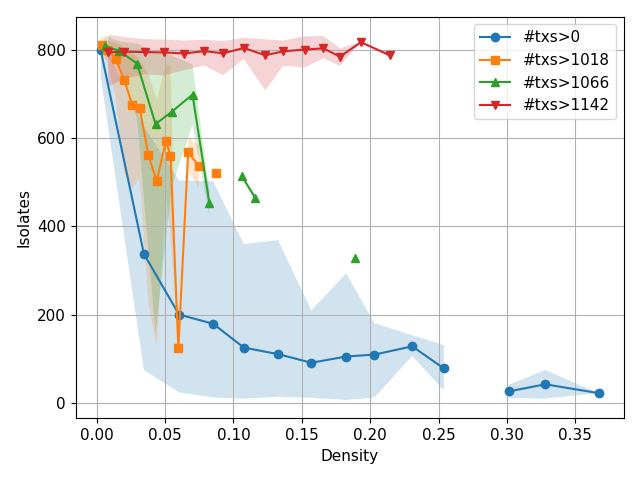}}
\caption{isolates}
\label{fig:isolates}
\end{figure}
\begin{figure}
\centerline{\includegraphics[width=\columnwidth]{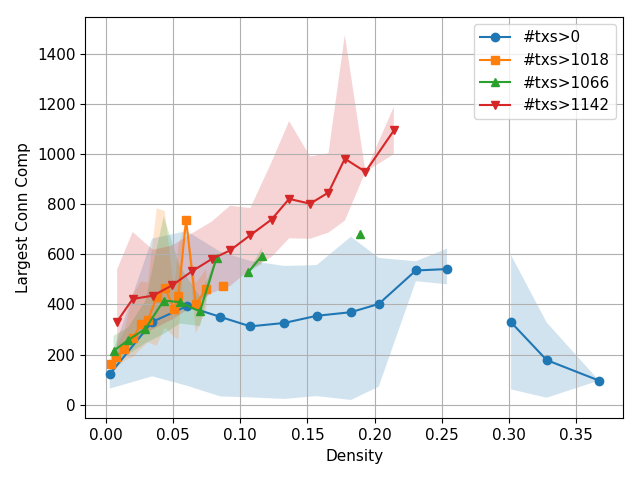}}
\caption{largest-conn-comp}
\label{fig:largest-conn-comp}
\end{figure}
\begin{figure}
\centerline{\includegraphics[width=\columnwidth]{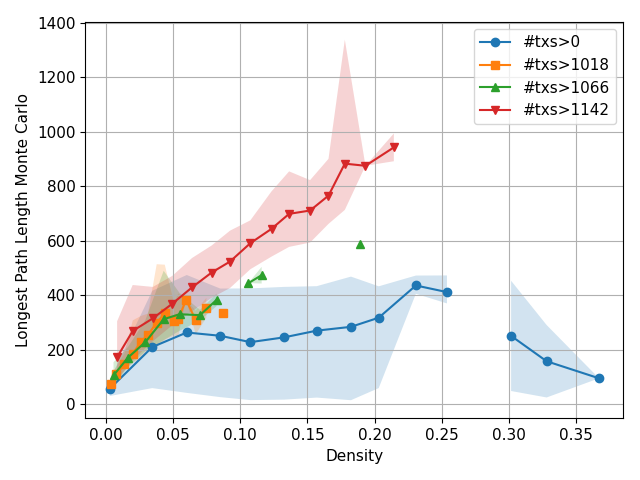}}
\caption{longest-path-length-monte-carlo}
\label{fig:longest-path-length-monte-carlo}
\end{figure}
\begin{figure}
\centerline{\includegraphics[width=\columnwidth]{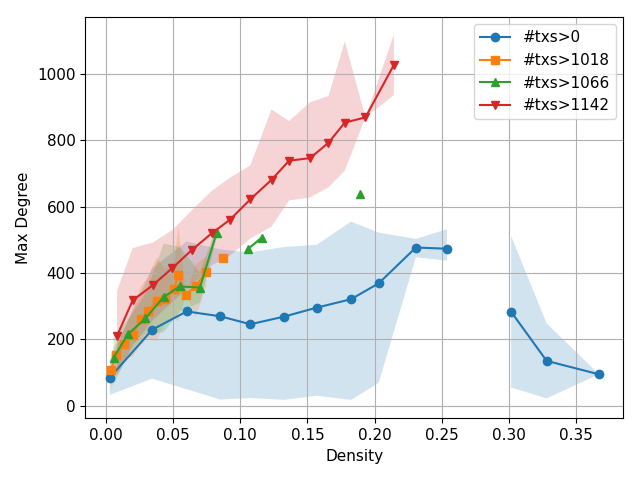}}
\caption{max-degree}
\label{fig:max-degree}
\end{figure}
\begin{figure}
\centerline{\includegraphics[width=\columnwidth]{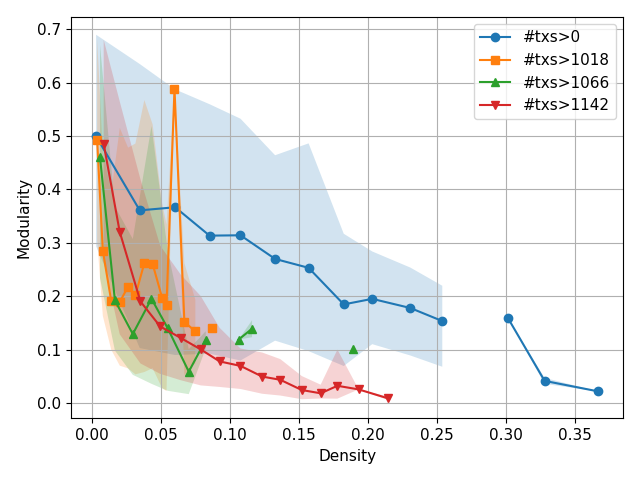}}
\caption{modularity}
\label{fig:modularity}
\end{figure}
\begin{figure}
\centerline{\includegraphics[width=\columnwidth]{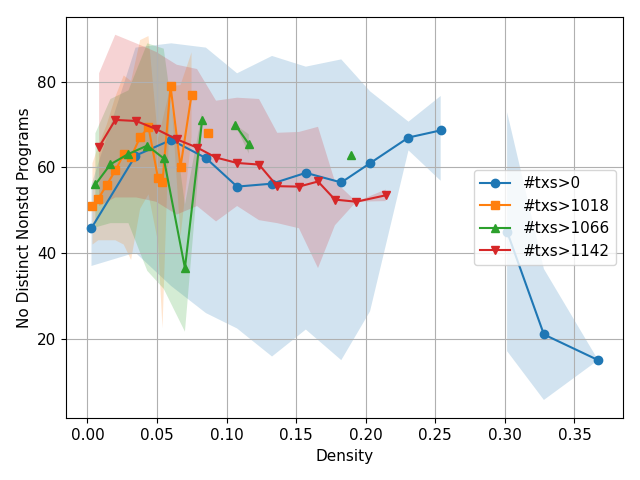}}
\caption{no-distinct-nonstd-programs}
\label{fig:no-distinct-nonstd-programs}
\end{figure}
\begin{figure}
\centerline{\includegraphics[width=\columnwidth]{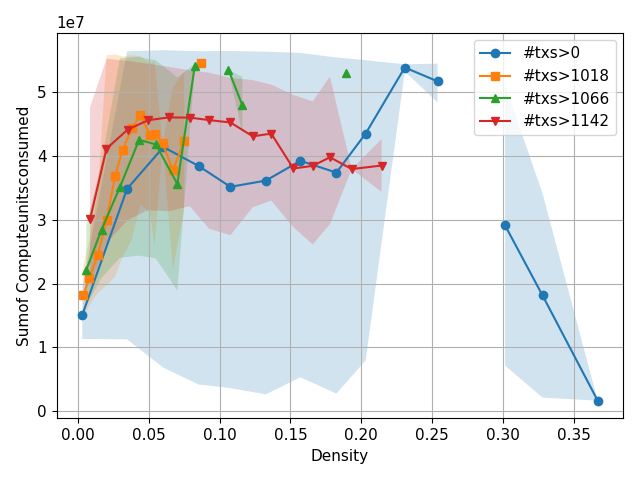}}
\caption{sumof-computeUnitsConsumed}
\label{fig:sumof-computeUnitsConsumed}
\end{figure}
\begin{figure}
\centerline{\includegraphics[width=\columnwidth]{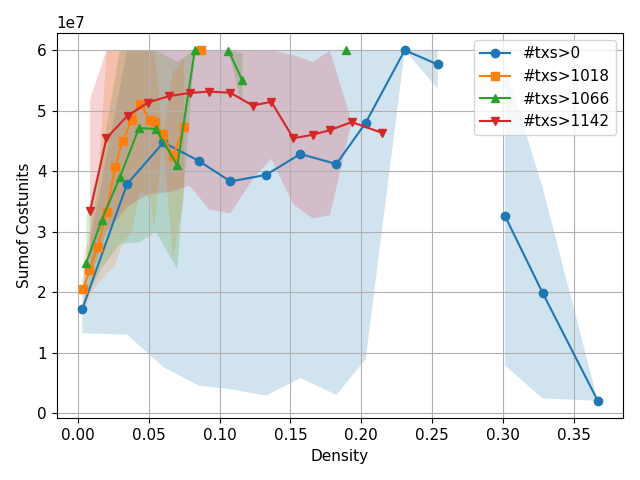}}
\caption{sumof-costUnits}
\label{fig:sumof-costUnits}
\end{figure}
\begin{figure}
\centerline{\includegraphics[width=\columnwidth]{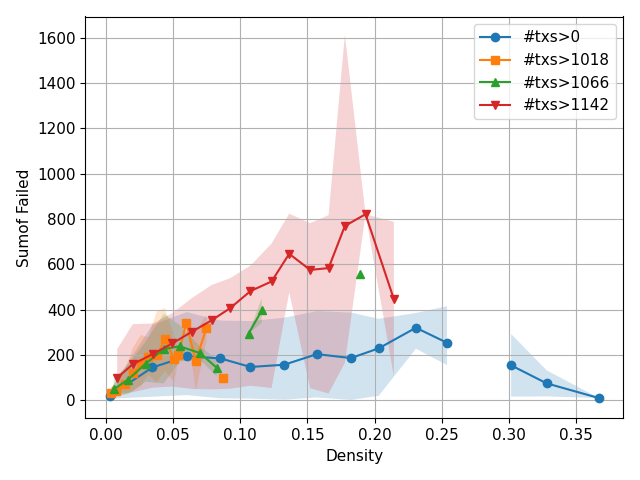}}
\caption{sumof-failed}
\label{fig:sumof-failed}
\end{figure}
\begin{figure}
\centerline{\includegraphics[width=\columnwidth]{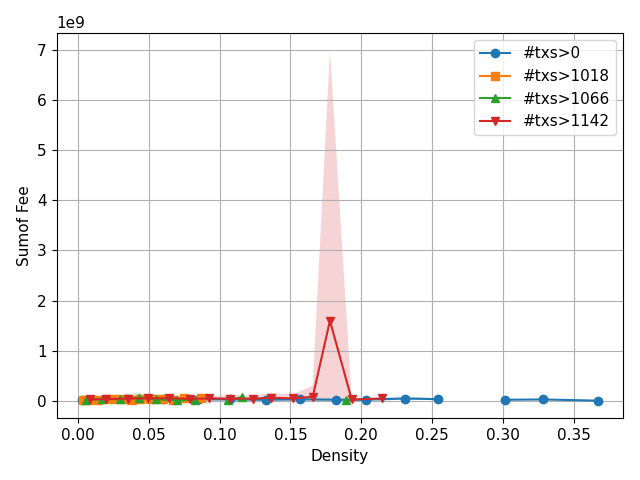}}
\caption{sumof-fee}
\label{fig:sumof-fee}
\end{figure}
\begin{figure}
\centerline{\includegraphics[width=\columnwidth]{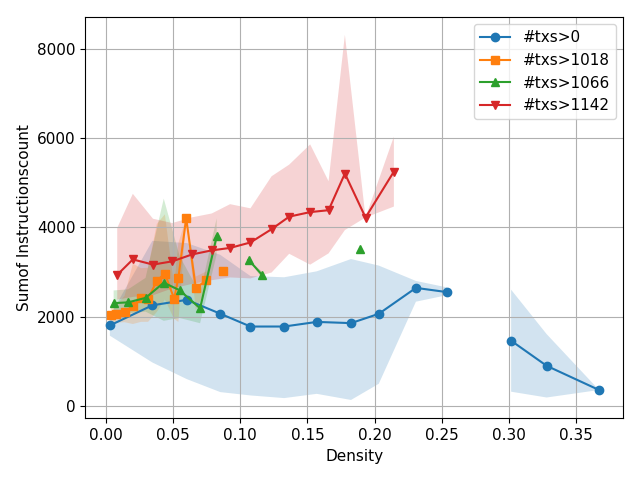}}
\caption{sumof-instructionsCount}
\label{fig:sumof-instructionsCount}
\end{figure}
\begin{figure}
\centerline{\includegraphics[width=\columnwidth]{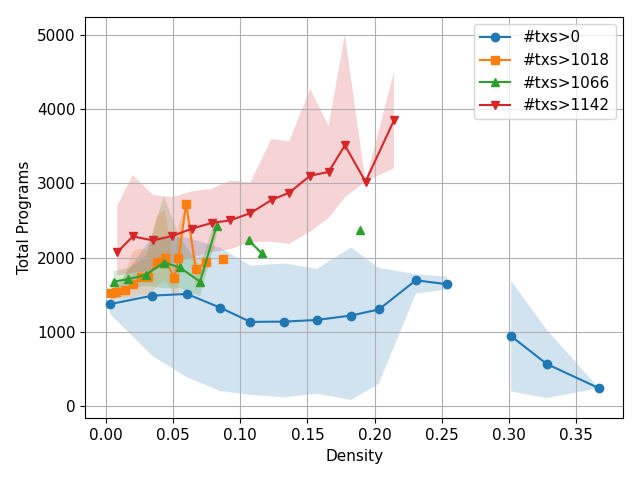}}
\caption{total-programs}
\label{fig:total-programs}
\end{figure}
\begin{figure}
\centerline{\includegraphics[width=\columnwidth]{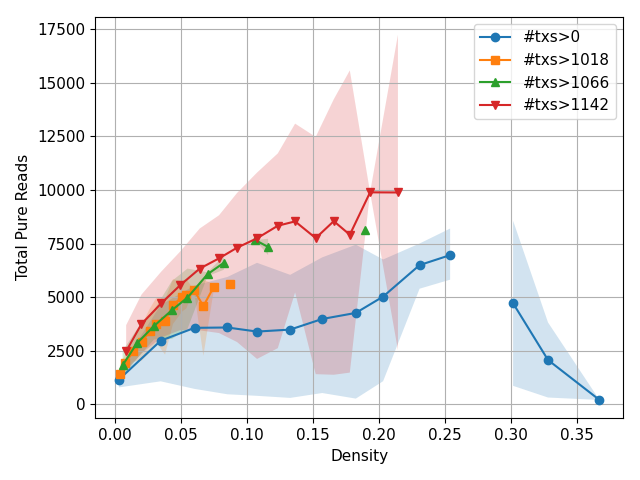}}
\caption{total-pure-reads}
\label{fig:total-pure-reads}
\end{figure}
\begin{figure}
\centerline{\includegraphics[width=\columnwidth]{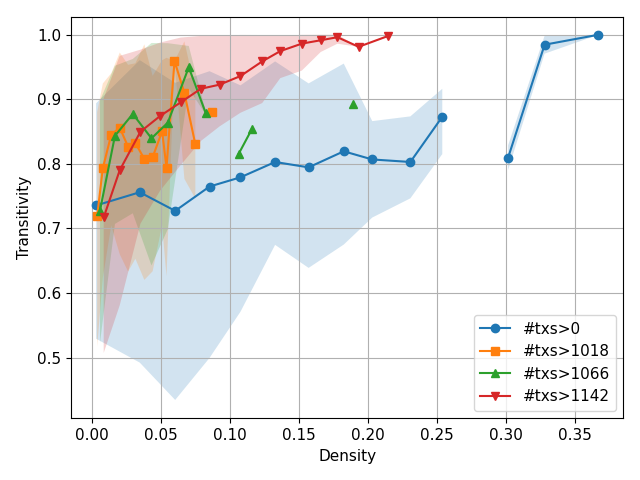}}
\caption{transitivity}
\label{fig:transitivity}
\end{figure}
\begin{figure}
\centerline{\includegraphics[width=\columnwidth]{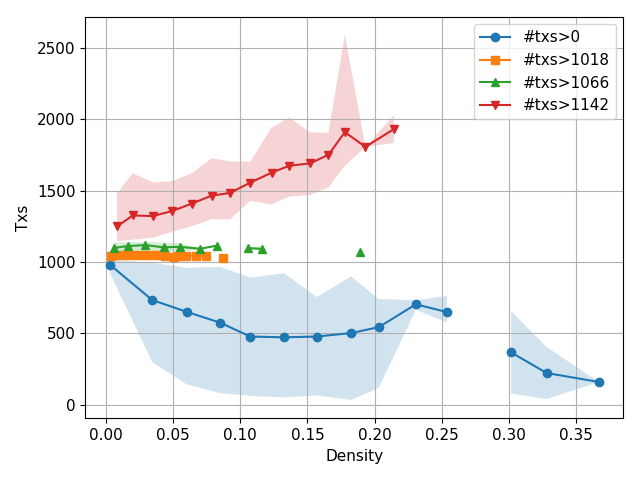}}
\caption{txs}
\label{fig:txs}
\end{figure}
\begin{figure}
\centerline{\includegraphics[width=\columnwidth]{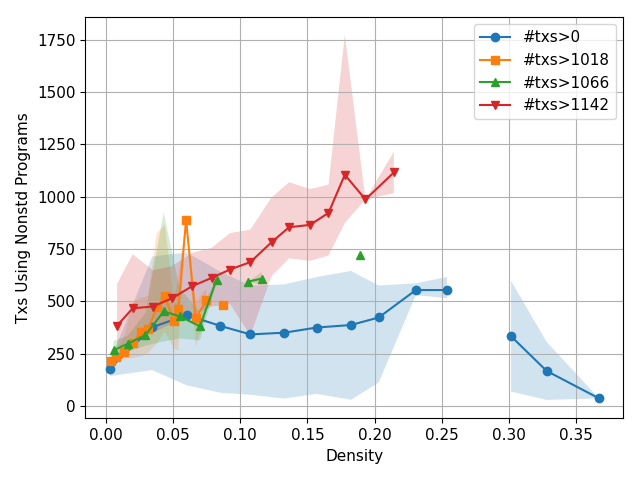}}
\caption{txs-using-nonstd-programs}
\label{fig:txs-using-nonstd-programs}
\end{figure}
\begin{figure}
\centerline{\includegraphics[width=\columnwidth]{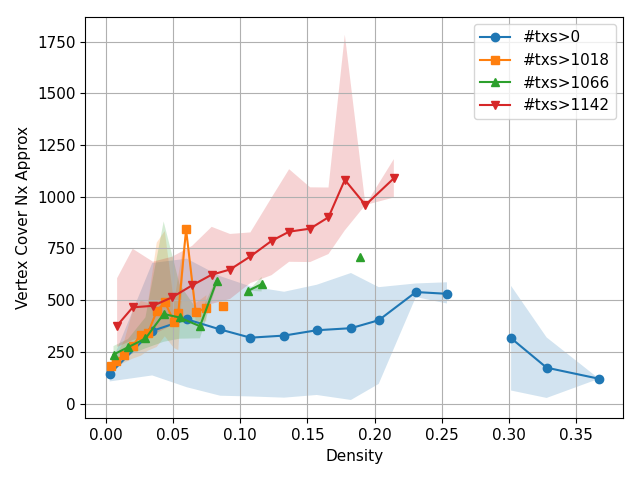}}
\caption{vertex-cover-nx-approx}
\label{fig:vertex-cover-nx-approx}
\end{figure}
\begin{figure}
\centerline{\includegraphics[width=\columnwidth]{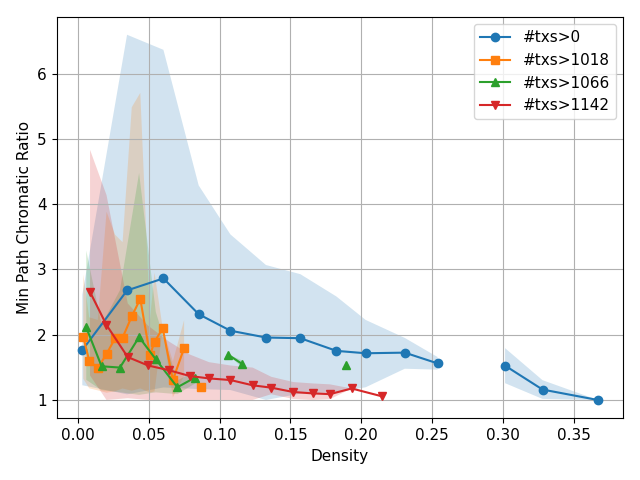}}
\caption{min-path-chromatic-ratio}
\label{fig:min-path-chromatic-ratio}
\end{figure}
\begin{figure}
\centerline{\includegraphics[width=\columnwidth]{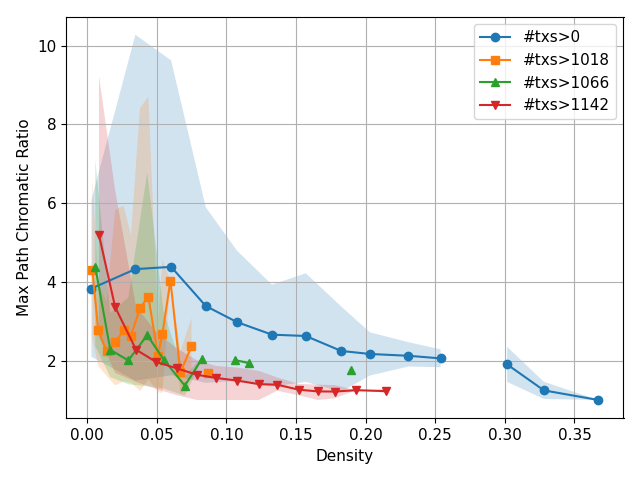}}
\caption{max-path-chromatic-ratio}
\label{fig:max-path-chromatic-ratio}
\end{figure}
\begin{figure}
\centerline{\includegraphics[width=\columnwidth]{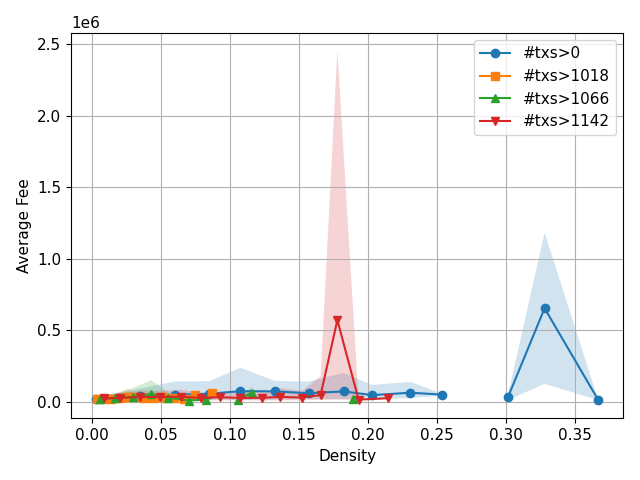}}
\caption{avg-fee}
\label{fig:avg-fee}
\end{figure}
\begin{figure}
\centerline{\includegraphics[width=\columnwidth]{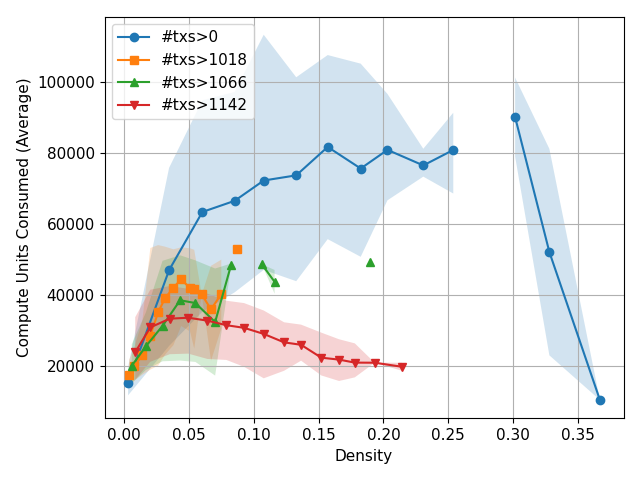}}
\caption{avg-computeUnitsConsumed}
\label{fig:avg-computeUnitsConsumed}
\end{figure}
\begin{figure}
\centerline{\includegraphics[width=\columnwidth]{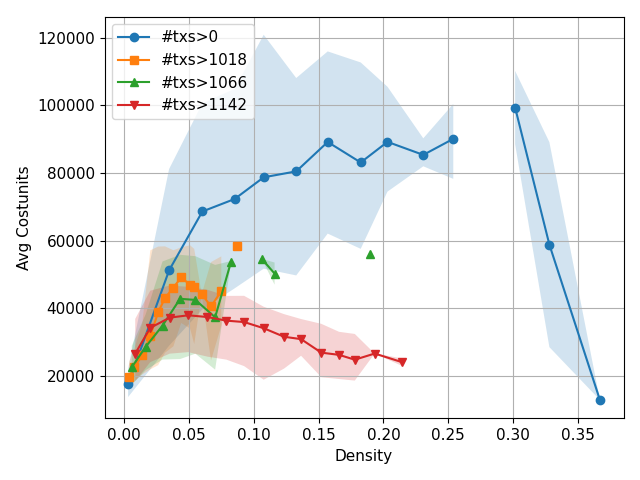}}
\caption{avg-costUnits}
\label{fig:avg-costUnits}
\end{figure}
\begin{figure}
\centerline{\includegraphics[width=\columnwidth]{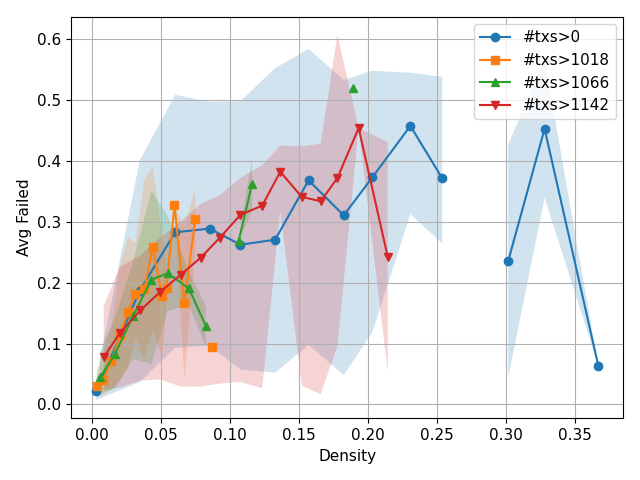}}
\caption{avg-failed}
\label{fig:avg-failed}
\end{figure}
\begin{figure}
\centerline{\includegraphics[width=\columnwidth]{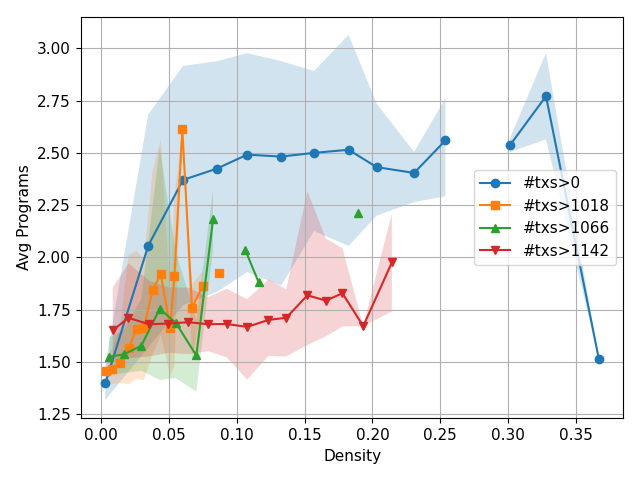}}
\caption{avg-programs}
\label{fig:avg-programs}
\end{figure}
\begin{figure}
\centerline{\includegraphics[width=\columnwidth]{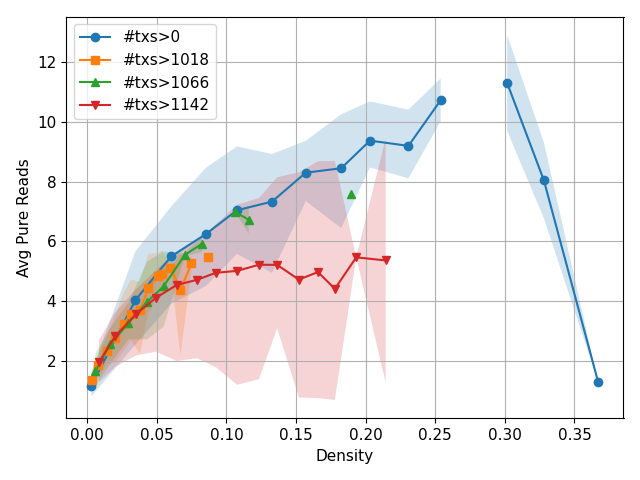}}
\caption{avg-pure-reads}
\label{fig:avg-pure-reads}
\end{figure}
\begin{figure}
\centerline{\includegraphics[width=\columnwidth]{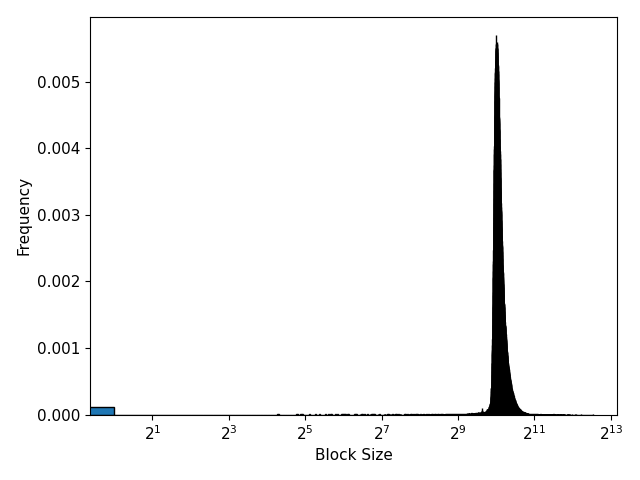}}
\caption{block-size-dist}
\label{fig:block-size-dist}
\end{figure}
\begin{figure}
\centerline{\includegraphics[width=\columnwidth]{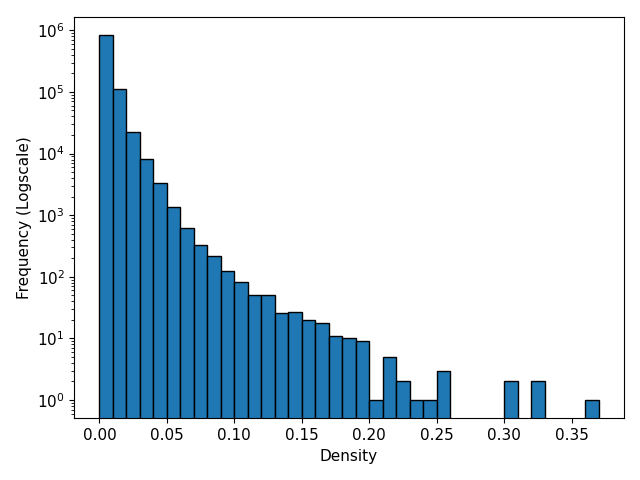}}
\caption{density-dist}
\label{fig:density-dist}
\end{figure}
\begin{figure}
\centerline{\includegraphics[width=\columnwidth]{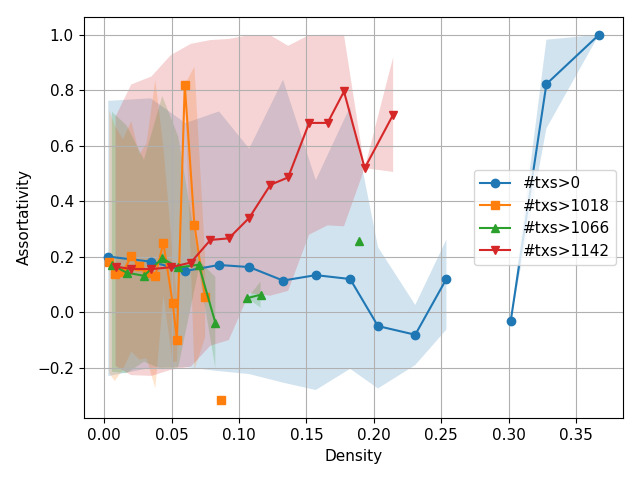}}
\caption{assortativity}
\label{fig:assortativity}
\end{figure}
\begin{figure}
\centerline{\includegraphics[width=\columnwidth]{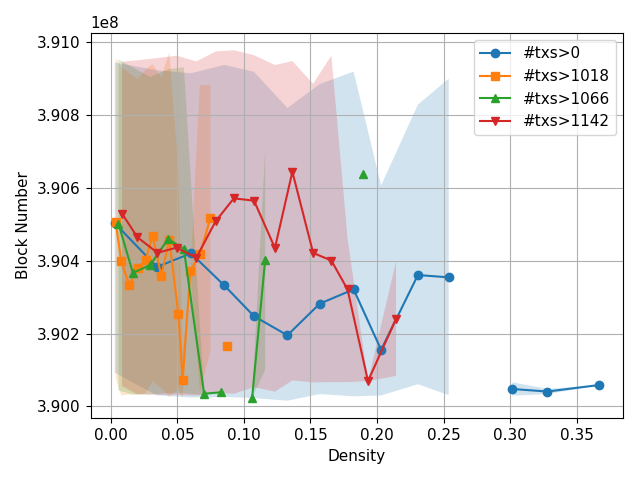}}
\caption{block-number}
\label{fig:block-number}
\end{figure}
\begin{figure}
\centerline{\includegraphics[width=\columnwidth]{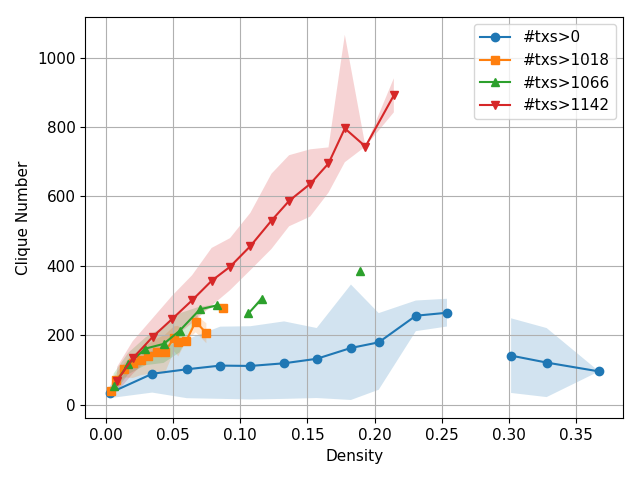}}
\caption{clique-number}
\label{fig:clique-number}
\end{figure}
\begin{figure}
\centerline{\includegraphics[width=\columnwidth]{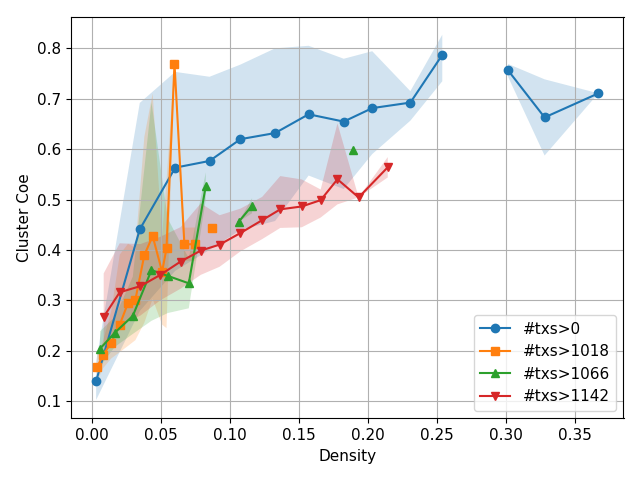}}
\caption{cluster-coe}
\label{fig:cluster-coe}
\end{figure}
\begin{figure}
\centerline{\includegraphics[width=\columnwidth]{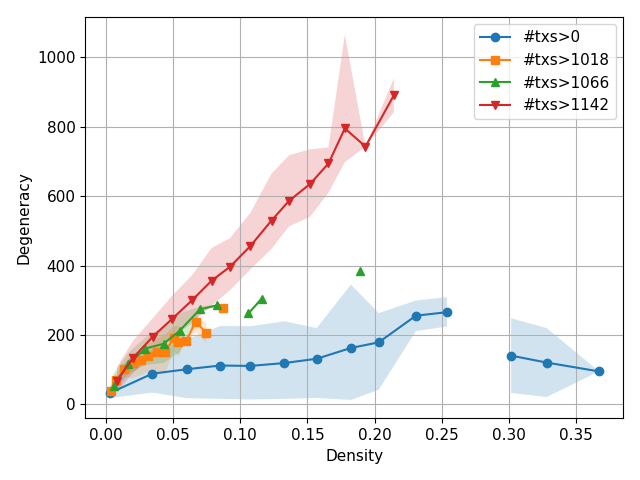}}
\caption{degeneracy}
\label{fig:degeneracy}
\end{figure}
\begin{figure}
\centerline{\includegraphics[width=\columnwidth]{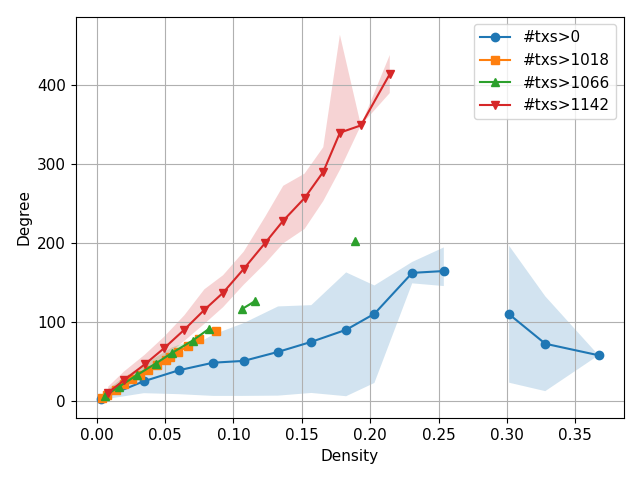}}
\caption{degree}
\label{fig:degree}
\end{figure}
\begin{figure}
\centerline{\includegraphics[width=\columnwidth]{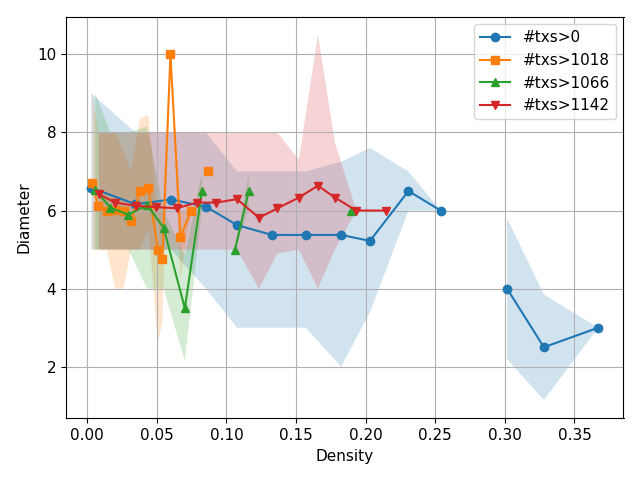}}
\caption{diameter}
\label{fig:diameter}
\end{figure}
\begin{figure}
\centerline{\includegraphics[width=\columnwidth]{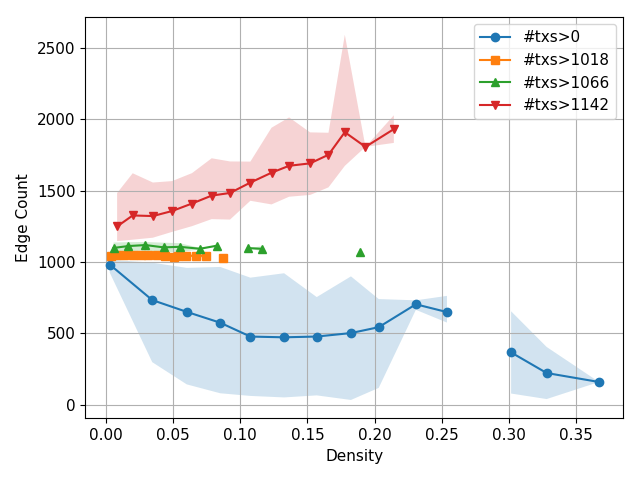}}
\caption{edge-count}
\label{fig:edge-count}
\end{figure}
\begin{figure}
\centerline{\includegraphics[width=\columnwidth]{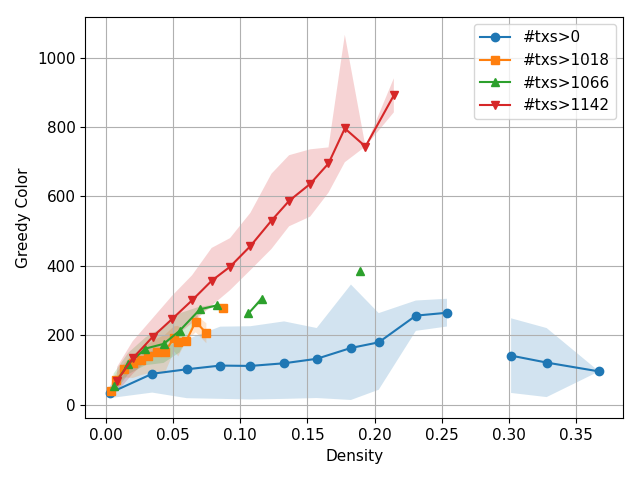}}
\caption{greedy-color}
\label{fig:greedy-color}
\end{figure}
\begin{figure}
\centerline{\includegraphics[width=\columnwidth]{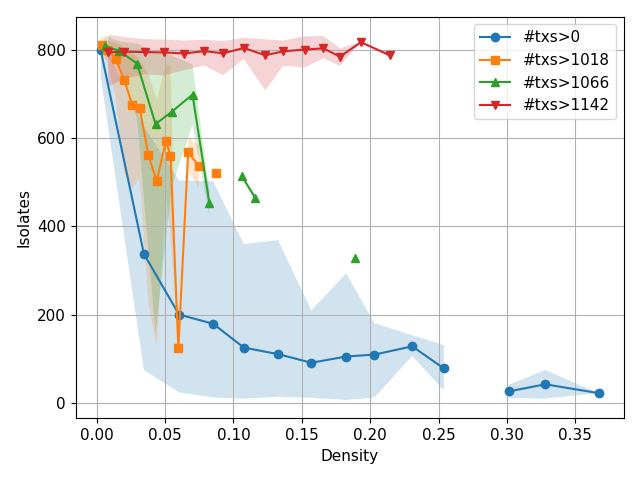}}
\caption{isolates}
\label{fig:isolates}
\end{figure}
\begin{figure}
\centerline{\includegraphics[width=\columnwidth]{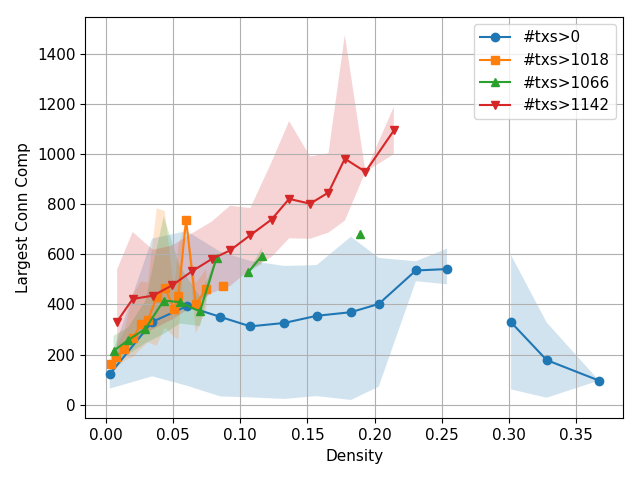}}
\caption{largest-conn-comp}
\label{fig:largest-conn-comp}
\end{figure}
\begin{figure}
\centerline{\includegraphics[width=\columnwidth]{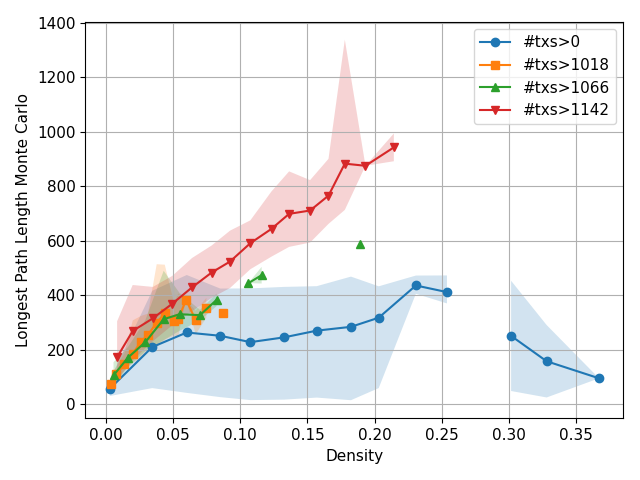}}
\caption{longest-path-length-monte-carlo}
\label{fig:longest-path-length-monte-carlo}
\end{figure}
\begin{figure}
\centerline{\includegraphics[width=\columnwidth]{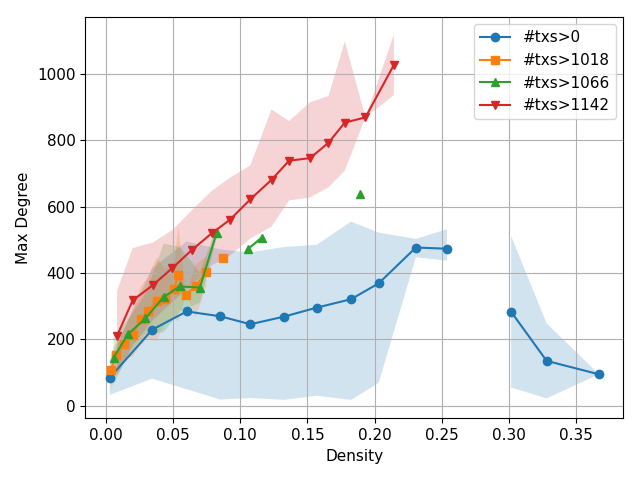}}
\caption{max-degree}
\label{fig:max-degree}
\end{figure}
\begin{figure}
\centerline{\includegraphics[width=\columnwidth]{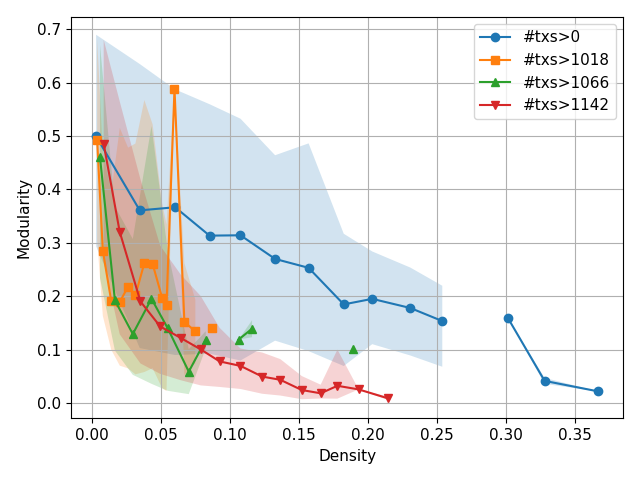}}
\caption{modularity}
\label{fig:modularity}
\end{figure}
\begin{figure}
\centerline{\includegraphics[width=\columnwidth]{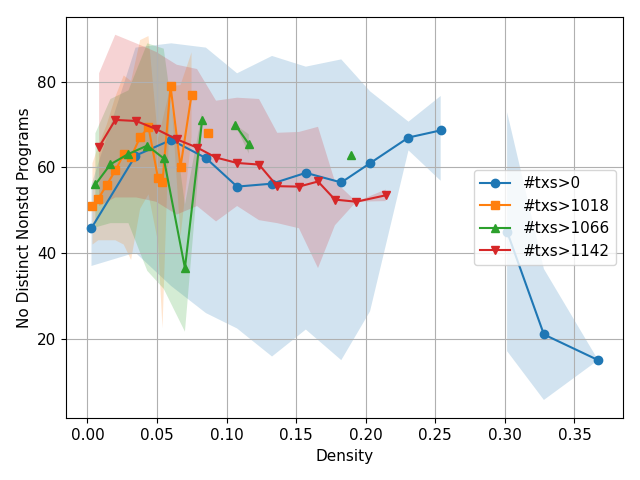}}
\caption{no-distinct-nonstd-programs}
\label{fig:no-distinct-nonstd-programs}
\end{figure}
\begin{figure}
\centerline{\includegraphics[width=\columnwidth]{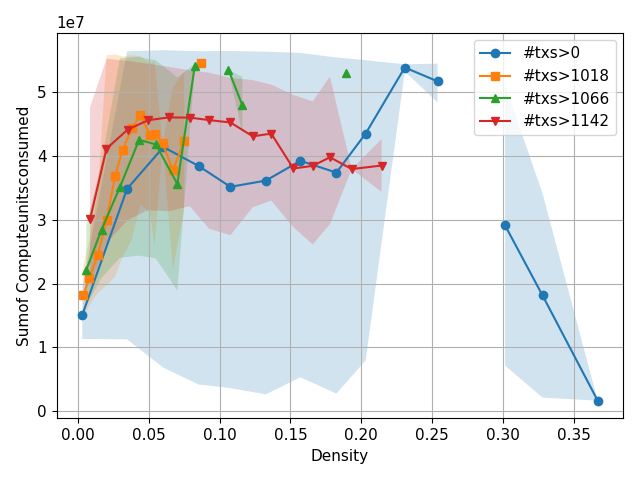}}
\caption{sumof-computeUnitsConsumed}
\label{fig:sumof-computeUnitsConsumed}
\end{figure}
\begin{figure}
\centerline{\includegraphics[width=\columnwidth]{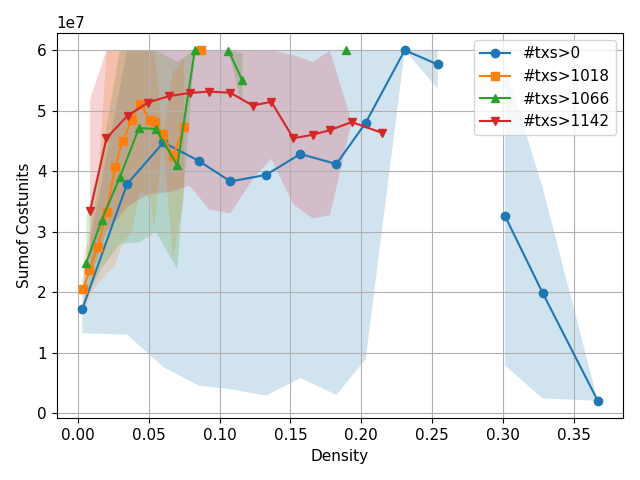}}
\caption{sumof-costUnits}
\label{fig:sumof-costUnits}
\end{figure}
\begin{figure}
\centerline{\includegraphics[width=\columnwidth]{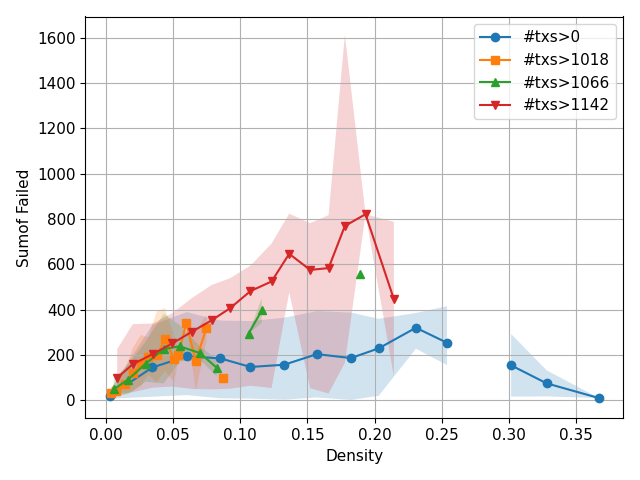}}
\caption{sumof-failed}
\label{fig:sumof-failed}
\end{figure}
\begin{figure}
\centerline{\includegraphics[width=\columnwidth]{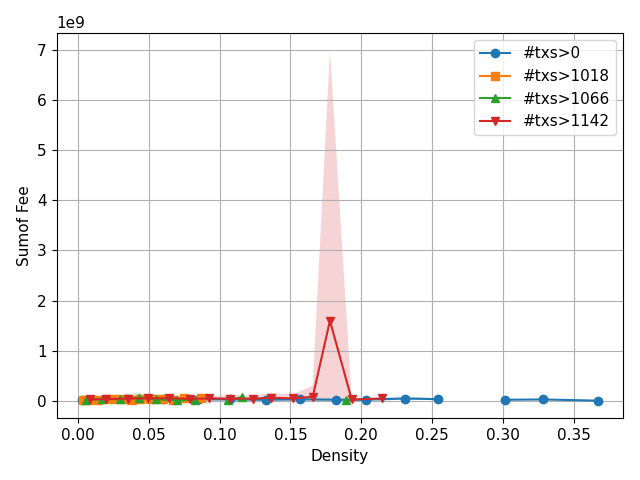}}
\caption{sumof-fee}
\label{fig:sumof-fee}
\end{figure}
\begin{figure}
\centerline{\includegraphics[width=\columnwidth]{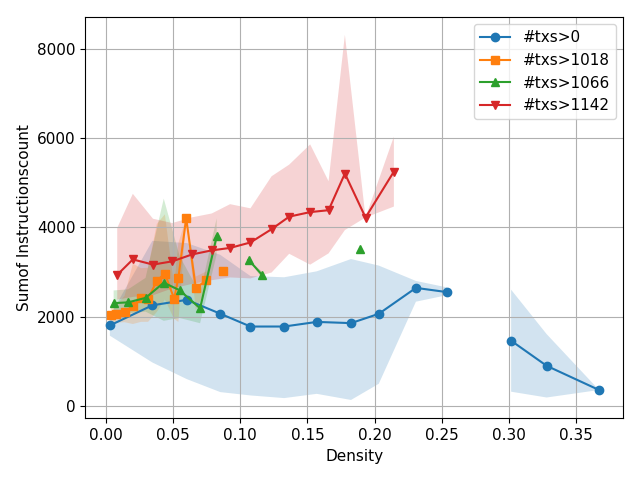}}
\caption{sumof-instructionsCount}
\label{fig:sumof-instructionsCount}
\end{figure}
\begin{figure}
\centerline{\includegraphics[width=\columnwidth]{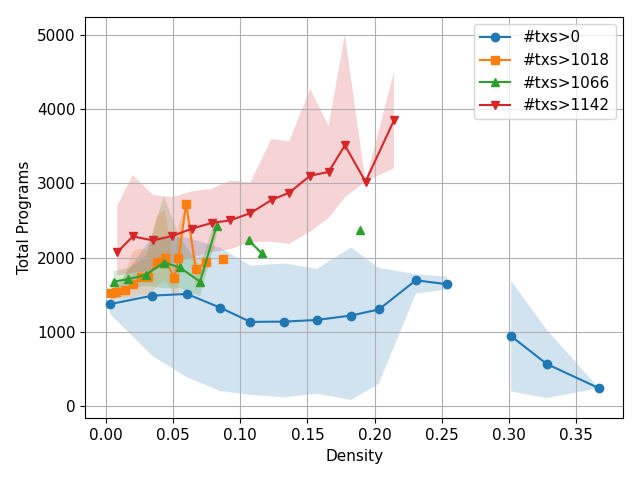}}
\caption{total-programs}
\label{fig:total-programs}
\end{figure}
\begin{figure}
\centerline{\includegraphics[width=\columnwidth]{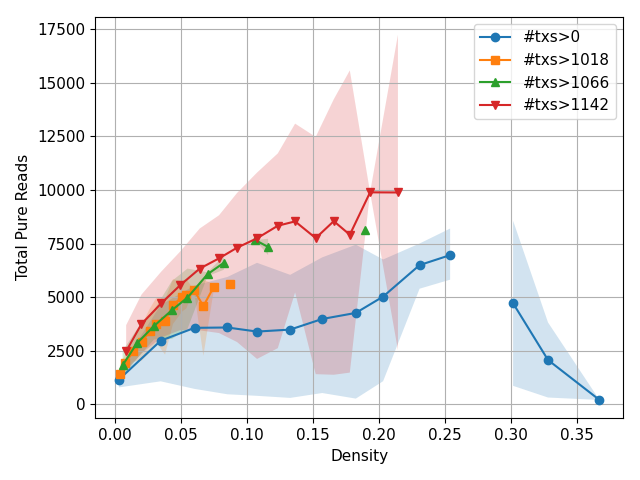}}
\caption{total-pure-reads}
\label{fig:total-pure-reads}
\end{figure}
\begin{figure}
\centerline{\includegraphics[width=\columnwidth]{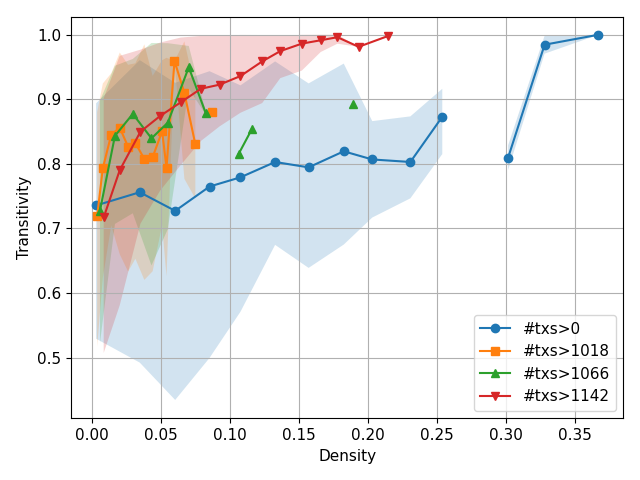}}
\caption{transitivity}
\label{fig:transitivity}
\end{figure}
\begin{figure}
\centerline{\includegraphics[width=\columnwidth]{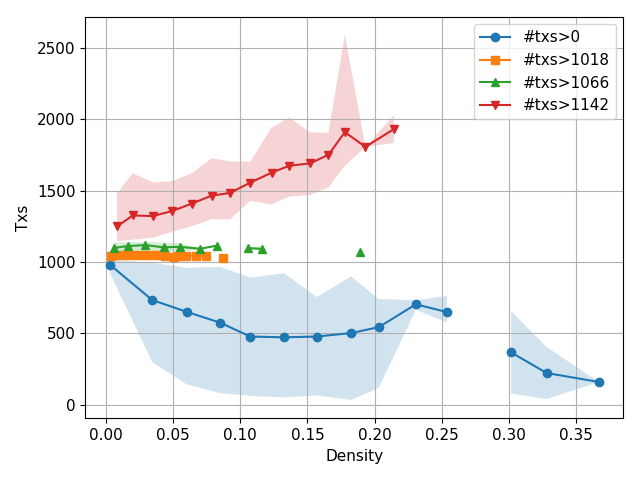}}
\caption{txs}
\label{fig:txs}
\end{figure}
\begin{figure}
\centerline{\includegraphics[width=\columnwidth]{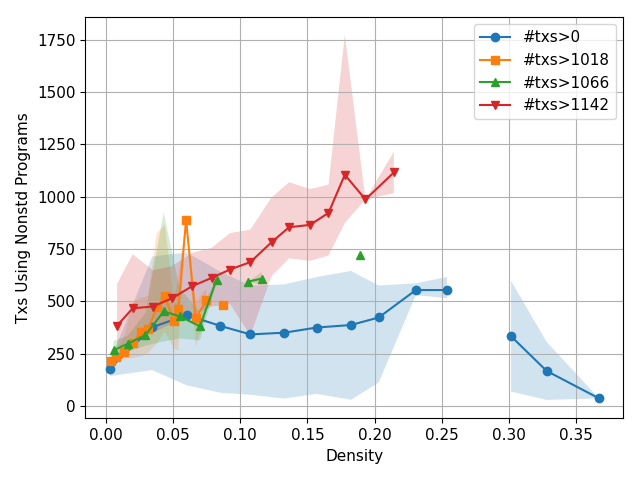}}
\caption{txs-using-nonstd-programs}
\label{fig:txs-using-nonstd-programs}
\end{figure}
\begin{figure}
\centerline{\includegraphics[width=\columnwidth]{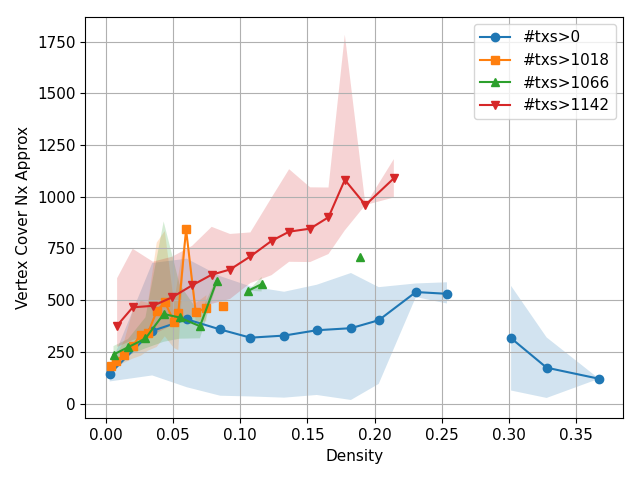}}
\caption{vertex-cover-nx-approx}
\label{fig:vertex-cover-nx-approx}
\end{figure}
\begin{figure}
\centerline{\includegraphics[width=\columnwidth]{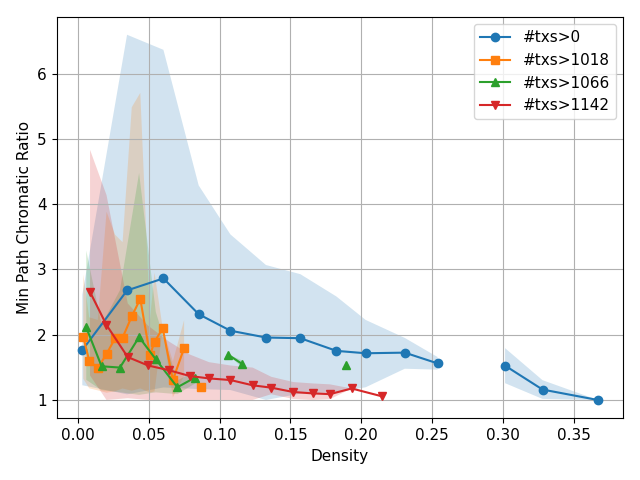}}
\caption{min-path-chromatic-ratio}
\label{fig:min-path-chromatic-ratio}
\end{figure}
\begin{figure}
\centerline{\includegraphics[width=\columnwidth]{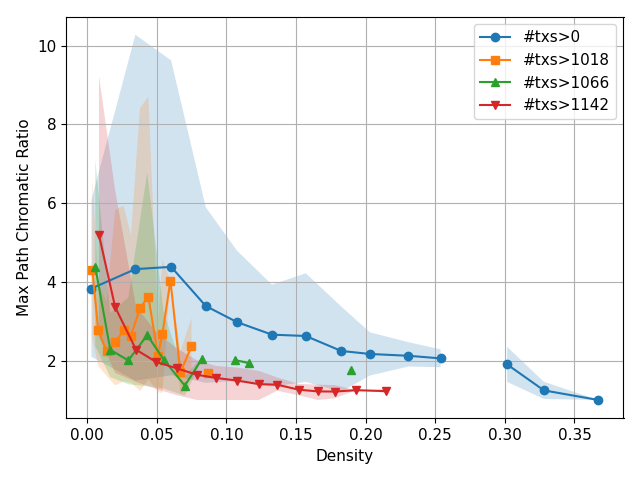}}
\caption{max-path-chromatic-ratio}
\label{fig:max-path-chromatic-ratio}
\end{figure}
\begin{figure}
\centerline{\includegraphics[width=\columnwidth]{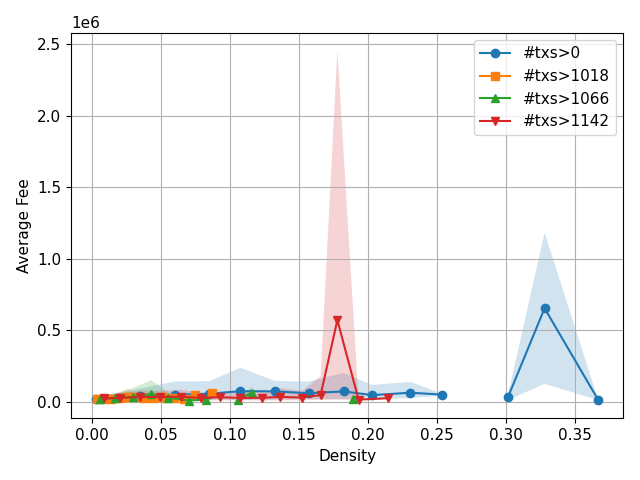}}
\caption{avg-fee}
\label{fig:avg-fee}
\end{figure}
\begin{figure}
\centerline{\includegraphics[width=\columnwidth]{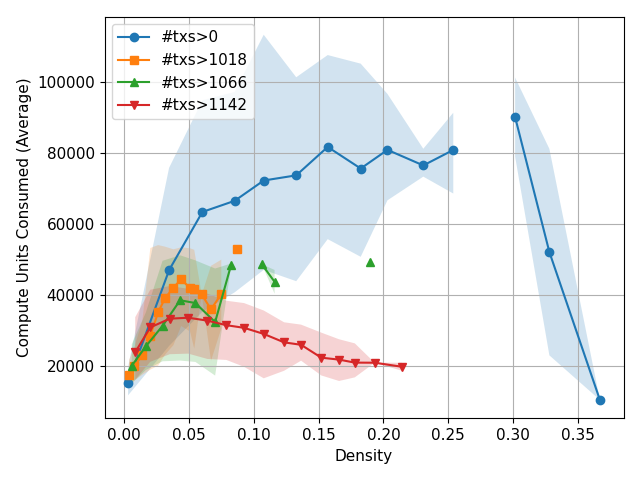}}
\caption{avg-computeUnitsConsumed}
\label{fig:avg-computeUnitsConsumed}
\end{figure}
\begin{figure}
\centerline{\includegraphics[width=\columnwidth]{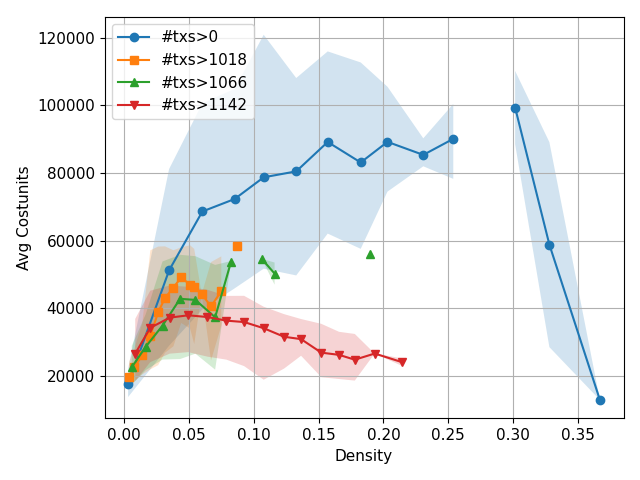}}
\caption{avg-costUnits}
\label{fig:avg-costUnits}
\end{figure}
\begin{figure}
\centerline{\includegraphics[width=\columnwidth]{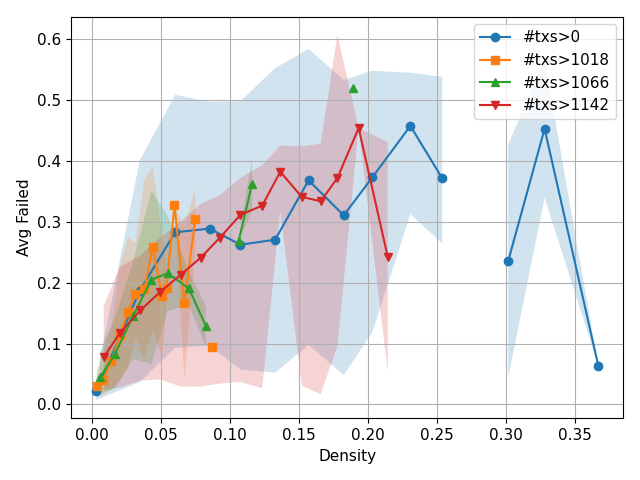}}
\caption{avg-failed}
\label{fig:avg-failed}
\end{figure}
\begin{figure}
\centerline{\includegraphics[width=\columnwidth]{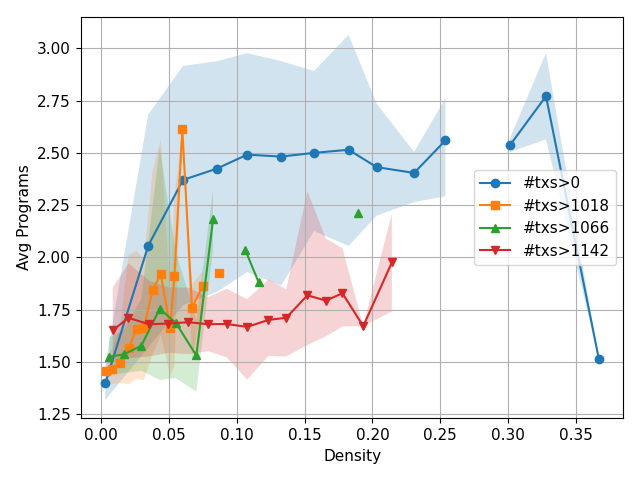}}
\caption{avg-programs}
\label{fig:avg-programs}
\end{figure}
\begin{figure}
\centerline{\includegraphics[width=\columnwidth]{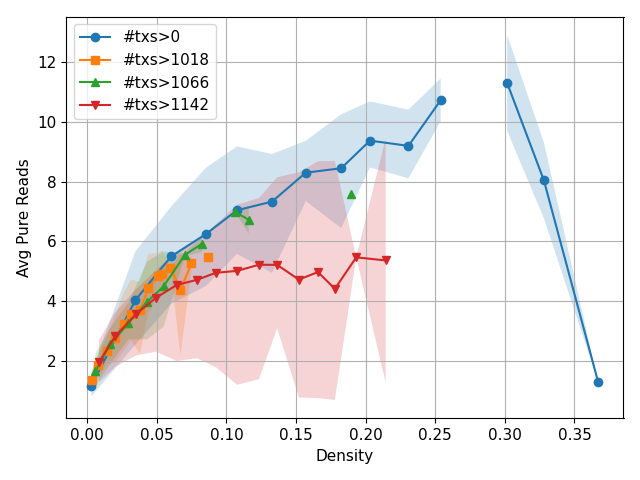}}
\caption{avg-pure-reads}
\label{fig:avg-pure-reads}
\end{figure}
\begin{figure}
\centerline{\includegraphics[width=\columnwidth]{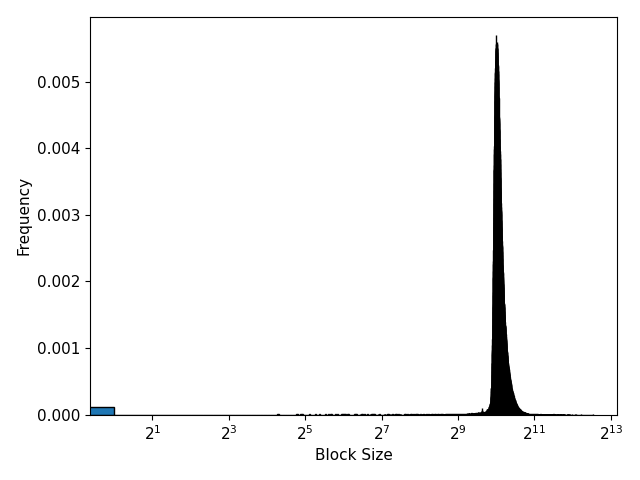}}
\caption{block-size-dist}
\label{fig:block-size-dist}
\end{figure}
\begin{figure}
\centerline{\includegraphics[width=\columnwidth]{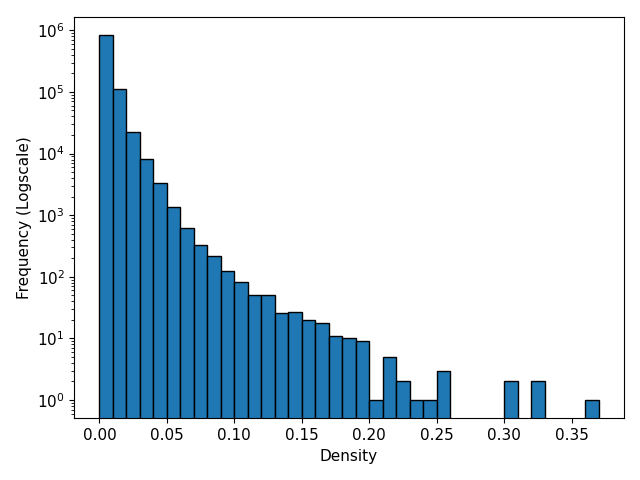}}
\caption{density-dist}
\label{fig:density-dist}
\end{figure}
\begin{figure}
\centerline{\includegraphics[width=\columnwidth]{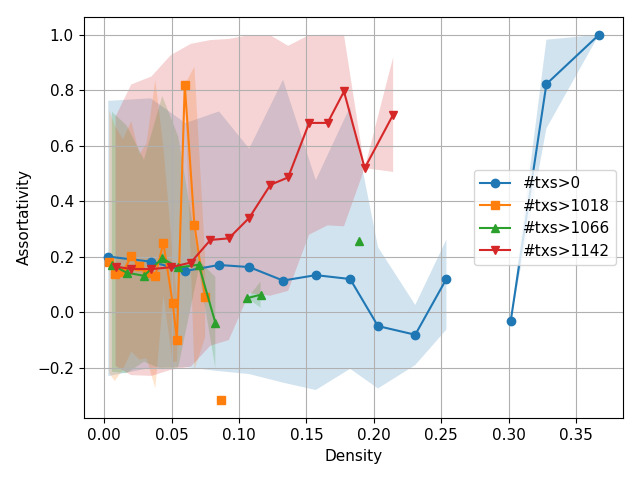}}
\caption{assortativity}
\label{fig:assortativity}
\end{figure}
\begin{figure}
\centerline{\includegraphics[width=\columnwidth]{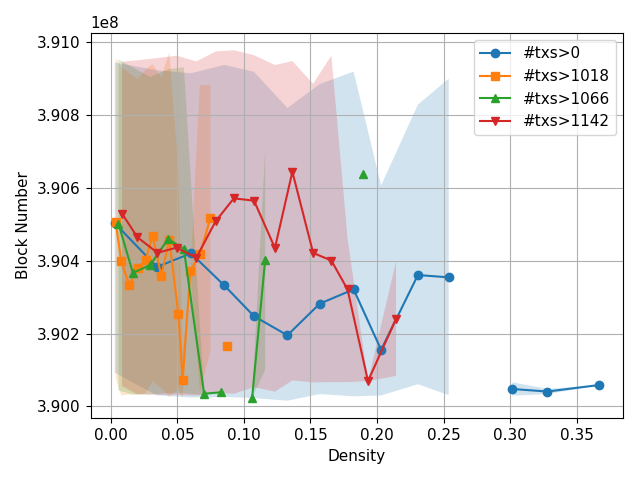}}
\caption{block-number}
\label{fig:block-number}
\end{figure}
\begin{figure}
\centerline{\includegraphics[width=\columnwidth]{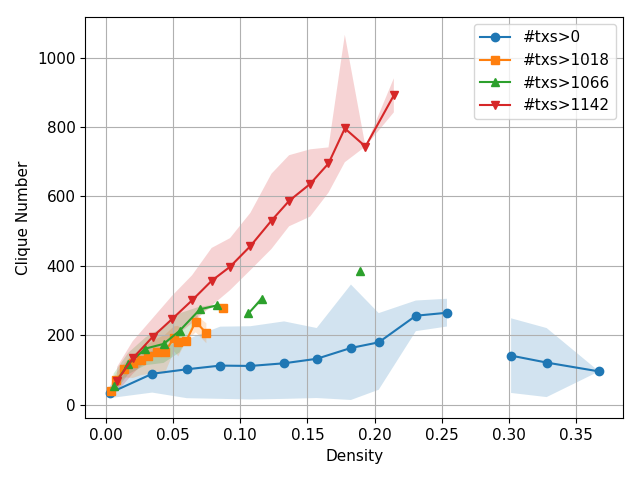}}
\caption{clique-number}
\label{fig:clique-number}
\end{figure}
\begin{figure}
\centerline{\includegraphics[width=\columnwidth]{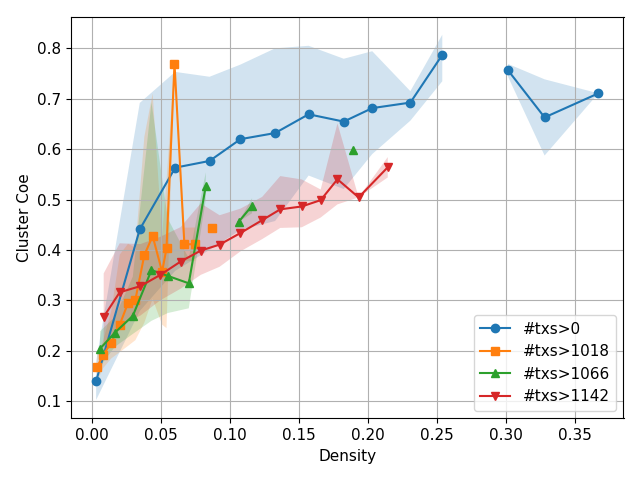}}
\caption{cluster-coe}
\label{fig:cluster-coe}
\end{figure}
\begin{figure}
\centerline{\includegraphics[width=\columnwidth]{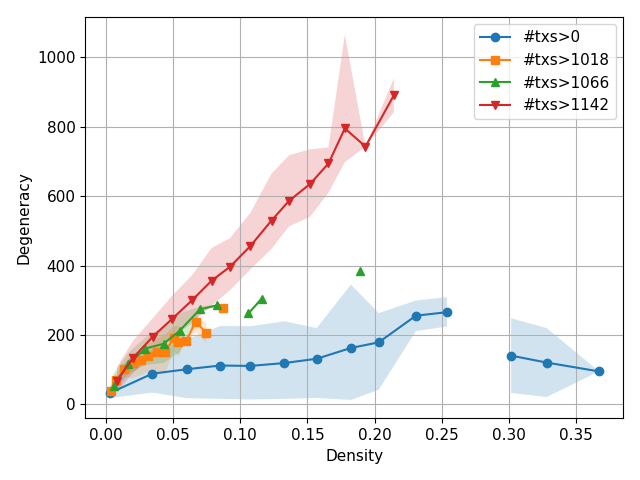}}
\caption{degeneracy}
\label{fig:degeneracy}
\end{figure}
\begin{figure}
\centerline{\includegraphics[width=\columnwidth]{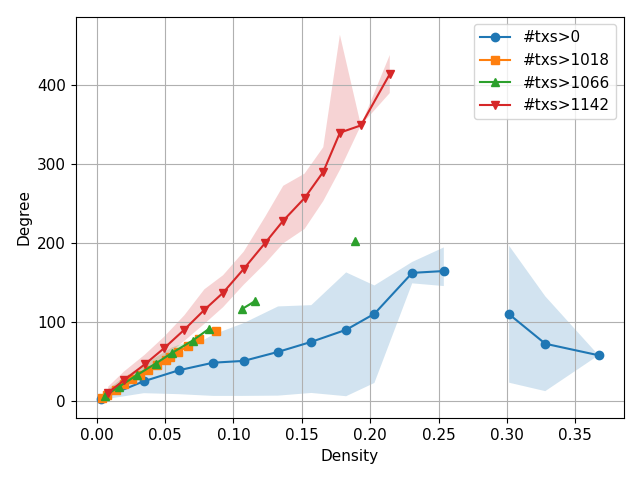}}
\caption{degree}
\label{fig:degree}
\end{figure}
\begin{figure}
\centerline{\includegraphics[width=\columnwidth]{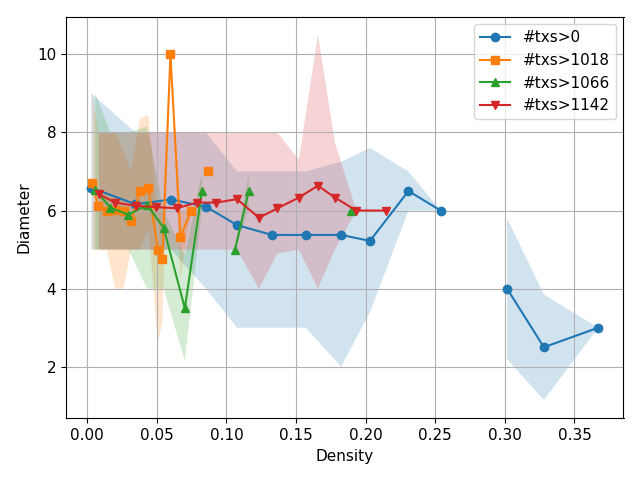}}
\caption{diameter}
\label{fig:diameter}
\end{figure}
\begin{figure}
\centerline{\includegraphics[width=\columnwidth]{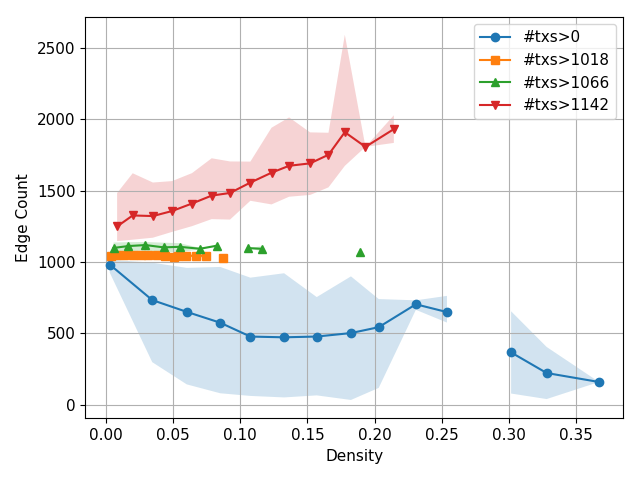}}
\caption{edge-count}
\label{fig:edge-count}
\end{figure}
\begin{figure}
\centerline{\includegraphics[width=\columnwidth]{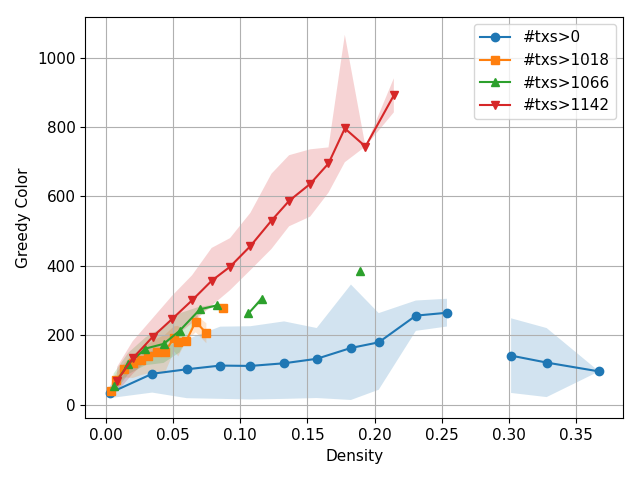}}
\caption{greedy-color}
\label{fig:greedy-color}
\end{figure}
\begin{figure}
\centerline{\includegraphics[width=\columnwidth]{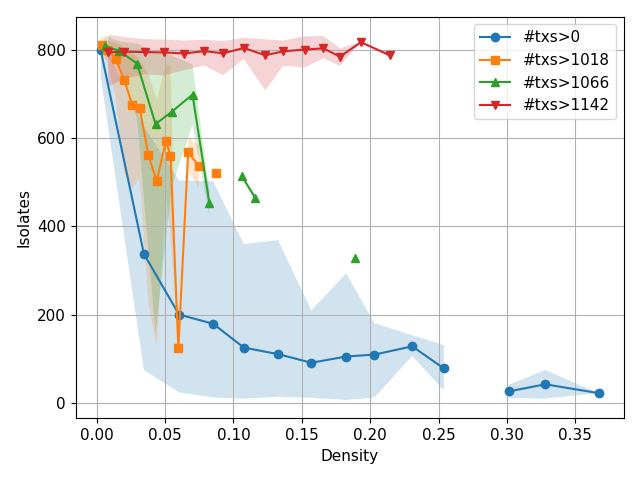}}
\caption{isolates}
\label{fig:isolates}
\end{figure}
\begin{figure}
\centerline{\includegraphics[width=\columnwidth]{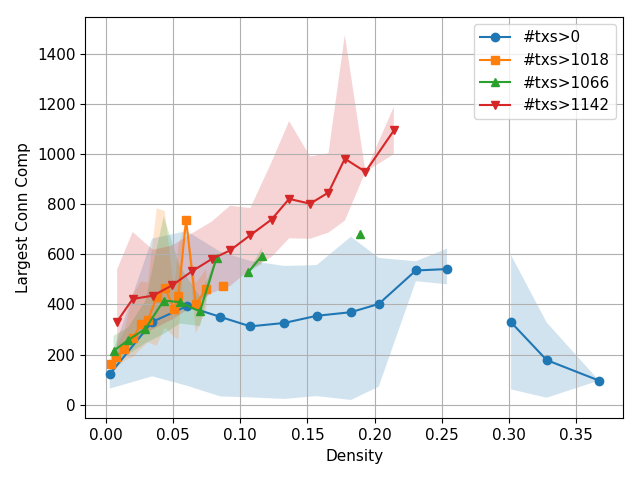}}
\caption{largest-conn-comp}
\label{fig:largest-conn-comp}
\end{figure}
\begin{figure}
\centerline{\includegraphics[width=\columnwidth]{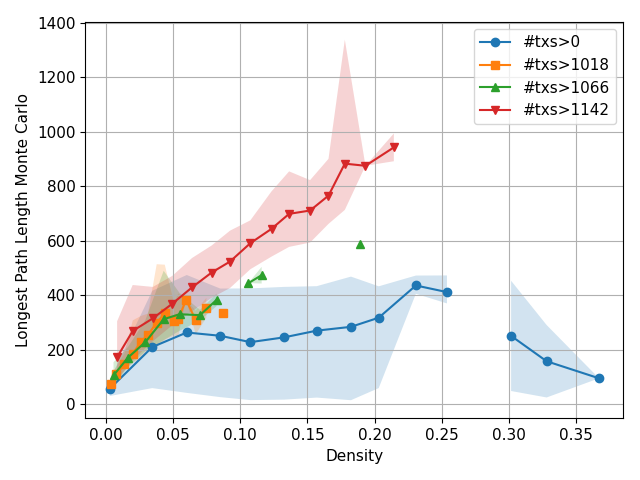}}
\caption{longest-path-length-monte-carlo}
\label{fig:longest-path-length-monte-carlo}
\end{figure}
\begin{figure}
\centerline{\includegraphics[width=\columnwidth]{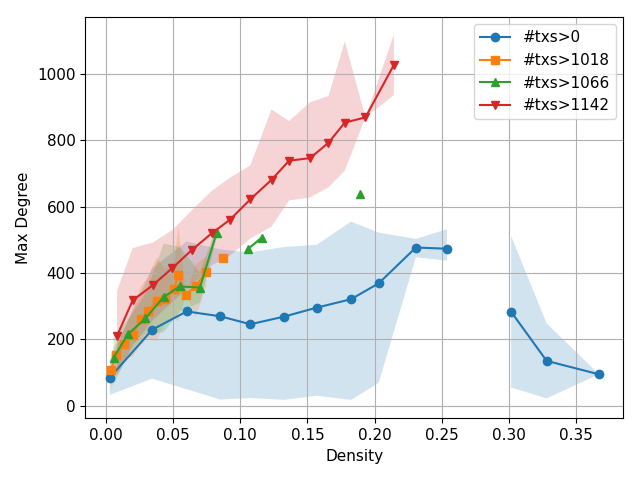}}
\caption{max-degree}
\label{fig:max-degree}
\end{figure}
\begin{figure}
\centerline{\includegraphics[width=\columnwidth]{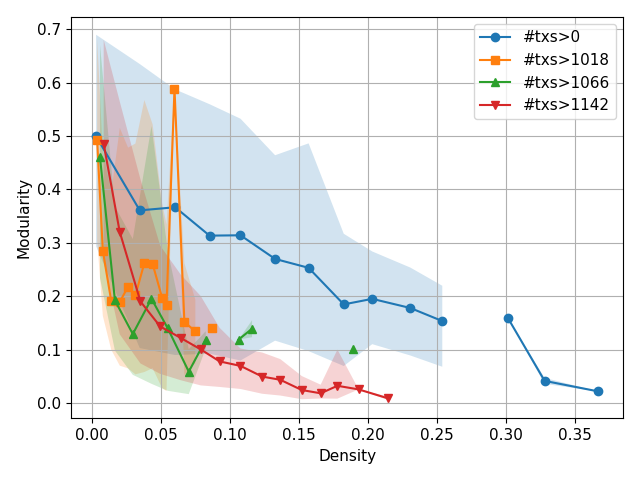}}
\caption{modularity}
\label{fig:modularity}
\end{figure}
\begin{figure}
\centerline{\includegraphics[width=\columnwidth]{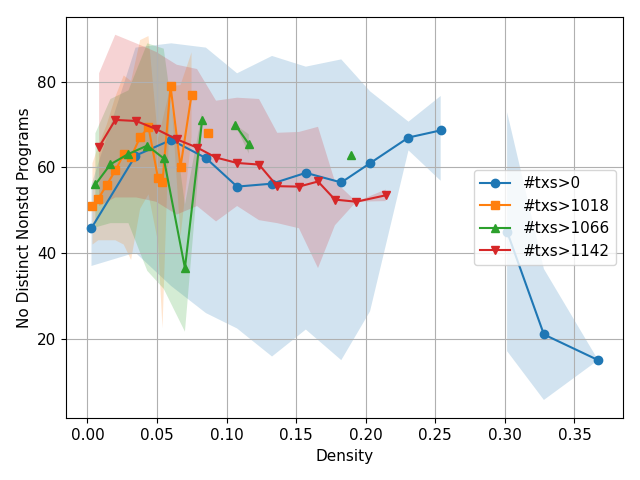}}
\caption{no-distinct-nonstd-programs}
\label{fig:no-distinct-nonstd-programs}
\end{figure}
\begin{figure}
\centerline{\includegraphics[width=\columnwidth]{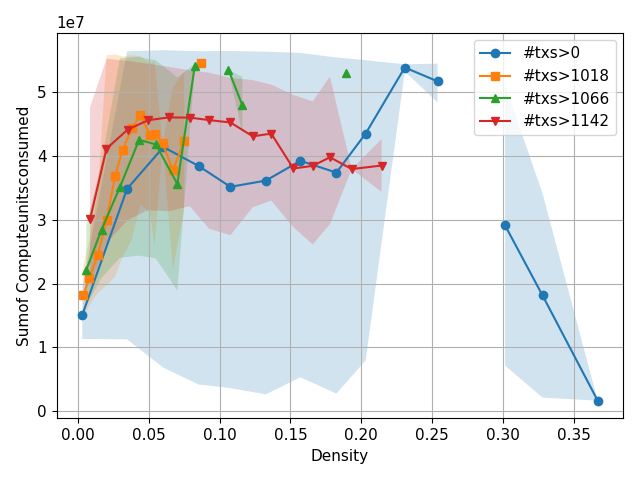}}
\caption{sumof-computeUnitsConsumed}
\label{fig:sumof-computeUnitsConsumed}
\end{figure}
\begin{figure}
\centerline{\includegraphics[width=\columnwidth]{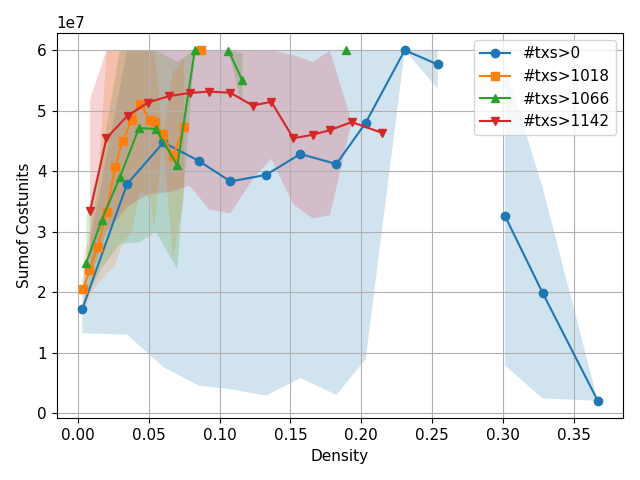}}
\caption{sumof-costUnits}
\label{fig:sumof-costUnits}
\end{figure}
\begin{figure}
\centerline{\includegraphics[width=\columnwidth]{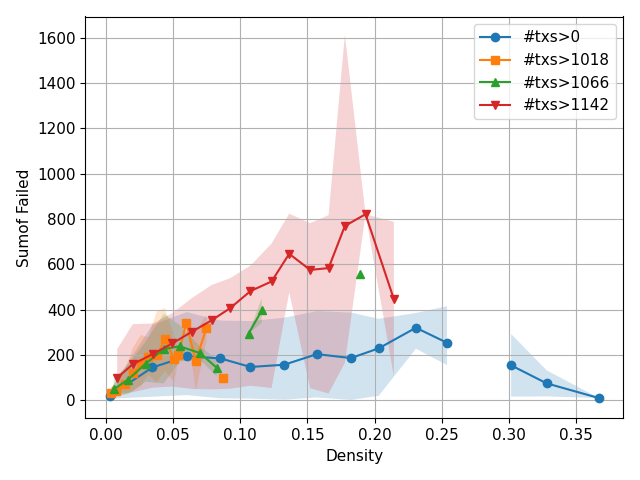}}
\caption{sumof-failed}
\label{fig:sumof-failed}
\end{figure}
\begin{figure}
\centerline{\includegraphics[width=\columnwidth]{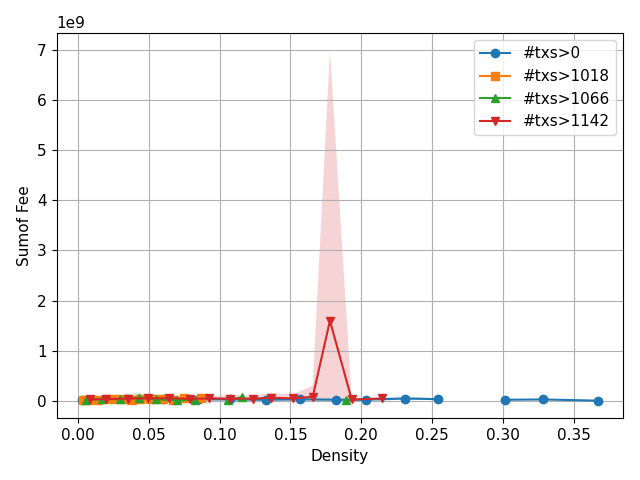}}
\caption{sumof-fee}
\label{fig:sumof-fee}
\end{figure}
\begin{figure}
\centerline{\includegraphics[width=\columnwidth]{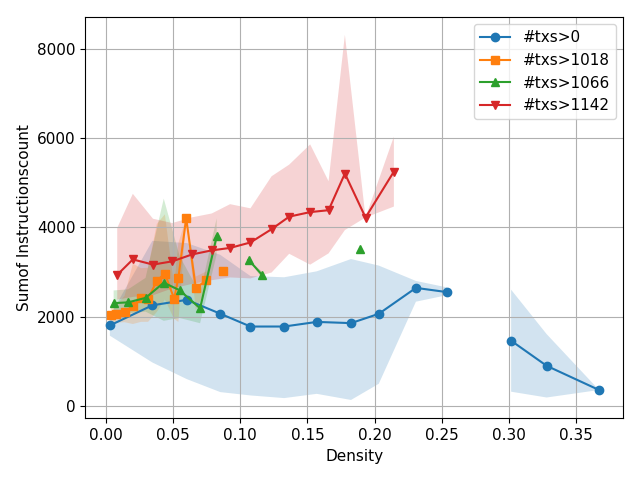}}
\caption{sumof-instructionsCount}
\label{fig:sumof-instructionsCount}
\end{figure}
\begin{figure}
\centerline{\includegraphics[width=\columnwidth]{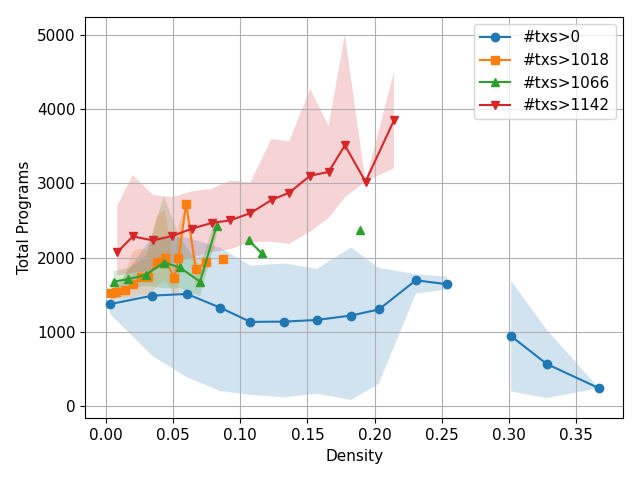}}
\caption{total-programs}
\label{fig:total-programs}
\end{figure}
\begin{figure}
\centerline{\includegraphics[width=\columnwidth]{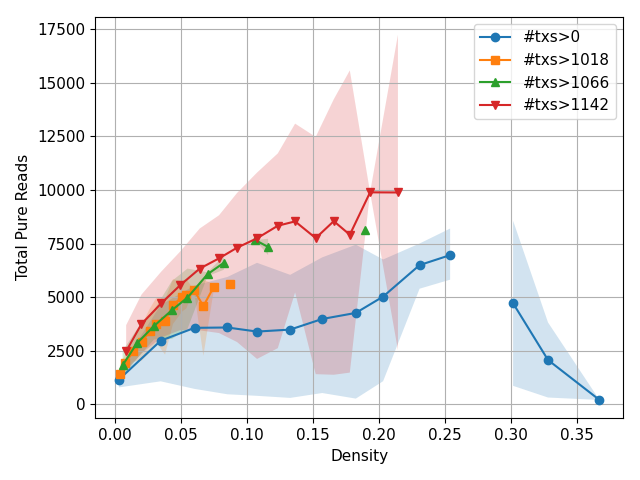}}
\caption{total-pure-reads}
\label{fig:total-pure-reads}
\end{figure}
\begin{figure}
\centerline{\includegraphics[width=\columnwidth]{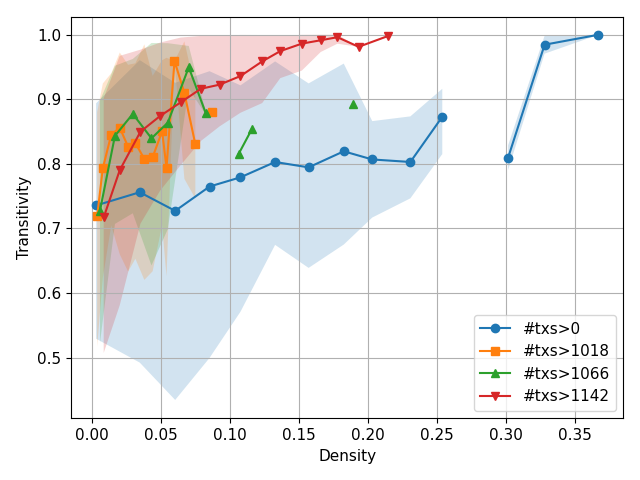}}
\caption{transitivity}
\label{fig:transitivity}
\end{figure}
\begin{figure}
\centerline{\includegraphics[width=\columnwidth]{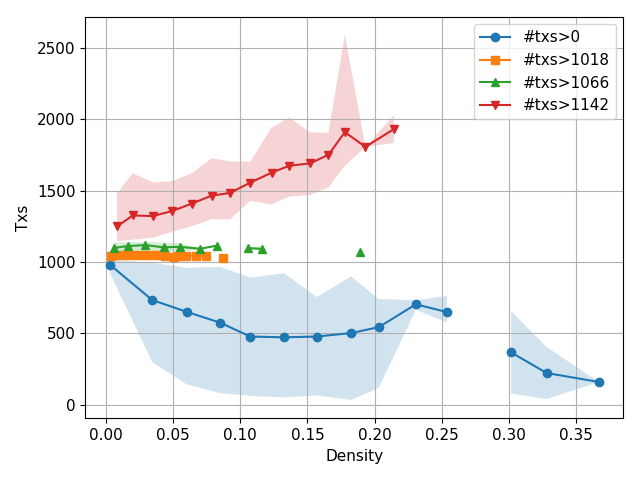}}
\caption{txs}
\label{fig:txs}
\end{figure}
\begin{figure}
\centerline{\includegraphics[width=\columnwidth]{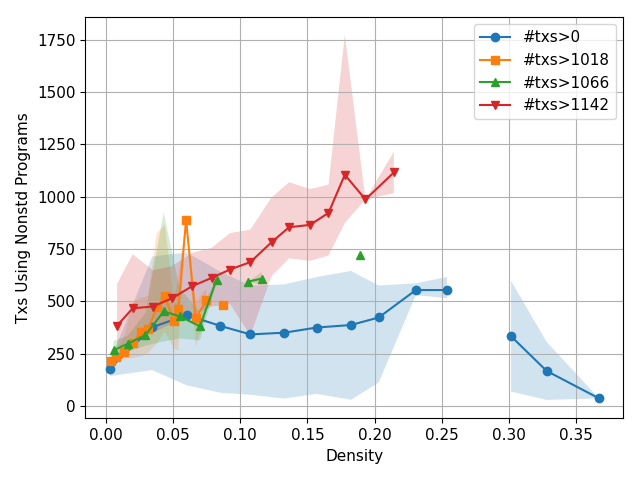}}
\caption{txs-using-nonstd-programs}
\label{fig:txs-using-nonstd-programs}
\end{figure}
\begin{figure}
\centerline{\includegraphics[width=\columnwidth]{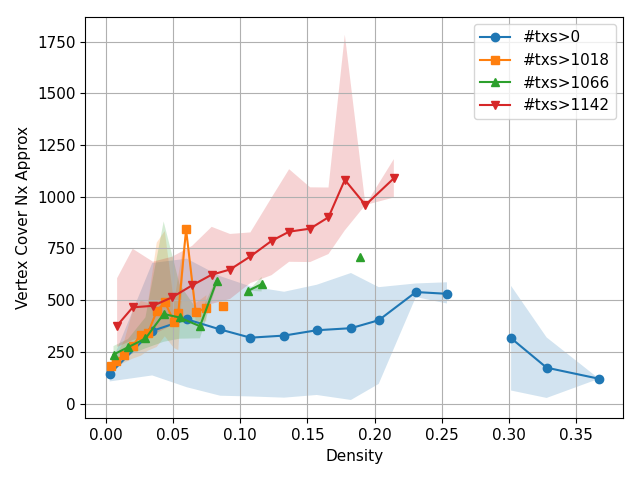}}
\caption{vertex-cover-nx-approx}
\label{fig:vertex-cover-nx-approx}
\end{figure}
\begin{figure}
\centerline{\includegraphics[width=\columnwidth]{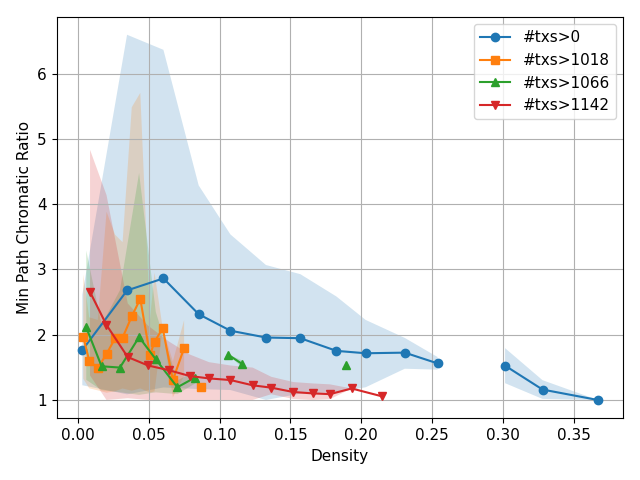}}
\caption{min-path-chromatic-ratio}
\label{fig:min-path-chromatic-ratio}
\end{figure}
\begin{figure}
\centerline{\includegraphics[width=\columnwidth]{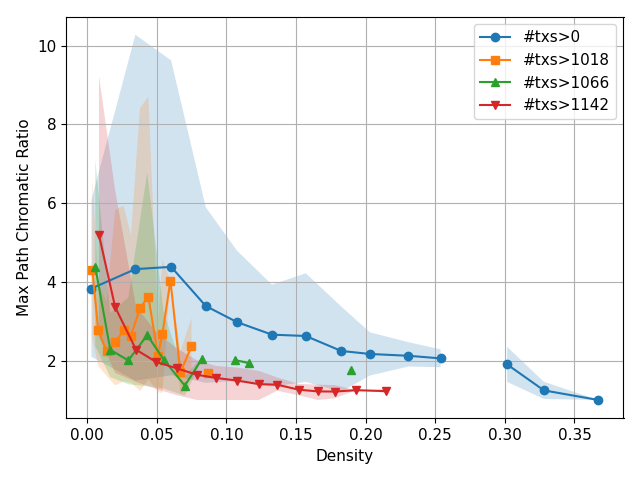}}
\caption{max-path-chromatic-ratio}
\label{fig:max-path-chromatic-ratio}
\end{figure}
\begin{figure}
\centerline{\includegraphics[width=\columnwidth]{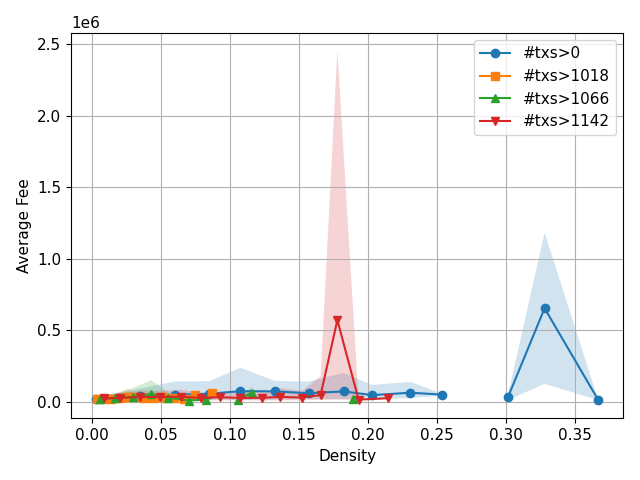}}
\caption{avg-fee}
\label{fig:avg-fee}
\end{figure}
\begin{figure}
\centerline{\includegraphics[width=\columnwidth]{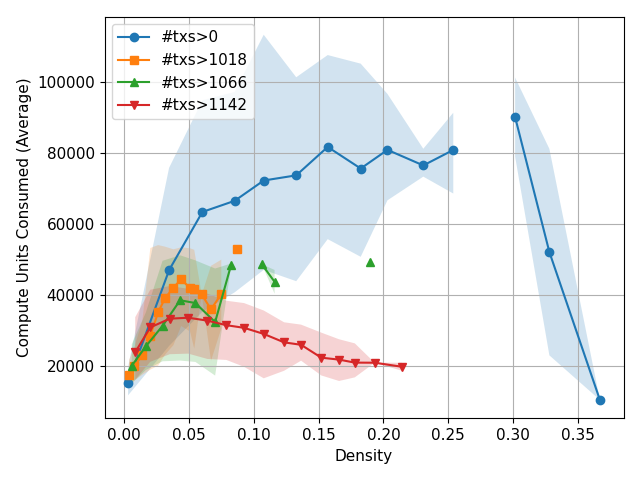}}
\caption{avg-computeUnitsConsumed}
\label{fig:avg-computeUnitsConsumed}
\end{figure}
\begin{figure}
\centerline{\includegraphics[width=\columnwidth]{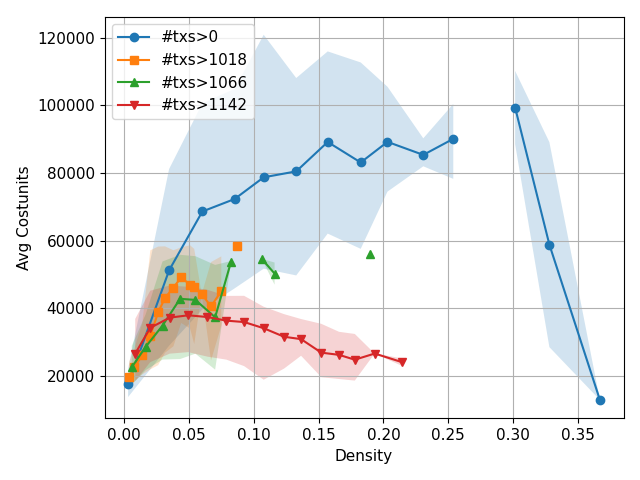}}
\caption{avg-costUnits}
\label{fig:avg-costUnits}
\end{figure}
\begin{figure}
\centerline{\includegraphics[width=\columnwidth]{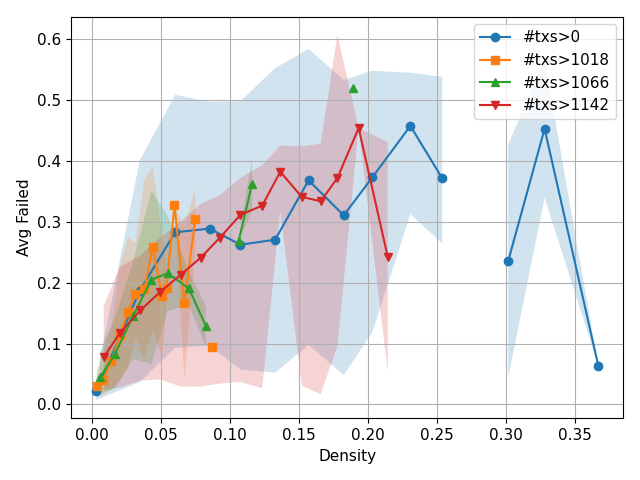}}
\caption{avg-failed}
\label{fig:avg-failed}
\end{figure}
\begin{figure}
\centerline{\includegraphics[width=\columnwidth]{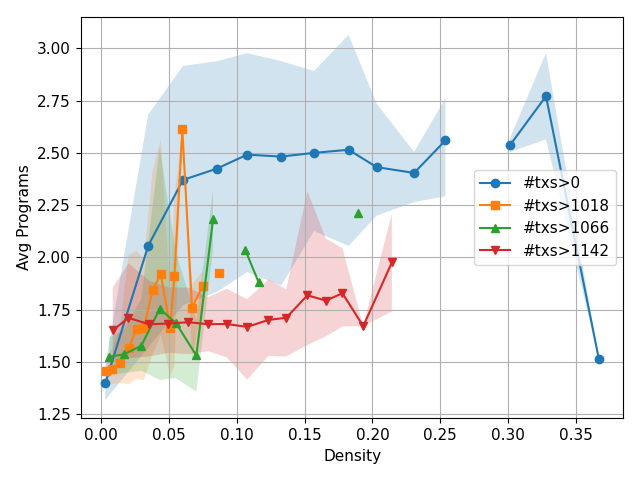}}
\caption{avg-programs}
\label{fig:avg-programs}
\end{figure}
\begin{figure}
\centerline{\includegraphics[width=\columnwidth]{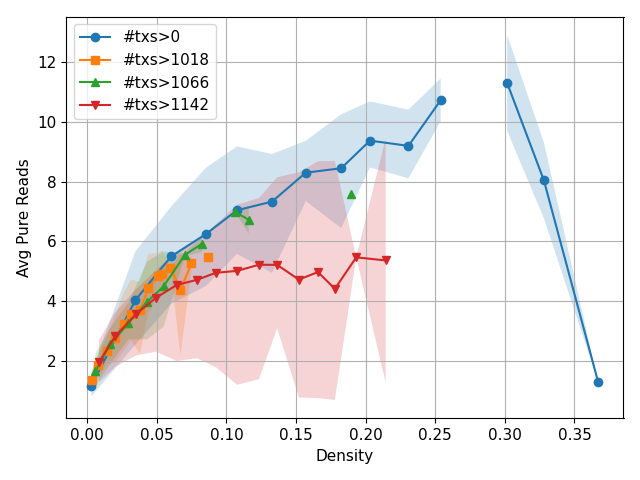}}
\caption{avg-pure-reads}
\label{fig:avg-pure-reads}
\end{figure}
\begin{figure}
\centerline{\includegraphics[width=\columnwidth]{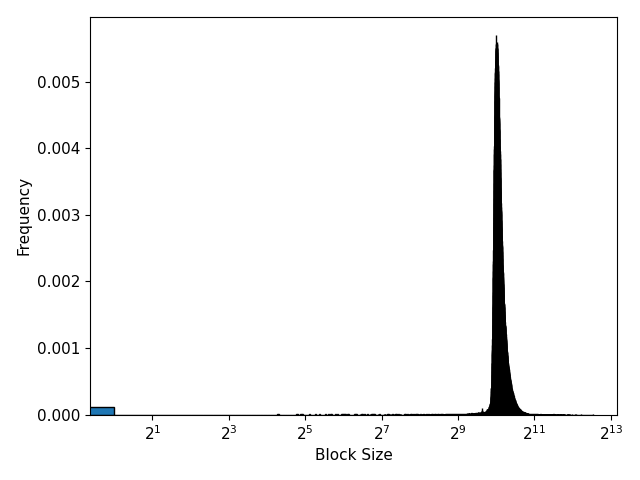}}
\caption{block-size-dist}
\label{fig:block-size-dist}
\end{figure}
\begin{figure}
\centerline{\includegraphics[width=\columnwidth]{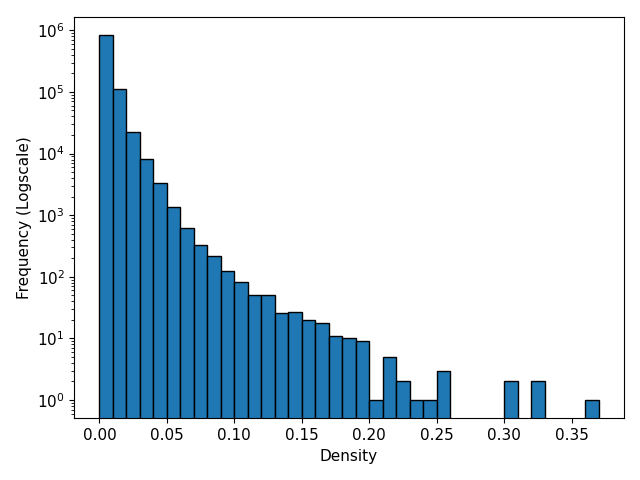}}
\caption{density-dist}
\label{fig:density-dist}
\end{figure}
\begin{figure}
\centerline{\includegraphics[width=\columnwidth]{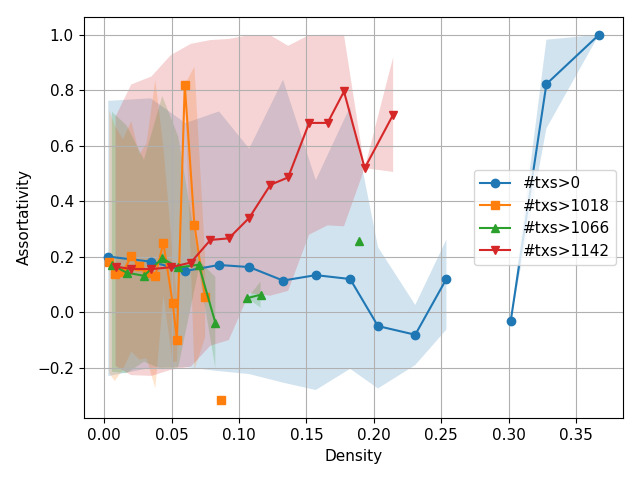}}
\caption{assortativity}
\label{fig:assortativity}
\end{figure}
\begin{figure}
\centerline{\includegraphics[width=\columnwidth]{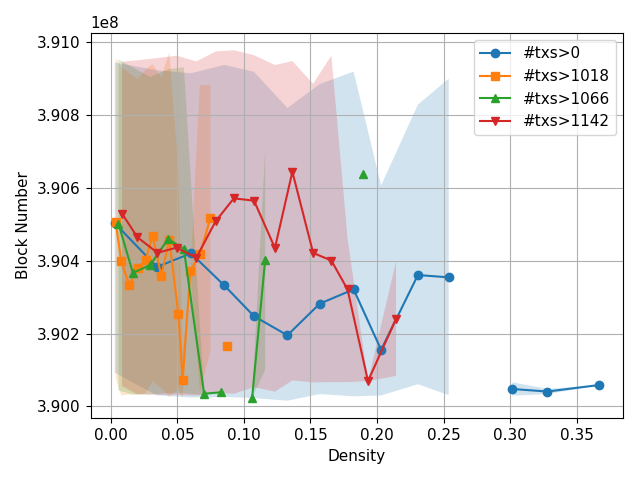}}
\caption{block-number}
\label{fig:block-number}
\end{figure}
\begin{figure}
\centerline{\includegraphics[width=\columnwidth]{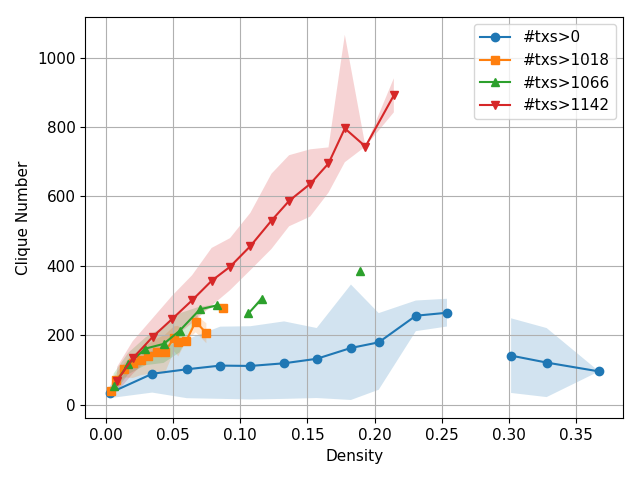}}
\caption{clique-number}
\label{fig:clique-number}
\end{figure}
\begin{figure}
\centerline{\includegraphics[width=\columnwidth]{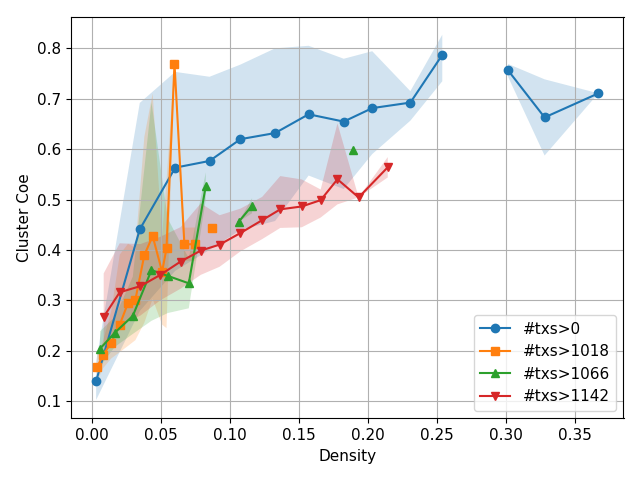}}
\caption{cluster-coe}
\label{fig:cluster-coe}
\end{figure}
\begin{figure}
\centerline{\includegraphics[width=\columnwidth]{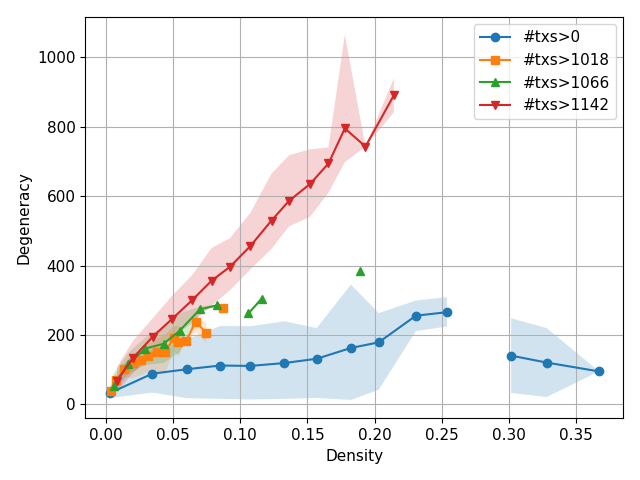}}
\caption{degeneracy}
\label{fig:degeneracy}
\end{figure}
\begin{figure}
\centerline{\includegraphics[width=\columnwidth]{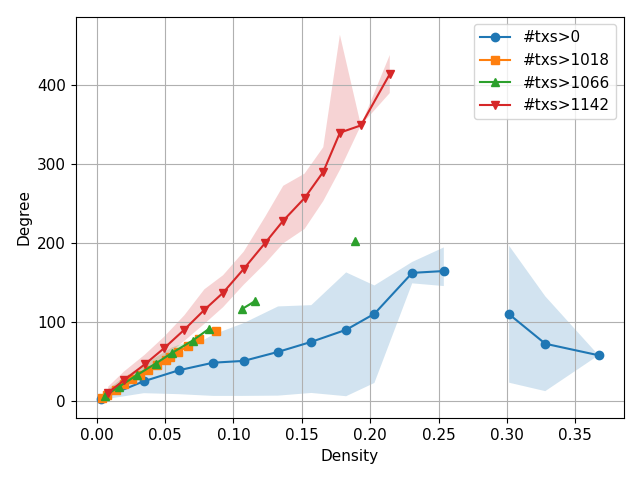}}
\caption{degree}
\label{fig:degree}
\end{figure}
\begin{figure}
\centerline{\includegraphics[width=\columnwidth]{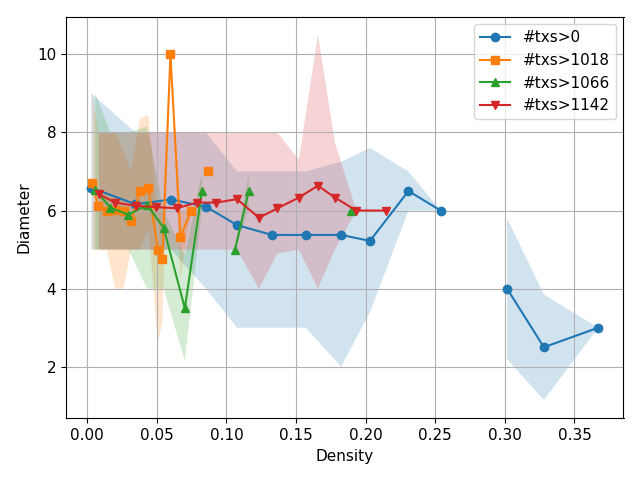}}
\caption{diameter}
\label{fig:diameter}
\end{figure}
\begin{figure}
\centerline{\includegraphics[width=\columnwidth]{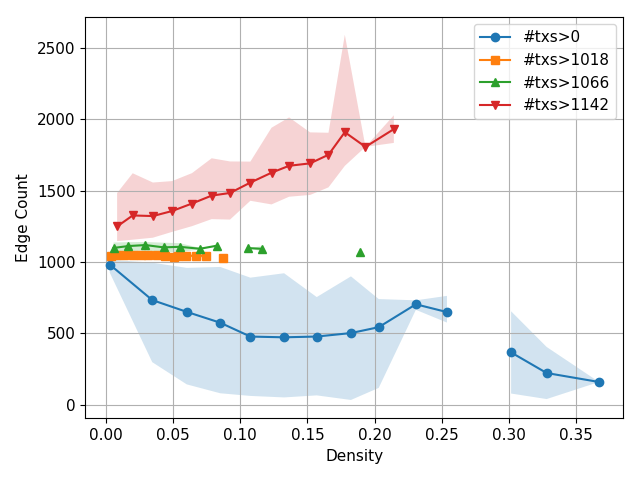}}
\caption{edge-count}
\label{fig:edge-count}
\end{figure}
\begin{figure}
\centerline{\includegraphics[width=\columnwidth]{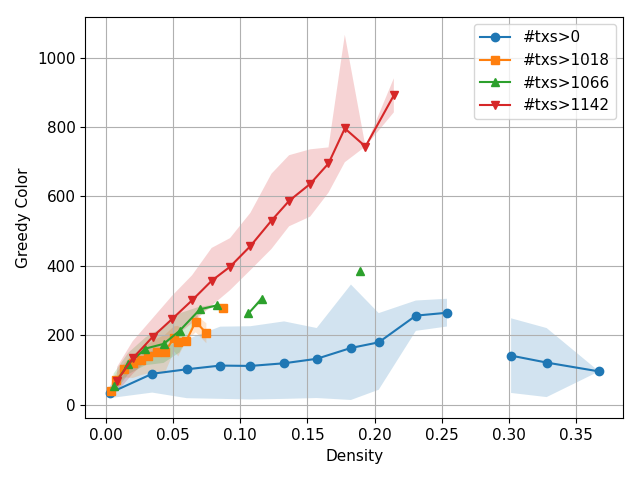}}
\caption{greedy-color}
\label{fig:greedy-color}
\end{figure}
\begin{figure}
\centerline{\includegraphics[width=\columnwidth]{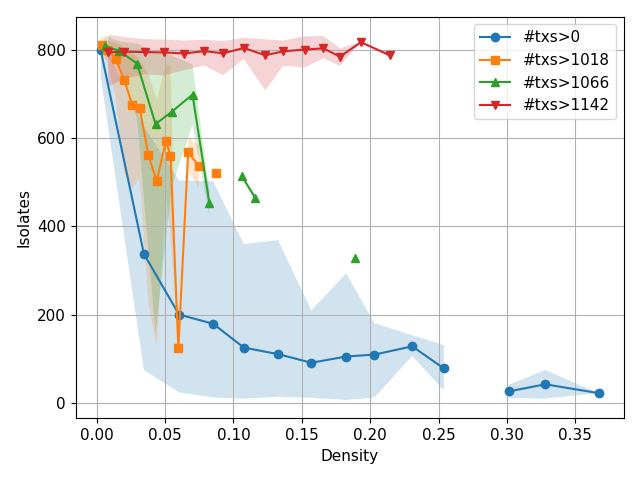}}
\caption{isolates}
\label{fig:isolates}
\end{figure}
\begin{figure}
\centerline{\includegraphics[width=\columnwidth]{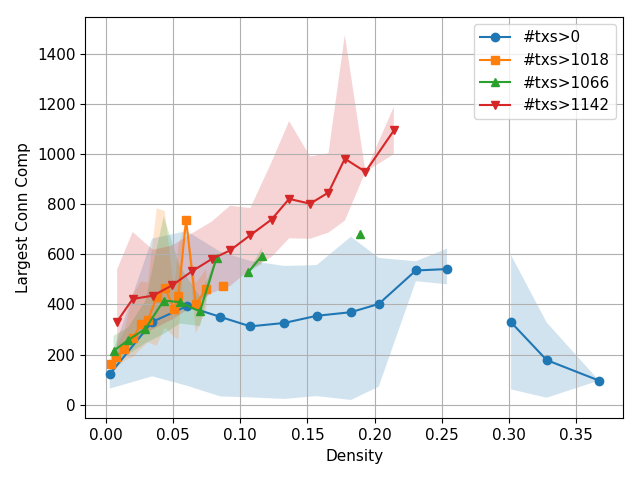}}
\caption{largest-conn-comp}
\label{fig:largest-conn-comp}
\end{figure}
\begin{figure}
\centerline{\includegraphics[width=\columnwidth]{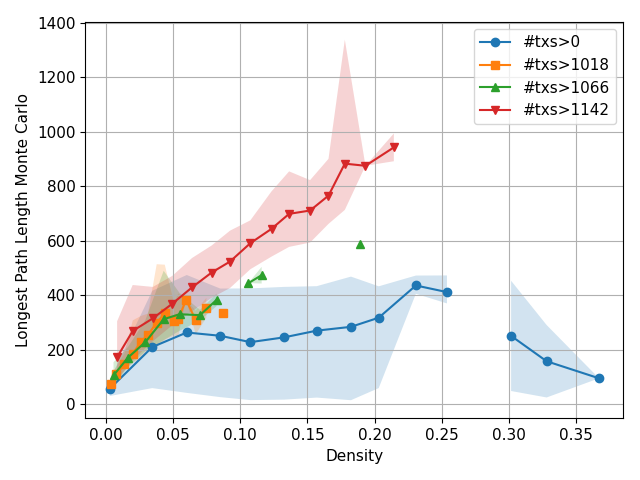}}
\caption{longest-path-length-monte-carlo}
\label{fig:longest-path-length-monte-carlo}
\end{figure}
\begin{figure}
\centerline{\includegraphics[width=\columnwidth]{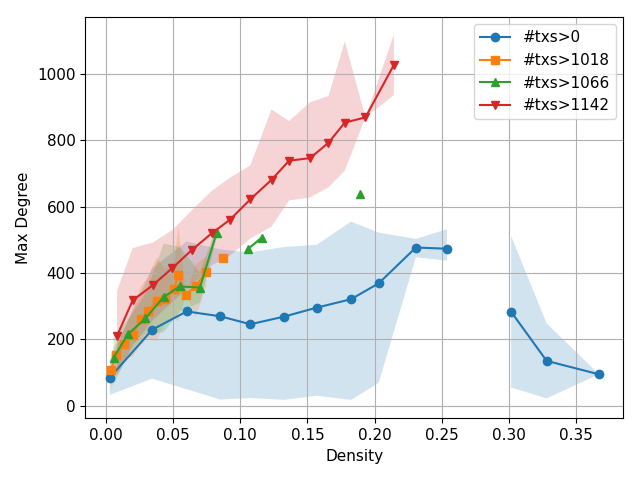}}
\caption{max-degree}
\label{fig:max-degree}
\end{figure}
\begin{figure}
\centerline{\includegraphics[width=\columnwidth]{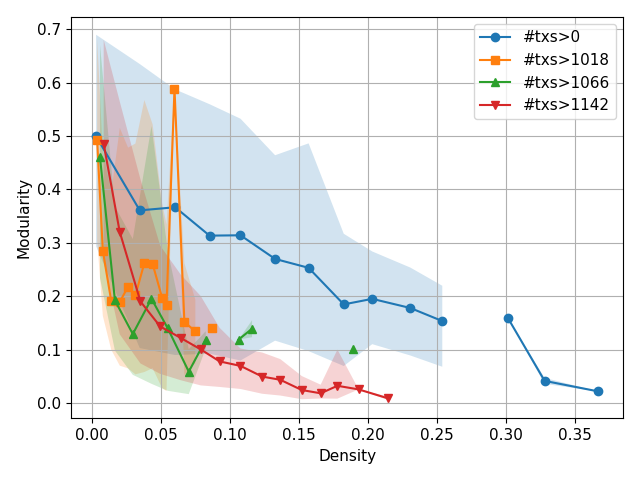}}
\caption{modularity}
\label{fig:modularity}
\end{figure}
\begin{figure}
\centerline{\includegraphics[width=\columnwidth]{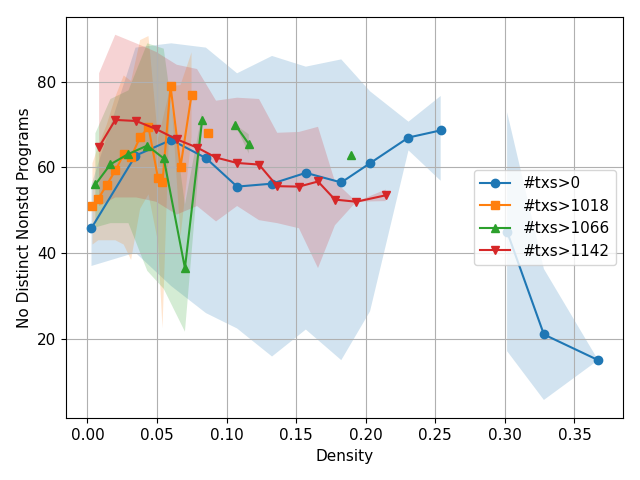}}
\caption{no-distinct-nonstd-programs}
\label{fig:no-distinct-nonstd-programs}
\end{figure}
\begin{figure}
\centerline{\includegraphics[width=\columnwidth]{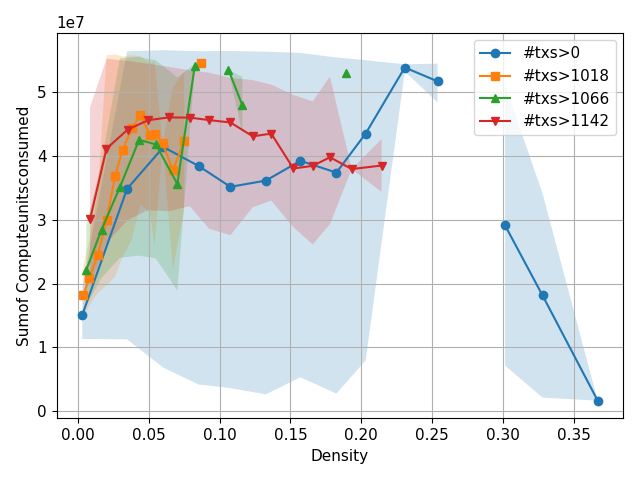}}
\caption{sumof-computeUnitsConsumed}
\label{fig:sumof-computeUnitsConsumed}
\end{figure}
\begin{figure}
\centerline{\includegraphics[width=\columnwidth]{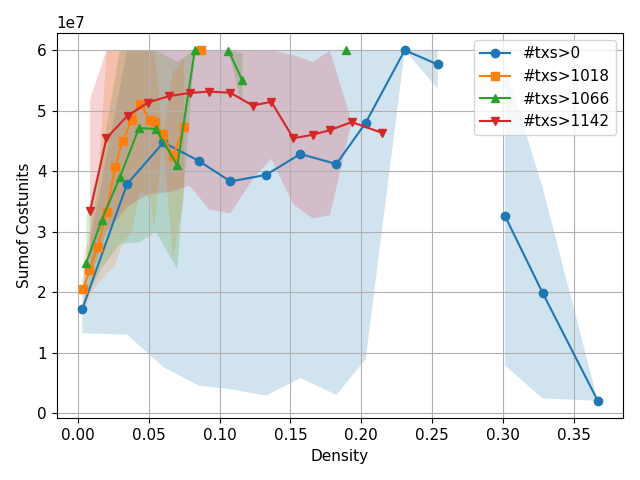}}
\caption{sumof-costUnits}
\label{fig:sumof-costUnits}
\end{figure}
\begin{figure}
\centerline{\includegraphics[width=\columnwidth]{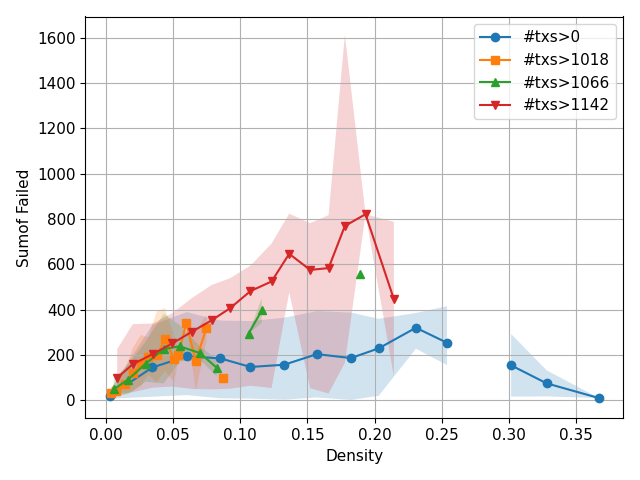}}
\caption{sumof-failed}
\label{fig:sumof-failed}
\end{figure}
\begin{figure}
\centerline{\includegraphics[width=\columnwidth]{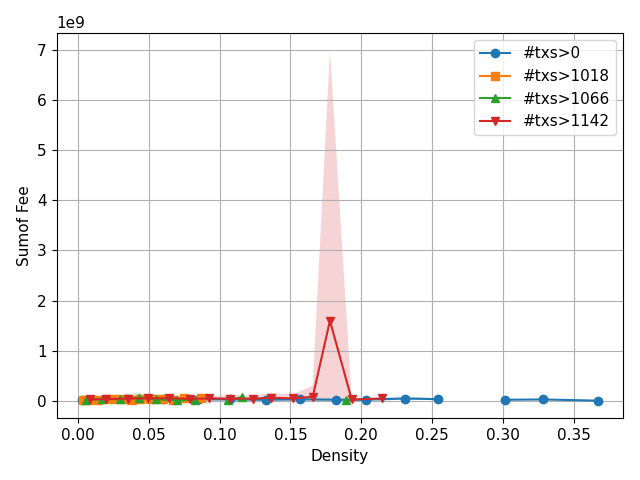}}
\caption{sumof-fee}
\label{fig:sumof-fee}
\end{figure}
\begin{figure}
\centerline{\includegraphics[width=\columnwidth]{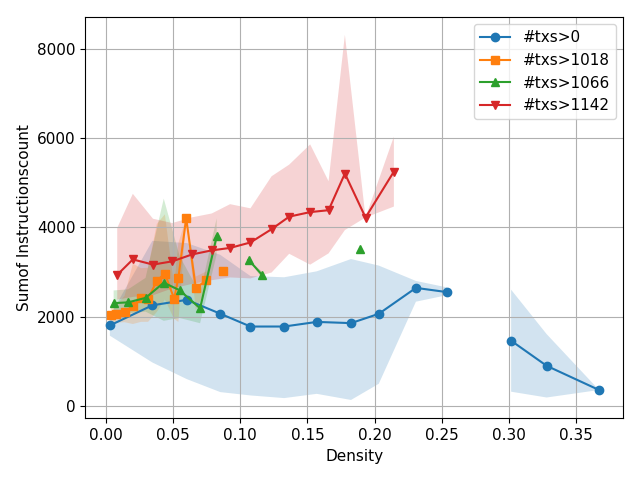}}
\caption{sumof-instructionsCount}
\label{fig:sumof-instructionsCount}
\end{figure}
\begin{figure}
\centerline{\includegraphics[width=\columnwidth]{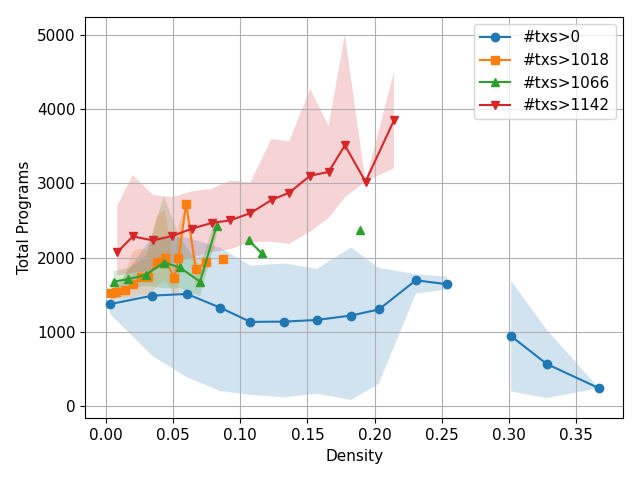}}
\caption{total-programs}
\label{fig:total-programs}
\end{figure}
\begin{figure}
\centerline{\includegraphics[width=\columnwidth]{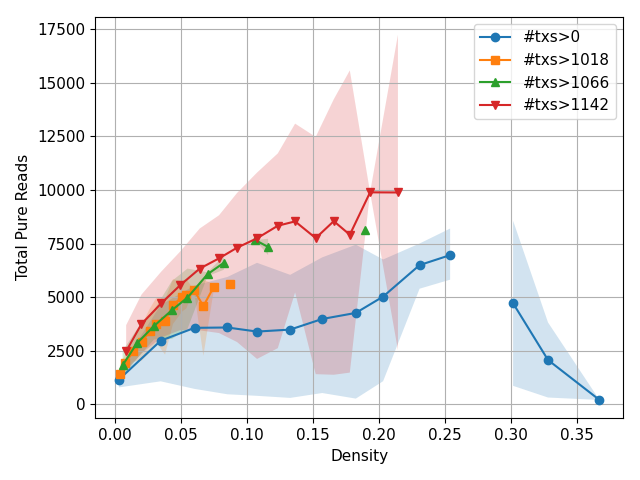}}
\caption{total-pure-reads}
\label{fig:total-pure-reads}
\end{figure}
\begin{figure}
\centerline{\includegraphics[width=\columnwidth]{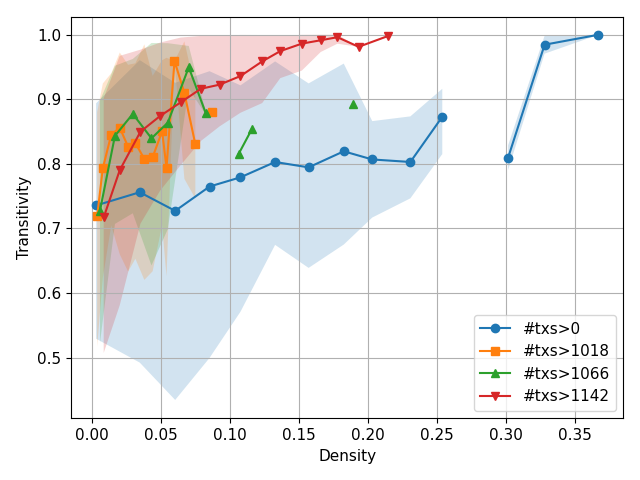}}
\caption{transitivity}
\label{fig:transitivity}
\end{figure}
\begin{figure}
\centerline{\includegraphics[width=\columnwidth]{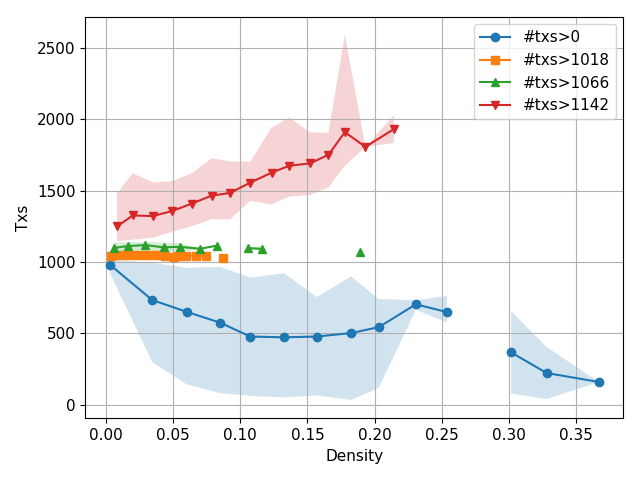}}
\caption{txs}
\label{fig:txs}
\end{figure}
\begin{figure}
\centerline{\includegraphics[width=\columnwidth]{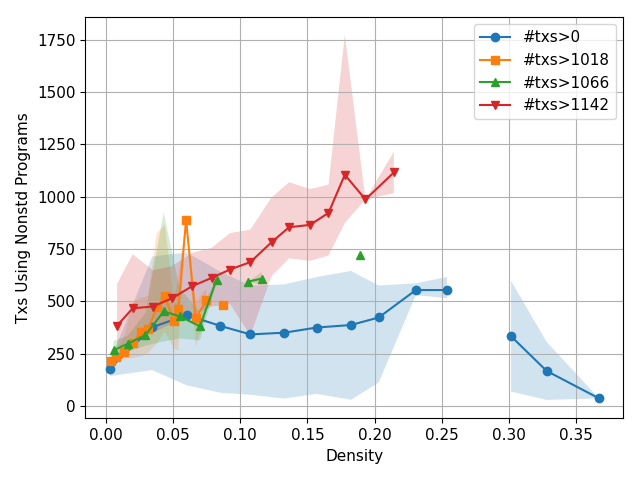}}
\caption{txs-using-nonstd-programs}
\label{fig:txs-using-nonstd-programs}
\end{figure}
\begin{figure}
\centerline{\includegraphics[width=\columnwidth]{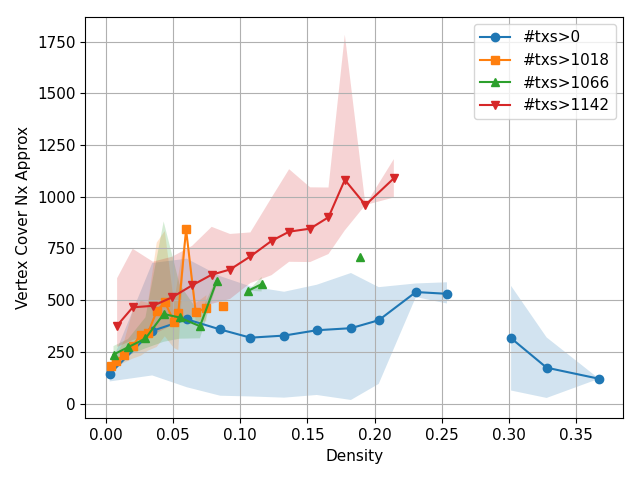}}
\caption{vertex-cover-nx-approx}
\label{fig:vertex-cover-nx-approx}
\end{figure}
\begin{figure}
\centerline{\includegraphics[width=\columnwidth]{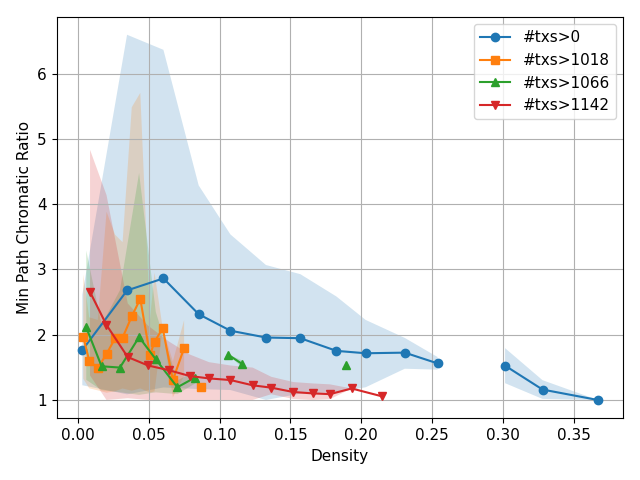}}
\caption{min-path-chromatic-ratio}
\label{fig:min-path-chromatic-ratio}
\end{figure}
\begin{figure}
\centerline{\includegraphics[width=\columnwidth]{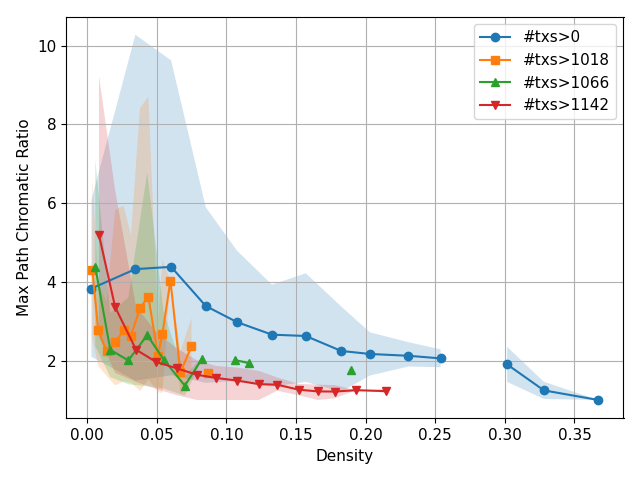}}
\caption{max-path-chromatic-ratio}
\label{fig:max-path-chromatic-ratio}
\end{figure}
\begin{figure}
\centerline{\includegraphics[width=\columnwidth]{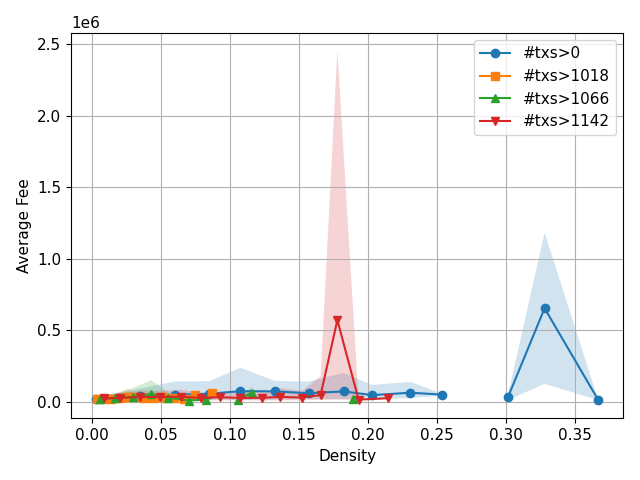}}
\caption{avg-fee}
\label{fig:avg-fee}
\end{figure}
\begin{figure}
\centerline{\includegraphics[width=\columnwidth]{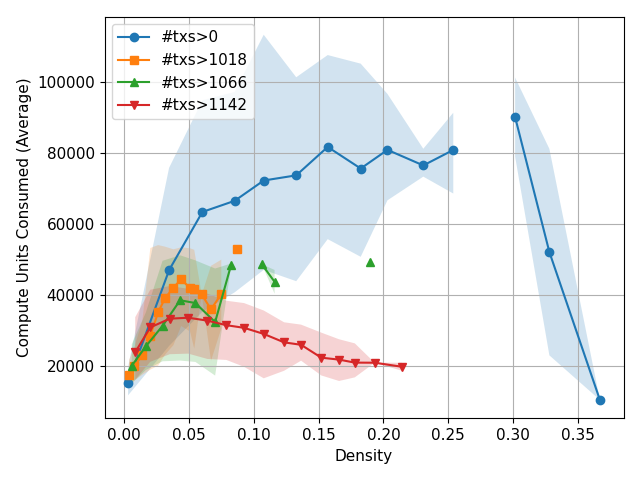}}
\caption{avg-computeUnitsConsumed}
\label{fig:avg-computeUnitsConsumed}
\end{figure}
\begin{figure}
\centerline{\includegraphics[width=\columnwidth]{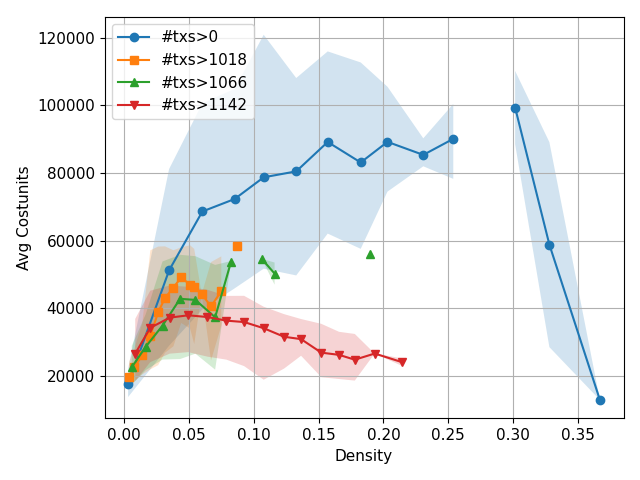}}
\caption{avg-costUnits}
\label{fig:avg-costUnits}
\end{figure}
\begin{figure}
\centerline{\includegraphics[width=\columnwidth]{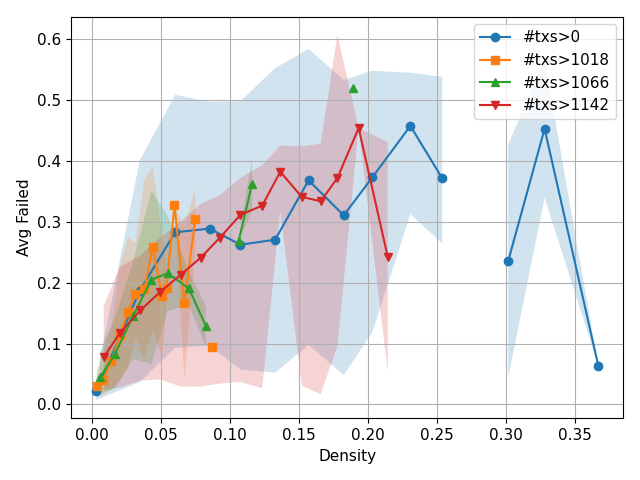}}
\caption{avg-failed}
\label{fig:avg-failed}
\end{figure}
\begin{figure}
\centerline{\includegraphics[width=\columnwidth]{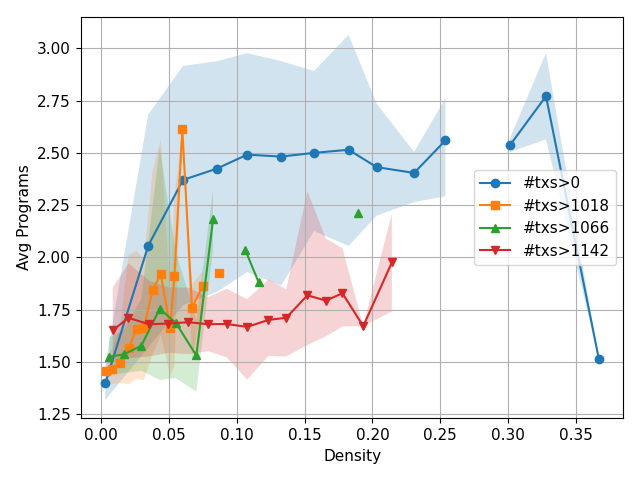}}
\caption{avg-programs}
\label{fig:avg-programs}
\end{figure}
\begin{figure}
\centerline{\includegraphics[width=\columnwidth]{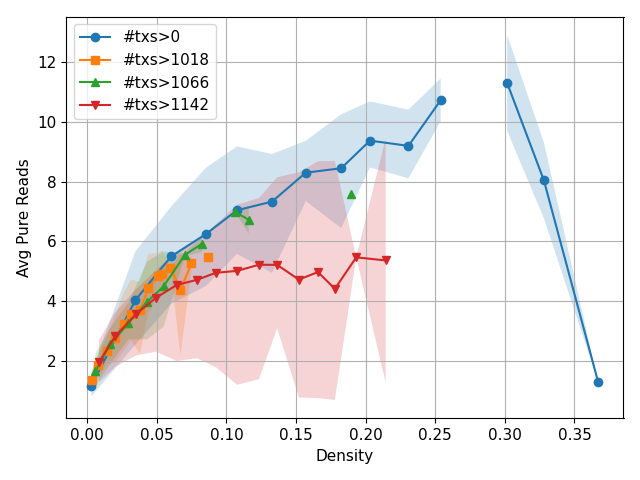}}
\caption{avg-pure-reads}
\label{fig:avg-pure-reads}
\end{figure}
\begin{figure}
\centerline{\includegraphics[width=\columnwidth]{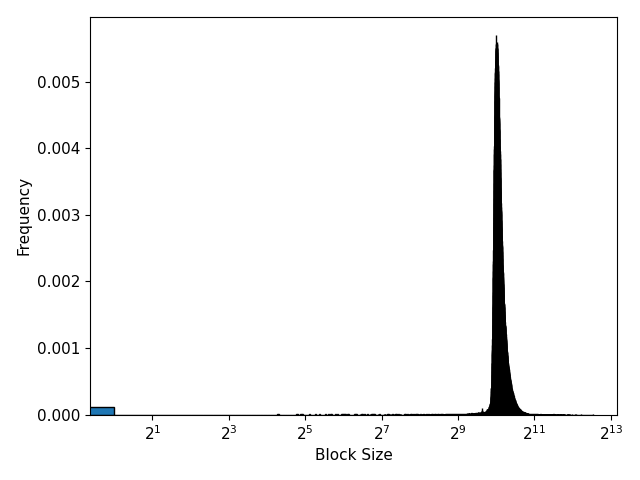}}
\caption{block-size-dist}
\label{fig:block-size-dist}
\end{figure}
\begin{figure}
\centerline{\includegraphics[width=\columnwidth]{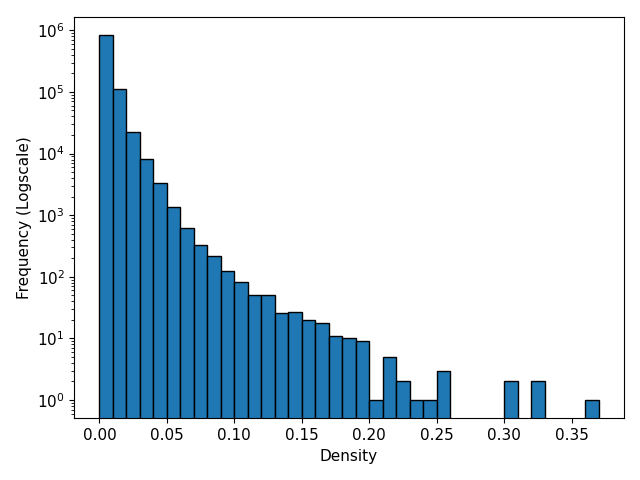}}
\caption{density-dist}
\label{fig:density-dist}
\end{figure}
\begin{figure}
\centerline{\includegraphics[width=\columnwidth]{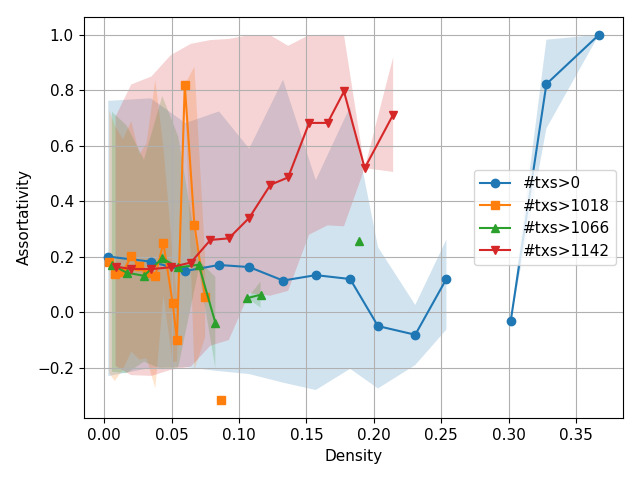}}
\caption{assortativity}
\label{fig:assortativity}
\end{figure}
\begin{figure}
\centerline{\includegraphics[width=\columnwidth]{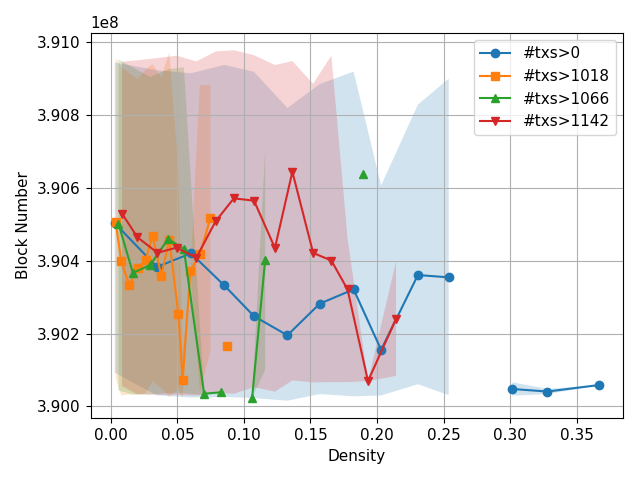}}
\caption{block-number}
\label{fig:block-number}
\end{figure}
\begin{figure}
\centerline{\includegraphics[width=\columnwidth]{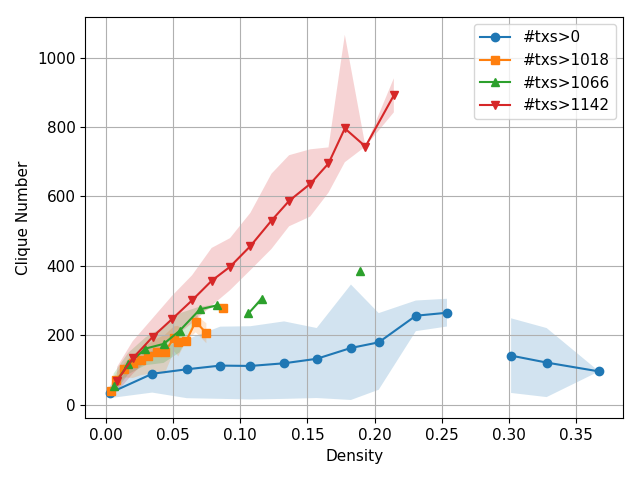}}
\caption{clique-number}
\label{fig:clique-number}
\end{figure}
\begin{figure}
\centerline{\includegraphics[width=\columnwidth]{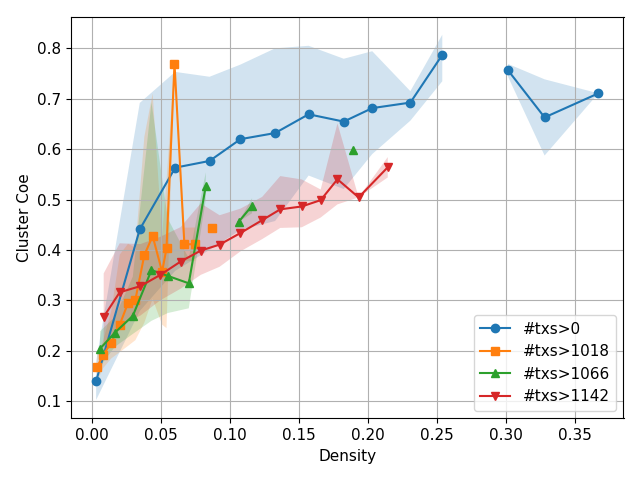}}
\caption{cluster-coe}
\label{fig:cluster-coe}
\end{figure}
\begin{figure}
\centerline{\includegraphics[width=\columnwidth]{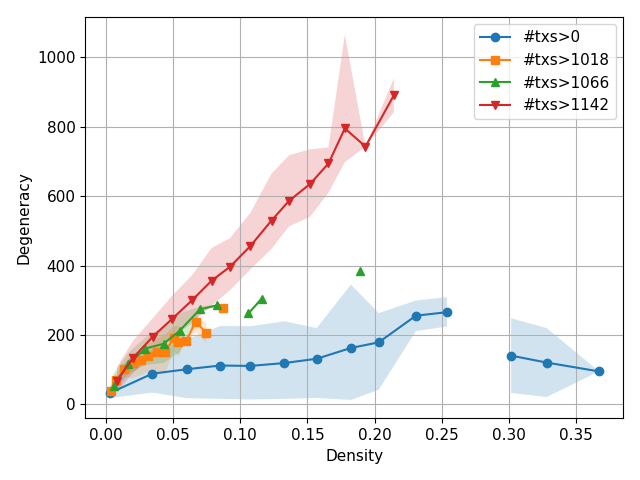}}
\caption{degeneracy}
\label{fig:degeneracy}
\end{figure}
\begin{figure}
\centerline{\includegraphics[width=\columnwidth]{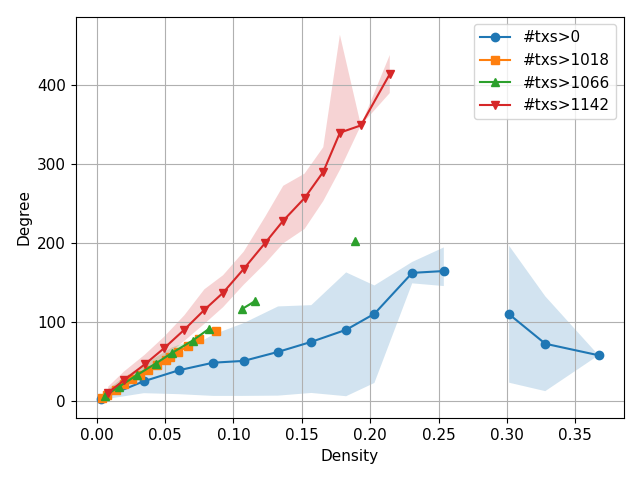}}
\caption{degree}
\label{fig:degree}
\end{figure}
\begin{figure}
\centerline{\includegraphics[width=\columnwidth]{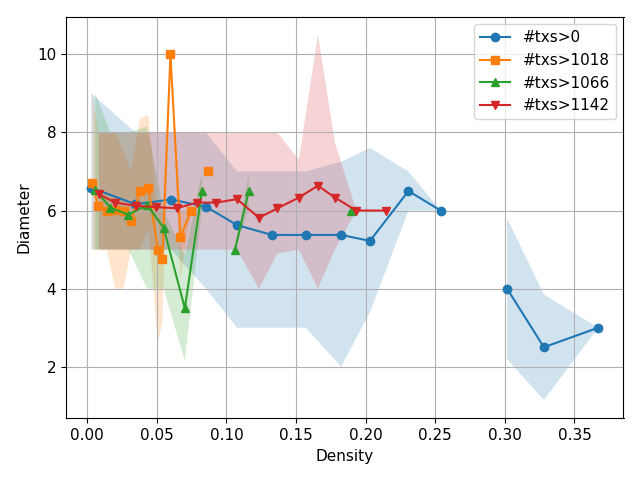}}
\caption{diameter}
\label{fig:diameter}
\end{figure}
\begin{figure}
\centerline{\includegraphics[width=\columnwidth]{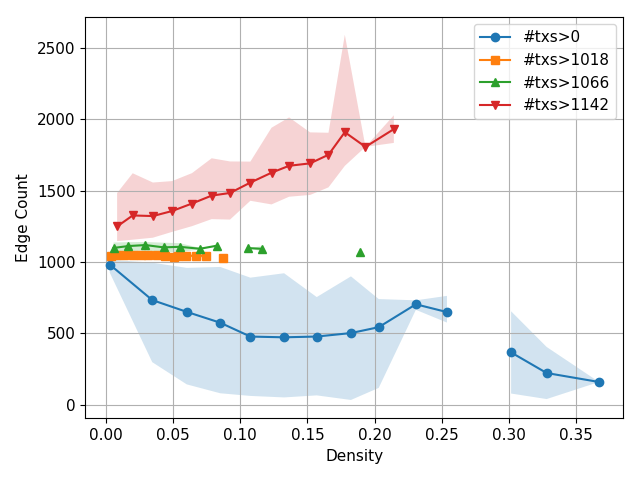}}
\caption{edge-count}
\label{fig:edge-count}
\end{figure}
\begin{figure}
\centerline{\includegraphics[width=\columnwidth]{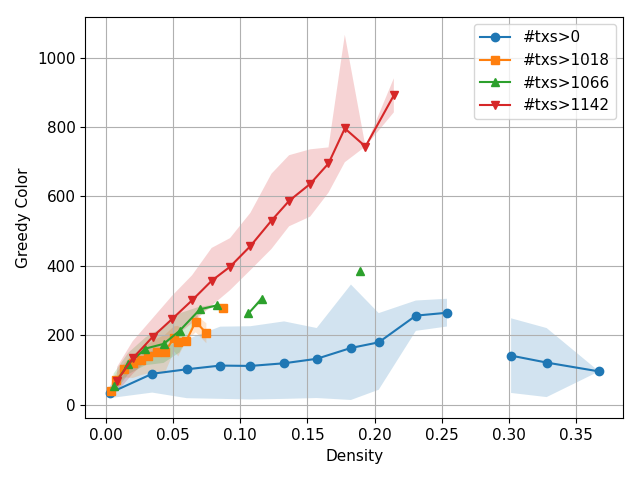}}
\caption{greedy-color}
\label{fig:greedy-color}
\end{figure}
\begin{figure}
\centerline{\includegraphics[width=\columnwidth]{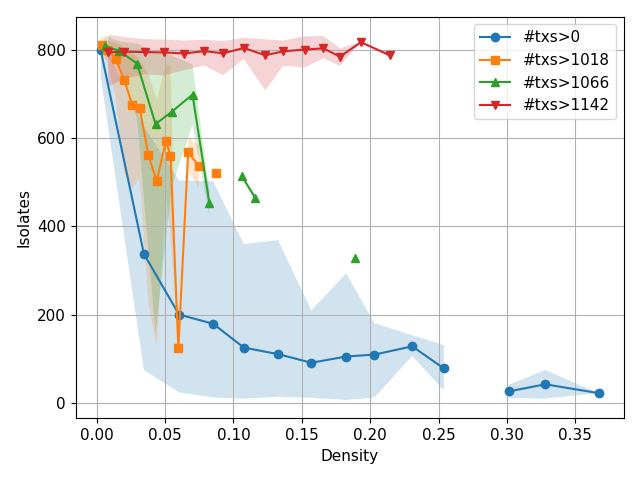}}
\caption{isolates}
\label{fig:isolates}
\end{figure}
\begin{figure}
\centerline{\includegraphics[width=\columnwidth]{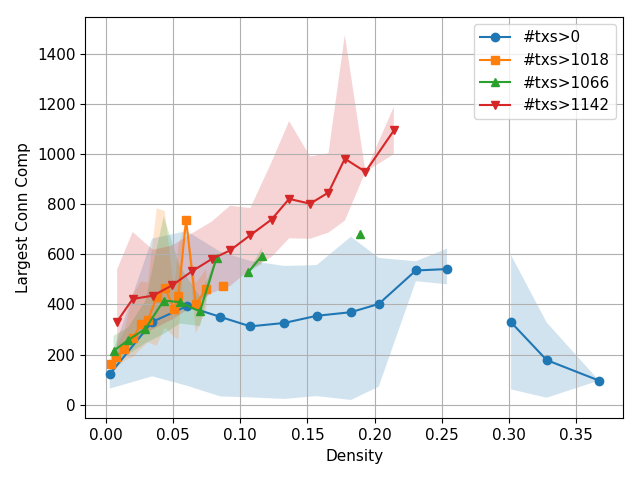}}
\caption{largest-conn-comp}
\label{fig:largest-conn-comp}
\end{figure}
\begin{figure}
\centerline{\includegraphics[width=\columnwidth]{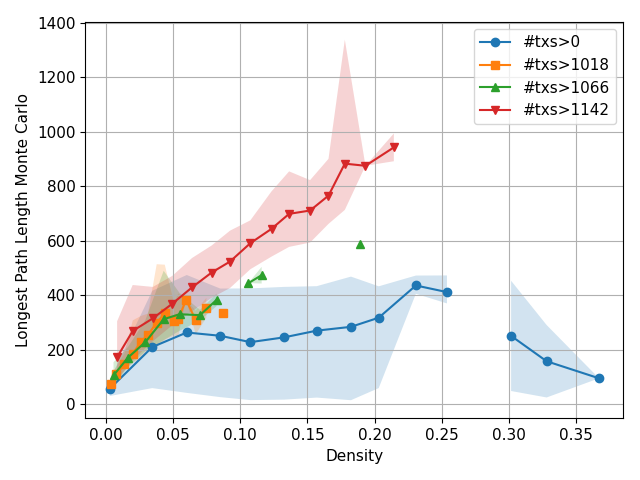}}
\caption{longest-path-length-monte-carlo}
\label{fig:longest-path-length-monte-carlo}
\end{figure}
\begin{figure}
\centerline{\includegraphics[width=\columnwidth]{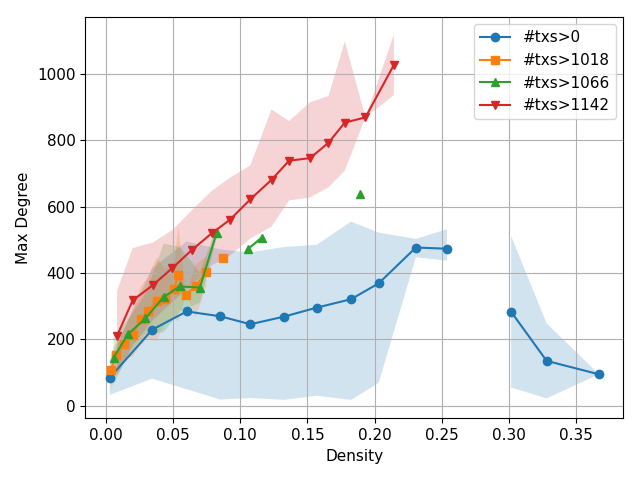}}
\caption{max-degree}
\label{fig:max-degree}
\end{figure}
\begin{figure}
\centerline{\includegraphics[width=\columnwidth]{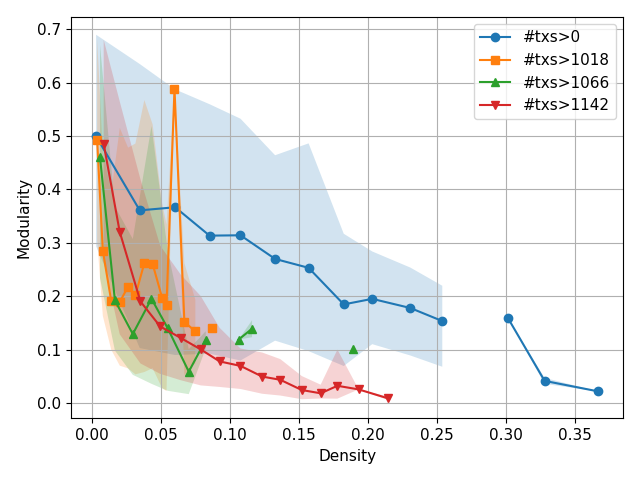}}
\caption{modularity}
\label{fig:modularity}
\end{figure}
\begin{figure}
\centerline{\includegraphics[width=\columnwidth]{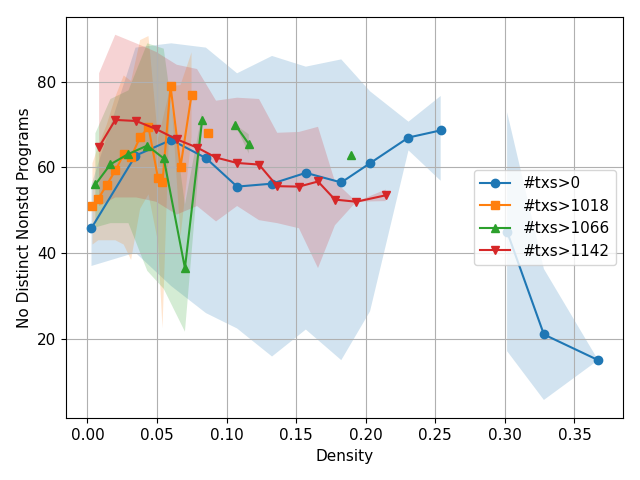}}
\caption{no-distinct-nonstd-programs}
\label{fig:no-distinct-nonstd-programs}
\end{figure}
\begin{figure}
\centerline{\includegraphics[width=\columnwidth]{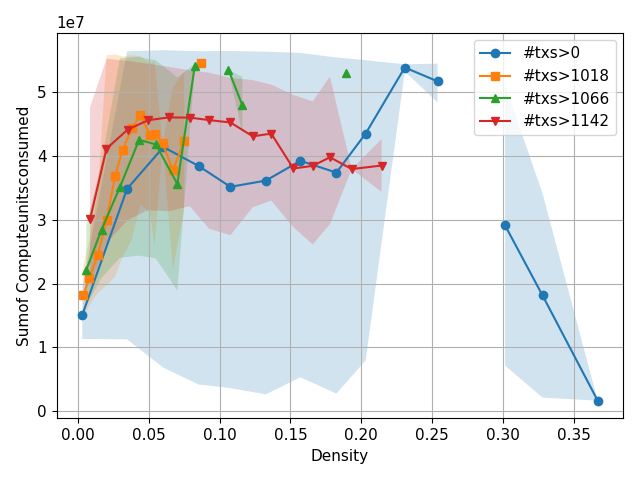}}
\caption{sumof-computeUnitsConsumed}
\label{fig:sumof-computeUnitsConsumed}
\end{figure}
\begin{figure}
\centerline{\includegraphics[width=\columnwidth]{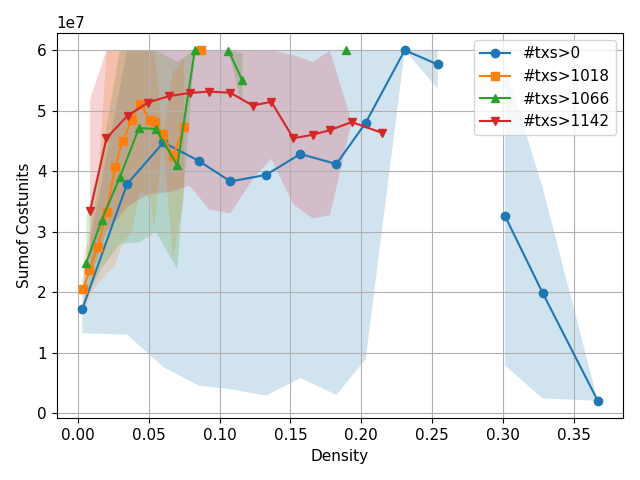}}
\caption{sumof-costUnits}
\label{fig:sumof-costUnits}
\end{figure}
\begin{figure}
\centerline{\includegraphics[width=\columnwidth]{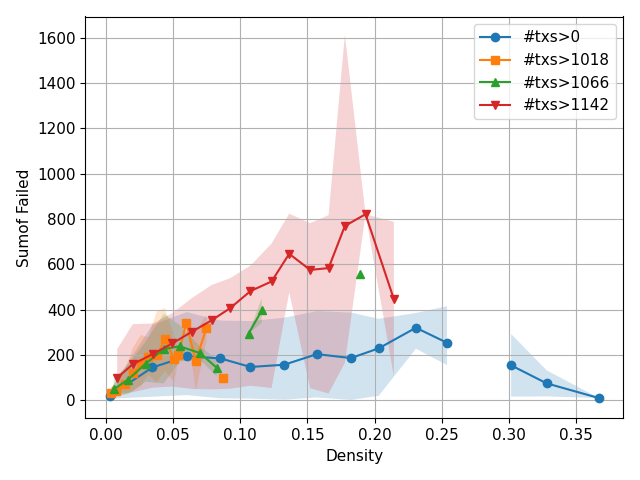}}
\caption{sumof-failed}
\label{fig:sumof-failed}
\end{figure}
\begin{figure}
\centerline{\includegraphics[width=\columnwidth]{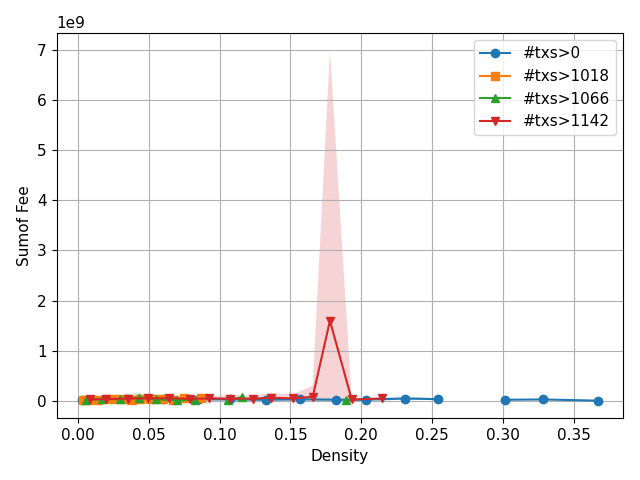}}
\caption{sumof-fee}
\label{fig:sumof-fee}
\end{figure}
\begin{figure}
\centerline{\includegraphics[width=\columnwidth]{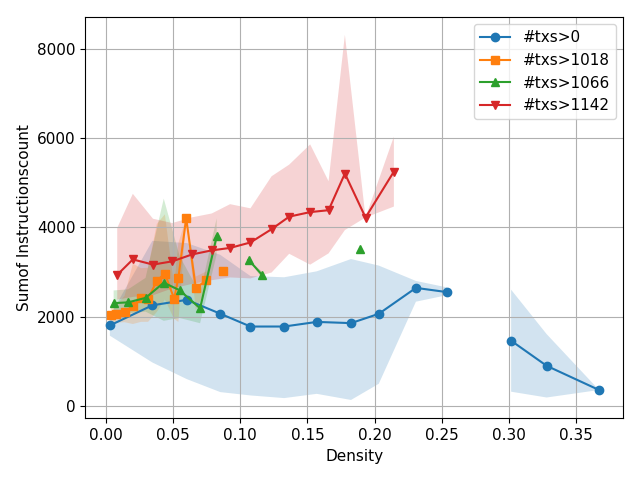}}
\caption{sumof-instructionsCount}
\label{fig:sumof-instructionsCount}
\end{figure}
\begin{figure}
\centerline{\includegraphics[width=\columnwidth]{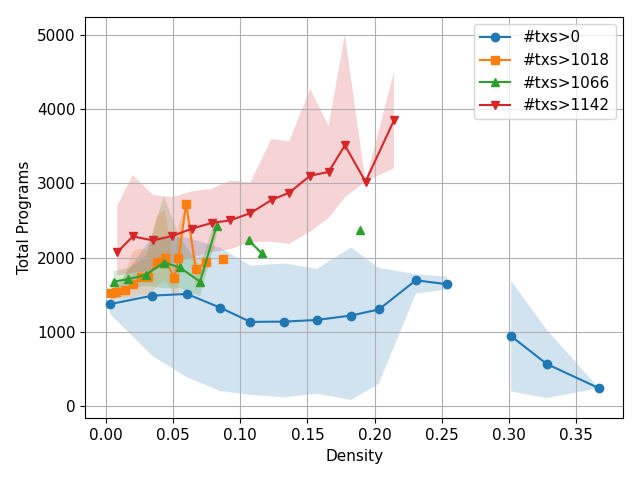}}
\caption{total-programs}
\label{fig:total-programs}
\end{figure}
\begin{figure}
\centerline{\includegraphics[width=\columnwidth]{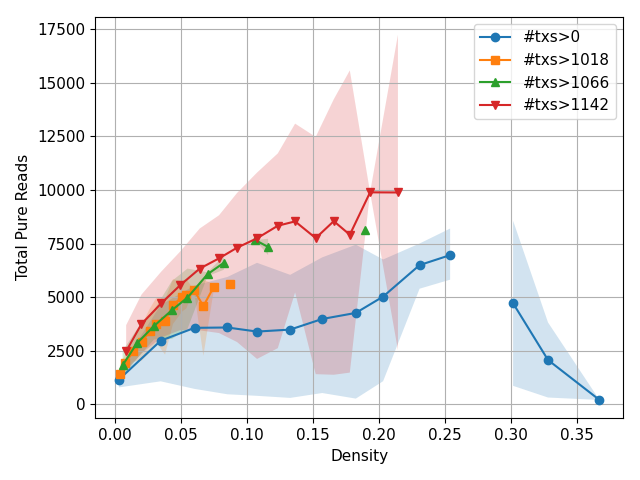}}
\caption{total-pure-reads}
\label{fig:total-pure-reads}
\end{figure}
\begin{figure}
\centerline{\includegraphics[width=\columnwidth]{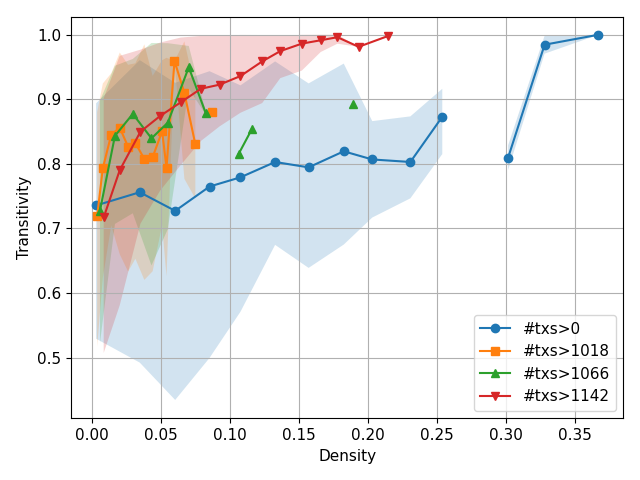}}
\caption{transitivity}
\label{fig:transitivity}
\end{figure}
\begin{figure}
\centerline{\includegraphics[width=\columnwidth]{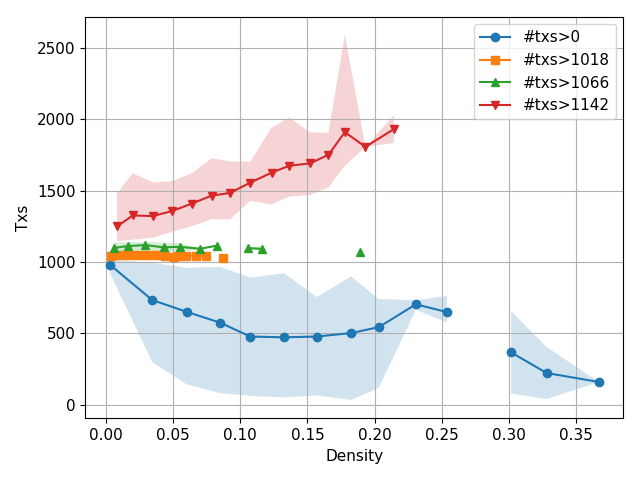}}
\caption{txs}
\label{fig:txs}
\end{figure}
\begin{figure}
\centerline{\includegraphics[width=\columnwidth]{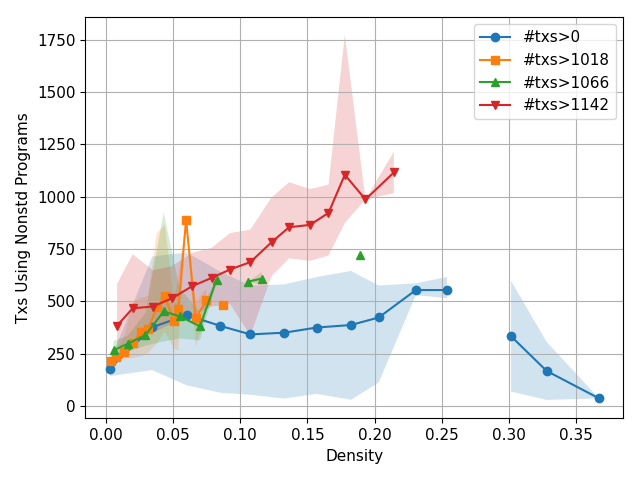}}
\caption{txs-using-nonstd-programs}
\label{fig:txs-using-nonstd-programs}
\end{figure}
\begin{figure}
\centerline{\includegraphics[width=\columnwidth]{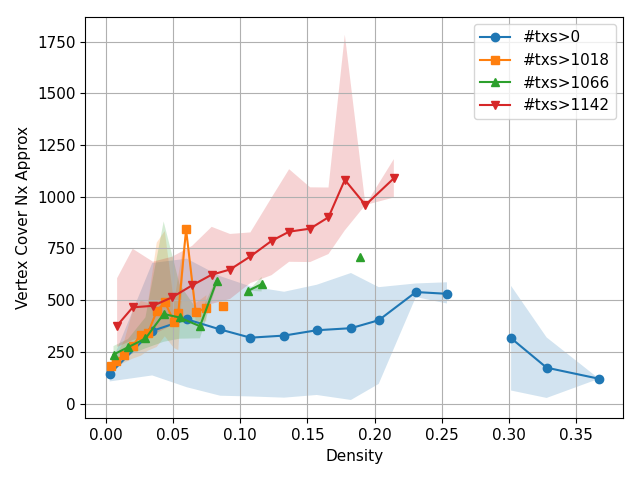}}
\caption{vertex-cover-nx-approx}
\label{fig:vertex-cover-nx-approx}
\end{figure}
\begin{figure}
\centerline{\includegraphics[width=\columnwidth]{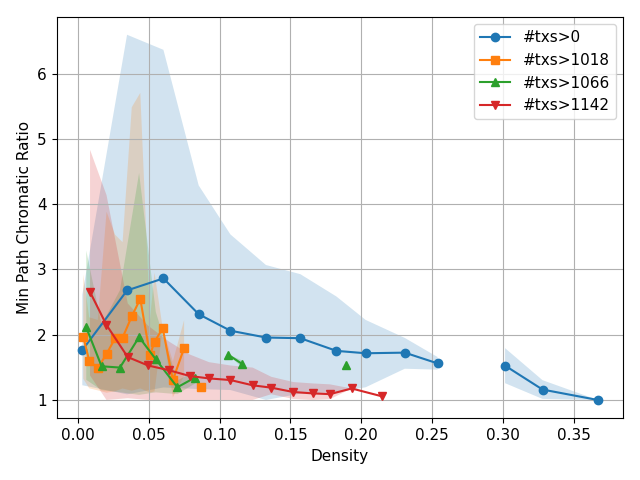}}
\caption{min-path-chromatic-ratio}
\label{fig:min-path-chromatic-ratio}
\end{figure}
\begin{figure}
\centerline{\includegraphics[width=\columnwidth]{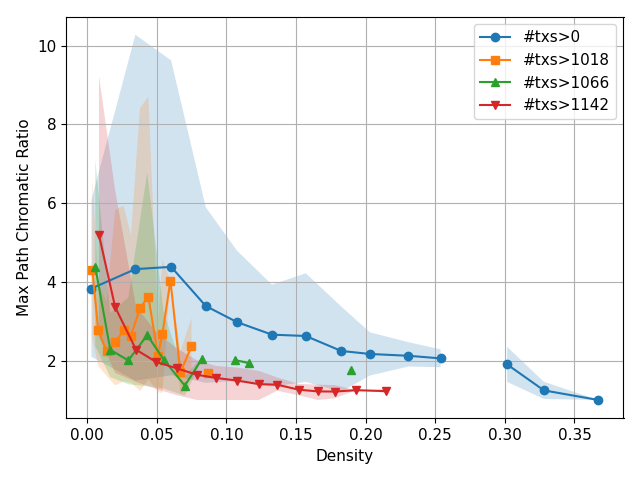}}
\caption{max-path-chromatic-ratio}
\label{fig:max-path-chromatic-ratio}
\end{figure}
\begin{figure}
\centerline{\includegraphics[width=\columnwidth]{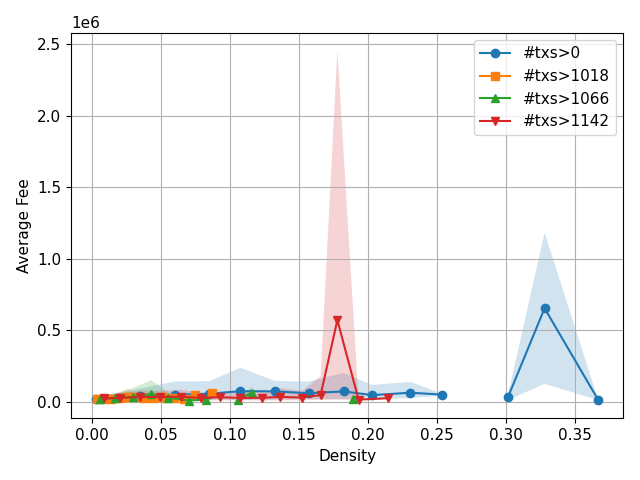}}
\caption{avg-fee}
\label{fig:avg-fee}
\end{figure}
\begin{figure}
\centerline{\includegraphics[width=\columnwidth]{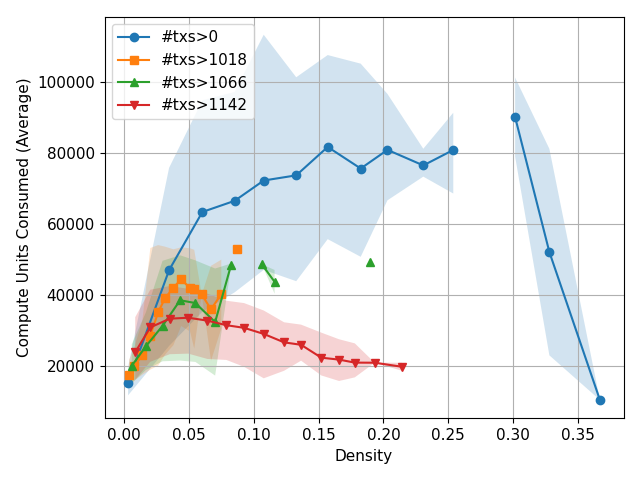}}
\caption{avg-computeUnitsConsumed}
\label{fig:avg-computeUnitsConsumed}
\end{figure}
\begin{figure}
\centerline{\includegraphics[width=\columnwidth]{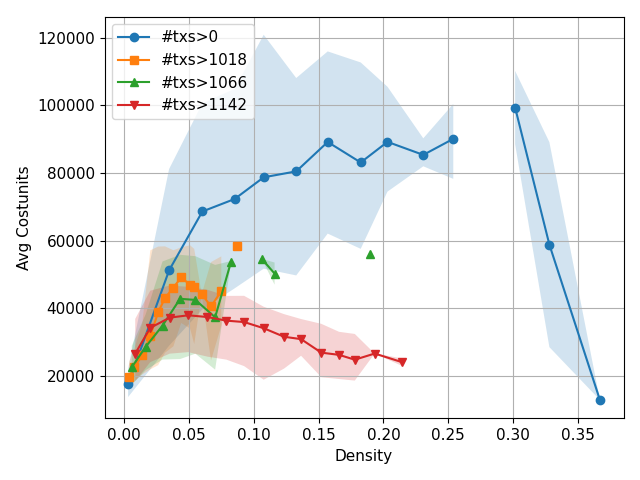}}
\caption{avg-costUnits}
\label{fig:avg-costUnits}
\end{figure}
\begin{figure}
\centerline{\includegraphics[width=\columnwidth]{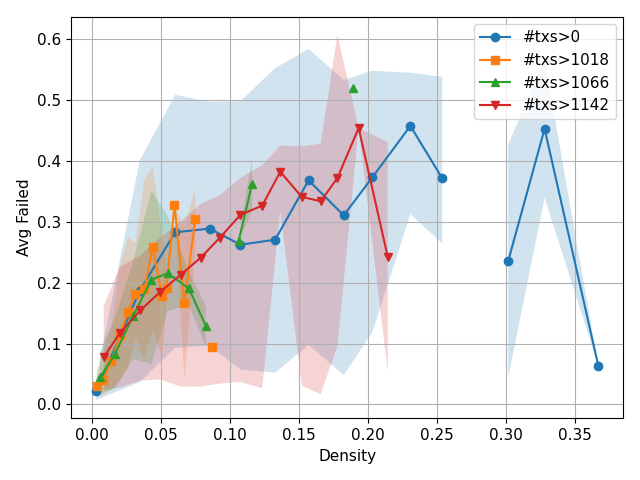}}
\caption{avg-failed}
\label{fig:avg-failed}
\end{figure}
\begin{figure}
\centerline{\includegraphics[width=\columnwidth]{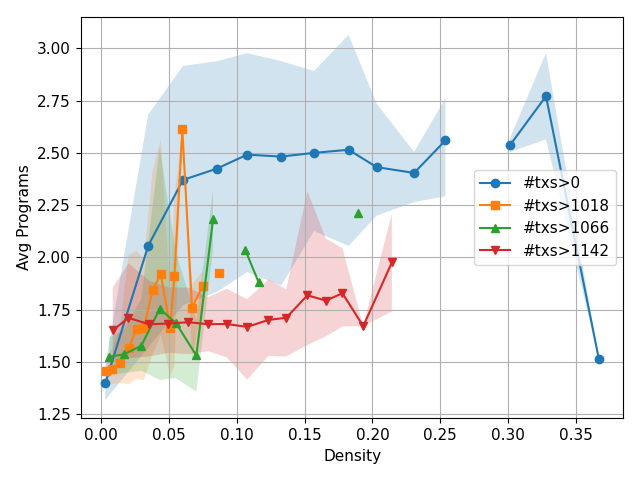}}
\caption{avg-programs}
\label{fig:avg-programs}
\end{figure}
\begin{figure}
\centerline{\includegraphics[width=\columnwidth]{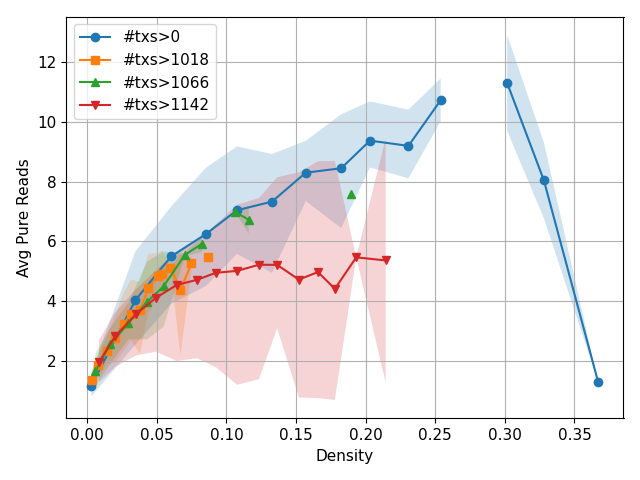}}
\caption{avg-pure-reads}
\label{fig:avg-pure-reads}
\end{figure}
\begin{figure}
\centerline{\includegraphics[width=\columnwidth]{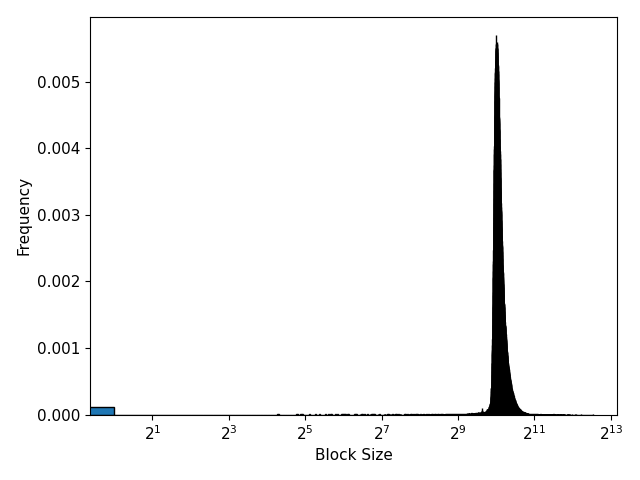}}
\caption{block-size-dist}
\label{fig:block-size-dist}
\end{figure}
\begin{figure}
\centerline{\includegraphics[width=\columnwidth]{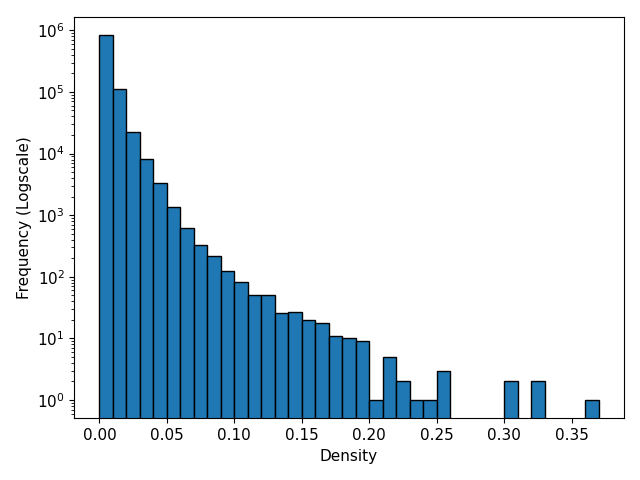}}
\caption{density-dist}
\label{fig:density-dist}
\end{figure}
\begin{figure}
\centerline{\includegraphics[width=\columnwidth]{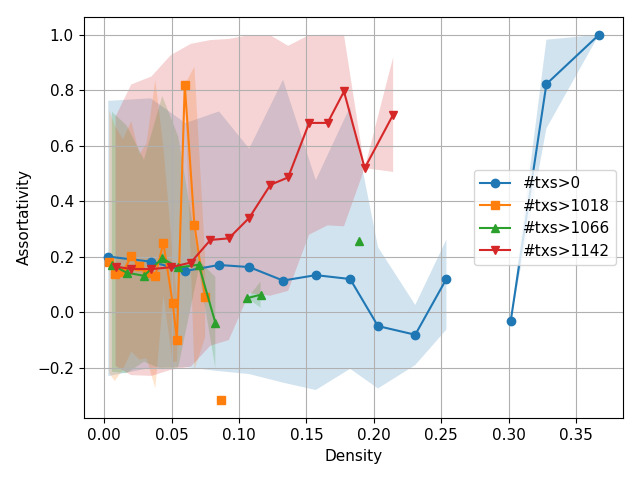}}
\caption{assortativity}
\label{fig:assortativity}
\end{figure}
\begin{figure}
\centerline{\includegraphics[width=\columnwidth]{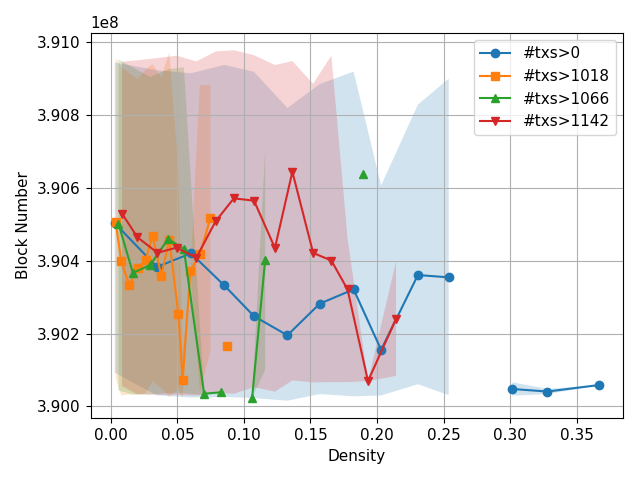}}
\caption{block-number}
\label{fig:block-number}
\end{figure}
\begin{figure}
\centerline{\includegraphics[width=\columnwidth]{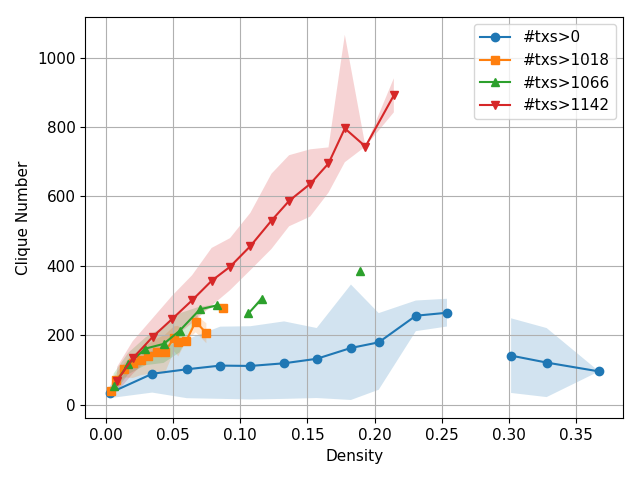}}
\caption{clique-number}
\label{fig:clique-number}
\end{figure}
\begin{figure}
\centerline{\includegraphics[width=\columnwidth]{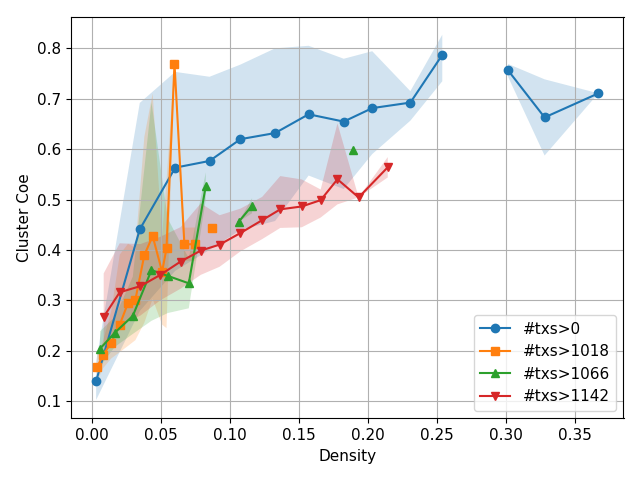}}
\caption{cluster-coe}
\label{fig:cluster-coe}
\end{figure}
\begin{figure}
\centerline{\includegraphics[width=\columnwidth]{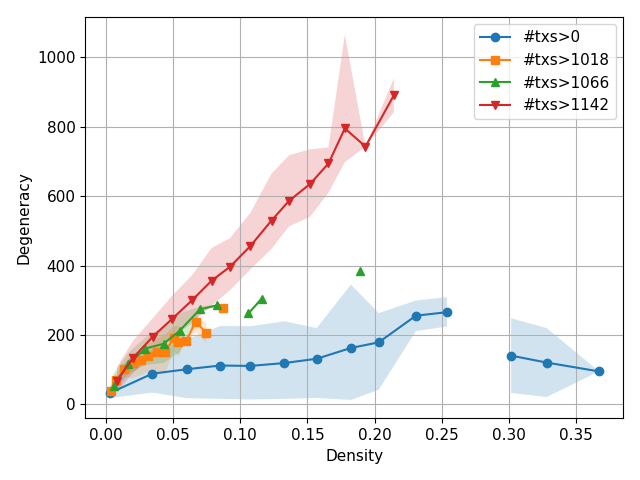}}
\caption{degeneracy}
\label{fig:degeneracy}
\end{figure}
\begin{figure}
\centerline{\includegraphics[width=\columnwidth]{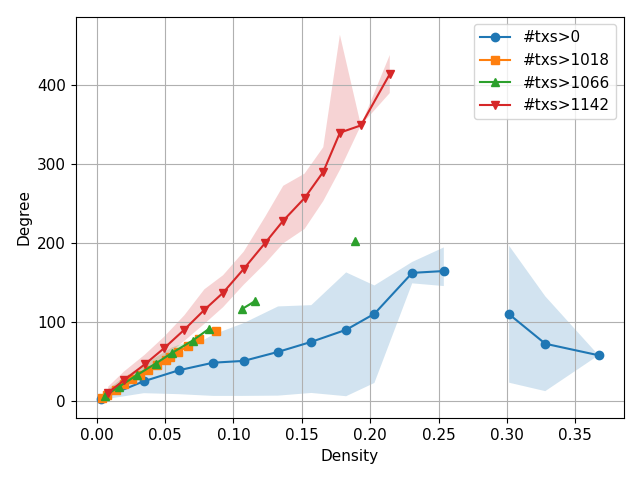}}
\caption{degree}
\label{fig:degree}
\end{figure}
\begin{figure}
\centerline{\includegraphics[width=\columnwidth]{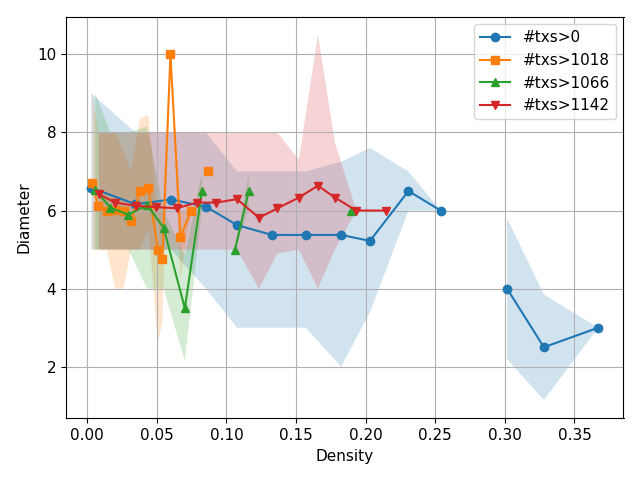}}
\caption{diameter}
\label{fig:diameter}
\end{figure}
\begin{figure}
\centerline{\includegraphics[width=\columnwidth]{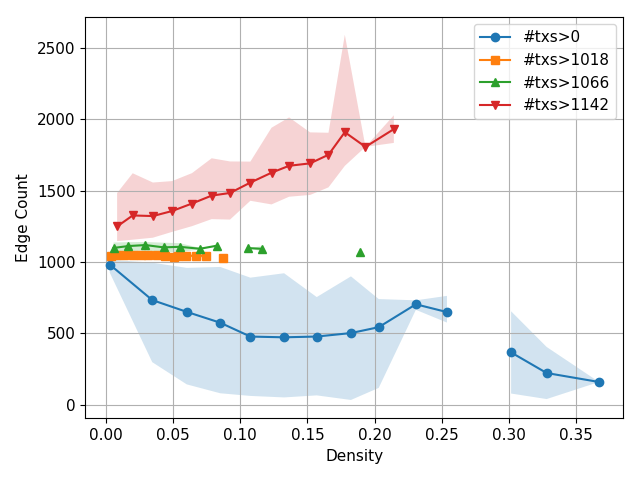}}
\caption{edge-count}
\label{fig:edge-count}
\end{figure}
\begin{figure}
\centerline{\includegraphics[width=\columnwidth]{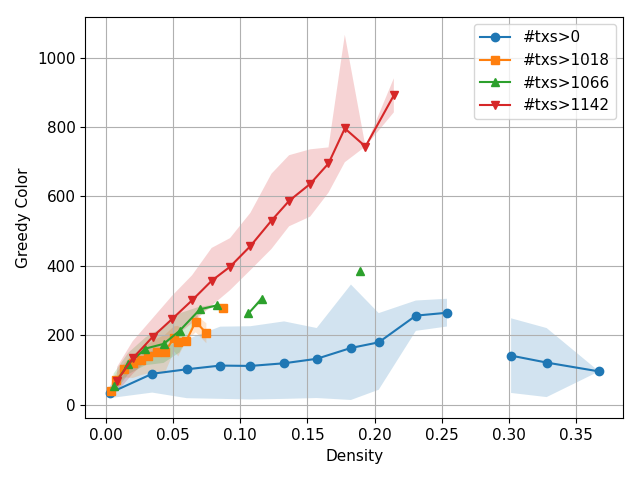}}
\caption{greedy-color}
\label{fig:greedy-color}
\end{figure}
\begin{figure}
\centerline{\includegraphics[width=\columnwidth]{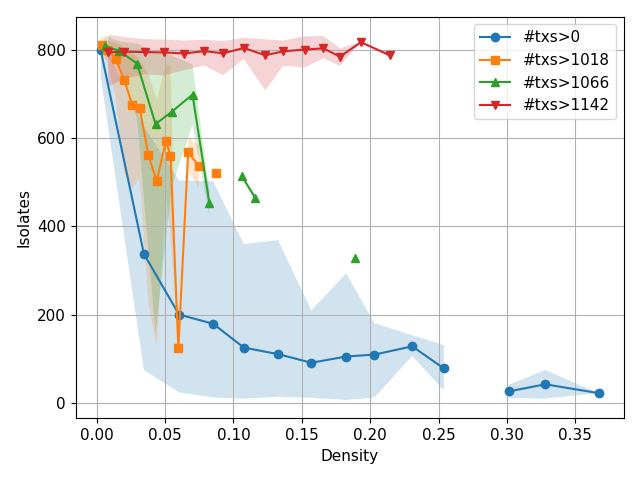}}
\caption{isolates}
\label{fig:isolates}
\end{figure}
\begin{figure}
\centerline{\includegraphics[width=\columnwidth]{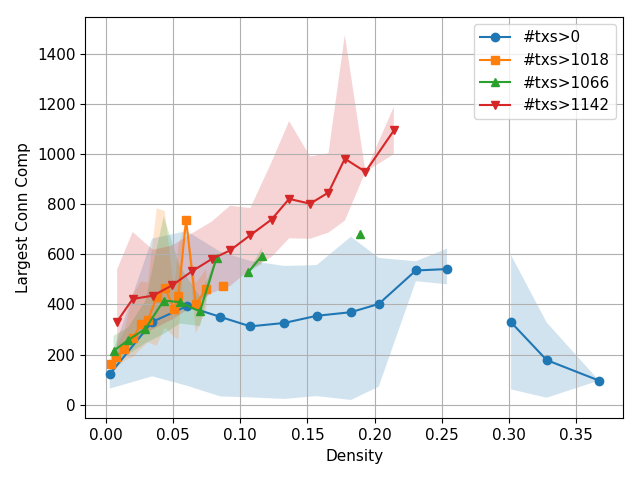}}
\caption{largest-conn-comp}
\label{fig:largest-conn-comp}
\end{figure}
\begin{figure}
\centerline{\includegraphics[width=\columnwidth]{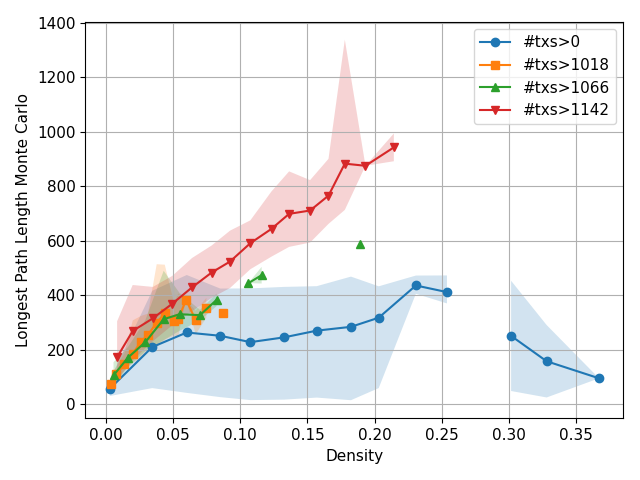}}
\caption{longest-path-length-monte-carlo}
\label{fig:longest-path-length-monte-carlo}
\end{figure}
\begin{figure}
\centerline{\includegraphics[width=\columnwidth]{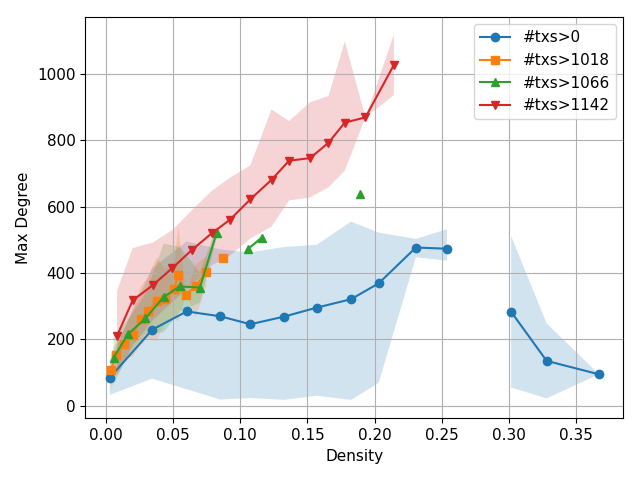}}
\caption{max-degree}
\label{fig:max-degree}
\end{figure}
\begin{figure}
\centerline{\includegraphics[width=\columnwidth]{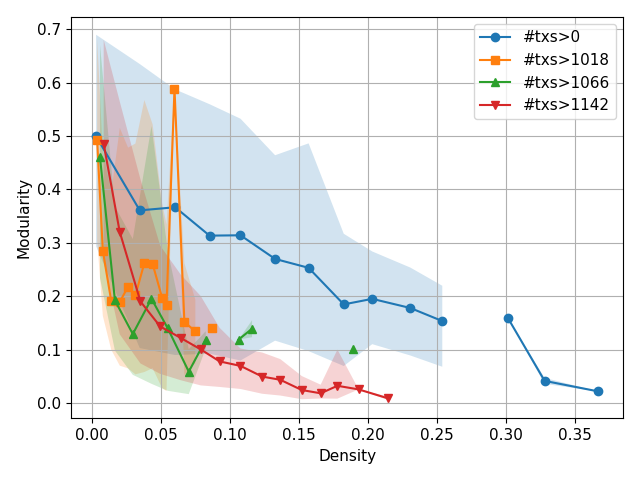}}
\caption{modularity}
\label{fig:modularity}
\end{figure}
\begin{figure}
\centerline{\includegraphics[width=\columnwidth]{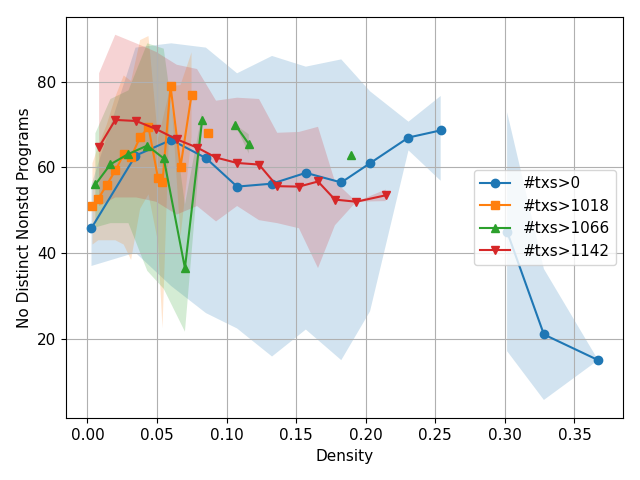}}
\caption{no-distinct-nonstd-programs}
\label{fig:no-distinct-nonstd-programs}
\end{figure}
\begin{figure}
\centerline{\includegraphics[width=\columnwidth]{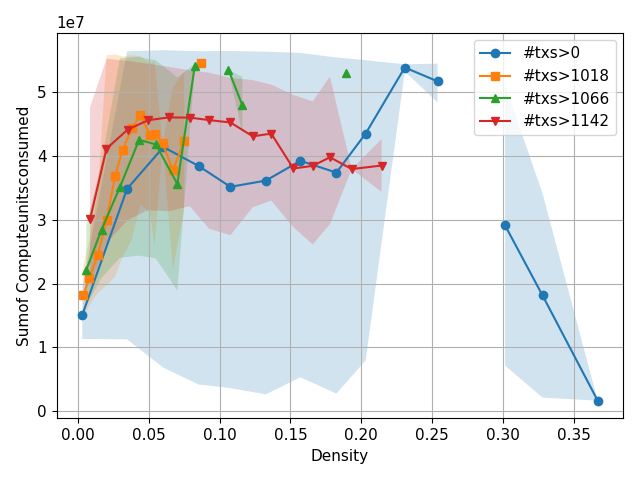}}
\caption{sumof-computeUnitsConsumed}
\label{fig:sumof-computeUnitsConsumed}
\end{figure}
\begin{figure}
\centerline{\includegraphics[width=\columnwidth]{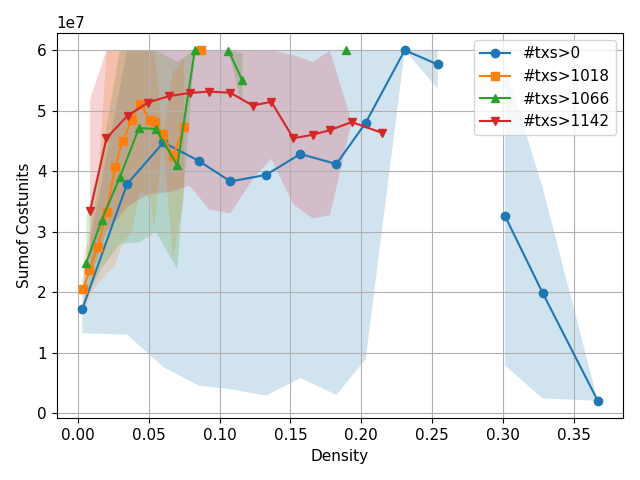}}
\caption{sumof-costUnits}
\label{fig:sumof-costUnits}
\end{figure}
\begin{figure}
\centerline{\includegraphics[width=\columnwidth]{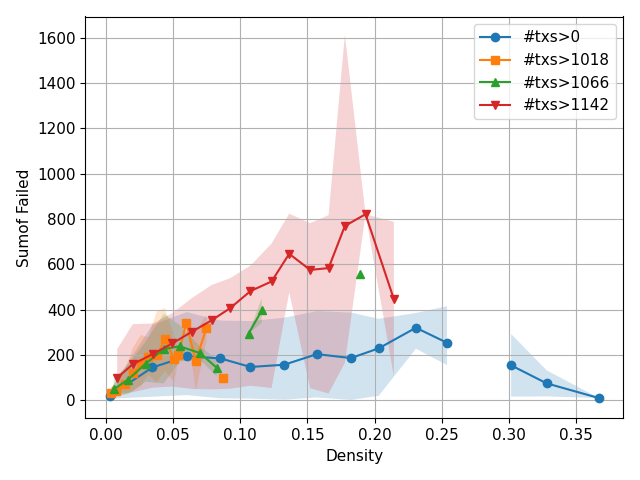}}
\caption{sumof-failed}
\label{fig:sumof-failed}
\end{figure}
\begin{figure}
\centerline{\includegraphics[width=\columnwidth]{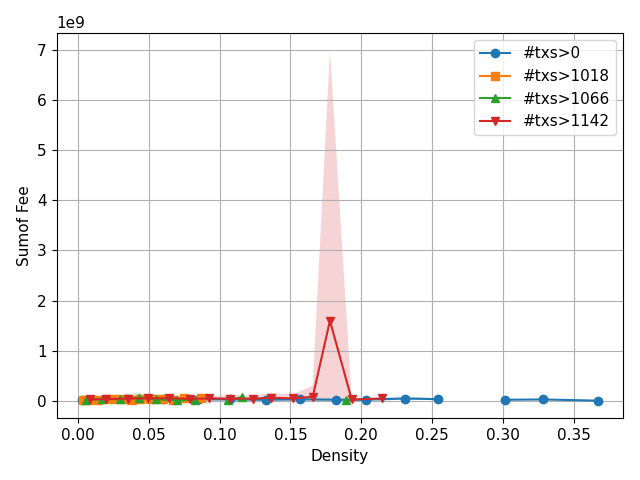}}
\caption{sumof-fee}
\label{fig:sumof-fee}
\end{figure}
\begin{figure}
\centerline{\includegraphics[width=\columnwidth]{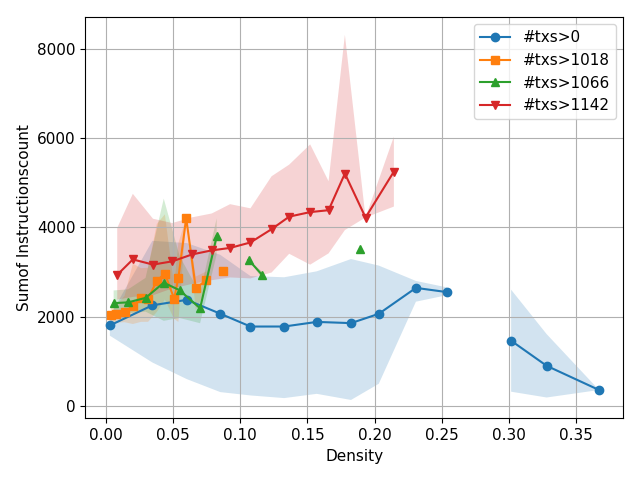}}
\caption{sumof-instructionsCount}
\label{fig:sumof-instructionsCount}
\end{figure}
\begin{figure}
\centerline{\includegraphics[width=\columnwidth]{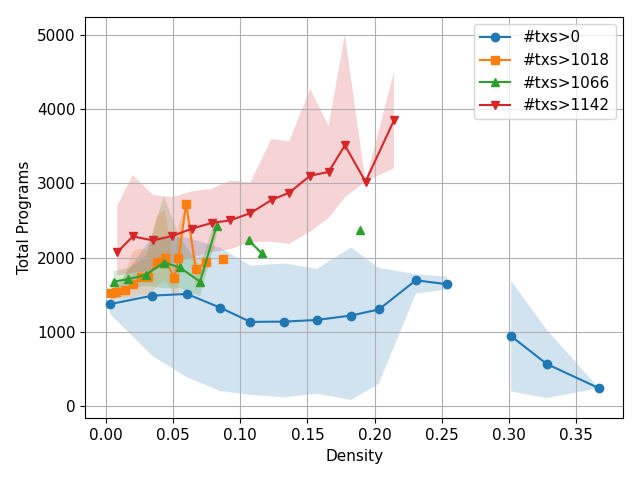}}
\caption{total-programs}
\label{fig:total-programs}
\end{figure}
\begin{figure}
\centerline{\includegraphics[width=\columnwidth]{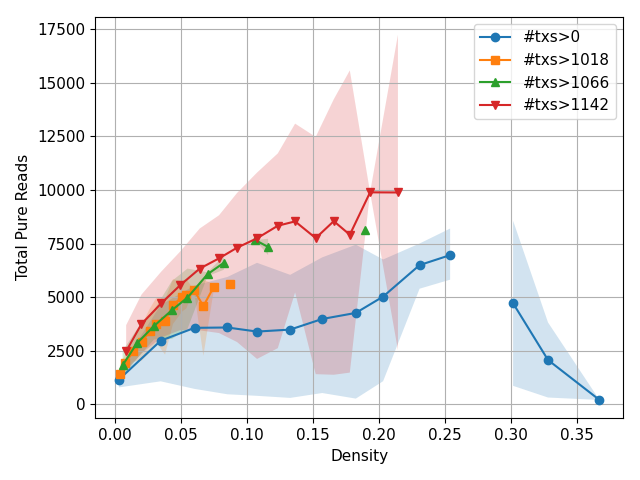}}
\caption{total-pure-reads}
\label{fig:total-pure-reads}
\end{figure}
\begin{figure}
\centerline{\includegraphics[width=\columnwidth]{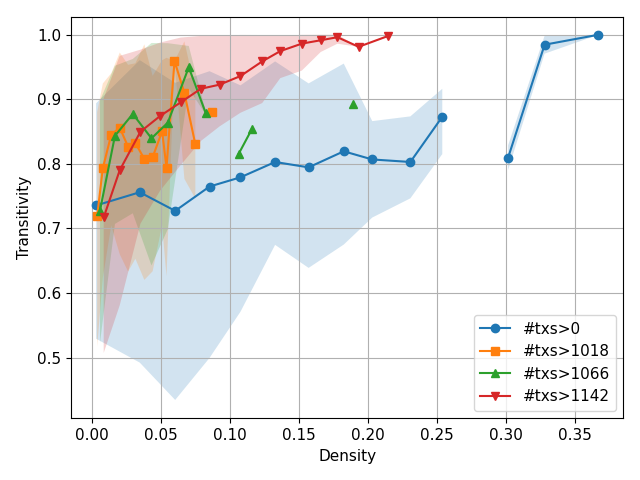}}
\caption{transitivity}
\label{fig:transitivity}
\end{figure}
\begin{figure}
\centerline{\includegraphics[width=\columnwidth]{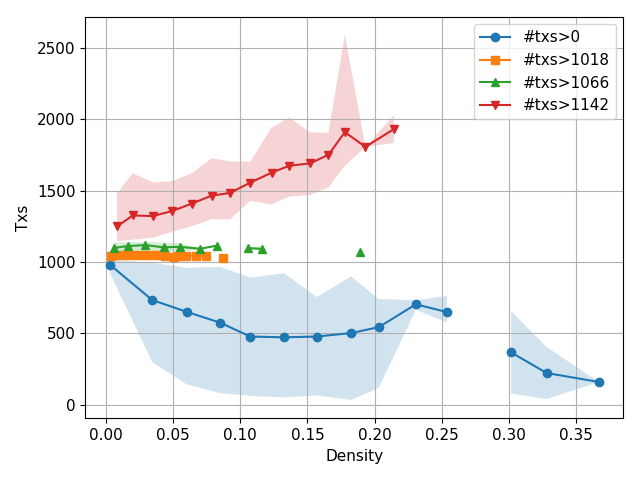}}
\caption{txs}
\label{fig:txs}
\end{figure}
\begin{figure}
\centerline{\includegraphics[width=\columnwidth]{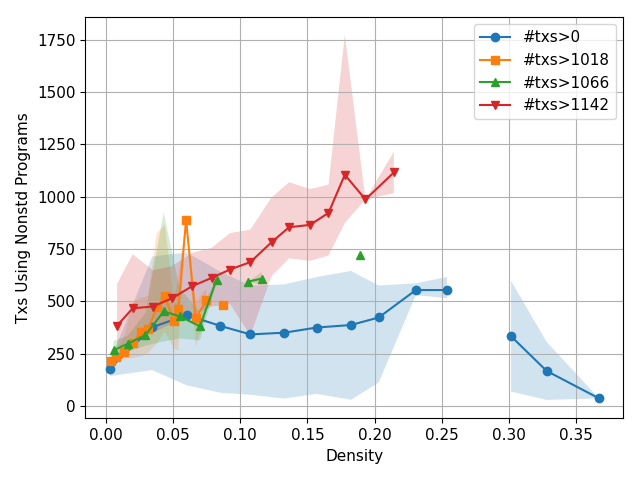}}
\caption{txs-using-nonstd-programs}
\label{fig:txs-using-nonstd-programs}
\end{figure}
\begin{figure}
\centerline{\includegraphics[width=\columnwidth]{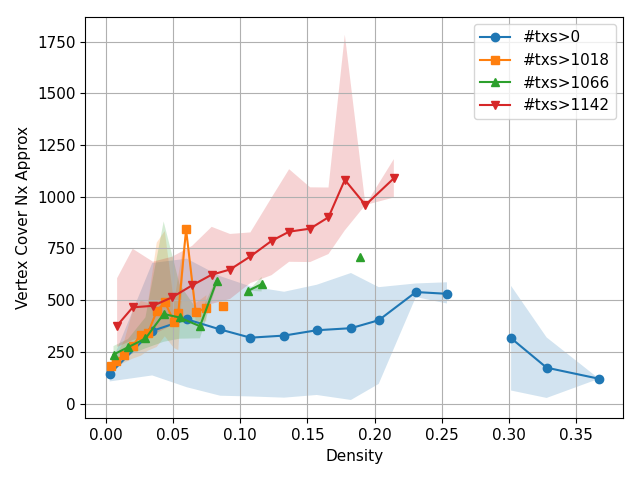}}
\caption{vertex-cover-nx-approx}
\label{fig:vertex-cover-nx-approx}
\end{figure}
\begin{figure}
\centerline{\includegraphics[width=\columnwidth]{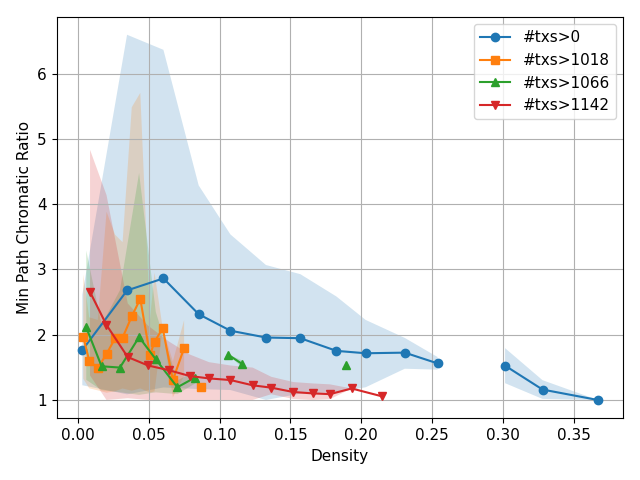}}
\caption{min-path-chromatic-ratio}
\label{fig:min-path-chromatic-ratio}
\end{figure}
\begin{figure}
\centerline{\includegraphics[width=\columnwidth]{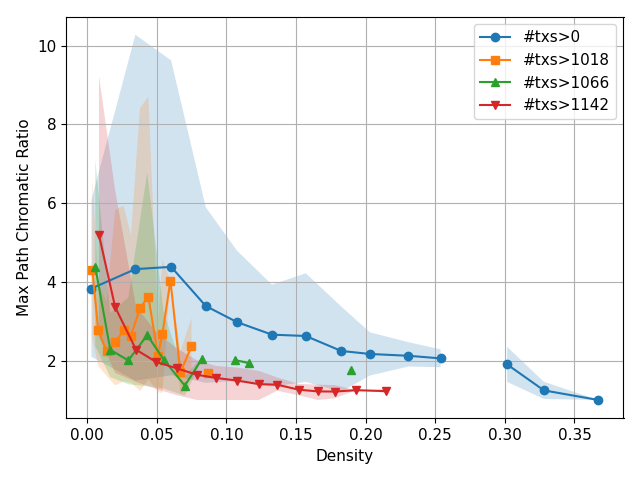}}
\caption{max-path-chromatic-ratio}
\label{fig:max-path-chromatic-ratio}
\end{figure}
\begin{figure}
\centerline{\includegraphics[width=\columnwidth]{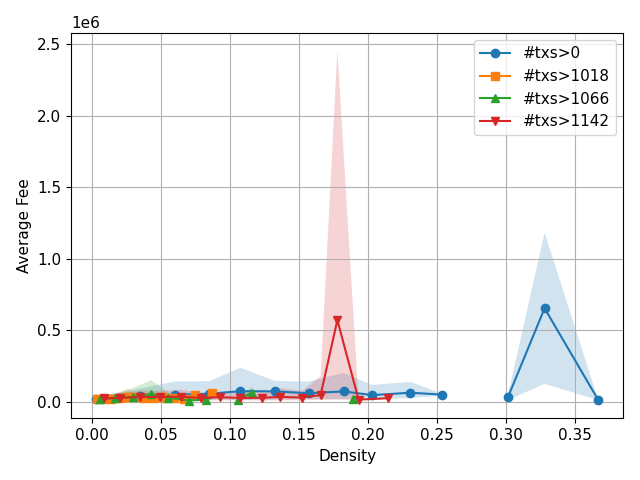}}
\caption{avg-fee}
\label{fig:avg-fee}
\end{figure}
\begin{figure}
\centerline{\includegraphics[width=\columnwidth]{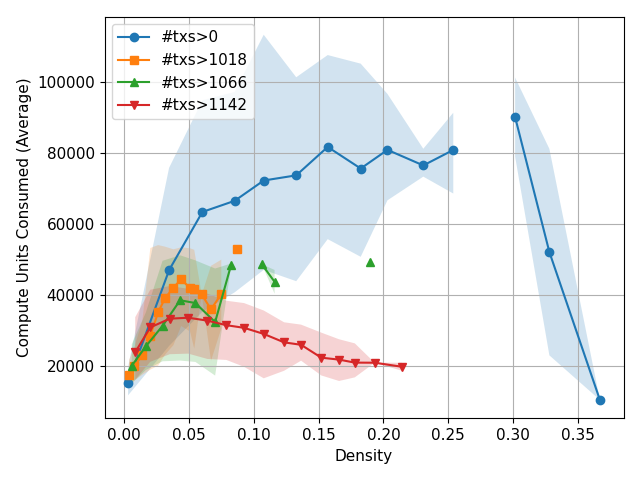}}
\caption{avg-computeUnitsConsumed}
\label{fig:avg-computeUnitsConsumed}
\end{figure}
\begin{figure}
\centerline{\includegraphics[width=\columnwidth]{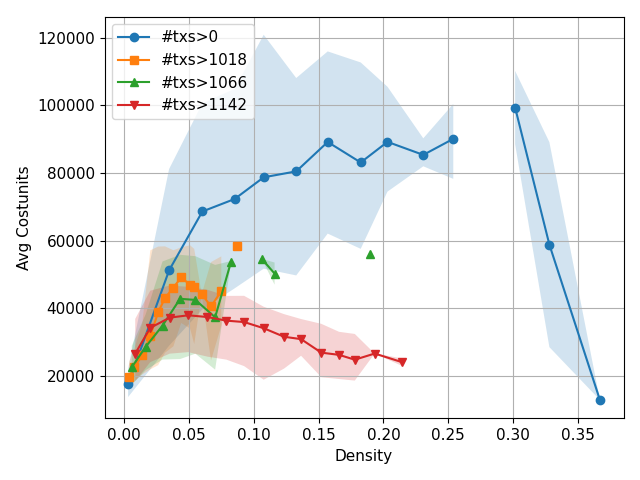}}
\caption{avg-costUnits}
\label{fig:avg-costUnits}
\end{figure}
\begin{figure}
\centerline{\includegraphics[width=\columnwidth]{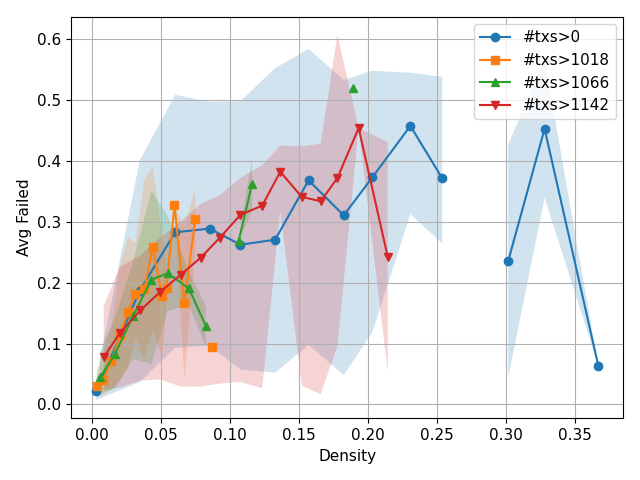}}
\caption{avg-failed}
\label{fig:avg-failed}
\end{figure}
\begin{figure}
\centerline{\includegraphics[width=\columnwidth]{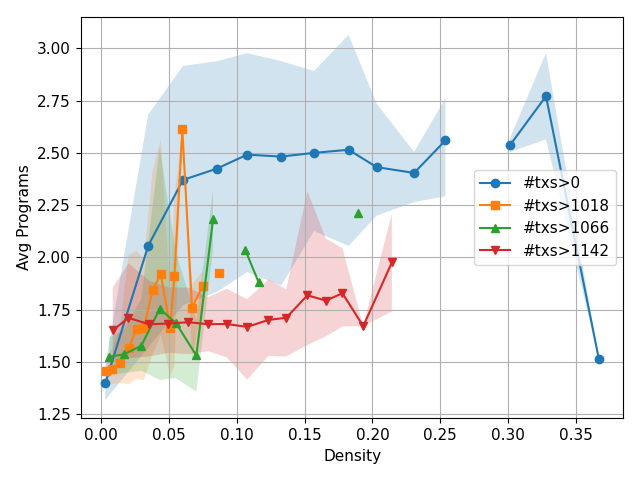}}
\caption{avg-programs}
\label{fig:avg-programs}
\end{figure}
\begin{figure}
\centerline{\includegraphics[width=\columnwidth]{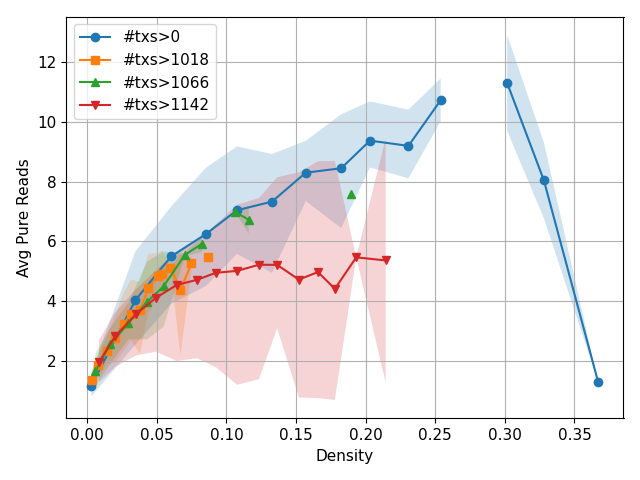}}
\caption{avg-pure-reads}
\label{fig:avg-pure-reads}
\end{figure}
\begin{figure}
\centerline{\includegraphics[width=\columnwidth]{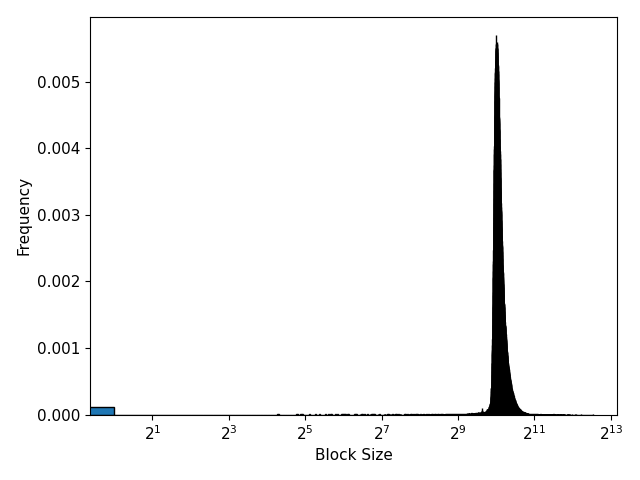}}
\caption{block-size-dist}
\label{fig:block-size-dist}
\end{figure}
\begin{figure}
\centerline{\includegraphics[width=\columnwidth]{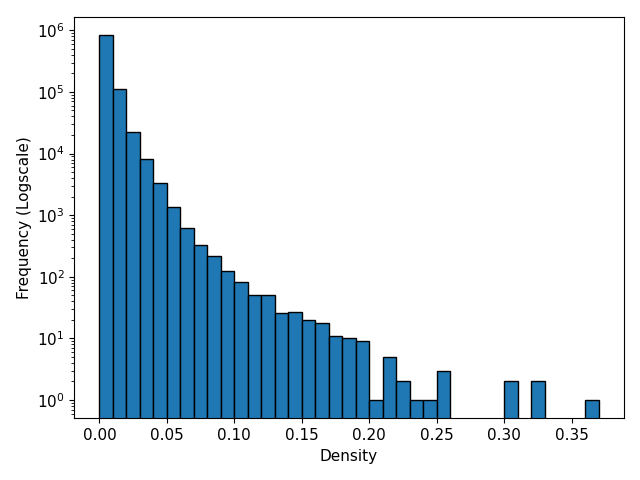}}
\caption{density-dist}
\label{fig:density-dist}
\end{figure}
\begin{figure}
\centerline{\includegraphics[width=\columnwidth]{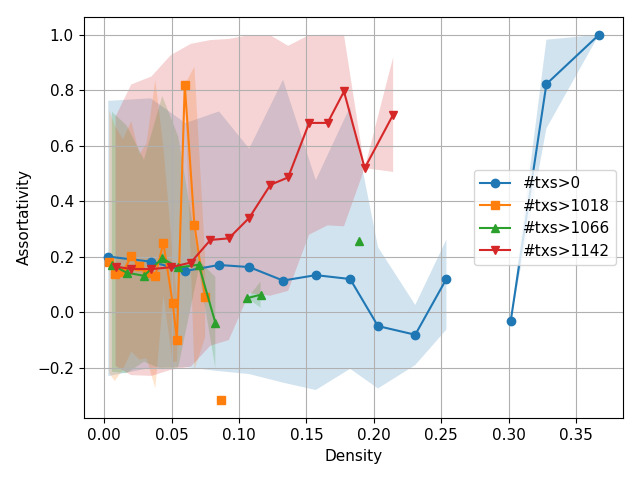}}
\caption{assortativity}
\label{fig:assortativity}
\end{figure}
\begin{figure}
\centerline{\includegraphics[width=\columnwidth]{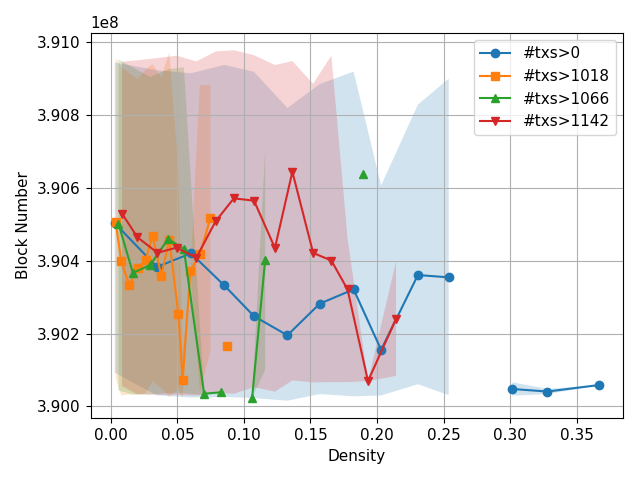}}
\caption{block-number}
\label{fig:block-number}
\end{figure}
\begin{figure}
\centerline{\includegraphics[width=\columnwidth]{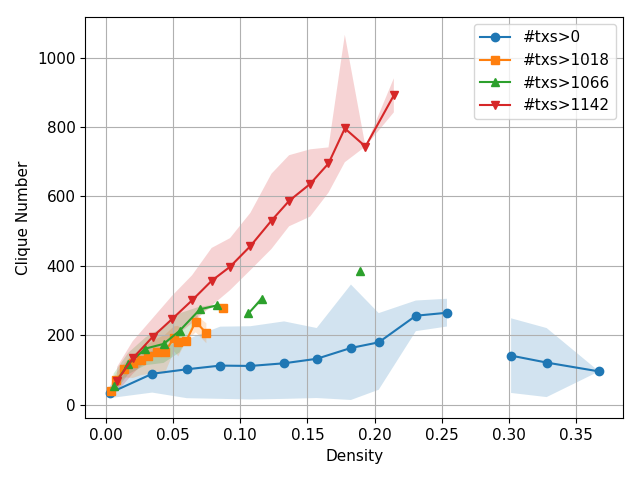}}
\caption{clique-number}
\label{fig:clique-number}
\end{figure}
\begin{figure}
\centerline{\includegraphics[width=\columnwidth]{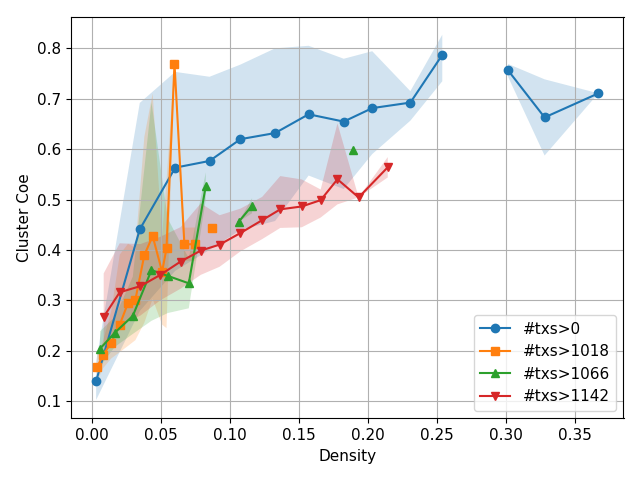}}
\caption{cluster-coe}
\label{fig:cluster-coe}
\end{figure}
\begin{figure}
\centerline{\includegraphics[width=\columnwidth]{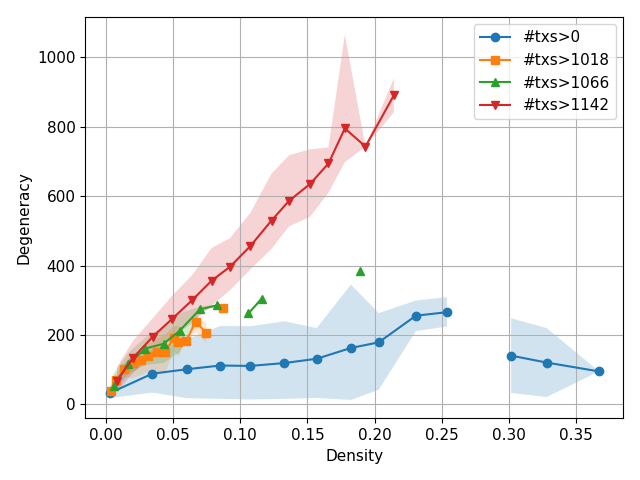}}
\caption{degeneracy}
\label{fig:degeneracy}
\end{figure}
\begin{figure}
\centerline{\includegraphics[width=\columnwidth]{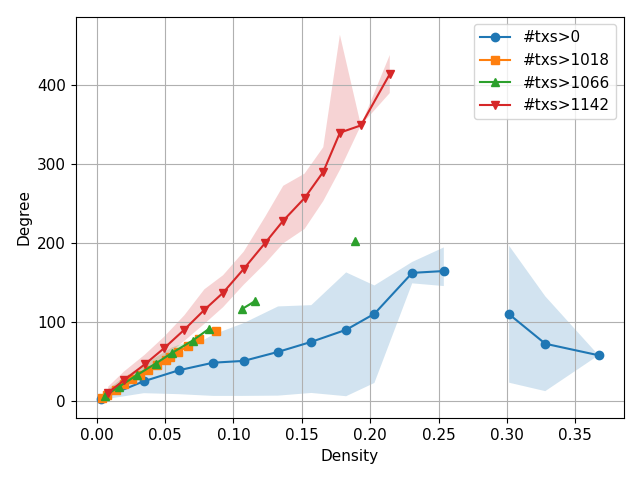}}
\caption{degree}
\label{fig:degree}
\end{figure}
\begin{figure}
\centerline{\includegraphics[width=\columnwidth]{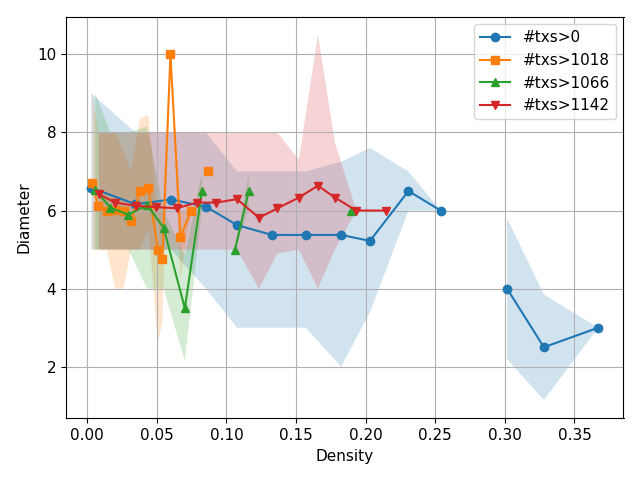}}
\caption{diameter}
\label{fig:diameter}
\end{figure}
\begin{figure}
\centerline{\includegraphics[width=\columnwidth]{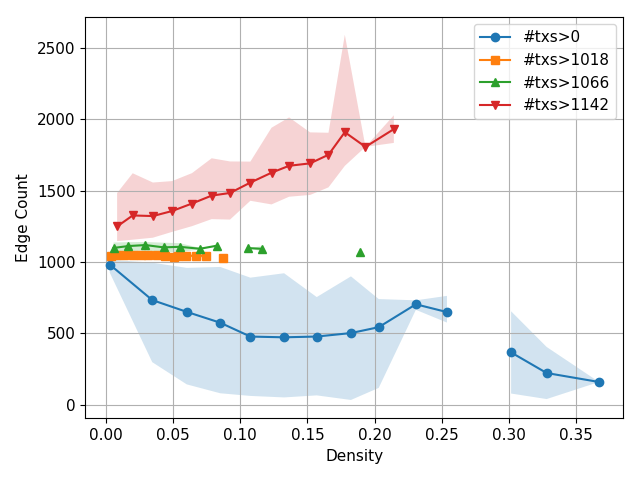}}
\caption{edge-count}
\label{fig:edge-count}
\end{figure}
\begin{figure}
\centerline{\includegraphics[width=\columnwidth]{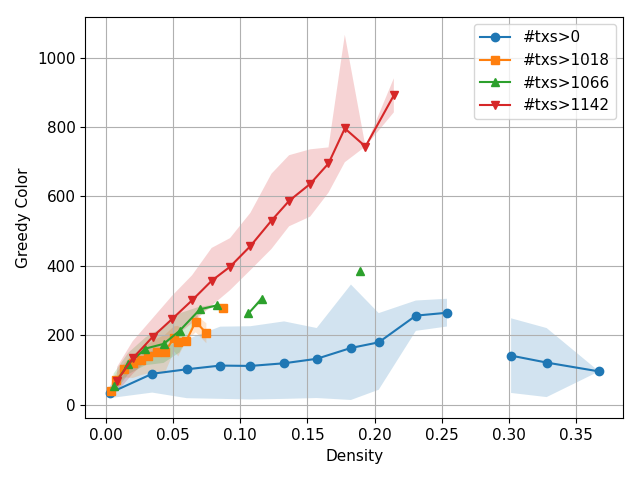}}
\caption{greedy-color}
\label{fig:greedy-color}
\end{figure}
\begin{figure}
\centerline{\includegraphics[width=\columnwidth]{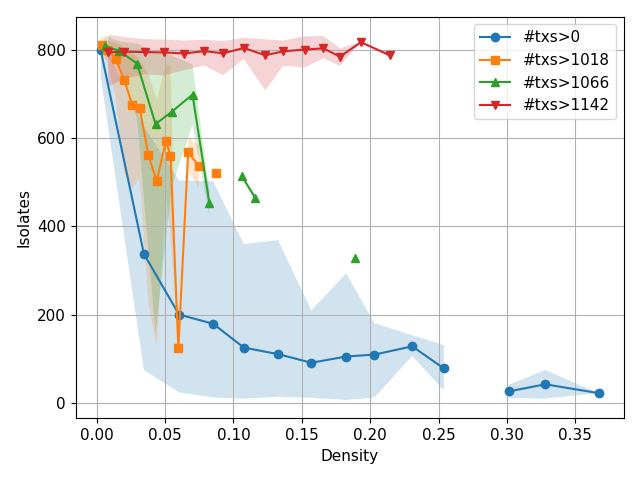}}
\caption{isolates}
\label{fig:isolates}
\end{figure}
\begin{figure}
\centerline{\includegraphics[width=\columnwidth]{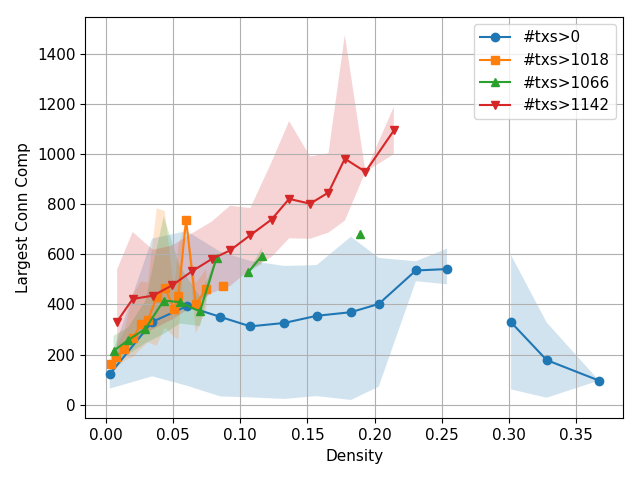}}
\caption{largest-conn-comp}
\label{fig:largest-conn-comp}
\end{figure}
\begin{figure}
\centerline{\includegraphics[width=\columnwidth]{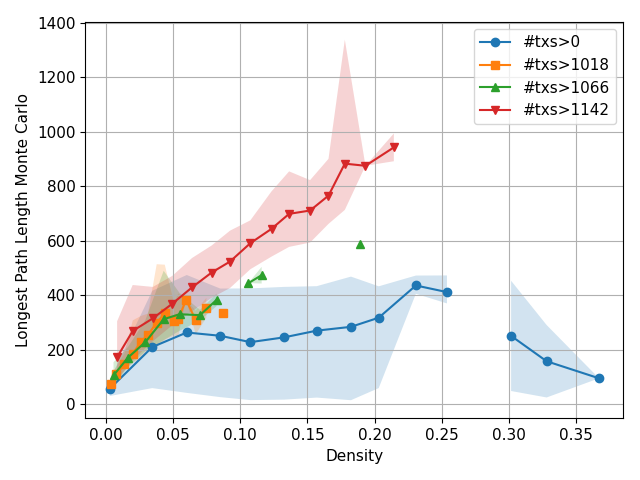}}
\caption{longest-path-length-monte-carlo}
\label{fig:longest-path-length-monte-carlo}
\end{figure}
\begin{figure}
\centerline{\includegraphics[width=\columnwidth]{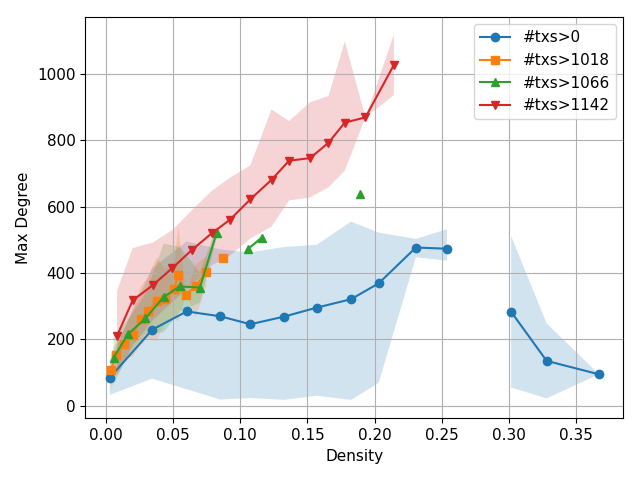}}
\caption{max-degree}
\label{fig:max-degree}
\end{figure}
\begin{figure}
\centerline{\includegraphics[width=\columnwidth]{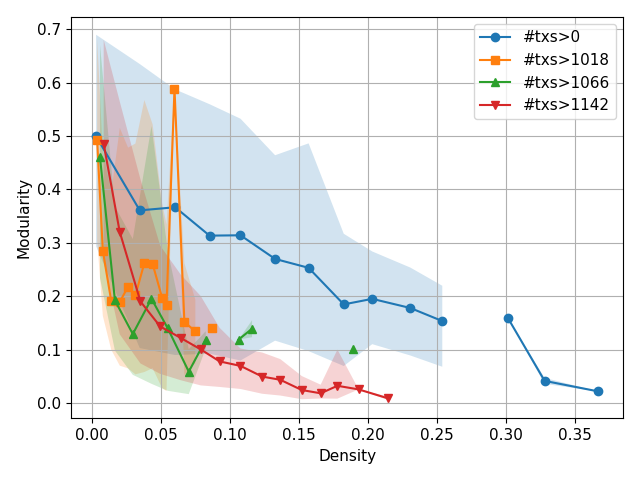}}
\caption{modularity}
\label{fig:modularity}
\end{figure}
\begin{figure}
\centerline{\includegraphics[width=\columnwidth]{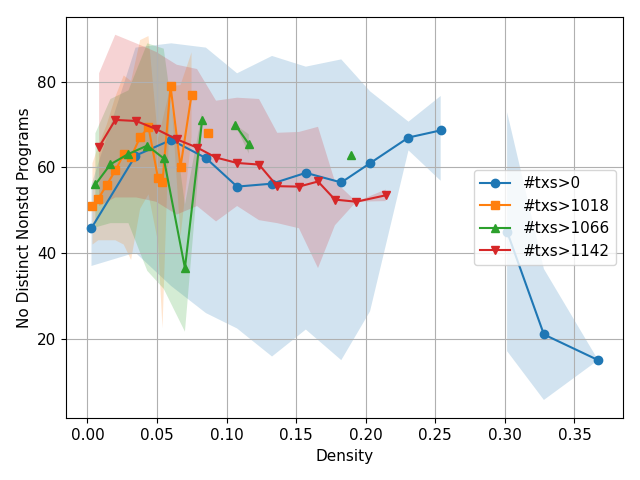}}
\caption{no-distinct-nonstd-programs}
\label{fig:no-distinct-nonstd-programs}
\end{figure}
\begin{figure}
\centerline{\includegraphics[width=\columnwidth]{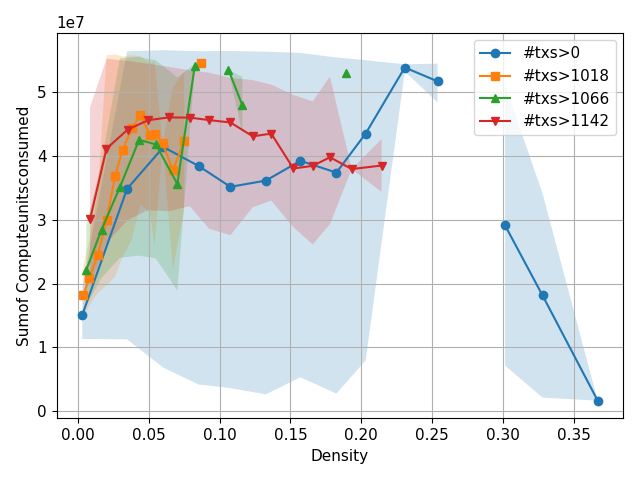}}
\caption{sumof-computeUnitsConsumed}
\label{fig:sumof-computeUnitsConsumed}
\end{figure}
\begin{figure}
\centerline{\includegraphics[width=\columnwidth]{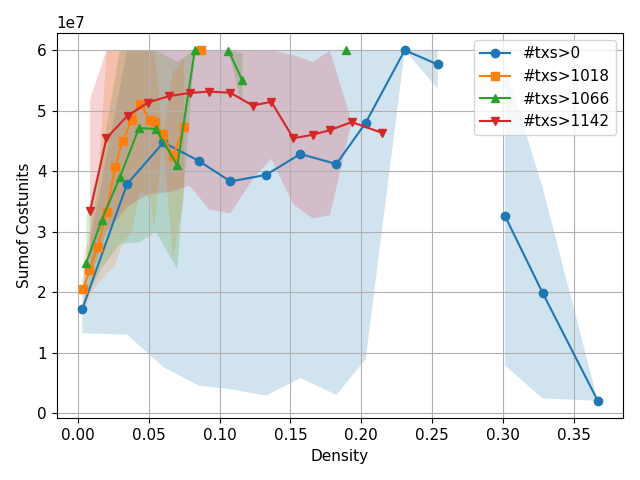}}
\caption{sumof-costUnits}
\label{fig:sumof-costUnits}
\end{figure}
\begin{figure}
\centerline{\includegraphics[width=\columnwidth]{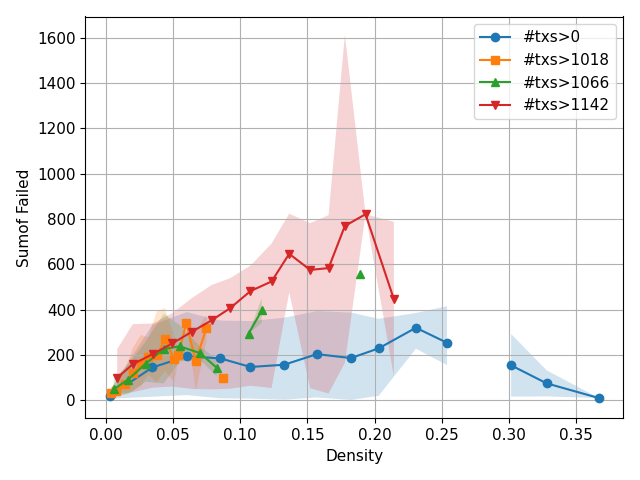}}
\caption{sumof-failed}
\label{fig:sumof-failed}
\end{figure}
\begin{figure}
\centerline{\includegraphics[width=\columnwidth]{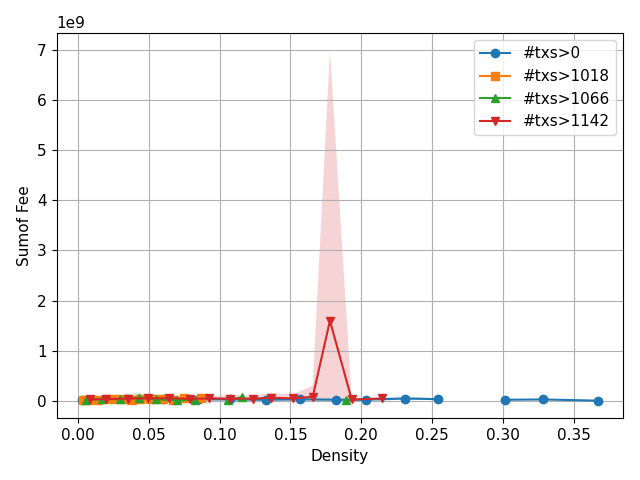}}
\caption{sumof-fee}
\label{fig:sumof-fee}
\end{figure}
\begin{figure}
\centerline{\includegraphics[width=\columnwidth]{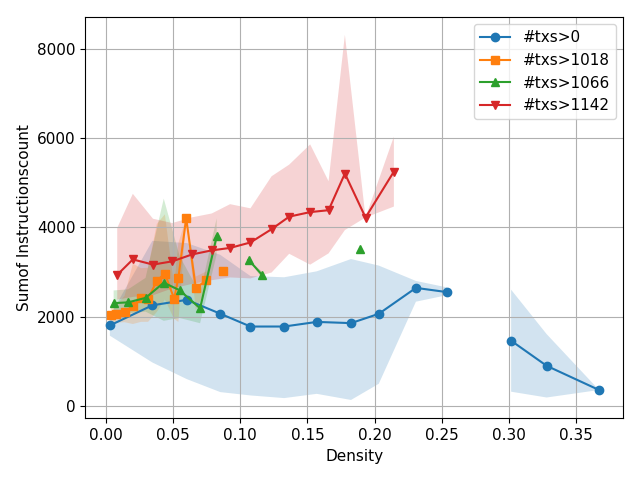}}
\caption{sumof-instructionsCount}
\label{fig:sumof-instructionsCount}
\end{figure}
\begin{figure}
\centerline{\includegraphics[width=\columnwidth]{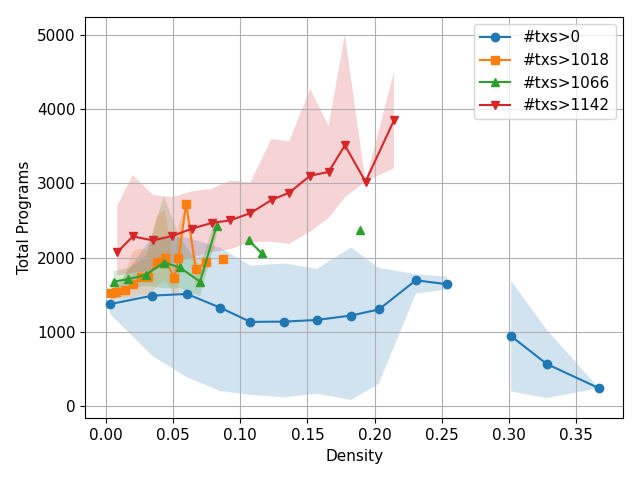}}
\caption{total-programs}
\label{fig:total-programs}
\end{figure}
\begin{figure}
\centerline{\includegraphics[width=\columnwidth]{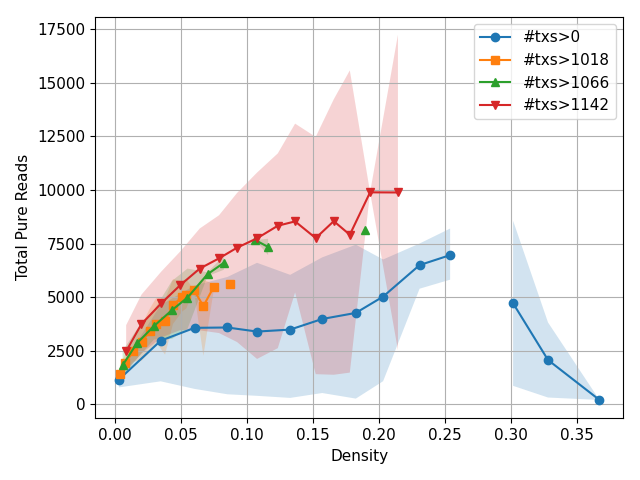}}
\caption{total-pure-reads}
\label{fig:total-pure-reads}
\end{figure}
\begin{figure}
\centerline{\includegraphics[width=\columnwidth]{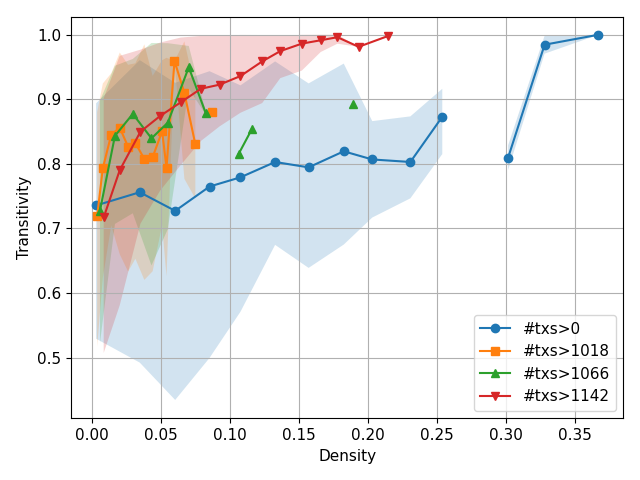}}
\caption{transitivity}
\label{fig:transitivity}
\end{figure}
\begin{figure}
\centerline{\includegraphics[width=\columnwidth]{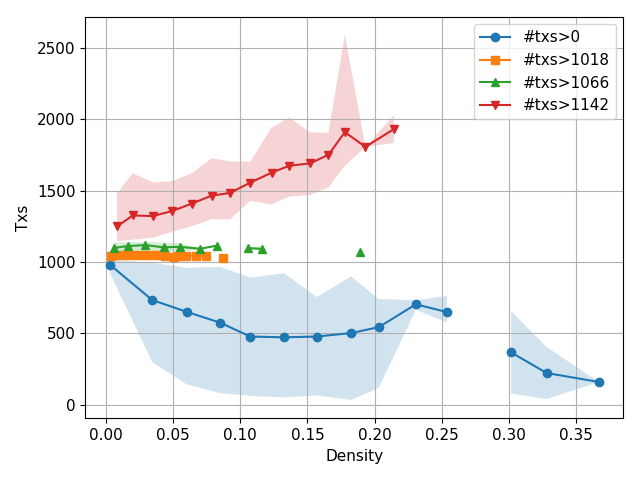}}
\caption{txs}
\label{fig:txs}
\end{figure}
\begin{figure}
\centerline{\includegraphics[width=\columnwidth]{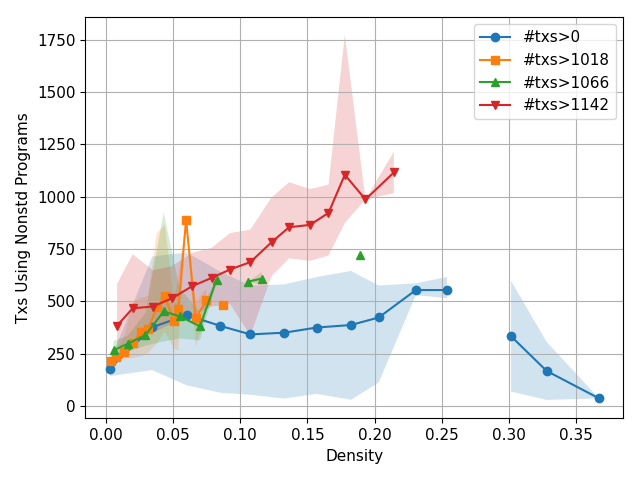}}
\caption{txs-using-nonstd-programs}
\label{fig:txs-using-nonstd-programs}
\end{figure}
\begin{figure}
\centerline{\includegraphics[width=\columnwidth]{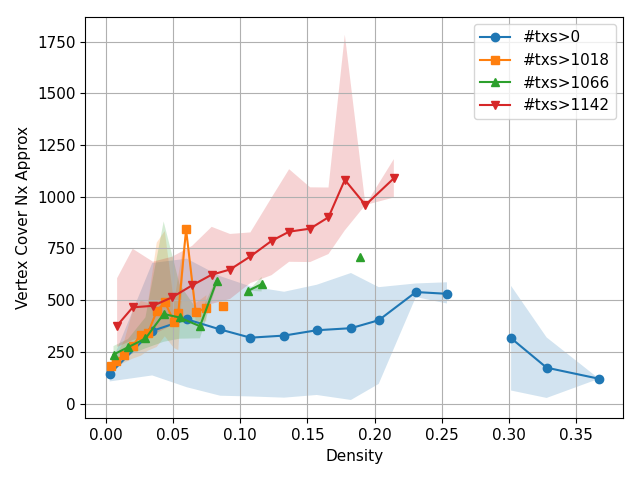}}
\caption{vertex-cover-nx-approx}
\label{fig:vertex-cover-nx-approx}
\end{figure}
\begin{figure}
\centerline{\includegraphics[width=\columnwidth]{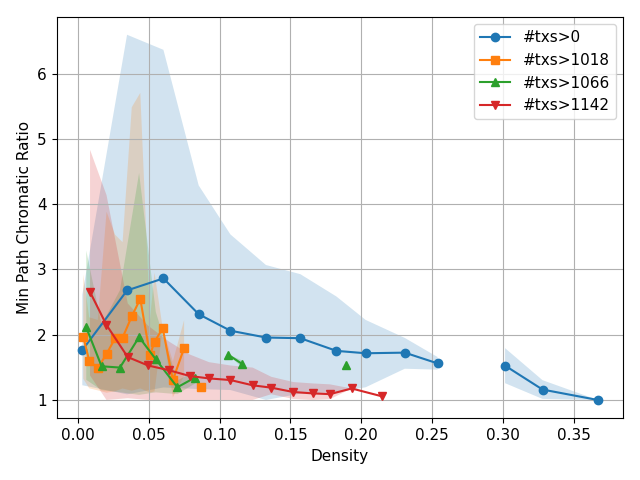}}
\caption{min-path-chromatic-ratio}
\label{fig:min-path-chromatic-ratio}
\end{figure}
\begin{figure}
\centerline{\includegraphics[width=\columnwidth]{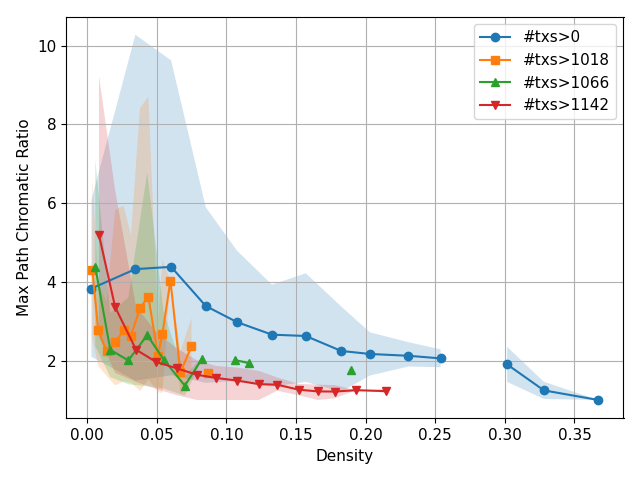}}
\caption{max-path-chromatic-ratio}
\label{fig:max-path-chromatic-ratio}
\end{figure}
\begin{figure}
\centerline{\includegraphics[width=\columnwidth]{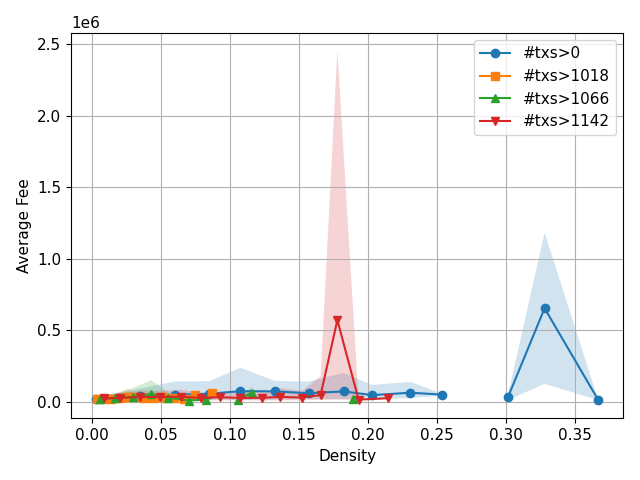}}
\caption{avg-fee}
\label{fig:avg-fee}
\end{figure}
\begin{figure}
\centerline{\includegraphics[width=\columnwidth]{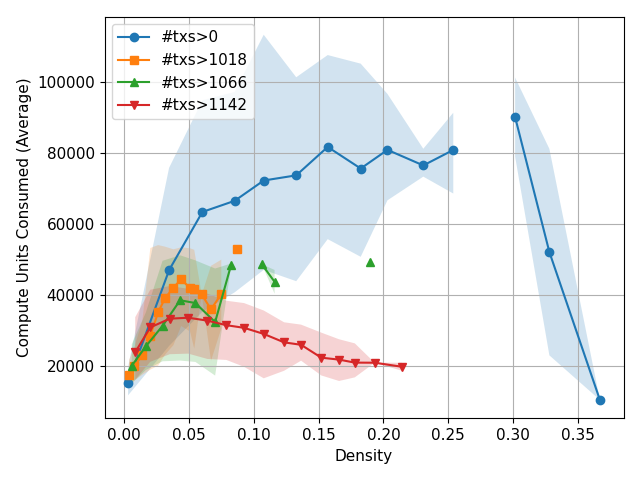}}
\caption{avg-computeUnitsConsumed}
\label{fig:avg-computeUnitsConsumed}
\end{figure}
\begin{figure}
\centerline{\includegraphics[width=\columnwidth]{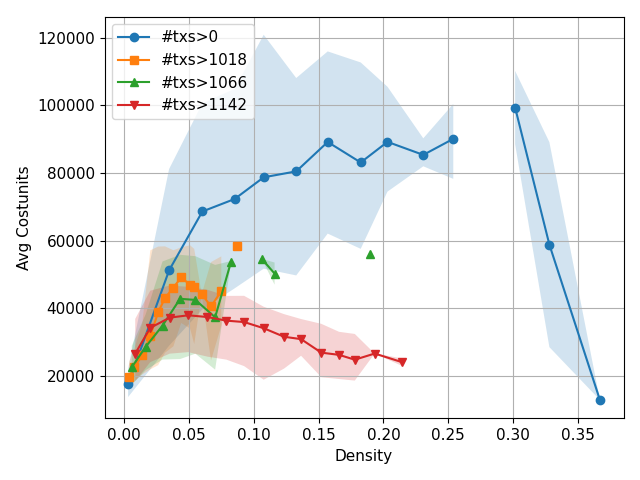}}
\caption{avg-costUnits}
\label{fig:avg-costUnits}
\end{figure}
\begin{figure}
\centerline{\includegraphics[width=\columnwidth]{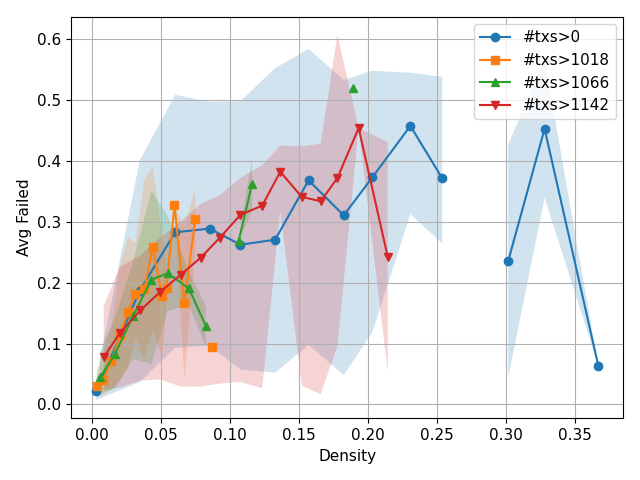}}
\caption{avg-failed}
\label{fig:avg-failed}
\end{figure}
\begin{figure}
\centerline{\includegraphics[width=\columnwidth]{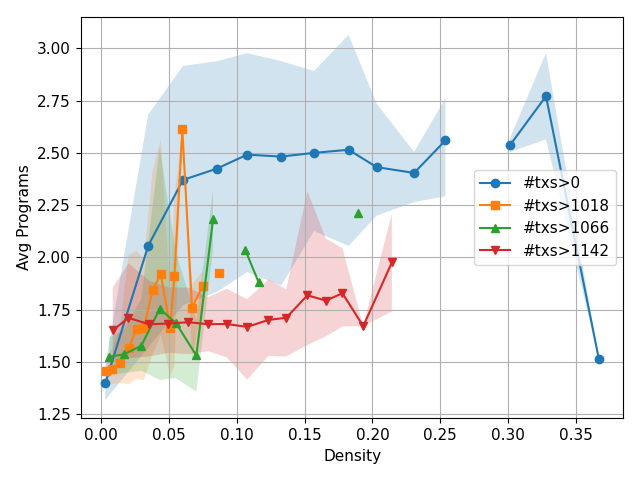}}
\caption{avg-programs}
\label{fig:avg-programs}
\end{figure}
\begin{figure}
\centerline{\includegraphics[width=\columnwidth]{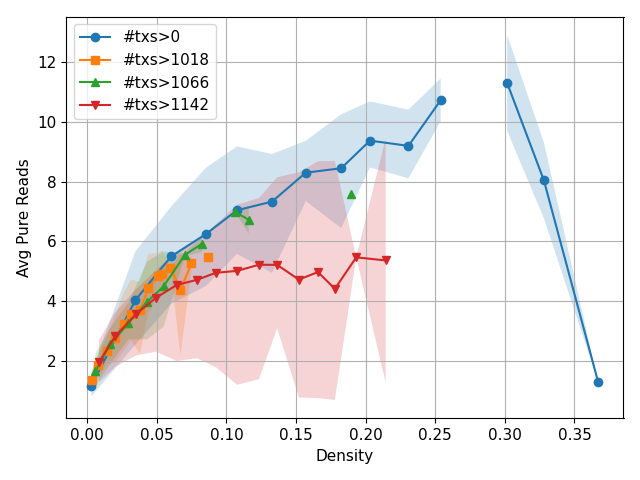}}
\caption{avg-pure-reads}
\label{fig:avg-pure-reads}
\end{figure}

\fi

\end{document}